ХАРЬКОВСКИЙ ГОСУДАРСТВЕННЫЙ УНИВЕРСИТЕТ

На правах рукописи

**Дуплий Степан Анатольевич**

# ПОЛУГРУППОВЫЕ МЕТОДЫ В СУПЕРСИММЕТРИЧНЫХ ТЕОРИЯХ ЭЛЕМЕНТАРНЫХ ЧАСТИЦ

**01.04.02 – Теоретическая физика**

**Диссертация на соискание ученой степени доктора физико-математических наук**

*Харьков* – **1999**

# СОДЕРЖАНИЕ

# СПИСОК УСЛОВНЫХ ОБОЗНАЧЕНИЙ

Латинскими наклонными буквами обозначены четные величины $a, b, Q$ и функции $f, g, F, G$; греческими буквами $\alpha, \beta, \theta, \mathit{\Delta}$ — нечетные;

$\mathrm{M}, \mathrm{S}, \mathrm{T}, \mathrm{A}, \mathrm{D} \ldots$ — матрицы и суперматрицы;

$\mathcal{M}, \mathcal{S}, \mathcal{P}, \mathcal{Q}, \mathcal{R} \ldots$ — полуматрицы и полуминоры;

$\mathfrak{M}, \mathfrak{S}, \mathfrak{T}, \mathfrak{A}, \mathfrak{D} \ldots$ — множества с умножением $(\star)$;

$\mathbf{M}, \mathbf{S}, \mathbf{T}, \mathbf{A}, \mathbf{D}, \mathbf{I} \ldots$ — абстрактные полугруппы с умножением $(*)$;

$\boldsymbol{S}, \boldsymbol{T}, \boldsymbol{A}, \boldsymbol{D} \ldots$ — полугруппы преобразований;

$\mathcal{T}, \mathcal{U}, \mathcal{A}, \mathcal{B}, \mathcal{H}, \mathcal{G} \ldots$ — преобразования с умножением $(\circ)$;

$\mathbb{C}^{n|m}, \mathbb{R}^{n|m}, \mathbb{D}^{n|m}, \mathbb{V}_{\alpha\beta} \ldots$ — (супер)пространства и области в них;

$\mathbf{\Lambda}^{n|m} \ldots$ — линейные суперпространства;

$T\mathbb{C}^{n|m} \left(T^*\mathbb{C}^{n|m}\right) \ldots$ — (ко)касательные суперпространства;

$\mathscr{M}, \mathscr{N}, \mathscr{X}, \mathscr{Y}, \mathscr{U}_\alpha \ldots$ — (супер)многообразия и области в них;

$\mathbb{Z}, \mathbb{K}, \mathbb{N}, \Lambda_0, \Lambda_1 \ldots$ — (супер)числовые поля;

$\mathscr{L}, \mathscr{R}, \mathscr{D}, \mathscr{H} \ldots$ — отношения Грина;

$\mathsf{L}, \mathsf{R}, \mathsf{D}, \mathsf{H} \ldots$ — классы эквивалентности с умножением $(\diamond)$;

$\mathbf{\Pi}, \mathbf{\Theta}_V, \mathbf{\Theta}_H, \mathbf{\Upsilon}_L, \mathbf{\Upsilon}_R, \mathsf{st} \ldots$ — транспонирования;

$\hat{\mathsf{S}}, \hat{\mathsf{T}}, \hat{\mathsf{A}}, \hat{\mathsf{D}} \ldots$ — операторы;

$\boldsymbol{r}, \boldsymbol{d}, \boldsymbol{D} \ldots$ — инвариванты;

$\Phi_{\alpha\beta}, \Lambda_{\alpha\beta} \ldots$ — функции перехода с композицией $(\circ)$;

$\varphi, \pi, \lambda \ldots$ — морфизмы и отображения;

$(\cdot)$ — умножение в грассмановой алгебре;

$(\odot_{\mathrm{X}}), (\circledcirc_{\mathrm{X}})$ — сэндвич умножения;

$\epsilon\left[x\right]$ — числовая часть величины $x$ (отбрасывание нильпотентов);

$\mathsf{SA}$ — суперsuccаналитический;

$\mathsf{SCf}$ — суперконформный;

$\mathsf{TPt}$ — сплетающий четность (twisting parity of tangent space).

# ВВЕДЕНИЕ

**Актуальность темы.** Построение единой теории всех фундаментальных взаимодействий — электромагнитного, слабого, сильного и гравитационного — является важнейшей теоретической проблемой современной физики элементарных частиц. Существенным достижением в этом направлении явилось развитие методов суперсимметрии и супергравитации, которые позволили разрешить такие трудности предшествующих суперсимметрии калибровочных теорий фундаментальных взаимодействий (квантовой электродинамики, квантовой хромодинамики и модели Вайнберга-Салама), как включение гравитации и рассмотрение процессов при планковских энергиях.

Нелокальное многомерное обобщение супергравитации – теория суперструн – дала ответ на многие открытые вопросы, связанные с неперенормируемостью и космологической постоянной, а также с последовательной унификацией всех фундаментальных взаимодействий. В теории суперструн осуществился синтез разнообразных методов теоретической и математической физики. Тем не менее, дальнейший прогресс в понимании глубинных физических основ строения материи, в свою очередь, требует интенсивных поисков нестандартных путей разрешения известных проблем и привлечения принципиально новых теоретических идей.

Наиболее фундаментальными и общими являются абстрактные алгебраические свойства теории, лежащей в основе физики элементарных частиц. Как правило, вначале исследований такие свойства вводятся с математической точки зрения и лишь затем формулируются на языке физических законов и предсказаний результатов эксперимента.

Так произошло и в случае суперсимметрии: антикоммутирующие величины рассматривались многими математиками еще начиная с прошлого столетия. Но лишь после открытия суперсимметрии физиками в начале 70-х годов она превратилась из чисто математической теории в "индустриальную" основу современного "моделестроения" с физическими конструкциями и конкретными предсказаниями новых элементарных частиц — суперпартнеров. Настоящий "бум суперсимметризации" потряс теоретическую физику 70-х и 80-х: все, что могло "суперсимметризоваться", незамедлительно "суперсимметризовалось". Основные ингредиенты теории после очевидных модификаций наделялись приставкой "супер", а затем построение уже суперсимметричной модели, исключая несущественные и не принимаемые в расчет моменты, копировались шаг за шагом из подобной несуперсимметричной версии, и последняя обязана была быть некоторым ее непрерывным пределом.

Однако, при этом абстрактные алгебраические свойства физической теории или вовсе не претерпевали изменений, либо влияние "суперсимметризации" было просто символичным. Так предполагалось, что именно супергруппы представляют собой адекватное суперобобщение соответствующих групп. И это удивительно, поскольку среди основных переменных суперсимметричной теории изначально присутствуют необратимые объекты и делители нуля. В частности, концепция суперпространства, допускающего унификацию описания бозонных и фермионных секторов теории, основана на введении дополнительных нильпотентных координат, тогда многие отображения и функции становятся необратимыми по определению. И все же, как это ни странно и ни парадоксально с математической точки зрения, они искусственно и необоснованно исключались из рассмотрения. Данная процедура была названа "факторизацией по нильпотентам" в физике (в теории полугрупп эта процедура хорошо известна и называется факторизацией Риса) и она (в основном неаргументированно) применялась или подразумевалась при суперсимметризациях.

На самом деле, все преобразования множества, содержащего нильпотенты, или все отображения суперпространства сохраняющего вид определенной структуры образуют полугруппу (а не группу) относительно композиции. Поэтому категория групп, в рамках которой строились несуперсимметричные теории элементарных частиц, должна быть обобщена до категории полугрупп при математически строгом включении суперсимметрии в основополагающие принципы теории.

Другими словами, переход от пространства к суперпространству должен сопровождаться одновременным переходом от групп к суперполугруппам, а не супергруппам — "супер" обобщение физической теории должно сопровождаться "полу" обобщением ее математики в целом. Тогда в глобальном теоретико-групповом смысле суперсимметричные модели элементарных частиц обязаны иметь структуру полугруппы, в то время, как наблюдаемый их сектор при настоящих энергиях может удовлетворительно описываться их обратимой групповой частью. Поэтому не следует ограничиваться исследованиями лишь последней, поскольку свойства идеальной и групповой частей взаимообусловлены и взаимозависимы. В этом контексте важным также является пересмотр стандартного анзаца "факторизации", а именно — "факторизовать по не-нильпотентам", т. е. изучать "негрупповые" (или идеальные) свойства суперсимметричных теорий.

Таким образом, построение и исследование таких суперсимметричных моделей элементарных частиц, которые, с одной стороны, обладали бы математической общностью и корректностью в рамках аппарата теории полугрупп, а с другой стороны, имели бы достаточную физическую предсказательную силу, представляет собой актуальную научно-теоретическую проблему.

Основной объект в теории суперструн — это мировая поверхность струны, следовательно построение и изучение необратимых и полугрупповых обобщений супермногообразий и суперконформной дифференциальной геометрии представляет собой первоочередную задачу.

В этой связи чрезвычайно актуальной является также проблема обратного влияния суперсимметрии на теорию полугрупп. Так, подробное исследование необратимых суперматриц приводит к новым и неожиданным результатам в идеальном строении и теории представлений суперматричных полугрупп, что, в свою очередь, может способствовать последовательному и корректному построению новых суперсимметричных моделей элементарных частиц, основанных на полугрупповых принципах.

**Связь работы с научными программами, планами, темами.** Диссертация выполнена как часть исследований, проводимых на кафедрах теоретической и экспериментальной ядерной физики ХГУ в рамках координационного плана Министерства образования Украины "Комплексные исследования ядерных процессов и создание на их основе ядерно-физических методов для использования в энергетике и радиационной безопасности ядерных энергетических установок и технологий радиационной модификации материалов и экологии".

Результаты диссертации вошли в отчеты госбюджетных тем "Исследования структуры атомных ядер и новых закономерностей в ядерных взаимодействиях" (тема №1-13-94, номер государственной регистрации 0194U018989) и "Исследования ядерных процессов с участием нуклонов и сложных частиц низких и средних энергий" (тема №1-13-97, номер государственной регистрации 0197U016494).

**Цель и задачи исследования.** Основной целью диссертационной работы является разработка и применение полугрупповых методов в суперсимметричных моделях элементарных частиц. Для этого решались такие задачи:

1. Подробный анализ необратимых свойств преобразований, возникающих при суперсимметризации физических теорий.
2. Поиск необратимых аналогов супермногообразий, расслоений и гомотопий.
3. Формулировка необратимой суперконформной дифференциальной геометрии и построение суперконформных полугрупп.
4. Классификация необратимых расширенных и нерасширенных суперконформных преобразований.
5. Нахождение нелинейных реализаций необратимых суперконформных преобразований.
6. Всесторонний анализ суперматричных полугрупп, поиск новых представлений и эквивалентностей.
7. Введение новых типов матриц, содержащих нильпотентные элементы и изучение их свойств.
8. Построение необратимого аналога гиперболической геометрии на суперплоскости.

**Научная новизна полученных результатов.** Научная новизна диссертационной работы состоит в построении нового направления в суперсимметричных моделях элементарных частиц, которое основано на включении полугрупп, идеалов и необратимых свойств в исследование математической структуры. Впервые определены необратимые аналоги супермногообразий, расслоений и гомотопий. Сформулирована новая необратимая суперконформная геометрия (и ее расширенные варианты), найдены новые типы суперконформных полугрупп и преобразований, которые сплетают четность касательного расслоения. Предложена альтернативная редукция суперматриц, которая приводит к новым абстрактным свойствам, полугруппам и супермодулям. Впервые суперматрицы используются для построения представлений полугрупп связок, при этом найдены новые обобщенные отношения Грина. Построен необратимый вариант гиперболической геометрии на суперплоскости, где найдены необратимые аналоги двойных отношений, инвариантов и расстояний.

**Практическое значение полученных результатов.** Диссертационная работа носит теоретический характер. Ее результаты могут быть использованы для построения новых математически корректных моделей элементарных частиц, основанных на теории суперструн, переосмысленного анализа необратимости в уже имеющихся моделях, а также для поиска новых полугрупповых свойств и структур в суперсимметричных объектах и пространствах.

**Личный вклад диссертанта.** Все результаты получены автором самостоятельно.

**Апробация результатов диссертации.** Основные результаты работы докладывались автором на 12 международных конференциях, 10 из которых проводились за рубежом:

1. Международная школа по теоретической и математической физике *(Гваделупа, Франция, 1993)*
2. Международный коллоквиум по теоретико-групповым методам в математической физике *(Париж, Франция, 1993)*
3. Международный конгресс по математической физике *(Париж, Франция, 1994)*
4. Международная краковская школа по теоретической физике *(Закопане, Польша, 1995)*
5. Международная конференция по калибровочным теориям, прикладной суперсимметрии и квантовой гравитации *(Леувен, Бельгия, 1995)*
6. Европейская школа по теории групп *(Валладолид, Испания, 1995)*

7. Международная конференция суперсимметрия-96 *(Коллеж Парк, США, 1996)*

8. Международная конференция по высшим гомотопическим структурам в математической физике *(Покипси, США, 1996)*

9. Международный семинар по суперсимметрии и квантовой теория поля памяти Д. В. Волкова *(Харьков, Украина, 1997)*

10. Международная конференция по суперсимметрии и квантовым симметриям памяти В. И. Огиевецкого *(Дубна, Россия, 1997)*

11. Международная алгебраическая конференция памяти Л. М. Глускина *(Славянск, Украина, 1997)*

12. Международный конгресс математиков *(Берлин, Германия, 1998)*

Материалы диссертационной работы представлялись и всесторонне обсуждались на многих семинарах в Украине, России, Германии, Англии, Франции, США и других странах.

**Публикации.** Основные результаты диссертации опубликованы в 21 работах (из них 12 в зарубежных изданиях), а также в трудах упомянутых конференций. Все работы выполнены без соавторов. Большинство работ предварительно опубликовано также в интернете и хранится в международных электронных архивах США, Англии, Италии, Японии. Прямой доступ к ним возможен с интернетовской страницы автора: `http://www-home.univer.kharkov.ua/~duplij`.

**Структура и объем работы.** Диссертация состоит из Введения, 5-ти основных разделов, раздела Выводы и приложений. Объем основного текста (без приложений и литературы) составляет 292 страницы. В работе имеется 3 рисунка, 3 таблицы и список литературы из 832 названий.

**Во Введении** обоснована актуальность проблемы, сформулирована цель работы, ее научная новизна, практическая ценность и апро-бация, кратко изложено ее содержание.

**В разделе “Теория необратимых супермногообразий”** подробно анализируются обобщения понятий супермногообразия, суперрасслоение и гомотопии на необратимый случай. На языке карт и функций перехода вводятся понятие полусупермногообразия как необратимого аналога супермногообразия. Префикс “полу-” отражает тот факт что лежащие в основе морфизмы формируют полугруппы состоящие из известной групповой части и новой идеальной необратимой части, т.е. рассматривается полугрупповое обобщение предыдущего формализма.

Полукарта определяется как пара из суперобласти $\mathscr{U}_{\alpha,}^{noninv}$ и необратимого морфизма $\varphi_{\alpha}^{noninv}$. Тогда полуатлас есть объединение стандартных обратимых карт $\left\{\mathscr{U}_{\alpha,}^{inv}\varphi_{\alpha}^{inv}\right\}$ и полукарт $\left\{\mathscr{U}_{\alpha,}^{noninv}\varphi_{\alpha}^{noninv}\right\}$. Полусупермногообразие $\mathscr{M}$ есть суперпространство, представленное в качестве полуатласа.

Функции перехода на полусупермногообразии находятся не из стандартных выражений $\Phi_{\alpha\beta} = \varphi_{\alpha} \circ \varphi_{\beta}^{-1}$ на пересечении суперобластей $\mathscr{U}_{\alpha} \cap \mathscr{U}_{\beta}$, а из системы уравнений

$$\Phi_{\alpha\beta} \circ \varphi_{\beta} = \varphi_{\alpha}, \qquad \Phi_{\beta\alpha} \circ \varphi_{\alpha} = \varphi_{\beta}.$$

В общем случае при нахождении $\Phi_{\alpha\beta}$ и $\Phi_{\beta\alpha}$ эти уравнения не могут быть решены с помощью $\Phi_{\alpha\beta} = \varphi_{\alpha} \circ \varphi_{\beta}^{-1}$ в силу необратимых $\varphi_{\alpha}$ и $\varphi_{\beta}$. Вместо этого ищутся искусственные приемы его решения, например, разложением в ряд по генераторам суперaлгебры, либо используя абстрактные методы теории полугрупп, которые рассматривают решения необратимых уравнений как классы эквивалентности.

Ослабление обратимости позволяет естественно обобщать условия коцикла для функций перехода полусупермногообразий. Они строятся аналогично условиям регулярности для элементов полугруппы. Так, вместо стандартного $n = 2$ условия взаимной обратности функций перехода $\Phi_{\alpha\beta}$ и $\Phi_{\beta\alpha}$ в виде $\Phi_{\alpha\beta} \circ \Phi_{\beta\alpha} = 1_{\alpha\alpha}$ (где $1_{\alpha\alpha}$ — тождественное отображение на $\mathscr{U}_\alpha$) имеем обобщенное условие

$$\Phi_{\alpha\beta} \circ \Phi_{\beta\alpha} \circ \Phi_{\alpha\beta} = \Phi_{\alpha\beta}$$

на пересечениях $\mathscr{U}_\alpha \cap \mathscr{U}_\beta$. А вместо известного $n = 3$ условия коцикла $\Phi_{\alpha\beta} \circ \Phi_{\beta\gamma} \circ \Phi_{\gamma\alpha} = 1_{\alpha\alpha}$ на пересечении трех суперобластей $\mathscr{U}_\alpha \cap \mathscr{U}_\beta \cap \mathscr{U}_\gamma$ получаем его необратимый аналог

$$\Phi_{\alpha\beta} \circ \Phi_{\beta\gamma} \circ \Phi_{\gamma\alpha} \circ \Phi_{\alpha\beta} = \Phi_{\alpha\beta}.$$

Аналогично строятся условия коцикла при произвольных $n$, которое мы называем $n$-регулярностью отображений. Понятно, что 3-регулярность совпадает с обыкновенной регулярностью.

Это позволяет сформулировать чрезвычайно общий анзац полукоммутативности для необратимых морфизмов, который при $n = 3$ описывается следующей коммутативной диаграммой

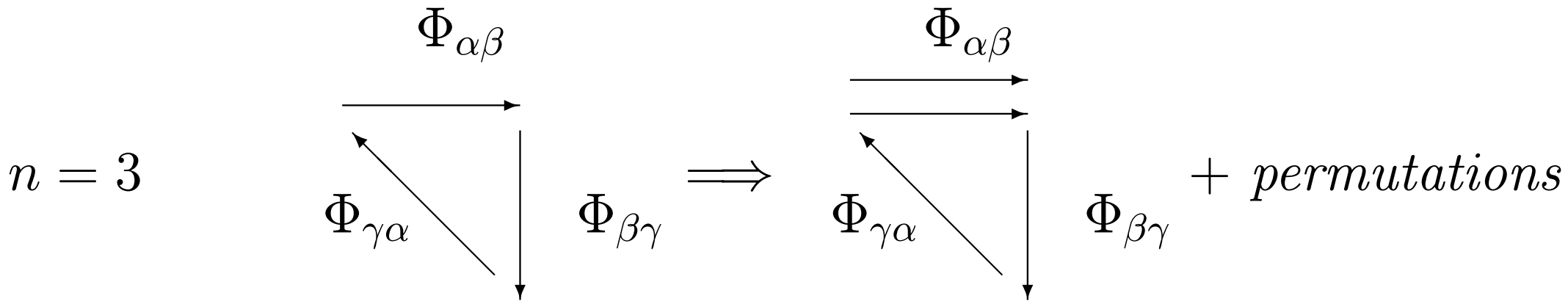

Обратимый морфизм Необратимый (регулярный) морфизм

Получены также необратимые аналоги коциклов для рефлексивных полусупермногообразий.

Найден дополнительный нильпотентный тип ориентируемости на полусупермногообразиях, который обусловлен нильпотентностью березиниана функций перехода. Индекс нильпотентности березиниана позволяет нам систематизировать полусупермногообразия имеющие нильпотентную ориентируемость. Вводятся также башенные тождества и препятственность, с помощью которых удается проклассифицировать полусупермногообразия. По аналогии с суперчислами имеем следующую классификацию:

- *Суперчисла.*

  **1.** Обыкновенные не равные нулю числа (обратимые).

  **2.** Суперчисла, имеющие ненулевую числовую часть (обратимые).

  **3.** Суперчисла, имеющие нулевую числовую часть (необратимые).

- *Полусупермногообразия.*

  **1.** Обыкновенные многообразия (функции перехода обратимы).

  **2.** Супермногообразия (функции перехода обратимы).

  **3.** Препятственные полусупермногообразия (функции перехода необратимы).

Аналогичным образом вводятся полурасслоения, в которых необратимость возникает за счет необратимости функций перехода, связанной с нильпотентами и дивизорами нуля в подстилающей супералгебре. Далее рассматриваются морфизмы и условия соответствия полурасслоений. Обобщенные условия коцикла для функций перехода полусупермногообразий и полурасслоений могут приводить к построению необратимых аналогов коциклов Чеха и спектральных последовательностей, что тесно связано с когомологическими методами теории полугрупп.

Для описания обобщенных морфизмов на полусупермногообразиях определяются четные и нечетные полугомотопии с необратимым четным или нечетным суперпараметром соответственно. Полугомотопии приводят к рассмотрению фундаментальных полугрупп и играют ту же роль в изучении свойств непрерывности и классификации полусупермногообразий, которую обыкновенные гомотопии играют для обыкновенных (супер)многообразий.

**Раздел "Необратимое обобщение $N = 1$ суперконформной геометрии"** посвящен построению необратимой суперконформной дифференциальной геометрии $(1|1)$-мерного комплексного суперпространства $Z = (z, \theta) \in \mathbb{C}^{1|1}$, которая исключительно важна в теории суперструн, суперримановых поверхностей и в двумерных суперконформных теориях поля.

Вначале строится полугруппа супераналитических преобразований $\mathbb{C}^{1|1} \to \mathbb{C}^{1|1}$ и проводится их классификация по необратимости. Вводятся локальные единицы и нули и анализируются их свойства. Приведены соотношения на тройных пересечениях $\mathscr{U}_\alpha \cap \mathscr{U}_\beta \cap \mathscr{U}_\gamma$ для супераналитических полусупермногообразий. Получено выражение для необратимого аналога березиниана и проведена классификация супераналитических преобразований $\mathbb{C}^{1|1} \to \mathbb{C}^{1|1}$ по индексу нильпотентности березиниана.

Далее подробно проанализированы все возможные редукции касательного $(1|1)$-мерного пространства без учета требования обратимости. Оказывается, что нетривиальных редукций имеется две, а не одна, как в обратимом случае. Это связано с фундаментальной формулой сложения березинианов редуцированных суперматриц касательного пространства

$$\operatorname{Ber} \mathrm{P}_A = \operatorname{Ber} \mathrm{P}_S + \operatorname{Ber} \mathrm{P}_T,$$

где $\mathrm{P}_A$ — полная суперматрица, $\mathrm{P}_S$ и $\mathrm{P}_T$ — треугольная и антитреугольная суперматрицы соответственно. Отсюда редуцированные (суперконформно-подобные) преобразования определяются проектированием березиниана на одно из слагаемых. Тогда в терминах преобразованных координат $\tilde{Z}=\left(\tilde{z},\tilde{\theta}\right)$ получаем два условия

$$\varDelta\left(z,\theta\right)=D\tilde{z}-D\tilde{\theta}\cdot\tilde{\theta}=0,\quad Q\left(z,\theta\right)=\partial\tilde{z}-\partial\tilde{\theta}\cdot\tilde{\theta}=0,$$

где $\partial$ и $D$ — обычная и суперпроизводная соответственно.

Первое из них определяет стандартные суперконформные преобразования $\mathcal{T}_{SCf}$ (обратимые и необратимые), а второе условие приводит к новым необратимым преобразованиям $\mathcal{T}_{TPt}$, сплетающим четность в касательном и кокасательном суперпространствах. Действительно, если суперконформные преобразования индуцируют ковариантные преобразования супердифференциалов $dZ=dz+\theta d\theta$ и суперпроизводных

$$D=D\tilde{\theta}\cdot\tilde{D},\quad d\tilde{Z}=Q\left(z,\theta\right)\cdot dZ,$$

то сплетающие четность преобразования также дают ковариантные преобразования в касательном суперпространстве, но с вращением четности

$$\partial=\partial\tilde{\theta}\cdot\tilde{D},\quad d\tilde{Z}=\varDelta\left(z,\theta\right)\cdot d\theta.$$

Первые два соотношения является ключевыми для построения теории распределения на суперримановых поверхностях, которые определяются уравнением $\varDelta\left(z,\theta\right)=0$. Другое условие $Q\left(z,\theta\right)=0$ определяет необратимый аналог суперримановых поверхностей, в которых четность касательного пространства не фиксирована. Такая конструкция с функциями перехода из $\mathcal{T}_{SCf}$ и $\mathcal{T}_{TPt}$ может рассматриваться как частный случай введенных ранее полусупермногообразий. Кроме того, новые сплетающие четность преобразования возможно могут приводить к дополнительным вкладам в амплитуду фермионных струн специальной конфигурации.

Рассмотрены также левые вырожденные редуцированные преобра-зования $\mathcal{T}_{Deg_L}$, для которых оба условия $\varDelta(z,\theta) = 0$ и $Q(z,\theta) = 0$ выполняются одновременно, а также правые вырожденные редуцированные преобразования $\mathcal{T}_{Deg_R}$, которые определяются условием $D\tilde{\theta} = 0$.

Найдено единое описание обоих типов редуцированных преобразований с помощью альтернативной параметризации, в котором различие между ними определяется проекцией некоторого "спина редукции" $n = \pm 1/2$, где знак $\pm$ соответствует преобразованиям $\mathcal{T}_{SCf}$ и $\mathcal{T}_{TPt}$ соответственно. Приведена таблица умножения для "спина редукции" и описаны его свойства. Если суперконформные преобразования $\mathcal{T}_{SCf}$ являются супераналогом голоморфных преобразований, то сплетающие четность преобразования $\mathcal{T}_{TPt}$ можно трактовать как супераналог антиголоморфных преобразований комплексной плоскости, которые обязаны быть необратимыми.

Другим важным свойством сплетающих четность преобразований $\mathcal{T}_{TPt}$ является незамкнутость композиции (как, впрочем, и антиголоморфных преобразований). Однако, на пересечении трех суперобластей $\mathscr{U} \cap \tilde{\mathscr{U}} \cap \tilde{\tilde{\mathscr{U}}}$ и $\mathcal{T} : \mathscr{U} \to \tilde{\mathscr{U}}$, $\tilde{\mathcal{T}} : \tilde{\mathscr{U}} \to \tilde{\tilde{\mathscr{U}}}$, $\tilde{\tilde{\mathcal{T}}} : \mathscr{U} \to \tilde{\tilde{\mathscr{U}}}$ выполняется следующий закон умножения преобразований $\tilde{\mathcal{T}}_{SCf} \circ \mathcal{T}_{TPt} = \tilde{\tilde{\mathcal{T}}}_{TPt}$. Отсюда видно, что множество сплетающих четность преобразований является правым идеалом для суперконформных преобразований. Кроме того, вместо стандартного условия коцикла на суперримановой поверхности $D\tilde{\tilde{\theta}} = D\tilde{\theta} \cdot \tilde{D}\tilde{\tilde{\theta}}$ мы определяем "сплетенный коцикл"

$$\partial\tilde{\tilde{\theta}} = \partial\tilde{\theta} \cdot \tilde{D}\tilde{\tilde{\theta}}$$

с множителями различной четности. Тогда возможно построение принципиально новых распределений и расслоений, которые не сохраняют четность, как в классическом случае.

Применяя анзац ослабления обратимости можно обобщить и сами суперконформные преобразования. Новая параметризация $N = 1$ супер-конформной группы позволила расширить ее до полугруппы $\mathbf{S}_{SCf}$ и унифицировать описание старых и новых преобразований. Мы нашли, что построенная полугруппа принадлежит к новому абстрактному типу полугрупп, удовлетворяющим необычному идеальному умножению

$$\mathbf{S}_{SCf} * \mathbf{I}_n \subseteq \mathbf{I}_n,\ \mathbf{I}_n * \mathbf{S}_{SCf} \subseteq \mathbf{I}_{n+1},\ \mathbf{S}_{SCf} * \mathbf{I}_n * \mathbf{S}_{SCf} \subseteq \mathbf{I}_{n+1},$$

где $\mathbf{I}_n$ — члены построенного идеального ряда, имеющего специфические свойства. Из этого умножения можно определить $\mathbf{I}_n$ как правый и двусторонний повышающий идеал, но обычный левый идеал, что говорит о нетривиальной идеальной структуре $N = 1$ суперконформной полугруппы.

Введены и изучены свойства обобщенных векторных и тензорных отношений Грина, также определены идеальные квазихарактеры в суперконформной полугруппе.

Исследование свойств дробно-линейных $N = 1$ редуцированных преобразований проводится в терминах нечетных аналогов миноров для суперматриц — полуминоров, которые являются полуматрицами вида $\mathcal{M} = \begin{pmatrix} a & b \\ \gamma & \delta \end{pmatrix}$ ($a, b$ — четные, $\gamma, \delta$ — нечетные) и описывают вращающие четность отображения линейных двумерных суперпространств $\mathbf{\Lambda}^{2|0} \to \mathbf{\Lambda}^{1|1}$ и $\mathbf{\Lambda}^{1|1} \to \mathbf{\Lambda}^{2|0}$. Определено отображение $\mathbf{\Theta}$ — полутранспонирование, связывающее полуматрицы с матрицами $\mathcal{M} \overset{\mathbf{\Theta}}{\leftrightarrow} \mathrm{M}$. Полутранспонирование можно трактовать как извлечение квадратного корня из хорошо известного оператора смены четности — $\mathbf{\Pi}$-транспонирования. Для описания сплетающих четность преобразований вводятся нечетные аналоги детерминанта и перманента от полуматриц — полудетерминант $\delta\mathrm{et}\mathcal{M} = a\delta - b\gamma$ и полуперманент $\pi\mathrm{er}\mathcal{M} = a\delta + b\gamma$, которые нильпотентны и удовлетворяют нетривиальным соотношениям. Полудетерминант дуален с детерминантом в том смысле, для необратимых преобразований полудетерминант $\delta\mathrm{et}\mathcal{M}$ играет роль, аналогичную той, которую корень из обычного детерминанта $\sqrt{\det \mathrm{M}}$ играет для обратимых преобразований. Найдена четно-нечетная симметрия дробно-линейных $N = 1$ суперконформных преобразований, которая состоит в симметрии относительно одновременной замены детерминанта на полудетерминант и четных координат на нечетные.

Найдены и исследованы необратимые суперaналоги расстояния в $(1|1)$-мерном суперпространстве. Введен необратимый TPt аналог метрики $ds$ по формулам

$$|ds| \operatorname{Im} \theta = |d\theta| , \quad |ds| \left( \operatorname{Im} \tilde{z} + \frac{1}{2} \tilde{\theta} \tilde{\bar{\theta}} \right) = \left| d\tilde{Z} \right|$$

и сформулирован необратимый аналог инвариантности — “полуинвариантность” введенной метрики.

Далее изучаются нелинейные реализации редуцированных суперконформно-подобных преобразований, и в дополнение к вышеупомянутым исследованиям, мы включаем в рассмотрение конечные преобразования и учитываем их необратимость. Рассмотрена трактовка нелинейных реализаций как движение нечетной кривой в суперпространстве $\mathbb{C}^{1|1}$ и получены представления для конечных обратимых и необратимых $N = 1$ суперконформных преобразований, а также для сплетающих четность преобразований как уравнений для SCf голдстино и TPt голдстино.

Соотношение между линейной и нелинейной реализациями изучены в рамках диаграммного подхода

$$\begin{array}{ccc} Z_A & \xrightarrow[\text{W-Z}]{\mathcal{G}} & \tilde{Z} \\ \mathcal{A}\uparrow & & \uparrow\mathcal{B} \\ Z & \xrightarrow[\text{A-V}]{\mathcal{H}} & Z_H \end{array}$$

Здесь преобразование $\mathcal{G}$ играет роль линейного преобразования, преобразование $\mathcal{H}$ является нелинейным (в обратимом случае — из подгруппы $\mathcal{G}$), в то время, как $\mathcal{A}$ и $\mathcal{B}$ соответствуют косетным преобразованиям с голдстоуновскими полями как параметрами. Для конечных редуцированных обратимых и необратимых преобразований с учетом их таблицы умножения получены следующие возможные представления

$$\mathcal{G}_{SCf} \circ \mathcal{A}_{SCf} = \mathcal{B}_{SCf} \circ \mathcal{H}_{SCf}, \quad \mathcal{G}_{TPt} \circ \mathcal{A}_{SCf} = \mathcal{B}_{TPt} \circ \mathcal{H}_{SCf}$$

(второе уравнение является новым) и соответствующие компонентные уравнения, которые решены в частных случаях.

**В разделе "Необратимая геометрия расширенных редуцированных преобразований"** рассмотрены $N = 2$ и $N = 4$ редуцированные обратимые и необратимые отображения. Получено общее выражение для березиниана расширенных преобразований в терминах полуминоров суперматриц касательного $(1|N)$-мерного пространства в комплексном базисе.

Сформулированы теоремы сложения $N = 2$ и $N = 4$ березинианов, откуда следуют возможные редукции $(1|N)$-мерных касательных пространств. Нетривиальных редукций оказывается $N + 1$, что приводит к $N$-обобщению понятия комплексной структуры: для $N$-редуцированных преобразований имеется 1 четный (обратимый или необратимый) суперконформный (SCf) супераналог голоморфных преобразований и $N$ нечетных необратимых сплетающих четность (TPt) супераналогов антиголоморфных преобразований.

Подробно классифицированы $N = 2$ и $N = 4$ суперконформные преобразования с использованием перманентов. Получен общий вид бе-резиниана для обратимых $N$-SCf преобразований

$$\mathrm{Ber}\,\left(\tilde{Z}/Z\right) = k\,(\det \mathrm{H})^{\frac{2-N}{N}}\,,$$

где H — матрица производных $D_i\tilde{\theta}_k$ в комплексном базисе и $k = \pm 1$.

В частном случае $N = 2$ получено выражение березиниана через перманент

$$\mathrm{Ber}\,\left(\tilde{Z}/Z\right) = \frac{\det \mathrm{H}}{\mathrm{per}\,\mathrm{H}}.$$

Проведена классификация в терминах перманентов обратимых и необратимых расщепленных суперконформных преобразований, описывающих спиновые структуры на обыкновенной римановой поверхности и играющих важную роль в расчете суперструнных амплитуд.

Построены $N = 2$ и $N = 4$ суперконформные полугруппы в альтернативной параметризации и подробно исследованы их свойства. Приведено компонентное представление. Определены и обсуждаются свойства сплетающих четность преобразований и соответствующих супердифференциалов, дуальных соответствующим суперпроизводным.

**Раздел “Суперматричные полугруппы, идеальное строение и редукции”** посвящен построению и исследованию идеальных свойств суперматриц. На примере $(1|1) \times (1|1)$ суперматриц изучено их необратимое строение и определяется два типа возможных редукций: четно-редуцированные (треугольные) суперматрицы S и нечетно-редуцированные (антитреугольные) суперматрицы T. Для них справедлива теорема сложения березинианов

$$\mathrm{Ber}\,\mathrm{M} = \mathrm{Ber}\,\mathrm{S} + \mathrm{Ber}\,\mathrm{T}.$$

Изучены мультипликативные свойства нечетно-редуцированных суперматриц, которые приводят к выводу о том, что нечетно-редуцирован-ный морфизм может представляться в качестве произведения нечетно- и четно-редуцированных морфизмов, таковых, что коммутативная диаграмма, которая ответственна также и за сплетенные коциклы в редуцированных суперконформных преобразованиях.

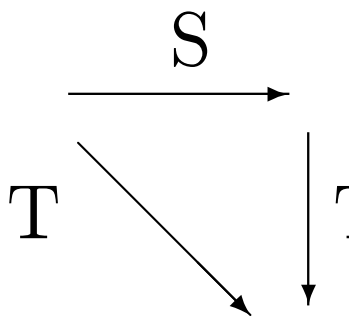

Построена полугруппа множеств редуцированных матриц. Множества четно- и нечетно-редуцированных суперматриц объединяются в некоторую сэндвич полугруппу с несимметричным умножением, зависящим от второго сомножителя. Полугруппа множеств редуцированных матриц изоморфна некоторой полугруппе правых нулей с сэндвич умножением.

Чтобы построить аналогичную сэндвич полугруппу с умножением не множеств, а самих суперматриц, вводится нечетный антикоммутирующий аналог $\mathcal{E}(\chi)$ (антискаляр) для скалярной суперматрицы $\mathrm{E}(x)$ (скаляра) по формулам $\mathrm{E}(x) = \begin{pmatrix} x & 0 \\ 0 & x \end{pmatrix}$, $\mathcal{E}(\chi) = \begin{pmatrix} 0 & \chi \\ \chi & 0 \end{pmatrix}$. Тогда прямая сумма скаляра и анти-скаляра совпадает со странной подалгеброй Березина $\mathrm{E}(x) \oplus \mathcal{E}(\chi) = \mathrm{Q}_\Lambda(1)$. Определяется в этой связи также правое $\boldsymbol{\Upsilon}_R$ и левое $\boldsymbol{\Upsilon}_L$ антитранспонирования, которые имеют смысл корня из оператора смены четности $\boldsymbol{\Pi}$, поскольку $\boldsymbol{\Upsilon}_R \boldsymbol{\Upsilon}_L = \boldsymbol{\Pi}$. Тогда конкретная реализация нечетного правого, левого и двустороннего модулей имеет вид

$$\mathcal{E}(\chi)\,\mathrm{M} = \chi \mathrm{M}^{\boldsymbol{\Upsilon}_R}, \quad \mathrm{M}\mathcal{E}(\chi) = \mathrm{M}^{\boldsymbol{\Upsilon}_L}\chi, \quad \mathcal{E}(\chi_1)\,\mathrm{M}\mathcal{E}(\chi_2) = \chi_1 \mathrm{M}^{\boldsymbol{\Pi}}\chi_2,$$

где, в отличие от стандартного супермодуля, в правой части появились антитранспонирования и оператор смены четности. Нахождение новых типов нечетных модулей является исключительно важным для построения и применения новых типов супермногообразий и полусупермногообразий.

Чтобы получить объединенное умножение четно- и нечетно-редуцированных суперматриц и построить соответствующую полугруппу, введенные антискаляры использовались наравне со скалярами. Если трактовать обычное умножение суперматриц как сэндвич-умножение со скаляром $\mathrm{E}\,(1)$, то сэндвич-умножение редуцированных суперматриц (с "суперполем" $X = (x, \chi)$) определится как

$$\mathrm{R}_1 \star_X \mathrm{R}_2 = \begin{cases} \mathrm{R}_1 \mathrm{E}\,(x)\,\mathrm{R}_2, & \mathrm{R}_2 = \mathrm{S}, \\ \mathrm{R}_1 \mathcal{E}\,(\chi)\,\mathrm{R}_2, & \mathrm{R}_2 = \mathrm{T}. \end{cases}$$

Поскольку сэндвич-умножение ассоциативно, редуцированные суперматрицы образуют полугруппу, которая изоморфна полугруппе правых нулей.

Рассмотрена также роль нечетных модулей и антискаляров в прямой сумме множеств редуцированных суперматриц, где введенны нечетные аналоги собственных чисел, характеристических функций (по формуле $\mathrm{Ber}\,(\mathcal{E}\,(\chi) - \mathrm{T})$ вместо $\mathrm{Ber}\,(\mathrm{E}\,(x) - \mathrm{S})$) и сформулирована обобщенная теорема Гамильтона-Якоби.

Важную роль в суперсимметричных теориях играют непрерывные полугруппы редуцированных суперматриц. Рассмотрена и подробно проанализирована идеальная структура многопараметрических полугрупп нечетно-редуцированных суперматриц. Показано, что общий вид суперматриц, образующих полугруппу ($\Gamma$-полугруппу), есть

$$\mathrm{T}^{\Gamma} = \begin{pmatrix} 0 & \Gamma \\ \mathrm{Ann}\,\Gamma & \mathrm{B} \end{pmatrix},$$

и их подмножество $\mathfrak{T}^{\Gamma} = \bigcup \mathrm{T}^{\Gamma}$ в множестве всех матриц $\mathfrak{M}$ является слабым идеалом, который для некоторого $\Gamma_1 \subseteq \Gamma$ определяется следующим соотношением $\mathfrak{T}^{\Gamma} \star \mathfrak{M} \star \mathfrak{T}^{\Gamma} \subseteq \mathfrak{T}^{\Gamma_1}$.

Обнаружено, что одно- и двухпараметрические полугруппы $\mathbf{P}_\alpha$ нечетно-редуцированных идемпотентных суперматриц вида $\begin{pmatrix} 0 & \alpha t \\ \alpha & 1 \end{pmatrix}$ и $\begin{pmatrix} 0 & \alpha t \\ \alpha u & 1 \end{pmatrix}$ ($\alpha^2 = 0$, $u, t$ — параметры) непрерывно представляют полугруппы левых нулей и прямоугольные связки соответственно. Это представление является неточным, поскольку нет редуктивности и сокращения. Поэтому стандартное отношение равенства $\mathbf{\Delta}$ заменяется на $\alpha$-отношение $\mathbf{\Delta}_\alpha \leftrightarrow t - u \in \operatorname{Ann} \alpha$. Полугруппа $\mathbf{P}_\alpha$ обладает необычным свойством — она является регулярной, но не инверсной. Для нее также найдены отношения Грина: $\mathscr{L}$-эквивалентность совпадает с универсальным отношением, а $\mathscr{R}$-эквивалентность равна $\alpha$-отношению $\mathbf{\Delta}_\alpha$ (а не $\mathbf{\Delta}$). Получено объединение однопараметрических полугрупп в некоторую нетривиальную полугруппу — скрученную прямоугольную связку, для которой выписана таблица Кэли и найдены все подполугруппы.

Рассматриваются суперматричные представления высших $(n|n)$-связок как обобщений прямоугольных связок, которые не могут быть сведены к произведению последних. Для них определяются высшие $\alpha$-отношения $\mathbf{\Delta}_\alpha^{n|n}$, которым равны соответствующие $\mathscr{R}$-эквивалентности. Вычислены отношения Грина для $(n|n)$-связок и установлен смысл стандартных $\mathscr{R}, \mathscr{L}, \mathscr{D}, \mathscr{H}$-классов для суперматриц. Далее мы определяем более общие отношения $\mathscr{R}^{(i)}, \mathscr{L}^{(i)}, \mathscr{D}^{(i)}, \mathscr{H}^{(i)}$ и называем их тонкими отношениями эквивалентности. Такие обобщенные отношения Грина необходимы для описания всех возможных классов элементов в $(n|n)$-связках, пропущенных в стандартном подходе. Из тонких эквивалентностей мы можем получать также и все известные отношения. Например, в случае $(2|2)$-связки, $\mathscr{R}^{(1)} \cap \mathscr{R}^{(2)} = \mathscr{R}$, $\mathscr{L}^{(1)} \cap \mathscr{L}^{(2)} = \mathscr{L}$, но дополнительно находим *смешанные* отношения вида $\mathscr{H}^{(i|j)} = \mathscr{R}^{(i)} \cap \mathscr{L}^{(j)}$, $\mathscr{D}^{(i|j)} = \mathscr{R}^{(i)} \vee \mathscr{L}^{(j)}$ и высших порядков

$$\mathscr{H}^{(ij|k)} = \left(\mathscr{R}^{(i)} \cap \mathscr{R}^{(j)}\right) \cap \mathscr{L}^{(k)}, \ \mathscr{D}^{(ij|k)} = \left(\mathscr{R}^{(i)} \cap \mathscr{R}^{(j)}\right) \vee \mathscr{L}^{(k)}.$$

Для каждого смешанного $\mathscr{D}$-класса мы можем построить смешанную eggbox диаграмму тонких $\mathscr{R}, \mathscr{L}$-классов, которая будет такой размерности, сколько слагаемых имеет в своей правой части заданное смешанное отношение. А именно, eggbox диаграммы $\mathscr{D}^{(i|j)}$-классов двумерны, а диаграммы $\mathscr{D}^{(ij|k)}$ и $\mathscr{D}^{(i|jk)}$-классов должны быть трехмерны. В случае $(n|n)$-связки необходимо рассматривать все возможные $k$-размерные eggbox диаграммы, где $2 \leq k \leq n-1$. Введенные тонкие отношения эквивалентности допускают подполугрупповую интерпретацию: стандартные отношения Грина на подполугруппе $\mathscr{U}$ полугруппы $\mathbf{S}$ имеют как свой аналог продолженные образы в $\mathbf{S}$, а именно тонкие отношения эквивалентности на $\mathbf{S}$.

**В разделе “Перманенты, scf-матрицы и необратимая гиперболическая геометрия”** детально исследованы свойства матриц, содержащих нильпотентные элементы и делители нуля, вполне определенный тип которых возникает при анализе $N$-расширенных редуцированных преобразований. Для таких матриц перманенты начинают играть дуальную (по отношению к детерминантам) роль, поэтому важно рассмотреть эти дуальные свойства в общем случае нильпотентных матриц, что может быть применено и в других моделях, использующих суперсимметрию в качестве основополагающего принципа.

Введено понятие scf-матрицы $\mathrm{A}_{\mathrm{scf}}$ из четных элементов, обладающих scf-свойством определенной ортогональности ее блоков между собой. В обратимом случае scf-матрицы подобны ортогональным матри-цам. Так, для $2 \times 2$ матрицы scf-свойство состоит в ортогональности элементов столбцов, и для них имеет место дуальность между перманентом и детерминантом и между минорами и алгебраическими дополнениями

$$\operatorname{per} \mathrm{A}_{\mathrm{scf}} \leftrightarrow \det \mathrm{A}_{\mathrm{scf}}, \quad \mathrm{A}_{\mathrm{scf}}^{M} \leftrightarrow \mathrm{A}_{\mathrm{scf}}^{D}.$$

Сформулирован критерий обратимости scf-матриц в терминах перманентов, а не детерминантов. Предложена новая формула для per-обратной scf-матрицы, которая в обратимом случае имеет вид

$$\mathrm{A}_{\mathrm{scf}}^{-1per} = \frac{\mathrm{A}_{\mathrm{scf}}^{MT}}{\operatorname{per} \mathrm{A}_{\mathrm{scf}}}.$$

Отличие от стандартного случая возникает лишь для необратимых scf-матриц. Получены формулы, связывающие след, перманент и детерминант, а также формула Бине-Коши для перманентов

$$\operatorname{per} \left(\mathrm{A}_{\mathrm{scf}} \cdot \mathrm{B}_{\mathrm{scf}}\right) = \operatorname{per} \mathrm{A}_{\mathrm{scf}} \cdot \operatorname{per} \mathrm{B}_{\mathrm{scf}},$$

которая совпадает с аналогичной формулой для детерминантов только в случае scf-матриц. Определяется полугруппа scf-матриц $SCF\,(N)$, подгруппа которой изоморфна $O\,(N)$ и для которой найдены идеалы и условия обратимости при $N = 2$ и $N = 4$.

Далее предлагается использовать scf-матрицы для изучения дробно-линейных (обратимых и необратимых) преобразований суперпространств $\mathbb{C}^{1|0} \to \mathbb{C}^{1|0}$, называемых per-отображениями. Показано, что для per-отображений имеет место симметрия $\operatorname{per} \leftrightarrow \det$, $\quad \operatorname{Re} \leftrightarrow \operatorname{Im}$ во всех основных соотношениях гиперболической геометрии.

Найден новый инвариант per-отображений— правое двойное отношение $\boldsymbol{D}^{+}\left(z_1, z_2, z_3, z_4\right)$, которое наряду с известным левым двойными отношениями $\boldsymbol{D}^{-}\left(z_1, z_2, z_3, z_4\right)$ является следующей функцией четырех

точек

$$\boldsymbol{D}^{\pm}\left(z_1, z_2, z_3, z_4\right) = \frac{\left(z_1 \pm z_3\right)\left(z_2 \pm z_4\right)}{\left(z_1 \pm z_4\right)\left(z_2 \pm z_3\right)}.$$

Это приводит к новым морфизмам группы перестановок, зеркальной per-гармонической последовательности точек и к per-аналогу классической формулы Лаггера, а также функция, которую можно трактовать как per-аналог производной Шварца. Два двойных отношения дают два — правое и левое — гиперболических расстояния

$$\boldsymbol{d}^{\pm}\left(z_1, z_2\right) = \ln \boldsymbol{D}^{\pm}\left(z_1, z_2, z_3, z_4\right).$$

В терминах правого двойного отношения $\boldsymbol{D}^{+}\left(z_1, z_2, z_3, z_4\right)$ и правого расстояния $\boldsymbol{d}^{+}\left(z_1, z_2\right)$ можно последовательно построить per-аналог гиперболической геометрии и тригонометрии на комплексной суперплоскости или в многомерных комплексных суперпространствах.

**В Заключении** сформулированы основные результаты диссертационной работы.

**В Приложениях** приведены необходимые сведения по супералгебрам, отдельные аспекты теории супермногообразий и суперримановых поверхностей, дополнительные факты из теории полугрупп, а также некоторые выкладки, не вошедшие в основной текст.

Основные результаты диссертации опубликованы в работах [1–21] и в трудах международных конференций, на которых докладывались работы автора.

# РАЗДЕЛ 1

# ТЕОРИЯ НЕОБРАТИМЫХ СУПЕРМНОГООБРАЗИЙ

В данном разделе рассматривается обобщение понятий супермногообразия, расслоения и гомотопии на необратимый случай. Используемый язык карт и функций перехода позволяет определить полусупермногообразие как необратимый аналог супермногообразия в общепринятом определении функционального подхода. Вводятся необратимые карты, атласы и функции перехода, для которых предлагаются соответствующие уравнения. Находятся обобщенные условия коцикла, а также новый нильпотентный тип ориентируемости полусупермногообразий. Формулируется общий принцип полукоммутативности для необратимых морфизмов. В терминах уравнений на функции перехода определяются морфизмы полурасслоений. Приводятся также условия рефлексивности для полусупермногообразий и полурасслоений. Вводятся четные и нечетные полугомотопии с необратимым четным или нечетным суперпараметром соответственно, которые играют важную роль в классификации полусупермногообразий и построении фундаментальных полугрупп.

Общепринятым считается [22, 23], что идея обратимых супермногообразий впервые была высказана неявно в работах [24, 25] в связи с обобщением классической динамики и дискуссией о классическом пределе для фермионов [26]. Математические аспекты групп и алгебр с антикоммутирующими переменными первоначально рассматривались в работах [27–29], но лишь в рамках формального правила “протаскивания знака” и предписания “о возможности обобщения всех основ-ных понятий анализа, при котором образующие грассмановой алгебры стали бы играть роль, равноправную с вещественными или комплексными переменными” ( [30, с. 9]). Именно в этой широко известной фразе и заключалось ограничение на дальнейшее развитие теории супермногообразий в абстрактном направлении: “равноправие” подразумевало в качестве “супераналогов” тривиально подобные (с точностью до замены некоторых знаков с минуса на плюс и четных величин на нечетные) объекты и не позволяло даже предполагать существования иных абстрактых алгебраических и геометрических структур.

В начале 70-х в конкретных моделях элементарных частиц [31,32] отечественными физиками был открыт новый тип симметрии [33–39] — между коммутирующими бозонами, которые описывают калибровочные взаимодействия, и антикоммутирующими фермионами, которые соответствуют взаимодействующим с их помощью частицам. Однако действительное признание это фундаментальное направление получило только через несколько лет *), когда такая же бозон-фермионная симметрия, но в других моделях, была названа западными учеными красивым и эффектным словом “суперсимметрия” [41–46]. К моменту появления суперсимметрии в физике оказалось, что математический аппарат для ее описания (супергруппы и супералгебры Ли) уже был создан [27, 47]. После чего количество работ по суперобобщениям физических теорий стремительно начало возрастать (см., например, обзоры [48–58] и книги [59–63]). Элементам рассматриваемых теорий присваивалась завораживающая приставка “супер”, но реальное “усовершенствование” опять-таки сводилось к заменам знаков и добавлению нечетных величин при неизменных основных абстрактных конструкциях, что, казалось бы, подтверждало математическую гипотезу “равноправия” антикоммутирующих величин [22, 30], но лишь на первый

---

*Примечание.* История этого периода подробно изложена в [40].

взгляд.

Важно, что суперсимметрия появилась благодаря ослаблению одного из условий теоремы Коулмена-Мандулы о числе симметрий $S$-матрицы [64] — ограничению только коммутирующими генераторами так, что "...можно представить себе, что дальнейшее ослабление условий может привести к новым симметриям" [65, с. 2] (например, неассоциативные генераторы рассматривались в [66, 67]). Именно введение антикоммутирующих генераторов [27, 29, 36, 68] позволило единым образом рассмотреть внутренние и пространственно-временные симметрии [34, 41, 43], т. е. объединить бозоны и фермионы в обобщенные мультиплеты – суперполя [44, 46, 69] и ввести суперпространство [70] как главную арену для "суперпревращений" элементарных частиц [51, 61, 71–73]. Так, согласно феноменологии объединенных суперсимметричных [74–76] и суперструнных [77, 78] теорий, каждой наблюдаемой частице должен соответствовать "суперпартнер" с противоположной статистикой (хотя и есть попытки включить в число суперпартнеров имеющиеся частицы [79] или вообще их не вводить [80]). Многочисленные экспериментальные поиски таких частиц (см. обзоры [14, 15, 81, 82]) пока не привели к их непосредственному обнаружению [*)] [87–90]. Это наводит на мысль о том, что, возможно, математические основания суперсимметричных теорий элементарных частиц нуждаются в дальнейшем внутреннем развитии.

Действительно, "равноправие" антикоммутирующих величин подразумевало однозначный ответ на вопрос "в каких категориях?" — в тех же, что и раньше: групп, топологических пространств и многообразий, хотя и "супер". Существенным оказывается то, что эта впечатляющая приставка не изменяла самого абстрактного и теоретико-категорного содержания понятий ("хотя ничего "супер" в суперматема-

---

*Примечание.* Удивительно, однако, что структура генетического кода человека описывается супералгебрами Ли [83–86].

тике нет..." [91, с. 6]). Например, супергруппа [92–94] является группой и не более того, т. е. принадлежит к категории групп [95–99], пусть с некоторыми дополнительными свойствами. То же касается суперпространств и супермногообразий. Добавление необратимых нильпотентных координат и направлений не позволяло изменить сами категории, а только несколько модифицировать уже имеющиеся в жестких рамках "гипотезы равноправия" [22].

Однако хорошо известно, что необратимые объекты описываются не группами, а *полугруппами**) [101–104], которые содержат группы как составную обратимую часть. Следовательно, категория групп [105] слишком узка для того, чтобы строить на ее основе суперсимметричные модели элементарных частиц (см. [4]). Основным и фундаментальным объектом таких моделей является понятие супермногообразия [47, 106–110] (см. **Приложение А**). Здесь мы построим необратимый аналог супермногообразий — полусупермногообразия, а также аналоги сопутствующих объектов — расслоений и гомотопий.

Необратимое расширение понятия супермногообразия представляется естественным в связи с предположениями, сделанными во многих работах относительно внутренней необратимости конкретных конструкций. Например, "... общая суперриманова поверхность не имеет числовой части" [111], "... возможно не существует обратимых операторов проектирования (числового отображения [112]) вообще" [113], или "... числовая часть даже может не существовать в самых экстремальных примерах" [114]. В частности, при исследовании свойств необратимости суперконформной симметрии [5, 8] предполагалось [1, 13] возможное существование суперсимметричных объектов, аналогичных суперримановым поверхностям, но без числовой части, и предварительно показано,

---

*Примечание.* Впервые полугруппы были введены харьковским математиком Сушкевичем еще в 30-х годах [100].

как их строить [9].

Необратимость в теории супермногообразий [115–118] в действительности является результатом добавления нечетных нильпотентных элементов [119,120] и делителей нуля [121–124], возникающих в алгебрах Грассмана-Банаха (см. [125, 126] для нетривиальных примеров). В бесконечномерном случае [127–131] имеются (топологически) квазинильпотентные нечетные элементы, которые на самом деле не нильпотентны [132], в некоторых суперaлгебрах можно построить чисто дỳховые элементы, которые не нильпотентны даже топологически [133] или ввести аналог обратимого нечетного символа [134], а также использовать методы нестандартного анализа [135, 136]. Высказывалась даже противоположная "равноправию" идея о том, что "четная геометрия = коллективному эффекту бесконечномерной нечетной геометрии" [137] (см. ее конкретную реализацию в [138]). Кстати, чисто нечетные многообразия рассматривались в [139–141], также вводились экзотические супермногообразия с нильпотентными четными координатами [142], супераналоги многообразий Фробениуса [143, 144] с нефиксированной метрикой [145, 146] и финслеровых пространств [147–151], рассматривалась гравитация [152] и супергравитация [153] с необратимым репером. Список общих проблем с нечетными направлениями (и, следовательно, связанных с необратимостью) для супермногообразий приведен в [154]. Отметим, что делались попытки абстрактного обобщения суперaлгебр и супермногообразий на тернарные структуры [155] и моноидальные категории [156,157], а также исследовать нильпотентность [158–160], обратимость [161] и полугруппы теоретико-категорными методами [162,163]. С другой стороны, полугруппы возникали в теории суперaлгебр Ли [164], градуированных алгебр [165,166] и алгебр Ли [167,168], топологической квантовой теории поля [169], свободная полугруппа возникала при обобщении фермионных и бозонных коммутационных соотношений [170], суперполугруппа трансляций в $\mathbb{R}_+^{1|1}$ применялась при суперполевой фор-мулировке характера Черна [171], полугруппа Брандта использовалась в тензорных конструкциях теории струн [172].

Необходимо также напомнить о возможности определения супермногообразия без введения понятия топологического пространства [173]. Здесь мы предлагаем пойти дальше в этом направлении и отказаться от рассмотрения конкретной внутренней структуры из подстилающих алгебр*) (грассмановых или более общих), а все определения необратимых супермногообразий давать в терминах абстрактной теории полугрупп [19].

## 1.1. Обратимые супермногообразия в терминах окрестностей

Рассмотрим стандартное определение супермногообразия $\mathscr{M}$ в терминах окрестностей [117, 175, 176], которое отличается от определения обычного многообразия [177, 178] лишь "супер" терминологией.

Следующее построение является общепринятым для описания многообразий [179] и супермногообразий [91, 174] в терминах окрестностей. Супермногообразие покрывается набором суперобластей $\mathscr{U}_\alpha$, таких,что $\mathscr{M} = \bigcup\limits_\alpha \mathscr{U}_\alpha$. Затем в каждой области выбираются некоторые функции (*координатные отображения*) $\varphi_\alpha : \mathscr{U}_\alpha \to D^{n|m} \subset \mathbb{R}^{n|m}$, где $\mathbb{R}^{n|m}$ представляет собой суперпространство,в котором существуют "супершары" и $D^{n|m}$ открытая область в $\mathbb{R}^{n|m}$. Далее, пара $\{\mathscr{U}_\alpha, \varphi_\alpha\}$ называется *локальной картой*, а объединение карт $\bigcup\limits_\alpha \{\mathscr{U}_\alpha, \varphi_\alpha\}$ объявляется атласом *супермногообразия.*

Затем вводятся склеивающие функций перехода следующим обра-

---

*Примечание.* Названной в [174] "скелетом" супермногообразия.

зом. Пусть $\mathscr{U}_{\alpha\beta} = \mathscr{U}_\alpha \cap \mathscr{U}_\beta \neq \varnothing$ и

$$\begin{aligned} \varphi_\alpha &: \mathscr{U}_\alpha \to V_\alpha \subset \mathbb{R}^{n|m}, \\ \varphi_\beta &: \mathscr{U}_\beta \to V_\beta \subset \mathbb{R}^{n|m}. \end{aligned} \tag{1.1}$$

К тому же, вышеупомянутые морфизмы ограничиваются $\varphi_\alpha : \mathscr{U}_{\alpha\beta} \to V_{\alpha\beta} = V_\alpha \cap \varphi_\alpha(\mathscr{U}_{\alpha\beta})$ и $\varphi_\beta : \mathscr{U}_{\alpha\beta} \to V_{\beta\alpha} = V_\beta \cap \varphi_\beta(\mathscr{U}_{\alpha\beta})$. Отображения $\Phi_{\alpha\beta} : V_{\beta\alpha} \to V_{\alpha\beta}$ , которые необходимы, чтобы сделать следующую диаграмму

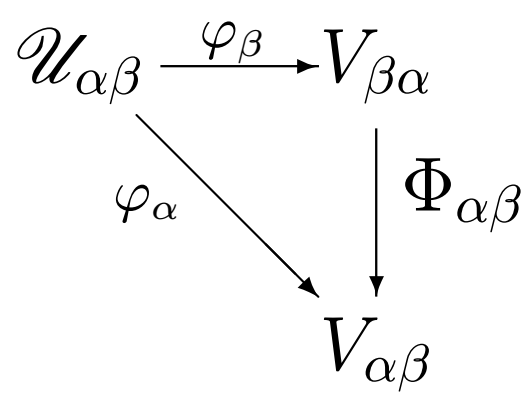

(1.2)

коммутирующей, называются *функциями перехода* многообразия в данном атласе.

Здесь мы подчеркиваем, во-первых, что $\mathscr{U}_{\alpha\beta} \subset \mathscr{M}$, а $V_{\alpha\beta}, V_{\beta\alpha} \subset \mathbb{R}^{n|m}$. Во-вторых, из (1.2) обычно делается вывод, что

$$\Phi_{\alpha\beta} = \varphi_\alpha \circ \varphi_\beta^{-1}. \tag{1.3}$$

(Супер) функции перехода $\Phi_{\alpha\beta}$ дают нам возможность к восстановить все (супер) многообразие $\mathscr{M}$ из индивидуальных карт и координатых отображений. В самом деле, они содержат всю информацию о (супер) многообразии. Они могут принадлежать к различным функциональным классам, что дает возможность уточнить более узкие классы многообразий и супермногообразий, например (супер) гладкие, аналитические, липшицевы и другие [179, 180]. В большинстве случаев “супер” только формально различает окрестностное определение многообразия и супермногообразия (что дает нам возможность записать его в скобках) и свойства функций $\Phi_{\alpha\beta}$, большинство же формул при этом остаются прежними [112,174,181]. Здесь мы не обсуждаем их подробно и пытаемся налагать минимум ограничений на $\Phi_{\alpha\beta}$, концентрируя наше внимание на их абстрактных свойствах и обобщениях, следующих из них.

Дополнительно, из (1.3) следует, что функции перехода удовлетворяют условиям коцикла

$$\Phi_{\alpha\beta}^{-1} = \Phi_{\beta\alpha} \tag{1.4}$$

на пересечениях $\mathscr{U}_\alpha \cap \mathscr{U}_\beta$ и

$$\Phi_{\alpha\beta} \circ \Phi_{\beta\gamma} \circ \Phi_{\gamma\alpha} = 1_{\alpha\alpha} \tag{1.5}$$

на тройных пересечениях $\mathscr{U}_\alpha \cap \mathscr{U}_\beta \cap \mathscr{U}_\gamma$, где $1_{\alpha\alpha} \stackrel{def}{=} \mathrm{id}\ (\mathscr{U}_\alpha)$. Обычно предполагается, что все отображения $\varphi_\alpha$ являются гомеоморфизмами, и они могут описываться взаимооднозначными обратимыми непрерывными (супер) гладкими функциями (т. е. происходит “переход” в обоих направлениях между любыми двумя пересекающимися областями $\mathscr{U}_\alpha \cap \mathscr{U}_\beta \neq \varnothing$). Можно было бы предположить, что логично не отличать $\mathscr{U}_\alpha$ и $D^{n|m}$, т. е. локально супермногообразия представляются как целостное суперпространство $\mathbb{R}^{n|m}$. Однако, дело не только в более богатой структуре расслоении [182–184] и пучка [106, 185, 186] из-за рассмотрения всех построений над алгеброй Грассмана (или над более общей алгеброй [117, 126, 133, 187]). Проблема заключается в ином абстрактном уровне построений, если условия обратимости в некоторой мере ослаблены.

## 1.2. Необратимые супермногообразия

Ранее существовал следующий общий рецепт: имеются готовые объекты (например вещественные многообразия, которые могут быть исследованы почти визуально), а затем, используя различные приемы и догадки, вычислялись ограничения на функции перехода (см., например, [177, 179, 188]). Несмотря на это, необратимые функции просто исключались из рассмотрения супермногообразий [110,189,190] (произнося магические слова “факторизуя по нильпотентам, мы опять получаем известный результат”), вследствие желания быть в наиболее близкой аналогии с интуитивно ясным и понимаемым несуперсимметричным случаем.

Здесь мы идем в обратном направлении: известно, что в суперматематике необратимые переменные и функции *существуют*. Какие объекты могут быть построены посредством них? Что дает “факторизация по ненильпотентам”, т. е. рассмотрение негрупповых особенностей теории? Как изменятся общий абстрактный смысл самых важных понятий, например многосвязных областей и расслоений? Мы сейчас попытаемся оставить в стороне внутреннее строение необратимых объектов, аналогичных супермногообразиям, и сконцентрируем наше внимание на общих абстрактных определениях.

Очевидно, что среди ординарных (несуперсимметричных) функций и отображений также существуют необратимые [191, 192] (и нереверсивные [193]), но тип необратимости, рассматриваемый здесь, весьма специальный: он возникает только из-за существования нильпотентов в подстилающей суперлгебре [125, 126, 132]. Здесь мы не рассматриваем конкретные уравнения и способы их решения, мы только используем факт их существования, чтобы переформулировать некоторые определения и расширить известные понятия.

**1.2.1.** Полусупермногообразия. Теперь мы сформулируем окрестностное определение объекта, аналогичного супермногообразию, т. е. попытаемся ослабить требование обратимости координатных отображений [6]. Рассмотрим некоторое обобщенное (в каком смысле, будет пояснено ниже) суперпространство $\mathscr{M}$, покрытое открытыми множествами $\mathscr{U}_\alpha$ как $\mathscr{M} = \bigcup_\alpha \mathscr{U}_\alpha$. Предположим, что отображения $\varphi_\alpha : \mathscr{U}_\alpha \to V_\alpha \subset \mathbb{R}^{n|m}$ не все обратимые гомеоморфизмы, т. е. среди них имеются необратимые отображения. Именно в этом смысле суперпространство $\mathscr{M}$ является необратимо обобщенным, и вместо $\mathbb{R}^{n|m}$ можно рассматривать также некоторое его необратимое обобщение.

**Определение 1.1.** *Карта есть пара $\left\{\mathscr{U}_\alpha^{inv}, \varphi_\alpha^{inv}\right\}$, где $\varphi_\alpha^{inv}$ — обратимый морфизм. Полукарта есть пара $\left\{\mathscr{U}_\alpha^{noninv}, \varphi_\alpha^{noninv}\right\}$, где $\varphi_\alpha^{noninv}$ — необратимые морфизмы.*

**Определение 1.2.** *Полуатлас $\{\mathscr{U}_\alpha, \varphi_\alpha\}$ есть объединение карт и полукарт $\left\{\mathscr{U}_\alpha^{inv}, \varphi_\alpha^{inv}\right\} \cup \left\{\mathscr{U}_\alpha^{noninv}, \varphi_\alpha^{noninv}\right\}$.*

**Определение 1.3.** *Полусупермногообразие есть суперпространство $\mathscr{M}$, представленное в качестве полуатласа $\mathscr{M} = \bigcup_\alpha \{\mathscr{U}_\alpha, \varphi_\alpha\}$.*

Определим аналог функций перехода полусупермногообразий *). Мы должны рассматривать ту же диаграмму (1.2), но мы не можем использовать (1.3) из-за необратимости некоторых $\varphi_\alpha$.

**Определение 1.4.** *Склеивающие функции полуперехода полусупермногообразия определяются уравнениями*

$$\Phi_{\alpha\beta} \circ \varphi_\beta = \varphi_\alpha \tag{1.6}$$

---

*) *Примечание.* Отметим, что имеется сходная терминология для других (отличных от рассматриваемого) обобщений многообразий: полуримановы многообразия [194–196], полупсевдоримановы пространства [197], полуинвариантные подмногообразия [198–200].

$$\Phi_{\beta\alpha} \circ \varphi_\alpha = \varphi_\beta. \tag{1.7}$$

*Замечание* **1.5.** Чтобы найти $\Phi_{\alpha\beta}$, уравнение (1.6) не может быть решено с помощью (1.3). Вместо этого мы должны искать искусственные приемы его решения, как в предыдущем подразделе, разложением в ряд по генераторам суперакгебры (см. например, [201–203]), либо используя абстрактные методы теории полугрупп [102, 204], которые рассматривают решения уравнений как классы эквивалентности.

Функции $\Phi_{\beta\alpha}$ теперь находятся не из (1.4), где левая часть не вполне определена, а из коммутативной диаграммы

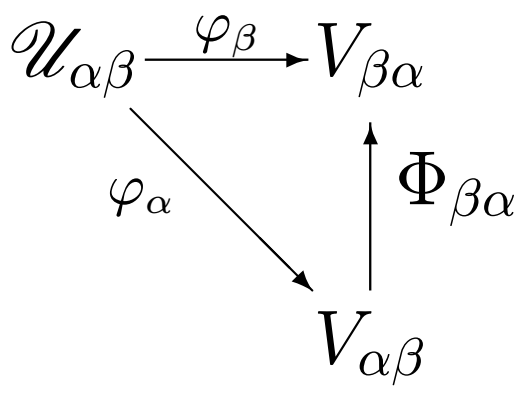

(1.8)

и уравнения (1.7), следующего из нее. Однако теперь функции $\Phi_{\beta\alpha}$ могут быть также необратимыми, и, следовательно, условия коцикла (1.4)–(1.5) должны быть модифицированы, чтобы не использовать обратимость [19].

*Замечание* **1.6.** Даже в стандартном случае условия коцикла (1.5) для супермногообразий автоматически не удовлетворяются, когда условие (1.3) имеет место, и поэтому они должны быть наложены искусственно дополнительными требованиями [189].

Таким образом, вместо (1.4) и (1.5) мы получаем

**Утверждение 1.7.** *Функции полуперехода полусупермногообразия удовлетворяют следующим отношениям*

$$\Phi_{\alpha\beta} \circ \Phi_{\beta\alpha} \circ \Phi_{\alpha\beta} = \Phi_{\alpha\beta} \tag{1.9}$$

*на* $\mathscr{U}_\alpha \cap \mathscr{U}_\beta$ *пересечениях и*

$$\Phi_{\alpha\beta} \circ \Phi_{\beta\gamma} \circ \Phi_{\gamma\alpha} \circ \Phi_{\alpha\beta} = \Phi_{\alpha\beta}, \tag{1.10}$$

$$\Phi_{\beta\gamma} \circ \Phi_{\gamma\alpha} \circ \Phi_{\alpha\beta} \circ \Phi_{\beta\gamma} = \Phi_{\beta\gamma}, \tag{1.11}$$

$$\Phi_{\gamma\alpha} \circ \Phi_{\alpha\beta} \circ \Phi_{\beta\gamma} \circ \Phi_{\gamma\alpha} = \Phi_{\gamma\alpha} \tag{1.12}$$

*на тройных пересечениях* $\mathscr{U}_\alpha \cap \mathscr{U}_\beta \cap \mathscr{U}_\gamma$ *и*

$$\Phi_{\alpha\beta} \circ \Phi_{\beta\gamma} \circ \Phi_{\gamma\rho} \circ \Phi_{\rho\alpha} \circ \Phi_{\alpha\beta} = \Phi_{\alpha\beta}, \tag{1.13}$$

$$\Phi_{\beta\gamma} \circ \Phi_{\gamma\rho} \circ \Phi_{\rho\alpha} \circ \Phi_{\alpha\beta} \circ \Phi_{\beta\gamma} = \Phi_{\beta\gamma}, \tag{1.14}$$

$$\Phi_{\gamma\rho} \circ \Phi_{\rho\alpha} \circ \Phi_{\alpha\beta} \circ \Phi_{\beta\gamma} \circ \Phi_{\gamma\rho} = \Phi_{\gamma\rho}, \tag{1.15}$$

$$\Phi_{\rho\alpha} \circ \Phi_{\alpha\beta} \circ \Phi_{\beta\gamma} \circ \Phi_{\gamma\rho} \circ \Phi_{\rho\alpha} = \Phi_{\rho\alpha} \tag{1.16}$$

*на* $\mathscr{U}_\alpha \cap \mathscr{U}_\beta \cap \mathscr{U}_\gamma \cap \mathscr{U}_\rho$.

Здесь первое соотношение (1.9) призвано обобщить первое условие коцикла (1.4), тогда как другие соотношения соответствуют (1.5). Мы называем соотношения (1.9)–(1.16) *башенными соотношениями* [19].

**Определение 1.8.** *Полусупермногообразие — рефлексивное, если, в дополнение к* (1.9)–(1.16), *функции полуперехода* $\Phi_{\alpha\beta}$ *удовлетворяют условиям рефлексивности*

$$\Phi_{\beta\alpha} \circ \Phi_{\alpha\beta} \circ \Phi_{\beta\alpha} = \Phi_{\beta\alpha} \tag{1.17}$$

*на* $\mathscr{U}_\alpha \cap \mathscr{U}_\beta$ *пересечениях и*

$$\Phi_{\alpha\gamma} \circ \Phi_{\gamma\beta} \circ \Phi_{\beta\alpha} \circ \Phi_{\alpha\gamma} = \Phi_{\alpha\gamma}, \tag{1.18}$$

$$\Phi_{\gamma\beta} \circ \Phi_{\beta\alpha} \circ \Phi_{\alpha\gamma} \circ \Phi_{\gamma\beta} = \Phi_{\gamma\beta}, \tag{1.19}$$

$$\Phi_{\beta\alpha} \circ \Phi_{\alpha\gamma} \circ \Phi_{\gamma\beta} \circ \Phi_{\beta\alpha} = \Phi_{\beta\alpha} \tag{1.20}$$

*на тройных пересечениях* $\mathscr{U}_\alpha \cap \mathscr{U}_\beta \cap \mathscr{U}_\gamma$ *и*

$$\Phi_{\alpha\rho} \circ \Phi_{\rho\gamma} \circ \Phi_{\gamma\beta} \circ \Phi_{\beta\alpha} \circ \Phi_{\alpha\rho} = \Phi_{\alpha\rho}, \tag{1.21}$$

$$\Phi_{\rho\gamma} \circ \Phi_{\gamma\beta} \circ \Phi_{\beta\alpha} \circ \Phi_{\alpha\rho} \circ \Phi_{\rho\gamma} = \Phi_{\rho\gamma}, \tag{1.22}$$

$$\Phi_{\gamma\beta} \circ \Phi_{\beta\alpha} \circ \Phi_{\alpha\rho} \circ \Phi_{\rho\gamma} \circ \Phi_{\gamma\beta} = \Phi_{\gamma\beta}, \tag{1.23}$$

$$\Phi_{\beta\alpha} \circ \Phi_{\alpha\rho} \circ \Phi_{\rho\gamma} \circ \Phi_{\gamma\beta} \circ \Phi_{\beta\alpha} = \Phi_{\beta\alpha} \tag{1.24}$$

*на* $\mathscr{U}_\alpha \cap \mathscr{U}_\beta \cap \mathscr{U}_\gamma \cap \mathscr{U}_\rho$.

*Замечание* **1.9.** Можно было бы считать, что условия рефлексивности (1.17)–(1.24) отличаются от (1.9)–(1.16) лишь индексом перестановки, однако, это так. Функции $\Phi_{\alpha\beta}$, входящие в эти две системы уравнений, являются теми же самыми, и, следовательно, последние представляют собой систему независимых уравнений, накладываемых на $\Phi_{\alpha\beta}$.

**Предложение 1.10.** *Соотношения, аналогичные* (1.9)–(1.24), *но имеющие два или более множителей в правой части, следуют из предыдущих.*

*Доказательство.* Например, рассмотрим

$$\Phi_{\alpha\beta} \circ \Phi_{\beta\gamma} \circ \Phi_{\gamma\alpha} \circ \Phi_{\alpha\beta} \circ \Phi_{\beta\gamma} = \Phi_{\alpha\beta} \circ \Phi_{\beta\gamma}. \tag{1.25}$$

Умножая справа на $\Phi_{\alpha\beta,}$ мы выводим

$$\Phi_{\alpha\beta} \circ \Phi_{\beta\gamma} \circ \Phi_{\gamma\alpha} \circ \Phi_{\alpha\beta} \circ \Phi_{\beta\gamma} \circ \Phi_{\alpha\beta} = \Phi_{\alpha\beta} \circ \Phi_{\beta\gamma} \circ \Phi_{\alpha\beta}. \tag{1.26}$$

Затем, используя (1.9), мы получаем

$$\Phi_{\alpha\beta} \circ \Phi_{\beta\gamma} \circ \Phi_{\gamma\alpha} \circ \Phi_{\alpha\beta} = \Phi_{\alpha\beta}, \tag{1.27}$$

что совпадает с (1.10). ■

*Замечание* **1.11.** В любых действиях с необратимыми функциями $\Phi_{\alpha\beta}$ мы не имеем права сокращать, поскольку полугруппа функций $\Phi_{\alpha\beta}$ представляет собой полугруппу без сокращений, и мы вынуждены использовать методы, подобные [205–208].

**Следствие 1.12.** *Соотношения* (1.9)–(1.24) *удовлетворяются тождественно в стандартном обратимом случае, т. е. когда условия* (1.3), (1.4) *и* (1.5) *выполняются.*

*Замечание* **1.13.** Уравнения (1.6)–(1.7), определяющие функции полуперехода $\Phi_{\alpha\beta}$, могут не иметь единственных решений, и в таком случае $\Phi_{\alpha\beta}$ должны рассматриваться, в качестве соответствующих множеств функций.

**Следствие 1.14.** *Функции* $\Phi_{\alpha\beta}$, *удовлетворяющие* (1.9)–(1.24), *могут быть рассмотрены как некоторое необратимое суперобобщение функций перехода для коциклов в чеховских когомологиях покрытий* [209, 210].

**1.2.2.** Ориентации полусупермногообразий. Известно, что ориентации обычных многообразий определяется знаком якобиана функций перехода $\Phi_{\alpha\beta}$, записанным в зависимости от локальных координат на $\mathcal{U}_\alpha \cap \mathcal{U}_\beta$ пересечениях [177, 178, 188]. Поскольку этот знак принадлежит $\mathbb{Z}_2$, существуют две ориентации на $\mathcal{U}_\alpha$. Две перекрывающиеся карты называются *согласованно ориентироваными* (или *сохраняющими ориентацию*), если $\Phi_{\alpha\beta}$ имеет положительный якобиан, и многообразие называется *ориентируемым*, если его можно покрыть такими картами. Следовательно, обычных многообразий имеется два типа: ориентируемый и неориентируемый [177, 211].

В суперсимметричном случае роль якобиана играет березиниан [22, 30], который имеет “знак”, принадлежащий к $\mathbb{Z}_2 \oplus \mathbb{Z}_2$ [110, 212], и таким образом здесь имеется четыре ориентации на $\mathscr{U}_\alpha$ и пять соответствующих типов ориентируемости супермногообразия [173, 213].

**Определение 1.15.** *В случае, если не обращающийся в нуль березиниан функций $\Phi_{\alpha\beta}$ является нильпотентным (и поэтому не имеет определенного знака в предыдущем смысле), существует дополнительная нильпотентная ориентация полусупермногообразия на $\mathscr{U}_\alpha$ и, соответственно, шестой (по классификации [173, 213]) тип ориентируемости — нильпотентная ориентируемость.*

Степень нильпотентности березиниана позволяет нам систематизировать полусупермногообразия, имеющие нильпотентую ориентируемость.

**1.2.3.** П р е п я т с т в е н н о с т ь  и  п о л у м н о г о о б р а з и я . Полусупермногообразия, определенные выше, представляют собой аналог так называемых препятственных полумногообразий [214–219]. Однако здесь мы определим *препятственность* в несколько ином смысле, чем определяется препятствие в [30], связав ее с необратимостью.

Запишем (1.3), (1.4) и (1.5) в виде следующего (бесконечного) ряда

$$n = 1: \ \Phi_{\alpha\alpha} = 1_{\alpha\alpha}, \tag{1.28}$$

$$n = 2: \ \Phi_{\alpha\beta} \circ \Phi_{\beta\alpha} = 1_{\alpha\alpha}, \tag{1.29}$$

$$n = 3: \ \Phi_{\alpha\beta} \circ \Phi_{\beta\gamma} \circ \Phi_{\gamma\alpha} = 1_{\alpha\alpha}, \tag{1.30}$$

$$n = 4: \ \Phi_{\alpha\beta} \circ \Phi_{\beta\gamma} \circ \Phi_{\gamma\delta} \circ \Phi_{\delta\alpha} = 1_{\alpha\alpha} \tag{1.31}$$

$$\dots \quad \dots$$

**Определение 1.16.** *Полумногообразие $\mathscr{M}$ — препятственное, если некоторые из условий коцикла* (1.28)–(1.31) *нарушаются.*

*Замечание* **1.17.** Введенное понятие препятственного многообразия не должно смешиваться с понятием препятствия для обыкновенных многообразий [214, 220] и супермногообразий [30, 221] или препятствием к расширению [209,222] и в теории характеристических классов [223,224].

Пусть, начиная с некоторого $n = n_\Pi$, все условия коцикла высшего порядка выполняются.

**Определение 1.18.** *Степень препятственности полумногообразия представляет собой максимальное $n_\Pi$, для которого условия коцикла* (1.28)–(1.31) *нарушаются. Если все из их выполняются, то $n_\Pi \stackrel{def}{=} 0$.*

**Следствие 1.19.** *Обычные многообразия (с обратимыми функциями перехода) имеют нулевую препятственность, и степень препятственности для них равна нулю, т. е. для них $n_\Pi = 0$.*

**Предположение 1.20.** *Препятственные полумногообразия могут также иметь ненулевое обычное препятствие которое может быть вычислено с помощью расширения общепринятых методов вычисления препятствий* [30, 215, 220] *на необратимый случай.*

Поэтому, используя степень препятственности $n_\Pi$, мы имеем возможность систематизировать полумногообразия должным образом.

В поиске аналогий мы можем сопоставить полусупермногообразия с суперчислами как в *Таблице* 1.1.

Далее учтем тот факт, что чистые духовые суперчисла существуют только при наличии нечетных направлений [174, 175, 225, 226].

*Таблица* 1.1

*Сравнительные типы суперчисел и полусупермногообразий*

| **Суперчисла** | **Полусупермногообразия** |
|---|---|
| Обыкновенные не равные нулю числа (обратимые) | Обыкновенные дифференцируемые многообразия (функции перехода обратимы) |
| Суперчисла, имеющие не обращающуюся в нуль числовую часть (обратимые) | Супермногообразия (функции перехода обратимы) |
| Чистые духовые суперчисла без числовой части (необратимые) | Препятственные полусупермногообразия (функции перехода необратимы) |

*Замечание* **1.21.** Препятственные полусупермногообразия имеют не равную нулю нечетную размерность.

Более того, очевидно чистые духовые суперчисла не содержат единицу.

*Замечание* **1.22.** Препятственные полусупермногообразия не могут иметь тождественных функций полуперехода.

Как возможные функции полуперехода для препятственных полусупермногообразий можно рассматривать преобразования, вращающие четность касательного пространства введенные в [1, 7, 9]. Объекты, полученные таким образом, могут быть рассмотрены как необратимые аналоги суперримановых поверхностей [111, 227, 228].

**1.2.4.** Полугруппа башенных тождеств. Рассмотрим ряд отображений $\mathrm{e}_{\alpha\alpha}^{(n)} : \mathscr{U}_\alpha \to \mathscr{U}_\alpha$ полумногообразия $\mathscr{M}$ в себя вида

$$\mathrm{e}_{\alpha\alpha}^{(1)} = \Phi_{\alpha\alpha}, \tag{1.32}$$

$$\mathrm{e}_{\alpha\alpha}^{(2)} = \Phi_{\alpha\beta} \circ \Phi_{\beta\alpha}, \tag{1.33}$$

$$\mathrm{e}_{\alpha\alpha}^{(3)} = \Phi_{\alpha\beta} \circ \Phi_{\beta\gamma} \circ \Phi_{\gamma\alpha}, \tag{1.34}$$

$$\mathrm{e}_{\alpha\alpha}^{(4)} = \Phi_{\alpha\beta} \circ \Phi_{\beta\gamma} \circ \Phi_{\gamma\delta} \circ \Phi_{\delta\alpha} \tag{1.35}$$

$$\dots \quad \dots$$

Мы будем называть $\mathrm{e}_{\alpha\alpha}^{(n)}$ *башенными тождествами*, которые вытекают из башенных соотношений (1.9)–(1.16).

Из формул (1.28)–(1.31) следует

**Утверждение 1.23.** *Для обычных супермногообразий все башенные тождества совпадают с обычным тождественным отображением*

$$\mathrm{e}_{\alpha\alpha}^{(n)} = 1_{\alpha\alpha}. \tag{1.36}$$

*Замечание* **1.24.** В тривиальном случае, когда все $\Phi_{\alpha\beta}$ являются тождественными отображениями, очевидно, что соотношения (1.32)–(1.35) удовлетворяются тождественно.

Степень препятственности может трактоваться в качестве максимального $n = n_{\Pi}$, для которой башенные тождества отличаются от тождества, т. е. соотношение (1.36) нарушено. Таким образом, башенные тождества задают меру отличия полусупермногообразия от обыкновенного супермногообразия. Будучи внутренней характеристикой, башенные тождества играют важную роль в описании полусупермногообразий [19].

Исследуем некоторый их свойства более подробно.

**Предложение 1.25.** *Башенные тождества являются единицами для функций полуперехода*

$$\mathrm{e}_{\alpha\alpha}^{(n)} \circ \Phi_{\alpha\beta} = \Phi_{\alpha\beta}, \tag{1.37}$$

$$\Phi_{\alpha\beta} \circ \mathrm{e}^{(n)}_{\beta\beta} = \Phi_{\alpha\beta}. \tag{1.38}$$

*Доказательство.* Следует прямо из соотношений (1.9)–(1.16) и определений (1.9)–(1.16). ■

**Предложение 1.26.** *Башенные тождества являются идемпотентами*

$$\mathrm{e}^{(n)}_{\alpha\alpha} \circ \mathrm{e}^{(n)}_{\alpha\alpha} = \mathrm{e}^{(n)}_{\alpha\alpha}. \tag{1.39}$$

*Доказательство.* Мы доказываем утверждение для $n = 2$ и для другого $n$ его можно доказать по индукции. Запишем (1.39) как

$$\mathrm{e}^{(2)}_{\alpha\alpha} \circ \mathrm{e}^{(2)}_{\alpha\alpha} = \mathrm{e}^{(2)}_{\alpha\alpha} \circ \Phi_{\alpha\beta} \circ \Phi_{\beta\alpha} = \left(\mathrm{e}^{(2)}_{\alpha\alpha} \circ \Phi_{\alpha\beta}\right) \circ \Phi_{\beta\alpha}.$$

Затем, используя (1.37), мы получаем

$$\left(\mathrm{e}^{(2)}_{\alpha\alpha} \circ \Phi_{\alpha\beta}\right) \circ \Phi_{\beta\alpha} = \Phi_{\alpha\beta} \circ \Phi_{\beta\alpha} = \mathrm{e}^{(2)}_{\alpha\alpha}.$$

■

Несуперсимметричные функциональные уравнения подобного вида были исследованы в [229].

**Определение 1.27.** *Сопряженные башенные тождества соответствуют тому же разбиению полусупермногообразия и состоят из функций полуперехода, взятых в противоположном порядке*

$$\tilde{\mathrm{e}}^{(1)}_{\alpha\alpha} = \mathrm{e}^{(1)}_{\alpha\alpha}, \tag{1.40}$$

$$\tilde{\mathrm{e}}^{(2)}_{\alpha\alpha} = \mathrm{e}^{(2)}_{\alpha\alpha}, \tag{1.41}$$

$$\tilde{\mathrm{e}}^{(3)}_{\alpha\alpha} = \Phi_{\alpha\gamma} \circ \Phi_{\gamma\beta} \circ \Phi_{\beta\alpha}, \tag{1.42}$$

$$\tilde{e}^{(4)}_{\alpha\alpha} = \Phi_{\alpha\delta} \circ \Phi_{\delta\gamma} \circ \Phi_{\gamma\beta} \circ \Phi_{\beta\alpha} \tag{1.43}$$

$$\dots \quad \dots$$

Сопряженные башенные тождества также играют роль башенных тождеств, но для условий рефлексивности (1.17)–(1.24).

По аналогии с (1.37)–(1.38) мы имеем

**Предложение 1.28.** *Сопряженныя башенные тождества являются рефлексивными единицами, но для функций полуперехода* $\Phi_{\beta\alpha}$

$$\tilde{e}^{(n)}_{\beta\beta} \circ \Phi_{\beta\alpha} = \Phi_{\beta\alpha}, \tag{1.44}$$

$$\Phi_{\beta\alpha} \circ \tilde{e}^{(n)}_{\alpha\alpha} = \Phi_{\beta\alpha}. \tag{1.45}$$

**Предложение 1.29.** *При одном и том же разбиении сопряженные башенные тождества аннулируют башенные тождества в следующем смысле*

$$e^{(n)}_{\alpha\alpha} \circ \tilde{e}^{(n)}_{\alpha\alpha} = e^{(2)}_{\alpha\alpha}. \tag{1.46}$$

*Доказательство.* Рассмотрим пример $n = 3$. Используя определения, мы выводим

$$e^{(3)}_{\alpha\alpha} \circ \tilde{e}^{(3)}_{\alpha\alpha} = \Phi_{\alpha\beta} \circ \Phi_{\beta\gamma} \circ \Phi_{\gamma\alpha} \circ \Phi_{\alpha\gamma} \circ \Phi_{\gamma\beta} \circ \Phi_{\beta\alpha}$$

$$= \Phi_{\alpha\beta} \circ \Phi_{\beta\gamma} \circ (\Phi_{\gamma\alpha} \circ \Phi_{\alpha\gamma}) \circ \Phi_{\gamma\beta} \circ \Phi_{\beta\alpha} = \Phi_{\alpha\beta} \circ \Phi_{\beta\gamma} \circ e^{(2)}_{\gamma\gamma} \circ \Phi_{\gamma\beta} \circ \Phi_{\beta\alpha}$$

$$= \Phi_{\alpha\beta} \circ (\Phi_{\beta\gamma} \circ \Phi_{\gamma\beta}) \circ \Phi_{\beta\alpha} = \Phi_{\alpha\beta} \circ e^{(2)}_{\beta\beta} \circ \Phi_{\beta\alpha} = \Phi_{\alpha\beta} \circ \Phi_{\beta\alpha} = e^{(2)}_{\alpha\alpha}.$$

Для остальных $n$ утверждение доказывается по индукции. ■

**Определение 1.30.** *Полусупермногообразие называется точным, если башенные тождества не зависят от разбиения.*

Умножение башенных тождеств для точного полусупермногообра-зия определяется следующим образом

$$\mathrm{e}_{\alpha\alpha}^{(n)} \circ \mathrm{e}_{\alpha\alpha}^{(m)} = \mathrm{e}_{\alpha\alpha}^{(n+m)}. \tag{1.47}$$

**Утверждение 1.31.** *Умножение* (1.47) *ассоциативно.*

Следовательно, мы можем дать

**Определение 1.32.** *Башенные тождества точного полусупермногообразия образуют башенную полугруппу относительно умножения* (1.47).

Таким образом, мы получили количественное описание внутренних свойств необратимости полусупермногообразий.

**Предположение 1.33.** *Введенная башенная полугруппа играет ту же роль для полусупермногообразий, что и фундаментальная группа для обыкновенных многообразий* [209, 230, 231].

**1.2.5.** Обобщенная регулярность и полукоммутативные диаграммы. Полученные выше построения имеют общее значение для любого числа необратимых отображений.

Расширение $n = 2$ коцикла, задаваемое (1.9), может быть рассмотрено как некоторая аналогия с регулярными [232–235] или псевдообратными [236] элементами в полугруппах [237–240] или обобщенными обратными в теории матриц [241–245] и в теории обобщенных инверсных морфизмов [246, 247].

Соотношения (1.10)–(1.16) с высшими $n$ могут рассматриваться как необратимый аналог регулярности для коциклов высшего порядка. Следовательно, по аналогии с (1.9)–(1.16), естественно сформулировать общее

**Определение 1.34.** *Отображение $\Phi_{\alpha\beta}$ называется n-регулярным, если*

*оно удовлетворяет условиям*

$$\overbrace{\Phi_{\alpha\beta} \circ \Phi_{\beta\gamma} \circ \ldots \circ \Phi_{\rho\alpha} \circ \Phi_{\alpha\beta}}^{n+1} = \Phi_{\alpha\beta} + permutations \tag{1.48}$$

*на пересечениях* $\overbrace{\mathscr{U}_\alpha \cap \mathscr{U}_\beta \cap \ldots \cap \mathscr{U}_\rho}^{n}$.

В этом определении формула (1.9) описывает 3-регулярные отображения, соотношения (1.10)–(1.12) соответствуют 4-регулярным отображениям, и (1.13)–(1.16) дают 5-регулярные отображения.

*Замечание* **1.35.** Очевидно, что 3-регулярность совпадает с обычной полугрупповой регулярностью [104, 204].

Иное определение $n$-регулярности может задаваться формулами (1.37)–(1.38). Условия регулярности высшего порядка существенно изменяют общий диаграммный метод для морфизмов, когда используются необратимые единицы*).

В самом деле, коммутативность диаграмм для обратимых морфизмов основана на зависимостях (1.28)–(1.31), т. е. на том факте, что башенные тождества являются в этом случае обычными тождествами (1.36). Когда морфизмы необратимы (полусупермногообразие имеет не обращающуюся в нуль препятственность), мы не можем “вернуться в ту же точку”, поскольку в общем случае $\mathrm{e}_{\alpha\alpha}^{(n)} \neq 1_{\alpha\alpha}$, и мы вынуждены рассматривать “незамкнутые” диаграммы из-за того факта, что соотношение $\mathrm{e}_{\alpha\alpha}^{(n)} \circ \Phi_{\alpha\beta} = \Phi_{\alpha\beta}$ теперь несократимо.

Подводя итог, мы предлагаем следующую интуитивно непротиворечивую замену стандартного диаграммного метода в применении к необратимым морфизмам [6, 19]. В каждом случае мы добавляем новую

---

*Примечание.* Отметим, что в несуперсимметричном случае похожая конструкция (“multiply wrapped cycles”) для многообразий Калаби-Яу рассматривалась в [248].

стрелку, которая соответствует дополнительному множителю в (1.37).

Таким образом, для $n = 2$ мы получаем обобщение диаграммного исчисления как на *Рис.* 1.1, что описывает переход от обратимого морфизма (1.29) к необратимому (1.9) и с абстрактной точки зрения представляет собой условие регулярности для морфизмов [246].

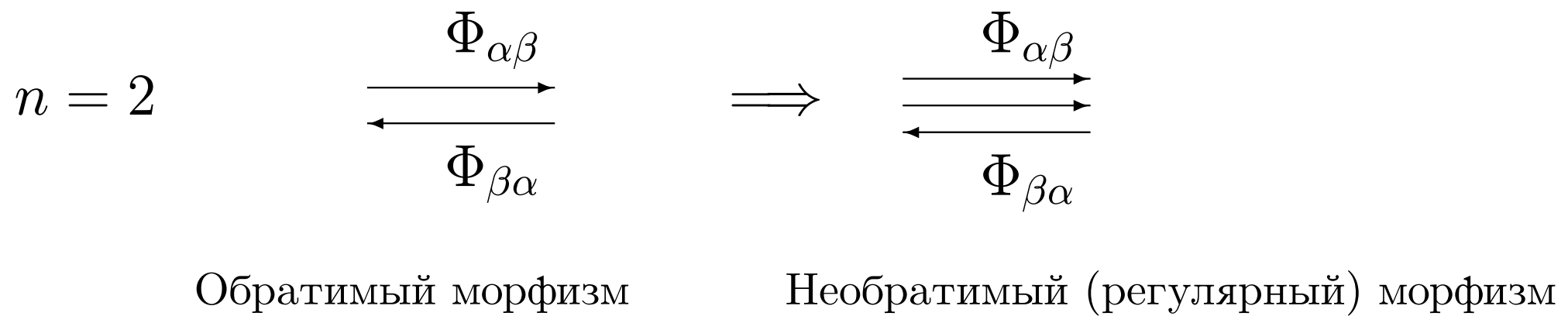

*Рис.* 1.1. *Переход от обратимого к необратимому морфизму при* $n = 2$

Более необычной полукоммутативной диаграммой является треугольная на *Рис.* 1.2, которая обобщает на необратимый случай условие коцикла (1.5).

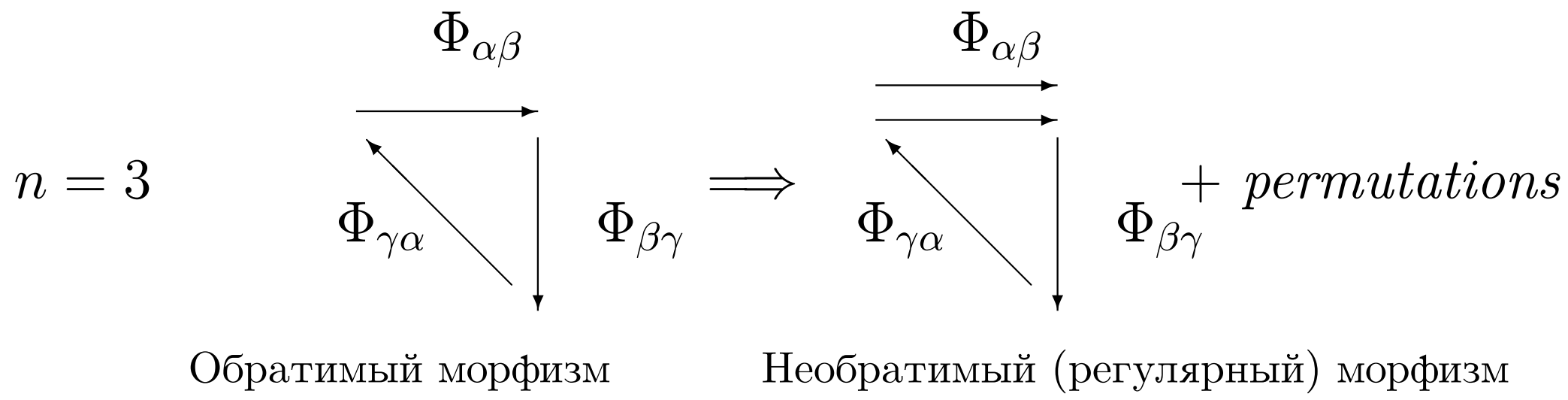

*Рис.* 1.2. *Обобщение условия коцикла на необратимый вариант составляющих морфизмов*

По аналогии мы можем представить полукоммутативные диаграммы для $n$-регулярности более высокого порядка, что можно рассмотреть также в рамках обобщенных категорий [154, 249–255].

## 1.3. Необратимость и полурасслоения

Подобный принцип замены обратимости морфизма на его регулярность может быть использован для необратимого расширения суперрасслоений [184, 256, 257], если определять их глобально на основе открытых покрытий и функций перехода [182, 258].

Следуя стандартным определениям расслоений [180, 223, 259], но ослабляя обратимость, построим новые объекты, аналогичные суперрасслоениям*).

**1.3.1.** Определение полурасслоений. Пусть $E$ и $\mathscr{M}$ представляют собой *полное* (*расслоенное*) суперпространство и *базовое* полусупермногообразие соответственно, и $\pi : E \to \mathscr{M}$ представляет собой *полупроективное отображение*, которое не обязательно обратимо (но может быть гладким). Обозначим $F_b$ множество точек $E$, которые отображаются в $b \in \mathscr{M}$ (*прообраз* $b$), т. е. *полуслой над* $b$ есть $F_b \stackrel{def}{=} \{x \in E \,|\, \pi(x) = b\}$. Тогда, $F = \cup F_b$ представляет собой *полуслой.*

**Определение 1.36.** *Полурасслоение определяется следующим набором* $\boldsymbol{\mathfrak{L}} \stackrel{def}{=} (E, \mathscr{M}, F, \pi)$.

*Сечение* $\boldsymbol{s} : \mathscr{M} \to F$ расслоения $(E, \mathscr{M}, F, \pi)$ обычно определяется соотношением $\pi(\boldsymbol{s}(b)) = b$, которое в виде $\pi \circ \boldsymbol{s} = 1_m$ весьма похоже на (1.4), (1.29) и выполняется тождественно только для обратимых отображений $\pi$ и $\boldsymbol{s}$. Следовательно, очень мало обыкновенных нетривиальных расслоений допускают соответствующие сечения [223].

Таким образом, используя аналогию с (1.9), мы приходим к следующему определению [6].

**Определение 1.37.** *Полусечение* $\boldsymbol{s}$ *полурасслоения* $\boldsymbol{\mathfrak{L}} = (E, \mathscr{M}, F, \pi)$

---

*Примечание.* В дальнейшем мы будем отбрасывать “супер”, если это не влияет существенно на ход рассуждений.

*определяется уравнением*

$$\pi \circ \boldsymbol{s} \circ \pi = \pi. \tag{1.49}$$

*Рефлексивное полусечение* $\boldsymbol{s}_{refl}$ *удовлетворяет дополнительному условию*

$$\boldsymbol{s}_{refl} \circ \pi \circ \boldsymbol{s}_{refl} = \boldsymbol{s}_{refl}. \tag{1.50}$$

Пусть $\tilde{\pi} : \mathscr{M} \times F \to \mathscr{M}$ представляет собой канонический индекс полуоператора проектирования на первый множитель $\tilde{\pi}(b, f) = b,\ f \in F$, тогда $\tilde{\pi}$ приводит к *расслоению-произведению.* Если $\lambda : E \to \mathscr{M} \times F$ представляет собой морфизм (называемый *тривиализацией*), тогда $\tilde{\pi} \circ \lambda = \pi$, и полурасслоение $\mathfrak{L} = (E, \mathscr{M}, F, \pi)$ является *тривиальным.* Если существует непрерывное отображение $\eta : \mathscr{M} \to F$, тогда полурасслоение $(\mathscr{M} \times F, \mathscr{M}, F, \tilde{\pi})$ допускает сечение $\boldsymbol{s} : \mathscr{M} \to \mathscr{M} \times F$ заданное формулой $\boldsymbol{s}(b) = (b, \eta(b))$.

Пусть для заданной суперобласти $\mathscr{U}_\alpha$ в полусупермногообразии имеем соответственную суперобласть в базе

$$E_\alpha \stackrel{def}{=} \{x \in E \,|\, \pi_\alpha(x) = b,\, b \in \mathscr{U}_\alpha \subset \mathscr{M}\}$$

(здесь мы намеренно не используем стандартное обозначение $\pi^{-1}(\mathscr{U}_\alpha)$ для $E_\alpha$, так как теперь допускается, чтобы $\pi_\alpha$ было необратимым), где $\pi_\alpha : E_\alpha \to \mathscr{U}_\alpha$ представляет собой сужение отображения $\pi$ на суперобласть $\mathscr{U}_\alpha$, т. е. $\pi_\alpha \stackrel{def}{=} \pi\,|_{\mathscr{U}_\alpha}$.

**Определение 1.38.** *Полурасслоение, определяемое* $\mathfrak{L} = (E, M, F, \pi)$, *называется локально тривиальным, если* $\forall b \in \mathscr{M}$ *существуют суперобласти* $\mathscr{U}_\alpha \ni b$ *такие, что можно найти тривиализирующие морфизмы* $\lambda_\alpha : E_\alpha \to \mathscr{U}_\alpha \times F$ *удовлетворяющие* $\tilde{\pi}_\alpha \circ \lambda_\alpha = \pi_\alpha$.

Так что диаграмма

$$
\begin{array}{ccc}
E_\alpha & \xrightarrow{\lambda_\alpha} & \mathscr{U}_\alpha \times F \\
 & \searrow^{\pi_\alpha} & \downarrow \tilde{\pi}_\alpha \\
 & & \mathscr{U}_\alpha
\end{array}
\tag{1.51}
$$

коммутирует.

**Определение 1.39.** *Полусечение локально тривиального полурасслоения* $\mathfrak{L}$ *дается отображениями* $\boldsymbol{s}_\alpha : \mathscr{U}_\alpha \to E$, *которые удовлетворяют условиям совместимости*

$$
\lambda_\alpha \circ \boldsymbol{s}_\alpha \mid_b = \lambda_\beta \circ \boldsymbol{s}_\beta \mid_b, \;\; b \in \mathscr{U}_\alpha \cap \mathscr{U}_\beta. \tag{1.52}
$$

Теперь пусть $\{\mathscr{U}_\alpha, \lambda_\alpha\}$ представляет собой тривиализирующее покрытие такое, что $\cup \mathscr{U}_\alpha = \mathscr{M}$ и $\mathscr{U}_\alpha \cap \mathscr{U}_\beta \neq \varnothing \Rightarrow E_\alpha \cap E_\beta \neq \varnothing$. Тогда мы требуем, чтобы тривиализирующие морфизмы $\lambda_\alpha$ находились в соответствии, и это значит, что диаграммы

$$
\begin{array}{ccc}
E_\alpha \cap E_\beta & \xrightarrow{\lambda_\beta} & \mathscr{U}_\alpha \cap \mathscr{U}_\beta \times F \\
 & \searrow^{\lambda_\alpha} & \downarrow \Lambda_{\alpha\beta} \\
 & & \mathscr{U}_\alpha \cap \mathscr{U}_\beta \times F
\end{array}
\tag{1.53}
$$

и

$$
\begin{array}{ccc}
E_\alpha \cap E_\beta & \xrightarrow{\lambda_\beta} & \mathscr{U}_\alpha \cap \mathscr{U}_\beta \times F \\
 & \searrow^{\lambda_\alpha} & \uparrow \Lambda_{\beta\alpha} \\
 & & \mathscr{U}_\alpha \cap \mathscr{U}_\beta \times F
\end{array}
\tag{1.54}
$$

должны коммутировать,где $\Lambda_{\alpha\beta}$ и $\Lambda_{\beta\alpha}$ — отображения, действующие вдоль полуслоя $F$.

**Определение 1.40.** *Склеивающие функции полуперехода локально тривиального полурасслоения* $\mathfrak{L} = (E, \mathscr{M}, F, \pi)$ *определяются уравнени-*

*ями*

$$\Lambda_{\alpha\beta} \circ \lambda_\beta = \lambda_\alpha, \tag{1.55}$$

$$\Lambda_{\beta\alpha} \circ \lambda_\alpha = \lambda_\beta. \tag{1.56}$$

**Утверждение 1.41.** *Функции полуперехода полурасслоения $\mathcal{L}$ удовлетворяют следующим соотношениям*

$$\Lambda_{\alpha\beta} \circ \Lambda_{\beta\alpha} \circ \Lambda_{\alpha\beta} = \Lambda_{\alpha\beta} \tag{1.57}$$

*на $\mathcal{U}_\alpha \cap \mathcal{U}_\beta$ пересечениях и*

$$\Lambda_{\alpha\beta} \circ \Lambda_{\beta\gamma} \circ \Lambda_{\gamma\alpha} \circ \Lambda_{\alpha\beta} = \Lambda_{\alpha\beta}, \tag{1.58}$$

$$\Lambda_{\beta\gamma} \circ \Lambda_{\gamma\alpha} \circ \Lambda_{\alpha\beta} \circ \Lambda_{\beta\gamma} = \Lambda_{\beta\gamma}, \tag{1.59}$$

$$\Lambda_{\gamma\alpha} \circ \Lambda_{\alpha\beta} \circ \Lambda_{\beta\gamma} \circ \Lambda_{\gamma\alpha} = \Lambda_{\gamma\alpha} \tag{1.60}$$

*на тройных пересечениях $\mathcal{U}_\alpha \cap \mathcal{U}_\beta \cap \mathcal{U}_\gamma$ и*

$$\Lambda_{\alpha\beta} \circ \Lambda_{\beta\gamma} \circ \Lambda_{\gamma\rho} \circ \Lambda_{\rho\alpha} \circ \Lambda_{\alpha\beta} = \Lambda_{\alpha\beta}, \tag{1.61}$$

$$\Lambda_{\beta\gamma} \circ \Lambda_{\gamma\rho} \circ \Lambda_{\rho\alpha} \circ \Lambda_{\alpha\beta} \circ \Lambda_{\beta\gamma} = \Lambda_{\beta\gamma}, \tag{1.62}$$

$$\Lambda_{\gamma\rho} \circ \Lambda_{\rho\alpha} \circ \Lambda_{\alpha\beta} \circ \Lambda_{\beta\gamma} \circ \Lambda_{\gamma\rho} = \Lambda_{\gamma\rho}, \tag{1.63}$$

$$\Lambda_{\rho\alpha} \circ \Lambda_{\alpha\beta} \circ \Lambda_{\beta\gamma} \circ \Lambda_{\gamma\rho} \circ \Lambda_{\rho\alpha} = \Lambda_{\rho\alpha} \tag{1.64}$$

*на $\mathcal{U}_\alpha \cap \mathcal{U}_\beta \cap \mathcal{U}_\gamma \cap \mathcal{U}_\rho$.*

**Определение 1.42.** *Полурасслоение $\mathfrak{L}$ называется рефлексивным, если, в дополнение к* (1.57)–(1.64)*, функции полуперехода удовлетворяют усло-*

*виям рефлексивности*

$$\Lambda_{\beta\alpha} \circ \Lambda_{\alpha\beta} \circ \Lambda_{\beta\alpha} = \Lambda_{\beta\alpha} \tag{1.65}$$

*на* $\mathscr{U}_\alpha \cap \mathscr{U}_\beta$ *пересечениях и*

$$\Lambda_{\alpha\gamma} \circ \Lambda_{\gamma\beta} \circ \Lambda_{\beta\alpha} \circ \Lambda_{\alpha\gamma} = \Lambda_{\alpha\gamma}, \tag{1.66}$$

$$\Lambda_{\gamma\beta} \circ \Lambda_{\beta\alpha} \circ \Lambda_{\alpha\gamma} \circ \Lambda_{\gamma\beta} = \Lambda_{\gamma\beta}, \tag{1.67}$$

$$\Lambda_{\beta\alpha} \circ \Lambda_{\alpha\gamma} \circ \Lambda_{\gamma\beta} \circ \Lambda_{\beta\alpha} = \Lambda_{\beta\alpha} \tag{1.68}$$

*на тройных пересечениях* $\mathscr{U}_\alpha \cap \mathscr{U}_\beta \cap \mathscr{U}_\gamma$ *и*

$$\Lambda_{\alpha\rho} \circ \Lambda_{\rho\gamma} \circ \Lambda_{\gamma\beta} \circ \Lambda_{\beta\alpha} \circ \Lambda_{\alpha\rho} = \Lambda_{\alpha\rho}, \tag{1.69}$$

$$\Lambda_{\rho\gamma} \circ \Lambda_{\gamma\beta} \circ \Lambda_{\beta\alpha} \circ \Lambda_{\alpha\rho} \circ \Lambda_{\rho\gamma} = \Lambda_{\rho\gamma}, \tag{1.70}$$

$$\Lambda_{\gamma\beta} \circ \Lambda_{\beta\alpha} \circ \Lambda_{\alpha\rho} \circ \Lambda_{\rho\gamma} \circ \Lambda_{\gamma\beta} = \Lambda_{\gamma\beta}, \tag{1.71}$$

$$\Lambda_{\beta\alpha} \circ \Lambda_{\alpha\rho} \circ \Lambda_{\rho\gamma} \circ \Lambda_{\gamma\beta} \circ \Lambda_{\beta\alpha} = \Lambda_{\beta\alpha} \tag{1.72}$$

*на* $\mathscr{U}_\alpha \cap \mathscr{U}_\beta \cap \mathscr{U}_\gamma \cap \mathscr{U}_\rho$.

Для заданного $b \in \mathscr{U}_\alpha \cap \mathscr{U}_\beta$ склеивающие функции перехода $\Lambda_{\alpha\beta}$ описывают морфизмы полуслоя $F$ в себя условием

$$\Lambda_{\alpha\beta} : (b, f) \to (b, L_{\alpha\beta} f)\,, \tag{1.73}$$

где $L_{\alpha\beta} : \mathscr{U}_\alpha \cap \mathscr{U}_\beta \to F$ и $f \in F$. Функции $L_{\alpha\beta}$ удовлетворяют обобщенным условиям коцикла аналогичного (1.57)–(1.72).

*Замечание* **1.43.** Сечения и функции перехода расслоения необратимы даже в стандартном случае [260, 261]. Но такой вид необратимости имеет природу, отличную от той, которая может иметь место в супер-симметричных объектах.

Это можно сравнить с необратимостью обычных функций [192, 262] и необратимостью суперфункций, что имеет место из-за присутствия нильпотентов и делителей нуля. Подразумевается, что стандартные функции перехода должны быть гомеоморфизмами, а сечения должны быть во взаимооднозначном соответствии[*)] с отображениями из базы в слой [267,268]. Наши определения (1.9)–(1.24) и (1.49)–(1.72) расширяют их, допуская включение в рассмотрение должным образом также и необратимые суперфункции.

**1.3.2.** Морфизмы полурасслоений. Пусть мы имеем два полурасслоения $\mathfrak{L} = (E, \mathscr{M}, F, \pi)$ и $\mathfrak{L}' = (E', \mathscr{M}', F', \pi')$.

**Определение 1.44.** *Морфизм полурасслоения* $\mathbf{f} : \mathfrak{L} \to \mathfrak{L}'$ *состоит из двух морфизмов* $\mathbf{f} = (\mathrm{f}_E, \mathrm{f}_{\mathscr{M}})$, *где* $\mathrm{f}_E : E \to E'$ *и* $\mathrm{f}_{\mathscr{M}} : \mathscr{M} \to \mathscr{M}'$, *удовлетворяют* $\mathrm{f}_{\mathscr{M}} \circ \pi = \pi' \circ \mathrm{f}_E$, *так что диаграмма*

$$\begin{array}{ccc} E & \xrightarrow{\mathrm{f}_E} & E' \\ \pi \downarrow & & \downarrow \pi' \\ \mathscr{M} & \xrightarrow{\mathrm{f}_{\mathscr{M}}} & \mathscr{M}' \end{array} \qquad (1.74)$$

*коммутативна.*

Пусть

$$E_b = \{x \in E \mid \pi(x) = b,\, b \in \mathscr{U} \subset \mathscr{M}\},$$

тогда $\mathrm{f}_E(E_b) \subset E'_{\mathrm{f}_{\mathscr{M}}(b)}$ для каждого $b$, и полуслой над $b \in \mathscr{M}$ переносится в полуслой над $\mathrm{f}_{\mathscr{M}}(b) \in \mathscr{M}'$, так, что $\mathrm{f}_E$ представляет собой морфизм слоя. Если полурасслоение имеет сечение (что может быть не все-

---

*Примечание.* Интересные примеры невзаимооднозначных (несуперсимметричных) отображений и диффеоморфизмов приведены в [263–266].

гда), то морфизм $\mathrm{f}_E$ действует следующим образом $\boldsymbol{s}(b) \to \boldsymbol{s}'(\mathrm{f}_{\mathscr{M}}(b))$.

В большинстве приложений расслоенных пространств морфизм $\mathrm{f}_{\mathscr{M}}$ есть тождество, и $\mathbf{f}_0 \stackrel{def}{=} (\mathrm{f}_E, \mathrm{id})$ называется *b-морфизмом* [260]. Тем не менее, в случае полурасслоений может иметь место обратная ситуация, когда $\mathrm{f}_{\mathscr{M}}$ представляет собой необратимый морфизм.

Для каждого заданного $b \in \mathscr{M}$ существуют тривиализирующие отображения $\lambda : E_b \to \mathscr{U} \times F$ и $\lambda' : E_{\mathrm{f}_{\mathscr{M}}(b)} \to \mathscr{U}' \times F'$, $\mathrm{f}_{\mathscr{M}}(\mathscr{U}) \subset \mathscr{U}'$, которые приводят к отображению полуслоя $h_b$, определяемого коммутативной диаграммой

$$\begin{array}{ccc} E_b & \xrightarrow{\mathrm{f}_E(b)} & E'_{\mathrm{f}_{\mathscr{M}}(b)} \\ \lambda \downarrow & & \downarrow \lambda' \\ \mathscr{U} \times F & \xrightarrow{h_b} & \mathscr{U}' \times F' \end{array} \qquad (1.75)$$

Чтобы локально описать морфизм полурасслоений $\mathfrak{L} \xrightarrow{\mathbf{f}} \mathfrak{L}'$, мы выбираем открытые покрытия $\mathscr{M} = \cup \mathscr{U}_\alpha$ и $\mathscr{M}' = \cup \mathscr{U}'_{\alpha'}$ наряду с тривиализациями $\lambda_\alpha$ и $\lambda'_{\alpha'}$ (см. (1.51)). Тогда связь между функциями полуперехода $\Lambda_{\alpha\beta}$ и $\Lambda'_{\alpha'\beta'}$ (1.55)–(1.56) двух полурасслоений $\mathfrak{L}$ и $\mathfrak{L}'$ может быть найдена из коммутативной диаграммы

$$\begin{array}{ccc} \mathscr{U}_{\alpha\beta} \times F & \xrightarrow{\Lambda_{\alpha\beta}} & \mathscr{U}_{\alpha\beta} \times F \\ h_\alpha \downarrow & & \downarrow h_\beta \\ \mathscr{U}'_{\alpha'\beta'} \times F' & \xrightarrow{\Lambda'_{\alpha'\beta'}} & \mathscr{U}'_{\alpha'\beta'} \times F' \end{array} \qquad (1.76)$$

где морфизмы $h_\alpha$ определяются диаграммой

$$\begin{array}{ccc} E & \xrightarrow{\mathrm{f}_E} & E' \\ \lambda_\alpha \downarrow & & \downarrow \lambda'_{\alpha'} \\ \mathscr{U}_\alpha \times F & \xrightarrow{h_\alpha} & \mathscr{U}'_{\alpha'} \times F' \end{array} \qquad (1.77)$$

Из (1.76) мы имеем соотношение между функциями полуперехода

$$h_\alpha \circ \Lambda_{\alpha\beta} = \Lambda'_{\alpha'\beta'} \circ h_\beta \tag{1.78}$$

которое выполняется тождественно также и для необратимых $h_\alpha$, тогда как в обратимом случае [260,261] уравнение (1.78) решается относительно $\Lambda'_{\alpha'\beta'}$ стандартным образом $\Lambda'_{\alpha'\beta'} = h_\alpha \circ \Lambda_{\alpha\beta} \circ h_\beta^{-1}$, что может рассматриваться как эквивалентность коциклов. Однако в общем случае (1.78) представляет собой систему суперуравнений, которые должны решаться стандартными [30] либо расширенными [269] методами суперанализа [91].

Предположим, $\mathscr{M}$ допускает тривиализирующие покрытия $\{\mathscr{U}_\alpha, \lambda_\alpha\}$ и $\{\mathscr{U}'_{\alpha'}, \lambda'_{\alpha'}\}$. В общем случае они не связаны между собой, и функции полуперехода $\Lambda_{\alpha\beta}$ и $\Lambda'_{\alpha'\beta'}$ независимы. Однако, если $\mathscr{M}$ представляет собой базовое суперпространство для двух полурасслоений $\boldsymbol{\mathfrak{L}}$ и $\boldsymbol{\mathfrak{L}}'$, которые связаны $b$-морфизмом $\boldsymbol{\mathfrak{L}} \stackrel{\mathbf{f}_0}{\to} \boldsymbol{\mathfrak{L}}'$, тогда $\Lambda_{\alpha\beta}$ и $\Lambda'_{\alpha'\beta'}$ должны находиться в соответствии.

**Предложение 1.45.** *Функции полуперехода $\Lambda_{\alpha\beta}$ и $\Lambda'_{\alpha'\beta'}$ двух полурасслоений находятся в соответствии, если существуют дополнительные отображения $\tilde{\Lambda}_{\alpha'\beta} : \mathscr{U}'_{\alpha'} \cap \mathscr{U}_\beta$ и $\tilde{\Lambda}_{\alpha\beta'} : \mathscr{U}_\alpha \cap \mathscr{U}'_{\beta'}$ связанные между собой соотношениями*

$$\tilde{\Lambda}_{\alpha'\beta} \circ \tilde{\Lambda}_{\beta\alpha'} \circ \tilde{\Lambda}_{\alpha'\beta} = \tilde{\Lambda}_{\alpha'\beta} \tag{1.79}$$

*на $\mathscr{U}'_{\alpha'} \cap \mathscr{U}_\beta$ и*

$$\tilde{\Lambda}_{\alpha\beta'} \circ \tilde{\Lambda}_{\beta'\alpha} \circ \tilde{\Lambda}_{\alpha\beta'} = \tilde{\Lambda}_{\alpha\beta'} \tag{1.80}$$

*на $\mathscr{U}_\alpha \cap \mathscr{U}'_{\beta'}$ пересечениях.*

*Условия соответствия для* $\Lambda_{\alpha\beta}$ *и* $\Lambda'_{\alpha'\beta'}$ *имеют вид*

$$\tilde{\Lambda}_{\alpha'\beta} \circ \Lambda_{\beta\gamma} \circ \tilde{\Lambda}_{\gamma\alpha'} \circ \tilde{\Lambda}_{\alpha'\beta} = \tilde{\Lambda}_{\alpha'\beta}, \tag{1.81}$$

$$\Lambda_{\beta\gamma} \circ \tilde{\Lambda}_{\gamma\alpha'} \circ \tilde{\Lambda}_{\alpha'\beta} \circ \Lambda_{\beta\gamma} = \Lambda_{\beta\gamma}, \tag{1.82}$$

$$\tilde{\Lambda}_{\gamma\alpha'} \circ \tilde{\Lambda}_{\alpha'\beta} \circ \Lambda_{\beta\gamma} \circ \tilde{\Lambda}_{\gamma\alpha'} = \tilde{\Lambda}_{\gamma\alpha'} \tag{1.83}$$

*на тройных пересечениях* $\mathscr{U}'_{\alpha'} \cap \mathscr{U}_{\beta} \cap \mathscr{U}_{\gamma}$ *и*

$$\Lambda'_{\alpha'\beta'} \circ \tilde{\Lambda}_{\beta'\gamma} \circ \tilde{\Lambda}_{\gamma\alpha'} \circ \Lambda'_{\alpha'\beta'} = \Lambda'_{\alpha'\beta'}, \tag{1.84}$$

$$\tilde{\Lambda}_{\beta'\gamma} \circ \tilde{\Lambda}_{\gamma\alpha'} \circ \Lambda'_{\alpha'\beta'} \circ \tilde{\Lambda}_{\beta'\gamma} = \tilde{\Lambda}_{\beta'\gamma}, \tag{1.85}$$

$$\tilde{\Lambda}_{\gamma\alpha'} \circ \Lambda'_{\alpha'\beta'} \circ \tilde{\Lambda}_{\beta'\gamma} \circ \tilde{\Lambda}_{\gamma\alpha'} = \tilde{\Lambda}_{\gamma\alpha'} \tag{1.86}$$

*на* $\mathscr{U}'_{\alpha'} \cap \mathscr{U}'_{\beta'} \cap \mathscr{U}_{\gamma}$ *пересечениях.*

*Тогда*

$$\tilde{\Lambda}_{\alpha'\beta} \circ \Lambda_{\beta\gamma} \circ \Lambda_{\gamma\rho} \circ \tilde{\Lambda}_{\rho\alpha'} \circ \tilde{\Lambda}_{\alpha'\beta} = \tilde{\Lambda}_{\alpha'\beta}, \tag{1.87}$$

$$\Lambda_{\beta\gamma} \circ \Lambda_{\gamma\rho} \circ \tilde{\Lambda}_{\rho\alpha'} \circ \tilde{\Lambda}_{\alpha'\beta} \circ \Lambda_{\beta\gamma} = \Lambda_{\beta\gamma}, \tag{1.88}$$

$$\Lambda_{\gamma\rho} \circ \tilde{\Lambda}_{\rho\alpha'} \circ \tilde{\Lambda}_{\alpha'\beta} \circ \Lambda_{\beta\gamma} \circ \Lambda_{\gamma\rho} = \Lambda_{\gamma\rho}, \tag{1.89}$$

$$\tilde{\Lambda}_{\rho\alpha'} \circ \tilde{\Lambda}_{\alpha'\beta} \circ \Lambda_{\beta\gamma} \circ \Lambda_{\gamma\rho} \circ \tilde{\Lambda}_{\rho\alpha'} = \tilde{\Lambda}_{\rho\alpha'} \tag{1.90}$$

*на* $\mathscr{U}'_{\alpha'} \cap \mathscr{U}_{\beta} \cap \mathscr{U}_{\gamma} \cap \mathscr{U}_{\rho}$ *и*

$$\Lambda'_{\alpha'\beta'} \circ \tilde{\Lambda}_{\beta'\gamma} \circ \Lambda_{\gamma\rho} \circ \tilde{\Lambda}_{\rho\alpha'} \circ \tilde{\Lambda}_{\alpha\beta'} = \Lambda'_{\alpha'\beta'}, \tag{1.91}$$

$$\tilde{\Lambda}_{\beta'\gamma} \circ \Lambda_{\gamma\rho} \circ \tilde{\Lambda}_{\rho\alpha'} \circ \Lambda'_{\alpha'\beta'} \circ \tilde{\Lambda}_{\beta'\gamma} = \tilde{\Lambda}_{\beta'\gamma}, \tag{1.92}$$

$$\Lambda_{\gamma\rho} \circ \tilde{\Lambda}_{\rho\alpha'} \circ \Lambda'_{\alpha'\beta'} \circ \tilde{\Lambda}_{\beta'\gamma} \circ \Lambda_{\gamma\rho} = \Lambda_{\gamma\rho}, \tag{1.93}$$

$$\tilde{\Lambda}_{\rho\alpha'} \circ \Lambda'_{\alpha'\beta'} \circ \tilde{\Lambda}_{\beta'\gamma} \circ \Lambda_{\gamma\rho} \circ \tilde{\Lambda}_{\rho\alpha'} = \tilde{\Lambda}_{\rho\alpha'} \tag{1.94}$$

*на* $\mathscr{U}'_{\alpha'} \cap \mathscr{U}'_{\beta'} \cap \mathscr{U}_\gamma \cap \mathscr{U}_\rho$ *и*

$$\Lambda'_{\alpha'\beta'} \circ \Lambda'_{\beta'\gamma'} \circ \tilde{\Lambda}_{\gamma'\rho} \circ \tilde{\Lambda}_{\rho\alpha'} \circ \Lambda'_{\alpha'\beta'} = \Lambda'_{\alpha'\beta'}, \tag{1.95}$$

$$\Lambda'_{\beta'\gamma'} \circ \tilde{\Lambda}_{\gamma'\rho} \circ \tilde{\Lambda}_{\rho\alpha'} \circ \Lambda'_{\alpha'\beta'} \circ \tilde{\Lambda}_{\beta'\gamma} = \tilde{\Lambda}_{\beta'\gamma}, \tag{1.96}$$

$$\tilde{\Lambda}_{\gamma'\rho} \circ \tilde{\Lambda}_{\rho\alpha'} \circ \Lambda'_{\alpha'\beta'} \circ \Lambda'_{\beta'\gamma'} \circ \tilde{\Lambda}_{\gamma'\rho} = \tilde{\Lambda}_{\gamma'\rho}, \tag{1.97}$$

$$\tilde{\Lambda}_{\rho\alpha'} \circ \Lambda'_{\alpha'\beta'} \circ \Lambda'_{\beta'\gamma'} \circ \tilde{\Lambda}_{\gamma'\rho} \circ \tilde{\Lambda}_{\rho\alpha'} = \tilde{\Lambda}_{\rho\alpha'} \tag{1.98}$$

*на* $\mathscr{U}'_{\alpha'} \cap \mathscr{U}'_{\beta'} \cap \mathscr{U}'_{\gamma'} \cap \mathscr{U}_\rho$.

*Доказательство.* Конструируем сумму тривиализирующих покрытий $\{\mathscr{U}_\alpha, \lambda_\alpha\}$ и $\{\mathscr{U}'_{\alpha'}, \lambda'_{\alpha'}\}$, а затем используем (1.57)–(1.64). ■

**Предложение 1.46.** *Функции полуперехода* $\Lambda_{\alpha\beta}$ *и* $\Lambda'_{\alpha'\beta'}$ *рефлексивно находятся в соответствии, если существуют дополнительные рефлексивные отображения* $\tilde{\Lambda}_{\alpha'\beta} : \mathscr{U}'_{\alpha'} \cap \mathscr{U}_\beta$ *и* $\tilde{\Lambda}_{\alpha\beta'} : \mathscr{U}_\alpha \cap \mathscr{U}'_{\beta'}$ *связанные между собой (в дополнение к* (1.79)–(1.80)) *рефлексивными отношениями*

$$\tilde{\Lambda}_{\beta\alpha'} \circ \tilde{\Lambda}_{\alpha'\beta} \circ \tilde{\Lambda}_{\beta\alpha'} = \tilde{\Lambda}_{\beta\alpha'} \tag{1.99}$$

*на* $\mathscr{U}'_{\alpha'} \cap \mathscr{U}_\beta$ *и*

$$\tilde{\Lambda}_{\beta'\alpha} \circ \tilde{\Lambda}_{\alpha\beta'} \circ \tilde{\Lambda}_{\beta'\alpha} = \tilde{\Lambda}_{\beta'\alpha} \tag{1.100}$$

*на* $\mathscr{U}_\alpha \cap \mathscr{U}'_{\beta'}$ *пересечениях.*

*Рефлексивные функции полуперехода* $\Lambda_{\alpha\beta}$ *и* $\Lambda'_{\alpha'\beta'}$ *должны удовлетворять (в дополнение к* (1.81)–(1.98)) *следующим соотношениям рефлексивной согласованности*

$$\tilde{\Lambda}_{\alpha'\gamma} \circ \Lambda_{\gamma\beta} \circ \tilde{\Lambda}_{\beta\alpha'} \circ \tilde{\Lambda}_{\alpha'\gamma} = \tilde{\Lambda}_{\alpha'\gamma}, \tag{1.101}$$

$$\Lambda_{\gamma\beta} \circ \tilde{\Lambda}_{\beta\alpha'} \circ \tilde{\Lambda}_{\alpha'\gamma} \circ \Lambda_{\gamma\beta} = \Lambda_{\gamma\beta}, \tag{1.102}$$

$$\tilde{\Lambda}_{\beta\alpha'} \circ \tilde{\Lambda}_{\alpha'\gamma} \circ \Lambda_{\gamma\beta} \circ \tilde{\Lambda}_{\beta\alpha'} = \tilde{\Lambda}_{\beta\alpha'} \tag{1.103}$$

*на* $\mathscr{U}'_{\alpha'} \cap \mathscr{U}_\beta \cap \mathscr{U}_\gamma$ *и*

$$\tilde{\Lambda}_{\alpha'\gamma} \circ \tilde{\Lambda}_{\gamma\beta'} \circ \Lambda'_{\beta'\alpha'} \circ \tilde{\Lambda}_{\alpha'\gamma} = \tilde{\Lambda}_{\alpha'\gamma}, \tag{1.104}$$

$$\tilde{\Lambda}_{\gamma\beta'} \circ \Lambda'_{\beta'\alpha'} \circ \tilde{\Lambda}_{\alpha'\gamma} \circ \tilde{\Lambda}_{\gamma\beta'} = \tilde{\Lambda}_{\gamma\beta'}, \tag{1.105}$$

$$\Lambda'_{\beta'\alpha'} \circ \tilde{\Lambda}_{\alpha'\gamma} \circ \Lambda'_{\gamma\beta'} \circ \Lambda'_{\beta'\alpha'} = \Lambda'_{\beta'\alpha'} \tag{1.106}$$

*на* $\mathscr{U}'_{\alpha'} \cap \mathscr{U}'_{\beta'} \cap \mathscr{U}_\gamma$ *пересечениях.*

*Тогда*

$$\tilde{\Lambda}_{\alpha'\rho} \circ \Lambda_{\rho\gamma} \circ \Lambda_{\gamma\beta} \circ \tilde{\Lambda}_{\beta\alpha'} \circ \tilde{\Lambda}_{\alpha'\rho} = \tilde{\Lambda}_{\alpha'\rho}, \tag{1.107}$$

$$\Lambda_{\rho\gamma} \circ \Lambda_{\gamma\beta} \circ \tilde{\Lambda}_{\beta\alpha'} \circ \tilde{\Lambda}_{\alpha'\rho} \circ \Lambda_{\rho\gamma} = \Lambda_{\rho\gamma}, \tag{1.108}$$

$$\Lambda_{\gamma\beta} \circ \tilde{\Lambda}_{\beta\alpha'} \circ \tilde{\Lambda}_{\alpha'\rho} \circ \Lambda_{\rho\gamma} \circ \Lambda_{\gamma\beta} = \Lambda_{\gamma\beta}, \tag{1.109}$$

$$\tilde{\Lambda}_{\beta\alpha'} \circ \tilde{\Lambda}_{\alpha'\rho} \circ \Lambda_{\rho\gamma} \circ \Lambda_{\gamma\beta} \circ \tilde{\Lambda}_{\beta\alpha'} = \tilde{\Lambda}_{\beta\alpha'} \tag{1.110}$$

*на* $\mathscr{U}'_{\alpha'} \cap \mathscr{U}_\beta \cap \mathscr{U}_\gamma \cap \mathscr{U}_\rho$ *и*

$$\tilde{\Lambda}_{\alpha'\rho} \circ \Lambda_{\rho\gamma} \circ \tilde{\Lambda}_{\gamma\beta'} \circ \Lambda'_{\beta'\alpha'} \circ \tilde{\Lambda}_{\alpha'\rho} = \tilde{\Lambda}_{\alpha'\rho}, \tag{1.111}$$

$$\Lambda_{\rho\gamma} \circ \tilde{\Lambda}_{\gamma\beta'} \circ \Lambda'_{\beta'\alpha'} \circ \tilde{\Lambda}_{\alpha'\rho} \circ \Lambda_{\rho\gamma} = \Lambda_{\rho\gamma}, \tag{1.112}$$

$$\tilde{\Lambda}_{\gamma\beta'} \circ \Lambda'_{\beta'\alpha'} \circ \tilde{\Lambda}_{\alpha'\rho} \circ \Lambda_{\rho\gamma} \circ \tilde{\Lambda}_{\gamma\beta'} = \tilde{\Lambda}_{\gamma\beta'}, \tag{1.113}$$

$$\Lambda'_{\beta'\alpha'} \circ \tilde{\Lambda}_{\alpha'\rho} \circ \Lambda_{\rho\gamma} \circ \tilde{\Lambda}_{\gamma\beta'} \circ \Lambda'_{\beta'\alpha'} = \Lambda'_{\beta'\alpha'} \tag{1.114}$$

*на* $\mathscr{U}'_{\alpha'} \cap \mathscr{U}'_{\beta'} \cap \mathscr{U}_\gamma \cap \mathscr{U}_\rho$ *и*

$$\tilde{\Lambda}_{\alpha'\rho} \circ \tilde{\Lambda}_{\rho\gamma'} \circ \Lambda'_{\gamma'\beta'} \circ \Lambda'_{\beta'\alpha'} \circ \tilde{\Lambda}_{\alpha'\rho} = \tilde{\Lambda}_{\alpha'\rho}, \tag{1.115}$$

$$\tilde{\Lambda}_{\rho\gamma'} \circ \Lambda'_{\gamma'\beta'} \circ \Lambda'_{\beta'\alpha'} \circ \tilde{\Lambda}_{\alpha'\rho} \circ \tilde{\Lambda}_{\rho\gamma'} = \tilde{\Lambda}_{\rho\gamma'}, \tag{1.116}$$

$$\Lambda'_{\gamma'\beta'} \circ \Lambda'_{\beta'\alpha'} \circ \tilde{\Lambda}_{\alpha'\rho} \circ \tilde{\Lambda}_{\rho\gamma'} \circ \Lambda'_{\gamma'\beta'} = \Lambda'_{\gamma'\beta'}, \tag{1.117}$$

$$\Lambda'_{\beta'\alpha'} \circ \tilde{\Lambda}_{\alpha'\rho} \circ \tilde{\Lambda}_{\rho\gamma'} \circ \Lambda'_{\gamma'\beta'} \circ \Lambda'_{\beta'\alpha'} = \Lambda'_{\beta'\alpha'} \tag{1.118}$$

*на* $\mathscr{U}'_{\alpha'} \cap \mathscr{U}'_{\beta'} \cap \mathscr{U}'_{\gamma'} \cap \mathscr{U}_{\rho}$.

Аналогично мы можем определять и исследовать главные и ассоциированные полурасслоения со структурной полугруппой.

## 1.4. Необратимость и полугомотопии

Здесь мы кратко остановимся на некоторых возможностях расширения понятия гомотопии на необратимые непрерывные отображения [19].

*Гомотопия* [188,209,230,231] представляет собой непрерывное отображение между двумя отображениями пространств $f : \mathscr{X} \to \mathscr{Y}$ и $g : \mathscr{X} \to \mathscr{Y}$ в пространстве $\boldsymbol{C}(\mathscr{X}, \mathscr{Y})$ отображений $\mathscr{X} \to \mathscr{Y}$ таковых, что $\gamma_{t=0}(x) = f(x)$, $\gamma_{t=1}(x) = g(x)$, $x \in \mathscr{X}$. Отображения $f(x)$ и $g(x)$ называются *гомотопными.* Другими словами [210] гомотопия из $\mathscr{X}$ в $\mathscr{Y}$ представляет собой непрерывную функцию $\gamma : \mathscr{X} \times I \to \mathscr{Y}$, где $I = [0.1]$ единичный интервал. Для заданного $t \in I$ имеются *шаги* $\gamma_t : \mathscr{X} \to \mathscr{Y}$ определяемые, как $\gamma_t(x) = \gamma(x, t)$.

Гомотопическое отношение, делящее $\boldsymbol{C}(\mathscr{X}, \mathscr{Y})$ на множество классов эквивалентности $\pi(\mathscr{X}, \mathscr{Y})$ , называется *гомотопическими классами,* которые представляют собой множество связных компонент из $\boldsymbol{C}(\mathscr{X}, \mathscr{Y})$. Поэтому для $\pi(\bullet, \mathscr{Y})$ (где $\bullet$ представляет собой точку) гомотопические классы соответствуют связным компонентам $\mathscr{Y}$. Если $\boldsymbol{C}(\mathscr{X}, \mathscr{Y})$ связны, тогда гомотопия между $f(x)$ и $g(x)$ может выбираться как их среднее, т. е.

$$\gamma_t(x) = tf(x) + (1 - t)\, g(x). \tag{1.119}$$

Два отображения $f$ и $g$ *гомотопически эквивалентны*, если $f \circ g$ и $g \circ f$ гомотопны тождественному отображению.

Теперь предположим $\mathscr{X}$ и $\mathscr{Y}$ — супермногообразия в некотором из определений [111, 112, 117, 181] или полусупермногообразие в нашей формулировке (см. **Определение 1.3**), тогда существует возможность расширения понятия гомотопии$^{*)}$ [19]. Идея заключается в том, чтобы расширить определение параметра $t$. В стандартном случае единичный интервал $I = [0, 1]$ выбирался для простоты, поскольку любые два отрезка на оси вещественных чисел гомеоморфны, и поэтому они топологически эквивалентны [231].

В случае супермногообразий [273–275], а особенно полусупермногообразий [19] ситуация существенно отличается. Мы имеем три топологически разделенных случая:

**1.** Параметр $t \in \Lambda_0$ четный и имеет числовую часть, т.е. $\epsilon(t) \neq 0$.

**2.** Параметр $t \in \Lambda_0$ четный и не имеет числовой части, т.е. $\epsilon(t) = 0$.

**3.** Параметр $\tau \in \Lambda_1$ нечетный (любой нечетный элемент не имеет числовой части).

Первая возможность может быть сведена стандартному случаю посредством соответствующего гомеоморфизма, и такой $t$ может всегда рассматриваться в единичном интервале $I = [0, 1]$. Однако следующие две возможности топологически не связаны с первой и между собой.

**Определение 1.47.** *Четная полугомотопия между двумя отображениями полусуперпространств $f : \mathscr{X} \to \mathscr{Y}$ и $g : \mathscr{X} \to \mathscr{Y}$ представляет собой необратимое (в общем случае) отображение $\mathscr{X} \to \mathscr{Y}$, зависящее от нильпотентного четного параметра $t \in \Lambda_0$ без числовой части и двух четных констант $a, b \in \Lambda_0$ без числовой части та-*

---

*) *Примечание.* Для различных несуперсимметричных обобщений гомотопии см. [270–272].

*ких, что*

$$\Delta I^{ab}\gamma^{even}_{t=a} = \Delta I^{ab} f(x)\,,\qquad \Delta I^{ab}\gamma^{even}_{t=b} = \Delta I^{ab} g(x)\,, \tag{1.120}$$

*где*

$$\gamma^{even}_{t}(x) = \Gamma^{even}(x,t)\,,\ \Gamma^{even} : \mathscr{X} \times I^{ab} \to \mathscr{Y}\,,\qquad I^{ab} = [a,b]\,,\ \Delta I^{ab} = b - a. \tag{1.121}$$

**Определение 1.48.** *Нечетная полугомотопия между двумя отображениями $f : \mathscr{X} \to \mathscr{Y}$ и $g : \mathscr{X} \to \mathscr{Y}$ представляет собой необратимое (в общем случае) отображение $\mathscr{X} \to \mathscr{Y}$, зависящее на нильпотентного нечетного параметра $\tau \in \Lambda_1$ и двух нечетных констант $\mu, \nu \in \Lambda_1$ таких, что*

$$\Delta \mathcal{I}^{\alpha\beta}\gamma^{odd}_{\tau=\alpha} = \Delta \mathcal{I}^{\alpha\beta} f(x)\,,\qquad \Delta \mathcal{I}^{\alpha\beta}\gamma^{odd}_{\tau=\beta} = \Delta \mathcal{I}^{\alpha\beta} g(x)\,, \tag{1.122}$$

*где*

$$\gamma^{odd}_{\tau}(x) = \Gamma^{odd}(x,\tau)\,,\Gamma^{odd} : \mathscr{X} \times \mathcal{I}^{\alpha\beta} \to \mathscr{Y}\,,\qquad \mathcal{I}^{\alpha\beta} = [\alpha,\beta]\,,\ \Delta \mathcal{I}^{\alpha\beta} = \beta - \alpha. \tag{1.123}$$

*Замечание* **1.49.** В (1.121) и (1.123) величины $I^{ab}$ и $\mathcal{I}^{\alpha\beta}$ не являются отрезками в обычном смысле, так как среди переменных без числовой части нет возможности устанавливить отношение упорядоченности [181, 268, 276], и поэтому $\Delta I^{ab}$ и $\Delta \mathcal{I}^{\alpha\beta}$ только формальные обозначения обозначения.

Тем не менее, мы можем привести пример аналога среднего (1.119) для нечетной полугомотопии

$$(\beta - \alpha)\,\gamma^{odd}_{\tau}(x) = (\beta - \tau)\, f(x) + (\tau - \alpha)\, g(x)\,, \tag{1.124}$$

который может удовлетворять условиям супергладкости.

*Замечание* **1.50.** В (1.120) и (1.122) нельзя сокращать левую и правую части на $I^{ab}$ и $\mathcal{I}^{\alpha\beta}$ соответственно, потому что решения для полугомо-топий $\gamma_t^{even}$ и $\gamma_\tau^{odd}$ рассмотриваются как отношения эквивалентности. Это отчетливо видно из (1.124), где деление на $(\beta - \alpha)$ невозможно, тем не менее решение для $\gamma_\tau^{odd}(x)$ существует.

Наиболее важное свойство полугомотопий — это их возможная необратимость, которая следует из нильпотентности $t$ и $\tau$ и определений (1.120) и (1.122). Поэтому, $\mathscr{Y}$ не может быть супермногообразием, оно может быть только полусупермногообразием [6, 19].

**Предположение 1.51.** *Полугомотопии играют ту же роль в изучении свойств непрерывности и классификации полусупермногообразий, какую обычные гомотопии играют для обычных многообразий.*

## 1.5. Основные результаты и выводы

1. Сформулирована теория полусупермногообразий в терминах атласов и функций перехода.

2. Найдены обобщенные условия коцикла и рефлексивности.

3. Предложен новый тип ориентируемости — нильпотентная ориентируемость.

4. Сформулирован общий принцип полукоммутативности для необратимых морфизмов.

5. Проведена классификация полусупермногообразий в терминах новой характеристики — препятственности.

6. Построены необратимые аналоги расслоений — полурасслоения — в терминах уравнений на функции перехода.

7. Изучены морфизмы полурасслоений и рефлексивность.

8. Введены полугомотопии с необратимым четным или нечетным суперпараметром.

# РАЗДЕЛ 2

# НЕОБРАТИМОЕ ОБОБЩЕНИЕ $N = 1$ СУПЕРКОНФОРМНОЙ ГЕОМЕТРИИ

В этом разделе формулируется необратимая $N = 1$ суперконформная геометрия на суперплоскости, играющая важную роль в теории суперструн и в двумерных суперконформных теориях поля. Прежде всего, строится полугруппа супераналитических преобразований, проводится их классификация по необратимости, дается формулировка супераналитических полусупермногообразий в терминах необратимых функций перехода. Далее анализируются все возможные редукции касательного суперпространства при ослаблении требования обратимости, что приводит к новым редукциям и необратимым аналогам антиголоморфных преобразований — сплетающим четность касательного пространства преобразованиям, которые характеризуются нильпотентным березинианом и наличием нового типа коциклов с разными стрелками. Единое описание обоих типов редуцированных преобразований проводится с помощью альтернативной параметризации, и переключение между ними происходит с помощью введенного спина редукции, равного 1/2 для $N = 1$ преобразований. В альтернативной параметризации строится суперконформная полугруппа, которая принадлежит к новому абстрактному типу полугрупп, удовлетворяющим необычному идеальному умножению. Для нее определяются обобщенные векторные и тензорные отношения Грина, а также идеальные квазихарактеры.

Исследование дробно-линейных необратимых редуцированных преобразований проводится в терминах полуминоров и полуматриц — нечетных аналогов обычных. Для них определяются функции полупер-манента и полудетерминанта, которые дуальны стандартным матричным функциям в рамках введенной четно-нечетной симметрии дробно-линейных $N = 1$ суперконформных преобразований. Находятся необратимые супераналоги расстояния в $N = 1$ суперпространстве и формулируется необратимый аналог инвариантности — "полуинвариантность" введенного необратимого аналога метрики для сплетающих четность преобразований.

Нелинейные реализации редуцированных преобразований рассматриваются в рамках двух подходов — как движение нечетной кривой в суперпространстве и диаграммное описание необратимого аналога индуцированного представления. Находятся уравнения для двух типов голдстино и для связи между линейной и нелинейной реализациями.

Идея суперконформной симметрии [277–280] играет ключевую роль в построении суперструнных [281] моделей элементарных частиц [282–286], в рамках которых удается объединить*) непротиворечивым образом все фундаментальные взаимодействия [288–291]. В последнее время значение суперконформной симметрии было переосмыслено из-за ее исключительной роли в построении $M$-теории [292–300], описании $D$-бран [301–305] и черных дыр [306–308], а также в ее связи с предельными теоремами в пространствах анти-Де Ситтера [309–318].

С одной стороны, суперконформная симметрия исключительно важна в теории суперримановых поверхностей [111, 176, 227, 319–322] как локального подхода для вычисления древесных [323–325] и многопетлевых [326–332] фермионных амплитуд в формализме Полякова [333–336]. С другой стороны, двумерные суперконформные теории поля [337–340] описывают квантовую геометрию мировой поверхности струны [341–345] и позволяют свести вычисление струнных амплитуд в критиче-

---

*) *Примечание.* Впервые использование струн для построения фундаментальной теории, описывающей в низкоэнергетическом пределе все существующие взаимодействия, было предложено в [287].

ской размерности [346] к интегрированию по суперконформному пространству модулей [347–355]. Возникшие здесь трудности с нечетными модулями [356–358] (а фактически, с нильпотентными направлениями [359–361]), несмотря на то, что некоторые многопетлевые вклады и были заново получены в [362–364], позволяют предположить*) возможность необратимого обобщения суперконформной геометрии [9, 18].

## 2.1. Необратимость и $N = 1$ суперконформные преобразования

Основным ингредиентом суперконформной симметрии является специальный класс редуцированных отображений двумерного $(1|1)$-мерного комплексного суперпространства, суперконформные преобразования [111, 345, 355, 366]. В локальном подходе к построению суперримановых поверхностей, представленных как семейства открытых суперобластей, суперконформные преобразования используются как склеивающие функции перехода [324, 341, 343]. С другой стороны, они возникают в результате специальной редукции структурной супергруппы [367, 368]. Аналогичный подход применяется и для клейновых поверхностей [369] и суперповерхностей [370–373].

Здесь мы рассматриваем альтернативную редукцию касательного пространства, что приводит к новым преобразованиям (см. также [1,8]). Мы используем функциональный подход к суперпространству [91, 112, 117] (см. также **Приложения Б.2** и **Б.4**), который допускает существование нетривиальной топологии в четных и нечетных нильпотентных направлениях [175, 268] и может быть подходящим для физических при-

---

*) *Примечание.* В связи с этими трудностями было высказано такое предположение: “может случиться, что основные конструкции должны быть модифицированы...” [365].

ложений [374, 375].

Кроме того, необратимые преобразования (см. также, [263, 376]) могут служить аналогом функций перехода для полусупермногообразий, введенных в **Разделе 1**, что позволяет последовательным образом сформулировать необратимый аналог суперримановой поверхности [9]. Отметим, что исследование четных нильпотентных направлений в суперсимметричной механике [16, 377] и квантовой механике [378–380] играет важную роль в прояснении общих механизмов нарушения суперсимметрии; они также возникают в контракциях групп [381–383] и в конкретных полевых моделях [384–387].

**2.1.1.** Супераналитические преобразования. Локально суперпространство $\mathbb{C}^{1|1}$, имеющее размерность $(1|1)$, на координатном языке описывается парой $Z = (z, \theta)$, где $z$ четная координата и $\theta$ нечетная.

В функциональном определении суперпространства существуют ду́ховые части в четной координате $z = z_{body} + z_{soul}$, $z_{body} = \epsilon(z)$, $z_{soul} \stackrel{def}{=} z - z_{body}$, где $\epsilon$ представляет собой числовое отображение [112], зануляющее все нильпотентные генераторы подстилающей суперoccurrences.

Числовое отображение действует на координатах следующим образом $\epsilon(z) = z_{body}$, $\epsilon(\theta) = 0$ (см. также **Пункт Б.2**). Это позволяет нам рассматривать нетривиальную ду́ховую топологию в четных направлениях на равных началах с нечетными [175, 181, 268].

Используя условия голоморфности, общее супераналитическое преобразование $\mathcal{T}_{SA} : \mathbb{C}^{1|1} \to \mathbb{C}^{1|1}$ можно представить (см., например, [388]) в виде

$$\begin{cases} \tilde{z} = \tilde{z}(z, \theta), \\ \tilde{\theta} = \tilde{\theta}(z, \theta), \end{cases} \tag{2.1}$$

где нет зависимости от комплексно сопряженной координаты.

Учитывая нильпотентность нечетной координаты $\theta^2 = 0$, мы по-

лучаем

$$\begin{cases} \tilde{z} &= f(z) + \theta \cdot \chi(z), \\ \tilde{\theta} &= \psi(z) + \theta \cdot g(z), \end{cases} \tag{2.2}$$

где четыре координатные функции $f(z), g(z) : \mathbb{C}^{1|0} \to \mathbb{C}^{1|0}$ и $\psi(z), \chi(z) : \mathbb{C}^{1|0} \to \mathbb{C}^{0|1}$ удовлетворяют супергладким условиям, обобщающим $C^{\infty}$ (см. [112, 226, 389] и **Пункт Б.2**).

Очевидно, что нечетные функции $\psi(z), \chi(z)$ по определению необратимы (см. [120], хотя имеются и некоторые контрпримеры [132–134]). Таким образом, обратимость супераналитического преобразования $\mathcal{T}_{SA}$ (2.1) контролируется четными функциями $f(z), g(z)$. Обычно они выбираются обратимыми [111, 355]. Здесь мы не будем ограничивать их обратимость и рассмотрим оба случая на равных началах.

**Определение 2.1.** *Множества обратимых и необратимых преобразований* $\mathbb{C}^{1|1} \to \mathbb{C}^{1|1}$ (2.2) *образуют полугруппу супераналитических преобразований* $\boldsymbol{T}_{SA}$ *относительно композиции.*

Обратимые преобразования принадлежат подгруппе этой полугруппы, тогда как необратимые преобразования принадлежат ее идеалу [1, 5]. Будем классифицировать все преобразования следующим образом [7].

**Определение 2.2.** *Обратимые супераналитические преобразования определяются условиями*

$$\epsilon[f(z)] \neq 0, \; \epsilon[g(z)] \neq 0. \tag{2.3}$$

**Определение 2.3.** *Полунеобратимые супераналитические преобразования определяются условиями*

$$\epsilon[f(z)] = 0, \; \epsilon[g(z)] \neq 0. \tag{2.4}$$

**Определение 2.4.** *Необратимые суператалитические преобразования определяются условием*

$$\epsilon\left[f\left(z\right)\right]=0,\ \epsilon\left[g\left(z\right)\right]=0. \tag{2.5}$$

*Замечание* **2.5.** Полунеобратимые суператалитические преобразования *могут разрешаться*, но только лишь относительно $\theta$, а не относительно $z$.

Очевидно, можно использовать координатные функции из (2.2) для соответствующей параметризации полугруппы суператалитических преобразований $\boldsymbol{T}_{SA}$.

**Определение 2.6.** *Элемент* $\mathbf{s}$ *суператалитической полугруппы* $\mathbf{S}_{SA}$ *может быть параметризован четверкой функций*

$$\left\{\begin{array}{cc} f & \chi \\ \psi & g \end{array}\right\} \stackrel{def}{=} \mathbf{s}\in\mathbf{S}_{SA}, \tag{2.6}$$

*и действие в* $\mathbf{S}_{SA}$ *есть*

$$\left\{\begin{array}{cc} f_1 & \chi_1 \\ \psi_1 & g_1 \end{array}\right\} * \left\{\begin{array}{cc} f_2 & \chi_2 \\ \psi_2 & g_2 \end{array}\right\} =$$

$$\left\{\begin{array}{cc} f_1\circ f_2+\psi_2\cdot\chi_1\circ f_2 & f_1'\circ f_2\cdot\chi_2+g_2\cdot\chi_1\circ f_2 \\ & +\chi_1'\circ f_2\cdot\chi_2\cdot\psi_2 \\ \psi_1\circ f_2+\psi_2\cdot g_1\circ f_2 & \psi_1'\circ f_2\cdot\chi_2+g_2\cdot g_1\circ f_2 \\ & +g_1'\circ f_2\cdot\chi_2\cdot\psi_2 \end{array}\right\}. \tag{2.7}$$

*где*

$$f_1\circ f_2=f_1\left(f_2\left(z\right)\right) \tag{2.8}$$

*и штрих* $(')$ *означает дифференцирование по аргументу.*

Ассоциативность в $\mathbf{S}_{SA}$

$$\mathbf{s}_1 * (\mathbf{s}_2 * \mathbf{s}_3) = (\mathbf{s}_1 * \mathbf{s}_2) * \mathbf{s}_3 \tag{2.9}$$

нетривиальна для (2.7) и требует проверки.

**Предложение 2.7.** *Умножение* (2.7) *ассоциативно.*

*Доказательство.* Соотношение (2.9) состоит из четырех уравнений, соответствующих четырем функциям в (2.6).

Используя (2.7) для 1-1 элемента, мы находим

$$\begin{aligned}
\mathbf{s}_1 * (\mathbf{s}_2 * \mathbf{s}_3)\,|_{1-1} \;&=\; f_1 \circ (f_2 \circ f_3 + \psi_3 \cdot \chi_2 \circ f_3) \\
&\quad + (\psi_2 \circ f_3 + \psi_3 \cdot g_2 \circ f_3) \cdot \chi_1 \circ (f_2 \circ f_3 + \psi_3 \cdot \chi_2 \circ f_3)\,.
\end{aligned}$$

Открывая скобки, раскладывая в ряд Тэйлора и учитывая нильпотентность входящих нечетных функций, мы имеем

$$\begin{aligned}
\mathbf{s}_1 * (\mathbf{s}_2 * \mathbf{s}_3)\,|_{1-1} \;&=\; f_1 \circ f_2 \circ f_3 + \psi_3 \cdot \chi_2 \circ f_3 \cdot f_1' \circ f_2 \circ f_3 \\
&\quad + \psi_2 \circ f_3 \cdot \chi_1 \circ f_2 \circ f_3 \\
&\quad + \psi_3 \cdot g_2 \circ f_3 \cdot \chi_1 \circ f_2 \circ f_3 \\
&\quad + \psi_2 \circ f_3 \cdot \chi_1' \circ f_2 \circ f_3 \cdot \psi_3 \cdot \chi_2 \circ f_3.
\end{aligned}$$

Далее группируем элементы различным способом и получаем

$$\begin{aligned}
\mathbf{s}_1 * (\mathbf{s}_2 * \mathbf{s}_3)\,|_{1-1} \;&=\; (f_1 \circ f_2 + \psi_2 \cdot \chi_1 \circ f_2) \circ f_3 \\
&\quad + \psi_3 \cdot (f_1' \circ f_2 \cdot \chi_2 + \chi_1' \circ f_2 \cdot \chi_2 \cdot \psi_2 + g_2 \cdot \chi_1 \circ f_2) \circ f_3 \\
&=\; (\mathbf{s}_1 * \mathbf{s}_2) * \mathbf{s}_3|_{1-1}.
\end{aligned}$$

Аналогичные вычисления могут быть проведены и для других элементов, это доказывает ассоциативность (2.7) и тот факт, что параметризация (2.6) задает действительно полугруппу. ■

*Замечание* **2.8.** Умножение (2.7) содержит два произведения: суперпозицию (2.8) и произведение в подстилающей алгебре Грассмана $(\cdot)$. Поэтому супераналитическая полугруппа не принадлежит ни к классу полугрупп непрерывных функций [262,390,391], ни к классу мультипликативных полугрупп [205, 392–394].

Наличие двух умножений, делителей нуля и нильпотентов делает анализ абстрактных свойств супераналитической полугруппы*) (и суперконформной полугруппы, рассматриваемой ниже) гораздо более сложным по сравнению с хорошо исследованными полугруппами функций [191, 208, 262, 399].

**Предложение 2.9.** *Двусторонняя единица в* $\mathbf{S}_{SA}$ *есть*

$$\mathbf{e} = \begin{Bmatrix} z & 0 \\ 0 & 1 \end{Bmatrix}, \tag{2.10}$$

*и двусторонний нуль представляет собой матрицу* (2.6), *имеющую нулевые элементы.*

*Доказательство.* Это можно легко проверить, используя (2.7). ■

Рассмотрим гомоморфизм $\varphi$ супераналитической полугруппы $\mathbf{S}_{SA}$ в полугруппу $\boldsymbol{T}_{SA}$ супераналитических преобразований $\varphi : \mathbf{S}_{SA} \to \boldsymbol{T}_{SA}$.

**Предложение 2.10.** *Как это и должно быть* $\ker \varphi = \mathbf{e}$.

При изучении суперчисловых систем, содержащих делители нуля

---

*) *Примечание.* Полугруппы несуперсимметричных аналитических эндоморфизов рассматривались в [395–398].

и нильпотенты, обычно говорят магические слова “факторизация по нильпотентам” или “по модулю нильпотентов” и исключают дополнительные экзотические свойства [117, 174, 187], являющиеся результатом тщательного рассмотрения последних. В системах рассматриваемых функций ситуация более тонкая и требует дополнительных абстрактных исследований.

Например, в супераналитической полугруппе $\mathbf{S}_{SA}$ наряду со стандартными элементами $\mathbf{e}$ и $\mathbf{z}$ мы можем вводить элементнозависимые “локальные” единицы и нули.

**Определение 2.11.** *Для заданного элемента* $\mathbf{s}$ *супераналитической полугруппы* $\mathbf{S}_{SA}$ *локальные левая, правая и двусторонняя единицы определяются равенствами*

$$\mathbf{e}_{\mathbf{s}}^{left} * \mathbf{s} = \mathbf{s}, \tag{2.11}$$

$$\mathbf{s} * \mathbf{e}_{\mathbf{s}}^{right} = \mathbf{s}, \tag{2.12}$$

$$\mathbf{e}_{\mathbf{s}} * \mathbf{s} * \mathbf{e}_{\mathbf{s}} = \mathbf{s}, \tag{2.13}$$

*где* $\mathbf{e}_{\mathbf{s}}^{left}, \mathbf{e}_{\mathbf{s}}^{right}, \mathbf{e}_{\mathbf{s}} \in \mathbf{S}_{SA}$.

**Определение 2.12.** *Для заданного элемента* $\mathbf{s}$ *супераналитической полугруппы* $\mathbf{S}_{SA}$ *локальные левый, правый и двусторонний нули определяются равенствами*

$$\mathbf{z}_{\mathbf{s}}^{left} * \mathbf{s} = \mathbf{z}_{\mathbf{s}}^{left}, \tag{2.14}$$

$$\mathbf{s} * \mathbf{z}_{\mathbf{s}}^{right} = \mathbf{z}_{\mathbf{s}}^{right}, \tag{2.15}$$

$$\mathbf{z}_{\mathbf{s}} * \mathbf{s} * \mathbf{z}_{\mathbf{s}} = \mathbf{z}_{\mathbf{s}}, \tag{2.16}$$

*где* $\mathbf{z}_{\mathbf{s}}^{left}, \mathbf{z}_{\mathbf{s}}^{right}, \mathbf{z}_{\mathbf{s}} \in \mathbf{S}_{SA}$.

Локальные единицы и нули являются множествами элементов из $\mathbf{S}_{SA}$ и могут найдены из соответствующих систем функционально-дифференциальных уравнений. Например, для $\mathbf{e}_{\mathbf{s}}^{left}$ из (2.11) в компонент-ном виде мы имеем систему

$$\begin{gathered} f_1 \circ f_2 + \psi_2 \cdot \chi_1 \circ f_2 = f_2, \\ \psi_1 \circ f_2 + \psi_2 \cdot g_1 \circ f_2 = \psi_2, \\ f_1' \circ f_2 \cdot \chi_2 + \chi_1' \circ f_2 \cdot \chi_2 \cdot \psi_2 + g_2 \cdot \chi_1 \circ f_2 = \chi_2, \\ \psi_1' \circ f_2 \cdot \chi_2 + g_1' \circ f_2 \cdot \chi_2 \cdot \psi_2 + g_2 \cdot g_1 \circ f_2 = g_2. \end{gathered} \tag{2.17}$$

*Пример* **2.13.** Пусть $\mathbf{s} = \left\{ \begin{array}{cc} z^2 & \beta \\ \alpha & z^{-1} \end{array} \right\}$, тогда $\mathbf{e}_{\mathbf{s}}^{left} = \left\{ \begin{array}{cc} z^2 & \beta \\ \alpha & z^{-1} \end{array} \right\}$.

Чтобы подчеркнуть отличие от полугрупп функций [208, 390, 399], рассмотрим левые нули. Из закона умножения (2.8) следует

**Утверждение 2.14.** *Для полугрупп функций роль левых нулей играют константные отображения*

$$f_0(z) : z \to c_f = const, \tag{2.18}$$

*поскольку* $\forall g(z),\ f_0 \circ g = f_0(g(z)) = c_f = f_0$.

Возьмем элемент $\mathbf{s}_0$ супераналитической полугруппы $\mathbf{S}_{SA}$, который имеет вид, аналогичный (2.18), т.е.

$$\mathbf{s}_0 = \left\{ \begin{array}{cc} f_0 & \chi_0 \\ \psi_0 & g_0 \end{array} \right\}. \tag{2.19}$$

Тогда из (2.7) имеем

$$\mathbf{s}_0 * \mathbf{s} = \left\{ \begin{array}{cc} f_0 & \chi_0 \\ \psi_0 & g_0 \end{array} \right\} * \left\{ \begin{array}{cc} f & \chi \\ \psi & g \end{array} \right\} = \left\{ \begin{array}{cc} c_f + c_\chi \cdot g & c_\chi \cdot g \\ c_\psi + c_g \cdot \psi & c_g \cdot g \end{array} \right\}, \tag{2.20}$$

и, таким образом, $\mathbf{s}_0 * \mathbf{s} \neq const$ в противоположость полугруппам

функций [192, 262].

*Замечание* **2.15.** Сопоставляя супераналитическое умножение (2.7) с матричным полугрупповым умножением [400, 401], мы обращаем внимание на то, что множеству нижнетреугольных суперматриц (2.6), т. е. с элементами $\chi = 0$, формируют подполугруппу, как обычно. Однако множество верхнетреугольных матриц, имеющих $\psi = 0$, не формируют подполугруппу из-за наличия среднего члена в 2-2 элементе (2.7).

Посредством супераналитических преобразований (2.2) можно построить супераналитическое полусупермногообразие $\mathscr{M}_{SA}$ стандартным способом (см. [112,117,174] и **Раздел 1**), в котором координатные функции играют роль склеивающих функций перехода.

Таким образом, пусть $\mathscr{M}_{SA} = \bigcup_{\alpha} \mathscr{U}_{\alpha}$, где $\mathscr{U}_{\alpha}$ — суперобласти, накрывающие полусупермногообразие $\mathscr{M}_{SA}$. Его строение определяется четырьмя функциями перехода $f_{\alpha\beta}(z_{\beta}), \chi_{\alpha\beta}(z_{\beta}), \; g_{\alpha\beta}(z_{\beta}), \psi_{\alpha\beta}(z_{\beta})$, описывающих супераналитическое преобразование $Z_{\beta} \to Z_{\alpha}$ на пересечении $\mathscr{U}_{\alpha} \cap \mathscr{U}_{\beta}$.

**Предложение 2.16.** *На тройных пересечениях $\mathscr{U}_{\alpha} \cap \mathscr{U}_{\beta} \cap \mathscr{U}_{\gamma}$ функциях перехода супераналитического супермногообразия удовлетворяют условиям согласованности*

$$\begin{aligned} f_{\alpha\gamma} &= f_{\alpha\beta} \circ f_{\beta\gamma} + \psi_{\beta\gamma} \cdot \chi_{\alpha\beta} \circ f_{\beta\gamma}, \\ \chi_{\alpha\gamma} &= f'_{\alpha\beta} \circ f_{\beta\gamma} \cdot \chi_{\beta\gamma} + g_{\beta\gamma} \cdot \chi_{\alpha\beta} \circ f_{\beta\gamma} + \chi'_{\alpha\beta} \circ f_{\beta\gamma} \cdot \chi_{\beta\gamma} \cdot \psi_{\beta\gamma}, \\ g_{\alpha\gamma} &= f'_{\alpha\beta} \circ f_{\beta\gamma} \cdot \chi_{\beta\gamma} + g_{\beta\gamma} \cdot g_{\alpha\beta} \circ f_{\beta\gamma} + g'_{\alpha\beta} \circ f_{\beta\gamma} \cdot \chi_{\beta\gamma} \cdot \psi_{\beta\gamma}, \\ \psi_{\alpha\gamma} &= \psi_{\alpha\beta} \circ f_{\beta\gamma} + \psi_{\beta\gamma} \cdot g_{\alpha\beta} \circ f_{\beta\gamma}. \end{aligned} \tag{2.21}$$

*Доказательство.* Непосредственно следует из умножения (2.7). ■

Дальнейшие коциклические свойства $N = 1$ преобразований изложены в **Приложении З**.

**2.1.2.** Касательное суперпространство и варианты его редукций. Рассмотрим действие супераналитических преобразований в некотором необратимом аналоге касательного суперпространства [7, 18] и возможные его редукции (см. обратимый вариант редукций в [367, 402–404]). Здесь мы покажем, что среди редуцированных необратимых преобразований имеются новые преобразования, которые в некотором смысле дуальны суперконформным преобразованиям [1, 9], и сконцентрируем внимание на новых свойствах, связанных с необратимостью, для ясности пытаясь останавливаться на рассмотрении нетривиальных моментов..

Касательное суперпространство $T\mathbb{C}^{1|1}$ определяется стандартным суперсимметричным базисом $\{\partial,\ D\}$, где $D = \partial_\theta + \theta\partial$, $\partial_\theta = \partial/\partial\theta$, $\partial = \partial/\partial z$. Дуальное кокасательное пространство $T^*\mathbb{C}^{1|1}$ определяется 1-формами $\{dz,\ d\theta\}$, где $dZ = dz + \theta d\theta$ (знаки как в [111]). В этих обозначениях соотношения суперсимметрии есть $D^2 = \partial$, $dZ^2 = dz$.

Полугруппа супераналитических преобразований $\boldsymbol{T}_{SA}$ действует в касательном и кокасательном суперпространствах посредством матрицы $\mathrm{P}_{SA}$ как

$$\begin{pmatrix} \partial \\ D \end{pmatrix} = \mathrm{P}_{SA} \begin{pmatrix} \tilde{\partial} \\ \tilde{D} \end{pmatrix}, \tag{2.22}$$

$$\begin{pmatrix} d\tilde{Z}, & d\tilde{\theta} \end{pmatrix} = \begin{pmatrix} dz, & d\theta \end{pmatrix} \mathrm{P}_{SA}, \tag{2.23}$$

где

$$\mathrm{P}_{SA} = \begin{pmatrix} \partial\tilde{z} - \partial\tilde{\theta}\cdot\tilde{\theta} & \partial\tilde{\theta} \\ D\tilde{z} - D\tilde{\theta}\cdot\tilde{\theta} & D\tilde{\theta} \end{pmatrix}. \tag{2.24}$$

Рассмотрим суперобобщения (включая и необратимые) внешнего дифференциала де Рама [211].

**Предложение 2.17.** *Внешний дифференциал* $d = dZ\partial + d\theta D$ *является*

*инвариантом супераналитических преобразований.*

*Доказательство.* Мы имеем

$$\begin{aligned} d &= \begin{pmatrix} dZ, & d\theta \end{pmatrix} \begin{pmatrix} \partial \\ D \end{pmatrix} = \begin{pmatrix} dZ, & d\theta \end{pmatrix} \mathrm{P}_{SA} \begin{pmatrix} \tilde{\partial} \\ \tilde{D} \end{pmatrix} \\ &= \begin{pmatrix} d\tilde{Z}, & d\tilde{\theta} \end{pmatrix} \begin{pmatrix} \tilde{\partial} \\ \tilde{D} \end{pmatrix} = \tilde{d} \end{aligned} \tag{2.25}$$

■

*Замечание* **2.18.** Важно отметить, что в (2.25) обратимость не использована.

**Предложение 2.19.** $\mathrm{Ber}\left(\tilde{Z}/Z\right) = \mathrm{Ber}\,\mathrm{P}_{SA}$.

*Доказательство.* Видим, что

$$\begin{pmatrix} \dfrac{\partial \tilde{z}}{\partial z} & \dfrac{\partial \tilde{\theta}}{\partial z} \\ \dfrac{\partial \theta}{\partial z} & \dfrac{\partial \theta}{\partial \theta} \end{pmatrix} = \begin{pmatrix} 1 & 0 \\ -\theta & 1 \end{pmatrix} \cdot \begin{pmatrix} \partial \tilde{z} - \partial \tilde{\theta} \cdot \tilde{\theta} & \partial \tilde{\theta} \\ D\tilde{z} - D\tilde{\theta} \cdot \tilde{\theta} & D\tilde{\theta} \end{pmatrix} \begin{pmatrix} 1 & 0 \\ \tilde{\theta} & 1 \end{pmatrix}. \tag{2.26}$$

Тогда из (2.26), (E.9), (E.10) и (2.24) следует

$$\begin{aligned} \mathrm{Ber}\left(\tilde{Z}/Z\right) &= \mathrm{Ber}\,\mathrm{P}^0_{SA} = \mathrm{Ber}\left(\begin{pmatrix} 1 & 0 \\ -\theta & 1 \end{pmatrix} \cdot \mathrm{P}_{SA} \cdot \begin{pmatrix} 1 & 0 \\ \tilde{\theta} & 1 \end{pmatrix}\right) \\ &= \mathrm{Ber}\begin{pmatrix} 1 & 0 \\ -\theta & 1 \end{pmatrix} \cdot \mathrm{Ber}\,\mathrm{P}_{SA} \cdot \mathrm{Ber}\begin{pmatrix} 1 & 0 \\ \tilde{\theta} & 1 \end{pmatrix} \\ &= \mathrm{Ber}\,\mathrm{P}_{SA}. \end{aligned}$$

■

В случае обратимых супераналитических преобразований матрица

$\mathrm{P}_{SA}$ определяет структуру супермногообразия, для которого эти преобразования играют роль функций перехода [367]. Поэтому различные редукции матрицы $\mathrm{P}_{SA}$ приводят к различным дополнительным структурам. Но только один из них обычно рассматривается [367, 368] поскольку лишь он может быть обратимым.

Учитывая необратимость, мы проанализируем *все* редукции [7] посредством зануления *каждого* элемента из $\mathrm{P}_{SA}$ поочередно, что дает в общем четыре возможности:

$$1)\ D\tilde{\theta} = 0, \tag{2.27}$$

$$2)\ \partial\tilde{\theta} = 0, \tag{2.28}$$

$$3)\ \varDelta\left(z,\theta\right) \equiv D\tilde{z} - D\tilde{\theta}\cdot\tilde{\theta} = 0, \tag{2.29}$$

$$4)\ Q\left(z,\theta\right) \equiv \partial\tilde{z} - \partial\tilde{\theta}\cdot\tilde{\theta} = 0, \tag{2.30}$$

которые упорядочены соответственно возрастанию их нетривиальности. Первые два случая (2.27) и (2.28) являются наиболее простыми, но они также имеют некоторые интересные особенности и будут рассмотрены отдельно.

**2.1.3.** Редуцированные $N=1$ преобразования . Здесь мы рассмотрим две остальные возможные редукции (2.29) и (2.30). В **Подразделе 4.1** показано, что существуют *две* нетривиальные редукции любой суперматрицы (а не одна, треугольная, как в обратимом случае). Мы применяем этот результат к $\mathrm{P}_{SA}$ (2.24).

**Утверждение 2.20.** *Условие* $\epsilon\left[D\tilde{\theta}\right] \neq 0$ *совпадает с полунеобратимостью суперапалитического преобразования* (2.4), *а не с его полной обратимостью.*

*Доказательство.* В самом деле, мы замечаем из (2.26), что березиниан может быть представлен в виде двух слагаемых

$$\operatorname{Ber} P_A = \frac{\partial \tilde{z} - \partial \tilde{\theta} \cdot \tilde{\theta}}{D\tilde{\theta}} + \frac{\left(D\tilde{z} - D\tilde{\theta} \cdot \tilde{\theta}\right) \partial \tilde{\theta}}{\left(D\tilde{\theta}\right)^2} = \tag{2.31}$$

$$\frac{Q(z,\theta)}{D\tilde{\theta}} + \frac{\Delta(z,\theta) \cdot \partial \tilde{\theta}}{\left(D\tilde{\theta}\right)^2}. \tag{2.32}$$

только, если $\epsilon\left[D\tilde{\theta}\right] \neq 0$. Тогда из компонентного вида (2.2) мы выводим $D\tilde{\theta} = g(z) + \theta \cdot \psi(z)$ и так $\epsilon\left[D\tilde{\theta}\right] = \epsilon\left[g(z)\right]$, поэтому $\epsilon\left[D\tilde{\theta}\right] \neq 0 \Rightarrow \epsilon\left[g(z)\right] \neq 0$, что действительно является условием полунеобратимости преобразования (2.4). ■

**Предложение 2.21.** *В случае $D\tilde{\theta} \neq 0$ березиниан супераналитических преобразований описывается выражением*

$$\operatorname{Ber} \mathrm{P}_{SA} = D\left(\frac{D\tilde{z}}{D\tilde{\theta}}\right). \tag{2.33}$$

*Доказательство.* После дифференцирования правой части, используя $D^2 = \partial$, мы получаем

$$D\left(\frac{D\tilde{z}}{D\tilde{\theta}}\right) = \frac{\partial \tilde{z} \cdot D\tilde{\theta} + D\tilde{z} \cdot \partial \tilde{\theta}}{\left(D\tilde{\theta}\right)^2} =$$
$$\frac{\left(\partial \tilde{z} + \tilde{\theta} \cdot \partial \tilde{\theta}\right) \cdot D\tilde{\theta} - \left(D\tilde{z} - \tilde{\theta} \cdot D\tilde{\theta}\right) \cdot \partial \tilde{\theta}}{\left(D\tilde{\theta}\right)^2} =$$
$$\frac{Q(z,\theta) \cdot D\tilde{\theta} - \Delta(z,\theta) \cdot \partial \tilde{\theta}}{\left(D\tilde{\theta}\right)^2},$$

что совпадает с (2.32). ■

По теореме сложения березинианов (4.7) имеем

$$\operatorname{Ber} \mathrm{P}_A = \operatorname{Ber} \mathrm{P}_S + \operatorname{Ber} \mathrm{P}_T, \tag{2.34}$$

где

$$\mathrm{P}_S \stackrel{def}{=} \begin{pmatrix} \partial\tilde{z} - \partial\tilde{\theta}\cdot\tilde{\theta} & \partial\tilde{\theta} \\ 0 & D\tilde{\theta} \end{pmatrix} = \begin{pmatrix} Q\left(z,\theta\right) & \partial\tilde{\theta} \\ 0 & D\tilde{\theta} \end{pmatrix}, \tag{2.35}$$

$$\mathrm{P}_T \stackrel{def}{=} \begin{pmatrix} 0 & \partial\tilde{\theta} \\ D\tilde{z} - D\tilde{\theta}\cdot\tilde{\theta} & D\tilde{\theta} \end{pmatrix} = \begin{pmatrix} 0 & \partial\tilde{\theta} \\ \varDelta\left(z,\theta\right) & D\tilde{\theta} \end{pmatrix}. \tag{2.36}$$

Обозначим множества матриц (2.35) и (2.36) за $\mathcal{P}_S$ и $\mathcal{P}_T$ соответственно. Подчеркиваем, что до сих пор на вид преобразований мы не налагали никаких ограничений, и они общие супераналитические (2.1).

**Определение 2.22.** *Редуцированные преобразования определяются проектированием березиниана на одно из слагаемых в* (2.32).

Другими словами, мы проектируем множество супераналитических матриц $\mathcal{P}_{SA}$ на $\mathcal{P}_S$ или $\mathcal{P}_T$.

Следовательно, имеется *два* (!) вида редуцированных (суперконформно-подобных) преобразований [1, 7, 21].

**Определение 2.23.** *Обратимые, полунеобратимые и необратимые суперконформные преобразования определяются условием*

$$\varDelta\left(z,\theta\right) = D\tilde{z} - D\tilde{\theta}\cdot\tilde{\theta} = 0. \tag{2.37}$$

**Определение 2.24.** *Полунеобратимые и необратимые преобразования, сплетающие четность, определяются условием*

$$Q\left(z,\theta\right) = \partial\tilde{z} - \partial\tilde{\theta}\cdot\tilde{\theta} = 0. \tag{2.38}$$

Такое определение понятно из следующих рассуждений. Если мы применим условия (2.38) и (2.37) к матрицам $\mathrm{P}_S$ и $\mathrm{P}_T$, то получим

$$\mathrm{P}_{SCf} \overset{def}{=} \mathrm{P}_S|_{\varDelta=0} = \begin{pmatrix} Q_{SCf}(z,\theta) & \partial\tilde{\theta}_{SCf} \\ 0 & D\tilde{\theta}_{SCf} \end{pmatrix}, \tag{2.39}$$

$$\mathrm{P}_{TPt} \overset{def}{=} \mathrm{P}_T|_{Q=0} = \begin{pmatrix} 0 & \partial\tilde{\theta}_{TPt} \\ \varDelta_{TPt}(z,\theta) & D\tilde{\theta}_{TPt} \end{pmatrix}, \tag{2.40}$$

где

$$Q_{SCf}(z,\theta) \overset{def}{=} Q(z,\theta)\,|_{\varDelta(z,\theta)=0}, \tag{2.41}$$

$$\varDelta_{TPt}(z,\theta) \overset{def}{=} \varDelta(z,\theta)\,|_{Q(z,\theta)=0}. \tag{2.42}$$

Отсюда следуют преобразования касательного и кокасательного пространств в стандартном базисе

$$\mathsf{SCf} \quad : \quad \begin{cases} D &= \quad D\tilde{\theta}_{SCf}\cdot\tilde{D}, \\ d\tilde{Z} &= \quad Q_{SCf}(z,\theta)\cdot dZ, \end{cases} \tag{2.43}$$

$$\mathsf{TPt} \quad : \quad \begin{cases} \partial &= \quad \partial\tilde{\theta}_{TPt}\cdot\tilde{D}, \\ d\tilde{Z} &= \quad \varDelta_{TPt}(z,\theta)\cdot d\theta. \end{cases} \tag{2.44}$$

Условие $\varDelta(z,\theta)=0$ (2.37) в обратимом случае задает обычные суперконформные преобразования $\mathcal{T}_{SCf}$ [111, 332, 345, 405], и приведенная матрица $\mathrm{P}_{SCf}$ (2.39) представляет собой результат стандартной редукции структурной супергруппы (см., например, [367]).

Другое условие $Q(z,\theta)=0$ (2.38) приводит к необратимым преобразованиям $\mathcal{T}_{TPt}$ (см. [1]). Из (2.44) следует, что они приводят к изменению четности касательного пространства, и поэтому определение (2.24) имеет смысл.

*Замечание* **2.25.** Альтернативная редукция [8] суперматрицы $\mathrm{P}_A$ каса-тельного пространства приводит к антитреугольной суперматрице $\mathrm{P}_{TPt}$ (2.40).

Дуальная роль суперконформных и сплетающих четность преобразований отчетливо видна из теоремы сложения березинианов (2.34) (см. [8]) и операторов проекций (2.39) и (2.40).

**Предположение 2.26.** *Поскольку суперконформные преобразования могут быть рассмотрены в качестве супераналога комплексной структуры* [406, 407], *мы можем трактовать сплетающие четность преобразования как иной нечетный $N = 1$ супераналог комплексной структуры* [9].

Более естественно называть сплетающие четность преобразования *антисуперконформными* из-за следующей аналогии с несуперсимметричным случаем. Для обыкновенной $2 \times 2$ матрицы $\mathrm{P} = \begin{pmatrix} a & b \\ c & d \end{pmatrix}$ мы, очевидно, имеем следующее тождество

$$\det P = \det \begin{pmatrix} a & 0 \\ 0 & d \end{pmatrix} + \det \begin{pmatrix} 0 & b \\ c & 0 \end{pmatrix} = \det \mathrm{P}_{diag} + \det \mathrm{P}_{antidiag}, \qquad (2.45)$$

которое можно назвать “формулой сложения детерминантов”. В теории комплексных функций первая матрица описывает матрицу касательного пространства для голоморфных отображений, а вторая – антиголоморфных отображений.

*Замечание* **2.27.** В суперсимметричном случае *треугольная и антитреугольная* суперматрицы $\mathrm{P}_S$ и $\mathrm{P}_T$ играют роль, подобную несуперсимметричным *диагональной и антидиагональной* матрицам в обычной теории матриц, как это видно из (2.34). Поэтому, если $\mathrm{P}_{SCf}$ обобщает матрицу касательного пространства для голоморфных отображений, суперматрицы $\mathrm{P}_{TPt}$ могут рассматриваться как соответственное обобще-ние для антиголоморфных отображений.

**Следствие 2.28.** *Очевидно, что*

$$\operatorname{Ber} \mathrm{P}_T|_{\varDelta(z,\theta)=0} = \operatorname{Ber} \begin{pmatrix} 0 & \partial\tilde{\theta}_{SCf} \\ 0 & D\tilde{\theta}_{SCf} \end{pmatrix} = 0, \tag{2.46}$$

$$\operatorname{Ber} \mathrm{P}_S|_{Q(z,\theta)=0} = \operatorname{Ber} \begin{pmatrix} 0 & \partial\tilde{\theta}_{TPt} \\ 0 & D\tilde{\theta}_{TPt} \end{pmatrix} = 0. \tag{2.47}$$

*Замечание* **2.29.** Отметим, что вырожденные суперматрицы в (2.46)–(2.47) различны $\mathrm{P}_S|_{Q(z,\theta)=0} \neq \mathrm{P}_T|_{\varDelta(z,\theta)=0}$, поскольку различны условия, налагаемые на их ненулевые элементы, $\partial\tilde{\theta}_{SCf} \neq \partial\tilde{\theta}_{TPt}$ и $D\tilde{\theta}_{SCf} \neq D\tilde{\theta}_{TPt}$.

Используя данные соотношения наряду с (2.39) и (2.40), мы можем спроектировать формулу сложения березинианов (2.34) на редуцированные преобразования $\mathcal{T}_{SCf}$ и $\mathcal{T}_{TPt}$ следующим образом

$$\begin{aligned} \operatorname{Ber} \mathrm{P}_A \;&=\; \begin{cases} \operatorname{Ber} \mathrm{P}_S + \operatorname{Ber} \mathrm{P}_T, & \varDelta(z,\theta) = 0, \\ \operatorname{Ber} \mathrm{P}_S + \operatorname{Ber} \mathrm{P}_T, & Q(z,\theta) = 0. \end{cases} = \\ \begin{cases} \operatorname{Ber} \mathrm{P}_{SCf} + 0, \\ 0 + \operatorname{Ber} \mathrm{P}_{TPt}, \end{cases} \;&=\; \begin{cases} \operatorname{Ber} \mathrm{P}_{SCf}, & (\mathsf{SCf}) \\ \operatorname{Ber} \mathrm{P}_{TPt}, & (\mathsf{TPt}) \end{cases} \end{aligned} \tag{2.48}$$

После соответствующих проекций для $Q(z,\theta)$ и $\varDelta(z,\theta)$ мы имеем

$$Q_{SCf}(z,\theta) \overset{def}{=} \left(\partial\tilde{z} - \partial\tilde{\theta}\cdot\tilde{\theta}\right)|_{\varDelta(z,\theta)=0} = \left(D\tilde{\theta}_{SCf}\right)^2, \tag{2.49}$$

$$\varDelta_{TPt}(z,\theta) \overset{def}{=} \left(D\tilde{z} - D\tilde{\theta}\cdot\tilde{\theta}\right)|_{Q(z,\theta)=0} = \partial_\theta\tilde{z}_{TPt} - \partial_\theta\tilde{\theta}_{TPt}\cdot\tilde{\theta}_{TPt}. \tag{2.50}$$

*Замечание* **2.30.** Примечательно отметить сходство формул (2.49) и (2.50), что доказывает нам еще раз *дуальность* между суперконформными и сплетающими четность преобразованиями.

Используя (2.49), можно получить [367]

$$\mathrm{P}_{SCf} = \begin{pmatrix} \left(D\tilde{\theta}_{SCf}\right)^2 & \partial\tilde{\theta}_{SCf} \\ 0 & D\tilde{\theta}_{SCf} \end{pmatrix}. \tag{2.51}$$

Если $\epsilon\left[D\tilde{\theta}_{SCf}\right] \neq 0$, тогда $\operatorname{Ber}\mathrm{P}_{SCf}$ может быть просто вычислен из (2.51) (см. [111, 341])

$$\operatorname{Ber}\mathrm{P}_{SCf} = D\tilde{\theta}_{SCf}. \tag{2.52}$$

В необратимом случае $\epsilon\left[D\tilde{\theta}_{SCf}\right] = 0$ березиниан не может быть определен, но мы принимаем формулу (2.52) в качестве определения якобиана необратимых суперконформных преобразований (см. [1, 13]).

**Определение 2.31.** *Березиниан полунеобратимых суперконформных преобразований есть*

$$\operatorname{Ber}\mathrm{P}_{SCf}^{noninv} = D\tilde{\theta}_{SCf}. \tag{2.53}$$

Рассмотрим березиниан для сплетающих четность преобразований. Из (2.50) мы получаем антитреугольную матрицу

$$\mathrm{P}_{TPt} = \begin{pmatrix} 0 & \partial\tilde{\theta}_{TPt} \\ \varDelta_{TPt}(z,\theta) & D\tilde{\theta}_{TPt} \end{pmatrix}. \tag{2.54}$$

Если $\epsilon\left[D\tilde{\theta}_{TPt}\right] \neq 0$, то березиниан суперматрицы $\mathrm{P}_{TPt}$ (2.54) есть

$$\operatorname{Ber}\mathrm{P}_{TPt} = \frac{\varDelta_{TPt}(z,\theta) \cdot \partial\tilde{\theta}_{TPt}}{\left(D\tilde{\theta}_{TPt}\right)^2}. \tag{2.55}$$

Из (2.50) следует, что $D\varDelta_{TPt}(z,\theta) = -\left(D\tilde{\theta}_{TPt}\right)^2$, и поэтому

$$\partial\varDelta_{TPt}(z,\theta) = -2\cdot D\tilde{\theta}_{TPt}\cdot\partial\tilde{\theta}_{TPt}, \tag{2.56}$$

что дает для березиниана

$$\operatorname{Ber}\mathrm{P}_{TPt} = \frac{\partial\varDelta_{TPt}(z,\theta)\cdot\varDelta_{TPt}(z,\theta)}{2\left(D\tilde{\theta}_{TPt}\right)^3}. \tag{2.57}$$

*Замечание* **2.32.** Поскольку $\varDelta_{TPt}$ является нечетным и нильпотентным, березиниан $\operatorname{Ber}\mathrm{P}_{TPt}$ также нильпотентен и чисто ду́ховый.

Четные и нечетные суперфункции $Q(z,\theta)$ и $\varDelta(z,\theta)$ играют важную роль в возможных редукциях супераналитического структуры, и поэтому стоит исследовать их подробнее. Общее соотношение между $Q(z,\theta)$ и $\varDelta(z,\theta)$ есть

$$Q(z,\theta) - D\varDelta(z,\theta) = \left(D\tilde{\theta}\right)^2. \tag{2.58}$$

Из этой связи и (2.32) мы получаем другое полезное выражение для березиниана общего супераналитического преобразования (если $\epsilon\left[D\tilde{\theta}\right] \neq 0$)

$$\operatorname{Ber}\mathrm{P}_{SA} = D\tilde{\theta} + D\left(\frac{\varDelta(z,\theta)}{D\tilde{\theta}}\right) = D\left(\tilde{\theta} + \frac{\varDelta(z,\theta)}{D\tilde{\theta}}\right), \tag{2.59}$$

в котором суперконформное условие $\varDelta(z,\theta) = 0$ явно прослеживается явным образом.

В дальнейшем будет полезно иметь компонентные выражения

$$\begin{aligned}
\varDelta(z,\theta) &= \chi(z) - \psi(z)\cdot g(z) + \theta\cdot\left(f'(z) - \psi'(z)\cdot\psi(z) - g^2(z)\right),\\
Q(z,\theta) &= f'(z) - \psi'(z)\cdot\psi(z) + \theta\cdot\left(\chi(z) - \psi(z)\cdot g'(z) + \psi'(z)\cdot g(z)\right).
\end{aligned} \tag{2.60}$$

Из этих величин можно построить нечетную суперфункцию

$$\mathbf{\Sigma}(z,\theta) = \mathit{\Delta}(z,\theta) - \theta \cdot Q(z,\theta) = \chi(z) - \psi(z) \cdot g(z) - \theta \cdot g^2(z), \quad (2.61)$$

которая представляет собой важную характеристику преобразования. В частности, $\mathbf{\Sigma}(z,\theta) = 0$ для суперконформных преобразований с нильпотентной функцией $g(z)$, которые будут рассматриваться ниже.

Известно, что различные редукции матрицы касательного расслоения приводят к различным связям на кручение и различным $G$-структурам [408–412] в гравитации и супергравитации [413, 414]. Так, суперматрицы $\mathrm{P}_{SA}$ соответствуют различным вариантам наложения связей на кручение в двумерной супергравитации [415–418].

**Предположение 2.33.** *Аналогично тому, как треугольная редукция суперматрицы* $\mathrm{P}_{SA} \to \mathrm{P}_{SCf}$ (2.51) *отвечает суперконформной двумерной супергравитации* [368, 415] *и суперримановым поверхностям* [367], *склеенным с помощью суперконформных преобразований* [111], *можно предположить, что альтернативная редукция* $\mathrm{P}_{SA} \to \mathrm{P}_{TPt}$ (2.54) *отвечает нечетному необратимому аналогу двумерной супергравитации и, соответственно, нечетному аналогу суперримановых поверхностей* [9], *склеенных с помощью сплетающих четность преобразований* (*см.* **Определение 2.24**).

Рассмотрим более подробнее преобразование производных (2.24) при общем супераналитическом отображении

$$\begin{aligned} \partial &= \partial\tilde{\theta} \cdot \tilde{D} + Q(z,\theta) \cdot \tilde{\partial}, \\ D &= D\tilde{\theta} \cdot \tilde{D} + \mathit{\Delta}(z,\theta) \cdot \tilde{\partial}. \end{aligned} \quad (2.62)$$

Исключая первые слагаемые в правой части (2.62), определим чет-ный дифференциальный оператор $\hat{\mathsf{R}}$ по формуле

$$\hat{\mathsf{R}} \overset{def}{=} D\tilde{\theta} \cdot \partial - \partial\tilde{\theta} \cdot D = \left(D\tilde{\theta} \cdot Q\left(z,\theta\right) - \partial\tilde{\theta} \cdot \varDelta\left(z,\theta\right)\right)\tilde{\partial}. \qquad (2.63)$$

Если $\epsilon\left[D\tilde{\theta}\right] \neq 0$, то, используя (2.32), для $\hat{\mathsf{R}}$ в общем случае суперaналитических преобразований получаем

$$\hat{\mathsf{R}} = \left(D\tilde{\theta}\right)^2 \cdot \mathrm{Ber}\left(\tilde{Z}/Z\right) \cdot \tilde{\partial}. \qquad (2.64)$$

Тогда для суперконформно-подобных преобразований имеем

$$\hat{\mathsf{R}} = \begin{cases} \left(D\tilde{\theta}_{SCf}\right)^3 \cdot \tilde{\partial}, & (\mathsf{SCf}), \\ \dfrac{\partial\varDelta_{TPt}\left(z,\theta\right) \cdot \varDelta_{TPt}\left(z,\theta\right)}{2\left(D\tilde{\theta}_{TPt}\right)} \cdot \tilde{\partial}, & (\mathsf{TPt}). \end{cases} \qquad (2.65)$$

Отсюда видно, что, как и в (2.57), оператор $\hat{\mathsf{R}}$ для сплетающих четность преобразований нильпотентен.

**2.1.4.** Вырожденные преобразования. Очевидно, что вырожденным преобразованиям соответствует нулевой дифференциальный оператор $\hat{\mathsf{R}} = 0$, а следовательно, и нулевой необратимый якобиан (E.12), но не березиниан (E.9), который в данном случае не определен вообще.

**Определение 2.34.** *Вырожденные преобразования определяются нулевым якобианом $\boldsymbol{J}^{noninv} = 0$ и оператором $\hat{\mathsf{R}} = 0$.*

В терминах компонентных функций (2.2) уравнения вырожденных

преобразований имеют вид

$$g(z) \cdot f'(z) = \psi'(z) \cdot \chi(z), \\ g(z) \cdot \chi'(z) = g'(z) \cdot \chi(z). \tag{2.66}$$

После алгебраических преобразований можно получить следствие

$$g'(z) \cdot f'(z) = \psi'(z) \cdot \chi'(z). \tag{2.67}$$

Имеется два типа вырожденных преобразований, *левые* и *правые*, в соответствии с тем, какой из столбцов суперматрицы $\mathrm{P}_{SA}$ в (2.24) зануляется.

Пересечение множеств суперматриц $\mathcal{P}_{D_L} = \mathcal{P}_S \cap \mathcal{P}_T$ представляет собой множество *левых* вырожденных матриц $\mathrm{P}_{D_{L:}} \in \mathcal{P}_{D_L}$ формы

$$\mathrm{P}_{D_L} \stackrel{def}{=} \begin{pmatrix} 0 & \partial\tilde{\theta} \\ 0 & D\tilde{\theta} \end{pmatrix}. \tag{2.68}$$

Отсюда видно, что $\mathrm{P}_{D_L}$ зависит от преобразования только нечетной координаты $\theta$. Вырожденная матрица вида (2.68) может получаться из $\mathrm{P}_S$ и $\mathrm{P}_T$ матриц соответствующими проекциями (2.46). Это означает, что, если преобразование нечетного сектора задано, т. е. фиксированы функции $\psi(z)$ и $g(z)$, то условия (2.38) и (2.37) определяют поведение четного сектора (функции $f(z)$ и $\chi(z)$). При этом, поскольку вырожденная матрица $\mathrm{P}_{D_L}$ зависит только от нечетного сектора преобразования, мы получаем

$$\mathrm{P}_{D_L} = \mathrm{P}_{SCf}|_{Q(z,\theta)=0} = \mathrm{P}_{TPt}|_{\Delta(z,\theta)=0} \tag{2.69}$$

(ср. *Замечание* **2.29**).

*Левые* вырожденные преобразования характеризуются только одной нечетной функцией $\psi(z)$ и отсутствием $\theta$-зависимости преобразования $Z \to \tilde{Z}$ (см. (2.50)), так что

$$\begin{cases} \tilde{z}_{Deg_L} = f(z), \\ \tilde{\theta}_{Deg_L} = \psi(z), \end{cases} \tag{2.70}$$

где

$$f'(z) = \psi'(z)\,\psi(z). \tag{2.71}$$

Решение последнего уравнения можно представить в виде бесконечного ряда [1,3]

$$f(z) = \sum_{n=0}^{\infty} \frac{z^{n+1}}{(n+1)!} \left(-\frac{\partial}{\partial z}\right)^n (\psi'(z) \cdot \psi(z)) + c, \tag{2.72}$$

где $c = const$.

Поскольку суперматрицы с левым нулевым столбцом замкнуты относительно умножения, то левые вырожденные преобразования образуют полугруппу $\boldsymbol{T}_{Deg_L}$. Из явного вида (2.70) следует, что полугруппе преобразований $\boldsymbol{T}_{Deg_L}$ соответствует полугруппа функций $\mathbf{S}_{Deg_L}$, элемент которой $\mathbf{S}_{Deg_L} \ni \mathbf{d}_L = \{\psi\}$ определяется одной нечетной функцией $\psi(z)$, а левое умножение имеет вид

$$\begin{aligned} \{\psi_1\} *_L \{\psi_2\} &= \{\psi_1 \circ f_2\}, \\ f_2'(z) &= \psi_2'(z) \cdot \psi_2(z). \end{aligned} \tag{2.73}$$

**Утверждение 2.35.** *Левое умножение* (2.73) *замкнуто и ассоциативно, и поэтому* $\mathbf{S}_{Deg_L}$ *действительно — полугруппа.*

*Замечание* **2.36.** Преобразование (2.70) является $1 \to 2$ преобразова-

нием и поэтому представляет собой вложение*).

Рассмотрим по аналогии *правые* вырожденные преобразования.

**Утверждение 2.37.** *Правые вырожденные преобразования описываются уравнением*

$$D\tilde{\theta} = 0. \tag{2.74}$$

*Доказательство.* Если $D\tilde{\theta} = 0$, тогда $\tilde{\theta} = \alpha = const$, а также $\partial\tilde{\theta} = D\left(D\tilde{\theta}\right) = 0$, что соответствует суперматрице $\mathrm{P}_{SA}$ с правым нулевым столбцом. ■

Таким образом, учитывая условие (2.74) и выражения для $Q(z,\theta)$ (2.37) и $\varDelta(z,\theta)$ (2.38), получаем

$$\mathrm{P}_{D_R} = \begin{pmatrix} Q(z,\theta)\,|_{\partial\tilde{\theta}=0} & 0 \\ \varDelta(z,\theta)\,|_{D\tilde{\theta}=0} & 0 \end{pmatrix} = \begin{pmatrix} \partial\tilde{z} & 0 \\ D\tilde{z} & 0 \end{pmatrix}. \tag{2.75}$$

В этом случае нечетный сектор становится вырожденным, представляя собой левые нули и константные отображения аналогично (2.18). Такие отображения формируют ограничительные полугруппы (см., например, ( [191, 399, 421])).

В несуперсимметричном случае различные отображения $2 \to 1$ изучались в [422], а голоморфные отображения между пространствами различных размерностей рассматривались в [423, 424].

Тем не менее, полное супераналитическое преобразование (2.2) не является левым нулем из-за (2.20) и имеет следующий вид

$$\begin{cases} \tilde{z} = f(z) + \theta\cdot\chi(z), \\ \tilde{\theta} = \alpha. \end{cases} \tag{2.76}$$

---

*) *Примечание.* Общие вопросы вложения суперпространств и супермногообразий изложены в [419, 420].

Эти преобразования необратимы (из-за вырожденного нечетного сектора) и формируют полугруппу правых вырожденных преобразований $\boldsymbol{T}_{Deg_R}$, которая является подполугруппой в $\boldsymbol{T}_{SA}$, вследствие $\mathcal{P}_{D_R} \cdot \mathcal{P}_{D_R} \subseteq \mathcal{P}_{D_R}$.

Элемент соответствующей полугруппы функций $\mathbf{S}_{Deg_R}$ запишем в виде

$$\mathbf{S}_{Deg_R} \ni \mathbf{d}_R = \left\{ \begin{array}{c} f \\ \chi \\ \alpha \end{array} \right\}, \tag{2.77}$$

а умножение в $\mathbf{S}_{Deg_R}$ имеет

$$\left\{ \begin{array}{l} f_1 \\ \chi_1 \\ \alpha_1 \end{array} \right\} *_R \left\{ \begin{array}{l} f_2 \\ \chi_2 \\ \alpha_2 \end{array} \right\} = \left\{ \begin{array}{l} f_1 \circ f_2 + \alpha_1 \cdot \chi_1 \circ f_2 \\ \chi_2 \cdot f_1' \circ f_2 + \chi_1' \circ f_2 \cdot \alpha_2 \\ \alpha_1 \end{array} \right\}. \tag{2.78}$$

**Утверждение 2.38.** *Правое умножение* (2.78) *замкнуто и ассоциативно.*

Схематически умножение вырожденных и рассмотренных ранее преобразований можно представить в виде *Таблицы* 2.1. Отсюда следует

*Таблица* 2.1

*Умножение обратимых и необратимых редуцированных $N = 1$ преобразований, включая вырожденные*

| | SCf | TPt | $\mathrm{Deg}_L$ | $\mathrm{Deg}_R$ |
|---|---|---|---|---|
| SCf | SCf | SA | $\mathrm{Deg}_L$ | $\mathrm{Deg}_R$ |
| TPt | TPt | SA | $\mathrm{Deg}_L$ | $\mathrm{Deg}_R$ |
| $\mathrm{Deg}_L$ | $\mathrm{Deg}_L$ | $\mathrm{Deg}_L$ | $\mathrm{Deg}_L$ | $\mathrm{Deg}_R$ |
| $\mathrm{Deg}_R$ | $\mathrm{Deg}_R$ | $\mathrm{Deg}_R$ | $\mathrm{Deg}_L$ | $\mathrm{Deg}_R$ |

**Утверждение 2.39.** *Множества преобразований $\mathfrak{T}_{Deg_{L,R}}$ (не рассматриваемые как полугруппы) есть идеалы в $\mathfrak{T}_{SA}$, $\mathfrak{T}_{SCf}$ и $\mathfrak{T}_{TPt}$, а $\mathfrak{T}_{SCf}$, $\mathfrak{T}_{Deg_R}$ и $\mathfrak{T}_{Deg_L}$ — замкнутые подмножества в $\mathfrak{T}_{SA}$.*

**2.1.5.** Альтернативная параметризация. Условия редукции (2.38) и (2.37) определяют 2 из 4 компонентных функций в (2.2) в каждом случае. Обычно [111, 176, 405] суперконформные преобразования $\mathcal{T}_{SCf}$ параметризуются парой функций

$$\mathbf{s}_{old} = \begin{pmatrix} f \\ \psi \end{pmatrix}, \tag{2.79}$$

тогда, как остальные функции находятся из (2.38) и (2.37). Однако очевидно, последнее можно сделать только для обратимых преобразований. Чтобы избежать этой трудности, мы вводим альтернативную параметризацию другой парой [9, 13]

$$\mathbf{s} = \begin{pmatrix} g \\ \psi \end{pmatrix}, \tag{2.80}$$

что позволяет нам исследовать редуцированные преобразования объединенным образом и естественно включить в рассмотрение необратимость [1, 21].

В самом деле, фиксируя $g(z)$ и $\psi(z)$, мы получаем из (2.37) и (2.38) для остальных компонентных функций из (2.2) уравнения

$$\begin{cases} f'_{\mathrm{m}}(z) &= \psi'(z)\cdot\psi(z) + \frac{1+\mathrm{m}}{2} g^2(z), \\ \chi'_{\mathrm{m}}(z) &= g'(z)\cdot\psi(z) + \mathrm{m} g(z)\cdot\psi'(z), \end{cases} \tag{2.81}$$

где $\mathrm{m} = \begin{cases} +1, & \mathsf{SCf}, \\ -1, & \mathsf{TPt}, \end{cases}$ может трактоваться в качестве проекции некоторого "спина редукции", который переключает тип преобразования.

Таким образом, редуцированное преобразование четной координаты (см. (2.2)) должно содержать данный добавочный индекс, т. е. $z \to \tilde{z}_{\mathrm{m}}$ (в этом месте дополнительно к (2.34) становится прозрачной аналогия с комплексной структурой).

Поскольку $f'_{-1}(z) = \psi'(z) \cdot \psi(z)$ является нильпотентным, TPt преобразования всегда необратимы и вырождены после числового отображения [9].

Объединенный закон умножения суперконформных и сплетающих четность преобразований имеет вид

$$\begin{pmatrix} g_1 \\ \psi_1 \end{pmatrix}_{\mathrm{m}_1} * \begin{pmatrix} g_2 \\ \psi_2 \end{pmatrix}_{\mathrm{m}_2} = \begin{pmatrix} g_2 \cdot g_1 \circ f_{2\mathrm{m}} + \chi_{2\mathrm{m}} \cdot \psi_2 \cdot g'_1 \circ f_{2\mathrm{m}} + \chi_{2\mathrm{m}} \cdot \psi'_1 \circ f_{2\mathrm{m}} \\ \psi_1 \circ f_{2\mathrm{m}} + \psi_2 \cdot g_1 \circ f_{2\mathrm{m}} \end{pmatrix}, \tag{2.82}$$

где $(*)$ есть композиция преобразований и $(\circ)$ –композиция функции.

Для проекции "спина редукции" мы имеем только два определенных произведения $(+1) * (+1) = (+1)$ и $(+1) * (-1) = (-1)$ (см. также **Приложение З** и диаграммы (З.21) и (4.13)). Первое выражение представляет собой следствие умножения множеств матриц $\mathcal{P}_S \star \mathcal{P}_S \subseteq \mathcal{P}_S$ (см. (2.35)), это есть проявление того факта, что суперконформные преобразования $\mathcal{T}_{SCf}$ формируют подструктуру [367], т. е. подполугруппу $\boldsymbol{T}_{SCf}$ супераналитической полугруппы $\boldsymbol{T}_{SA}$ (в обратимом случае – подгруппу [111, 341, 367]).

## 2.2. Суперконформные полугруппы

Исследование новых абстрактных типов полугрупп и их идеалов [204, 425–428] представляет само по себе важную теоретико-категорную задачу. Интерес к изучению суперконформных полугрупп обусловлен прежде всего тем, что они имеют необычные идеальные (негрупповые) свойства [5], которые можно использовать в приложениях к теоретическим моделям элементарных частиц.

**2.2.1.** Локальное строение $N=1$ суперконформной полугруппы. Рассмотрим свойства обратимости $N=1$ суперконформных преобразований, связанные с нильпотентностью компонентных функций $g(z)$, входящих в альтернативную параметризацию (2.80). Так, обобщенный суперякобиан (E.12) суперконформных (обратимых и необратимых) преобразований в терминах компонент элемента $\mathbf{s}$ (2.80) имеет вид

$$\boldsymbol{J}_{SCf} = D\tilde{\theta}_{SCf} = g\left(z\right) + \theta \cdot \psi'\left(z\right), \tag{2.83}$$

что следует из (2.53) и (2.2).

**Предложение 2.40.** *Индекс необратимости* (E.15) *общего суперконформного преобразования и его степень необратимости* (E.16) *связаны с индексом функции* $g(z)$ *формулой*

$$\operatorname{ind} \boldsymbol{J}_{SCf} = \frac{1}{\boldsymbol{m}_{SCf}} = \operatorname{ind} g(z) + 1. \tag{2.84}$$

*Доказательство.* Возведем обе части равенства (2.83) в степень $n$ и воспользуемся тем, что грассманов индекс нильпотентности второго слагаемого в нем минимален и равен двум, тогда получим

$$\boldsymbol{J}_{SCf}^{n} = g^{n}(z) + n \cdot g^{n-1}(z) \cdot \theta \cdot \psi'(z). \tag{2.85}$$

Отсюда и следует соотношение (2.84). ■

Из (2.85) видно, что имеется другая возможность в зависимости от присутствия последнего слагаемого. Среди необратимых суперконформных преобразований с $\operatorname{ind} g\,(z) = n$ можно выделить следующие преобразования, имеющие существенно отличные от общего случая абстрактные свойства.

**Определение 2.41.** Ann-*преобразования, имеющие индекс необратимости* $n$*, определяются формулами*

$$\operatorname{ind} g(z) = n,\ g^{n-1}(z) \in \operatorname{Ann} \psi'(z). \tag{2.86}$$

**Предложение 2.42.** *Индекс необратимости* Ann-*преобразования равен индексу функции* $g(z)$

$$\operatorname{ind} \boldsymbol{J}^{A}_{SCf} = \operatorname{ind} g(z). \tag{2.87}$$

*Доказательство.* Из (2.86) следует, что второе слагаемое в (2.85) равно нулю. Отсюда получаем (2.87). ■

Соотношения (2.84) и (2.87) справедливы лишь для суперконформных преобразований, т. е. они являются условиями суперконформности, записанными через индексы нильпотентности [13].

Элементы суперконформной полугруппы с $\operatorname{ind} g(z) = 1$ являются обратимыми, а элементы с $\operatorname{ind} g(z) = 0$ – необратимыми. Обратимые элементы полугруппы $\mathbf{g}$ составляют подгруппу $\mathbf{G}_{SCf} = \cup\, \mathbf{g}$ суперконформной полугруппы, а необратимые элементы $\mathbf{i}$ с нулем $\mathbf{z}$ — ее идеал $\mathbf{I}_{SCf} = \cup\, \mathbf{i} \cup \mathbf{z}$. Из закона умножения (2.82) следует, что если хотя бы один из сомножителей необратим, то и результирующее преобразование также необратимо, т.е. $\mathbf{I}_{SCf} * \mathbf{S}_{SCf} \subseteq \mathbf{I}_{SCf},\ \ \mathbf{S}_{SCf} * \mathbf{I}_{SCf} \subseteq \mathbf{I}_{SCf}$, поэтому $\mathbf{I}_{SCf}$ – изолированный идеал, а подгруппа $\mathbf{G}_{SCf}$ — фильтр (см. определения в **Приложении А**). Обратимые суперконформные преобразо-вания, соответствующие элементам $\mathbf{G}_{SCf}$, рассматривались в [111, 227, 429]. Поэтому подробнее остановимся на необратимых преобразованиях и структуре идеала $\mathbf{I}_{SCf}$.

Выделим в идеале $\mathbf{I}_{SCf}$ следующие подмножества элементов:

$$\mathbf{I}_n \stackrel{def}{=} \left\{\mathbf{i} \in \mathbf{I}_{SCf} \mid g^n(z) = 0\right\}, \tag{2.88}$$

$$\mathbf{J}_n \stackrel{def}{=} \left\{\mathbf{i} \in \mathbf{I}_n \mid \operatorname{ind} g(z) = n\right\}, \tag{2.89}$$

$$\mathbf{J}_n^A \stackrel{def}{=} \left\{\mathbf{i} \in \mathbf{J}_n \mid g^{n-1}(z) \in \operatorname{Ann} \psi'(z)\right\}, \tag{2.90}$$

которые связаны очевидными соотношениями $\mathbf{J}_n = \mathbf{I}_n \setminus \mathbf{I}_{n-1}$, причем $\mathbf{I}_0 = \mathbf{J}_0 = \mathbf{z}$.

Пусть $\mathbf{s}_3 = \mathbf{s}_1 * \mathbf{s}_2$, тогда из (2.82) при $\mathrm{m}_1 = \mathrm{m}_2 = \mathrm{m}_3 = +1$ имеем

$$g_3(z) = g_1\left(f_2(z)\right) \cdot g_2(z) + \psi_2(z) \cdot \psi_1'\left(f_2(z)\right) \cdot g_2(z), \tag{2.91}$$

где $f_2'(z) = g_2^2(z) + \psi_2'(z) \cdot \psi_2(z)$. Возводя (2.91) в степень $n$ в грассмановой алгебре, получаем

$$\begin{aligned} g_3^n(z) \;=\; & g_1^n\left(f_2(z)\right) \cdot g_2^n(z) + \\ & + n \cdot g_1^{n-1}\left(f_2(z)\right) \cdot \psi_2(z) \cdot \psi_1'\left(f_2(z)\right) \cdot g_2^n(z). \end{aligned} \tag{2.92}$$

Отсюда следует, что здесь условием обращения в нуль второго слагаемого по-прежнему является (2.86), и это снова выделяет необратимые Ann-преобразования.

**Теорема 2.43.** *Множество элементов $\mathbf{J}_n^A \subseteq \mathbf{I}_n$ является правым идеалом для $\mathbf{I}_n$ относительно* (2.86).

*Доказательство.* Пусть $\mathbf{s}_3 = \mathbf{s}_1 * \mathbf{s}_2$, $\mathbf{s}_i \in \mathbf{I}_n$, и для $\mathbf{s}_1$ выполняется (2.86), т. е. $g_1^{n-1}(z) \cdot \psi_1'(z) = 0$. Покажем, что $g_3^{n-1}(z) \cdot \psi_3(z) = 0$.

Из (2.91) имеем

$$\omega_3(z) = \omega_1(z) \cdot g_2^{n+1}(z)$$
$$+\omega_2(z) \cdot g_1^n \left(h_2(z)\right) + n \cdot \psi_2(z) \cdot \omega_1(z) \cdot \omega_2(z) +$$
$$+g_1^{n-1} \left(h_2(z)\right) \cdot g_1' \left(h_2(z)\right) \cdot g_2^{n-1}(z) \cdot \psi_2(z) = 0,$$

где

$$\begin{aligned} \omega_1(z) &= g_1^{n-1} \left(h_2(z)\right) \cdot \psi_1' \left(h_2(z)\right), \\ \omega_2(z) &= g_2^{n-1}(z) \cdot \psi_2'(z), \omega_3(z) = g_3^{n-1}(z) \cdot \psi_3'(z), \end{aligned}$$

и в последнем равенстве использована очевидная импликация $g^n(z) = 0 \Rightarrow g^{n-1}(z) \cdot g'(z) = 0$. Поэтому $\mathbf{J}_n^A * \mathbf{I}_n \subseteq \mathbf{J}_n^A$. ■

Отсюда следует, что $\mathbf{J}_n^A * \mathbf{J}_n^A \subseteq \mathbf{J}_n^A$, т.е. множество $\mathbf{J}_n^A$ замкнуто относительно свойства (2.86), поэтому объединение $\bigcup\limits_n \mathbf{J}_n^A = \mathbf{A}_{SCf}$ есть подполугруппа в $\mathbf{S}_{SCf}$, которую будем называть Ann-*полугруппой*.

**2.2.2.** Ann - п о л у г р у п п а . Свойства идеалов в Ann-полугруппе существенно отличаются от таковых в оставшейся части суперконформной полугруппы, поэтому рассмотрим их отдельно.

**Предложение 2.44.** *Все элементы из* Ann-*полугруппы необратимы, следовательно, групповая часть в* $\mathbf{A}_{SCf}$ *отсутствует.*

*Доказательство.* Из (2.86) следует, что

$$g^{n-1}(z) \cdot \psi'(z) = 0, \tag{2.93}$$

поэтому $\operatorname{ind} g(z) < \infty$ (считаем, что $\psi'(z) \neq 0$). ■

Чтобы изучить свойства нильпотентности Ann-преобразований, возведем (2.91) в степень $n$ при учете (2.93), тогда получим Ann-аналог

соотношения (2.92)

$$g_3^n(z) = g_1^n\left(h_2(z)\right) \cdot g_2(z). \tag{2.94}$$

Отсюда видно, что множества элементов

$$\mathbf{A}_n \stackrel{def}{=} \{\mathbf{s} \in \mathbf{A}_{SCf} \mid g^n(z) = 0\} \tag{2.95}$$

являются двухсторонними идеалами в $\mathbf{A}_{SCf}$ и, кроме того, имеют место строгие включения $\mathbf{A}_{n-1} \subset \mathbf{A}_n$. Следовательно, идеалу Ann-полугруппы $\mathbf{I}^A \equiv \mathbf{A}_{SCf}$ можно поставить в соответствие бесконечную двусторонне-идеальную цепь

$$\mathbf{z} \subset \mathbf{A_1} \subset \mathbf{A_2} \ldots \subset \mathbf{A_n} \subset \ldots \mathbf{I}^A \equiv \mathbf{A}_{SCf}, \tag{2.96}$$

начинающуюся с тривиального минимального идеала – нуля $\mathbf{z}$ Ann-полугруппы – и заканчивающуюся самой полугруппой $\mathbf{A}_{SCf}$. Идеальные цепи различных полугрупп рассматривались в [430–432].

Из закона умножения (2.94) следует, что каждый идеал $\mathbf{A}_n$ содержит нильидеал (см., например, [160, 433])

$$\mathbf{N}_n \stackrel{def}{=} \left\{\mathbf{s} \in \mathbf{A}_n \mid \mathbf{s}^{*n} = \mathbf{z}\right\}, \tag{2.97}$$

причем реализуется строгое включение $\mathbf{N}_n \subset \mathbf{A}_n$. Можно показать, что разность $\mathbf{A}_n \setminus \mathbf{N}_n$ содержит только нильэлементы более высокого полугруппового индекса и, следовательно, принадлежит к соответствующим нильидеалам. Поэтому объединение всех нильидеалов совпадает с Ann-полугруппой. Таким образом, Ann-полугруппа является нильполугруппой [434–438]. Поскольку $\mathbf{A}_{n-1}$ есть идеал в $\mathbf{A}_n$, то, как это следует из (2.94), идеальная цепь (2.96) представляет собой идеальный ряд Ann-полугруппы. Факторами этого ряда являются фактор-полугруппы Риса $\mathbf{A}_n/\mathbf{A}_{n-1}$, и для них коидеал $\mathbf{A}_n \setminus \mathbf{A}_{n-1}$ совпадает с $\mathbf{J}_n^A$ (2.90). Кроме того, $\mathbf{A}_{n+1}/\mathbf{A}_n$ является идеалом фактор-полугруппы $\mathbf{I}^A/\mathbf{A}_n$, и выполняется следующее соотношение:

$$\mathbf{I^A}/\mathbf{A_{n+1}} \cong (\mathbf{I^A}/\mathbf{A_n})/(\mathbf{A_{n+1}}/\mathbf{A_n}).$$

Однако идеальный ряд (2.96) не является аннуляторным ни справа, ни слева, как этого следовало бы ожидать для нильполугруппы [435, 437, 439, 440].

Пользуясь (2.94) и очевидными свойствами нильпотентных элементов, для множеств $\mathbf{A}_n$ и $\mathbf{J}_n^A$ из Ann-полугруппы построим таблицу умножения

$$\begin{aligned} &\mathbf{A}_n * \mathbf{A}_m \subseteq \mathbf{A}_k, \quad \mathbf{J}_n^A * \mathbf{J}_m^A \subseteq \mathbf{A}_k, \\ &\mathbf{J}_n^A * \mathbf{A}_m \subseteq \mathbf{A}_k, \quad \mathbf{A}_n * \mathbf{J}_m^A \subseteq \mathbf{A}_k, \end{aligned} \tag{2.98}$$

где $k = \min(n, m)$.

Множество $\mathbf{A}_{SCf}$ представляет собой объединение взаимно непересекающихся множеств: $\mathbf{A}_{SCf} = \bigcup_n \mathbf{J}_n^A$, $\mathbf{J}_n^A \cap \mathbf{J}_m^A = \varnothing$, однако $\mathbf{J}_n^A$ не является подполугруппой ни для $\mathbf{A}_n$, ни для $\mathbf{A}_{SCf}$. Но с $\mathbf{J}_n^A$ можно связать полугруппу $\mathbf{U}_n^A \overset{def}{=} \{\mathbf{A}_n \cup \mathbf{z}, \circledast\}$, в которой умножение определяется формулой

$$\mathbf{s} \circledast \mathbf{t} \overset{def}{=} \begin{cases} \mathbf{s} * \mathbf{t}, & \mathbf{s} * \mathbf{t} \in \mathbf{J}_n^A, \\ \mathbf{z}, & \mathbf{s} * \mathbf{t} \notin \mathbf{J}_n^A. \end{cases} \tag{2.99}$$

Отметим, что полугруппа $\mathbf{U}_n^A$ может быть построена также и с помощью характеристической функции

$$\mathbf{c}_n(\mathbf{s}) \overset{def}{=} \begin{cases} \mathbf{e}, \mathbf{s} \in \mathbf{J}_n^A, \\ \mathbf{z}, \mathbf{s} \notin \mathbf{J}_n^A. \end{cases} \tag{2.100}$$

Тогда умножение в (2.99) можно представить следующим образом:

$$\mathbf{s} \circledast \mathbf{t} = \mathbf{c}_n(\mathbf{s} * \mathbf{t}) * \mathbf{s} * \mathbf{t}. \tag{2.101}$$

При одинаковых индексах из (2.98) имеем $\mathbf{A}_n * \mathbf{A}_n \subseteq \mathbf{A}_n$. Поэтому представляется естественным выделить в $\mathbf{A}_n$ подмножество $\mathbf{A}_n^{(k)} \subset \mathbf{A}_n$, обладающее свойством

$$\mathbf{A}_n^{(k)} * \mathbf{A}_n^{(k)} \subseteq \mathbf{J}_k^A,\ 0 \leq k \leq n, \tag{2.102}$$

что можно трактовать как извлечение квадратного корня из $\mathbf{J}_k^A$. При $k = n$ получаем $\mathbf{U}_n^A = \mathbf{A}_n^{(n)} \cup \mathbf{z}$. В другом предельном случае, при $k = 0$, имеем $\mathbf{A}_n^{(0)} = \mathbf{A}_n \cap \mathbf{N}_2$. Но поскольку умножение (2.100) снова не замыкается, подмножество $\mathbf{A}_n^{(k)}$ не является полугруппой.

Из соотношений (2.98) получаем для главных идеалов (см. определения в **Приложении А**)

$$\mathbf{R}(\mathbf{s}) \subseteq \mathbf{A}_n,\ \ \mathbf{L}(\mathbf{s}) \subseteq \mathbf{A}_n,\ \ \mathbf{J}(\mathbf{s}) \subseteq \mathbf{A}_n, \tag{2.103}$$

где $\mathbf{s} \in \mathbf{J}_n^A$. Поскольку $\mathbf{A}_{SCf}$ – нильполугруппа, все отношения эквивалентности Грина (см. [103, 104, 428] и **Приложение А**) совпадают между собой и с отношением равенства $\mathbf{\Delta}$. По аналогии с [436, 441, 442] для Ann-полугруппы можно доказать следующую теорему.

**Теорема 2.45.** Ann-*полугруппа является* $\mathscr{J}$ *-тривиальной.*

*Доказательство.* Пусть $\mathbf{s} \in \mathbf{R}(\mathbf{s}) \wedge \mathbf{s} \neq \mathbf{z}$, тогда найдется элемент $\mathbf{t} \neq \mathbf{s}$ такой, что $\mathbf{s} = \mathbf{s} * \mathbf{t}$, а следовательно, и*) $\mathbf{s} = \mathbf{s} * \mathbf{t}^{*\mathbf{k}}$, где $k$ произвольно.

---

*Примечание.* Звездочка в степени означает умножение в рассматриваемой полугруппе, т. е. $\mathbf{t}^{*\mathbf{2}} = \mathbf{t} * \mathbf{t}$.

Но полугруппа $\mathbf{A}_{SCf}$ содержит по определению только нильэлементы, поэтому $\exists n$, $\mathbf{t}^{*\mathbf{n}} = \mathbf{z}$. Выберем $k = n$ и получим

$$\mathbf{s} = \mathbf{s} * \mathbf{t}^{*\mathbf{n}} = \mathbf{s} * \mathbf{z} = \mathbf{z},$$

что противоречит условию $\mathbf{s} \neq \mathbf{z}$. Наоборот, пусть $\mathbf{R}(\mathbf{s}) = \mathbf{R}(\mathbf{t})$, $\mathbf{s} \neq \mathbf{z}$, тогда из определения главных идеалов [104] получаем

$$\mathbf{s} = \mathbf{t} * \mathbf{x} = \mathbf{s} * (\mathbf{y} * \mathbf{x}) = \mathbf{s} * (\mathbf{y} * \mathbf{x})^{*\mathbf{k}}, \ \mathbf{x}, \mathbf{y} \in \mathbf{A}_{SCf}.$$

Снова в силу, того что $\mathbf{A}_{SCf}$ – нильполугруппа, найдется такая степень $n$, что $(\mathbf{y} * \mathbf{x})^{*\mathbf{n}} = \mathbf{z}$, поэтому $\mathbf{s} = \mathbf{s} * \mathbf{z} = \mathbf{z}$ — противоречие. Отсюда следует требуемая импликация $\mathbf{R}(\mathbf{s}) = \mathbf{R}(\mathbf{t}) \Rightarrow \mathbf{s} = \mathbf{t}$. Аналогично и для других отношений Грина. ■

**Следствие 2.46.** *$\mathscr{L}$, $\mathscr{R}$, $\mathscr{G}$-классы Ann-полугруппы содержат ровно по одному элементу.*

**2.2.3.** К в а з и и д е а л ь н ы й р я д . Переходим теперь к анализу идеального строения суперконформной полугруппы $\mathbf{S}_{SCf}$ в общем случае. В отличие от (2.86), полагаем, что $g^{n-1}(z) \notin \operatorname{Ann} \psi'(z)$. Такая полугруппа может содержать, кроме необратимых, также и обратимые элементы, а следовательно, подгруппу $\mathbf{G}_{SCf} \subset \mathbf{S}_{SCf}$, которая определяется преобразованиями с ненильпотентными и обратимыми $g(z)$. Если положить для обратимых элементов индекс нильпотентности равным бесконечности, то в терминах величин, введенных в (2.88–(2.90), имеем $\mathbf{G}_{SCf} = \mathbf{J}_\infty$, $\mathbf{S}_{SCf} = \mathbf{I}_\infty$, что позволяет в некоторых случаях формально включить $\mathbf{G}_{SCf}$ в закон умножения, аналогичный (2.98). Очевидно, что множество $\mathbf{G}_{SCf} \cup \{\mathbf{z}\}$ является фактор-полугруппой Риса $\mathbf{S}_{SCf}/\mathbf{I}_{SCf}$ [104]. Тогда суперконформную полугруппу $\mathbf{S}_{SCf}$ можно трак-товать как идеальное расширение [443–445] суперконформной группы $\mathbf{G}_{SCf}$ при помощи идеала $\mathbf{I}_{SCf}$.

Рассмотрим множества (2.88–(2.90) в случае полной суперконформной полугруппы $\mathbf{S}_{SCf}$. Очевидно, что строгие включения $\mathbf{I}_{n-1} \subset \mathbf{I}_n$ сохраняются. Поэтому идеалу суперконформной полугруппы $\mathbf{I}_{SCf}$ можно поставить в соответствие цепь множеств $\mathbf{I}_n$, аналогичную (2.96), следующим образом:

$$\mathbf{z} \subset \mathbf{I}_1 \subset \mathbf{I}_2 \subset \ldots \subset \mathbf{I}_n \subset \ldots \subset \mathbf{I}_{SCf}. \tag{2.104}$$

Однако в данном случае вместо (2.98) имеет место

**Предложение 2.47.** *Множества* $\mathbf{I}_n$ *удовлетворяют соотношениям*

$$\mathbf{S}_{SCf} * \mathbf{I}_n \subseteq \mathbf{I}_n, \tag{2.105}$$

$$\mathbf{I}_n * \mathbf{S}_{SCf} \subseteq \mathbf{I}_{n+1}, \tag{2.106}$$

$$\mathbf{S}_{SCf} * \mathbf{I}_n * \mathbf{S}_{SCf} \subseteq \mathbf{I}_{n+1}. \tag{2.107}$$

*Доказательство.* Действительно, если в (2.92) $g_1^n(z) = 0$ и $g_2^n(z) \neq 0$, то найдется такое $n = \operatorname{ind} g_1(z)$, что $g_1^{n-1}(z)$ может быть отлично от нуля, в то время как $g_3^{n+1}(z)= 0$ за счет обращения в нуль уже второго слагаемого в (2.91). ■

**Следствие 2.48.** *Множество* $\mathbf{I}_n$ *является только левым идеалом суперконформной полугруппы, но не правым и двухсторонним.*

**Предложение 2.49.** $\mathbf{I}_n$ *– квазиидеал* [446–448] *и одновременно биидеал* [449–452].

*Доказательство.* Из формул (2.94) и (2.105)–(2.107) непосредственно получаем свойства $\mathbf{I}_n$ как квазиидеала $\mathbf{S}_{SCf} * \mathbf{I}_n \cap \mathbf{I}_n * \mathbf{S}_{SCf} \subseteq \mathbf{I}_n$ и как биидеала $\mathbf{I}_n * \mathbf{S}_{SCf} * \mathbf{I}_n \subseteq \mathbf{I}_n$. ■

В соотношениях (2.106)–(2.107) происходит подъем лишь в соседнее множество $\mathbf{I}_{n+1}$ (в цепи (2.104)), поэтому $\mathbf{I}_n$ можно определить как правый и двухсторонний *повышающий идеал*. Таким образом, цепь (2.104) представляет собой левоидеальную цепь или цепь правых и двухсторонних повышающих идеалов $\mathbf{I}_n$. Поскольку из (2.107) следует, что $\mathbf{S}_{SCf} * \mathbf{I}_n \cup \mathbf{I}_n * \mathbf{S}_{SCf} \subseteq \mathbf{I}_{n+1}$, цепь (2.104) естественно назвать *антианнуляторным возрастающим рядом*, длина которого равна бесконечности. Можно предположить, что многие свойства антианнуляторного ряда (2.104) обусловлены нильпотентностью нильидеала $\mathbf{I}_{SCf}$, рассматриваемого как самостоятельная полугруппа (для аннуляторных рядов подобные связи установлены в [453–455]).

Непосредственно из (2.92) следует таблица умножения множеств $\mathbf{I}_n$ и $\mathbf{J}_n$ в общем случае:

$$\begin{aligned}
\mathbf{I}_n * \mathbf{I}_{n+k} &\subseteq \mathbf{I}_{n+1}, \\
\mathbf{I}_{n+k-1} * \mathbf{I}_n &\subseteq \mathbf{I}_n, \\
\mathbf{J}_n * \mathbf{J}_{n+k} &\subseteq \mathbf{I}_{n+1}, \\
\mathbf{J}_{n+k-1} * \mathbf{J}_n &\subseteq \mathbf{I}_n, \\
\mathbf{I}_n * \mathbf{J}_{n+k} &\subseteq \mathbf{I}_{n+1}, \\
\mathbf{I}_{n+k-1} * \mathbf{J}_n &\subseteq \mathbf{I}_n, \\
\mathbf{J}_n * \mathbf{I}_{n+k} &\subseteq \mathbf{I}_{n+1}, \\
\mathbf{J}_{n+k-1} * \mathbf{I}_n &\subseteq \mathbf{I}_n, \\
\mathbf{I}_n * \mathbf{G}_{SCf} &\subseteq \mathbf{I}_{n+1}, \\
\mathbf{G}_{SCf} * \mathbf{I}_n &\subseteq \mathbf{I}_n, \\
\mathbf{J}_n * \mathbf{G}_{SCf} &\subseteq \mathbf{I}_{n+1}, \\
\mathbf{G}_{SCf} * \mathbf{J}_n &\subseteq \mathbf{J}_n, \qquad (2.108)
\end{aligned}$$

где $k > 0$. Отсюда видно, что $\mathbf{I}_n$ является подполугруппой, так как $\mathbf{I}_n * \mathbf{I}_n \subseteq \mathbf{I}_n$, а множество $\mathbf{J}_n$ не является таковой, как и в случае Ann-полугруппы, что есть следствие наличия делителей нуля [121–123] и нильпотентов [159, 456–460] в суперконформной полугруппе.

Отметим, что из предпоследнего включения в (2.108) следует, что с помощью действия подгруппы $\mathbf{G}_{SCf}$ справа можно попасть в любое множество $\mathbf{I}_n$ с б´ольшим индексом, начиная с любого ненулевого члена левоидеального ряда (2.104). Из последних двух соотношений (2.108) имеем

$$\mathbf{G}_{SCf} * \mathbf{J}_n * \mathbf{G}_{SCf} \subseteq \mathbf{I}_{n+1}, \tag{2.109}$$

т. е. некоторые из элементов множества $\mathbf{J}_{n+1}$ оказываются сопряженными по подгруппе $\mathbf{G}_{SCf}$ с элементами предыдущего множества. По аналогии с [461–464] назовем два подмножества суперконформной полугруппы $\mathbf{A} \subseteq \mathbf{S}_{SCf}$ и $\mathbf{B} \subseteq \mathbf{S}_{SCf}$ *взаимно-$G$-нормальными*, если

$$\mathbf{g}^{-1} * \mathbf{A} * \mathbf{g} \subseteq \mathbf{B}, \ \mathbf{g} \in \mathbf{G}_{SCf}.$$

Тогда из (2.109) следует, что любые два соседние множества $\mathbf{J}_n$ из (2.108) содержат взаимно-$G$-нормальные элементы. Общие свойства классов сопряженных элементов в абстрактных полугруппах исследовались в [465], а в полугруппах преобразований — в работах [466–469].

**2.2.4.** Обобщенные отношения Грина. В случае суперконформной полугруппы стандартных отношений Грина [103, 433] недостаточно для описания всех классов элементов, что связано с (2.108). Чтобы обойти трудность, связанную с появлением $\mathbf{I}_{n+1}$ в правой части соотношения (2.108), построим при фиксированном $n$ разбиение суперконформной полугруппы на непересекающиеся части

$$\mathbf{S}_{SCf} \;=\; \mathbf{V}_1^{(n)} \cup \mathbf{V}_2^{(n)} \cup \mathbf{V}_3^{(n)} \cup \mathbf{V}_4, \tag{2.110}$$

$$\begin{aligned}
\mathbf{V}_i^{(n)} \cap \mathbf{V}_j^{(n)} &= \varnothing,\ i \neq j,\ \ \mathbf{V}_i^{(n)} \cap \mathbf{V}_4 = \varnothing, \\
\mathbf{V}_1^{(n)} &= \mathbf{I}_{n-1},\ \ \mathbf{V}_2^{(n)} = \mathbf{J}_n,\ \ \mathbf{V}_3^{(n)} = \mathbf{I}_{SCf} \setminus \mathbf{I}_n,
\end{aligned}$$

причем $\mathbf{V}_1^{(n)} \cup \mathbf{V}_2^{(n)} = \mathbf{V}_1^{(n+1)} = \mathbf{I}_n$.

Тогда для некоторых из введенных множеств будут справедливы стандартные соотношения [104], а для остальных появятся новые. Введем индекс $\mu = 1 \div 4$, тогда разбиение (2.110) запишется в виде $\mathbf{S}_{SCf} = \bigcup\limits_{\mu} \mathbf{V}_\mu^{(n)}$. Используя (2.108), можно построить таблицу умножения компонент "векторов" $\mathbf{V}_\mu^{(n)}$ в виде

$$\begin{aligned}
\mathbf{V}_\mu^{(n)} * \mathbf{V}_1^{(n)} &\subseteq \mathbf{V}_1^{(n)}, \quad \mathbf{V}_\mu^{(n)} * \mathbf{V}_2^{(n)} \subseteq \mathbf{V}_1^{(n+1)}, \\
\mathbf{V}_1^{(n)} * \mathbf{V}_3^{(n)} &\subseteq \mathbf{V}_1^{(n+1)}, \quad \mathbf{V}_2^{(n)} * \mathbf{V}_3^{(n)} \subseteq \mathbf{V}_1^{(n+2)}, \\
\mathbf{V}_1^{(n)} * \mathbf{V}_4 &\subseteq \mathbf{V}_1^{(n+1)}, \quad \mathbf{V}_3^{(n)} * \mathbf{V}_3^{(n)} \subseteq \mathbf{I}_{SCf}, \\
\mathbf{V}_4 * \mathbf{V}_3^{(n)} &\subseteq \mathbf{I}_{SCf}, \quad \mathbf{V}_3^{(n)} * \mathbf{V}_4 \subseteq \mathbf{I}_{SCf}, \\
\mathbf{V}_2^{(n)} * \mathbf{V}_4 &\subseteq \mathbf{V}_1^{(n+2)}, \quad \mathbf{V}_4 * \mathbf{V}_4 \subseteq \mathbf{V}_4.
\end{aligned} \tag{2.111}$$

Отсюда следует, что только два множества $\mathbf{V}_1^{(n)}$ и $\mathbf{V}_4$ являются подполугруппами (последнее — подгруппа) полугруппы $\mathbf{S}_{SCf}$, а для остальных множеств умножение незамкнуто. Тем не менее изучение свойств подобных разбиений представляет значительный интерес с абстрактно-алгебраической точки зрения.

**Определение 2.50.** *Главные векторные левый, правый идеалы и двусторонний тензорный идеал определяются формулами*

$$\begin{aligned}
&\mathbf{L}_\mu^{(n)}(\mathbf{s}) \stackrel{def}{=} \mathbf{s} * \mathbf{V}_\mu^{(n)}, \\
&\mathbf{R}_\mu^{(n)}(\mathbf{s}) \stackrel{def}{=} \mathbf{V}_\mu^{(n)} * \mathbf{s}, \\
&\mathbf{J}_{\mu\nu}^{(n)}(\mathbf{s}) \stackrel{def}{=} \mathbf{V}_\mu^{(n)} * \mathbf{s} * \mathbf{V}_\nu^{(n)},
\end{aligned} \tag{2.112}$$

*где* $\mathbf{s} \in \mathbf{J}_n$.

Из (2.108) и (2.111) следуют включения

$$\begin{aligned}
\mathbf{L}_\mu^{(n)}(\mathbf{s}) &\subseteq \mathbf{V}_1^{(n+1)},\ \mathbf{R}_2^{(n)}(\mathbf{s}) \subseteq \mathbf{V}_1^{(n+1)},\ \mathbf{R}_1^{(n)}(\mathbf{s}) \subseteq \mathbf{V}_1^{(n)}, \\
\mathbf{R}_3^{(n)}(\mathbf{s}) &\subseteq \mathbf{V}_1^{(n+2)},\ \mathbf{J}_{\mu 1}^{(n)}(\mathbf{s}) \subseteq \mathbf{V}_1^{(n)},\ \mathbf{J}_{13}^{(n)}(\mathbf{s}) \subseteq \mathbf{V}_1^{(n+1)}, \\
\mathbf{J}_{\mu 3}^{(n)}(\mathbf{s}) &\subseteq \mathbf{V}_1^{(n+2)},\ \mu > 1,\ \mathbf{J}_{\mu 4}^{(n)}(\mathbf{s}) \subseteq \mathbf{V}_1^{(n+2)},\ \mu > 1, \\
\mathbf{R}_4^{(n)}(\mathbf{s}) &\subseteq \mathbf{V}_1^{(n+2)},\ \mathbf{J}_{\mu 2}^{(n)}(\mathbf{s}) \subseteq \mathbf{V}_1^{(n+1)},\ \mathbf{J}_{14}^{(n)}(\mathbf{s}) \subseteq \mathbf{V}_1^{(n+1)}.
\end{aligned} \tag{2.113}$$

Выясним свойства векторных (2.112) и тензорных (2.112) идеалов по отношению к $\mathbf{L}_\mu^{(n)}(\mathbf{s})$. Так, левый векторный идеал является обычным левым идеалом множества $\mathbf{L}_\mu^{(n)}(\mathbf{s})$, поскольку

$$\mathbf{V}_\mu^{(n)} * \mathbf{L}_\mu^{(n)}(\mathbf{s}) \subseteq \mathbf{L}_\mu^{(n)}(\mathbf{s}). \tag{2.114}$$

Однако для правого векторного идеала подобное включение реализуется только при следующих комбинациях индексов:

$$\begin{aligned}
&\mathbf{R}_\mu^{(n)}(\mathbf{s}) * \mathbf{V}_1^{(n)} \subseteq \mathbf{R}_\mu^{(n)}(\mathbf{s}), \\
&\mathbf{R}_\mu^{(n)}(\mathbf{s}) * \mathbf{V}_2^{(n)} \subseteq \mathbf{R}_\mu^{(n)}(\mathbf{s}),\ \mu \neq 1, \\
&\mathbf{R}_3^{(n)}(\mathbf{s}) * \mathbf{V}_3^{(n)} \subseteq \mathbf{R}_3^{(n)}(\mathbf{s}),
\end{aligned} \tag{2.115}$$

причем последнее справедливо, если $\mathbf{V}_3^{(n)} \cap \mathbf{V}_1^{(n+2)} \neq \emptyset$.

Укажем также на соотношения, в которых $\mathbf{R}_\mu^{(n)}(\mathbf{s})$ ведет себя как *$\mu$-повышающий идеал*:

$$\mathbf{R}_1^{(n)}(\mathbf{s}) * \mathbf{V}_\mu^{(n)} \subseteq \mathbf{R}_2^{(n)}(\mathbf{s}),\ \mathbf{R}_2^{(n)}(\mathbf{s}) * \mathbf{V}_\mu^{(n)} \subseteq \mathbf{R}_3^{(n)}(\mathbf{s}). \tag{2.116}$$

**Определение 2.51.** *Обобщенные отношения Грина определяются формулами*

$$\begin{aligned} \mathbf{s}\mathscr{L}^{(nm)}_{\mu\nu}\mathbf{t} &\Leftrightarrow \mathbf{L}^{(n)}_{\mu}(\mathbf{s}) = \mathbf{L}^{(m)}_{\nu}(\mathbf{t}), \\ \mathbf{s}\mathscr{R}^{(nm)}_{\mu\nu}\mathbf{t} &\Leftrightarrow \mathbf{R}^{(n)}_{\mu}(\mathbf{s}) = \mathbf{R}^{(m)}_{\nu}(\mathbf{t}), \\ \mathbf{s}\mathscr{G}^{(nm)}_{\mu\nu\rho\sigma}\mathbf{t} &\Leftrightarrow \mathbf{J}^{(n)}_{\mu\nu}(\mathbf{s}) = \mathbf{J}^{(m)}_{\rho\sigma}(\mathbf{t}), \end{aligned} \tag{2.117}$$

*где* $\mathbf{s} \in \mathbf{J}_n, \quad \mathbf{t} \in \mathbf{J}_m$*.*

Классы эквивалентности по векторным и тензорным отношениям Грина имеют вид

$$\begin{aligned} \mathsf{L}^{(nm)}_{\mathbf{s},\mu\nu} &\stackrel{def}{=} \left\{\mathbf{t} \in \mathbf{J}_m \mid \mathbf{L}^{(n)}_{\mu}(\mathbf{s}) = \mathbf{L}^{(m)}_{\nu}(\mathbf{t})\right\}, \\ \mathsf{R}^{(nm)}_{\mathbf{s},\mu\nu} &\stackrel{def}{=} \left\{\mathbf{t} \in \mathbf{J}_m \mid \mathbf{R}^{(n)}_{\mu}(\mathbf{s}) = \mathbf{R}^{(m)}_{\nu}(\mathbf{t})\right\}, \\ \mathsf{J}^{(nm)}_{\mathbf{s},\mu\nu\rho\sigma} &\stackrel{def}{=} \left\{\mathbf{t} \in \mathbf{J}_m \mid \mathbf{J}^{(n)}_{\mu\nu}(\mathbf{s}) = \mathbf{J}^{(m)}_{\rho\sigma}(\mathbf{t})\right\}. \end{aligned} \tag{2.118}$$

Задание частичного порядка на множествах классов (2.118) превращает фактор-множества $\mathbf{S}_{SCf}/\mathscr{L}$, $\mathbf{S}_{SCf}/\mathscr{R}$, $\mathbf{S}_{SCf}/\mathscr{G}$ в частично упорядоченные множества: *правый, левый* и (просто) *остов* [470–472] суперконформной полугруппы, причем мощность каждого остова равна бесконечности [13].

**Предложение 2.52.** *Суперконформная полугруппа* $\mathbf{S}_{SCf}$ *не является устойчивой* [473] *ни справа, ни слева.*

*Доказательство.* Из (2.84) и определений (2.86)–(2.90) следует

$$\begin{aligned} \forall \mathbf{s}, \mathbf{t} &\in \mathbf{S}_{SCf}, \mathbf{s} \in \mathbf{S}_{SCf} * \mathbf{s} * \mathbf{t} \not\Rightarrow \mathbf{L}^{(n)}_{\mu}(\mathbf{s}) = \mathbf{L}^{(m)}_{\nu}(\mathbf{s} * \mathbf{t}), \\ \mathbf{s} &\in \mathbf{t} * \mathbf{s} * \mathbf{S}_{SCf} \not\Rightarrow \mathbf{R}^{(n)}_{\mu}(\mathbf{s}) = \mathbf{R}^{(m)}_{\nu}(\mathbf{t} * \mathbf{s}). \end{aligned}$$

■

**2.2.5.** Квазихарактеры. Рассмотрим подробнее свойства нильпотентности элементов полугруппы $\mathbf{S}_{SCf}$.

**Определение 2.53.** *Идеальный индекс элемента* $\mathbf{s}$ *суперконформной полугруппы определяется формулой*

$$\operatorname{ind}_{ideal}\mathbf{s} \stackrel{def}{=} \operatorname{ind} g(z), \tag{2.119}$$

*причем* $\operatorname{ind}_{ideal}\mathbf{g} = \infty$.

Отметим, что все элементы, обладающие конечным идеальным индексом (2.119), нильпотентны в смысле полугруппового умножения, т. е. $\forall \mathbf{s} \in \mathbf{S}_{SCf}\ \exists n \in \mathbb{N}$ такое, что $\mathbf{s}^{*\mathbf{n}} = \mathbf{z}$.

Для произведения элементов суперконформной полугруппы из формул (2.119) имеем

$$\max\operatorname{ind}_{ideal}(\mathbf{s}*\mathbf{t}) = \operatorname{ind}_{ideal}\mathbf{t},\ \operatorname{ind}_{ideal}\mathbf{s} \geq \operatorname{ind}_{ideal}\mathbf{t}, \tag{2.120}$$

$$\operatorname{ind}_{ideal}\mathbf{s}+1,\ \operatorname{ind}_{ideal}\mathbf{s} < \operatorname{ind}_{ideal}\mathbf{t}. \tag{2.121}$$

В частности,

$$\operatorname{ind}_{ideal}(\mathbf{g}*\mathbf{s}) \leq \operatorname{ind}_{ideal}\mathbf{s}, \tag{2.122}$$

$$\operatorname{ind}_{ideal}(\mathbf{s}*\mathbf{g}) \leq \operatorname{ind}_{ideal}\mathbf{s}+1. \tag{2.123}$$

Аналогично определяются индексы соответствующих множеств элементов (2.110). Для них получаем

$$\max\operatorname{ind}_{ideal}\mathbf{V}_1^{(n)} = n-1,\ \operatorname{ind}_{ideal}\mathbf{V}_2^{(n)} = n,\ \min\operatorname{ind}_{ideal}\mathbf{V}_3^{(n)} = n+1. \tag{2.124}$$

Из соотношений (2.120)–(2.121) и (2.122)–(2.123) следует, что величина

$$|\mathrm{ind}_{ideal}\,(\mathbf{s}*\mathbf{t})-\mathrm{ind}_{ideal}\,\mathbf{s}-\mathrm{ind}_{ideal}\,\mathbf{t}| \tag{2.125}$$

ограничена, поэтому отличие отображения $\mathbf{s}\to\mathrm{ind}_{ideal}\,\mathbf{s}$ от гомоморфизма конечно, что позволяет определить квазихарактер [474–478] по формуле $\chi(\mathbf{s})\stackrel{def}{=}\mathrm{ind}_{ideal}\,\mathbf{s}$, который мы назовем *идеальным квазихарактером*. Отметим некоторые свойства идеального квазихарактера: $\chi(\mathbf{s}^{*2})\leq\chi(\mathbf{s})$, $\chi(\mathbf{g})=\infty$. Из того факта, что множества $\mathbf{J}_n$, на которых определен идеальный квазихарактер, не пересекаются: $\mathbf{J}_n\cap\mathbf{J}_k=\varnothing$, $n\neq k$, следует вывод о том, что $\chi(\mathbf{s})$ действительно разделяет элементы полугруппы [479–482], а отношение $\pi_\chi$, заданное формулой $\mathbf{s}\stackrel{\pi_\chi}{\sim}\mathbf{t}\Leftrightarrow\chi(\mathbf{s})=\chi(\mathbf{t})$, является отношением эквивалентности в суперконформной полугруппе $\mathbf{S}_{SCf}$.

## 2.3. Сплетающие четность преобразования

Рассмотрим более подробно сплетающие четность $N=1$ преобразования, задаваемые уравнением $Q\left(z,\theta\right)=0$ (2.38).

Прежде всего обратим внимание на дуальную роль таких преобразований с суперконформными преобразованиями при определении порядка $\mathrm{sord}D$ дифференциального оператора $D$ (см., например, [483] и применения в [484–487]).

**Предложение 2.54.** *При* $Q\left(z,\theta\right)=0$ *(как и при* $\varDelta\left(z,\theta\right)=0$ *в* [483]*) для некоторого целого* $k\geq0$ *имеем*

$$\mathrm{sord}\left(\tilde{D}^{2k+1}\right)=\mathrm{sord}\left(D^{2k+1}\right)=\frac{2k+1}{2}. \tag{2.126}$$

*Доказательство.* Учитывая соотношение суперсимметрии $D^2=\partial$, не-

посредственно из (2.24) имеем

$$D = D\tilde{\theta} \cdot \tilde{D} + \Delta(z,\theta) \cdot \tilde{D}^2 \tag{2.127}$$

и

$$D^2 = D^2\tilde{\theta} \cdot \tilde{D} + Q(z,\theta) \cdot \tilde{D}^2. \tag{2.128}$$

После возведения оператора $D$ в степень $(2k+1)$ получаем

$$\begin{aligned} D^{2k+1} &= \left(D^2\right)^k D = \\ &\left(D^2\tilde{\theta} \cdot \tilde{D} + Q(z,\theta) \cdot \tilde{D}^2\right)^k \cdot \left(D\tilde{\theta} \cdot \tilde{D} + \Delta(z,\theta) \cdot \tilde{D}^2\right). \end{aligned}$$

Видно, что наибольшая степень $D$ есть $(2k+2)$, и нечетный коэффициент при ней равен

$$\Xi(z,\theta) = Q^k(z,\theta) \cdot \Delta(z,\theta). \tag{2.129}$$

Отсюда следует, что $\Xi(z,\theta) = 0$ в случаях

**1.** $\Delta(z,\theta) = 0$ — SCf (суперконформные преобразования, как в [483]);

**2.** $Q(z,\theta) = 0$ — TPt (сплетающие четность преобразования).

**3.** $Q^k(z,\theta) = 0$, $Q(z,\theta) \neq 0$ — нередуцированные преобразования с нильпотентным $Q(z,\theta)$ (которые мы здесь не рассматриваем).

■

Отметим, что формула (2.126) и соотношение $\Xi(z,\theta) = 0$ играют ключевую роль при построении интегрируемых иерархий в $(1|1)$-мерном суперпространстве [484, 486–488]. Отсюда заключаем, что сплетающие четность преобразования могут дать нечетный вариант иерархий и соответствующих нелинейных уравнений.

**2.3.1.** Касательное суперпространство и кручение четности. Действие сплетающих четность преобразований в касательном и кокасательном $(1|1)$-пространствах определяется суперматрицей $\mathrm{P}_{TPt}$ (2.54). Из (2.22), (2.23) и (2.40) получаем для суперпроизводных

$$\partial = \partial\tilde{\theta} \cdot \tilde{D}, \tag{2.130}$$

$$D = \mathit{\Delta}_{TPt}(z,\theta) \cdot \tilde{\partial} + D\tilde{\theta} \cdot \tilde{D} \tag{2.131}$$

и диффереациалов

$$d\tilde{Z} = d\theta \cdot \mathit{\Delta}_{TPt}(z,\theta), \tag{2.132}$$

$$d\tilde{\theta} = dZ \cdot \partial\tilde{\theta} + d\theta \cdot D\tilde{\theta}, \tag{2.133}$$

где $\mathit{\Delta}_{TPt}(z,\theta)$ определена в (2.50).

Соотношения (2.130) и (2.132) свидетельствуют о том, что преобразования, удовлетворяющие условию $Q(z,\theta) = 0$, изменяют четность касательного и кокасательного суперпространств, действуя, как $T\mathbb{C}^{1|0} \to T\mathbb{C}^{0|1}$ и $T^*\mathbb{C}^{0|1} \to T^*\mathbb{C}^{1|0}$. Поэтому можно переформулировать **Определение 2.24** в виде

**Определение 2.55.** *Назовем сплетающими четность (касательного пространства) преобразованиями* (TPt – *twisting parity of tangent space transformations*) *такие преобразования, действующие в касательном пространстве как* $T\mathbb{C}^{1|0} \to T\mathbb{C}^{0|1}$ *и* $T^*\mathbb{C}^{0|1} \to T^*\mathbb{C}^{1|0}$ *(с кручением четности), которые удовлетворяют условию* $Q(z,\theta) = 0$ (2.38).

Тем не менее, необратимый аналог инвариантности (2.25) имеет место и для сплетающих четность преобразований.

Отметим некоторые сэндвич-соотношения, следующие из нильпо-тентности $\varDelta_{TPt}(z,\theta)$ и $\partial\tilde{\theta}$. Поскольку березиниан сплетающих четность преобразований (2.55) пропорционален $\partial\tilde{\theta}$, то из (2.133) следует

$$d\tilde{\theta}\cdot \mathrm{Ber}_{\,TPt}\left(\tilde{Z}/Z\right)\cdot\tilde{D}=d\theta\cdot \mathrm{Ber}_{\,TPt}\left(\tilde{Z}/Z\right)\cdot D, \tag{2.134}$$

что можно трактовать как ортогональность березиниана изменению оператора $d\theta D$ под действием сплетающих четность преобразований.

Интересно отметить и другую ортогональность, следующую из (2.132)

$$\mathrm{Ber}_{\,TPt}\left(\tilde{Z}/Z\right)\cdot d\tilde{Z}=0. \tag{2.135}$$

Кроме того, умножая обе части уравнения (2.131) на $\varDelta_{TPt}(z,\theta)$ и пользуясь ее нильпотентностью, получаем

$$\varDelta_{TPt}(z,\theta)\cdot D=\varDelta_{TPt}(z,\theta)\cdot D\tilde{\theta}\cdot\tilde{D}, \tag{2.136}$$

что интересно сравнить с суперконформным условием (2.43).

Подобное соотношение имеет место, если умножить обе части уравнения (2.133) на $\varDelta'_{TPt}(z,\theta)$ и воспользоваться соотношением (2.56)

$$d\tilde{\theta}\cdot\varDelta'_{TPt}(z,\theta)=d\theta\cdot D\tilde{\theta}\cdot\varDelta'_{TPt}(z,\theta)\,. \tag{2.137}$$

Далее, из уравнений (2.130) и (2.132) получаем

$$d\tilde{Z}\tilde{D}=d\theta\varDelta_{TPt}(z,\theta)\cdot\partial\tilde{\theta}D=\left(D\tilde{\theta}\right)^2\mathrm{Ber}_{\,TPt}\left(\tilde{Z}/Z\right)\cdot d\theta\partial. \tag{2.138}$$

**Утверждение 2.56.** *При* $\left(D\tilde{\theta}\right)^2=1$ *равенство* (2.138) *определяет ковариантный объект, преобразующийся с помощью березиниана как множителя.*

**2.3.2.** Обобщенное редуцированное расслоение с кручением четности. Для построения TPt аналога линейного расслоения на суперримановой поверхности [183, 365, 405, 489] необходимо построить инвариантный объект, но не с помощью суперконформного дифференциала $d\tau_{SCf}$, а с помощью его аналога для TPt преобразований.

В отличие от случая суперконформных преобразований (см. **Подраздел 2.1**) объект $d\tau_{SCf} = dZD + d\theta$ (введенный в [489] в качестве SCf супердифференциала) при сплетающих четность преобразованиях (с условием $\epsilon\left(D\tilde{\theta}\right) \neq 0$) не преобразуется ковариантно. Действительно,

$$d\tilde{\tau}_{SCf} = d\tilde{Z}\tilde{D} + d\tilde{\theta} = dZ \cdot \partial\tilde{\theta} + d\theta \cdot \left(D\tilde{\theta} + \Delta_{TPt}\left(z,\theta\right) \cdot \tilde{D}\right).$$

Пользуясь (2.136) и (2.56), преобразуем это выражение в

$$d\tilde{\tau}_{SCf} = -\frac{1}{2D\tilde{\theta}}\left[dZ\Delta'_{TPt}\left(z,\theta\right) + 2d\theta D\Delta_{TPt}\left(z,\theta\right)\right], \tag{2.139}$$

что невозможно выразить через $d\tau_{SCf}$. Поэтому необходимо ввести TPt аналог суперконформного дифференциала.

**Определение 2.57.** *Дифферециалом с кручением четности назовем такой объект $d\tau_{TPt}$, который преобразуется при сплетающих четность преобразованиях по закону*

$$d\tilde{\tau}^{even}_{TPt} = d\tau^{odd}_{TPt} \cdot \partial\tilde{\theta}. \tag{2.140}$$

*Замечание* **2.58.** Четности $d\tilde{\tau}^{even}_{TPt}$ и $d\tau^{odd}_{TPt}$ противоположны*).

*Примечание.* Здесь можно проследить некоторую аналогию с кручением квантовых дифференциалов [490].

Тогда можно построить TPt аналог линейного расслоения, если ввести TPt аналог $\delta_{TPt}$ суперконформного дифференциала по формулам

$$\delta_{TPt} = d\tau_{TPt}^{odd}\partial, \qquad (2.141)$$

$$\tilde{\delta}_{TPt} = d\tilde{\tau}_{TPt}^{even}\tilde{D}. \qquad (2.142)$$

*Замечание* **2.59.** Четность TPt дифференциала $\delta_{TPt}$ фиксирована, он — нечетный при любых сплетающих четность преобразованиях.

**Предложение 2.60.** *Дифференциал $\delta_{TPt}$ инвариантен относительно сплетающих четность преобразований.*

*Доказательство.* Пользуясь формулами (2.130) и (2.140), получаем

$$\delta_{TPt} = d\tau_{TPt}^{odd}\partial = d\tau_{TPt}^{odd}\partial\tilde{\theta}\tilde{D} = d\tilde{\tau}_{TPt}^{even}\tilde{D} = \tilde{\delta}_{TPt}.$$

■

Таким образом, величины $\partial\tilde{\theta}$ и $d\tau_{TPt}$ играют такую же фундаментальную роль для сплетающих четность преобразований [9, 13], как и $D\tilde{\theta}$ и $d\tau_{SCf}$ — для суперконформных преобразований [341, 491, 492].

*Замечание* **2.61.** Рассматриваемое здесь сплетение четности (2.130) и (2.132) существенно отличается от другого подобного объекта, существующего в литературе — $Q$-многообразия [493–497], где изменение четности касательного пространства делается искуственно из начальных определений.

*Замечание* **2.62.** Следует также отличать “сплетение четности” от скрученных кокасательных расслоений и дифференциальных операторов на многообразиях, рассмотренных в [498, 499], скрученных представлений [500–502], скрученных кокасательных расслоений в механических системах с точной пуассоновской симметрией [503, 504], а также от скрученных комплексов де Рама [505–509].

С помощью TPt супердифференциалов $d\tau_{TPt}$ можно определить TPt аналоги с кручением четности для линейных [183, 405] и векторных [510] расслоений , линейных интегралов [489] и соответствующих спектральных последовательностей [269, 332, 511].

**2.3.3.** Компонентный анализ . Условие сплетающих четность преобразований $Q(z,\theta)=0$ имеет в компонентах следующий вид

$$\begin{cases} f'(z)=\psi'(z)\cdot\psi(z), \\ \chi'(z)=g'(z)\cdot\psi(z)-g(z)\cdot\psi'(z), \end{cases} \tag{2.143}$$

который получается из (2.81) проекцией спина редукции $\mathrm{m}=-1$.

Решение первого уравнения в (2.143) можно представить в виде бесконечного ряда (2.72). Из второго уравнения можно получить

$$\chi'(z)=-g^2(z)\left(\frac{\psi(z)}{g(z)}\right)',$$

тогда

$$\chi(z)=2\int g'(z)\cdot\psi(z)\,dz-g(z)\cdot\psi(z). \tag{2.144}$$

Отсюда следуют сплетающие четность преобразования в интегральной форме

$$\begin{cases} f(z)=\int\psi'(z)\cdot\psi(z)\,dz, \\ \chi(z)=2\int g'(z)\cdot\psi(z)\,dz-g(z)\cdot\psi(z). \end{cases} \tag{2.145}$$

Видно, что при условии (аналогичном тому, которое выделяет Ann-преобразования (2.86))

$$g(z)=g_{nil}(z)\in\operatorname{Ann}\psi(z) \tag{2.146}$$

функция $\chi(z) = 0$ (поскольку из (2.146) следует, что и $g'_{nil}(z) \in \mathrm{Ann}\,\psi(z)$), поэтому преобразование четного сектора отщепляется и становится конформным необратимым преобразованием $\tilde{z} = f_{nil}(z)$ с нильпотентной правой частью в том смысле, что $f_{nil}(z) \cdot f_{nil}(z) = 0$.

**Определение 2.63.** *Преобразования, удовлетворяющие* (2.146), *назовем* Nil *преобразованиями.*

Далее, из условия (2.146) следует нильпотентность функции $g(z) = g_{nil}(z)$. Если при этом индекс нильпотентности функции $g_{nil}(z)$ равен двум, т. е. $g_{nil}(z) \cdot g_{nil}(z) = 0$, то уравнения для суперконформных и вращающих четность преобразований (2.81) совпадают между собой и с первым уравнением в (2.143). Поэтому имеет место

**Утверждение 2.64.** *Преобразования, выделяемые условием* (2.146), *представляют собой нильпотентное расширение левых вырожденных преобразований* (2.70).

**Утверждение 2.65.** *Для сплетающих четность преобразований с условием* (2.146) *нечетная характеристическая функция* (2.61) *зануляется, т. е.* $\mathbf{\Sigma}(z,\theta) = 0$.

Поскольку $\epsilon[g_{nil}(z)] = 0$, то березиниан (E.8) таких преобразований не определен и можно пользоваться доопределенной формулой (2.53), которая в данном случае имеет вид

$$\boldsymbol{J}_{nil}^{noninv} = g_{nil}(z) + \theta \cdot \psi'(z). \tag{2.147}$$

Однако, если $\epsilon[g(z)] \neq 0$, то березиниан сплетающих четность преобразований определен формулой (2.57), хотя и необратим (нильпотентен), т. е. такие преобразования полунеобратимы (см. (2.4)). В этом случае, пользуясь нильпотентностью функции $\psi(z)$, получаем компо-

нентный вид березиниана

$$\mathrm{Ber}_{TPt}\left(\tilde{Z}/Z\right) =$$
$$\left(\psi\left(z\right)\cdot\int g\left(z\right)\cdot\psi'\left(z\right)dz+\theta\cdot\left(\frac{2}{g\left(z\right)}\int g\left(z\right)\cdot\psi'\left(z\right)dz-\psi\left(z\right)\right)\right)'. \tag{2.148}$$

## 2.4. Нелинейная реализация $N = 1$ редуцированных преобразований

Изучение нелинейных реализаций редуцированных преобразований (см. [1, 7, 9, 21] и **Подраздел 2.1**) представляет интерес по многим причинам. С одной стороны, первые статьи по суперсимметрии [34, 35, 38, 512] были написаны в терминах нелинейных реализаций (несуперсимметричный вариант этого метода изложен в [31, 32, 513, 514], внутренние суперсимметрии рассматривались в [515, 516], а различные обобщения представлены в [517]). Позднее появилась надежда, что с помощью метода нелинейных реализаций можно решить проблему суперпартнеров [518] и спонтанного нарушения суперсимметрии [519–523] в реалистичных [524, 525] и суперконформных четырехмерных моделях [526–529]. С другой стороны, нелинейно реализованная двумерная суперконформная симметрия [530, 531] была использована в теории суперструн [532] для построения иерархий и вложений [533–535] с различным количеством суперсимметрий на мировом листе [536–538], нелинейных $W$ симметрий [539–541], а также в (расширенной) суперконформной механике [306, 308, 542] и теории супермембран [419, 420, 543–549].

В данном подразделе, в дополнение к этим исследованиям, мы включаем в рассмотрение конечные преобразования и учитываем их необратимость [20]. Мы также рассматриваем связь между "линейными" и нелинейными реализациями [550–554], но с чисто кинематической точки зрения и предлагаем прозрачное диаграммное описание, которое можно применять и в общем случае.

**2.4.1.** Движение нечетной кривой в $\mathbb{C}^{1|1}$. Напомним, что $N = 1$ суперааналитические преобразования в $\mathbb{C}^{1|1}$ имеют вид (см. (2.2) и [111, 189])

$$\begin{cases} \tilde{z} &= f(z) + \theta \cdot \chi(z), \\ \tilde{\theta} &= \psi(z) + \theta \cdot g(z), \end{cases} \tag{2.149}$$

Согласно интерпретации Весса [555] мы можем изучать движение кривой $\theta = \lambda(z)$ в $\mathbb{C}^{1|1}$. Тогда получаем

$$\tilde{z} = f(z) + \lambda(z) \cdot \chi(z), \tag{2.150}$$

$$\tilde{\lambda}(\tilde{z}) = \psi(z) + \lambda(z) \cdot g(z), \tag{2.151}$$

где второе уравнение отражает эйнштейновский тип преобразований.

В четырехмерном случае функция $\lambda(z)$ обычно называется полем Акулова-Волкова [524, 555] и в физических приложениях играет роль голдстоуновского фермиона [35, 38, 512] (и поэтому называемого также голдстино ).

Как это видно из (2.151) преобразование функции $\lambda(z)$ является существенно нелинейным. Соотношения такого типа являюся стандартными для нелинейных реализаций, и голдстино $\lambda(z)$ описывает нарушение суперсимметрии [518, 556, 557].

Чтобы найти преобразование голдстино, разложим функцию $\tilde{\lambda}(\tilde{z})$ в ряд и использум нильпотентность нечетных функций

$$\tilde{\lambda}(f(z)) = \psi(z) + \lambda(z) \cdot g(z) - \tilde{\lambda}'(f(z)) \cdot \lambda(z) \cdot \chi(z) . \tag{2.152}$$

В случае, если $f^{-1}$ существует, мы можем записать искомые преобразования в явном виде [7, 20]

$$\tilde{\lambda} = \psi \circ f^{-1} + \lambda \circ f^{-1} \cdot g \circ f^{-1} - \tilde{\lambda}' \cdot \lambda \circ f^{-1} \cdot \chi \circ f^{-1}, \tag{2.153}$$

где $f \circ g = f(g(z))$.

Найти общее решение уравнения (2.152) не представляется возможным, поэтому рассмотрим различные частные случаи.

**2.4.2.** Глобальная суперсимметрия в $\mathbb{C}^{1|1}$. В этом случае компонентные функции в (2.149) имеют вид

$$f(z) = z,\ g(z) = 1,\ \chi(z) = \varepsilon,\ \psi(z) = \varepsilon, \tag{2.154}$$

где $\varepsilon$ постоянный нечетный параметр. Тогда из (2.150) и (2.151) имеем

$$\tilde{\lambda}_{Glob}(z) = \varepsilon + \lambda(z) - \tilde{\lambda}'_{Glob}(z) \cdot \lambda(z) \cdot \varepsilon\,. \tag{2.155}$$

Эти уравнения также достаточно сложны для явного решения. Однако, в случае инфенитезимальных преобразований получаем решение

$$\delta_\varepsilon \lambda_{Glob}(z) = \tilde{\lambda}_{Glob}(z) - \lambda(z) = \varepsilon \cdot [1 + \lambda(z) \cdot \lambda'(z)]\,, \tag{2.156}$$

которое удовлетворяет стандартной алгебре суперсимметрии в двух измерениях

$$[\delta_\varepsilon, \delta_\eta]\, \lambda_{Glob}(z) = 2\varepsilon\eta \cdot \lambda(z) \cdot \lambda'(z) \tag{2.157}$$

в соответствие с [38, 512].

*Замечание* **2.66.** В конечном глобальном случае имеем

$$\tilde{\lambda}^{fin}_{Glob}(z) = \tilde{\lambda}_{Glob}(z) + \sigma(z), \tag{2.158}$$

где $\tilde{\lambda}_{Glob}(z)$ дается в (2.156). Подставляя (2.158) в (2.155), для $\sigma(z)$ получаем следующее уравнение

$$\sigma'(z) \cdot \varepsilon \cdot \lambda(z) = \sigma(z), \tag{2.159}$$

которое может быть решено разложение по нильпотентам.

**2.4.3.** Редуцированные преобразования. Рассмотрим $N = 1$ редуцированные преобразования, параметризованные функциями $g(z)$, $\psi(z)$ (см **Подраздел 2.1** и [9, 13]). В терминах той же нечетной функции $\lambda(z)$ мы можем в общем случае найти преобразованную функцию $\tilde{\lambda}_{\mathrm{m}}(z)$ из (2.151) в виде двух решений (соответствующих различным проекциям проекции “спина редукции” m (2.81)) следующей системы уравнений [20]

$$\begin{cases} \tilde{\lambda}_{\mathrm{m}}\left(f^{(g\psi)}_{\mathrm{m}}(z)\right) & = \psi(z) + \lambda(z) \cdot g(z) - \tilde{\lambda}'_{\mathrm{m}}\left(f^{(g\psi)}_{\mathrm{m}}(z)\right) \cdot \lambda(z) \cdot \chi^{(g\psi)}_{\mathrm{m}}(z), \\ f^{(g\psi)\prime}_{\mathrm{m}}(z) & = \psi'(z) \cdot \psi(z) + \dfrac{1+\mathrm{m}}{2} \cdot g^2(z), \\ \chi^{(g\psi)\prime}_{\mathrm{m}}(z) & = g'(z) \cdot \psi(z) + \mathrm{m} \cdot g(z) \cdot \psi'(z), \end{cases} \tag{2.160}$$

где штрих означает дифференцирование по аргументу, $\mathrm{m} = +1$ соответствует суперконформным преобразованиям и $\mathrm{m} = -1$ - преобразованиям, сплетающим четность касательного пространства (см. [1, 7, 9]).

**Определение 2.67.** *Соответственно спину редукции назовем решения* $\tilde{\lambda}_{SCf}(z) = \tilde{\lambda}_{\mathrm{m}=+1}(z)$ — *SCf голдстино, и* $\tilde{\lambda}_{TPt}(z) = \tilde{\lambda}_{\mathrm{m}=-1}(z)$ — *TPt голдстино.*

Как и ранее, уравнение (2.160) невозможно решить явно в общем случае.

**Утверждение 2.68.** *Аналог кривизны кривой нильпотентен и совпадает со второй производной голдстино.*

*Доказательство.* По стандартной формуле $\varkappa_\lambda(z) = \dfrac{\tilde{\lambda}''(z)}{\left(1+\left(\tilde{\lambda}'(z)\right)^2\right)^{3/2}}$, и после подстановки $\left(\tilde{\lambda}'(z)\right)^2 = 0$ получаем $\varkappa_\lambda(z) = \tilde{\lambda}''(z)$. ■

*Замечание* **2.69.** Необходимо подчеркнуть, что уравнения (2.160) не зависят от свойств обратимости суперконформно-подобных преобразований [1,13], и только они, а не диаграммный метод, изложенный ниже, могут быть использованы для нахождения эволюции голдстино для преобразований, сплетающих четность касательного пространства (m = −1 случай).

*Пример* **2.70.** Параметризуем инфинитезимальные суперконформные преобразования следующим образом

$$f(z) = z + r(z),\ g(z) = 1 + \frac{1}{2}r'(z),\ \chi(z) = \varepsilon(z),\ \psi(z) = \varepsilon(z), \quad (2.161)$$

где $r(z), \varepsilon(z)$ бесконечно малые четная и нечетная функции. Тогда из (2.160) получаем

$$\delta_{r,\varepsilon}\lambda_{SCf}(z) = \varepsilon(z)\cdot[1+\lambda(z)\cdot\lambda'(z)] + \frac{1}{2}r'(z)\cdot\lambda(z) - r(z)\cdot\lambda'(z) \quad (2.162)$$

в полном соответствии с [531].

**2.4.4.** Диаграммный подход к связи между линейной и нелинейной реализациями. Соотношение между линейной и нелинейной реализациями [550,552,554] играет важную роль в понимании механизмов спонтанного нарушения суперсимметрии [551]. Интерес к изучению $N = 1$ суперконформных и редуцированных преобразований обусловлен тем фактом, что нелинейно реализованные инфинитезимальные суперконформные преобразования [531, 537, 538] широко используются в методе погружения суперструн [532, 536], а также в их иерархиях [533, 535].

Здесь мы исследуем двумерные конечные (в общем случае необратимые) редуцированные преобразования (см. [11,20] и **Подразделы 2.1** и **2.3**), что с очевидными модификациями применимо и к многомерному случаю.

Рассмотрим следующую диаграмму

$$\begin{array}{ccc} Z_A & \xrightarrow[\text{W-Z}]{\mathcal{G}} & \tilde{Z} \\ \mathcal{A} \uparrow & & \uparrow \mathcal{B} \\ Z & \xrightarrow[\text{A-V}]{\mathcal{H}} & Z_H \end{array} \tag{2.163}$$

где $\mathcal{A} : Z \to Z_A$, $\mathcal{G} : Z_A \to \tilde{Z}$, $\mathcal{B} : Z_H \to \tilde{Z}$, $\mathcal{H} : Z \to Z_H$ (и $Z = (z, \theta)$) супераналитические преобразования (2.149).

Преобразование $\mathcal{G}$ играет роль линейного преобразования вессзуминовского типа, а нелинейное преобразование $\mathcal{H}$ является преобразованием акулов-волковского типа, в то время, как $\mathcal{A}$ и $\mathcal{B}$ соответствуют косетным преобразованиям с голдстоуновскими полями как параметрами [31, 513].

**2.4.5.** Глобальная двумерная суперсимметрия в терминах нелинейных реализаций. В соответствие с [32, 550] мы можем рассмотреть $\mathcal{G}$ как глобальные линейные двумерные суперсимметричные преобразования

$$\mathcal{G}:\begin{cases}\tilde{z} &= z_A+\theta_A\cdot\varepsilon,\\ \tilde{\theta} &= \varepsilon+\theta_A,\end{cases} \tag{2.164}$$

тогда $\mathcal{H}$ — обычные конформные преобразования с составными параметрами, которые должны быть найдены из соответствующих уравнений, а $\mathcal{A}$ and $\mathcal{B}$ можно интерпретировать как косетные преобразования с локальными нечетными параметрами $\lambda(z)$ и $\tilde{\lambda}_{Glob}(z_H)$.

$$\mathcal{A}:\begin{cases}z_A &= z+\theta\cdot\lambda(z),\\ \theta_A &= \lambda(z)+\theta,\end{cases}\qquad \mathcal{B}:\begin{cases}\tilde{z} &= z_H+\theta_H\cdot\tilde{\lambda}_{Glob}(z_H),\\ \tilde{\theta} &= \tilde{\lambda}_{Glob}(z_H)+\theta_H,\end{cases} \tag{2.165}$$

Именно коммутативность диаграммы (2.163) задает эволюцию голдстино $\lambda(z)$ подобно (2.151) и (2.155) и уравнения для составных параметров преобразования $\mathcal{H}$ следующим образом [11, 20].

**Определение 2.71.** *Будем считать, что "линейное" редуцированное преобразование $\mathcal{G}$ представимо "нелинейным" преобразованием $\mathcal{H}$, если диаграмма* (2.163) *коммутативна*

$$\mathcal{G}\circ\mathcal{A}=\mathcal{B}\circ\mathcal{H}. \tag{2.166}$$

*Замечание* **2.72.** В теории групп эта конструкция связана с индуцированным представлением [558, 559]. Однако, здесь мы не требуем обратимости составляющих преобразований (2.166) и включаем в рассмотрение также конечные преобразования.

Используя (2.166), мы получаем соотношения

$$
\begin{aligned}
\tilde{z}_{\mathcal{G}\circ\mathcal{A}} &= \tilde{z}_{\mathcal{B}\circ\mathcal{H}}, \\
\tilde{\theta}_{\mathcal{G}\circ\mathcal{A}} &= \tilde{\theta}_{\mathcal{B}\circ\mathcal{H}},
\end{aligned}
\tag{2.167}
$$

которые являются условиями представимости (2.166) в координатном виде (как 4 компонентных уравнения после разложения по $\theta$).

В частном случае глобальной суперсимметрии (2.164) уравнения (2.167) имеют вид

$$
\begin{aligned}
& z_A + \theta_A \cdot \varepsilon = z_H + \theta_H \cdot \tilde{\lambda}_{Glob}\left(z_H\right), \\
& \theta_A + \varepsilon = \tilde{\lambda}_{Glob}\left(z_H\right) + \theta_H .
\end{aligned}
\tag{2.168}
$$

Используя (2.165), мы получаем составные параметры преобразования $\mathcal{H}$ в виде

$$
\mathcal{H}: \begin{cases} z_H = z + \lambda\left(z\right) \cdot \varepsilon, \\ \theta_H = \theta, \end{cases}
\tag{2.169}
$$

а также уравнение для эволюции голдстино

$$
\tilde{\lambda}_{Glob}\left(z_H\right) = \varepsilon + \lambda\left(z\right).
\tag{2.170}
$$

После разложения по нильпотентам получаем

$$
\varepsilon + \lambda\left(z\right) = \tilde{\lambda}_{Glob}\left(z\right) + \tilde{\lambda}'_{Glob}\left(z\right) \cdot \lambda\left(z\right) \cdot \varepsilon,
\tag{2.171}
$$

что совпадает с (2.155).

Таким образом, именно из соотношений (2.166) и (2.167) определяется эволюция голдстино. Если $\mathcal{A}$ обратимо, условие представимости

(2.166) принимает следующий вид

$$\mathcal{G} = \mathcal{B} \circ \mathcal{H} \circ \mathcal{A}^{-1}. \tag{2.172}$$

В глобальном случае обратимость $\mathcal{A}$ очевидна, тогда из (2.165) получаем

$$\mathcal{A}^{-1} : \begin{cases} z &= z_A - \theta_A \cdot \lambda(z_A), \\ \theta &= -\lambda(z_A) + \theta_A \left[1 + \lambda(z_A) \cdot \lambda'(z_A)\right]. \end{cases} \tag{2.173}$$

Это объясняет хорошо известное "$-\lambda$ правило" [550,560] при сравнении суперполей в линейной и нелинейной реализациях [561].

Соотношение (2.172) представляет собой общий вид "расщепляющего трюка" ("splitting trick") [550, 551], в соответствии с которым любое линейное суперполе может быть представлено как суперпозиция нелинейно преобразующихся компонент. Аналогом этой процедуры в необратимом случае является условие представимости (2.166), которое не должно разрешаться относительно $\mathcal{A}$.

Таким образом, для суперполя $\Phi(z,\theta)$ мы можем записать

$$\delta_{\mathcal{H}}\Phi(z,\theta) = \Phi(z + \lambda(z) \cdot \varepsilon, \theta) - \Phi(z,\theta) = \varepsilon \cdot \lambda(z) \cdot \frac{\partial \Phi(z,\theta)}{\partial z}, \tag{2.174}$$

где $\delta_{\mathcal{H}}$ — инфинитезимальное нелинейное преобразование $\mathcal{H}$, соответствующее $\mathcal{G}$. Если использовать (2.173) и положить

$$\begin{aligned} \Phi(z,\theta) &= \Phi\left(z_A - \theta_A \cdot \lambda(z_A), -\lambda(z_A) + \theta_A \left[1 + \lambda(z_A) \cdot \lambda'(z_A)\right]\right) \\ &\overset{def}{=} \Phi_A(z_A, \theta_A), \end{aligned} \tag{2.175}$$

тогда для инфинитезимального линейного преобразования $\mathcal{G}$ мы полу-чаем стандартное соотношение суперсимметрии

$$\delta_{\mathcal{G}}\Phi_A(z_A,\theta_A)=\Phi(z_A+\varepsilon\cdot\theta_A,\theta_A+\varepsilon)-\Phi_A(z_A,\theta_A)=\varepsilon\cdot Q_A\Phi_A(z_A,\theta_A), \tag{2.176}$$

где $Q_A$ - обыкновенные супертрансляции (см. [550]).

Теперь мы можем доказать обратный расщепляющий трюк, который явно следует из условия представимости (2.166), примененного к глобальной двумерной суперсимметрии.

**Предложение 2.73.** *Любое суперполе* $\Phi(z,\theta)$, *преобразующееся нелинейно, как в* (2.174), *вместе с голдстино* $\lambda(z)$, *преобразующегося, как в* (2.156), *задает линейное глобально преобразующееся суперполе* (2.176).

*Доказательство.* Мы должны доказать, что $\Delta\Phi(z,\theta)=\delta_{\mathcal{G}}\Phi_A(z_A,\theta_A)$, где $\Delta\Phi(z,\theta)\stackrel{def}{=}\delta_{\mathcal{H}}\Phi(z,\theta)+\delta_{\mathcal{B}}\Phi(z,\theta)-\delta_{\mathcal{A}}\Phi(z,\theta)$ и $\delta_{\mathcal{H}}$ дается формулой (2.174). Из (2.165) следует, что $\delta_{\mathcal{B}}-\delta_{\mathcal{A}}$ описывает изменения $\lambda(z)$, поэтому $\delta_{\mathcal{B}}\Phi(z,\theta)-\delta_{\mathcal{A}}\Phi(z,\theta)=\delta_\varepsilon\lambda_{Glob}(z)\cdot\frac{\partial\Phi(z,\theta)}{\partial\lambda}$. Так, что из (2.156) мы имеем

$$\Delta\Phi(z,\theta)=\varepsilon\cdot\left(\lambda(z)\cdot\frac{\partial\Phi(z,\theta)}{\partial z}+\left(1+\lambda(z)\cdot\lambda'(z)\right)\cdot\frac{\partial\Phi(z,\theta)}{\partial\lambda}\right).$$

Делая замену переменных $(z,\theta)\to(z_A,\theta_A)$ и используя соотношения

$$\frac{\partial\Phi(z,\theta)}{\partial z}=\left(1+\theta\cdot\lambda'(z)\right)\cdot\frac{\partial\Phi_A(z_A,\theta_A)}{\partial z_A}+\lambda'(z)\cdot\frac{\partial\Phi_A(z_A,\theta_A)}{\partial\theta_A}$$

и

$$\frac{\partial\Phi(z,\theta)}{\partial\lambda}=-\theta\cdot\frac{\partial\Phi_A(z_A,\theta_A)}{\partial z_A}+\frac{\partial\Phi_A(z_A,\theta_A)}{\partial\theta_A},$$

следующие из (2.165), мы получаем

$$\begin{aligned}\Delta\Phi(z,\theta) = (\theta+\lambda(z))\cdot\varepsilon\cdot\frac{\partial\Phi_A(z_A,\theta_A)}{\partial z_A} + \varepsilon\cdot\frac{\partial\Phi_A(z_A,\theta_A)}{\partial\theta_A} &= \\ \delta_{\mathcal{G}} z_A\cdot\frac{\partial\Phi_A(z_A,\theta_A)}{\partial z_A} + \delta_{\mathcal{G}}\theta_A\cdot\frac{\partial\Phi_A(z_A,\theta_A)}{\partial\theta_A} &= \\ \delta_{\mathcal{G}}\Phi_A(z_A,\theta_A)\,. &\end{aligned}$$

■

**2.4.6.** Нелинейная реализация конечных редуцированных преобразований. Рассмотрим условие представимости (2.166) для общих $N=1$ редуцированных преобразований $Z_A \to \tilde{Z}$, которые играют роль "линейных". В соответствии с [7, 9, 13] они могут быть параметризованы двумя функциями $g(z_A)$ и $\psi(z_A)$ и имеют вид

$$\mathcal{G}: \begin{cases} \tilde{z} &= f_{\mathrm{m}}^{(g\psi)}(z_A) + \theta_A\cdot\chi_{\mathrm{m}}^{(g\psi)}(z_A)\,, \\ \tilde{\theta} &= \psi(z_A) + \theta_A\cdot g(z_A)\,, \end{cases} \tag{2.177}$$

где

$$\begin{aligned} f_{\mathrm{m}}^{(g\psi)\prime}(z_A) &= \psi'(z_A)\,\psi(z_A) + \tfrac{1+\mathrm{m}}{2}\cdot g^2(z_A)\,, \\ \chi_{\mathrm{m}}^{(g\psi)\prime}(z_A) &= g'(z_A)\,\psi(z_A) + \mathrm{m}\cdot g(z_A)\,\psi'(z_A)\,, \end{aligned} \tag{2.178}$$

где $\mathrm{m} = \begin{cases} +1, & \textsf{SCf}\ \text{преобразования} \\ -1, & \textsf{TPt}\ \text{преобразования} \end{cases}$ — проекция "спина редукции", отвечающего за тип преобразований (см. **Подраздел 2.1** и [9]).

В попытках представить $\mathcal{G}$ в терминах нелинейных составляющих, подобно диаграмме (2.163), мы сталкиваемся со следующим ограничением, которое является следствием закона умножения $N=1$ суперконформно-подобных преобразований [7, 9].

Если $\mathcal{T}$ — суперконформно-подобное преобразование, то в компо-зиции $z \xrightarrow{\mathcal{T}} \tilde{z} \xrightarrow{\tilde{\mathcal{T}}} \tilde{\tilde{z}}$ имеется лишь две возможности

$$\begin{aligned} \tilde{\mathcal{T}}_{SCf} \circ \mathcal{T}_{SCf} &= \widetilde{\tilde{\mathcal{T}}}_{SCf}, \\ \tilde{\mathcal{T}}_{TPt} \circ \mathcal{T}_{SCf} &= \widetilde{\tilde{\mathcal{T}}}_{TPt}. \end{aligned} \tag{2.179}$$

Соответственно в терминах составляющих преобразований из диаграммы (2.163) имеем

$$\mathcal{G}_{SCf} \circ \mathcal{A}_{SCf} = \mathcal{B}_{SCf} \circ \mathcal{H}_{SCf}, \tag{2.180}$$

$$\mathcal{G}_{TPt} \circ \mathcal{A}_{SCf} = \mathcal{B}_{TPt} \circ \mathcal{H}_{SCf}. \tag{2.181}$$

Первое соотношение представляет собой аналог нелинейного представления $N = 1$ суперконформной группы (см. инфинитезимальный обратимый четырехмерный случай в [550, 562] and (2.166)), в котором $\mathcal{A}_{SCf}$ и $\mathcal{B}_{SCf}$ играют роль косетных преобразований.

Рассмотрим уравнение (2.180) в компонентах. Выберем косетные преобразования $\mathcal{A}_{SCf}$ and $\mathcal{B}_{SCf}$ в виде

$$\mathcal{A}_{SCf} : \begin{cases} z_A = z + \theta \cdot \lambda(z), \\ \theta_A = \lambda(z) + \theta \sqrt{1 + \lambda(z) \cdot \lambda'(z)}, \end{cases} \tag{2.182}$$

$$\mathcal{B}_{SCf} : \begin{cases} \tilde{z} = z_H + \theta_H \cdot \tilde{\lambda}(z_H), \\ \tilde{\theta} = \tilde{\lambda}(z_H) + \theta_H \sqrt{1 + \tilde{\lambda}(z_H) \cdot \tilde{\lambda}'(z_H)}, \end{cases} \tag{2.183}$$

и $\mathcal{H}$ параметризуем следующим образом

$$\mathcal{H}_{SCf} : \begin{cases} z_H = p(z), \\ \theta_H = \rho(z) + \theta \cdot q(z) \end{cases} \tag{2.184}$$

Тогда, разлагая координатные уравнения (2.167) на компоненты,

мы получаем четыре уравнения для четырех неизвестных составных функций $p(z), q(z), \rho(z), \tilde{\lambda}(z)$ в следующем виде

$$p(z) + \rho(z) \cdot \tilde{\lambda}(p(z)) = f_{+1}^{(g\psi)}(z) + g(z) \cdot \lambda(z) \cdot \psi(z), \tag{2.185}$$

$$\tilde{\lambda}(p(z)) + \rho(z) \cdot \sqrt{1 + \tilde{\lambda}(p(z)) \cdot \tilde{\lambda}'(p(z))} = \psi(z) + g(z) \cdot \lambda(z), \tag{2.186}$$

$$\begin{aligned} q(z) \cdot \tilde{\lambda}(p(z)) = {} & \lambda(z) \cdot f_{+1}^{(g\psi)\prime}(z) + \\ & g(z) \cdot \psi(z) \cdot \sqrt{1 + \lambda(z) \cdot \lambda'(z)}, \end{aligned} \tag{2.187}$$

$$\begin{aligned} q(z) \cdot \sqrt{1 + \tilde{\lambda}(p(z)) \cdot \tilde{\lambda}'(p(z))} = {} & \lambda(z) \cdot \psi'(z) + \\ & g(z) \cdot \sqrt{1 + \lambda(z) \cdot \lambda'(z)}, \end{aligned} \tag{2.188}$$

где $f_{+1}^{(g\psi)}(z)$ определяется в (2.178).

В случае, если $q(z)$ and $g(z)$ обратимы, эти уравнения имеют следующее решение для параметров преобразования $\mathcal{H}$ в терминах параметров “линейных” преобразований $\mathcal{G}$

$$p(z) = f_{+1}^{(g\psi)}(z) + g(z) \cdot \lambda(z) \cdot \psi(z), \tag{2.189}$$

$$q(z) = \sqrt{p'(z)}, \tag{2.190}$$

$$\rho(z) = 0, \tag{2.191}$$

и для эволюции голдстино

$$\tilde{\lambda}(p(z)) = \psi(z) + g(z) \cdot \lambda(z), \tag{2.192}$$

что естественно совпадает с предыдущим подходом (2.152), если подставить $f(z) = f_{+1}^{(g\psi)}(z)$ и $\chi(z) = g(z) \cdot \psi(z)$.

Следовательно, преобразование $\mathcal{H}$ есть расщепленное $N = 1$ су-перконформное преобразование [491, 563]

$$\mathcal{H}_{SCf} : \begin{cases} z_H &= p(z), \\ \theta_H &= \theta \cdot \sqrt{p'(z)} \end{cases} \tag{2.193}$$

с составным параметром $p(z)$ из (2.189).

Это может быть представлено в виде следующей диаграммы

$$\begin{array}{ccc} Z_A & \xrightarrow[\text{full}]{\mathcal{G}_{SCf}} & \tilde{Z} \\ \mathcal{A}_{SCf}\uparrow & & \uparrow\mathcal{B}_{SCf} \\ Z & \xrightarrow[\text{split}]{\mathcal{H}_{SCf}} & Z_H \end{array} \tag{2.194}$$

Второе соотношение (2.181) не имеет столь прозрачного смысла, поскольку аналог косетного преобразования $\mathcal{B}_{TPt}$ является теперь необратимым в отличие от стандартного косета [559]. Соответствующая коммутативная диаграмма имеет следующий вид

$$\begin{array}{ccc} Z_A & \xrightarrow{\mathcal{G}_{TPt}} & \tilde{Z} \\ \mathcal{A}_{SCf}\uparrow & & \uparrow\mathcal{B}_{TPt} \\ Z & \xrightarrow{\mathcal{H}_{SCf}} & Z_H \end{array} \tag{2.195}$$

Тем не менее, если предположить, что предыдущий подход дает правильное выражение (2.160) для составных компонент “нелинейного” преобразования $\mathcal{H}_{SCf}$, то необратимый аналог косетных преобразований $\mathcal{B}_{TPt}$ может быть в принципе найден из уравнений, аналогичных (2.185)–(2.188) [7, 11].

Запишем преобразование $\mathcal{B}_{TPt}$ в виде

$$\mathcal{B}_{SCf}:\begin{cases}\tilde{z} &= f_{-1}^{\left(b\widetilde{\lambda}\right)}\left(z_H\right)+\theta_H\cdot\chi_{-1}^{\left(b\widetilde{\lambda}\right)}\left(z_H\right),\\ \tilde{\theta} &= \tilde{\lambda}\left(z_H\right)+\theta_H\cdot b\left(z_H\right),\end{cases}\tag{2.196}$$

где

$$\begin{aligned}f_{\mathrm{m}}^{\left(b\widetilde{\lambda}\right)\prime}\left(z_H\right) &= \quad\tilde{\lambda}'\left(z_H\right)\cdot\tilde{\lambda}\left(z_H\right)+\tfrac{1+\mathrm{m}}{2}\cdot b^2\left(z_H\right),\\ \chi_{\mathrm{m}}^{\left(b\widetilde{\lambda}\right)\prime}\left(z_H\right) &= b'\left(z_H\right)\cdot\tilde{\lambda}\left(z_H\right)+\mathrm{m}\cdot b\left(z_H\right)\cdot\tilde{\lambda}'\left(z_H\right),\end{aligned}\tag{2.197}$$

и штрих означает производную по аргументам. Тогда соответствующая система уравнений имеет вид

$$f_{-1}^{\left(b\widetilde{\lambda}\right)}\left(p\left(z\right)\right)+\rho\left(z\right)\cdot\chi_{-1}^{\left(b\widetilde{\lambda}\right)}\left(p\left(z\right)\right) = f_{+1}^{\left(g\psi\right)}\left(z\right)+\lambda\left(z\right)\cdot\chi_{+1}^{\left(g\psi\right)}\left(z\right),\tag{2.198}$$

$$\tilde{\lambda}\left(p\left(z\right)\right)+\rho\left(z\right)\cdot b\left(p\left(z\right)\right) = \psi\left(z\right)+g\left(z\right)\cdot\lambda\left(z\right),\tag{2.199}$$

$$\begin{aligned}\rho\left(z\right)\cdot f_{-1}^{\left(b\widetilde{\lambda}\right)\prime}\left(p\left(z\right)\right)+q\left(z\right)\cdot\chi_{-1}^{\left(b\widetilde{\lambda}\right)}\left(p\left(z\right)\right) &= \lambda\left(z\right)\cdot f_{+1}^{\left(g\psi\right)\prime}\left(z\right)+\\ &\quad\chi_{+1}^{\left(g\psi\right)}\left(z\right)\cdot\sqrt{1+\lambda\left(z\right)\cdot\lambda'\left(z\right)},\end{aligned}\tag{2.200}$$

$$\begin{aligned}\rho\left(z\right)\cdot q\left(z\right)\cdot\tilde{\lambda}'\left(p\left(z\right)\right)+q\left(z\right)\cdot b\left(p\left(z\right)\right) &= \lambda\left(z\right)\cdot\psi'\left(z\right)+\\ &\quad g\left(z\right)\cdot\sqrt{1+\lambda\left(z\right)\cdot\lambda'\left(z\right)}.\end{aligned}\tag{2.201}$$

Если преобразование $\mathcal{A}_{SCf}$ обратимо, то мы получаем

$$\mathcal{G}_{TPt}=\mathcal{B}_{TPt}\circ\mathcal{H}_{SCf}\circ\mathcal{A}_{SCf}^{-1}\tag{2.202}$$

что дает диаграммных аналог нелинейной реализации для необратимых сплетающих четность преобразований [7, 20].

## 2.5. Дробно-линейные преобразования

Выясним, какие из преобразований,удовлетворяющих (2.37) или (2.38), могут быть реализованы как линейные преобразования в суперпроективном пространстве $\mathbb{C}P^{1|1}$ после перехода к однородным координатам.

Предположим, что $\epsilon\left[y\right] \neq 0$, тогда, вводя однородные координаты [342, 429]

$$\mathrm{X} = \begin{pmatrix} x \\ y \\ \eta \end{pmatrix} \in \mathbb{C}^{2|1}, \tag{2.203}$$

неоднородные координаты можно записать в виде $z = x/y,\ \theta = \eta/y$.

Поставим в соответствие общему $(5|4)$-мерному линейному отображению в $\mathbb{C}P^{1,1}$ преобразование однородных координат

$$\tilde{\mathrm{X}} = \mathrm{M} \cdot \mathrm{X}, \tag{2.204}$$

где

$$\mathrm{M} = \begin{pmatrix} a & b & \alpha \\ c & d & \beta \\ \gamma & \delta & e \end{pmatrix}. \tag{2.205}$$

Соответствующее дробно-линейное преобразование в неоднородных координатах выражается через элементы суперматрицы $\mathrm{M}$

$$\begin{aligned} \tilde{z} &= \frac{az+b}{cz+d} + \theta \cdot \frac{\left(\beta a - \alpha c\right)z + \beta b - \alpha d}{\left(cz+d\right)^2}, \\ \tilde{\theta} &= \frac{\gamma z + \delta}{cz+d} + \theta \cdot \frac{\left(\beta\gamma + ec\right)z + \beta\delta + ed}{\left(cz+d\right)^2}. \end{aligned} \tag{2.206}$$

Исследование свойств дробно-линейных преобразований удобно про-водить в терминах нечетных аналогов миноров для суперматриц — полуминоров и полуматриц, которые введены в **Приложении Д.2**.

**2.5.1.** Суперконформные преобразования. В терминах полуминоров преобразования (2.206) имеют вид

$$\begin{aligned} \tilde{z} &= \frac{az+b}{cz+d} + \theta \cdot \frac{\delta\mathrm{et}\mathcal{M}_{\delta} \cdot z + \delta\mathrm{et}\mathcal{M}_{\gamma}}{(cz+d)^2}, \\ \tilde{\theta} &= \frac{\gamma z+\delta}{cz+d} + \theta \cdot \frac{\det M_b \cdot z + \det M_a}{(cz+d)^2}. \end{aligned} \tag{2.207}$$

Из этого выражения видно, зачем были введены полуминоры и аналоги матричных функций от них.

**Определение 2.74.** *Четно-нечетная симметрия дробно-линейных преобразований определяется как*

$$\begin{aligned} a &\leftrightarrow \gamma, \\ b &\leftrightarrow \delta, \\ \det &\leftrightarrow \delta\mathrm{et}. \end{aligned} \tag{2.208}$$

Суперконформные условия (2.37) дают четыре уравнения на параметры суперматрицы M (2.205) в виде

$$\begin{aligned} & e\beta \cdot \delta\mathrm{et}\mathcal{M}_{\alpha} = 0, \\ & e \cdot \delta\mathrm{et}\mathcal{M}_{\alpha} = \beta \cdot \det M_e, \\ & \beta \cdot \mathrm{per}\, M_e - e \cdot \pi\mathrm{er}\mathcal{M}_{\alpha} = 2\alpha cd, \\ & \det M_e = e^2 + \gamma\delta + \frac{\beta e}{2cd} \cdot \pi\mathrm{er}\mathcal{M}_{\alpha}. \end{aligned} \tag{2.209}$$

Здесь предполагается, что $\epsilon\,[cd] \neq 0$.

Рассмотрим возможные решения системы уравнений (2.209), учитывая также и необратимый вариант. Первое уравнение в (2.209) задает три типа соответствующих решений по количеству вариантов его решения

$$\beta \cdot \delta\text{et}\mathcal{M}_\alpha = 0, \tag{2.210}$$

$$e\beta \cdot \delta\text{et}\mathcal{M}_\alpha = 0, \tag{2.211}$$

$$e\beta = 0. \tag{2.212}$$

Рассмотрим первое уравнение (2.210) более детально. при ненулевых сомножителях оно имеет следующее решение

$$\beta = const \cdot \delta\text{et}\mathcal{M}_\alpha. \tag{2.213}$$

Тогда получаем решение для M в виде матрицы суперпроективных преобразований [342, 429]

$$\mathrm{M}^{inv}_{SCf} = \begin{pmatrix} a & b & \dfrac{\delta\text{et}\mathcal{M}_\beta}{\sqrt{\det \mathrm{M}_e}} \\ c & d & \dfrac{\delta\text{et}\mathcal{M}_\alpha}{\sqrt{\det \mathrm{M}_e}} \\ \gamma & \delta & \sqrt{\det \mathrm{M}_e} - \dfrac{3}{2}\gamma\delta \end{pmatrix}. \tag{2.214}$$

Березиниан суперматрицы $\mathrm{M}^{inv}_{SCf}$ имеет вид

$$\mathrm{Ber}\,^{inv}_{SCf}\mathrm{M} = \sqrt{\det \mathrm{M}_e} + \frac{3}{2}\gamma\delta - \frac{2\gamma\delta}{\sqrt{\det \mathrm{M}_e}}.$$

Обратимые матрицы (2.214) с единичным березинианом образуют $(3|2)$-мерную супергруппу $OSp_{\mathbb{C}}(2|1)$, свойства которой применяются для расчета многопетлевых амплитуд в суперструнных теориях [320, 344, 362]. Описание классов суперконформных многообразий [341, 343] может быть проведено с помощью различных дискретных подгрупп этой супергруппы [353].

В неоднородных координатах для обратимого дробно-линейного суперконформного преобразования $\mathbb{C}^{1,1} \to \mathbb{C}^{1,1}$ получаем

$$\begin{aligned} \tilde{z} &= \frac{az+b}{cz+d} + \theta \cdot \frac{\gamma z + \delta}{(cz+d)^2} \sqrt{\det \mathrm{M}_e}, \\ \tilde{\theta} &= \frac{\gamma z + \delta}{cz+d} - \theta \cdot \frac{\sqrt{\det \mathrm{M}_e - \gamma\delta}}{cz+d}, \end{aligned} \tag{2.215}$$

и березиниан преобразования $Z \to \tilde{Z}$ имеет вид

$$\boldsymbol{J}_{SCf}^{inv} = \mathrm{Ber}\,\left(\tilde{Z}/Z\right) = \frac{\sqrt{\det \mathrm{M}_e - \gamma\delta}}{cz+d} + \theta \cdot \frac{\delta \mathrm{et} \mathcal{M}_\alpha}{(cz+d)^2}. \tag{2.216}$$

*Замечание* **2.75.** Здесь мы видим явно смысл введения полудетерминантов: если $\sqrt{\det \mathrm{M}_e}$ контролирует числовую часть березиниана, то $\delta\mathrm{et}\mathrm{M}_\alpha$ отвечает за его $\theta$-зависимость [1]. Это позволяет также трактовать полудетерминат как квадратный корень из обычного детерминанта.

Если включить в рассмотрение и полунеобратимые (2.4) суперконформные преобразования, то можно воспользоваться другими уравнениями (2.211) и (2.212).

Так, условие (2.211) может быть выполено с помощью подстановки $e = \mu \cdot \delta\mathrm{et}\mathrm{M}_\alpha$, что для суперматрицы M дает

$$\mathrm{M}_{SCf}^{halfinv} = \begin{pmatrix} a & b & \mu\delta\gamma + \dfrac{\beta}{2cd} \mathrm{per}\,\mathrm{M}_e \\ c & d & \beta \\ \gamma & \delta & \delta\mathrm{et}\mathcal{M}_\alpha \end{pmatrix}, \tag{2.217}$$

где выполняются дополнительные условия

$$\det \mathrm{M}_e = \gamma\delta, \tag{2.218}$$

$$\beta \det \mathrm{M}_e = 0. \tag{2.219}$$

Если считать, что $\beta \neq 0$ и $\det \mathrm{M}_e \neq 0$, то должно быть $\beta \in \mathrm{Ann}\,[\det \mathrm{M}_e]$ или $\beta\gamma\delta = 0$. Для выполнения (2.218) и (2.219) положим $a = a_0\gamma\delta$ и $b = b_0\gamma\delta$, где $\det \begin{pmatrix} a_0 & b_0 \\ c & d \end{pmatrix} = 1$. Тогда получаем полунеобратимое преобразование

$$\begin{aligned} \tilde{z} &= \frac{a_0 z + b_0}{cz + d} + \theta \cdot \frac{\mu\gamma\delta}{cz + d}, \\ \tilde{\theta} &= \frac{\gamma z + \delta}{cz + d} - \theta \cdot \frac{\mu \cdot \delta\mathrm{et}\mathcal{M}_\alpha}{cz + d}, \end{aligned} \tag{2.220}$$

для которого необратимый аналог якобиана (E.8) имеет вид

$$\boldsymbol{J}_{SCf}^{halfinv} = \frac{\delta\mathrm{et}\mathcal{M}_\alpha}{cz + d} \left( \mu + \frac{\theta}{cz + d} \right). \tag{2.221}$$

*Замечание* **2.76.** Сравнивая (2.216) и (2.221), можем убедиться, что для полунеобратимых преобразований полудетерминант $\delta\mathrm{et}\,\mathrm{M}_\alpha$ играет роль, аналогичную той, которую корень из обычного детерминанта $\sqrt{\det \mathrm{M}_e}$ играет для обратимых преобразований.

Остальные возможные случаи перечислены в [1].

**2.5.2.** Сплетающие четность преобразования. Применяя условие (2.38) к общим дробно-линейным преобразованиям, получаем уравнения на параметры суперматрицы M

$$\begin{aligned} &\det \mathrm{M}_e = \gamma\delta, \\ &\delta\mathrm{et}\mathcal{M}_\delta = \gamma e, \\ &c \cdot \delta\mathrm{et}\mathcal{M}_\gamma = \gamma \cdot \det \mathrm{M}_a. \end{aligned} \tag{2.222}$$

Видно, что первое уравнение в совпадает с полуобратимым супер-конформным условием (2.218).

Тогда в одном из возможных вариантов решения системы (2.222) для матрицы M получаем

$$\mathrm{M}_{TPt}^{noninv} = \begin{pmatrix} a & b & -ad\eta \\ c & d & -cd\eta \\ \gamma & \delta & \eta \cdot \pi\mathrm{er}\mathcal{M}_{\alpha} \end{pmatrix}, \tag{2.223}$$

где $\eta$ — нечетный параметр.

Отметим, что среди всех рассмотренных преобразований групповыми свойствами обладают лишь обратимые суперконформные преобразования (2.215).

**2.5.3.** Супераналоги расстояния в $\mathbb{C}^{1|1}$. Расстояние между двумя точками в $\mathbb{C}^{1|1}$ определяется как $|Z_{12}|$, где

$$Z_{12} = z_1 - z_2 - \theta_1\theta_2 \tag{2.224}$$

(см. [111, 491, 564]). Относительно суперконформных обратимых преобразований (2.215) величина $Z_{12}$ преобразуется ковариантно [565]

$$\tilde{Z}_{12} = \boldsymbol{J}_{SCf}^{inv}(Z_1)\,\boldsymbol{J}_{SCf}^{inv}(Z_2)\,Z_{12}, \tag{2.225}$$

где $\boldsymbol{J}_{SCf}^{inv}(Z)$ - якобиан преобразования, который в данном случае равен березиниану (2.216). Чтобы рассмотреть, как преобразуется $Z_{12}$ в необратимом случае, остановимся на соотношении более подробно. Используя (2.2), представим левую часть (2.225) в виде

$$\begin{aligned} \tilde{Z}_{12} &= f(z_1) - f(z_2) - \psi(z_1)\cdot\psi(z_2) + \\ &\quad (\theta_1 g(z_1) + \theta_2 g(z_2))\cdot(\psi(z_1) - \psi(z_2)) - \theta_1\theta_2 g(z_1)\,g(z_2)\,. \end{aligned} \tag{2.226}$$

Здесь мы использовали суперконформное условие $\chi(z) = \psi(z)\, g(z)$ (см. (2.81) при $\mathrm{m} = +1$).

Отметим, для любых дробно-линейных функций

$$f(z) = \frac{az+b}{cz+d},\ \psi(z) = \frac{\gamma z+\delta}{cz+d},$$

что соответствует фиксации элементов из первых двух столбцов суперматрицы M (2.205), можно получить

$$\begin{aligned} & f(z_1) - f(z_2) = R\cdot(z_1 - z_2)\cdot \det \mathrm{M}_e, \\ & \psi(z_1) - \psi(z_2) = R\cdot(z_1 - z_2)\cdot \delta\mathrm{et}\mathcal{M}_\alpha, \\ & \psi(z_1)\, f(z_2) - \psi(z_2)\, f(z_1) = R\cdot(z_1 - z_2)\cdot \delta\mathrm{et}\mathcal{M}_\beta, \\ & \psi(z_1)\cdot\psi(z_2) = R\cdot(z_1 - z_2)\,\gamma\delta, \end{aligned} \tag{2.227}$$

где $R^{-1} = (cz_1 + d)(cz_2 + d)$ (см. также (Д.23) и (Д.24)).

Тогда для преобразованного расстояния $\tilde{Z}_{12}$ из (2.226) получаем

$$\begin{aligned} \tilde{Z}_{12} \;=\; & R\cdot(z_1 - z_2)\cdot(\det \mathrm{M}_e - \gamma\delta) + \\ & R\cdot(z_1 - z_2)\cdot(\theta_1 g(z_1) + \theta_2 g(z_2))\cdot \delta\mathrm{et}\mathcal{M}_\alpha - \theta_1\theta_2 g(z_1)\, g(z_2)\,. \end{aligned} \tag{2.228}$$

Отсюда следует, что от функции $g(z)$ зависят дальнейшие свойства $Z_{12}$. Так, в случае обратимых суперконформных преобразований (2.215)

$$g(z) = \frac{\sqrt{\det \mathrm{M}_e - \gamma\delta}}{cz+d}. \tag{2.229}$$

Все необратимые преобразования содержат уравнение $\det \mathrm{M}_e = \gamma\delta$ (см. (2.218) и (2.222)), поэтому из (2.228) мы имеем

**Утверждение 2.77.** *Для необратимых дробно-линейных преобразований $\tilde{Z}_{12}$ не содержит $\theta$-независимых слагаемых, и поэтому является* *чисто дýховой величиной.*

Из рассмотрения последнего слагаемого в (2.228) следует

**Утверждение 2.78.** *Нильпотентность $g(z)$ приводит к линейности $\tilde{Z}_{12}$ по четным координатам.*

Окончательно можно сформулировать

**Предложение 2.79.** *Для полунеобратимых суперконформных преобразований*

$$\tilde{Z}_{12} = 0. \tag{2.230}$$

*Доказательство.* В всех вариантах полунеобратимых преобразований

$$g(z) \subset \mathrm{Ann}\left[\delta\mathrm{et}\mathcal{M}_{\alpha}\right], \tag{2.231}$$

поэтому второе слагаемое в (2.228) равно нулю. Первое слагаемое зануляется вследствие **Утверждения 2.77**. Поскольку выполняется (2.231), функция $g(z)$ нильпотентна, и последнее слагаемое в (2.228) также равно нулю по **Утверждению 2.78**. ■

Соотношение (2.230) можно трактовать также и как определение полунеобратимых преобразований.

Отметим, что для всех полунеобратимых преобразований выполняется аналогичное (2.225) соотношение

$$\delta\mathrm{et}\tilde{\mathcal{N}} = R \cdot (z_1 - z_2) \cdot \delta\mathrm{et}\mathcal{M}_{\beta}, \tag{2.232}$$

где

$$\mathcal{N} = \begin{pmatrix} z_1 & z_2 \\ \theta_1 & \theta_2 \end{pmatrix}.$$

**Предложение 2.80.** *Для полуобратимых суперконформных преобра-*

*зований разности преобразованных четных и нечетных координат пропорциональны, т. е.* $\tilde{z}_1 - \tilde{z}_2 \sim \tilde{\theta}_1 - \tilde{\theta}_2$.

*Доказательство.* Непосредственно из (2.220) получаем

$$\begin{aligned} \tilde{z}_1 - \tilde{z}_2 &= R \cdot Z_{12}^{hinv} \cdot \det \mathrm{M}_e, \\ \tilde{\theta}_1 - \tilde{\theta}_2 &= R \cdot Z_{12}^{hinv} \cdot \delta\mathrm{et}\mathcal{M}_\alpha, \end{aligned} \tag{2.233}$$

где

$$Z_{12}^{hinv} = z_1 - z_2 + \delta\mathrm{et}\mathcal{N} \cdot \mu c + (\theta_1 - \theta_2) \cdot \mu d \tag{2.234}$$

и содержит все типы расстояний из (2.232) и (2.233). ■

Другие подобные соотношения для суперконформных преобразований приведены в [1].

В случае вращающих четность преобразований (2.223) суперрасстояние между двумя точками является образом нечетного расстояния

$$\tilde{Z}_{12}^{TPt} = R \cdot (\theta_1 - \theta_2) \cdot (\delta\mathrm{et}\mathcal{M}_\gamma - e\delta) \,. \tag{2.235}$$

Здесь параметры выбраны таким образом, чтобы занулить квадратичное по $\theta$ слагаемое в (2.226).

**2.5.4.** Необратимый аналог метрики в $\mathbb{C}^{1|1}$ . Если положить элементы суперматрицы M (2.205) действительными, а расстояние между точками в (2.235) бесконечно малым, то получаем

$$d\tilde{Z}^{TPt} = \frac{\delta\mathrm{et}\mathcal{M}_\gamma - e\delta}{|cz+d|^2} d\theta, \tag{2.236}$$

что может рассматриваться как ключевое соотношение для нахождения необратимого TPt аналога [1] суперконформной метрики на верхней полуплоскости [111, 330, 566]. Очевидно, что (2.236) является “дробно-линейным” следствием общего соотношения между дифференциалами при вращающих четность преобразованиях (2.44). Далее из (2.235) находим

$$\operatorname{Im}\tilde{z} + \frac{1}{2}\tilde{\theta}\tilde{\theta} = \frac{\delta \mathrm{et} \mathcal{M}_\gamma - e\delta}{|cz+d|^2}\operatorname{Im}\theta. \tag{2.237}$$

Поэтому

$$\left|d\tilde{Z}^{TPt}\right|\operatorname{Im}\theta = |d\theta|\left(\operatorname{Im}\tilde{z} + \frac{1}{2}\tilde{\theta}\tilde{\bar{\theta}}\right). \tag{2.238}$$

Заметим, что здесь нет деления, поскольку некоторые сомножители могут быть необратимыми. Отсюда получаем

**Определение 2.81.** *Необратимый* TPt *аналог метрики на верхней* $\mathbb{C}^{1|1}$ *полуплоскости* $\left|ds^{TPt}\right|$ *удовлетворяет одновременно соотношениям*

$$\begin{aligned} &\left|ds^{TPt}\right|\operatorname{Im}\theta = |d\theta|, \\ &\left|ds^{TPt}\right|\left(\operatorname{Im}\tilde{z} + \frac{1}{2}\tilde{\theta}\tilde{\bar{\theta}}\right) = \left|d\tilde{Z}^{TPt}\right|. \end{aligned} \tag{2.239}$$

Таким образом, приведенные соотношения могут трактоваться как необратимый аналог инвариантности — “полуинвариантность” введенной метрики [1].

## 2.6. Основные результаты и выводы

1. Построена супераналитическая полугруппа и дано определение супераналитических полусупермногообразий.
2. Введены дополнительные редукции касательного суперпространства с учетом необратимости, которые приводят к более глубокому обобщению понятия суперсимметричной комплексной структуры.
3. Построена $N = 1$ суперконформная полугруппа, принадлежащая к новому абстрактному типу полугрупп, которые имеют необычную идеальную структуру.

**4.** Изучена локальная структура суперконформной полугруппы, определены обобщенные векторные и тензорные отношения Грина, новая характеристика — идеальный квазихарактер.

**5.** Найден необратимый супераналог антиголоморфных преобразований — сплетающие четность преобразования, которые дуальны суперконформным преобразованиям в смысле приведенной формулы сложения березинианов.

**6.** Получена новая категория коциклов — сплетенные коциклы с различным типом стрелок.

**7.** Определены редуцированные преобразования, объединяющие старые и новые преобразования. Введен спин редукции, проекция которого различает их между собой.

**8.** Изучены дробно-линейные необратимые редуцированные преобразования в терминах введенных полуминоров и полуматриц,для которых определены функции полуперманента и полудетерминанта.

**9.** Найдена четно-нечетная симметрия дробно-линейных $N = 1$ суперконформных преобразований, которая состоит в симметрии относительно одновременной замены детерминанта на полудетерминант и четных координат на нечетные.

**10.** Построены необратимые супераналоги расстояния в $(1|1)$-мерном суперпространстве. Введен необратимый аналог метрики и показана ее полуинвариантность.

**11.** Изучены нелинейные реализации редуцированных $N = 1$ редуцированных преобразований и найден новый тип голдстино как решение, которое соответствует сплетающим преобразованиям.

# РАЗДЕЛ 3

# НЕОБРАТИМАЯ ГЕОМЕТРИЯ РАСШИРЕННЫХ РЕДУЦИРОВАННЫХ ПРЕОБРАЗОВАНИЙ

В разделе исследуются свойства $N = 2$ и $N = 4$ редуцированных обратимых и необратимых отображений суперплоскости. Нетривиальные редукции расширенных касательных суперпространств приводят к $N$-обобщению понятия комплексной структуры, к полной классификации $N = 2$ и $N = 4$ преобразований в терминах перманентов и полуминоров. Строятся и анализируются $N = 2$ и $N = 4$ суперконформные полугруппы в альтернативной параметризации и приводится их компонентное представление. Обсуждаются свойства сплетающих четность расширенных преобразований и соответствующих супердифференциалов.

Хорошо известно, что многие пространственно-временные свойства суперструнных теорий элементарных частиц тесно связаны со свойствами мирового листа струны [282, 283, 286, 567]. Так, в работах [568–571] было показано, что необходимым условием $N = 1$ $(N = 2)$ пространственно-временной суперсимметрии является $N = 2$ $(N = 4)$ расширенная суперконформная симметрия на мировом листе, которая была впервые рассмотрена в контексте фермионных струн [572–574]. Общие свойства $N$-расширенных суперконформных алгебр изучались в работах [575–581], а $N$-расширенные суперконформные теории поля исследовались в [565, 582–587]. Исключительно важной является также связь расширенных суперконформных алгебр с геометрией пространства анти-Де Ситтера при $N = 2$ [588–590] и $N = 4$ [591, 592]. Однако как было показано в [575], при $N \geq 5$ не существует центральных расширений, а суперконформные теории поля, хотя и могут быть сформулированы при произвольных $N$ [593,594], но становятся тривиальными и не имеют осмысленной квантовой физики [565]. Поэтому здесь мы рассмотрим необратимые обобщения суперконформной геометрии только на $N = 2$ и $N = 4$ [2, 3, 12, 17].

## 3.1. $N = 2$ суперконформная геометрия

Исследования различных вариантов $N = 2$ суперконформной теории поля [581, 595–601] и $N = 2$ суперконформной геометрии [602] явилось чрезвычайно важным инструментом для построения как гипотетических теорий критических $N = 2$ струн*) в пространстве-времени с сигнатурой $(2, 2)$ [610–613], так и последовательных реалистичных моделей, основанных на суперструнных компактификациях методом косетов $G/H$ [614–619]. С геометрической точки зрения мировой лист суперструны (в моделях с пространственно-временной суперсимметрией) представляет собой $N = 2$ суперриманову поверхность [563, 620–624], склеенную $N = 2$ суперконформными преобразованиями [596, 625].

В этом подразделе мы подробно изучим аналоги этих преобразований — обратимые и необратимые $N = 2$ редуцированные преобразования, используя также и несуперконформные редукции, аналогичные введенным в **Подразделе 2.3** и формализм перманентов (см. [626] и **Раздел 5**).

В стандартном базисе суперпространство $\mathbb{C}^{1|2}$ локально описывается голоморфными суперкоординатами $Z = \left(z, \theta^1, \theta^2\right)$, где $\left(\theta^i\right)^2 = 0$, $\left\{\theta^1, \theta^2\right\} = 0$. При четных $N$ удобнее пользоваться комплексным ба-

---

*) *Примечание.* Тем не менее, недавно [603, 604] обнаружена тесная связь $N = 2$ струн с $M$-теорией [298, 299, 605] и $D$-бранами [302, 606–609].

зисом в нечетном секторе [563] $\theta^{\pm} = \dfrac{\theta^1 \pm i\theta^2}{\sqrt{2}}$. (для произвольных $N$ см. **Приложение E.2**). Тогда касательное суперпространство также можно рассматривать в комплексном базисе $\left(\partial, D^+, D^-\right)^T$, где

$$D^{\pm} = \frac{D_1 \pm iD_2}{\sqrt{2}} = \partial_{\mp} + \theta^{\pm}\partial, \; \partial_{\mp} = \frac{\partial}{\partial\theta^{\mp}} \tag{3.1}$$

и $D_i$ определены в (E.18), кроме того, для соотношения суперсимметрии вместо (E.19) имеем

$$\left(D^{\pm}\right)^2 = 0, \tag{3.2}$$

$$\left\{D^+, D^-\right\} = 2\partial. \tag{3.3}$$

Аналогично для кокасательного $(1|2)$ суперпространства вместо (E.28) получаем

$$dZ = dz + \theta^+ d\theta^- + \theta^- d\theta^+, \quad d\theta^{\pm} = \frac{d\theta^1 \pm id\theta^2}{\sqrt{2}}. \tag{3.4}$$

В комплексном базисе суперполевое разложение (E.17) имеет вид

$$F\left(z, \theta^+, \theta^-\right) = F_0\left(z\right) + \theta^+ F_-\left(z\right) + \theta^- F_+\left(z\right) + \theta^+\theta^- F_{+-}\left(z\right), \tag{3.5}$$

где $F_0$ и $F_{+-}$ — одной четности с $F$, а $F_+$ и $F_-$ — противоположной.

**3.1.1.** К л а с с и ф и к а ц и я $N=2$ р а с ш и р е н н ы х с у п е р а н а л и т и ч е с к и х п р е о б р а з о в а н и й . Для классификации по необратимости $N=2$ супераналитических преобразований необходимо получить выражение для $N=2$ березиниана $\mathrm{Ber}^{N=2}\left(\tilde{Z}/Z\right)$ (или его необратимого аналога) через суперматрицу $\mathrm{P}_{SA}^{(N=2)}$ подобно $N=1$ преобразованиям (E.9)–(E.10) (см. **Приложение E.3**).

Запишем суперматрицу производных $\mathrm{P}_{SA}^{(N=2)}$ (E.30) в следующей форме, к удобной для рассмотрения дальшейших редукций,

$$\mathrm{P}_{SA}^{(N=2)} = \left( \begin{array}{cc} Q\left(z,\theta^{+},\theta^{-}\right) & \partial\tilde{\theta}^{+} \;\; \partial\tilde{\theta}^{-} \\ \mathit{\Delta}^{-}\left(z,\theta^{+},\theta^{-}\right) & \\ \mathit{\Delta}^{+}\left(z,\theta^{+},\theta^{-}\right) & \mathrm{H} \end{array} \right), \tag{3.6}$$

где

$$Q\left(z,\theta^{+},\theta^{-}\right) = \partial\tilde{z} - \partial\tilde{\theta}^{+}\cdot\tilde{\theta}^{-} - \partial\tilde{\theta}^{-}\cdot\tilde{\theta}^{+}, \tag{3.7}$$

$$\mathit{\Delta}^{\pm}\left(z,\theta^{+},\theta^{-}\right) = D^{\pm}\tilde{z} - D^{\pm}\tilde{\theta}^{-}\cdot\tilde{\theta}^{+} - D^{\pm}\tilde{\theta}^{+}\cdot\tilde{\theta}^{-}, \tag{3.8}$$

$$\mathrm{H} = \left( \begin{array}{cc} D^{-}\tilde{\theta}^{+} & D^{-}\tilde{\theta}^{-} \\ D^{+}\tilde{\theta}^{+} & D^{+}\tilde{\theta}^{-} \end{array} \right). \tag{3.9}$$

Тогда $N=2$ березиниан в случае $\epsilon\left[\det\mathrm{H}\right]\neq 0$ равен

$$\mathrm{Ber}^{\,N=2}\left(\tilde{Z}/Z\right) = \mathrm{Ber}\,\mathrm{P}_{SA}^{(N=2)} =$$

$$\frac{Q\left(z,\theta^{+},\theta^{-}\right) - \left(\begin{array}{cc}\partial\tilde{\theta}^{+}, & \partial\tilde{\theta}^{-}\end{array}\right)\cdot\mathrm{H}^{-1}\cdot\left(\begin{array}{c}\mathit{\Delta}^{-}\left(z,\theta^{+},\theta^{-}\right)\\ \mathit{\Delta}^{+}\left(z,\theta^{+},\theta^{-}\right)\end{array}\right)}{\det\mathrm{H}} =$$

$$\frac{Q\left(z,\theta^{+},\theta^{-}\right)}{\det\mathrm{H}} - \frac{\partial\tilde{\theta}^{+}\cdot D^{+}\tilde{\theta}^{-} - \partial\tilde{\theta}^{-}\cdot D^{+}\tilde{\theta}^{+}}{\det^{2}\mathrm{H}}\cdot\mathit{\Delta}^{-}\left(z,\theta^{+},\theta^{-}\right) -$$

$$\frac{\partial\tilde{\theta}^{-}\cdot D^{-}\tilde{\theta}^{+} - \partial\tilde{\theta}^{+}\cdot D^{-}\tilde{\theta}^{-}}{\det^{2}\mathrm{H}}\cdot\mathit{\Delta}^{+}\left(z,\theta^{+},\theta^{-}\right). \tag{3.10}$$

Отсюда следует классификация по необратимости общих $N=2$ супераналитических преобразований.

**Определение 3.1.** *Обратимые* $N=2$ *супераналитические преобра-*

*зования определяются условиями*

$$\epsilon\left[Q\left(z,\theta^{+},\theta^{-}\right)\right]\neq 0,\ \epsilon\left[\det \mathrm{H}\right]\neq 0. \tag{3.11}$$

**Определение 3.2.** *Полунеобратимые* $N=2$ *супераналитические преобразования определяются условиями*

$$\epsilon\left[Q\left(z,\theta^{+},\theta^{-}\right)\right]=0,\ \epsilon\left[\det \mathrm{H}\right]\neq 0. \tag{3.12}$$

**Определение 3.3.** *Необратимые* $N=2$ *супераналитические преобразования определяются условиями*

$$\epsilon\left[Q\left(z,\theta^{+},\theta^{-}\right)\right]=0,\ \epsilon\left[\det \mathrm{H}\right]=0. \tag{3.13}$$

**3.1.2.** Компонентное представление и $N=2$ супераналитическая полугруппа. Супераналитическое отображение $Z\left(z,\theta^{+},\theta^{-}\right)\to\tilde{Z}\left(\tilde{z},\tilde{\theta}^{+},\tilde{\theta}^{-}\right)$ суперпространства $\mathbb{C}^{1|2}$ после разложения в ряд по нечетным координатам (как (3.5)) определяется 12 функциями на $\mathbb{C}^{1|0}$ (6 четных $f,h,g_{ab}:\mathbb{C}^{1|0}\to\mathbb{C}^{1|0}$ и 6 нечетных $\psi_a,\chi_a,\lambda_a:\mathbb{C}^{1|0}\to\mathbb{C}^{0|1}$, где $a,b=\pm$) следующим образом

$$\begin{cases}\tilde{z}=f\left(z\right)+\theta^{+}\chi_{-}\left(z\right)+\theta^{-}\chi_{+}\left(z\right)+\theta^{+}\theta^{-}h\left(z\right),\\ \tilde{\theta}^{\pm}=\psi_{\pm}\left(z\right)+\theta^{\pm}g_{\pm\mp}\left(z\right)+\theta^{\mp}g_{\pm\pm}\left(z\right)+\theta^{\pm}\theta^{\mp}\lambda_{\pm}\left(z\right).\end{cases} \tag{3.14}$$

**Определение 3.4.** *Множество обратимых и необратимых преобразований* $\mathbb{C}^{1|2}\to\mathbb{C}^{1|2}$ (3.14) *образует полугруппу относительно композиции преобразований, которую мы назовем полугруппой* $N=2$ *супераналитических преобразований* $\boldsymbol{T}_{SA}^{(N=2)}$.

*Замечание* **3.5.** Обратимые преобразования очевидно образуют подгруппу $\boldsymbol{G}_{SA}^{(N=2)}$ полугруппы $\boldsymbol{T}_{SA}^{(N=2)}$.

**Определение 3.6.** *Необратимые преобразования* $\mathbb{C}^{1|2} \to \mathbb{C}^{1|2}$ (3.14) *входят в идеал* $\boldsymbol{I}_{SA}^{(N=2)}$ *полугруппы* $\boldsymbol{T}_{SA}^{(N=2)}$.

*Замечание* **3.7.** Согласно абстрактной теории полугрупп [102–104] все преобразования некоторого множества образут полугруппу относительно композиции.

Поскольку нечетные функции $\psi_{\pm}(z)\,, \chi_{\pm}(z)\,, \lambda_{\pm}(z)$ необратимы по определению [120], а функция $h(z)$ входит с коэффициентом $\theta^{+}\theta^{-}$, то мы имеем

**Утверждение 3.8.** *Обратимость всего преобразования будет определяться только функциями* $f(z)$ *и* $g_{ab}(z)$.

*Доказательство.* Действительно, в терминах компонентных функций $f(z)$ и $g_{ab}(z)$ для $Q(z,\theta^{+},\theta^{-})$ (3.7) и $\det \mathrm{H}$ (3.9) получаем

$$\epsilon\left[Q\left(z,\theta^{+},\theta^{-}\right)\right] = \epsilon\left[f'(z)\right], \tag{3.15}$$

$$\epsilon\left[\det \mathrm{H}\right] = \epsilon\left[\det \mathrm{G}\right], \tag{3.16}$$

где

$$\mathrm{G} = \begin{pmatrix} g_{+-}(z) & g_{++}(z) \\ g_{--}(z) & g_{-+}(z) \end{pmatrix}. \tag{3.17}$$

■

Поэтому определения (3.11)–(3.13) могут быть переформулированы в терминах функций $f(z)$ и $g_{ab}(z)$ с очевидными заменами (3.15).

Для соответствующей параметризации полугруппы $N=2$ супераналитических преобразований $\boldsymbol{T}_{SA}^{(N=2)}$ мы используем компонентные функции на $\mathbb{C}^{1|0}$, входящие в (3.14).

**Определение 3.9.** *Суперaналитическая полугруппа* $\mathbf{S}_{SA}^{(N=2)} \ni \mathbf{s}$ *параметризуется функциональной матрицей*

$$\left\{ \begin{array}{cccc} f & h & \chi_- & \chi_+ \\ \psi_+ & \lambda_+ & g_{+-} & g_{++} \\ \psi_- & \lambda_- & g_{--} & g_{-+} \end{array} \right\} \stackrel{def}{=} \mathbf{s} \in \mathbf{S}_{SA}^{(N=2)}, \tag{3.18}$$

*а действие*

$$\mathbf{s}_1 * \mathbf{s}_2 = \mathbf{s}_3 \tag{3.19}$$

*определяется композицией преобразований* $Z \to \tilde{Z} \to \tilde{\tilde{Z}}$.

*Замечание* **3.10.** Умножение в полугруппе $\mathbf{S}_{SA}^{(N=2)}$ не связано с обычным матричным умножением*), а определяется композицией $N = 2$ супераналитических преобразований, записанных в компонентном виде (3.14), поэтому функциональная матрица, определяющая элемент $\mathbf{s}$, не обязана быть квадратной, как, например, в случае $N = 1$ (2.6).

Ассоциативность умножения (3.19)

$$\mathbf{s}_1 * (\mathbf{s}_2 * \mathbf{s}_3) = (\mathbf{s}_1 * \mathbf{s}_2) * \mathbf{s}_3 \tag{3.20}$$

следует из ассоциативности преобразований относительно композиции (для $N = 1$ см. **Предложение 2.7**).

Двусторонняя единица в полугруппе $\mathbf{S}_{SA}^{(N=2)}$ определяется функци-

---

*Примечание.* Для этого и использованы фигурные скобки вместо матричных круглых.

ональной матрицей следующего вида

$$\mathbf{e} = \left\{ \begin{array}{cccc} z & 0 & 0 & 0 \\ 0 & 0 & 1 & 0 \\ 0 & 0 & 0 & 1 \end{array} \right\}, \tag{3.21}$$

а двусторонний нуль определяется нулевой такой матрицей.

Рассмотрим гомоморфизм $\varphi$ $N=2$ супераналитической полугруппы в полугруппу $N=2$ супераналитических преобразований $\varphi : \mathbf{S}_{SA}^{(N=2)} \to \boldsymbol{T}_{SA}^{(N=2)}$. Тогда легко проверить, что, как и должно быть, $\ker\varphi = \mathbf{e}$.

Приведенная процедура представляет собой специальную "нелинейную реализацию" $N=2$ супераналитической полугруппы функциональными матрицами*), умножение в которых задается композицией $N=2$ супераналитических преобразований.

**3.1.3.** Редукции $N=2$ касательного суперпространства и перманенты. Сначала найдем соотношение между суперфункциями $Q\left(z,\theta^{+},\theta^{-}\right)$ и $\varDelta^{\pm}\left(z,\theta^{+},\theta^{-}\right)$, аналогичное $N=1$ случаю (2.58). Для этого продифференцируем $\varDelta^{\pm}\left(z,\theta^{+},\theta^{-}\right)$, применим (3.3) и получим

$$Q\left(z,\theta^{+},\theta^{-}\right) - \frac{D^{+}\varDelta^{-}\left(z,\theta^{+},\theta^{-}\right) + D^{-}\varDelta^{+}\left(z,\theta^{+},\theta^{-}\right)}{2} = \operatorname{per}\mathrm{H}, \tag{3.22}$$

где

$$\operatorname{per}\mathrm{H} = D^{-}\tilde{\theta}^{+}\cdot D^{+}\tilde{\theta}^{-} + D^{+}\tilde{\theta}^{+}\cdot D^{-}\tilde{\theta}^{-} \tag{3.23}$$

— перманент обычной матрицы H с четными (и возможно нильпотент-

---

*Примечание.* Это название не связано с нелинейными реализациями, обусловленными индуцированными представлениями, которые рассмотрены в **Подразделе 2.4**.

ными) элементами (см. **Раздел 5**).

*Замечание* **3.11.** Обе (!) матричные функции — *перманент и детерминант* — матрицы H играют существенную роль в $N = 2$ геометрии и редукциях касательного суперпространства.

Введем в рассмотрение следующие $2 \times 2$ матрицы с чисто нильпотентными элементами

$$\mathrm{Q} = \begin{pmatrix} \partial\tilde{\theta}^{+} & \partial\tilde{\theta}^{-} \\ \tilde{\theta}^{+} & \tilde{\theta}^{-} \end{pmatrix}, \tag{3.24}$$

$$\mathrm{D} = \begin{pmatrix} \varDelta^{+}\left(z, \theta^{+}, \theta^{-}\right) & \varDelta^{-}\left(z, \theta^{+}, \theta^{-}\right) \\ \partial\tilde{\theta}^{+} & \partial\tilde{\theta}^{-} \end{pmatrix}, \tag{3.25}$$

а также горизонтальные полуматрицы (см. **Пункт Д.2**)

$$\mathcal{D}^{\pm} = \begin{pmatrix} D^{\pm}\tilde{\theta}^{+} & D^{\pm}\tilde{\theta}^{-} \\ \tilde{\theta}^{+} & \tilde{\theta}^{-} \end{pmatrix}, \tag{3.26}$$

$$\mathcal{R}^{\pm} = \begin{pmatrix} \partial\tilde{\theta}^{+} & \partial\tilde{\theta}^{-} \\ D^{\pm}\tilde{\theta}^{+} & D^{\pm}\tilde{\theta}^{-} \end{pmatrix}. \tag{3.27}$$

Используя (3.22), для $N = 2$ березиниана имеем

$$\begin{aligned} \mathrm{Ber}\left(\tilde{Z}/Z\right) = \frac{\mathrm{per}\,\mathrm{H}}{\det \mathrm{H}} + \frac{D^{+}\varDelta^{-}\left(z, \theta^{+}, \theta^{-}\right) + D^{-}\varDelta^{+}\left(z, \theta^{+}, \theta^{-}\right)}{2\det \mathrm{H}} + \\ \frac{\varDelta^{-}\left(z, \theta^{+}, \theta^{-}\right)}{\det \mathrm{H}} \cdot \frac{\delta\mathrm{et}\mathcal{R}^{+}}{\det \mathrm{H}} - \frac{\varDelta^{+}\left(z, \theta^{+}, \theta^{-}\right)}{\det \mathrm{H}} \cdot \frac{\delta\mathrm{et}\mathcal{R}^{-}}{\det \mathrm{H}}. \end{aligned} \tag{3.28}$$

В то же время функции (3.7) и (3.8) можно выразить через ма-тричные функции и полуматричные функции симметричным образом

$$Q\left(z, \theta^{+}, \theta^{-}\right) = \partial \tilde{z} - \operatorname{per} \mathrm{Q}, \tag{3.29}$$

$$\Delta^{\pm}\left(z, \theta^{+}, \theta^{-}\right) = D^{\pm} \tilde{z} - \pi \mathrm{er} \mathcal{D}^{\pm}, \tag{3.30}$$

где полуматричные функции $\pi\mathrm{er}$ и $\delta\mathrm{et}$ определены в (Д.24) и (Д.23) (см. **Пункт Д.2**).

*Замечание* **3.12.** В формулах (3.29) явно прослеживается четно-нечетная симметрия, аналогичная (2.208).

Чтобы выяснить, какие редукции $N = 2$ касательного суперпространства возможны, докажем теорему сложения березинианов в случае $N = 2$ (см. для $N = 1$ (2.34) и (4.7)).

**Теорема 3.13.** (Теорема сложения $N = 2$ березинианов) *Для $N = 2$ супераналитических преобразований $Z\left(z, \theta^{+}, \theta^{-}\right) \rightarrow \tilde{Z}\left(\tilde{z}, \tilde{\theta}^{+}, \tilde{\theta}^{-}\right)$ полный $N = 2$ березиниан в обратимом* (3.11) *и полунеобратимом* (3.12) *случаях представляется в виде суммы трех березинианов*

$$\operatorname{Ber}\left(\tilde{Z}/Z\right) = \operatorname{Ber} \mathrm{P}_{S}^{(N=2)} + \operatorname{Ber} \mathrm{P}_{T+}^{(N=2)} + \operatorname{Ber} \mathrm{P}_{T-}^{(N=2)}. \tag{3.31}$$

*Доказательство.* С этой целью представим березиниан (3.10) (или (3.28) в виде трех слагаемых

$$\operatorname{Ber}\left(\tilde{Z}/Z\right) = \frac{Q\left(z, \theta^{+}, \theta^{-}\right)}{\det \mathrm{H}} + \tag{3.32}$$

$$\frac{\Delta^{-}\left(z, \theta^{+}, \theta^{-}\right)}{\det \mathrm{H}} \cdot \frac{\delta \mathrm{et} \mathcal{R}^{+}}{\det \mathrm{H}} - \frac{\Delta^{+}\left(z, \theta^{+}, \theta^{-}\right)}{\det \mathrm{H}} \cdot \frac{\delta \mathrm{et} \mathcal{R}^{-}}{\det \mathrm{H}}.$$

Легко видеть, что каждое из этих слагаемых представляет собой березиниан суперматрицы, которая получается из общей суперматрицы

$\mathrm{P}_{SA}^{(N=2)}$ (3.6) занулением некоторых ее элементов. Отсюда получаем вид суперматриц и их березинианов, входящих в правую часть (3.31)

$$\mathrm{P}_{S}^{(N=2)} = \begin{pmatrix} Q\left(z,\theta^{+},\theta^{-}\right) & \partial\tilde{\theta}^{+} \; \partial\tilde{\theta}^{-} \\ 0 & \\ 0 & \mathrm{H} \end{pmatrix}, \tag{3.33}$$

$$\mathrm{Ber}\,\mathrm{P}_{S}^{(N=2)} = \frac{Q\left(z,\theta^{+},\theta^{-}\right)}{\det \mathrm{H}}, \tag{3.34}$$

$$\mathrm{P}_{T+}^{(N=2)} = \begin{pmatrix} 0 & \partial\tilde{\theta}^{+} \; \partial\tilde{\theta}^{-} \\ \varDelta^{-}\left(z,\theta^{+},\theta^{-}\right) & \\ 0 & H \end{pmatrix}, \tag{3.35}$$

$$\mathrm{Ber}\,\mathrm{P}_{T+}^{(N=2)} = \frac{\varDelta^{-}\left(z,\theta^{+},\theta^{-}\right)}{\det \mathrm{H}} \cdot \frac{\delta\mathrm{et}\mathcal{R}^{+}}{\det \mathrm{H}}, \tag{3.36}$$

$$\mathrm{P}_{T-}^{(N=2)} = \begin{pmatrix} 0 & \partial\tilde{\theta}^{+} \; \partial\tilde{\theta}^{-} \\ 0 & \mathrm{H} \\ \varDelta^{+}\left(z,\theta^{+},\theta^{-}\right) & \end{pmatrix}, \tag{3.37}$$

$$\mathrm{Ber}\,\mathrm{P}_{T-}^{(N=2)} = -\frac{\varDelta^{+}\left(z,\theta^{+},\theta^{-}\right)}{\det \mathrm{H}} \cdot \frac{\delta\mathrm{et}\mathcal{R}^{-}}{\det \mathrm{H}}. \tag{3.38}$$

■

Из формул (3.33), (3.36) и (3.38) следует, что при $N = 2$ имеется не одна (как в обратимом случае [563,565,625]), не две, как в $N = 1$ случае (см. **Пункт 2.1.3**), а три возможные редукции, соответствующие трем различным типам преобразований.

**Определение 3.14.** *Обратимые, полунеобратимые и необратимые редуцированные $N = 2$ суперконформные преобразования определяются двумя условиями*

$$\varDelta^{+}\left(z,\theta^{+},\theta^{-}\right) = D^{+}\tilde{z} - D^{+}\tilde{\theta}^{-}\cdot\tilde{\theta}^{+} - D^{+}\tilde{\theta}^{+}\cdot\tilde{\theta}^{-} = 0, \tag{3.39}$$

$$\varDelta^-\left(z,\theta^+,\theta^-\right) = D^-\tilde{z} - D^-\tilde{\theta}^-\cdot\tilde{\theta}^+ - D^-\tilde{\theta}^+\cdot\tilde{\theta}^- = 0. \tag{3.40}$$

Определение полунеобратимых и необратимых преобразований для $N=1$ дано в (2.4) и (2.5), а для $N=2$ — в (3.12) и (3.13).

**Определение 3.15.** *Полунеобратимые и необратимые левые $N=2$ редуцированные вращающие четность*[*)] *касательного пространства преобразования определяются двумя условиями*

$$Q\left(z,\theta^+,\theta^-\right) = \partial\tilde{z} - \partial\tilde{\theta}^+\cdot\theta^- - \partial\tilde{\theta}^-\cdot\theta^+ = 0, \tag{3.41}$$
$$\varDelta^-\left(z,\theta^+,\theta^-\right) = D^-\tilde{z} - D^-\tilde{\theta}^-\cdot\tilde{\theta}^+ - D^-\tilde{\theta}^+\cdot\tilde{\theta}^- = 0. \tag{3.42}$$

**Определение 3.16.** *Полунеобратимые и необратимые правые $N=2$ редуцированные сплетающие четность касательного пространства преобразования определяются двумя условиями*

$$Q\left(z,\theta^+,\theta^-\right) = \partial\tilde{z} - \partial\tilde{\theta}^+\cdot\theta^- - \partial\tilde{\theta}^-\cdot\theta^+ = 0, \tag{3.43}$$
$$\varDelta^+\left(z,\theta^+,\theta^-\right) = D^+\tilde{z} - D^+\tilde{\theta}^-\cdot\tilde{\theta}^+ - D^+\tilde{\theta}^+\cdot\tilde{\theta}^- = 0. \tag{3.44}$$

Будем называть условия (3.39)–(3.40) SCf условиями, условия (3.41)–(3.42) — $\mathsf{TPt}^-$ условиями и (3.43)–(3.44) — $\mathrm{TPt}^+$ условиями. В терминах перманентов и полуперманентов они приобретают вид

$$D^\pm\tilde{z} = \pi\mathrm{er}\mathcal{D}^\pm, \quad (\mathsf{SCf}) \tag{3.45}$$

$$\begin{cases} \partial\tilde{z} = \mathrm{per}\,\mathrm{Q}, \\ D^-\tilde{z} = \pi\mathrm{er}\mathcal{D}^-, \end{cases} \quad (\mathsf{TPt}^-) \tag{3.46}$$

---

*Примечание.* Причина такого названия будет пояснена ниже (для $N=1$ вращающих четность преобразований см. **Подраздел 2.3**).

$$\begin{cases} \partial \tilde{z} = \mathrm{per}\,\mathrm{Q}, \\ D^{+}\tilde{z} = \pi \mathrm{er} \mathcal{D}^{+}. \end{cases} \quad (\mathsf{TPt}^{\,+}) \tag{3.47}$$

Исходя из этих условий, можно определить три соответствующие редуцированные суперматрицы по формулам

$$\mathrm{P}_{SCf}^{(N=2)} \stackrel{def}{=} \mathrm{P}_{S}^{(N=2)}|_{\Delta^{+}(z,\theta^{+},\theta^{-})=0,\, \Delta^{-}(z,\theta^{+},\theta^{-})=0} =$$

$$\begin{pmatrix} Q_{SCf}\left(z,\theta^{+},\theta^{-}\right) & \partial\tilde{\theta}_{SCf}^{+} \; \partial\tilde{\theta}_{SCf}^{-} \\ 0 & \\ 0 & \mathrm{H}_{SCf} \end{pmatrix}, \tag{3.48}$$

$$\mathrm{P}_{TPt+}^{(N=2)} \stackrel{def}{=} \mathrm{P}_{T+}^{(N=2)}|_{Q(z,\theta^{+},\theta^{-})=0,\, \Delta^{+}(z,\theta^{+},\theta^{-})=0} =$$

$$\begin{pmatrix} 0 & \partial\tilde{\theta}_{TPt+}^{+} \; \partial\tilde{\theta}_{TPt+}^{-} \\ \Delta_{TPt+}^{-}\left(z,\theta^{+},\theta^{-}\right) & \\ 0 & \mathrm{H}_{TPt+} \end{pmatrix}, \tag{3.49}$$

$$\mathrm{P}_{TPt-}^{(N=2)} \stackrel{def}{=} \mathrm{P}_{T-}^{(N=2)}|_{Q(z,\theta^{+},\theta^{-})=0,\, \Delta^{-}(z,\theta^{+},\theta^{-})=0} =$$

$$\begin{pmatrix} 0 & \partial\tilde{\theta}_{TPt-}^{+} \; \partial\tilde{\theta}_{TPt-}^{-} \\ 0 & \\ \Delta_{TPt-}^{+}\left(z,\theta^{+},\theta^{-}\right) & \mathrm{H}_{TPt-} \end{pmatrix}, \tag{3.50}$$

где

$$Q_{SCf}\left(z,\theta^{+},\theta^{-}\right) \stackrel{def}{=} Q\left(z,\theta^{+},\theta^{-}\right)|_{\Delta^{+}(z,\theta^{+},\theta^{-})=0,\, \Delta^{-}(z,\theta^{+},\theta^{-})=0}, \tag{3.51}$$

$$\Delta_{TPt\mp}^{\pm}\left(z,\theta^{+},\theta^{-}\right) \stackrel{def}{=} \Delta^{\pm}\left(z,\theta^{+},\theta^{-}\right)|_{Q(z,\theta^{+},\theta^{-})=0,\, \Delta^{\mp}(z,\theta^{+},\theta^{-})=0}, \tag{3.52}$$

$$\begin{aligned}
\partial\tilde{\theta}^{\pm}_{SCf} &= \partial\tilde{\theta}^{\pm}|_{\Delta^{+}(z,\theta^{+},\theta^{-})=0,\,\Delta^{-}(z,\theta^{+},\theta^{-})=0},\\
\partial\tilde{\theta}^{+}_{TPt\pm} &= \partial\tilde{\theta}^{+}|_{Q(z,\theta^{+},\theta^{-})=0,\,\Delta^{\pm}(z,\theta^{+},\theta^{-})=0},\\
\partial\tilde{\theta}^{-}_{TPt\pm} &= \partial\tilde{\theta}^{-}|_{Q(z,\theta^{+},\theta^{-})=0,\,\Delta^{\pm}(z,\theta^{+},\theta^{-})=0},
\end{aligned} \tag{3.53}$$

$$\begin{aligned}
\mathrm{H}_{SCf} &= \mathrm{H}|_{\Delta^{+}(z,\theta^{+},\theta^{-})=0,\,\Delta^{-}(z,\theta^{+},\theta^{-})=0},\\
\mathrm{H}_{TPt\pm} &= \mathrm{H}|_{Q(z,\theta^{+},\theta^{-})=0,\,\Delta^{\pm}(z,\theta^{+},\theta^{-})=0},
\end{aligned} \tag{3.54}$$

$$\mathcal{R}_{TPt\pm} = \mathcal{R}|_{Q(z,\theta^{+},\theta^{-})=0,\,\Delta^{\pm}(z,\theta^{+},\theta^{-})=0}. \tag{3.55}$$

Для нахождения функции $Q_{SCf}\left(z,\theta^{+},\theta^{-}\right)$ воспользуемся также (3.22), тогда получаем

$$Q_{SCf}\left(z,\theta^{+},\theta^{-}\right) = \operatorname{per}\mathrm{H}_{SCf}. \tag{3.56}$$

Отсюда следует окончательный вид SCf редуцированной суперматрицы

$$\mathrm{P}^{(N=2)}_{SCf} = \begin{pmatrix} \operatorname{per}\mathrm{H}_{SCf} & \partial\tilde{\theta}^{+}_{SCf} & \partial\tilde{\theta}^{-}_{SCf}\\ 0 & & \\ & \multicolumn{2}{c}{\mathrm{H}_{SCf}} \\ 0 & & \end{pmatrix} \tag{3.57}$$

и фундаментальная *тройная формула*, связывающая березиниан, перманент и детерминант

$$\operatorname{Ber}\mathrm{P}^{(N=2)}_{SCf} = \frac{\operatorname{per}\mathrm{H}_{SCf}}{\det\mathrm{H}_{SCf}}. \tag{3.58}$$

**Предложение 3.17.** *Композиция двух $N=2$ SCf преобразований есть $N=2$ SCf преобразование, а композиция $N=2$ SCf преобразования и $N=2$ $\mathsf{TPt}^{\pm}$ преобразования есть $N=2$ $\mathsf{TPt}^{\pm}$ преобразование.*

*Доказательство.* Следует из умножения суперматриц (3.48)–(3.50)

$$\mathrm{P}^{(N=2)}_{SCf_1}\cdot\mathrm{P}^{(N=2)}_{SCf_2} = \mathrm{P}^{(N=2)}_{SCf_3}, \tag{3.59}$$

$$\mathrm{P}^{(N=2)}_{TPt^{\pm}_1}\cdot\mathrm{P}^{(N=2)}_{SCf_2} = \mathrm{P}^{(N=2)}_{TPt^{\pm}_3}. \tag{3.60}$$

■

Покажем, что березинианы суперконформно-подобных и сплетающих четность преобразований выражаются через введенные суперматрицы (3.48), (3.49) и (3.50). Для этого нам понадобится

**Утверждение 3.18.** *Применение условий редукции* (3.39)–(3.44) *к суперматрицам* $P_S^{(N=2)}$ *и* $P_{T\pm}^{(N=2)}$ *в порядке, обратном, чем в* (3.48)–(3.50), *приводит к вырожденным суперматрицам с нулевым березинианом.*

*Доказательство.* Применяя $\mathsf{TPt}^{\pm}$ условия (3.41)–(3.44) к суперматрице $\mathrm{P}_S^{(N=2)}$ (3.33), получаем вырожденные суперматрицы следующего вида

$$\mathrm{P}_{D\pm}^{(N=2)} = \mathrm{P}_S^{(N=2)}|_{Q(z,\theta^+,\theta^-)=0,\, \Delta^\pm(z,\theta^+,\theta^-)=0} = \begin{pmatrix} 0 & \partial\tilde{\theta}^+_{TPt\pm} & \partial\tilde{\theta}^-_{TPt\pm} \\ 0 & & \\ 0 & \multicolumn{2}{c}{\mathrm{H}_{TPt\pm}} \end{pmatrix}, \tag{3.61}$$

березиниан которых, очевидно, равен нулю $\mathrm{Ber}\,\mathrm{P}_{D\pm}^{(N=2)} = 0$. С другой стороны, если применить $\mathsf{SCf}$ условия (3.39)–(3.40) к суперматрицам $\mathrm{P}_{T\pm}^{(N=2)}$ (3.36)–(3.38), то получим в обоих случаях одну и ту же вырожденную суперматрицу

$$\mathrm{P}_{D}^{(N=2)} = \mathrm{P}_{T\pm}^{(N=2)}|_{\Delta^+(z,\theta^+,\theta^-)=0,\, \Delta^-(z,\theta^+,\theta^-)=0} = \begin{pmatrix} 0 & \partial\tilde{\theta}^+_{SCf} & \partial\tilde{\theta}^-_{SCf} \\ 0 & & \\ 0 & \multicolumn{2}{c}{\mathrm{H}_{SCf}} \end{pmatrix}, \tag{3.62}$$

березиниан которой также равен нулю $\mathrm{Ber}\,\mathrm{P}_{D}^{(N=2)} = 0$. ■

Важно отметить, что все три вырожденные суперматрицы (3.61) и (3.62), несмотря на подобный внешний вид, не совпадают между собой $\mathrm{P}_{D+}^{(N=2)} \neq \mathrm{P}_{D-}^{(N=2)} \neq \mathrm{P}_{D}^{(N=2)}$, поскольку на их оставшиеся ненулевые элементы $\partial\tilde{\theta}^\pm$ и H наложены различные условия — $\mathsf{TPt}^+$ (3.41)–(3.42), $\mathsf{TPt}^-$ (3.43)–(3.44) и $\mathsf{SCf}$ (3.39)–(3.40).

Для того, чтобы найти березинианы редуцированных преобразований, необходимо спроектировать формулу сложения $N = 2$ березинианов (3.32) на различные вырианты редукции, пользуясь SCf и TPt$^{\pm}$ условиями (3.39)–(3.44), а также **Предложением E.12** и **Утверждением 3.18**. Тогда получим

$$\operatorname{Ber}\left(\tilde{Z}/Z\right) = \operatorname{Ber}\mathrm{P}_{SA}^{(N=2)} =$$

$$\begin{cases} \left(\operatorname{Ber}\mathrm{P}_{S}^{(N=2)} + \operatorname{Ber}\mathrm{P}_{T+}^{(N=2)} + \operatorname{Ber}\mathrm{P}_{T-}^{(N=2)}\right)|_{\Delta^{\pm}(z,\theta^{+},\theta^{-})=0} \\ \left(\operatorname{Ber}\mathrm{P}_{S}^{(N=2)} + \operatorname{Ber}\mathrm{P}_{T+}^{(N=2)} + \operatorname{Ber}\mathrm{P}_{T-}^{(N=2)}\right)|_{Q(z,\theta^{+},\theta^{-})=0,\,\Delta^{+}(z,\theta^{+},\theta^{-})=0} \\ \left(\operatorname{Ber}\mathrm{P}_{S}^{(N=2)} + \operatorname{Ber}\mathrm{P}_{T+}^{(N=2)} + \operatorname{Ber}\mathrm{P}_{T-}^{(N=2)}\right)|_{Q(z,\theta^{+},\theta^{-})=0,\,\Delta^{-}(z,\theta^{+},\theta^{-})=0} \end{cases} =$$

$$\begin{cases} \operatorname{Ber}\mathrm{P}_{SCf}^{(N=2)} + 0 + 0 \\ 0 + \operatorname{Ber}\mathrm{P}_{TPt+}^{(N=2)} + 0 \\ 0 + 0 + \operatorname{Ber}\mathrm{P}_{TPt-}^{(N=2)} \end{cases} = \begin{cases} \operatorname{Ber}\mathrm{P}_{SCf}^{(N=2)}, & (\mathsf{SCf}) \\ \operatorname{Ber}\mathrm{P}_{TPt+}^{(N=2)}, & (\mathsf{TPt}^{+}) \\ \operatorname{Ber}\mathrm{P}_{TPt-}^{(N=2)}, & (\mathsf{TPt}^{-}) \end{cases} =$$

$$\begin{cases} \dfrac{\operatorname{per}\mathrm{H}_{SCf}}{\det\mathrm{H}_{SCf}}, & (\mathsf{SCf}) \\ \dfrac{\Delta_{TPt+}^{-}(z,\theta^{+},\theta^{-})}{\det\mathrm{H}_{TPt+}} \cdot \dfrac{\delta\mathrm{et}\mathcal{R}_{TPt+}^{+}}{\det\mathrm{H}_{TPt+}}, & (\mathsf{TPt}^{+}) \\ -\dfrac{\Delta_{TPt-}^{+}(z,\theta^{+},\theta^{-})}{\det\mathrm{H}_{TPt-}} \cdot \dfrac{\delta\mathrm{et}\mathcal{R}_{TPt-}^{-}}{\det\mathrm{H}_{TPt-}}, & (\mathsf{TPt}^{-}) \end{cases} \tag{3.63}$$

где суперматрицы $\mathrm{P}_{SCf}^{(N=2)}$ и $\mathrm{P}_{TPt\pm}^{(N=2)}$ определены в (3.48)–(3.50), и мы воспользовались тройной формулой (3.58) для березиниана $N = 2$ суперконформно-подобных преобразований.

**3.1.4.** Классификация $N = 2$ SCf преобразований. Рассмотрим более подробно $N = 2$ преобразования, определяемые SCf условиями (3.39)–(3.40) или (3.45).

В обратимом случае они называются $N = 2$ суперконформными преобразованиями [563, 565, 602] и используются для описания скрытой

$N = 2$ суперконформной симметрии в суперструнной теории [337, 347], $N = 2$ суперримановых поверхностей [622, 624, 625] и $N = 2$ суперконформной теории поля [595, 596]. Необратимые случае $N = 2$ редуцированные преобразования рассматривались в [3, 12, 17].

Из (3.57) и тройной формулы (3.58) следует, что редуцированные $N = 2$ суперконформно-подобные преобразования полностью определяются элементами обычной матрицы $\mathrm{H}_{SCf}$ (3.9) с возможно нильпотентными элементами (см. [3, 12] и ниже).

Так, для преобразования суперпроизводных $D^{\pm}$ и дифференциала $dZ$ из (3.57) и (3.9) имеем

$$\begin{pmatrix} D^- \\ D^+ \end{pmatrix} = \mathrm{H}_{SCf} \cdot \begin{pmatrix} \tilde{D}^- \\ \tilde{D}^+ \end{pmatrix}, \tag{3.64}$$

$$d\tilde{Z} = dZ \cdot \operatorname{per} \mathrm{H}_{SCf}. \tag{3.65}$$

Последняя формула (3.65), в частности, может трактоваться так, что $\operatorname{per} \mathrm{H}_{SCf}$ играет роль якобиана в комплексном базисе [563].

*Замечание* **3.19.** Матрица $\mathrm{H}_{SCf}$ является полуминором (см. **Пункт Д.2**) четного элемента $Q_{SCf}(z, \theta^+, \theta^-)$ в суперматрице (3.48).

Из (3.64) видно, что нечетные суперпроизводные $D^{\pm}$ образуют $(0|2)$-мерное подпространство в $(1|2)$-мерном касательном пространстве. Другими словами, они преобразуются только друг через друга как

$$D^{\pm} = D^{\pm}\tilde{\theta}^- \cdot \tilde{D}^+ + D^{\pm}\tilde{\theta}^+ \cdot \tilde{D}^-. \tag{3.66}$$

Соответствующее кокасательное $(0|2)$-мерное пространство строится с помощью $N = 2$ супердифференциалов, преобразующихся ду-ально с помощью той же матрицы $\mathrm{H}_{SCf}$ следующим образом

$$\left( \begin{array}{cc} d\tilde{\tau}^+_{SCf} & d\tilde{\tau}^-_{SCf} \end{array} \right) = \left( \begin{array}{cc} d\tau^+_{SCf} & d\tau^-_{SCf} \end{array} \right) \cdot \mathrm{H}_{SCf}. \tag{3.67}$$

**Определение 3.20.** *Внешний* SCf *супердифференциал* $\delta^{(N=2)}_{SCf}$ *определяется формулой*

$$\delta^{(N=2)}_{SCf} = d\tau^+_{SCf} \cdot D^- + d\tau^-_{SCf} \cdot D^+. \tag{3.68}$$

**Утверждение 3.21.** *Внешний дифференциал* $(0|2)$*-мерного подпространства инвариантен относительно* $N = 2$ *суперконформно-подобных преобразований.*

*Доказательство.* Из (3.64), (3.67) и (3.68) имеем

$$\delta^{(N=2)}_{SCf} = \left( \begin{array}{cc} d\tau^+_{SCf} & d\tau^-_{SCf} \end{array} \right) \cdot \left( \begin{array}{c} D^- \\ D^+ \end{array} \right) =$$

$$\left( \begin{array}{cc} d\tau^+_{SCf} & d\tau^-_{SCf} \end{array} \right) \cdot \mathrm{H}_{SCf} \cdot \left( \begin{array}{c} \tilde{D}^- \\ \tilde{D}^+ \end{array} \right) =$$

$$\left( \begin{array}{cc} d\tilde{\tau}^+_{SCf} & d\tilde{\tau}^-_{SCf} \end{array} \right) \cdot \left( \begin{array}{c} \tilde{D}^- \\ \tilde{D}^+ \end{array} \right) = \tilde{\delta}^{(N=2)}_{SCf}. \tag{3.69}$$

■

Введенные $N = 2$ супердифференциалы $d\tau^\pm_{SCf}$ дуальны к суперпроизводным и в смысле соотношения (см. (3.3))

$$\left\{ d\tau^+_{SCf}, d\tau^-_{SCf} \right\} = 2dZ. \tag{3.70}$$

Применим оператор $D^+$ к SCf условию (3.39)

$$D^+\left(D^+\tilde{z} - D^+\tilde{\theta}^- \cdot \tilde{\theta}^+ - D^+\tilde{\theta}^+ \cdot \tilde{\theta}^-\right) = 0. \tag{3.71}$$

Используя нильпотентность суперпроизводных в комплексном базисе (3.2) $(D^+)^2 = 0$, получаем

$$\begin{aligned} D^+\tilde{\theta}^- \cdot D^+\tilde{\theta}^+ &= 0, \\ D^-\tilde{\theta}^+ \cdot D^-\tilde{\theta}^- &= 0. \end{aligned} \tag{3.72}$$

Отсюда следует, что матрица $\mathrm{H}^T_{SCf}$ является $2 \times 2$ scf-матрицей (см. **Подраздел 5.1**), т. е. элементы в столбцах $\mathrm{H}^T_{SCf}$ взаимно ортогональны. Для таких матриц справедливо общее соотношение

$$(\det \mathrm{H}_{SCf})^2 = (\mathrm{per}\, \mathrm{H}_{SCf})^2 . \tag{3.73}$$

Тогда в случае обратимых преобразований, которые удовлетворяют условию $\epsilon\left[\det \mathrm{H}_{SCf}\right] \neq 0$ (3.11)–(3.12) (с очевидностью, также и $\epsilon\left[\mathrm{per}\, \mathrm{H}_{SCf}\right] \neq 0$), для березиниана (3.63) имеем

$$\mathrm{Ber}_{\,SCf}\left(\tilde{Z}/Z\right) = \frac{\det \mathrm{H}_{SCf}}{\mathrm{per}\, \mathrm{H}_{SCf}} = \frac{\mathrm{per}\, \mathrm{H}_{SCf}}{\det \mathrm{H}_{SCf}}, \tag{3.74}$$

поэтому

$$\left(\mathrm{Ber}_{\,SCf}\left(\tilde{Z}/Z\right)\right)^2 = 1. \tag{3.75}$$

Следовательно, для обратимых $N = 2$ суперконформных преобразований березиниан равен

$$\mathrm{Ber}_{\,SCf}\left(\tilde{Z}/Z\right) = k = \pm 1, \tag{3.76}$$

где $k = +1$ отвечает подгруппе $SO_{\Lambda_0}(2)$ преобразований (в координатном базисе (E.24)), описывающих $N = 2$ суперримановы поверхности без твиста, а $k = -1$ соответствует общим $O_{\Lambda_0}(2)$ преобразованиям, описывающих $N = 2$ суперримановы поверхности [563, 565, 622] или $N = 2$ суперконформные алгебры [502, 627, 628] с твистом. Этого и естестеноо, поскольку группой внешних автоморфизмов здесь является $Z_2 = \epsilon\left[O_{\Lambda_0}(2)/SO_{\Lambda_0}(2)\right]$ [563, 565].

Если $\epsilon\left[\det \mathrm{H}_{SCf}\right] \neq 0$, то легко видеть, что элемент матрицы может быть либо ненулевым с ненулевой числовой частью, либо нулем, поэтому верхнему и нижнему знакам отвечает диагональная и антидиагональная матрица $\mathrm{H}_{SCf}$ соответственно

$$\mathrm{H}_{SCf}^{(k=+1)} = \begin{pmatrix} D^-\tilde{\theta}^+ & 0 \\ 0 & D^+\tilde{\theta}^- \end{pmatrix}, \; D^-\tilde{\theta}^- = D^+\tilde{\theta}^+ = 0, \tag{3.77}$$

$$\mathrm{H}_{SCf}^{(k=-1)} = \begin{pmatrix} 0 & D^-\tilde{\theta}^- \\ D^+\tilde{\theta}^+ & 0 \end{pmatrix}, \; D^-\tilde{\theta}^+ = D^+\tilde{\theta}^- = 0. \tag{3.78}$$

Это следует и из общего соотношения между $2 \times 2$ матрицами в координатном $\mathrm{H}_0$ (E.21) и комплексном H (3.9) базисах (см. [3] и **Подраздел 5.1**)

$$\mathrm{H}_0^T \cdot \mathrm{H}_0 = \operatorname{per}\mathrm{H} \cdot \mathrm{I} + \operatorname{scf}_-\mathrm{H}^T \cdot \sigma^+ + \operatorname{scf}_+\mathrm{H}^T \cdot \sigma^-, \tag{3.79}$$

где I — единичная $2 \times 2$ матрица, $\sigma^\pm = \sigma_3 \pm i\sigma_1$, $\sigma_i$ — матрицы Паули и

$$\operatorname{scf}_\pm\mathrm{H}^T = D^\pm\tilde{\theta}^+ \cdot D^\pm\tilde{\theta}^-. \tag{3.80}$$

Из (3.79) видно, что условие для матрицы $\mathrm{H}_0$ в координатном базисе быть пропорциональной ортогональной матрице совпадает с условием для матрицы H в комплексном базисе быть scf-матрицей (см. **Под-** **раздел 5.1**) $\operatorname{scf}_{\pm}\mathrm{H}_{SCf}^{T}=0$, что совпадает с условиями (3.72) и соответственно с SCf условиями (3.39)–(3.40).

Таким образом, при $\epsilon\left[\det\mathrm{H}_{SCf}\right]\neq 0$ scf-матрица $\mathrm{H}_{SCf}$ является диагональной (3.77) или антидиагональной (3.78), и тогда из (3.66) имеем

$$D^{\pm}=\begin{cases} D^{\pm}\tilde{\theta}^{-}\cdot\tilde{D}^{\pm}, \ k=+1 \\ D^{\pm}\tilde{\theta}^{+}\cdot\tilde{D}^{\mp}, \ k=-1 \end{cases}. \tag{3.81}$$

Первое условие в (3.81) приводит к возможности глобального определения $D^{\pm}$, и такие преобразования могут применяться как функции перехода на $N=2$ суперримановых поверхностях без твиста, допускающих $U_{\Lambda_0}(1)$ $\left(=SO_{\Lambda_0}(2)\right)$ группу голономии и линейное расслоение над обычными римановыми поверхностями в то время, как второе условие не позволяет глобально определить $D^{\pm}$,что приводит к поверхностям с твистом [563, 565, 622].

В необратимом случае $\epsilon\left[\det\mathrm{H}_{SCf}\right]=0$, хотя может оказаться, что $\det\mathrm{H}_{SCf}\neq 0$ из-за наличия нильпотентных элементов в подстилающей алгебре и соответствующих нильпотентных функций, входящих в $\mathrm{H}_{SCf}$ [3, 12]. Тогда условие (3.72) может выполняться не за счет зануления сомножителей, а за счет делителей нуля в компонентных функциях. И для суперпроизводных $D^{\pm}$ будет выполняться соотношение (3.66) с двумя ненулевыми членами в правой части, несмотря на выполнение (3.72).

Таким образом, в полунеобратимом и необратимом случаях мы будем будем избегать деления в (3.74) и пользоваться необратимым аналогом (см. для $N=1$ **Подраздел 2.1**) якобиана (который назван в [3] доопределенным березинианом) в виде

$$\operatorname{per}\mathrm{H}_{SCf}\cdot\boldsymbol{J}_{SCf}^{noninv}=\det\mathrm{H}_{SCf}, \tag{3.82}$$

Здесь $\operatorname{per} \mathrm{H}_{SCf} \neq 0$ и $\det \mathrm{H}_{SCf} \neq 0$, хотя и $\epsilon [\operatorname{per} \mathrm{H}_{SCf}] = \epsilon [\det \mathrm{H}_{SCf}] = 0$. Тогда решение (3.76) не является единственным за счет нильпотентности $\operatorname{per} \mathrm{H}_{SCf}$ и $\det \mathrm{H}_{SCf}$. Примеры таких необратимых преобразований, удовлетворяющих SCf условиям (3.39)–(3.40), приведены в [3, 12] (см. ниже).

**Утверждение 3.22.** *Для преобразований, удовлетворяющих* SCf *условиям* (3.39)–(3.40)

$$\epsilon [\operatorname{per} \mathrm{H}_{SCf}] = \epsilon [\det \mathrm{H}_{SCf}] . \tag{3.83}$$

*Доказательство.* Непосредственно следует из (3.72) и (3.73). ∎

Таким образом, классификация обратимых и необратимых преобразований, удовлетворяющих SCf условиям (3.39)–(3.40), может быть проведена в терминах перманента матрицы $\mathrm{H}_{SCf}$ и имеет вид:

**1.** Обратимые $N = 2$ суперконформные преобразования, удовлетворяющие условию $\epsilon [\operatorname{per} \mathrm{H}_{SCf}] \neq 0$.

   **а)** $U_{\Lambda_0}(1)$ преобразования $\operatorname{per} \mathrm{H}_{SCf} = \det \mathrm{H}_{SCf}$ (матрица $\mathrm{H}_{SCf}$ диагональна);

   **б)** $O_{\Lambda_0}(2)$ преобразования $\operatorname{per} \mathrm{H}_{SCf} = -\det \mathrm{H}_{SCf}$ (матрица $\mathrm{H}_{SCf}$ антидиагональна).

**2.** Необратимые $N = 2$ суперконформные преобразования, удовлетворяющие условию $\epsilon [\operatorname{per} \mathrm{H}_{SCf}] = 0$.

   **а)** $\operatorname{per} \mathrm{H}_{SCf} \neq 0$ (матрица $\mathrm{H}_{SCf}$ состоит из нильпотентных элементов);

   **б)** $\operatorname{per} \mathrm{H}_{SCf} = 0$ (матрица $\mathrm{H}_{SCf}$ мономиальна, биномиальна или состоит из взаимноортогональных элементов).

Особый случай представляют собой расщепленные $N = 2$ SCf преобразования, которые рассмотрены в **Приложении Ж.2**.

**3.1.5.** Конечные обратимые и необратимые SCf преобразования и $N=2$ SCf полугруппа. Чтобы получить и проклассифицировать нерасщепленные $N = 2$ SCf преобразования, применим SCf условия (3.39)–(3.40) к полным $N = 2$ супераналитическим преобразованиям вида (3.14). Для этого удобно воспользоваться соотношениями, следующими из (3.14) и (3.1),

$$D^{\pm}\tilde{z} = \chi_{\pm}(z) + \theta^{\pm} \cdot \left(f'(z) \mp h(z)\right) + \theta^{\pm}\theta^{\mp} \cdot \chi'_{\pm}(z), \tag{3.84}$$

$$D^{\pm}\tilde{\theta}^{\mp} = g_{\mp\pm}(z) + \theta^{\pm} \cdot \left(\psi'_{\mp}(z) + \lambda_{\mp}(z)\right) + \theta^{\pm}\theta^{\mp} \cdot g'_{\mp\pm}(z), \tag{3.85}$$

$$D^{\pm}\tilde{\theta}^{\pm} = g_{\pm\pm}(z) + \theta^{\pm} \cdot \left(\psi'_{\pm}(z) - \lambda_{\pm}(z)\right) + \theta^{\pm}\theta^{\mp} \cdot g'_{\pm\pm}(z). \tag{3.86}$$

Тогда непосредственно из SCf условий (3.39)–(3.40) получаем систему уравнений на компонентные функции

$$\begin{aligned} f'(z) \mp h(z) &= g_{+-}(z)\, g_{-+}(z) + g_{++}(z)\, g_{--}(z) + \psi'_{\pm}(z)\, \psi_{\mp}(z) + \\ &\quad \psi'_{\mp}(z)\, \psi_{\pm}(z) + \lambda_{\mp}(z)\, \psi_{\pm}(z) - \lambda_{\pm}(z)\, \psi_{\mp}(z), \end{aligned}$$

$$\begin{aligned} \chi'_{\pm} &= g'_{\mp\pm}(z)\, \psi_{\pm}(z) - g_{\mp\pm}(z)\, \psi'_{\pm}(z) + g'_{\pm\pm}(z)\, \psi_{\mp}(z) - \\ &\quad g_{\pm\pm}(z)\, \psi'_{\mp}(z) + 2g_{\mp\pm}(z)\, \lambda_{\pm}(z) - 2g_{\pm\pm}(z)\, \lambda_{\mp}(z), \end{aligned}$$

$$\chi_{\pm} = g_{\mp\pm}(z)\, \psi_{\pm}(z) + g_{\pm\pm}(z)\, \psi_{\mp}(z),$$

$$g_{\mp\pm}(z)\, g_{\pm\pm}(z) = 0,$$

которую можно привести к следующему виду

$$f'(z) = g_{+-}(z)\, g_{-+}(z) + g_{++}(z)\, g_{--}(z) + \psi'_{+}(z)\, \psi_{-}(z) + \psi'_{-}(z)\, \psi_{+}(z), \tag{3.87}$$

$$h(z) = \lambda_{+}(z)\, \psi_{-}(z) - \lambda_{-}(z)\, \psi_{+}(z), \tag{3.88}$$

$$\chi_\pm = g_{\mp\pm}(z)\,\psi_\pm(z) + g_{\pm\pm}(z)\,\psi_\mp(z)\,, \tag{3.89}$$

$$g_{\mp\pm}(z)\left[\lambda_\pm(z) - \psi'_\pm(z)\right] = g_{\pm\pm}(z)\left[\lambda_\mp(z) + \psi'_\mp(z)\right], \tag{3.90}$$

$$g_{\mp\pm}(z)\,g_{\pm\pm}(z) = 0. \tag{3.91}$$

В терминах матричных функций и их нечетных аналогов (см. 2.5) первые 4 уравнения можно представить в компактном четно-нечетно симметричном виде

$$f'(z) = \operatorname{per}\mathrm{G}_{SCf} + \operatorname{per}\begin{pmatrix} \psi'_+(z) & \psi'_-(z) \\ \psi_+(z) & \psi_-(z) \end{pmatrix}, \tag{3.92}$$

$$h(z) = \det\begin{pmatrix} \lambda_+(z) & \lambda_+(z) \\ \psi_+(z) & \psi_-(z) \end{pmatrix}, \tag{3.93}$$

$$\chi_\pm = \pi\mathrm{er}\begin{pmatrix} g_{\mp\pm}(z) & g_{\pm\pm}(z) \\ \psi_\mp(z) & \psi_\pm(z) \end{pmatrix}, \tag{3.94}$$

$$\pi\mathrm{er}\begin{pmatrix} g_{\mp\pm}(z) & g_{\pm\pm}(z) \\ \psi'_\mp(z) & \psi'_\pm(z) \end{pmatrix} = \delta\mathrm{et}\begin{pmatrix} g_{\mp\pm}(z) & g_{\pm\pm}(z) \\ \lambda_\mp(z) & \lambda_\pm(z) \end{pmatrix}. \tag{3.95}$$

Отсюда следует, что число независимых функций, которыми определяется $N = 2$ SCf преобразование равно 12 (супераналитических компонентных функций (3.14)) - 8 (уравнений) = 4. Остальные функции могут быть найдены из 8 уравнений (3.87)–(3.91).

В частности, в обратимом случае, если $\epsilon\left[\det\mathrm{G}_{SCf}\right] \neq 0$, то функции $h(z)$ и $\lambda_\pm(z)$ можно получить из уравнений (3.93) и (3.94) в явном виде

$$h(z) \;=\; \frac{\operatorname{per}\mathrm{G}_{SCf}}{\det\mathrm{G}_{SCf}}\left(\psi_+(z)\,\psi_-(z)\right)' + \tag{3.96}$$

$$\frac{2g_{-+}(z)\,g_{--}(z)}{\det \mathrm{G}_{SCf}}\psi'_{+}(z)\,\psi_{+}(z)+\frac{2g_{+-}(z)\,g_{++}(z)}{\det \mathrm{G}_{SCf}}\psi'_{-}(z)\,\psi_{-}(z)\,,$$

$$\lambda_{\pm}(z)=\frac{\operatorname{per}\mathrm{G}_{SCf}}{\det \mathrm{G}_{SCf}}\psi'_{\pm}(z)+\frac{2g_{\pm\mp}(z)\,g_{\pm\pm}(z)}{\det \mathrm{G}_{SCf}}\psi'_{\mp}(z)\,. \tag{3.97}$$

Кроме того, при $\epsilon\left[\det \mathrm{G}_{SCf}\right]\neq 0$ имеется лишь два решения уравнений (3.91): матрица $\mathrm{G}_{SCf}$ — диагональна (Ж.16) или антидиагональна (Ж.17), что снова соответствует $U_{\Lambda_0}(1)$ и $O_{\Lambda_0}(2)$ случаям. При этом

$$\frac{\operatorname{per}\mathrm{G}_{SCf}}{\det \mathrm{G}_{SCf}}=k=\begin{cases}+1, & U_{\Lambda_0}(1)\\ -1, & O_{\Lambda_0}(2)\end{cases}. \tag{3.98}$$

Тогда получаем для $U_{\Lambda_0}(1)$ и $O_{\Lambda_0}(2)$ преобразования в выбранной параметризации

$$\begin{cases}\tilde{z}=f(z)+\theta^{+}g_{+-}(z)\,\psi_{-}(z)+\theta^{-}g_{-+}(z)\,\psi_{+}(z)+\theta^{+}\theta^{-}\left(\psi_{+}(z)\,\psi_{-}(z)\right)',\\ \tilde{\theta}^{\pm}=\psi_{\pm}(z)+\theta^{\pm}g_{\pm\mp}(z)+\theta^{\pm}\theta^{\mp}\psi'_{\pm}(z)\,,\end{cases} \tag{3.99}$$

где $f'(z)=g_{+-}(z)\,g_{-+}(z)+\psi'_{+}(z)\,\psi_{-}(z)+\psi'_{-}(z)\,\psi_{+}(z)$, и

$$\begin{cases}\tilde{z}=f(z)+\theta^{+}g_{--}(z)\,\psi_{+}(z)+\theta^{-}g_{++}(z)\,\psi_{-}(z)-\theta^{+}\theta^{-}\left(\psi_{+}(z)\,\psi_{-}(z)\right)',\\ \tilde{\theta}^{\pm}=\psi_{\pm}(z)+\theta^{\mp}g_{\pm\pm}(z)-\theta^{\pm}\theta^{\mp}\psi'_{\pm}(z)\,,\end{cases} \tag{3.100}$$

где $f'(z)=g_{++}(z)\,g_{--}(z)+\psi'_{+}(z)\,\psi_{-}(z)+\psi'_{-}(z)\,\psi_{+}(z)$.

Как и в случае расщепленной $N=2$ полугруппы (Ж.25), представление полной $N=2$ SCf полугруппы функциональными матрицами (см. **Определение 3.9**) будет некоторым сужением представле-

ния (3.18), а именно

$$\left\{\begin{array}{cccc} f & h & \chi_- & \chi_+ \\ \psi_+ & \lambda_+ & g_{+-} & g_{++} \\ \psi_- & \lambda_- & g_{--} & g_{-+} \end{array}\right\}|_{SCf\ nonsplit} \longrightarrow \left\{\begin{array}{ccc} \psi_+ & g_{+-} & g_{++} \\ \psi_- & g_{--} & g_{-+} \end{array}\right\}. \qquad (3.101)$$

Поэтому можно определить $N = 2$ SCf полугруппу следующим образом.

**Определение 3.23.** *Элемент* $\mathbf{s}$ *полной* $N = 2$ *суперконформной полугруппы* $\mathbf{S}_{SCf}^{(N=2)}$ *параметризуется функциональной матрицей*

$$\left\{\begin{array}{ccc} \psi_+ & g_{+-} & g_{++} \\ \psi_- & g_{--} & g_{-+} \end{array}\right\}|_{g_{\mp\pm}(z)g_{\pm\pm}(z)=0} \stackrel{def}{=} \mathbf{s} \in \mathbf{S}_{SCf}^{(N=2)}, \qquad (3.102)$$

*а действие* $\mathbf{s}^{(1)} * \mathbf{s}^{(2)} = \mathbf{s}^{(3)}$ *определяется композицией полных преобразований* $Z \to \tilde{Z} \to \tilde{\tilde{Z}}$.

Остальные функции $f(z), h(z), \lambda_\pm(z), \chi_\pm(z)$ определяются из уравнений (3.87)–(3.90), а в обратимом случае — из (3.96)–(3.97).

В необратимом случае при $\epsilon[\det \mathrm{G}_{SCf}] = 0$ (и, следовательно, $\epsilon[\mathrm{per}\, \mathrm{G}_{SCf}] = 0$) число возможных решений системы (3.87)–(3.91) резко увеличивается за счет делителей нуля и нильпотентов среди компонентных функций. Фактически необходимо решить систему двух уравнений над расширенным кольцом, содержащим нильпотенты.

Во-первых, как и в расщепленном случае (см. **Приложение Ж.2**), ортогональность элементов столбца матрицы $\mathrm{G}_{SCf}$ теперь уже означает не зануление одного из сомножителей, а их возможную пропорциональность одной и той же нильпотентной нечетной функции (подобно (Ж.24)). Более того, уравнения (3.90), используя тот же подход, могут решаться многими способами, например, путем взаимной ортогональ-ности каждого из сомножителей в правой и левой части. Тогда удобно параметризовать все преобразование только нечетными функциями, а элементы матрицы $\mathrm{G}_{SCf}$ находить из уравнений (3.90)–(3.91).

Отметим, что возможен и половинный случай, когда в одном столбце матрицы $\mathrm{G}_{SCf}$ элементы ортогональны за счет нильпотентности, а в другом — за счет зануления одного из них. Следует учесть и "самый необратимый" вариант, когда все элементы матрицы $\mathrm{G}_{SCf}$ равны нулю.

Суммируя, можно проклассифицировать возможные преобразования, удовлетворяющие системе (3.87)–(3.91), следующим образом:

**1.** Обратимые преобразования с $\epsilon\,[\mathrm{per}\,\mathrm{G}] \neq 0$.

   **а)** $U_{\Lambda_0}(1)$ преобразования $\mathrm{per}\,\mathrm{G}_{SCf} = \det \mathrm{G}_{SCf}$ (матрица $\mathrm{G}_{SCf}$ диагональна);

   **б)** $O_{\Lambda_0}(2)$ преобразования $\mathrm{per}\,\mathrm{G}_{SCf} = -\det \mathrm{G}_{SCf}$ (матрица $\mathrm{G}_{SCf}$ антидиагональна).

**2.** Необратимые преобразования с $\epsilon\,[\mathrm{per}\,\mathrm{G}_{SCf}] = 0$.

   **а)** "Половинный" вариант, когда одно уравнение из (3.91) выполняется, как в обратимом случае, за счет зануления одного или двух сомножителей, а другое — за счет нильпотентной ортогональности;

   **б)** "Полный" необратимый вариант, когда все элементы матрицы $\mathrm{G}_{SCf}$ не равны нулю, но взаимонильпотентны.

**3.** "Самый необратимый" вариант $\mathrm{G}_{SCf} = 0$.

Обратимые случаи рассматривались выше (3.99)–(3.100), поэтому мы рассмотрим необратимые.

Последний "самый необратимый" вариант **3** получается из большинства необратимых вариантов **2** путем зануления некоторых компо-нентных функций. Соответствующие $N=2$ необратимые SCf преобразования вообще не содержат четных функций и имеют вид

$$\begin{cases} \tilde{z} = f(z) + \theta^+\theta^-\left(\psi_+(z)\,\lambda_-(z) - \psi_-(z)\,\lambda_+(z)\right), \\ \tilde{\theta}^{\pm} = \psi_{\pm}(z) + \theta^{\pm}\theta^{\mp}\lambda_{\pm}(z), \end{cases} \tag{3.103}$$

где $f'(z) = \psi'_+(z)\,\psi_-(z) + \psi'_-(z)\,\psi_+(z)$.

Матрица $\mathrm{G}_{SCf}$ в "половинных вариантах" **2a** содержит один нуль и один элемент, который может быть ненильпотентным, а уравнения (3.90)–(3.91) имеют следующие возможные решения

$$1)\ \mathrm{G}_{SCf} = \begin{pmatrix} g_{+-}(z) & \eta(z)\,\psi'_-(z) \\ 0 & \eta(z)\left(\lambda_+(z) - \psi'_+(z)\right) \end{pmatrix},\ \lambda_-(z) = \psi'_-(z), \tag{3.104}$$

$$2)\ \mathrm{G}_{SCf} = \begin{pmatrix} \eta(z)\left(\lambda_-(z) - \psi'_-(z)\right) & 0 \\ \eta(z)\,\psi'_+(z) & g_{-+}(z) \end{pmatrix},\ \lambda_+(z) = \psi'_+(z), \tag{3.105}$$

$$3)\ \mathrm{G}_{SCf} = \begin{pmatrix} 0 & \eta(z)\left(\lambda_-(z) + \psi'_-(z)\right) \\ g_{--}(z) & \eta(z)\,\psi'_+(z) \end{pmatrix},\ \lambda_+(z) = -\psi'_+(z), \tag{3.106}$$

$$4)\ \mathrm{G}_{SCf} = \begin{pmatrix} \eta(z)\,\psi'_-(z) & g_{++}(z) \\ \eta(z)\left(\lambda_+(z) + \psi'_+(z)\right) & 0 \end{pmatrix},\ \lambda_-(z) = -\psi'_-(z). \tag{3.107}$$

Остальные функции $f(z), h(z), \chi_{\pm}(z)$ находятся из уравнений (3.87)–(3.89). Параметризация таких преобразований проводится с помощью одной четной функции $g_{ab}(z)$ (ненильпотентный элемент матрицы $\mathrm{G}_{SCf}$) и четырех нечетных функций $\psi_+(z)$, $\psi_-(z)$, $\eta(z)$, $\lambda_+(z)$ (или $\lambda_-(z)$). Например, для варианта 1) (3.104) получаем необратимое $N=2$ пре-

образование

$$\begin{cases} \tilde{z} = f(z) + \theta^- \left[\eta(z)\left(\lambda_+(z) - \psi'_+(z)\right)\psi_+(z) + \eta(z)\psi'_-(z)\psi_-(z)\right] + \\ \qquad \theta^+ g_{+-}(z)\psi_-(z) + \theta^+\theta^-\left[\lambda_+(z)\psi_-(z) - \psi'_-(z)\psi_+(z)\right], \\ \tilde{\theta}^+ = \psi_+(z) + \theta^+ g_{+-}(z) + \theta^-\eta(z)\psi'_-(z) + \theta^+\theta^-\lambda_+(z), \\ \tilde{\theta}^- = \psi_-(z) + \theta^-\eta(z)\left(\lambda_+(z) - \psi'_+(z)\right) - \theta^+\theta^-\psi'_-(z), \end{cases} \tag{3.108}$$

где

$$f'(z) = g_{+-}(z)\eta(z)\left(\lambda_+(z) - \psi'_+(z)\right) + \psi'_+(z)\psi_-(z) + \psi'_-(z)\psi_+(z).$$

В случае **2б** “полных” необратимых преобразований все элементы матрицы $\mathrm{G}_{SCf}$ отличны от нуля, но нильпотентны. Тогда решение уравнений (3.90)–(3.91) дает

$$1)\ \mathrm{G}_{SCf} = \begin{pmatrix} \eta(z)\left(\lambda_-(z) - \psi'_-(z)\right) & \rho(z)\left(\lambda_-(z) + \psi'_-(z)\right) \\ \eta(z)\left(\lambda_+(z) + \psi'_+(z)\right) & \rho(z)\left(\lambda_+(z) - \psi'_+(z)\right) \end{pmatrix}, \tag{3.109}$$

$$2)\ \mathrm{G}_{SCf} = \begin{pmatrix} \eta(z)\left(\lambda_+(z) + \psi'_+(z)\right) & \rho(z)\left(\lambda_+(z) - \psi'_+(z)\right) \\ \eta(z)\left(\lambda_-(z) - \psi'_-(z)\right) & \rho(z)\left(\lambda_-(z) + \psi'_-(z)\right) \end{pmatrix}. \tag{3.110}$$

Такие преобразования параметризуются шестью нечетными функциями $\psi_\pm(z)$, $\lambda_\pm(z)$, $\eta(z)$, $\rho(z)$ и, например, в первом варианте (3.109) имеют вид

$$\begin{aligned} &\tilde{z} = f(z) + \theta^+\theta^-\left[\lambda_+(z)\psi_-(z) - \lambda_-(z)\psi_+(z)\right] + \\ &\theta^+\left[\eta(z)\lambda_+(z) + \eta(z)\lambda_-(z) + \left(\psi_+(z)\psi_-(z)\right)'\right] + \\ &\theta^-\left[\rho(z)\lambda_-(z) + \rho(z)\lambda_+(z) - \left(\psi_+(z)\psi_-(z)\right)'\right], \end{aligned}$$

$$\tilde{\theta}^{+} = \psi_{+}(z) + \theta^{+}\eta(z)\left(\lambda_{+}(z) + \psi'_{+}(z)\right) +$$
$$\theta^{-}\rho(z)\left(\lambda_{+}(z) - \psi'_{+}(z)\right) + \theta^{+}\theta^{-}\lambda_{+}(z),$$

$$\tilde{\theta}^{-} = \psi_{-}(z) + \theta^{-}\rho(z)\left(\lambda_{-}(z) + \psi'_{-}(z)\right) +$$
$$\theta^{+}\eta(z)\left(\lambda_{-}(z) - \psi'_{-}(z)\right) - \theta^{+}\theta^{-}\lambda_{-}(z), \tag{3.111}$$

где

$$f'(z) = 2\eta(z)\rho(z)\left[\lambda_{-}(z)\psi'_{+}(z) - \lambda_{+}(z)\psi'_{-}(z)\right] +$$
$$\psi'_{+}(z)\psi_{-}(z) + \psi'_{-}(z)\psi_{+}(z).$$

Иной вариант решения уравнений (3.90)–(3.91) возникает, когда обе функции $\lambda_{\pm}(z)$ приравниваются к $\psi'_{\pm}(z)$ с различными знаками, что приводит еще к четырем решениям

$$1)\ \mathrm{G}_{SCf} = \begin{pmatrix} \eta(z)\sigma(z) & \rho(z)\psi'_{-}(z) \\ \eta(z)\psi'_{+}(z) & \rho(z)\mu(z) \end{pmatrix},\ \lambda_{\pm}(z) = \psi'_{\pm}(z), \tag{3.112}$$

$$2)\ \mathrm{G}_{SCf} = \begin{pmatrix} \eta(z)\psi'_{-}(z) & \rho(z)\mu(z) \\ \eta(z)\sigma(z) & \rho(z)\psi'_{+}(z) \end{pmatrix},\ \lambda_{\pm}(z) = -\psi'_{\pm}(z), \tag{3.113}$$

$$3)\ \mathrm{G}_{SCf} = \begin{pmatrix} \eta(z)\psi'_{-}(z) & \rho(z)\mu(z) \\ \eta(z)\psi'_{+}(z) & \rho(z)\sigma(z) \end{pmatrix},\ \lambda_{+}(z) = \psi'_{+}(z),\ \lambda_{-}(z) = -\psi'_{-}(z), \tag{3.114}$$

$$4)\ \mathrm{G}_{SCf} = \begin{pmatrix} \eta(z)\mu(z) & \rho(z)\psi'_{-}(z) \\ \eta(z)\sigma(z) & \rho(z)\psi'_{+}(z) \end{pmatrix},\ \lambda_{+}(z) = -\psi'_{+}(z),\ \lambda_{-}(z) = \psi'_{-}(z). \tag{3.115}$$

Необратимые $N = 2$ преобразования, соответствующие, напри-мер, варианту (3.112), имеют вид

$$\tilde{z} = f(z) + \theta^+ \left[\eta(z)\,\sigma(z)\,\psi_-(z) + \eta(z)\,\psi'_+(z)\,\psi_+(z)\right] +$$
$$\theta^- \left[\rho(z)\,\mu(z)\,\psi_+(z) + \rho(z)\,\psi'_-(z)\,\psi_-(z)\right] + \theta^+\theta^- \left(\psi_+(z)\,\psi_-(z)\right)',$$

$$\tilde{\theta}^+ = \psi_+(z) + \theta^+\eta(z)\,\sigma(z) + \theta^-\rho(z)\,\psi'_-(z) + \theta^+\theta^-\psi'_+(z)\,,$$

$$\tilde{\theta}^- = \psi_-(z) + \theta^-\rho(z)\,\mu(z) + \theta^+\eta(z)\,\psi'_+(z) - \theta^+\theta^-\psi'_-(z)\,, \qquad (3.116)$$

где

$$f'(z) = \rho(z)\,\eta(z)\left[\sigma(z)\,\mu(z) + \psi'_+(z)\,\psi'_-(z)\right] + \psi'_+(z)\,\psi_-(z) + \psi'_-(z)\,\psi_+(z)\,. \qquad (3.117)$$

Оставшийся вариант — это биномиальная матрица G, которая содержит два нулевых элемента в одном из столбцов и два ненильпотентных элемента. При этом одна из нечетных координат вырождается (как в (2.76) для $N = 1$), а другая сохраняет общий суперагалитический вид (3.14). При этом возможны два решения

$$1)\ \mathrm{G}_{SCf} = \begin{pmatrix} 0 & g_{++}(z) \\ 0 & g_{-+}(z) \end{pmatrix},\ \lambda_+(z) = 0,\ \psi'_+(z) = 0, \qquad (3.118)$$

$$2)\ \mathrm{G}_{SCf} = \begin{pmatrix} g_{+-}(z) & 0 \\ g_{--}(z) & 0 \end{pmatrix},\ \lambda_-(z) = 0,\ \psi'_-(z) = 0, \qquad (3.119)$$

первое из которых приводит к следующим вырожденным преобразованиям

$$\begin{cases} \tilde{z} = \tilde{\theta}^-\alpha + c = \psi_-(z)\,\alpha + \theta^+ g_{--}(z)\,\alpha + \theta^- g_{-+}(z)\,\alpha - \theta^+\theta^-\lambda_-(z)\,\alpha + c, \\ \tilde{\theta}^+ = \alpha = const, \\ \tilde{\theta}^- = \psi_-(z) + \theta^- g_{-+}(z) + \theta^+ g_{--}(z) - \theta^+\theta^-\lambda_-(z)\,. \end{cases} \qquad (3.120)$$

Для выяснения полугрупповых свойств всех приведенных преобразований необходимо построить их таблицу умножения, подобную приведенной в **Пункте 2.1.4**. Однако, это не представляется возможным из-за неимоверного размера формул и количества различных вариантов. Ограничися лишь замечанием, что $U_{\Lambda_0}(1)$ преобразования (3.99) представляют собой очевидную подполугруппу (или подгруппу в обратимом случае [563, 565, 622, 625]). Также подполугруппы (но не подгруппы) представляют собой вырожденные преобразования с биномиальными матрицами $\mathrm{G}_{SCf}$ (3.118)–(3.119) и "самые необратимые" преобразования варианта **3** с нулевой матрицей G (3.103).

**3.1.6.** Сплетающие четность $N=2$ преобразования. Рассмотрим здесь другие типы редукций (3.49)–(3.50) и соответствующие преобразования, определяемые условиями (3.41)–(3.44).

Сначала воспользуемся некоторыми соотношениями, следующими из общей формулы (3.22) и $\mathsf{TPt}^{\pm}$ условий (3.41)–(3.44).

Отметим такое соотношение

$$\operatorname{per}\mathrm{H}_{TPt\pm} = -\frac{1}{2} D^{\pm} \varDelta^{\mp}_{TPt\pm}\left(z,\theta^{+},\theta^{-}\right), \tag{3.121}$$

следующее из (3.22). Отсюда $D^{\pm}\left(\operatorname{per}\mathrm{H}_{TPt\pm}\right) = 0$ или

$$D^{\pm}D^{\mp}\tilde{\theta}^{\pm}_{TPt\pm}\cdot D^{\pm}\tilde{\theta}^{\mp}_{TPt\pm} = -D^{\pm}D^{\mp}\tilde{\theta}^{\mp}_{TPt\pm}\cdot D^{\pm}\tilde{\theta}^{\pm}_{TPt\pm}. \tag{3.122}$$

Кроме того, из условий $D^{\pm}\varDelta^{\pm}_{TPt\mp}\left(z,\theta^{+},\theta^{-}\right) = 0$ (3.42), (3.44) находим

$$D^{\pm}\tilde{\theta}^{+}_{TPt\mp}\cdot D^{\pm}\tilde{\theta}^{-}_{TPt\mp} = 0, \tag{3.123}$$

и это свидетельствует о том, что теперь элементы лишь одного столбца матрицы $\mathrm{H}^{T}_{TPt\pm}$ ортогональны (ср. $\mathsf{SCf}$ (3.72)), и поэтому $\mathrm{H}^{T}_{TPt\pm}$ более не является scf-матрицей (см. **Подраздел 5.1**).

Далее выясним действие $\mathsf{TPt}^{\pm}$ преобразований в касательном пространстве. Из (3.49)–(3.50) следуют законы преобразования производных и дифференциалов для $\mathsf{TPt}^{+}$ преобразований (3.41)–(3.42)

$$\begin{pmatrix} \partial \\ D^- \end{pmatrix} = \mathcal{R}_{TPt+} \cdot \begin{pmatrix} \tilde{D}^- \\ \tilde{D}^+ \end{pmatrix}, \tag{3.124}$$

$$d\tilde{Z} = d\theta^+ \cdot \varDelta^-_{TPt+}(z, \theta^+, \theta^-), \tag{3.125}$$

где $\mathcal{R}_{TPt+}$ — полуматрица (см. **Пункт Д.2**) из (3.55).

Соответственно для $\mathsf{TPt}^{-}$ преобразований (3.43)–(3.44)

$$\begin{pmatrix} \partial \\ D^+ \end{pmatrix} = \mathcal{R}_{TPt-} \cdot \begin{pmatrix} \tilde{D}^- \\ \tilde{D}^+ \end{pmatrix}, \tag{3.126}$$

$$d\tilde{Z} = d\theta^- \cdot \varDelta^+_{TPt-}(z, \theta^+, \theta^-). \tag{3.127}$$

*Замечание* **3.24.** Полуматрицы $\mathcal{R}_{TPt+}$ и $\mathcal{R}_{TPt-}$ являются полуминорами (см. **Пункт Д.2**) следующих нечетных элементов $\varDelta^-_{TPt+}(z, \theta^+, \theta^-)$ и $\varDelta^+_{TPt-}(z, \theta^+, \theta^-)$ в суперматрицах (3.49) и (3.50) соответственно (см. *Замечание* **3.19**).

Из сравнения SCf преобразований касательного пространства (3.64) и формул (3.124) и (3.126) следует, что здесь имеется некоторая аналогия с $(0|2)$ мерным подпространством $(1|2)$ мерного касательного пространства, где матрица $\mathrm{H}_{SCf}$ оставляла его инвариантным $T\mathbb{C}^{0|2} \to T\mathbb{C}^{0|2}$.

*Замечание* **3.25.** В данном нечетном случае полуматрицы $\mathcal{R}_{TPt\pm}$ действуют также в двумерном подпространстве, однако меняют его четность, а именно $T\mathbb{C}^{1|1} \to T\mathbb{C}^{0|2}$.

Поэтому, по аналогии с $N = 1$ (см. **Определение 2.55**) можно сформулировать

**Определение 3.26.** *Редуцированные $N = 2$ преобразования, удовлетворяющие следующим условиям $Q(z, \theta^+, \theta^-) = 0$, $\varDelta^+(z, \theta^+, \theta^-) = 0$ или $Q(z, \theta^+, \theta^-) = 0$, $\varDelta^-(z, \theta^+, \theta^-) = 0$* (3.41)–(3.44)*, действующие в касательном пространстве как $T\mathbb{C}^{1|1} \to T\mathbb{C}^{0|2}$ назовем сплетающими четность (касательного пространства) $N = 2$ преобразованиями* (TPt *— twisting parity of tangent space transformations*).

Происхождение этого определения ясно из выражения для четной производной

$$\partial = \partial\tilde{\theta}^+_{TPt\pm} \cdot \tilde{D}^- + \partial\tilde{\theta}^-_{TPt\pm} \cdot \tilde{D}^+, \tag{3.128}$$

следующего из (3.124) и (3.126) (ср. SCf (3.66)), а также из TPt формул для четного дифференциала (3.125) и (3.127) (ср. $N = 1$ (2.130)–(2.132)).

По аналогии с $N = 2$ суперконформными дифференциалами (3.67), которые дуальны суперпроизводным $D^\pm$ в смысле формулы (3.64), определим $N = 2$ TPt дифференциалы, исходя из (3.124) следующим образом (ср. $N = 1$ (2.140)).

**Определение 3.27.** *Назовем $N = 2$* TPt *супердифференциалами с кручением четности такие объекты $d\tau_{TPt\pm}$, которые преобразуются при сплетающих четность преобразованиях $Z \to \tilde{Z}$ (см.* **Определение 3.26**) *по закону*

$$\left( d\tilde{\tau}^{even+}_{TPt\pm} \;\; d\tilde{\tau}^{even-}_{TPt\pm} \right) = \left( d\tau^{odd}_{TPt\pm} \;\; d\tau^{even\pm}_{TPt\pm} \right) \cdot \mathcal{R}_{TPt\pm}, \tag{3.129}$$

*где полуматрицы $\mathcal{R}_{TPt\pm}$ определены в* (3.27).

В явном виде имеем

$$d\tilde{\tau}^{even+}_{TPt\pm} = d\tau^{odd}_{TPt\pm} \cdot \partial\tilde{\theta}^{+}_{TPt\pm} + d\tau^{even\pm}_{TPt\pm} \cdot D^{\pm}\tilde{\theta}^{+}_{TPt\pm}, \tag{3.130}$$

$$d\tilde{\tau}^{even-}_{TPt\pm} = d\tau^{odd}_{TPt\pm} \cdot \partial\tilde{\theta}^{-}_{TPt\pm} + d\tau^{even\pm}_{TPt\pm} \cdot D^{\pm}\tilde{\theta}^{-}_{TPt\pm}. \tag{3.131}$$

*Замечание* **3.28.** Четности $d\tilde{\tau}^{even+}_{TPt\pm}$, $d\tilde{\tau}^{even-}_{TPt\pm}$, $d\tau^{even+}_{TPt\pm}$ и $d\tau^{odd-}_{TPt\pm}$ противоположны, поэтому в кокасательном пространстве мы имеем отображение с кручением четности $T^*\mathbb{C}^{1|1} \to T^*\mathbb{C}^{2|0}$ (ср. *Замечание* **3.25**).

По аналогии с суперконформным случаем (3.68) определим внешние $\mathsf{TPt}^{\pm}$ дифференциалы

$$\delta_{TPt+} = d\tau^{odd}_{TPt+} \cdot \partial + d\tau^{even+}_{TPt+} \cdot D^{-}. \tag{3.132}$$

$$\delta_{TPt-} = d\tau^{odd}_{TPt-} \cdot \partial + d\tau^{even-}_{TPt-} \cdot D^{+}. \tag{3.133}$$

$$\tilde{\delta}_{TPt\pm} = d\tilde{\tau}^{even+}_{TPt\pm} \cdot \tilde{D}^{-} + d\tilde{\tau}^{even-}_{TPt\pm} \cdot \tilde{D}^{+}. \tag{3.134}$$

*Замечание* **3.29.** Четность внешних $N = 2$ $\mathsf{TPt}$ дифференциалов (3.132)–(3.134) фиксирована, они — нечетны при любых сплетающих четность преобразованиях.

**Предложение 3.30.** *Внешние $N = 2$ $\mathsf{TPt}$ дифференциалы инвариантны относительно $N = 2$ $\mathsf{TPt}$ преобразований.*

*Доказательство.* Из определений (3.132)–(3.134) и законов преобразования (3.124), (3.126) и (3.129) имеем, например, для $\mathsf{TPt}^{+}$ преобразований

$$\delta_{TPt+} = \begin{pmatrix} d\tau^{odd}_{TPt\pm} & d\tau^{even\pm}_{TPt\pm} \end{pmatrix} \cdot \begin{pmatrix} \partial \\ D^{-} \end{pmatrix} =$$

$$\left( \begin{array}{cc} d\tau^{odd}_{TPt\pm} & d\tau^{even\pm}_{TPt\pm} \end{array} \right) \cdot \mathcal{R}_{TPt+} \cdot \left( \begin{array}{c} \tilde{D}^- \\ \tilde{D}^+ \end{array} \right) =$$

$$\left( \begin{array}{cc} d\tilde{\tau}^{even+}_{TPt\pm} & d\tilde{\tau}^{even-}_{TPt\pm} \end{array} \right) \cdot \left( \begin{array}{c} \tilde{D}^- \\ \tilde{D}^+ \end{array} \right) = \tilde{\delta}_{TPt+}.$$

И аналогично для $\mathsf{TPt}^-$ преобразований. ■

Таким образом, необратимый аналог $N = 2$ дифференциальной геометрии при $\mathsf{TPt}$ преобразованиях оказывается не столь прост и прозрачен, как в $\mathsf{SCf}$ случае. Это дает возможность построения $N = 2$ расслоений с кручением четности (см. для $N = 1$ **Пункт 2.3.2**).

Исходя из (3.63), а также из теоремы сложения $N = 2$ березинианов, можно трактовать $N = 2$ преобразования следующим образом.

**Предположение 3.31.** *Если считать $N = 2$ $\mathsf{SCf}$ преобразования $N = 2$ супераналогом обычных голоморфных преобразований* [563, 565]*, то для антиголоморфных преобразований, в отличие от $N = 1$* (*см.* **Подраздел 2.3**)*, имеется два* (!) *нечетных супераналога:* $\mathsf{TPt}^+$ *и* $\mathsf{TPt}^-$ *преобразования.*

**3.1.7.** Дуальные суперaналитические $N=1$ преобразования и редуцированные $N=2$ преобразования. Необходимость рассмотрения связи $N = 1$ и $N = 2$ редуцированных преобразований обусловлена, прежде всего, обнаружением скрытой $N = 2$ суперконформной симметрии в суперструнной теории [337, 629]. Более того, из расширения аксиоматики [630–632] конформной теории поля [633–636] на $N = 2$ делался вывод о том, что "$N = 2$ суперконформная симметрия более фундаментальна, чем $N = 1$ суперконформная симметрия" [407].

Здесь мы обобщим с учетом необратимости получение дуальных $N = 1$ преобразований из редуцированных $N = 2$ преобразований подобно [269, 629]. Кроме того, в **Приложении Ж.3** мы рассмотрим вложения $N = 1 \hookrightarrow N = 2$, играющие важную роль в суперструнных вычислениях [637].

Пусть мы имеем $U(1)$ SCf преобразование (3.99), определяемое двумя четными $g_{\pm\mp}(z)$ и двумя нечетными функциями $\psi_{\pm}(z)$, записанное в виде

$$\mathcal{T}_{SCf}^{(N=2)}: \begin{cases} \tilde{z} = f(z) + \theta^+ g_{-+}(z)\,\psi_+(z) + \theta^- g_{+-}(z)\,\psi_-(z) + \\ \qquad\qquad \theta^+\theta^- \left(\psi_+(z)\,\psi_-(z)\right)', \\ \tilde{\theta}^+ = \psi_+(z) + \theta^+ g_{+-}(z) + \theta^+\theta^-\psi'_+(z), \\ \tilde{\theta}^- = \psi_-(z) + \theta^- g_{-+}(z) - \theta^+\theta^-\psi'_-(z), \end{cases} \tag{3.135}$$

где

$$f'(z) = g_{+-}(z)\,g_{-+}(z) + \psi'_+(z)\,\psi_-(z) + \psi'_-(z)\,\psi_+(z)\,. \tag{3.136}$$

Обратим внимание на то, что правая часть второго уравнения в (3.135) зависит от $\theta^-$ только в комбинации $z+\theta^+\theta^-$, а зависимость от $\theta^+$ в третьем уравнении — в комбинации $z-\theta^+\theta^-$. Поэтому естественным является введение новых $N = 2$ координат $(Z_A, \eta_A)$ и $(Z_B, \eta_B)$, где $N = 1$ координаты равны $Z_A = (z_A, \theta_A)$ и $Z_B = (z_B, \theta_B)$, по формулам

$$\mathcal{U}_A: \begin{cases} z_A = z + \theta^+\theta^-, \\ \theta_A = \dfrac{\theta^+}{\sqrt{2}}, \\ \eta_A = \dfrac{\theta^-}{\sqrt{2}}, \end{cases} \qquad \mathcal{U}_B: \begin{cases} z_B = z - \theta^+\theta^-, \\ \eta_B = \dfrac{\theta^+}{\sqrt{2}}, \\ \theta_B = \dfrac{\theta^-}{\sqrt{2}}. \end{cases} \tag{3.137}$$

Очевидно, что $\mathrm{Ber}\left((Z_A, \eta_A)/Z\right) = \mathrm{Ber}\left((Z_B, \eta_B)/Z\right) = 2$. Тогда из (3.135)–(3.137) получаем $N = 2$ SCf преобразования $\mathcal{T}_A^{(N=2)}: (Z_A, \eta_A) \to$ $\left(\tilde{Z}_A, \tilde{\eta}_A\right)$ и $\mathcal{T}_B^{(N=2)} : (Z_B, \eta_B) \to \left(\tilde{Z}_B, \tilde{\eta}_B\right)$ в виде*)

$$\mathcal{T}_A^{(N=1)} : \begin{cases} \tilde{z}_A = F_A(z_A, \theta_A) = f(z_A) + \psi_+(z_A)\psi_-(z_A) + \\ \qquad\qquad \theta_A \cdot g_{+-}(z_A)\sqrt{2}\psi_-(z_A), \\ \tilde{\theta}_A = \Psi_A(z_A, \theta_A) = \sqrt{2}\psi_+(z_A) + \theta_A \cdot g_{+-}(z_A), \end{cases} \tag{3.138}$$

$$\mathcal{T}_{\eta_A}^{(N=1)} : \begin{array}{l} \tilde{\eta}_A = \eta_A \cdot H_A(z_A, \theta_A) + \Phi_A(z_A, \theta_A) = \\ \sqrt{2}\psi_-(z_A) + \eta_A \cdot g_{-+}(z_A) - \theta_A\eta_A \cdot \sqrt{2}\psi'_-(z_A), \end{array} \tag{3.139}$$

$$\mathcal{T}_B^{(N=1)} : \begin{cases} \tilde{z}_B = F_B(z_B, \theta_B) = f(z_B) + \psi_+(z_B)\psi_-(z_B) + \\ \qquad\qquad \theta_B \cdot g_{-+}(z_B)\sqrt{2}\psi_+(z_B), \\ \tilde{\theta}_B = \Psi_B(z_B, \theta_B) = \sqrt{2}\psi_-(z_B) + \theta_B \cdot g_{-+}(z_B), \end{cases} \tag{3.140}$$

$$\mathcal{T}_{\eta_B}^{(N=1)} : \begin{array}{l} \tilde{\eta}_B = \eta_B \cdot H_B(z_B, \theta_B) + \Phi_B(z_B, \theta_B) = \\ \sqrt{2}\psi_+(z_B) + \eta_B \cdot g_{+-}(z_B) - \theta_B\eta_B \cdot \sqrt{2}\psi'_+(z_B). \end{array} \tag{3.141}$$

Обратим внимание на то, что преобразование переменных $\eta_A, \eta_B$ в (3.138)–(3.141) "отщепляется", т. е. не входит в первые 2 уравнения, и поэтому можно схематически записать $\mathcal{T}_A^{(N=2)} = \mathcal{T}_A^{(N=1)} \otimes \mathcal{T}_{\eta_A}^{(N=1)}$ и $\mathcal{T}_B^{(N=2)} = \mathcal{T}_B^{(N=1)} \otimes \mathcal{T}_{\eta_B}^{(N=1)}$.

Таким образом, мы получаем следующее

**Утверждение 3.32.** *Каждому $N = 2$ SCf преобразованию без твиста (или $U(1)$) $\mathcal{T}_{\mathrm{SCf}}^{(N=2)} : (z, \theta^+, \theta^-) \to \left(\tilde{z}, \tilde{\theta}^+, \tilde{\theta}^-\right)$ (3.135) можно поставить в соответствие пару дуальных $N = 1$ (в общем случае не суперконформных, а супераналитических (2.2)) преобразований $\mathcal{T}_A^{(N=1)} : (z_A, \theta_A) \to \left(\tilde{z}_A, \tilde{\theta}_A\right)$ и $\mathcal{T}_B^{(N=1)} : (z_B, \theta_B) \to \left(\tilde{z}_B, \tilde{\theta}_B\right)$ по формулам (3.138)–(3.141).*

---

*) *Примечание.* По повторяющимся индексам нет суммирования, и нижеследующие уравнения являются одновременно определением функций $F_A(z_A, \theta_A)$, $\Psi_A(z_A, \theta_A)$, $H_A(z_A, \theta_A)$, $\Psi_A(z_A, \theta_A)$ и $F_B(z_B, \theta_B)$, $\Psi_B(z_B, \theta_B)$, $H_B(z_B, \theta_B)$, $\Psi_B(z_B, \theta_B)$.

Тогда легко видеть, что диграмма преобразований

$$\begin{array}{ccc}
\eta_A, Z_A & \xrightarrow{\mathcal{T}_A^{(N=1)}} & \tilde{Z}_A, \tilde{\eta}_A \\
\uparrow \mathcal{U}_A & & \uparrow \tilde{\mathcal{U}}_A \\
Z & \xrightarrow{\mathcal{T}_{\mathrm{SCf}}^{(N=2)}} & \tilde{Z} \\
\downarrow \mathcal{U}_B & & \downarrow \tilde{\mathcal{U}}_B \\
\eta_B, Z_B & \xrightarrow{\mathcal{T}_B^{(N=1)}} & \tilde{Z}_B, \tilde{\eta}_B
\end{array} \qquad (3.142)$$

коммутативна.

Важно отметить фундаментальные равенства

$$H_A(z_A, \theta_A) = \mathrm{Ber}^{\,N=1}\left(\tilde{Z}_A / Z_A\right), \qquad (3.143)$$

$$\Phi_A(z_A, \theta_A) = \frac{\dfrac{\partial F_A(z_A, \theta_A)}{\partial \theta_A}}{\dfrac{\partial \Psi_A(z_A, \theta_A)}{\partial \theta_A}}, \qquad (3.144)$$

(и аналогичные для $A \to B$), которые следуют непосредственно из $N = 2$ SCf условий и требования ковариантности преобразования дифференциалов $dZ = dz + \theta^+ d\theta^- + \theta^- d\theta^+ = dz_A + \eta_A d\theta_A = dz_B + \eta_B d\theta_B$.

В обратимом случае, если использовать преобразования $\mathcal{T}_{SCf}^{(N=2)}$ как функции перехода на $N = 2$ суперримановой поверхности, а преобразования $\mathcal{T}_A^{(N=1)}$ и $\mathcal{T}_B^{(N=1)}$ — как функции перехода для $(1|1)$ мерных супермногообразий, то получаем

**Утверждение 3.33.** *Каждой $N = 2$ суперримановой поверхности без твиста соответствует пара дуальных $(1|1)$-мерных супермногообразий, компонентные функции перехода которых (см. (2.2)) равны*

$$f_A(z) = f_B(z) = f(z) + \psi_+(z)\,\psi_-(z), \qquad (3.145)$$

$$g_A(z) = g_{+-}(z), \qquad g_B(z) = g_{-+}(z), \qquad (3.146)$$

$$\psi_A(z) = \sqrt{2}\psi_+(z), \qquad \psi_B(z) = \sqrt{2}\psi_-(z), \tag{3.147}$$

$$\chi_A(z) = \sqrt{2}\psi_-(z)\, g_{+-}(z), \ \chi_A(z) = \sqrt{2}\psi_+(z)\, g_{-+}(z). \tag{3.148}$$

*Доказательство.* Следует из вида преобразований (3.138)–(3.141). ■

Отсюда можно получить

**Предложение 3.34.** *Компонентные функции дуальных $N = 1$ преобразований (и функции перехода дуальных $(1|1)$ супермногообразий) связаны между собой соотношениями*

$$f'_A(z) = f'_B(z) = g_A(z)\, g_B(z) + \psi'_A(z)\, \psi_B(z), \tag{3.149}$$

$$\chi_A(z) = g_A(z)\, \psi_B(z), \tag{3.150}$$

$$\chi_B(z) = g_B(z)\, \psi_A(z). \tag{3.151}$$

*Доказательство.* Следует непосредственно из (3.135)–(3.136) и выражений (3.145)–(3.148). ■

Рассмотрим расщепленные дуальные $N = 1$ преобразования, которые не содержат нечетных компонентных функций.

**Утверждение 3.35.** *Березинианы расщепленных дуальных преобразований взаимообратны относительно $f'(z)$.*

*Доказательство.* По общей формуле для березиниана $N = 1$ суперааналитических преобразований (E.13) имеем

$$\mathrm{Ber}^{\,N=1}\left(\tilde{Z}_A/Z_A\right) = \frac{f'_A(z)}{g_A(z)}, \quad \mathrm{Ber}^{\,N=1}\left(\tilde{Z}_B/Z_B\right) = \frac{f'_B(z)}{g_B(z)},$$

тогда, пользуясь (3.149) и (3.145), получаем

$$\mathrm{Ber}^{\,N=1}\left(\tilde{Z}_A/Z_A\right) \mathrm{Ber}^{\,N=1}\left(\tilde{Z}_B/Z_B\right) =$$

$$\frac{f'_A(z)}{g_A(z)}\frac{f'_B(z)}{g_B(z)} = \frac{(f'(z))^2}{f'(z)} = f'(z)\,. \quad (3.152)$$

■

В терминах введенной дуальности $N = 1$ суперконформные преобразования (и в обратимом случае соответствующие им суперримановы поверхности) можно определить следующим образом.

**Утверждение 3.36.** $N = 1$ *суперконформные преобразования самодуальны.*

*Доказательство.* Если приравнять функции с индексами $A$ и $B$ в уравнениях дуальности (3.149)–(3.151), то получим $N = 1$ суперконформные условия (2.81). ■

Аналогичные конструкции можно построить и для различных типов необратимых $N = 2$ суперконформных преобразований, рассмотренных в **Пункте 3.1.5** и допускающих "отщепление" одной из нечетных координат (например, (3.108) и (3.116)).

## 3.2. Редуцированные $N = 4$ преобразования

Суперструнные теории, имеющие $N = 4$ суперсимметрию на мировом листе, после компактификации предсказывают нефизические значения размерности пространства-времени [568, 638–640], тем не менее, интерес к $N = 4$ суперконформной теории поля [641, 642] и $N = 4$ суперконформным алгебрам [643–650] (включая алгебры с твистом [651]) обусловлен применимостью к $\sigma$-моделям [652–655], суперконформной топологической теории поля [656, 657] и компактификациям в шесть измерений [658], а в последнее время — к нетривиальным решениям для $D$-бран [659–663] и к геометрии трехмерного пространства анти-де-Ситтера [315, 588, 664]. Общие вопросы $N = 4$ суперконформной гео-метрии изучались в работах [665–667].

В этом разделе мы обратимся к $N=4$ преобразованиям и приведем возможные редукции касательного пространства и соответствующие типы преобразований с учетом необратимости [2].

При описании суперпространства $\mathbb{C}^{1|4}$ мы также воспользуемся комплексным базисом. Если $Z=\left(z,\theta^1,\theta^2,\theta^3,\theta^4\right)\in\mathbb{C}^{1|4}$, то в комплексном базисе имеем

$$\theta_1^{\pm}=\frac{\theta^1\pm i\theta^2}{\sqrt{2}},\;\;\theta_2^{\pm}=\frac{\theta^3\pm i\theta^3}{\sqrt{2}}, \tag{3.153}$$

$$D_1^{\pm}=\frac{D_1\pm iD_2}{\sqrt{2}}=\frac{\partial}{\partial\theta_1^{\mp}}+\theta_1^{\pm}\partial,\; D_2^{\pm}=\frac{D_3\pm iD_4}{\sqrt{2}}=\frac{\partial}{\partial\theta_2^{\mp}}+\theta_2^{\pm}\partial, \tag{3.154}$$

где $D_i$ определены в (E.18) и удовлетворяют соотношениям $N=4$ суперсимметрии

$$\left\{D_i^+,D_j^-\right\}=2\delta_{ij}\partial,\;\left\{D_i^+,D_j^+\right\}=\left\{D_i^-,D_j^-\right\}=0. \tag{3.155}$$

Аналогично (3.4) базис в $(1|4)$ кокасательном пространстве имеет вид *)

$$dZ=dz+\theta_i^+d\theta_i^-+\theta_i^-d\theta_i^+, \tag{3.156}$$

$$d\theta_1^{\pm}=\frac{d\theta^1\pm id\theta^2}{\sqrt{2}},\;\;d\theta_2^{\pm}=\frac{d\theta^3\pm id\theta^4}{\sqrt{2}}. \tag{3.157}$$

При действии общих $N=4$ супераналитических преобразований $Z\left(z,\theta_i^+,\theta_i^-\right)\to\tilde{Z}\left(\tilde{z},\tilde{\theta}_i^+,\tilde{\theta}_i^-\right)$ суперпроизводные (3.154) и дифференциалы

---

*Примечание.* В этом подразделе $i,j,k\ldots\in\{1,2\}$, и по повторяющимся латинским индексам производится суммирование.

преобразуются как

$$\begin{pmatrix} \partial \\ D_1^- \\ D_1^+ \\ D_2^- \\ D_2^+ \end{pmatrix} = \mathrm{P}_{SA}^{(N=4)} \cdot \begin{pmatrix} \tilde{\partial} \\ \tilde{D}_1^- \\ \tilde{D}_1^+ \\ \tilde{D}_2^- \\ \tilde{D}_2^+ \end{pmatrix}, \tag{3.158}$$

$$\begin{pmatrix} d\tilde{Z} & d\tilde{\theta}_1^+ & d\tilde{\theta}_1^- & d\tilde{\theta}_2^+ & d\tilde{\theta}_2^- \end{pmatrix} = \begin{pmatrix} dZ & d\theta_1^+ & d\theta_1^- & d\theta_2^+ & d\theta_2^- \end{pmatrix} \cdot \mathrm{P}_{SA}^{(N=4)}, \tag{3.159}$$

где $\mathrm{P}_{SA}^{(N=4)}$ — суперматрица касательного $N = 4$ суперпространства.

Независимо от конретного вида суперматрицы $\mathrm{P}_{SA}^{(N=4)}$ имеем

**Предложение 3.37.** *Внешний $N = 4$ дифференциал*

$$\mathrm{d}^{(N=4)} = dz\partial + d\theta_i^+ \frac{\partial}{\partial\theta_i^+} + d\theta_i^- \frac{\partial}{\partial\theta_i^-} \tag{3.160}$$

*инвариантен относительно общих $N = 4$ супераналитических преобразований $Z\left(z, \theta_i^+, \theta_i^-\right) \to \tilde{Z}\left(\tilde{z}, \tilde{\theta}_i^+, \tilde{\theta}_i^-\right)$.*

*Доказательство.* Пользуясь определениями (3.158) и (3.159), внешний $N = 4$ дифференциал (3.160) можно представить в виде

$$\begin{aligned} \mathrm{d}^{(N=4)} &= \left(dz - d\theta_i^+\theta_i^- - d\theta_i^-\theta_i^+\right)\partial + d\theta_i^+ \left(\frac{\partial}{\partial\theta_i^+} + \theta_i^-\partial\right) + \\ d\theta^- \left(\frac{\partial}{\partial\theta_i^-} + \theta_i^+\partial\right) &= dZ\partial + d\theta_i^+ D_i^- + d\theta_i^- D_i^+. \end{aligned} \tag{3.161}$$

Далее доказательство полностью совпадает с доказательством подобного $N = 2$ **Предложения E.11**. ■

Представим суперматрицу $\mathrm{P}_{SA}^{(N=4)}$, входящую в (3.158) и (3.159), в удобном для дальнейших редукций виде (ср. (3.6))

$$\mathrm{P}_{SA}^{(N=4)} = \begin{pmatrix} Q\left(z,\theta_i^+,\theta_i^-\right) & \partial\tilde{\theta}_1^+ \ \partial\tilde{\theta}_1^- \ \partial\tilde{\theta}_2^+ \ \partial\tilde{\theta}_2^- \\ \Delta_1^-\left(z,\theta_i^+,\theta_i^-\right) & \\ \Delta_1^+\left(z,\theta_i^+,\theta_i^-\right) & \mathrm{H} \\ \Delta_2^-\left(z,\theta_i^+,\theta_i^-\right) & \\ \Delta_2^+\left(z,\theta_i^+,\theta_i^-\right) & \end{pmatrix}, \tag{3.162}$$

где

$$Q\left(z,\theta_i^+,\theta_i^-\right) = \partial\tilde{z} - \partial\tilde{\theta}_i^+ \cdot \tilde{\theta}_i^- - \partial\tilde{\theta}_i^- \cdot \tilde{\theta}_i^+, \tag{3.163}$$

$$\Delta_i^\pm\left(z,\theta_k^+,\theta_k^-\right) = D_i^\pm\tilde{z} - D_i^\pm\tilde{\theta}_j^- \cdot \tilde{\theta}_j^+ - D_i^\pm\tilde{\theta}_j^+ \cdot \tilde{\theta}_j^-, \tag{3.164}$$

а $4\times 4$ матрица H имеет блочный вид

$$\mathrm{H} = \begin{pmatrix} \mathrm{H}_{11} & \mathrm{H}_{12} \\ \mathrm{H}_{21} & \mathrm{H}_{22} \end{pmatrix}, \ \ \mathrm{H}_{ij} = \begin{pmatrix} D_j^-\tilde{\theta}_i^+ & D_j^-\tilde{\theta}_i^- \\ D_j^+\tilde{\theta}_i^+ & D_j^+\tilde{\theta}_i^- \end{pmatrix}. \tag{3.165}$$

Березиниан общих $N = 4$ супераналитических преобразований $Z\left(z,\theta_i^+,\theta_i^-\right) \to \tilde{Z}\left(\tilde{z},\tilde{\theta}_i^+,\tilde{\theta}_i^-\right)$ определяется как [30]

$$\mathrm{Ber}^{\,N=4}\left(\tilde{Z}/Z\right) = \mathrm{Ber}\,\mathrm{P}_0^{(N=4)}, \tag{3.166}$$

где $\mathrm{P}_0^{(N=4)}$ — это $5\times 5$ суперматрица производных, аналогичная $3\times 3$ $N=2$ суперматрице (E.36).

Для $N = 4$ имеет место также подобное **Предложению E.12** следующее

**Предложение 3.38.** $\mathrm{Ber}^{\,N=4}\left(\tilde{Z}/Z\right) = \mathrm{Ber}\,\mathrm{P}_{SA}^{(N=4)}$.

*Доказательство.* Основывается на $N = 4$ аналоге формулы (E.38), но для $5\times 5$ суперматриц. ∎

Тогда в случае $\epsilon\,[\det \mathrm{H}] \neq 0$, для березиниана $N=4$ супераналитических преобразований получаем

$$\mathrm{Ber}^{\,N=4}\left(\tilde{Z}/Z\right) = \mathrm{Ber}\,\mathrm{P}_{SA}^{(N=4)} =$$

$$\frac{Q\left(z,\theta_i^+,\theta_i^-\right) - \left(\partial\tilde{\theta}_1^+ \;\; \partial\tilde{\theta}_1^- \;\; \partial\tilde{\theta}_2^+ \;\; \partial\tilde{\theta}_2^-\right)\cdot \mathrm{H}^{-1}\cdot \begin{pmatrix} \mathit{\Delta}_1^-\left(z,\theta_i^+,\theta_i^-\right) \\ \mathit{\Delta}_1^+\left(z,\theta_i^+,\theta_i^-\right) \\ \mathit{\Delta}_2^-\left(z,\theta_i^+,\theta_i^-\right) \\ \mathit{\Delta}_2^+\left(z,\theta_i^+,\theta_i^-\right) \end{pmatrix}}{\det \mathrm{H}}. \quad (3.167)$$

Легко видеть, что (при условии $\epsilon\,[\det \mathrm{H}] \neq 0$)

$$\epsilon\left[\mathrm{Ber}^{\,N=4}\left(\tilde{Z}/Z\right)\right] = \epsilon\left[Q\left(z,\theta_i^+,\theta_i^-\right)\right], \quad (3.168)$$

поскольку остальные слагаемые в числителе (3.167) не имеют числовой части по их определению. Отсюда следует классификация по необратимости общих $N=4$ супераналитических преобразований.

**Определение 3.39.** *Обратимые $N=4$ супераналитические преобразования определяются условиями*

$$\epsilon\left[Q\left(z,\theta_i^+,\theta_i^-\right)\right] \neq 0,\; \epsilon\,[\det \mathrm{H}] \neq 0. \quad (3.169)$$

**Определение 3.40.** *Полунеобратимые $N=4$ супераналитические преобразования определяются условиями*

$$\epsilon\left[Q\left(z,\theta_i^+,\theta_i^-\right)\right] = 0,\; \epsilon\,[\det \mathrm{H}] \neq 0. \quad (3.170)$$

**Определение 3.41.** *Необратимые $N=2$ супераналитические пре-*

*образования определяются условиями*

$$\epsilon\left[Q\left(z,\theta_i^+,\theta_i^-\right)\right]=0,\ \epsilon\left[\det \mathrm{H}\right]=0. \tag{3.171}$$

Обратимся к нахождению различных вариантов редукций суперматрицы $\mathrm{Ber}\,\mathrm{P}_{SA}^{(N=4)}$ (3.162) с учетом необратимости преобразований.

**3.2.1.** $N=4$ редукции в терминах перманентов. Для того, чтобы выяснить, какие возможны редукции $(1|4)$ касательного суперпространства, необходимо представить березиниан $N=4$ преобразований $\mathrm{Ber}^{\,N=4}\left(\tilde{Z}/Z\right)$ (3.162) в виде суммы некоторых слагаемых, подобно $N=2$ (3.32). Тогда можно сформулировать теорему сложения березинианов для $N=4$ (см. для $N=2$ (3.31) и $N=1$ (2.34)).

**Теорема 3.42.** (Теорема сложения $N=4$ березинианов) *Для* $N=4$ *супераналитических преобразований* $Z\left(z,\theta_i^+,\theta_i^-\right)\to\tilde{Z}\left(\tilde{z},\tilde{\theta}_i^+,\tilde{\theta}_i^-\right)$ *полный* $N=4$ *березиниан в обратимом* (3.169) *и полунеобратимом* (3.170) *случаях представляется в виде суммы пяти березинианов*

$$\begin{aligned}\mathrm{Ber}^{\,N=4}\left(\tilde{Z}/Z\right) &= \mathrm{Ber}\,\mathrm{P}_S^{(N=4)}+\mathrm{Ber}\,\mathrm{P}_{T_1^+}^{(N=4)}+\\ &\quad \mathrm{Ber}\,\mathrm{P}_{T_1^-}^{(N=4)}+\mathrm{Ber}\,\mathrm{P}_{T_2^+}^{(N=4)}+\mathrm{Ber}\,\mathrm{P}_{T_2^-}^{(N=4)}.\end{aligned} \tag{3.172}$$

*Доказательство.* Запишем столбец функций $\Delta_i^\pm\left(z,\theta_i^+,\theta_i^-\right)$ из (3.167) в виде суммы столбцов, в каждом из которых только один элемент не равен нулю. Тогда для суперматриц, входящих в правую часть (3.172)

получим

$$\mathrm{P}_S^{(N=4)} = \begin{pmatrix} Q\left(z, \theta_i^+, \theta_i^-\right) & \partial\tilde{\theta}_1^+ \;\; \partial\tilde{\theta}_1^- \;\; \partial\tilde{\theta}_2^+ \;\; \partial\tilde{\theta}_2^- \\ 0 & \\ 0 & \mathrm{H} \\ 0 & \\ 0 & \end{pmatrix}, \tag{3.173}$$

$$\mathrm{P}_{T_1^+}^{(N=4)} = \begin{pmatrix} 0 & \partial\tilde{\theta}_1^+ \;\; \partial\tilde{\theta}_1^- \;\; \partial\tilde{\theta}_2^+ \;\; \partial\tilde{\theta}_2^- \\ \varDelta_1^-\left(z, \theta_i^+, \theta_i^-\right) & \\ 0 & \mathrm{H} \\ 0 & \\ 0 & \end{pmatrix}, \tag{3.174}$$

$$\mathrm{P}_{T_1^-}^{(N=4)} = \begin{pmatrix} 0 & \partial\tilde{\theta}_1^+ \;\; \partial\tilde{\theta}_1^- \;\; \partial\tilde{\theta}_2^+ \;\; \partial\tilde{\theta}_2^- \\ 0 & \\ \varDelta_1^+\left(z, \theta_i^+, \theta_i^-\right) & \mathrm{H} \\ 0 & \\ 0 & \end{pmatrix}, \tag{3.175}$$

$$\mathrm{P}_{T_2^+}^{(N=4)} = \begin{pmatrix} 0 & \partial\tilde{\theta}_1^+ \;\; \partial\tilde{\theta}_1^- \;\; \partial\tilde{\theta}_2^+ \;\; \partial\tilde{\theta}_2^- \\ 0 & \\ 0 & \mathrm{H} \\ \varDelta_2^-\left(z, \theta_i^+, \theta_i^-\right) & \\ 0 & \end{pmatrix}, \tag{3.176}$$

$$\mathrm{P}_{T_2^-}^{(N=4)} = \begin{pmatrix} 0 & \partial\tilde{\theta}_1^+ \;\; \partial\tilde{\theta}_1^- \;\; \partial\tilde{\theta}_2^+ \;\; \partial\tilde{\theta}_2^- \\ 0 & \\ 0 & \mathrm{H} \\ 0 & \\ \varDelta_2^+\left(z, \theta_i^+, \theta_i^-\right) & \end{pmatrix}. \tag{3.177}$$

Из построения суперматриц (3.173)–(3.177) следует, что сумма их березинианов и дает полный березиниан, т. е. выполняется (3.172). ■

*Замечание* **3.43.** Такая процедура для обычных матриц аналогична разложению детерминанта по элементам столбца, умноженным на соответствующие алгебраические дополнения (теорема Лапласа). В суперсимметричном случае минорами нечетных элементов являются не суперматрицы общего положения*), а полуминоры, являющиеся полуматрицами, которые введены нами в **Приложении Д.2**.

Из (3.173) в случае $\epsilon\,[\det \mathrm{H}] \neq 0$ получаем

$$\mathrm{Ber}\, \mathrm{P}_S^{(N=4)} = \frac{Q\left(z, \theta_i^+, \theta_i^-\right)}{\det \mathrm{H}} \tag{3.178}$$

(ср. (3.34).

В остальных случаях выражения для березинианов громоздки и отличаются друг от друга лишь перестановками индексов. Поэтому мы приведем лишь один вариант

$$\mathrm{Ber}\, \mathrm{P}_{T_1^+}^{(N=4)} = \frac{\mathit{\Delta}_1^-\left(z, \theta_i^+, \theta_i^-\right)}{\det^2 \mathrm{H}} \cdot \boldsymbol{K}_{T_1^+}, \tag{3.179}$$

где

$$\begin{aligned} \boldsymbol{K}_{T_1^+} \;=\; & \delta\mathrm{et}\mathcal{R}_{11}^+ \cdot \det \mathrm{H}_{22} + \delta\mathrm{et}\mathcal{R}_{12}^+ \cdot \det \mathrm{H}_{21} + \\ & \delta\mathrm{et}\mathcal{R}_{21}^- \cdot \det \bar{\bar{\mathrm{H}}}_{22}^+ - \delta\mathrm{et}\mathcal{R}_{22}^- \cdot \det \bar{\bar{\mathrm{H}}}_{11}^+ - \\ & \delta\mathrm{et}\mathcal{R}_{22}^+ \cdot \det \bar{\mathrm{H}}_{11} - \delta\mathrm{et}\mathcal{R}_{21}^+ \cdot \det \bar{\mathrm{H}}_{21}. \end{aligned} \tag{3.180}$$

Здесь $\mathcal{R}_{ij}^{\pm}$ — горизонтальные полуматрицы, являющиеся полуми-

---

*) *Примечание.* Только такими суперматрицами и ограничено рассмотрение в [30, 106]

норами (см. **Приложение Д.2**) и определяемые как (ср. (3.27))

$$\mathcal{R}_{ij}^{\pm} = \begin{pmatrix} \partial\tilde{\theta}_j^{+} & \partial\tilde{\theta}_j^{-} \\ D_j^{\pm}\tilde{\theta}_i^{+} & D_j^{\pm}\tilde{\theta}_i^{-} \end{pmatrix}, \tag{3.181}$$

а $\mathrm{H}_{ij}$ определено в (3.165), $\bar{\mathrm{H}}_{ij}, \bar{\bar{\mathrm{H}}}_{ij}^{\pm}$ — миноры матрицы H (3.165) вида

$$\bar{\mathrm{H}}_{ij} = \begin{pmatrix} D_j^{-}\tilde{\theta}_i^{+} & D_j^{-}\tilde{\theta}_i^{-} \\ D_j^{+}\tilde{\theta}_{|i+1|_2}^{+} & D_j^{+}\tilde{\theta}_{|i+1|_2}^{-} \end{pmatrix}, \tag{3.182}$$

$$\bar{\bar{\mathrm{H}}}_{ij}^{\pm} = \begin{pmatrix} D_j^{\pm}\tilde{\theta}_i^{+} & D_j^{\pm}\tilde{\theta}_i^{-} \\ D_j^{\pm}\tilde{\theta}_{|i+1|_2}^{+} & D_j^{\pm}\tilde{\theta}_{|i+1|_2}^{-} \end{pmatrix}, \tag{3.183}$$

где $|i+1|_2$ означает по модулю 2, т.е. $|1+1|_2 = 2$, $|2+1|_2 = 1$ (здесь $i, j \in \mathrm{Z}_2$).

Из теоремы сложения $N = 4$ березинианов (3.172) следует, что число независимых редукций $(1|4)$ касательного пространства равно пяти (по сравнению с тремя при $N = 2$ (3.31) и двумя при $N = 1$ (2.34)).

**Определение 3.44.** *Обратимые, полунеобратимые и необратимые редуцированные $N = 4$ суперконформные преобразования определяются четырьмя условиями*

SCf:

$$\begin{aligned} \varDelta_1^{\pm}\left(z, \theta_i^{+}, \theta_i^{-}\right) &= D_1^{\pm}\tilde{z} - D_1^{\pm}\tilde{\theta}_i^{-} \cdot \tilde{\theta}_i^{+} - D_1^{\pm}\tilde{\theta}_i^{+} \cdot \tilde{\theta}_i^{-} = 0, && (3.184) \\ \varDelta_2^{\pm}\left(z, \theta_i^{+}, \theta_i^{-}\right) &= D_2^{\pm}\tilde{z} - D_2^{\pm}\tilde{\theta}_i^{-} \cdot \tilde{\theta}_i^{+} - D_2^{\pm}\tilde{\theta}_i^{+} \cdot \tilde{\theta}_i^{-} = 0. && (3.185) \end{aligned}$$

Определение полунеобратимых и необратимых преобразований для $N = 2$ дано в (3.12) и (3.13), а для $N = 4$ — в (3.170) и (3.171).

**Определение 3.45.** *Каждое из четырех полунеобратимых и необратимых $N = 4$ редуцированных сплетающих четность*[*)] *касательного пространства преобразований определяется четырьмя условиями*

1) $\mathsf{TPt}_1^-$:

$$Q\left(z, \theta_i^+, \theta_i^-\right) = \partial \tilde{z} - \partial \tilde{\theta}_i^+ \cdot \theta_i^- - \partial \tilde{\theta}_i^- \cdot \theta_i^+ = 0, \quad (3.186)$$

$$\varDelta_1^-\left(z, \theta_i^+, \theta_i^-\right) = D_1^- \tilde{z} - D_1^- \tilde{\theta}_i^- \cdot \tilde{\theta}_i^+ - D_1^- \tilde{\theta}_i^+ \cdot \tilde{\theta}_i^- = 0 \quad (3.187)$$

$$\varDelta_2^\pm\left(z, \theta_i^+, \theta_i^-\right) = D_2^\pm \tilde{z} - D_2^\pm \tilde{\theta}_i^- \cdot \tilde{\theta}_i^+ - D_2^\pm \tilde{\theta}_i^+ \cdot \tilde{\theta}_i^- = 0. \quad (3.188)$$

2) $\mathsf{TPt}_1^+$:

$$Q\left(z, \theta_i^+, \theta_i^-\right) = \partial \tilde{z} - \partial \tilde{\theta}_i^+ \cdot \theta_i^- - \partial \tilde{\theta}_i^- \cdot \theta_i^+ = 0, \quad (3.189)$$

$$\varDelta_1^+\left(z, \theta_i^+, \theta_i^-\right) = D_1^+ \tilde{z} - D_1^+ \tilde{\theta}_i^- \cdot \tilde{\theta}_i^+ - D_1^+ \tilde{\theta}_i^+ \cdot \tilde{\theta}_i^- = 0 \quad (3.190)$$

$$\varDelta_2^\pm\left(z, \theta_i^+, \theta_i^-\right) = D_2^\pm \tilde{z} - D_2^\pm \tilde{\theta}_i^- \cdot \tilde{\theta}_i^+ - D_2^\pm \tilde{\theta}_i^+ \cdot \tilde{\theta}_i^- = 0. \quad (3.191)$$

3) $\mathsf{TPt}_2^-$:

$$Q\left(z, \theta_i^+, \theta_i^-\right) = \partial \tilde{z} - \partial \tilde{\theta}_i^+ \cdot \theta_i^- - \partial \tilde{\theta}_i^- \cdot \theta_i^+ = 0, \quad (3.192)$$

$$\varDelta_2^-\left(z, \theta_i^+, \theta_i^-\right) = D_2^- \tilde{z} - D_2^- \tilde{\theta}_i^- \cdot \tilde{\theta}_i^+ - D_2^- \tilde{\theta}_i^+ \cdot \tilde{\theta}_i^- = 0 \quad (3.193)$$

$$\varDelta_1^\pm\left(z, \theta_i^+, \theta_i^-\right) = D_1^\pm \tilde{z} - D_1^\pm \tilde{\theta}_i^- \cdot \tilde{\theta}_i^+ - D_1^\pm \tilde{\theta}_i^+ \cdot \tilde{\theta}_i^- = 0. \quad (3.194)$$

4) $\mathrm{TPt}_2^+$:

$$Q\left(z, \theta_i^+, \theta_i^-\right) = \partial \tilde{z} - \partial \tilde{\theta}_i^+ \cdot \theta_i^- - \partial \tilde{\theta}_i^- \cdot \theta_i^+ = 0, \quad (3.195)$$

$$\varDelta_2^\pm\left(z, \theta_i^+, \theta_i^-\right) = D_2^\pm \tilde{z} - D_2^\pm \tilde{\theta}_i^- \cdot \tilde{\theta}_i^+ - D_2^\pm \tilde{\theta}_i^+ \cdot \tilde{\theta}_i^- = 0 \quad (3.196)$$

$$\varDelta_1^\pm\left(z, \theta_i^+, \theta_i^-\right) = D_1^\pm \tilde{z} - D_1^\pm \tilde{\theta}_i^- \cdot \tilde{\theta}_i^+ - D_1^\pm \tilde{\theta}_i^+ \cdot \tilde{\theta}_i^- = 0. \quad (3.197)$$

---

*Примечание.* Причина такого названия будет пояснена ниже (для $N = 2$ сплетающих четность преобразований см. **Пункт 3.1.6**).

Из (3.184)–(3.197) видно, что число уравнений во всех случаях одинаково и равно 4.

**Определение 3.46.** *Назовем условия* (3.184)–(3.185) SCf *условиями, условия* (3.186)–(3.187) — $\mathsf{TPt}_1^-$ *условиями,* (3.189)–(3.190) — $\mathsf{TPt}_1^+$ *условиями,* (3.192)–(3.193) — $\mathsf{TPt}_2^-$ *условиями и* (3.195)–(3.197) — $\mathsf{TPt}_2^+$ *условиями.*

Любой из этих индексов будет означать применение соответствующего условия к рассматриваемому объекту.

Найдем связь между функциями $Q\left(z,\theta_i^+,\theta_i^-\right)$ и $\varDelta_k^\pm\left(z,\theta_i^+,\theta_i^-\right)$. Для этого продифференцируем $\varDelta_k^\pm\left(z,\theta_i^+,\theta_i^-\right)$ и применим условия суперсимметрии (3.155), тогда получим

$$Q\left(z,\theta_i^+,\theta_i^-\right)-\frac{D_k^+\varDelta_k^-\left(z,\theta_i^+,\theta_i^-\right)+D_k^-\varDelta_k^+\left(z,\theta_i^+,\theta_i^-\right)}{4}=$$

$$\frac{\operatorname{per}\mathrm{H}_{11}+\operatorname{per}\mathrm{H}_{12}+\operatorname{per}\mathrm{H}_{21}+\operatorname{per}\mathrm{H}_{22}}{2}, \tag{3.198}$$

(ср. (3.22)).

Исходя из условий редукции (3.184)–(3.197), определим 5 редуцированных суперматриц $(1|4)$ касательного пространства

$$\mathrm{P}_{SCf}^{(N=4)}=\begin{pmatrix} Q_{SCf}\left(z,\theta_i^+,\theta_i^-\right) & \partial\tilde{\theta}_{1(SCf)}^+ & \partial\tilde{\theta}_{1(SCf)}^- & \partial\tilde{\theta}_{2(SCf)}^+ & \partial\tilde{\theta}_{2(SCf)}^- \\ 0 & & & & \\ 0 & & & & \\ 0 & & \mathrm{H}_{SCf} & & \\ 0 & & & & \end{pmatrix}, \tag{3.199}$$

$$\mathrm{P}_{TPt_1^+}^{(N=4)}=$$

$$\begin{pmatrix} 0 & \partial\tilde{\theta}^+_{1\left(TPt^+_1\right)} & \partial\tilde{\theta}^-_{1\left(TPt^+_1\right)} & \partial\tilde{\theta}^+_{2\left(TPt^+_1\right)} & \partial\tilde{\theta}^-_{2\left(TPt^+_1\right)} \\ \Delta^-_{1\left(TPt^+_1\right)}\left(z,\theta^+_i,\theta^-_i\right) & & & & \\ 0 & & & & \\ 0 & & \mathrm{H}_{TPt^+_1} & & \\ 0 & & & & \end{pmatrix}. \tag{3.200}$$

И аналогично для остальных трех редукций (3.175)–(3.177).

Если ввести матрицы

$$\mathrm{Q}_i = \begin{pmatrix} \partial\tilde{\theta}^+_i & \partial\tilde{\theta}^-_i \\ \tilde{\theta}^+_i & \tilde{\theta}^-_i \end{pmatrix}, \tag{3.201}$$

состоящие из нечетных элементов, и горизонтальные полуматрицы $\mathcal{D}^\pm_{ij}$ (см. **Приложение Д.2**)

$$\mathcal{D}^\pm_{ij} = \begin{pmatrix} D^\pm_i\tilde{\theta}^+_j & D^\pm_i\tilde{\theta}^-_j \\ \tilde{\theta}^+_j & \tilde{\theta}^-_j \end{pmatrix} \tag{3.202}$$

то

$$Q\left(z,\theta^+_i,\theta^-_i\right) = \partial\tilde{z} - \operatorname{per}\mathrm{Q}_1 - \operatorname{per}\mathrm{Q}_2, \tag{3.203}$$

$$\Delta^\pm_k\left(z,\theta^+_i,\theta^-_i\right) = D^\pm_k\tilde{z} - \pi\mathrm{er}\mathcal{D}^\pm_{k1} - \pi\mathrm{er}\mathcal{D}^\pm_{k2}, \tag{3.204}$$

(ср. (3.29)–(3.30)).

Тогда условия редукции (3.184)–(3.197) можно записать через перманенты и полуперманенты

$$D^\pm_k\tilde{z} = \pi\mathrm{er}\mathcal{D}^\pm_{k1} + \pi\mathrm{er}\mathcal{D}^\pm_{k2},\ k = 1,2\ \ (\mathsf{SCf}) \tag{3.205}$$

$$\begin{cases} \partial\tilde{z} = \operatorname{per} \mathrm{Q}_1 + \operatorname{per} \mathrm{Q}_2,\ D_1^- \tilde{z} = \pi\mathrm{er}\mathcal{D}_{11}^- + \pi\mathrm{er}\mathcal{D}_{12}^-, \\ D_2^\pm \tilde{z} = \pi\mathrm{er}\mathcal{D}_{21}^\pm + \pi\mathrm{er}\mathcal{D}_{22}^\pm, \qquad (\mathsf{TPt}_1^-) \end{cases} \tag{3.206}$$

$$\begin{cases} \partial\tilde{z} = \operatorname{per} \mathrm{Q}_1 + \operatorname{per} \mathrm{Q}_2,\ D_1^+ \tilde{z} = \pi\mathrm{er}\mathcal{D}_{11}^+ + \pi\mathrm{er}\mathcal{D}_{12}^+, \\ D_2^\pm \tilde{z} = \pi\mathrm{er}\mathcal{D}_{21}^\pm + \pi\mathrm{er}\mathcal{D}_{22}^\pm, \qquad (\mathsf{TPt}_1^+) \end{cases} \tag{3.207}$$

$$\begin{cases} \partial\tilde{z} = \operatorname{per} \mathrm{Q}_1 + \operatorname{per} \mathrm{Q}_2,\ D_2^- \tilde{z} = \pi\mathrm{er}\mathcal{D}_{21}^- + \pi\mathrm{er}\mathcal{D}_{22}^-, \\ D_1^\pm \tilde{z} = \pi\mathrm{er}\mathcal{D}_{11}^\pm + \pi\mathrm{er}\mathcal{D}_{12}^\pm, \qquad (\mathsf{TPt}_2^-) \end{cases} \tag{3.208}$$

$$\begin{cases} \partial\tilde{z} = \operatorname{per} \mathrm{Q}_1 + \operatorname{per} \mathrm{Q}_2,\ D_2^+ \tilde{z} = \pi\mathrm{er}\mathcal{D}_{21}^+ + \pi\mathrm{er}\mathcal{D}_{22}^+, \\ D_1^\pm \tilde{z} = \pi\mathrm{er}\mathcal{D}_{11}^\pm + \pi\mathrm{er}\mathcal{D}_{12}^\pm. \qquad (\mathsf{TPt}_2^+) \end{cases} \tag{3.209}$$

**Утверждение 3.47.** *Число редукций* $(1|N)$ *мерного касательного суперпространства равно* $N+1$, *среди которых лишь одна* SCf *редукция может быть обратимой, остальные* $N$ *являются необратимыми и полунеобратимыми, а число уравнений, определяющих редукции, равно* $N$ *в каждом случае.*

Следует также ожидать, что по аналогии с $N=1$ **Предположением 2.26** для произвольных $N$ имеет место

**Предположение 3.48.** *Среди редуцированных* $N$ *преобразований существует один четный* (SCf) *суперaналог голоморфных преобразований* (*среди которых могут быть обратимые*) *и* $N$ *нечетных необратимых и полунеобратимых* (TPt) *суперaналогов антиголоморфных преобразований.*

Березинианы обратимых и необратимых $N=4$ редуцированных преобразований получены в **Приложении E.4**.

**3.2.2.** Классификация $N=4$ SCf преобразований. Рассмотрим редуцированные $N=4$ преобразования, определяемые SCf условиями (3.184)–(3.185).

Запишем формулу (3.198) с учетом SCf условий $\varDelta_k^\pm\left(z,\theta_i^+,\theta_i^-\right)=0$

в виде

$$Q_{SCf}\left(z,\theta_i^+,\theta_i^-\right)=\frac{\operatorname{per}\mathrm{H}_{11}^{SCf}+\operatorname{per}\mathrm{H}_{12}^{SCf}+\operatorname{per}\mathrm{H}_{21}^{SCf}+\operatorname{per}\mathrm{H}_{22}^{SCf}}{2}, \qquad (3.210)$$

где матрицы $\mathrm{H}_{ij}$ определены в (3.165).

Далее, из (3.199) следует

$$\begin{pmatrix} D_1^- \\ D_1^+ \\ D_2^- \\ D_2^+ \end{pmatrix} = \mathrm{H}_{SCf}\cdot \begin{pmatrix} \tilde{D}_1^- \\ \tilde{D}_1^+ \\ \tilde{D}_2^- \\ \tilde{D}_2^+ \end{pmatrix}, \qquad (3.211)$$

Данная формула свидетельствует о том, что нечетные суперпроизводные $D_i^\pm$ образуют $(0|4)$ мерное подпространство в $(1|4)$ мерном касательном пространстве, т.е. $D_i^\pm$ преобразуются друг через друга

$$D_i^\pm = D_i^\pm\tilde{\theta}_j^-\cdot\tilde{D}_j^+ + D_i^\pm\tilde{\theta}_j^+\cdot\tilde{D}_j^-. \qquad (3.212)$$

Применим к SCf условиям (3.184)–(3.185) операторы суперпроизводной той же киральности $D_i^\pm\varDelta_{(SCf)k}^\pm\left(z,\theta_i^+,\theta_i^-\right)=0$ и воспользуемся нильпотентностью $D_i^\pm$, тогда получим [2]

$$\operatorname{scf}_\pm\mathrm{H}_{11}^{(SCf)T}+\operatorname{scf}_\pm\mathrm{H}_{21}^{(SCf)T} = 0, \qquad (3.213)$$

$$\operatorname{scf}_\pm\mathrm{H}_{12}^{(SCf)T}+\operatorname{scf}_\pm\mathrm{H}_{22}^{(SCf)T} = 0, \qquad (3.214)$$

$$\mathrm{H}_{12}^{(SCf)}\cdot\mathrm{H}_{11}^{(SCf)MT}+\mathrm{H}_{22}^{(SCf)}\cdot\mathrm{H}_{21}^{(SCf)MT} = 0, \qquad (3.215)$$

где $\mathrm{H}_{ij}^{(SCf)MT}$ обозначает транспонированную матрицу миноров и

$$\operatorname{scf}_\pm\mathrm{H}_{ij}^{(SCf)T}=D_j^\pm\tilde{\theta}_{i(SCf)}^+\cdot D_j^\pm\tilde{\theta}_{i(SCf)}^-. \qquad (3.216)$$

Кроме того, из SCf условий следует

$$\operatorname{per} \mathrm{H}_{11}^{SCf} + \operatorname{per} \mathrm{H}_{12}^{SCf} = \operatorname{per} \mathrm{H}_{21}^{SCf} + \operatorname{per} \mathrm{H}_{22}^{SCf}, \tag{3.217}$$

поэтому вместо (3.210) имеем

$$Q_{SCf}\left(z, \theta_i^+, \theta_i^-\right) = \operatorname{per} \mathrm{H}_{11}^{SCf} + \operatorname{per} \mathrm{H}_{12}^{SCf} = \operatorname{per} \mathrm{H}_{21}^{SCf} + \operatorname{per} \mathrm{H}_{22}^{SCf}. \tag{3.218}$$

*Замечание* **3.49.** Уравнения (3.213)–(3.217) совпадают с условиями того, что матрица $\mathrm{H}_{SCf}$ после нормировки на $\sqrt{\operatorname{per} \mathrm{H}_{11}^{SCf} + \operatorname{per} \mathrm{H}_{12}^{SCf}}$ (при условии $\epsilon\left[\operatorname{per} \mathrm{H}_{11}^{SCf} + \operatorname{per} \mathrm{H}_{12}^{SCf}\right] \neq 0$) подобна ортогональной матрице $O_{\Lambda_0}(4)$ (см. **Подраздел 5.1**).

Для березиниана $N = 4$ SCf преобразований (при $\epsilon\left[\det \mathrm{H}_{SCf}\right] \neq 0$) получаем [2]

$$\operatorname{Ber}_{SCf}^{N=4}\left(\tilde{Z}/Z\right) = \frac{\operatorname{per} \mathrm{H}_{11}^{SCf} + \operatorname{per} \mathrm{H}_{12}^{SCf}}{\det \mathrm{H}_{SCf}} = \frac{\operatorname{per} \mathrm{H}_{21}^{SCf} + \operatorname{per} \mathrm{H}_{22}^{SCf}}{\det \mathrm{H}_{SCf}}. \tag{3.219}$$

(ср. $N = 2$ (3.74)).

Из формул (3.213)–(3.215) следует , что матрица $\mathrm{H}_{SCf}$ является $N = 4$ scf-матрицей согласно **Определению 5.20**, т. е. $\mathrm{G} \in SCF_{\Lambda_0}(4)$ (см. **Пункт 5.1**). Поэтому детерминант $\mathrm{H}_{SCf}$ выражается через перманенты ее блоков (см. общую формулу (5.52))

$$\det \mathrm{H}_{SCf} = k\left(\operatorname{per} \mathrm{H}_{11}^{SCf} + \operatorname{per} \mathrm{H}_{12}^{SCf}\right)^2 = k\left(\operatorname{per} \mathrm{H}_{21}^{SCf} + \operatorname{per} \mathrm{H}_{22}^{SCf}\right)^2. \tag{3.220}$$

Тогда для березиниана $N = 4$ SCf преобразований окончательно

получаем

$$\operatorname{Ber}{}^{N=4}_{SCf}\left(\tilde{Z}/Z\right)=\frac{k}{\operatorname{per}\mathrm{H}^{SCf}_{11}+\operatorname{per}\mathrm{H}^{SCf}_{12}}=\frac{k}{\operatorname{per}\mathrm{H}^{SCf}_{21}+\operatorname{per}\mathrm{H}^{SCf}_{22}}, \qquad (3.221)$$

где $k=\pm 1$ (ср. $N=2$ (3.76)) и здесь подразумевается выполненными условия обратимости

$$\epsilon\left[\operatorname{per}\mathrm{H}^{SCf}_{11}+\operatorname{per}\mathrm{H}^{SCf}_{12}\right]\neq 0,\ \ \epsilon\left[\operatorname{per}\mathrm{H}^{SCf}_{21}+\operatorname{per}\mathrm{H}^{SCf}_{22}\right]\neq 0. \qquad (3.222)$$

В этом случае между матричными функциями блоков $\mathrm{H}^{SCf}_{ij}$ имеются соотношения (ср. (3.74))

$$\frac{\operatorname{per}\mathrm{H}^{SCf}_{11}}{\operatorname{per}\mathrm{H}^{SCf}_{21}}=\frac{\operatorname{per}\mathrm{H}^{SCf}_{22}}{\operatorname{per}\mathrm{H}^{SCf}_{12}}, \qquad (3.223)$$

$$\frac{\operatorname{per}\mathrm{H}^{SCf}_{ij}}{\det\mathrm{H}^{SCf}_{ij}}=\frac{\det\mathrm{H}^{SCf}_{ij}}{\operatorname{per}\mathrm{H}^{SCf}_{ij}}, \qquad (3.224)$$

$$\det\mathrm{H}^{SCf}_{11} = k\det\mathrm{H}^{SCf}_{22}, \qquad (3.225)$$

$$\det\mathrm{H}^{SCf}_{12} = k\det\mathrm{H}^{SCf}_{21}. \qquad (3.226)$$

Полезно также выразить березиниан (3.221) через детерминант матрицы $\mathrm{H}_{SCf}$ по формуле (при $\epsilon\left[\det\mathrm{H}_{SCf}\right]\neq 0$)

$$\operatorname{Ber}{}^{N=4}_{SCf}\left(\tilde{Z}/Z\right)=\frac{k}{\sqrt{\det\mathrm{H}_{SCf}}}. \qquad (3.227)$$

Отсюда следует

**Утверждение 3.50.** *Общее выражение для березиниана* SCf *преобра-*

*зований при произвольных $N$ имеет следующий вид*

$$\mathrm{Ber}_{SCf}\left(\tilde{Z}/Z\right) = k\left(\det \mathrm{H}_{SCf}\right)^{\frac{2-N}{N}}. \tag{3.228}$$

Как и в $N = 2$ (3.76), величина $k$ отличает между собой подгруппу $SO_{\Lambda_0}(4)$ ($\cong U_{\Lambda_0}(2)$) преобразований касательного пространства ($k = +1$) и общую $O_{\Lambda_0}(4)$ группу ($k = -1$) (в обратимом случае) [565, 668].

*Замечание* **3.51.** Из-за соотношений (3.220)–(3.226) при $N = 4$ не имеется полунеобратимых SCf преобразований (3.170).

Таким образом, мы приходим к следующей классификации $N = 4$ преобразований, удовлетворяющих SCf условиям (3.184)–(3.185):

**1.** Обратимые $N = 4$ суперконформные преобразования, удовлетворяющие условиям обратимости (3.222).

   **а)** $SO_{\Lambda_0}(4)$ преобразования с $k = +1$;

   **б)** $O_{\Lambda_0}(4)$ преобразования с $k = -1$.

**2.** Необратимые $N = 4$ SCf преобразования, удовлетворяющие соотношению $\epsilon\left[\mathrm{per}\,\mathrm{H}_{i1}^{SCf} + \mathrm{per}\,\mathrm{H}_{i2}^{SCf}\right] = 0$.

Относительно $N = 4$ SCf преобразований дифференциал $dZ$ преобразуется однородно, как это следует из (3.159), (3.199) и (3.218)

$$d\tilde{Z} = dZ \cdot \left(\mathrm{per}\,\mathrm{H}_{11}^{SCf} + \mathrm{per}\,\mathrm{H}_{12}^{SCf}\right) \tag{3.229}$$

(ср. $N = 2$ (3.65)). Используя (3.221), получаем

$$d\tilde{Z} = \frac{k\,dZ}{\mathrm{Ber}_{SCf}^{N=4}\left(\tilde{Z}/Z\right)}, \tag{3.230}$$

откуда следует определение $N$-SCf преобразований через (обратимый)

березиниан [2].

**Определение 3.52.** *При* $N \neq 2$ *общие* $N$-SCf *преобразования определяются ковариантным преобразованием дифференциала* $dZ$ *с конформным множителем, выражающимся через березиниан преобразований*

$$d\tilde{Z} = dZ \cdot \left[k\mathrm{Ber}_{SCf}\left(\tilde{Z}/Z\right)\right]^{\frac{2}{2-N}}. \tag{3.231}$$

*Замечание* **3.53.** Соотношение (3.231) является обобщением на $N$-SCf преобразования соотношения $d\tilde{z} = (\partial\tilde{z}/\partial z)\, dz$ (ср. [565, 625]).

Введем в рассмотрение $N = 4$ супердифференциалы $d\tau^{\pm}_{i(SCf)}$, преобразующиеся дуально к суперпроизводным $D^{\pm}_i$ (3.211), как

$$\begin{pmatrix} d\tilde{\tau}^+_1 & d\tilde{\tau}^-_1 & d\tilde{\tau}^+_2 & d\tilde{\tau}^-_2 \end{pmatrix} = \begin{pmatrix} d\tau^+_1 & d\tau^-_1 & d\tau^+_2 & d\tau^-_2 \end{pmatrix} \cdot \mathrm{H}_{SCf}. \tag{3.232}$$

Тогда по аналогии с $N = 2$ имеем

**Определение 3.54.** *Внешний* *N = 4 SCf супердифференциал* $\delta^{(N=4)}_{SCf}$ *определяется формулой*

$$\delta^{(N=4)}_{SCf} = d\tau^+_{i(SCf)} \cdot D^-_i + d\tau^-_{i(SCf)} \cdot D^+_i. \tag{3.233}$$

**Утверждение 3.55.** *Внешний дифференциал* $(0|4)$ *мерного подпространства инвариантен относительно* $N = 4$ SCf *преобразований.*

*Доказательство.* Совпадает с (3.69). ■

Введенные $N = 4$ супердифференциалы $d\tau^{\pm}_{i(SCf)}$ удовлетворяют дуальным по отношению к (3.155) соотношениям

$$\left\{d\tau^+_{i(SCf)}, d\tau^-_{j(SCf)}\right\} = 2\delta_{ij} dZ \tag{3.234}$$

и используются для построения действия фермионной струны [320], изучения линейных расслоений и линейных интегралов на суперримановых поверхностях [563].

**3.2.3.** Компонентное представление $N=4$ редуцированных преобразований. Рассмотрим произвольное $N = 4$ суперааналитическое отображение $Z\left(z, \theta_i^+, \theta_i^-\right) \to \tilde{Z}\left(\tilde{z}, \tilde{\theta}_i^+, \tilde{\theta}_i^-\right)$ суперпространства $\mathbb{C}^{1|4} \to \mathbb{C}^{1|4}$.

Раскладывая в ряд по нечетным координатам (как (3.5)), используя их нильпотентность, получаем общий вид $N = 4$ супераналитического преобразования

$$\begin{cases} \tilde{z} = f(z) + \theta_i^+ \chi_i^-(z) + \theta_i^- \chi_i^+(z) + \theta_i^+ \theta_j^- h_{ij}(z) + \theta_i^+ \theta_{3-i}^+ s_i^-(z) + \\ \theta_i^- \theta_{3-i}^- s_i^+(z) + \theta_i^+ \theta_i^- \left(\theta_{3-i}^+ \rho_{3-i}^-(z) + \theta_{3-i}^- \rho_{3-i}^+(z)\right) + \theta_1^+ \theta_1^- \theta_2^+ \theta_2^- v(z), \\ \tilde{\theta}_i^\pm = \psi_i^\pm(z) + \theta_j^\pm g_{ij}^{\pm\mp}(z) + \theta_j^\mp g_{ij}^{\pm\pm}(z) + \theta_i^\pm \theta_j^\mp \lambda_{ij}^\pm(z) + \theta_i^\pm \theta_j^\pm \sigma_{ij}^{\mp\pm}(z) + \\ \theta_i^\mp \theta_j^\mp \sigma_{ij}^{\pm\pm}(z) + \theta_i^\pm \theta_i^\mp \theta_{3-i}^\pm t_i^{\pm\mp}(z) + \theta_i^\pm \theta_i^\mp \theta_{3-i}^\mp t_i^{\pm\pm}(z) + \theta_{3-i}^\pm \theta_{3-i}^\mp \theta_i^\pm u_i^{\pm\mp}(z) \\ + \theta_{3-i}^\pm \theta_{3-i}^\mp \theta_i^\mp u_i^{\pm\pm}(z) + \theta_i^\pm \theta_i^\mp \theta_{3-i}^\pm \theta_{3-i}^\mp \mu^\pm(z), \end{cases} \tag{3.235}$$

где $i = 1, 2$ и по повторяющимся индексам подразумевается суммирование.

Отсюда следует, что $N = 4$ супераналитическое преобразование определяется 80 функциями на $\mathbb{C}^{1|0}$:

42 четных $f$, $h_{ij}$, $s_i^a$, $v$, $g_{ij}^{ab}$, $t_{ij}^{ab}$, $u_{ij}^{ab} : \mathbb{C}^{1|0} \to \mathbb{C}^{1|0}$ и

38 нечетных $\psi_i^a$, $\chi_i^a$, $\lambda_i^a$, $\rho_i^a$, $\sigma_{ij}^{ab}$, $\mu^a : \mathbb{C}^{1|0} \to \mathbb{C}^{0|1}$, где $a, b = \pm$.

**Определение 3.56.** *Множество обратимых и необратимых преобразований* $\mathbb{C}^{1|4} \to \mathbb{C}^{1|4}$ (3.235) *образует полугруппу относительно композиции преобразований, которую мы назовем полугруппой* $N = 4$ *супераналитических преобразований* $\boldsymbol{T}_{SA}^{(N=4)}$.

*Замечание* **3.57.** Обратимые преобразования, очевидно, образуют под-группу $\boldsymbol{G}_{SA}^{(N=4)}$ полугруппы $\boldsymbol{T}_{SA}^{(N=4)}$.

**Определение 3.58.** *Необратимые преобразования* $\mathbb{C}^{1|4} \to \mathbb{C}^{1|4}$ (3.235) *входят в идеал* $\boldsymbol{I}_{SA}^{(N=4)}$ *полугруппы* $\boldsymbol{T}_{SA}^{(N=4)}$.

Очевидно, что при $N = 4$ также имеет место

**Утверждение 3.59.** *Обратимость* $N = 4$ *супераналитического преобразования будет определяться только функциями* $f(z)$ *и* $g_{ij}^{ab}(z)$.

Компонентные функции, входящие в (3.235), могут быть использованы для параметризации $N = 4$ супераналитической полугруппы, элемент которой $\mathbf{s}$ есть функциональная матрица, аналогичная (3.18), но с 80 элементами. Поскольку действие $\mathbf{s}_1 * \mathbf{s}_2 = \mathbf{s}_3$ снова (как и в $N = 2$ случае (3.20)) определяется композицией $N = 4$ преобразований, ассоциативность умножения $N = 4$ функциональных матриц выполняется (см. **Замечание 3.10**).

Понятно, что рассматривать и решать условия редукции (3.205)–(3.209) как уравнения для 80 компонентных функций из (3.235) не представляется обозримым в общем виде. Однако всегда есть возможность исследовать частные случаи, что мы и сделаем в последующих пунктах. Так, обратимые и необратимые расщепленные $N = 4$ преобразования рассматриваются в **Приложении Ж.4**.

**3.2.4.** Киральные нерасщепленные преобразования. В общем случае нерасщепленных $N = 4$ преобразований (3.235) решить систему уравнений (3.213)–(3.215) относительно 80 компонентных функций, входящих в (3.235), не представляется возможным без дополнительных ограничений.

Наиболее естественными и необходимыми в приложениях являются киральные $N = 4$ SCf преобразования [2, 668], определяемые тем,

что суперпроизводные $D_i^{\pm}$ не меняют киральность, т. е.

$$D_1^{\pm} = D_1^{\pm}\tilde{\theta}_1^{\mp} \cdot \tilde{D}_1^{\pm} + D_1^{\pm}\tilde{\theta}_2^{\mp} \cdot \tilde{D}_2^{\pm}, \quad (3.236)$$

$$D_2^{\pm} = D_2^{\pm}\tilde{\theta}_1^{\mp} \cdot \tilde{D}_1^{\pm} + D_2^{\pm}\tilde{\theta}_2^{\mp} \cdot \tilde{D}_2^{\pm}. \quad (3.237)$$

Это приводит к условиям на преобразования

$$D_i^{\pm}\tilde{\theta}_j^{\pm} = 0. \quad (3.238)$$

В нашем формализме условия (3.238) соответствуют тому, что все матрицы $\mathrm{H}_{ij}^{SCf}$ являются $N = 2$ scf-матрицами и в случае $\epsilon\left[\mathrm{per}\,\mathrm{H}_{ij}^{SCf}\right] \neq 0$ принадлежат группе $GSCF\,(2, \Lambda_0)$ (см. **Пункт 5.1**), а матрица $\mathrm{H}_{SCf}$ (3.165) имеет вид

$$\mathrm{H}_{SCf}^{chiral} = \begin{pmatrix} D_1^{-}\tilde{\theta}_1^{+} & 0 & D_2^{-}\tilde{\theta}_1^{+} & 0 \\ 0 & D_1^{+}\tilde{\theta}_1^{-} & 0 & D_2^{+}\tilde{\theta}_1^{-} \\ D_1^{-}\tilde{\theta}_2^{+} & 0 & D_2^{-}\tilde{\theta}_2^{+} & 0 \\ 0 & D_1^{+}\tilde{\theta}_2^{-} & 0 & D_2^{+}\tilde{\theta}_2^{-} \end{pmatrix}. \quad (3.239)$$

В инфинитезимальном виде такие преобразования используются для описания $SU(2)$ расширенных суперконформных алгебр [668].

В наиболее общем случае выберем в качестве условий киральности следующие [2]

$$D_i^{n}\tilde{\theta}_j^{m_{ij}} = 0, \quad (3.240)$$

где $n, m_{ij} = \pm$ и по $i, j$ нет суммирования*).

Тогда закон преобразования суперпроизводных запишется в виде

---

*Примечание.* Только в этой формуле.

$$D_i^n = D_i^n \tilde{\theta}_j^{-m_{ij}} \cdot \tilde{D}_j^{m_{ij}}. \quad (3.241)$$

Решением условий (3.240) является следующая параметризация нечетного сектора 16 четными и 8 нечетными функциями [2]

$$\begin{aligned} \tilde{\theta}_1^n &= \psi_1^n \left(Z^{m_{11},m_{12}}\right) + \theta_1^{m_{11}} g_{11}^{n,-m_{11}} \left(Z^{m_{11},m_{12}}\right) + \\ &\theta_2^{m_{12}} g_{12}^{n,-m_{12}} \left(Z^{m_{11},m_{12}}\right) + \theta_1^{m_{11}} \theta_2^{m_{12}} \lambda_1^n \left(Z^{m_{11},m_{12}}\right), \quad (3.242) \\ \tilde{\theta}_2^n &= \psi_2^n \left(Z^{m_{21},m_{22}}\right) + \theta_1^{m_{21}} g_{21}^{n,-m_{21}} \left(Z^{m_{21},m_{22}}\right) + \\ &\theta_2^{m_{22}} g_{22}^{n,-m_{22}} \left(Z^{m_{21},m_{22}}\right) + \theta_1^{m_{21}} \theta_2^{m_{22}} \lambda_2^n \left(Z^{m_{21},m_{22}}\right), \quad (3.243) \end{aligned}$$

где $Z^{a,b} = z + \theta_1^a \theta_1^{-a} + \theta_2^b \theta_2^{-b}$, $a, b = \pm$.

Здесь при $m_{ij} = n$ киральность суперпроизводных сохраняется, и мы получаем предыдущий случай (3.236)–(3.239). Для простоты ограничимся в дальнейшем рассмотрением лишь киральных обратимых и необратимых конечных $N = 4$ SCf преобразований, поскольку остальные варианты формально отличаются только соответствующей перестановкой индексов.

**Утверждение 3.60.** *Киральные $N = 4$ SCf преобразования образуют подполугруппу $N = 4$ SCf полугруппы (см.* **Определение 3.56***).*

*Доказательство.* Следует из поведения $N = 4$ суперпроизводных (3.241) при композиции двух SCf преобразований, когда киральность сохраняется $m_{ij} = n$. ■

Отметим также, что различные подполугруппы образуют также преобразования с матрицей $\mathrm{H}_{SCf}$, у которой число антидиагональных блочных матриц является четным (см. (Ж.60)).

Применение SCf условий (3.184)–(3.185) к параметризации (3.242)–(3.243) приводит к тому, что, как и в расщепленном случае, матрица G (Ж.47)–(Ж.49) становится $N = 4$ scf-матрицей, т. е. $\mathrm{G} \in SCF(4, \Lambda_0)$ (см. **Подраздел 5.1**). Следовательно, для нее выполняются условия (Ж.44)–(Ж.45). Однако, дифференциальные условия на элементы матрицы G отличаются от расщепленного варианта (Ж.46) и имеют следующий вид

$$\begin{aligned} \mathrm{G}_{11}^{T\prime}\cdot\mathrm{G}_{11}^{M}+\mathrm{G}_{12}^{T\prime}\cdot\mathrm{G}_{12}^{M}+\mathrm{G}_{21}^{T\prime}\cdot\mathrm{G}_{21}^{M}+\mathrm{G}_{22}^{T\prime}\cdot\mathrm{G}_{22}^{M} &= \\ \mathrm{G}_{11}^{T}\cdot\mathrm{G}_{11}^{M\prime}+\mathrm{G}_{12}^{T}\cdot\mathrm{G}_{12}^{M\prime}+\mathrm{G}_{21}^{T}\cdot\mathrm{G}_{21}^{M\prime}+\mathrm{G}_{22}^{T}\cdot\mathrm{G}_{22}^{M\prime}, & \qquad (3.244) \\ \mathrm{G}_{12}^{T\prime}\cdot\mathrm{G}_{11}^{M\prime}+\mathrm{G}_{22}^{T\prime}\cdot\mathrm{G}_{21}^{M\prime} &= 0. \qquad (3.245) \end{aligned}$$

Связь четных и нечетных функций для киральных $N=4$ преобразований (3.238)–(3.239) определяется формулами

$$\begin{aligned} g_{11}^{\pm\mp}(z)\,\lambda_1^{\mp}(z)+g_{21}^{\pm\mp}(z)\,\lambda_2^{\mp}(z) &= \qquad (3.246) \\ & 2g_{12}^{\mp\pm}(z)\,\psi_1^{\pm\prime}(z)+2g_{22}^{\mp\pm}(z)\,\psi_2^{\pm\prime}(z), \\ g_{12}^{\pm\mp}(z)\,\lambda_1^{\mp}(z)+g_{22}^{\pm\mp}(z)\,\lambda_2^{\mp}(z) &= \qquad (3.247) \\ & 2g_{11}^{\mp\pm}(z)\,\psi_1^{\pm\prime}(z)+2g_{21}^{\mp\pm}(z)\,\psi_2^{\pm\prime}(z). \end{aligned}$$

Тогда из (3.244)–(3.247) для нечетных функций, входящих в (3.242)–(3.243), имеем

$$4\left[\psi_1^{+\prime}(z)\,\psi_1^{-\prime}(z)+\psi_2^{+\prime}(z)\,\psi_2^{-\prime}(z)\right]+\lambda_1^{+}(z)\,\lambda_1^{-}(z)+\lambda_2^{+}(z)\,\lambda_2^{-}(z)=0. \qquad (3.248)$$

Одним из возможных решений (3.248) является

$$\lambda_i^{\pm}(z)=2\psi_{3-i}^{\mp\prime}(z). \qquad (3.249)$$

Таким образом, разрешая остальные SCf условия, получаем общий вид конечных киральных $N=4$ SCf преобразований [2]

$$\begin{aligned}
\tilde{z} \;=\; & f(z)+\hat{\theta}_1^{+}\left[\psi_1^{-}\left(Z^{+}\right)u_1^{+}\left(Z^{+}\right)-\psi_2^{-}\left(Z^{+}\right)u_2^{-}\left(Z^{+}\right)\right]+ \\
& \hat{\theta}_1^{-}\left[\psi_1^{+}\left(Z^{-}\right)u_1^{-}\left(Z^{-}\right)-\psi_2^{+}\left(Z^{-}\right)u_2^{+}\left(Z^{-}\right)\right]+ \\
& \hat{\theta}_2^{+}\left[\psi_2^{-}\left(Z^{+}\right)u_1^{-}\left(Z^{+}\right)+\psi_1^{-}\left(Z^{+}\right)u_2^{+}\left(Z^{+}\right)\right]+ \\
& \hat{\theta}_2^{-}\left[\psi_2^{+}\left(Z^{-}\right)u_1^{+}\left(Z^{-}\right)+\psi_2^{+}\left(Z^{-}\right)u_2^{-}\left(Z^{-}\right)\right]+ \\
& \left(\hat{\theta}_1^{+}\hat{\theta}_1^{-}+\hat{\theta}_2^{+}\hat{\theta}_2^{-}\right)\left[\psi_1^{+}(z)\,\psi_1^{-}(z)+\psi_2^{+}(z)\,\psi_2^{-}(z)\right]'- \\
& 2\hat{\theta}_1^{+}\hat{\theta}_2^{+}\left[\psi_1^{-}(z)\,\psi_2^{-}(z)\right]'-2\hat{\theta}_1^{-}\hat{\theta}_2^{-}\left[\psi_1^{+}(z)\,\psi_2^{+}(z)\right]'+\hat{\theta}_1^{+}\hat{\theta}_1^{-}\hat{\theta}_2^{+}\hat{\theta}_2^{-}f''(z)\,,
\end{aligned} \tag{3.250}$$

$$\tilde{\theta}_1^{\pm} = \psi_1^{\pm}\left(Z^{\pm}\right)+\hat{\theta}_1^{\pm}u_1^{\pm}\left(Z^{\pm}\right)+\hat{\theta}_2^{\pm}u_2^{\pm}\left(Z^{\pm}\right)+2\hat{\theta}_1^{\pm}\hat{\theta}_2^{\pm}\psi_2^{\mp\prime}\left(Z^{\pm}\right), \tag{3.251}$$

$$\tilde{\theta}_2^{\pm} = \psi_2^{\pm}\left(Z^{\pm}\right)-\hat{\theta}_1^{\pm}u_2^{\mp}\left(Z^{\pm}\right)+\hat{\theta}_2^{\pm}u_1^{\mp}\left(Z^{\pm}\right)+2\hat{\theta}_2^{\pm}\hat{\theta}_1^{\pm}\psi_1^{\mp\prime}\left(Z^{\pm}\right), \tag{3.252}$$

где $Z^{\pm}=Z^{\pm\pm}$, $\hat{\theta}_i^{\pm}$ — определены в (Ж.67), и

$$f'(z)=u_1^{+}(z)\,u_1^{-}(z)+u_2^{+}(z)\,u_2^{-}(z)-\psi_i^{+}(z)\,\psi_i^{-\prime}(z)-\psi_i^{-}(z)\,\psi_i^{+\prime}(z)\,. \tag{3.253}$$

Легко видеть, что при занулении нечетных компонентных функций эти преобразования соответствуют расщепленным преобразованиям (Ж.69)–(Ж.72). За счет появления в правой части $\hat{\theta}_i^{\pm}$ они также имеют глобальную $SU_{global}(2,\Lambda_0)$ симметрию. Локальные $SU_{local}(2,\Lambda_0)$ вращения можно рассмотреть аналогично (Ж.73)–(Ж.78).

Представление киральной $N=4$ суперконформной полугруппы функциональными матрицами можно получить сужением представления $N=4$ супераналитической полугруппы, содержащего 80 функций из (3.235), на представление, содержащее только 8 функций (4 четных и 4 нечетных), входящих в (3.250)–(3.252). Таким образом, получаем

**Определение 3.61.** *Элемент* $\mathbf{s}$ *киральной* $N=4$ *суперконформной*

*полугруппы* $\mathbf{S}^{(N=4)}_{SCf(chiral)}$ *параметризуется функциональной матрицей*

$$\left\{ \begin{array}{cccc} \psi_1^+(z) & \psi_2^+(z) & u_1^+(z) & u_2^+(z) \\ \psi_1^-(z) & \psi_2^-(z) & u_1^-(z) & u_2^-(z) \end{array} \right\} \stackrel{def}{=} \mathbf{s}_{chiral} \in \mathbf{S}^{(N=4)}_{SCf(chiral)}, \tag{3.254}$$

*а действие* $\mathbf{s}^{(1)}_{chiral} *_{ch} \mathbf{s}^{(2)}_{chiral} = \mathbf{s}^{(3)}_{chiral}$ *определяется композицией расщепленных преобразований* $Z \to \tilde{Z} \to \tilde{\tilde{Z}}$ *(как при* $N = 2$ (Ж.28)).

*Замечание* **3.62.** Ассоциативность действия $*_{ch}$ следует из ассоциативности композиции киральных $N = 4$ преобразований.

Необратимые преобразования соответствуют идеалу полугруппы $\mathbf{I}^{(N=4)}_{SCf(chiral)} \trianglelefteq \mathbf{S}^{(N=4)}_{SCf(chiral)}$, а обратимые преобразования — ее подгруппе $\mathbf{G}^{(N=4)}_{SCf(chiral)} \subset \mathbf{S}^{(N=4)}_{SCf(chiral)}$.

Перейдем к рассмотрению необратимых киральных $N = 4$ преобразований (см. **Определение 3.41**), которые характеризуются условием $\epsilon[\det \mathrm{H}_{SCf}] = 0,\ \det \mathrm{H}_{SCf} \neq 0$. Наиболее экстремальный вариант — это отбрасывание всех четных функций в (3.242)–(3.243), т. е. $\mathrm{G} = 0$. Тогда условия (3.244)–(3.247) выполняются тождественно, и следовательно, такие преобразования параметризуются нечетными функциями $\psi_i^{\pm}(z)$, $\lambda_i^{\pm}(z)$ и имеют вид [2]

$$\begin{aligned} \tilde{z} &= f(z) + \left(\theta_1^+\theta_1^- + \theta_2^+\theta_2^-\right)\left[\psi_1^+(z)\,\psi_1^-(z) + \psi_2^+(z)\,\psi_2^-(z)\right]' - \\ &\quad 2\theta_1^+\theta_2^+\left[\psi_1^-(z)\,\lambda_1^+(z) + \lambda_2^+(z)\,\psi_2^-(z)\right] - \qquad (3.255)\\ &\quad 2\theta_1^-\theta_2^-\left[\psi_1^+(z)\,\lambda_1^-(z) + \lambda_2^-(z)\,\psi_2^+(z)\right] + \theta_1^+\theta_1^-\theta_2^+\theta_2^- f''(z), \\ \tilde{\theta}_1^{\pm} &= \psi_1^{\pm}\left(Z^{\pm}\right) + \theta_1^{\pm}\theta_2^{\pm}\lambda_1^{\pm}\left(Z^{\pm}\right), \qquad (3.256)\\ \tilde{\theta}_2^{\pm} &= \psi_2^{\pm}\left(Z^{\pm}\right) + \theta_2^{\pm}\theta_1^{\pm}\lambda_2^{\pm}\left(Z^{\pm}\right), \qquad (3.257) \end{aligned}$$

где

$$f'(z) = \psi_i^{+\prime}(z)\,\psi_i^-(z) + \psi_i^{-\prime}(z)\,\psi_i^+(z) \tag{3.258}$$

и выполняется условие (3.248).

Если воспользоваться решением (3.249), то получаем параметризацию необратимых преобразований 4 нечетными функциями

$$\begin{aligned}\tilde{z} &= f(z) + \left(\theta_1^+\theta_1^- + \theta_2^+\theta_2^-\right)\left[\psi_1^+(z)\,\psi_1^-(z) + \psi_2^+(z)\,\psi_2^-(z)\right]' - \\ &\quad 2\theta_1^+\theta_2^+\left[\psi_1^-(z)\,\psi_2^-(z)\right]' - 2\theta_1^-\theta_2^-\left[\psi_1^+(z)\,\psi_2^+(z)\right] + \theta_1^+\theta_1^-\theta_2^+\theta_2^- f''(z), \end{aligned} \qquad (3.259)$$

$$\tilde{\theta}_1^\pm = \psi_1^\pm\left(Z^\pm\right) + 2\theta_1^\pm\theta_2^\pm\psi_2^\mp\left(Z^\pm\right), \qquad (3.260)$$

$$\tilde{\theta}_2^\pm = \psi_2^\pm\left(Z^\pm\right) + 2\theta_2^\pm\theta_1^\pm\psi_1^\mp\left(Z^\pm\right), \qquad (3.261)$$

где $f(z)$ дается в (3.258).

Отметим, что уравнение (3.248) имеет и другое решение

$$\psi_1^\pm(z) = \psi_2^\mp(z) = \psi^\pm(z), \quad \lambda_1^\pm(z) = \lambda_2^\mp(z) = \lambda^\pm(z). \qquad (3.262)$$

Тогда из (3.255)–(3.257) вместо (3.259)–(3.261) имеем

$$\begin{aligned}\tilde{z} &= f(z) + 2\left(\theta_1^+\theta_2^+ - \theta_1^-\theta_2^-\right)\left[\lambda^+(z)\,\psi^-(z) + \psi^+(z)\,\lambda^-(z)\right] + \\ &\quad \theta_1^+\theta_1^-\theta_2^+\theta_2^- f''(z), \end{aligned} \qquad (3.263)$$

$$\tilde{\theta}_1^\pm = \psi^\pm\left(Z^\pm\right) + \theta_1^\pm\theta_2^\pm\lambda^\pm\left(Z^\pm\right), \qquad (3.264)$$

$$\tilde{\theta}_2^\pm = \psi^\mp\left(Z^\pm\right) + \theta_2^\pm\theta_1^\pm\lambda^\mp\left(Z^\pm\right), \qquad (3.265)$$

где $f'(z) = 2\left[\psi^{+\prime}(z)\,\psi^-(z) + \psi^{-\prime}(z)\,\psi^+(z)\right]$.

Если матрица G отлична от нуля, но содержит нильпотентные элементы, то одним из вариантов решения жесткого смешанного ограничения (3.246)–(3.247) является выбор элементов ее в виде $g_{ij}^{\pm\mp}(z) = \psi_i^{\mp\prime}(z)\,\psi_j^{\pm\prime}(z)$. Тогда SCf условия (3.244)–(3.247) выполняются за счет нильпотентности функций $\psi_i^\pm(z)$. В общем, количество различных типов необратимых преобразований велико и их таблица умножений, к со-жалению, является необозримой. Тем не менее, в конкретной задаче всегда можно обратиться к полученной здесь системе уравнений (3.244)–(3.248) и решить ее применительно к рассматриваемому случаю.

Например, чрезвычайно необходимыми в приложениях к суперконфомной теории поля и суперструне, обладающих расширенной суперсимметрией, являются конечные $N = 4$ киральные дробно-линейные преобразования, которые рассмотрены в **Приложении Ж.5**.

**3.2.5.** Сплетающие четность $N=4$ преобразования. Остановимся на других типах редуцированных $N = 4$ преобразований, которые удовлетворяют условиям редукции (3.186)–(3.197).

В касательном пространстве они действуют на суперпроизводные и дифференциалы следующим образом

$$\begin{pmatrix} \partial \\ D_1^+ \\ D_2^- \\ D_2^+ \end{pmatrix} = \mathcal{R}_{TPt_1^+}^{(N=4)} \cdot \begin{pmatrix} \tilde{D}_1^- \\ \tilde{D}_1^+ \\ \tilde{D}_2^- \\ \tilde{D}_2^+ \end{pmatrix}, \qquad (3.266)$$

$$d\tilde{Z} = d\theta_1^+ \cdot \varDelta_{TPt_1^+}^- \left(z, \theta_i^+, \theta_i^-\right), \qquad (3.267)$$

где нечетная функция*) $\varDelta_{TPt_1^+}^- \left(z, \theta_i^+, \theta_i^-\right)$ определена в (3.164) и $4 \times 4$ горизонтальная полуматрица (см. **Пункт Д.2**) $\mathcal{R}_{TPt_1^+}^{(N=4)}$ задается формулой

$$\mathcal{R}_{TPt_1^+}^{(N=4)} = \begin{pmatrix} \partial\tilde{\theta}_{1(TPt_1^+)}^+ & \partial\tilde{\theta}_{1(TPt_1^+)}^- & \partial\tilde{\theta}_{2(TPt_1^+)}^+ & \partial\tilde{\theta}_{2(TPt_1^+)}^- \\ D_1^+\tilde{\theta}_{1(TPt_1^+)}^+ & D_1^+\tilde{\theta}_{1(TPt_1^+)}^- & D_2^+\tilde{\theta}_{1(TPt_1^+)}^+ & D_2^+\tilde{\theta}_{1(TPt_1^+)}^- \\ D_1^-\tilde{\theta}_{2(TPt_1^+)}^+ & D_1^-\tilde{\theta}_{2(TPt_1^+)}^- & D_2^-\tilde{\theta}_{2(TPt_1^+)}^+ & D_2^-\tilde{\theta}_{2(TPt_1^+)}^- \\ D_1^+\tilde{\theta}_{2(TPt_1^+)}^+ & D_1^+\tilde{\theta}_{2(TPt_1^+)}^- & D_2^+\tilde{\theta}_{2(TPt_1^+)}^+ & D_2^+\tilde{\theta}_{2(TPt_1^+)}^- \end{pmatrix} \qquad (3.268)$$

---

*Примечание.* Относительно индексов см. **Определение 3.46**.

(ср. $N = 2$ (3.124)–(3.127)).

Здесь мы будем рассматривать только один из 4 вариантов TPt редукций (3.186)–(3.197), поскольку остальные отличаются лишь перестановкой индексов

*Замечание* **3.63.** Полуматрица $\mathcal{R}^{(N=4)}_{TPt_1^+}$ является полуминором нечетного элемента $\Delta^-_{TPt_1^+}\left(z, \theta_i^+, \theta_i^-\right)$ в суперматрице $\mathrm{P}^{(N=4)}_{TPt_1^+}$ (3.200).

Сравнивая суперконформные преобразования касательного пространства (3.158) и формулу (3.266), можно заметить, что здесь мы имеем аналогию с $(0|4)$-мерным подпространством $(1|4)$-мерного касательного пространства, когда матрица $\mathrm{H}_{SCf}$ оставляла его инвариантным $T\mathbb{C}^{0|4} \to T\mathbb{C}^{0|4}$ (см.(3.211)).

*Замечание* **3.64.** Полуматрицы $\mathcal{R}^{(N=4)}_{TPt_i^\pm}$ в данном нечетном случае действуют в четырехмерном подпространстве, однако меняют его четность, а именно $\mathbb{C}^{1|3} \to \mathbb{C}^{0|4}$.

Следовательно, по аналогии с $N = 2$ (см. **Определение 3.26**), получаем

**Определение 3.65.** *Назовем редуцированные $N = 4$ преобразования, которые удовлетворяют* $\mathsf{TPt}_i^\pm$ *условиям* (3.186)–(3.197) *и действуют в касательном пространстве $T\mathbb{C}^{1|3} \to T\mathbb{C}^{0|4}$, сплетающими четность (касательного пространства) $N = 4$ преобразованиями* (TPt – *twisting parity of tangent space transformations*).

Это определение становится ясным из выражения для четной производной для $\mathsf{TPt}_1^+$ преобразований

$$\partial = \partial\tilde{\theta}^+_{i\left(TPt_1^+\right)} \cdot \tilde{D}_i^- + \partial\tilde{\theta}^-_{i\left(TPt_1^+\right)} \cdot \tilde{D}_i^+, \tag{3.269}$$

которое следует из (3.266), а также из TPt формулы для четного дифференциала (3.267), вращающего четность в результате TPt преобразо-ваний.

Введем $N=4$ TPt супердифференциалы, дуальные производным из (3.266), по аналогии с суперконформными супердифференциалами (3.232).

**Определение 3.66.** *Назовем $N=4$ TPt супердифферециалами с кручением четности такие объекты**) $d\tau^{\pm}_{i\left(TPt_1^+\right)}$*, которые преобразуются при сплетающих четность $N=4$ редуцированных* TPt *преобразованиях $Z\to\tilde{Z}$ (см.* **Определение 3.65**) *по закону*

$$\begin{pmatrix} d\tilde{\tau}^{even+}_{1\left(TPt_1^+\right)} & d\tilde{\tau}^{even-}_{1\left(TPt_1^+\right)} & d\tilde{\tau}^{even+}_{2\left(TPt_1^+\right)} & d\tilde{\tau}^{even-}_{2\left(TPt_1^+\right)} \end{pmatrix} = \\ \begin{pmatrix} d\tau^{odd}_{\left(TPt_1^+\right)} & d\tau^{even-}_{1\left(TPt_1^+\right)} & d\tau^{even+}_{2\left(TPt_1^+\right)} & d\tau^{even-}_{2\left(TPt_1^+\right)} \end{pmatrix}\cdot\mathcal{R}_{TPt_1^+}, \tag{3.270}$$

*где полуматрица $\mathcal{R}_{TPt_1^+}$ определена в* (3.268).

*Замечание* **3.67.** Четности $d\tilde{\tau}^{even\pm}_{TPt_i^\pm}$ и $d\tau^{odd}_{TPt_i^\pm}$ противоположны, поэтому в кокасательном пространстве мы имеем отображение с кручением четности $T\mathbb{C}^{1|3}\to T\mathbb{C}^{4|0}$ (ср. *Замечание* **3.64**).

Определим внешние $N=4$ TPt дифференциалы по аналогии с внешними $N=4$ SCf дифференциалом (3.233) следующим образом (см. также $N=2$ TPt (3.132)–(3.134))

$$\delta_{TPt_1^+} = d\tau^{odd}_{TPt_1^+}\cdot\partial + d\tau^{even-}_{1\left(TPt_1^+\right)}\cdot D_1^+ + d\tau^{even-}_{2\left(TPt_1^+\right)}\cdot D_2^+ + d\tau^{even+}_{2\left(TPt_1^+\right)}\cdot D_2^-, \tag{3.271}$$

$$\tilde{\delta}_{TPt_i^\pm} = d\tilde{\tau}^{even+}_{j\left(TPt_i^\pm\right)}\cdot\tilde{D}_j^- + d\tilde{\tau}^{even-}_{j\left(TPt_i^\pm\right)}\cdot\tilde{D}^+. \tag{3.272}$$

Отметим, что четность внешних $N=4$ TPt дифференциалов, как и в суперконформном случае (см. также *Замечание* **3.29** относительно

---

*) *Примечание.* Для остальных индексов $\mathsf{TPt}_i^\pm$ справедливы те же определения и формулы с точностью до очевидных перестановок.

$N=2$ TPt дифференциалов), фиксирована, они — нечетны при $N=4$ сплетающих четность преобразованиях.

**Предложение 3.68.** *Внешние* $N = 4$ TPt *дифференциалы "инвариантны" относительно соответвтвующих* $N=4$ TPt *преобразований.*

*Доказательство.* Рассмотрим только $\mathsf{TPt}_1^+$ преобразования. Из законов преобразования (3.266), (3.270) и определений (3.271)–(3.272) получаем

$$\delta_{TPt_1^+} = \begin{pmatrix} d\tau^{odd}_{(TPt_1^+)} & d\tau^{even-}_{1(TPt_1^+)} & d\tau^{even+}_{2(TPt_1^+)} & d\tau^{even-}_{2(TPt_1^+)} \end{pmatrix} \begin{pmatrix} \partial \\ D_1^+ \\ D_2^- \\ D_2^+ \end{pmatrix} =$$

$$\begin{pmatrix} d\tau^{odd}_{(TPt_1^+)} & d\tau^{even-}_{1(TPt_1^+)} & d\tau^{even+}_{2(TPt_1^+)} & d\tau^{even-}_{2(TPt_1^+)} \end{pmatrix} \mathcal{R}^{(N=4)}_{TPt_1^+} \begin{pmatrix} \tilde{D}_1^- \\ \tilde{D}_1^+ \\ \tilde{D}_2^- \\ \tilde{D}_2^+ \end{pmatrix} =$$

$$\begin{pmatrix} d\tilde{\tau}^{even+}_{1(TPt_1^+)} & d\tilde{\tau}^{even-}_{1(TPt_1^+)} & d\tilde{\tau}^{even+}_{2(TPt_1^+)} & d\tilde{\tau}^{even-}_{2(TPt_1^+)} \end{pmatrix} \begin{pmatrix} \tilde{D}_1^- \\ \tilde{D}_1^+ \\ \tilde{D}_2^- \\ \tilde{D}_2^+ \end{pmatrix} = \tilde{\delta}_{TPt_1^+}.$$

И аналогично для остальных типов $\mathsf{TPt}_i^\pm$ преобразований. ■

Полученные соотношения дают возможность построения $N = 4$ расслоений с кручением четности (см. для $N = 2$ **Пункт 3.1.6**).

Используя (E.42) и теорему сложения $N = 4$ березинианов (3.172), можно трактовать $N = 4$ преобразования следующим образом.

**Предположение 3.69.** *Если считать* $N = 4$ SCf *преобразования* $N = 4$ *суперaналогом обычных голоморфных преобразований* [563, 565]*, то для антиголоморфных преобразований, в отличие от* $N = 1$ *(см.* **Под-** **раздел 2.3**), *имеется четыре* (!) *нечетных супераналога: редуцированные* $\mathsf{TPt}_1^\pm$ *и* $\mathsf{TPt}_2^\pm$ *преобразования.*

Из приведенных построений для частных случаев $N = 2$ и $N = 4$ следует ожидать, что в общем случае произвольных $N$ имеет место следующее

**Предположение 3.70.** *При ослаблении требования обратимости для редуцированных $N$ преобразований имеется 1 четный супераналог голоморфных преобразований и $N$ нечетных супераналогов антиголоморфных преобразований.*

## 3.3. Основные результаты и выводы

1. Подробно исследованы все редукции в расширенной суперконформной геометрии с учетом необратимости и проведена их классификация.
2. Альтернативная параметризация введена и использована для построения $N = 2$ и $N = 4$ суперконформных полугрупп.
3. Обобщается на произвольное $N$ понятие комплексной структуры на суперплоскости: имеется 1 супераналог голоморфных преобразований и $N$ необратимых супераналогов антиголоморфных преобразований.
4. Найдено, что переключение типа преобразования производится проекцией введенного спина редукции, который равен $N/2$.
5. Рассмотрены расщепленные $N$ расширенные преобразования и для них построена полугруппа и компонентное представление в альтернативной параметризации.
6. Получена общие формулы для березинианов редуцированных преобразований через перманенты и полуминоры.

**7.** Изучены сплетающие четность $N$ расширенные преобразования и получены компонентные представления.

**8.** Введены сплетающие четность дифференциалы как аналог супердифференциалов на суперримановых поверхностях.

**9.** Построены дуальные преобразования с половинным количеством суперсимметрий, подробно исследован случай $N = 2$.

**10.** Рассмотрены вложения $N = 1 \hookrightarrow N = 2$ и получены аналитические формулы для обратимого и необратимого случаев.

# РАЗДЕЛ 4

# СУПЕРМАТРИЧНЫЕ ПОЛУГРУППЫ, ИДЕАЛЬНОЕ СТРОЕНИЕ И РЕДУКЦИИ

Данный раздел посвящен исследованию идеальных свойств суперматриц и построению суперматричных полугрупп, важных с точки зрения их приложений к суперструнным теориям и к феноменологии суперсимметричных моделей элементарных частиц. Рассматриваются общие свойства и классифицируются возможные редукции суперматриц, вводится понятие нечетно-редуцированных суперматриц и показывается их существенная роль как новой категории в изучении суперматричных подструктур. Формулируется теорема сложения березинианов, в рамках которой видна дуальная роль нечетно-редуцированных суперматриц по отношению к четно-редуцированным (треугольным). Оба типа суперматриц объединяются в различные сэндвич-полугруппы с необычными свойствами. Вводятся новые типы супермодулей — нечетные супермодули, нечетное антитранспонирование, представления странной супералгебры Березина. Рассматривается прямая сумма редуцированных суперматриц, где определяются нечетные аналоги собственных чисел и характеристических функций, сформулирована обобщенная теорема Гамильтона-Якоби.

Подробно анализируется идеальная структура многопараметрических полугрупп нечетно-редуцированных суперматриц. Изучаются непрерывные представления полугрупповых связок нечетно-редуцированными суперматрицами антитреугольного вида и вводится новый тип связок — скрученная прямоугольная связка. Для высших связок определяются обобщения отношений Грина — тонкие и смешанные отноше-ния эквивалентности, которые приводят к обобщенным многомерным eggbox диаграммам и являются продолжением отношений Грина с подполугрупп на полугруппу.

## 4.1. Альтернативная редукция суперматриц

Согласно общей теории $G$-структур [412, 669–671] различные геометрии получаются редукцией структурной группы многообразия $\mathscr{M}$ к некоторой подгруппе $G$ эндоморфизмов касательного пространства $T\mathscr{M}$ [408, 411, 414]. В локальном подходе (используя координатное описание) это означает, что фактически необходимо преобразовать соответствующую матрицу производных в заданном представлении к некоторому редуцированному виду. В подавляющем большинстве случаев этот вид был треугольным [408, 669], и доводом этому было прозрачное наблюдение из обыкновенной теории матриц, что треугольные матрицы сохраняют форму и образуют подгруппу [672–674]. Кроме того, кольца верхнетреугольных матриц обладают нетривиальными алгебраическими свойствами [675].

В суперсимметричных теориях, несмотря на возникновение нечетных подпространств и антикоммутирующих величин, выбор формы редукции оставался тем же [109, 413, 414, 676]. Основанием этому было желание полностью отождествить умножение в подгруппах суперматриц с умножением обыкновенных матриц, и вытекающее из этого допущение, что вид матриц, образующих подструктуру, должен быть прежним [367, 368, 403, 620].

При рассмотрении вариантов нетривиальных суперсимметричных обобщений [9, 13] можно видеть, что замыкание умножения также может быть достигнуто и для других типов подструктур, не только треугольных, из-за существования дивизоров нуля в алгебре Грассмана или в кольце, над которым определяются суперпространства и супермного-образия [112, 117]. Более того, такие структуры можно объединить со стандартными треугольными в некоторую *более общую категорию*, которая может иметь дальнейшее применение, аналогичное подгруппам. Таким образом, абстрактный смысл собственно редукции [101, 677] может быть в принципе расширен и видоизменен, как это будет показано ниже.

В [1, 9] (см. **Разделы 2** и **3**) были рассмотрены варианты таких редукций в применении к аналогам суперконформных преобразований — редуцированным преобразованиям, которые имеют много необычных свойств. Например, они необратимы и сплетают четность касательного пространства в суперсимметричном базисе.

В данном подразделе мы изучаем общие свойства альтернативной редукции суперматриц с более абстрактной точки зрения без связи с конкретной физической моделью [8]. Однако многие полученные результаты могут быть использованы в теории суперструн [282] и в феноменологии суперсимметричных моделей элементарных частиц [74, 678].

Линейное суперпространство $\mathbf{\Lambda}^{p|q}$ размерности $(p|q)$ над $\Lambda = \Lambda_0 \oplus \Lambda_1$ определено в **Приложении B** (см. [30, 106]). Различные четные морфизмы $\mathrm{Hom}_0\left(\mathbf{\Lambda}^{p|q}, \mathbf{\Lambda}^{m|n}\right)$ между линейными суперпространствами $\mathbf{\Lambda}^{p|q} \to \mathbf{\Lambda}^{m|n}$ описываются посредством $(p|q) \times (m|n)$- суперматриц как операторов в некотором базисе (см. [30] и **Приложение B**).

В теории суперримановых поверхностей [111] $(1|1) \times (1|1)$-суперматрицы, описывающие голоморфные морфизмы касательного расслоения, имеют треугольный вид [367]. Здесь мы рассматриваем специальную альтернативную редукцию суперматриц. Для ясности мы ограничиваемся $(1|1) \times (1|1)$-суперматрицами, что позволит нам сосредоточиться на самих идеях, не скрывая их за громоздкими формулами. Обобщение на $(p|q) \times (m|n)$ случай понятно и может быть выполнено посредством элементарных блочных переобозначений.

**4.1.1.** Необратимое строение суперматриц. В стандартном базисе элементы из $\mathrm{Hom}_0\left(\mathbf{\Lambda}^{1|1},\mathbf{\Lambda}^{1|1}\right)$ описываются $(1|1)\times(1|1)$-суперматрицами [30]

$$\mathrm{M}\equiv\begin{pmatrix} a & \alpha \\ \beta & b \end{pmatrix}\in\mathrm{Mat}_{\Lambda}\left(1|1\right), \tag{4.1}$$

где $a,b\in\Lambda_0$, $\alpha,\beta\in\Lambda_1$ (мы полагаем здесь, что нечетные элементы имеют индекс нильпотентности, равный 2).

Для множеств суперматриц мы будем использовать соответствующие символы, например, $\mathfrak{M}\overset{def}{=}\left\{\mathrm{M}\in\mathrm{Mat}_{\Lambda}\left(1|1\right)\right\}$.

В данном $(1|1)$-мерном случае березиниан [30], определяемый как $\mathrm{Ber}:\mathrm{Mat}_{\Lambda}\left(1|1\right)\setminus\left\{\mathrm{M}|\,\epsilon\left(b\right)=0\right\}\to\Lambda_0$ имеет вид

$$\mathrm{Ber}\mathrm{M}=\frac{a}{b}+\frac{\beta\alpha}{b^2}. \tag{4.2}$$

Здесь мы предлагаем два типа возможных редукций суперматрицы M (в соответствие с двумя слагаемыми в (4.2)) и изучаем некоторые их свойства совместно [8].

**Определение 4.1.** *Четно-редуцированные суперматрицы есть элементы из* $\mathrm{Mat}_{\Lambda}\left(1|1\right)$*, имеющие вид*

$$\mathrm{S}\equiv\begin{pmatrix} a & \alpha \\ 0 & b \end{pmatrix}\in\mathrm{RMat}_{\Lambda}^{even}\left(1|1\right). \tag{4.3}$$

**Определение 4.2.** *Нечетно-редуцированные суперматрицы есть элементы из* $\mathrm{Mat}_{\Lambda}\left(1|1\right)$*, имеющие вид*

$$\mathrm{T}\equiv\begin{pmatrix} 0 & \alpha \\ \beta & b \end{pmatrix}\in\mathrm{RMat}_{\Lambda}^{odd}\left(1|1\right). \tag{4.4}$$

*Замечание* **4.3.** Причина обозначений происходит из нильпотентности березиниана $\mathrm{Ber}\,\mathrm{T}$ и из того факта, что четно-редуцированные суперматрицы $\mathrm{S}$ отвечают суперконформным преобразованиям, которые описывают морфизмы касательного расслоения над суперримановыми поверхностями [367], тогда как нечетно-редуцированные суперматрицы $\mathrm{T}$ приводят к преобразованиям, сплетающим четность касательного суперпространства $T\mathbb{C}^{1|1}$ в стандартном базисе (см. [1, 13] и **Подраздел 2.3.2**).

**Утверждение 4.4.** *Множество* $\mathfrak{M}$ *представляет собой прямую сумму диагональных* $\mathfrak{D}$ *и анти-диагональных* $\mathcal{A}$ *суперматриц* (*четные и нечетные суперматрицы в обозначениях* [30])

$$\mathfrak{M}=\mathfrak{D} \oplus \mathcal{A}, \tag{4.5}$$

$$\mathrm{D} \equiv \begin{pmatrix} a & 0 \\ 0 & b \end{pmatrix} \in \mathfrak{D} \equiv \mathrm{Mat}_{\Lambda}^{Diag}(1|1)\,,$$
$$\mathrm{A} \equiv \begin{pmatrix} 0 & \alpha \\ \beta & 0 \end{pmatrix} \in \mathcal{A} \equiv \mathrm{Mat}_{\Lambda}^{Adiag}(1|1)\,,$$

*где* $\mathfrak{D} \subset \mathcal{S}$, $\mathcal{A} \subset \mathcal{T}$.

Для редуцированных суперматриц находим

$$\mathcal{S} \cap \mathcal{T} = \begin{pmatrix} 0 & \alpha \\ 0 & b \end{pmatrix} \neq \varnothing. \tag{4.6}$$

Тем не менее, следующая теорема объясняет фундаментальную и дуальную роль четно-редуцированных суперматриц $\mathcal{S}$ и нечетно-редуцированных суперматриц $\mathcal{T}$.

**Теорема 4.5.** (Теорема сложения березинианов) *Березинианы четно- и нечетно-редуцированных суперматриц являются аддитивными компонентами березиниана соответствующей нередуцированной суперматрицы*

$$\mathrm{Ber}\,\mathrm{M} = \mathrm{Ber}\,\mathrm{S} + \mathrm{Ber}\,\mathrm{T}. \tag{4.7}$$

Первое слагаемое в (4.7) покрывает все подгруппы четно-редуцированных суперматриц из $\mathrm{Mat}_{\Lambda}(1|1)$, и только оно раньше рассматривалось в приложениях. Второе слагаемое в (4.7) дуально к первому в некотором смысле и соответствует нечетно- редуцированным суперматрицам из $\mathrm{Mat}_{\Lambda}(1|1)$ (см. **Определение 4.2**).

*Замечание* **4.6.** Соотношение (4.7) представляет собой суперсимметричный вариант очевидного равенства $\det \mathrm{M}_{nonsusy} = \det \mathrm{D}_{nonsusy} + \det \mathrm{A}_{nonsusy}$, где $\mathrm{D}_{nonsusy}$ и $\mathrm{A}_{nonsusy}$ — обыкновенные диагональная и антидиагональная матрицы.

Однако дело в том, что, если A из (4.5) — суперматрица, то BerA не определен вообще [30].

Обозначим множество обратимых элементов из $\mathfrak{M}$ за $\mathfrak{M}^{inv}$, и их разность за $\mathfrak{J} = \mathfrak{M} \setminus \mathfrak{M}^{inv}$. В [30] доказывается, что $\mathfrak{M}^{inv} = \{\mathrm{M} \in \mathfrak{M} \mid \epsilon(a) \neq 0 \wedge \epsilon(b) \neq 0\}$. Далее аналогично для редуцированных суперматриц

$$\mathfrak{S}^{inv} = \{\mathrm{S} \in \mathfrak{S} | \, \epsilon(a) \neq 0 \wedge \epsilon(b) \neq 0\}, \;\; \mathfrak{T}^{inv} = \varnothing, \tag{4.8}$$

т. е. получаем

**Утверждение 4.7.** *Нечетно-редуцированные суперматрицы* $T \in \mathfrak{T}$ *необратимы и* $\mathfrak{T} \subset \mathfrak{J}$.

Идеальная структура $(1|1)$-суперматриц подробно изложена в **Приложении B.5**.

**4.1.2.** Мультипликативные свойства нечетно-редуцированных суперматриц. Нечетно-редуцированные суперматрицы не образуют полугруппу в общем случае, поскольку

$$\mathrm{T}_1\mathrm{T}_2 = \begin{pmatrix} \alpha_1\beta_2 & \alpha_1 b_2 \\ b_1\beta_2 & b_1 b_2 + \beta_1\alpha_2 \end{pmatrix} \neq \mathrm{T}. \tag{4.9}$$

Однако,

$$\mathfrak{T} \star \mathfrak{T} \cap \mathfrak{T} \neq \varnothing \Rightarrow \alpha\beta = 0, \tag{4.10}$$

$$\mathfrak{T} \star \mathfrak{T} \cap \mathfrak{S} \neq \varnothing \Rightarrow \beta b = 0, \tag{4.11}$$

что может иметь место из-за наличия дивизоров нуля в $\Lambda$.

**Предложение 4.8.** 1) *Подмножество* $\mathfrak{T}^{SG} \subset \mathfrak{T}$ *нечетно-редуцированных суперматриц удовлетворяющих* $\alpha\beta = 0$ (4.10) *представляют нечетно-редуцированную подполугруппу* $\mathbf{T}^{SG} \stackrel{def}{=} \left\{\mathfrak{T}^{SG}; \cdot\right\}$ *полугруппы* $\mathbf{M}$.

2) *В нечетной-редуцированной подполугруппе* $\mathbf{T}^{SG}$ *подмножество суперматриц с* $\beta = 0$ *представляет собой левый идеал, и с* $\alpha = 0$ *представляет собой правый идеал, суперматрицы с* $b = 0$ *образуют двусторонний идеал.*

Другое условие $\beta b = 0$ (4.11) можно трактовать следующим образом.

**Утверждение 4.9.** *Подмножество* $\mathfrak{T}^{\sqrt{S}} \subset \mathfrak{T}$ *нечетно-редуцированных суперматриц удовлетворяющих* $\beta b = 0$ *представляет нечетную ветвь корня из четно-редуцированных суперматриц* $\mathfrak{S}$, *четная ветвь которого представляется всеми четно-редуцированными суперматрицами вследствие соотношения* $\mathfrak{S} \star \mathfrak{S} \subseteq \mathfrak{S}$.

**4.1.3.** Унификация редуцированных суперматриц.

Теперь мы объединим четно- и нечетно-редуцированные суперматрицы (4.3) и (4.4) в общий абстрактный объект. Сначала рассмотрим таблицу умножения всех введенных множеств

$$\begin{array}{rclrcl} \mathfrak{D}\star\mathfrak{D} & = & \mathfrak{D}, & \mathcal{A}\star\mathcal{A} & = & \mathfrak{D} \\ \mathfrak{D}\star\mathfrak{S} & = & \mathfrak{S}, & \mathfrak{T}\star\mathcal{A} & = & \mathfrak{S}^{\mathsf{st}}, \\ \mathfrak{S}\star\mathfrak{D} & = & \mathfrak{S}, & \mathfrak{S}\star\mathcal{A} & = & \mathfrak{T}^{\mathbf{\Pi}}, \\ \mathcal{A}\star\mathfrak{T} & = & \mathfrak{S}, & \mathfrak{S}\star\mathfrak{T} & = & \mathfrak{S}\cup\mathfrak{T} \\ \mathcal{A}\star\mathfrak{S} & = & \mathfrak{T}, & \mathfrak{T}\star\mathfrak{S} & = & \mathfrak{T}. \end{array} \tag{4.12}$$

Здесь $\mathsf{st}:\mathrm{Mat}_{\Lambda}(1|1)\to\mathrm{Mat}_{\Lambda}(1|1)$ представляет собой супертранспонирование [106], т. е. $\begin{pmatrix} a & \alpha \\ \beta & b \end{pmatrix}^{\mathsf{st}} = \begin{pmatrix} a & \beta \\ -\alpha & b \end{pmatrix}$.

Также мы употребляем $\mathbf{\Pi}$-транспонирование [679] определенное, как $\mathbf{\Pi}:\mathrm{Mat}_{\Lambda}(1|1)\to\mathrm{Mat}_{\Lambda}(1|1)$ и $\begin{pmatrix} a & \alpha \\ \beta & b \end{pmatrix}^{\mathbf{\Pi}} = \begin{pmatrix} b & \beta \\ \alpha & a \end{pmatrix}$.

*Замечание* **4.10.** Множества суперматриц $\mathfrak{S}$ и $\mathfrak{T}$ не замкнуты относительно $\mathsf{st}$ и $\mathbf{\Pi}$ операций, но $\mathfrak{S}^{\mathsf{st}}\cap\mathfrak{S}\subseteq\mathfrak{D}$ и $\mathfrak{T}^{\mathbf{\Pi}}\cap\mathfrak{T}\subseteq\mathcal{A}$.

Мы видим из первых двух соотношений в (4.12), что $\mathcal{A}$ в некотором базисе играет роль левого оператора $\hat{\mathsf{A}}$ изменения типа множества суперматриц (четно-редуцированный на нечетно- и наоборот) $\hat{\mathsf{A}}:\mathfrak{S}\to\mathfrak{T}$ и $\hat{\mathsf{A}}:\mathfrak{T}\to\mathfrak{S}$, тогда как оператор $\hat{\mathsf{D}}$, соответствующий множеству $\mathfrak{D}$, не изменяет тип.

Далее, из первых двух соотношений в (4.12) видно, что множества $\mathfrak{S}$ и $\mathfrak{D}$ представляют собой подполугруппы $\mathbf{S}\overset{def}{=}\{\mathfrak{S};\cdot\}$ и $\mathbf{D}\overset{def}{=}\{\mathfrak{D};\cdot\}$ полугруппы $\mathbf{M}$. К сожалению, из-за двух следующих соотношений в (4.12) множество $\mathfrak{T}$ не имеет такого отчетливого абстрактного смысла. Тем не менее, последняя зависимость $\mathfrak{T}\star\mathfrak{S}=\mathfrak{T}$ важна с иной точки зрения.

**Теорема 4.11.** *Любой нечетно-редуцированный морфизм* $\hat{\mathsf{T}} : \mathbf{\Lambda}^{1|1} \to \mathbf{\Lambda}^{1|1}$*, отвечающий множеству нечетно-редуцированных суперматриц* $\mathfrak{T}$*, может представляться в виде произведения нечетно- и четно-редуцированных морфизмов, таковых, что*

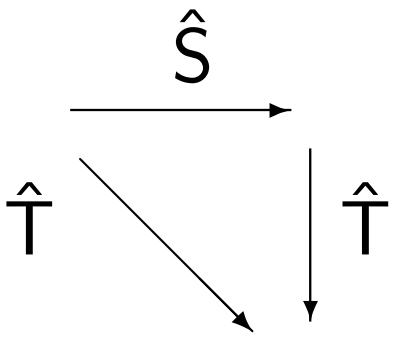

(4.13)

*представляет собой коммутативную диаграмму.*

Это разложение является решающим в приложениях к построению сплетающих четность преобразований — нечетных супераналогов антиголоморфных преобразований (см. [1] и **Подраздел 2.3**).

**4.1.4.** Скаляры, антискаляры, обобщенные модули и сэндвич-полугруппа редуцированных суперматриц. Введем аналог $\odot$-умножения для самих редуцированных матриц (не для множеств, как в **Приложении B.7**). Во-первых, определим строение обобщенного $\Lambda$-модуля в $\mathrm{Hom}_0\left(\mathbf{\Lambda}^{1|1}, \mathbf{\Lambda}^{1|1}\right)$ некоторым альтернативным способом, четная часть которого*) описана в [106].

**Определение 4.12.** *В* $\mathrm{Mat}_\Lambda(1|1)$ *скалярная матрица (скаляр)* $\mathrm{E}(x)$ *и антискалярная матрица (антискаляр)* $\mathcal{E}(\chi)$ *определяются формулами*

$$\mathrm{E}(x) \overset{def}{=} \begin{pmatrix} x & 0 \\ 0 & x \end{pmatrix} \in \mathfrak{D} = \mathrm{Mat}_\Lambda^{diag}(1|1), \; x \in \Lambda_0, \tag{4.14}$$

---

*Примечание.* В обыкновенной матричной теории — это тот факт, что произведение матрицы и числа равно произведению матрицы и диагональной матрицы, имеющей данное число на диагонали [680].

$$\mathcal{E}\left(\chi\right) \overset{def}{=} \begin{pmatrix} 0 & \chi \\ \chi & 0 \end{pmatrix} \in \boldsymbol{\mathcal{A}} = \mathrm{Mat}_{\Lambda}^{adiag}\left(1|1\right), \ \chi \in \Lambda_{1.} \tag{4.15}$$

**Утверждение 4.13.** *Странная супералгебра Березина* [30] (*см. также* **Приложение В.4**)

$$\boldsymbol{Q}_{\Lambda}\left(1\right) \equiv \begin{pmatrix} x & \chi \\ \chi & x \end{pmatrix} \subset \mathrm{Mat}_{\Lambda}\left(1|1\right) \tag{4.16}$$

*представляет собой прямую сумму скаляра и антискаляра*

$$\boldsymbol{Q}_{\Lambda}\left(1\right) = \mathrm{E}\left(x\right) \oplus \mathcal{E}\left(\chi\right). \tag{4.17}$$

Опишем некоторые свойства скаляров и антискаляров.

**Утверждение 4.14.** *Антискаляры между собой антикоммутируют* $\mathcal{E}\left(\chi_1\right)\mathcal{E}\left(\chi_2\right) + \mathcal{E}\left(\chi_2\right)\mathcal{E}\left(\chi_1\right) = 0$*, и поэтому они нильпотентны.*

**Предложение 4.15.** *Строение обобщенного* $\Lambda_0 \oplus \Lambda_1$*-модуля в* $\mathrm{Hom}_0\left(\boldsymbol{\Lambda}^{1|1}, \boldsymbol{\Lambda}^{1|1}\right)$ *определяется действием скаляров* (4.14) *и антискаляров* (4.15).

Это значит, что везде, где необходимо, мы заменяем умножение суперматриц четными и нечетными элементами из $\Lambda$ с умножением на скалярные и антискалярные суперматрицы (4.14)–(4.15). Соотношения, содержащие скаляры, уже известны [106], но для антискалярных величин мы получаем новые дуальные соотношения [8].

Рассмотрим подробнее их действие на элементах $\mathrm{M}{\in}\mathrm{Mat}_{\Lambda}\left(1|1\right)$. Во-первых, сформулируем следующее

**Определение 4.16.** *Левое* $\boldsymbol{\Upsilon}_L$ *и правое* $\boldsymbol{\Upsilon}_R$ *антитранспонирования — это отображения* $\mathrm{Hom}_0\left(\boldsymbol{\Lambda}^{1|1}, \boldsymbol{\Lambda}^{1|1}\right) \to \mathrm{Hom}_1\left(\boldsymbol{\Lambda}^{1|1}, \boldsymbol{\Lambda}^{1|1}\right)$*, действую-*

*щие на* $\mathrm{M} \in \mathfrak{M}$ *как*

$$\begin{pmatrix} a & \alpha \\ \beta & b \end{pmatrix}^{\Upsilon_L} = \begin{pmatrix} \beta & b \\ a & \alpha \end{pmatrix}, \tag{4.18}$$

$$\begin{pmatrix} a & \alpha \\ \beta & b \end{pmatrix}^{\Upsilon_R} = \begin{pmatrix} \alpha & a \\ b & \beta \end{pmatrix}. \tag{4.19}$$

**Следствие 4.17.** *Антитранспонирования являются квадратными корнями оператора смены четности* $\mathbf{\Pi}$ *в следующем смысле*

$$\Upsilon_L \Upsilon_R = \Upsilon_R \Upsilon_L = \mathbf{\Pi}. \tag{4.20}$$

Интересно сравнить (4.20) с полутранспонированиями, введенными в **Пункте Д.2**, и аналогичной формулой (Д.18).

**Утверждение 4.18.** *Антитранспонирования удовлетворяют соотношениям*

$$\begin{aligned} \left(\mathcal{E}\left(\chi\right)\mathrm{M}\right)^{\Upsilon_L} &= \chi\mathrm{M} \\ \left(\mathcal{E}\left(\chi\right)\mathrm{M}\right)^{\Upsilon_R} &= \chi\mathrm{M}^{\mathbf{\Pi}} \\ \left(\mathrm{M}\mathcal{E}\left(\chi\right)\right)^{\Upsilon_L} &= \mathrm{M}^{\mathbf{\Pi}}\chi \\ \left(\mathrm{M}\mathcal{E}\left(\chi\right)\right)^{\Upsilon_R} &= \mathrm{M}\chi \end{aligned} \tag{4.21}$$

Таким образом, конкретная реализация правого, левого и двустороннего обобщенных $\Lambda_0 \oplus \Lambda_1$- модулей в $\mathrm{Hom}_0\left(\mathbf{\Lambda}^{1|1}, \mathbf{\Lambda}^{1|1}\right)$ определяется новыми действиями

$$\begin{aligned} \mathcal{E}\left(\chi\right)\mathrm{M} &= \chi\mathrm{M}^{\Upsilon_L}, \\ \mathrm{M}\mathcal{E}\left(\chi\right) &= \mathrm{M}^{\Upsilon_R}\chi, \\ \mathcal{E}\left(\chi_1\right)\mathrm{M}\mathcal{E}\left(\chi_2\right) &= \chi_1\mathrm{M}^{\mathbf{\Pi}}\chi_2. \end{aligned} \tag{4.22}$$

Можно сравнить эти выражения со стандартной структурой $\Lambda$-

модуля [106]

$$\begin{aligned} \mathrm{E}\left(x\right)\mathrm{M} &= x\mathrm{M}, \\ \mathrm{M}\mathrm{E}\left(x\right) &= \mathrm{M}x, \\ \mathrm{E}\left(x_1\right)\mathrm{M}\mathrm{E}\left(x_2\right) &= x_1\mathrm{M}x_2. \end{aligned} \tag{4.23}$$

**Следствие 4.19.** *Обобщенные соотношения для* $\Lambda_0 \oplus \Lambda_1$*-модуля имеют следующий вид*

$$\begin{aligned} \left(\mathrm{E}\left(x\right)\mathrm{M}\right)\mathrm{N} &= \mathrm{E}\left(x\right)\left(\mathrm{M}\mathrm{N}\right) \\ \left(\mathrm{M}\mathrm{E}\left(x\right)\right)\mathrm{N} &= \mathrm{M}\left(\mathrm{E}\left(x\right)\mathrm{N}\right) \\ \mathrm{M}\left(\mathrm{N}\mathrm{E}\left(x\right)\right) &= \left(\mathrm{M}\mathrm{N}\right)\mathrm{E}\left(x\right) \\ \left(\mathcal{E}\left(\chi\right)\mathrm{M}\right)\mathrm{N} &= \mathcal{E}\left(\chi\right)\left(\mathrm{M}\mathrm{N}\right) \\ \left(\mathrm{M}\mathcal{E}\left(\chi\right)\right)\mathrm{N} &= \mathrm{M}\left(\mathcal{E}\left(\chi\right)\mathrm{N}\right) \\ \mathrm{M}\left(\mathrm{N}\mathcal{E}\left(\chi\right)\right) &= \left(\mathrm{M}\mathrm{N}\right)\mathcal{E}\left(\chi\right) \end{aligned} \tag{4.24}$$

*где* $\mathrm{M}, \mathrm{N} \in \mathrm{Mat}_\Lambda\left(1|1\right)$.

Таким же образом определяются и величины, дуальные относительно четности.

**Определение 4.20.** *Нечетные скаляр и антискаляр определяются формулами*

$$\mathrm{E}\left(\chi\right) \overset{def}{=} \begin{pmatrix} \chi & 0 \\ 0 & -\chi \end{pmatrix} \in \mathrm{Hom}_1\left(\mathbf{\Lambda}^{1|1}, \mathbf{\Lambda}^{1|1}\right), \tag{4.25}$$

$$\mathcal{E}\left(x\right) \overset{def}{=} \begin{pmatrix} 0 & x \\ x & 0 \end{pmatrix} \in \mathrm{Hom}_1\left(\mathbf{\Lambda}^{1|1}, \mathbf{\Lambda}^{1|1}\right). \tag{4.26}$$

**Предложение 4.21.** *Строение обобщенного* $\Lambda_0 \oplus \Lambda_1$*-модуля в* $\mathrm{Hom}_1\left(\mathbf{\Lambda}^{1|1}, \mathbf{\Lambda}^{1|1}\right)$ *определяется аналогичный действию нечетного скаляра и нечетного анти-скаляра* (4.24).

Одним способом объединения четно- (4.3) и нечетно-редуцированных (4.4) суперматриц в объект, аналогичный полугруппе, является рассмотрение сэндвич-умножения, подобного (B.49), но на уровне суперматриц (а не множеств), посредством скаляров и анти скаляров в качестве сэндвич-суперматриц.

В самом деле, обычное произведение суперматриц может быть записано, как $\mathrm{M}_1\mathrm{M}_2 = \mathrm{M}_1\mathrm{E}\left(1\right)\mathrm{M}_2$. Для антискаляра не существует аналога этого соотношения, потому, что среди нечетных величин $\chi \in \Lambda_1$ нет единицы. Следовательно, единственная возможность рассмотреть $\mathcal{E}\left(\chi\right)$ на равных началах с $\mathrm{E}\left(x\right)$ есть рассмотрение сэндвич-элементов (4.14)–(4.15), которые имеют в качестве аргументов $x$ и $\chi$ произвольно выбранные или фиксированные другими специальными условиями суперчисла.

**Определение 4.22.** *Сэндвич-произведение* $\Lambda_0 \oplus \Lambda_1$ *редуцированных суперматриц* $\mathrm{R} = \mathrm{S}, \mathrm{T} \in \mathfrak{R}$ *определяется формулой*

$$\mathrm{R}_1 \circledcirc_{\boldsymbol{X}} \mathrm{R}_2 \stackrel{def}{=} \begin{cases} \mathrm{R}_1\mathrm{E}\left(x\right)\mathrm{R}_2, & \mathrm{R}_2 = \mathrm{S}, \\ \mathrm{R}_1\mathcal{E}\left(\chi\right)\mathrm{R}_2, & \mathrm{R}_2 = \mathrm{T}, \end{cases} \tag{4.27}$$

*где* $\boldsymbol{X} = \{x, \chi\} \in \Lambda_0 \oplus \Lambda_1$ — *"суперполе" сэндвич-умножения.*

Введенное $\circledcirc_{\boldsymbol{X}}$-умножение ассоциативно, и его таблица совпадает с (B.50). Поэтому мы имеем

**Определение 4.23.** *Относительно* $\circledcirc_{\boldsymbol{X}}$*-умножения* (4.27) *редуцированные суперматрицы образуют полугруппу, которую мы будем называть сэндвич-полугруппой редуцированных матриц* $\mathbf{RMS}_{sandw}$ (**R***educed super***M***atrix sandwich* **S***emigroup*).

Из явного вида $\circledcirc_{\boldsymbol{X}}$-умножения следует

**Теорема 4.24.** *Введенная сэндвич-полугруппа редуцированных матриц*

$\mathbf{RMS}_{sandw}$ *изоморфна специальной полугруппе правых нулей*

$$\mathbf{RMS}_{sandw} \cong \mathbf{Z}_R = \{\mathcal{R} = \mathcal{S} \cup \mathcal{T}; \circledcirc_{\boldsymbol{X}}\}. \tag{4.28}$$

**4.1.5.** Прямая сумма редуцированных суперматриц. Иной способ объединить редуцированные суперматрицы — это рассмотреть связь между ними и обобщенными $\Lambda_0 \oplus \Lambda_1$-модулями, введенными в предыдущем пункте. Для этого необходимо определить прямую сумму пространств.

**Определение 4.25.** *Прямое пространство редуцированных суперматриц* $\mathbb{RMS}_{\oplus}$ (*Reduced superMatrix direct Superspace*) *представляет собой прямую сумму пространства четно-редуцированных суперматриц и пространства нечетно-редуцированных суперматриц.*

В терминах множеств имеем $\mathcal{R}_{\oplus} = \mathcal{S} \oplus \mathcal{T}$.

*Замечание* **4.26.** Отметим, что $\mathcal{R}_{\oplus} \neq \mathcal{M}$ из-за (4.6).

**Утверждение 4.27.** *В пространстве* $\mathbb{RMS}_{\oplus}$ *скаляр — это странная супералгебра Березина* $\boldsymbol{Q}_{\Lambda}(1)$ (*см.* (4.17)).

В пространстве $\mathbb{RMS}_{\oplus}$ скаляр играет ту же роль для четно-редуцированных суперматриц, как антискаляр — для нечетно-редуцированных суперматриц. Так, используя (4.3)–(4.4) и (4.14)–(4.15), легко проверить следующее

**Утверждение 4.28.** *В* $\mathbb{RMS}_{\oplus}$ *собственные значения четно-* $\mathrm{S}$ *и нечетно-редуцированных* $\mathrm{T}$ *суперматриц должны находиться из различных уравнений, а именно,*

$$\mathrm{S} \cdot \mathtt{V} = \mathrm{E}(x) \cdot \mathtt{V}, \tag{4.29}$$

$$\mathrm{T} \cdot \mathtt{V} = \mathcal{E}(\chi) \cdot \mathtt{V}, \tag{4.30}$$

*где* V *представляет собой вектор-столбец, а собственные значения равны*

$$x_1 = a, \ \ x_2 = b, \tag{4.31}$$

$$\chi_1 = \alpha, \ \ \chi_2 = \beta. \tag{4.32}$$

**Определение 4.29.** *Четная и нечетная характеристические функции для редуцированных суперматриц определяются в* $\mathbb{RMS}_{\oplus}$ *различными* (!) *формулами*

$$\boldsymbol{H}_{\mathrm{S}}^{even}(x) = \mathrm{Ber}\,(\mathrm{E}\,(x) - \mathrm{S})\,, \tag{4.33}$$

$$\boldsymbol{H}_{\mathrm{T}}^{odd}(\chi) = \mathrm{Ber}\,(\mathcal{E}\,(\chi) - \mathrm{T})\,. \tag{4.34}$$

*Замечание* **4.30.** В стандартном $\Lambda$-модуле над $\mathrm{Mat}_{\Lambda}(1|1)$ [30] характеристические функции и собственные значения для любой суперматрицы (включая и нечетно-редуцированные) получаются из уравнений (4.29) и (4.33), что дает в нечетном случае отличный от нашего результат (см. также [681]).

Используя (4.3)–(4.4) и (4.33)–(4.34), легко находим

$$\boldsymbol{H}_{\mathrm{S}}^{even}(x) = \frac{(x-b)(x-a)}{(x-b)^2}, \tag{4.35}$$

$$\boldsymbol{H}_{\mathrm{T}}^{odd}(\chi) = \frac{(\chi-\beta)(\chi-\alpha)}{b^2}. \tag{4.36}$$

Здесь мы замечаем полную симметрию между четно- и нечетно-редуцированными суперматрицами*), а также непротиворечивость с их $\Lambda_0 \oplus \Lambda_1$ собственными значениями (4.31)–(4.32).

---

*) *Примечание.* Чтобы это подчеркнуть, мы не проводили сокращения в равенстве (4.35).

В “четном” случае характеристический многочлен суперматрицы M определяется выражением $\boldsymbol{P}_{\mathrm{M}}(\mathrm{M}) = 0$ и в нетривиальных случаях [682–686] строится из частей характеристической функции $\boldsymbol{H}_{\mathrm{M}}(x)$ согласно особому алгоритму [681, 687, 688]. Для несуперсимметричной матрицы $\mathrm{M}_{nonsusy}$ он очевидно совпадает с характеристической функцией $\boldsymbol{P}_{\mathrm{M}_{nonsusy}}(x) = \boldsymbol{H}_{\mathrm{M}_{nonsusy}}(x) \equiv \det(\mathrm{I} \cdot x - \mathrm{M}_{nonsusy})$, где I представляет собой единичную матрицу. Однако в суперслучае из-за существования дивизоров нуля в $\Lambda$ степень характеристического многочлена $\boldsymbol{P}_{\mathrm{M}}(x)$ может быть меньше стандартной величины $n = p + q$, $\mathrm{M} \in \mathrm{Mat}_{\Lambda}(p|q)$ [681, 688]. Но этот алгоритм не может быть непосредственно применим для нечетно-редуцированных и антидиагональных суперматриц.

Поэтому, как и выше, мы рассматриваем два дуальных характеристических многочлена и, используя (4.35)–(4.36), получаем аналог теорему Кэли-Гамильтона для пространства $\mathbb{RMS}_{\oplus}$.

**Теорема 4.31.** (Обобщенная теорема Кэли-Гамильтона) *В* $\mathbb{RMS}_{\oplus}$ *характеристические многочлены имеют вид*

$$\boldsymbol{P}_{\mathrm{S}}^{even}(x) = (x - a)(x - b), \quad (4.37)$$

$$\boldsymbol{P}_{\mathrm{T}}^{odd}(\chi) = (\chi - \alpha)(\chi - \beta). \quad (4.38)$$

*и* $\boldsymbol{P}_{\mathrm{S}}^{even}(\mathrm{S}) = 0$ *для любого* S, *но* $\boldsymbol{P}_{\mathrm{T}}^{odd}(\mathrm{T}) = 0$ *только для нильпотентных* $b$.

## 4.2. Представление полугрупп связок суперматрицами

Матричные полугруппы [400,427,689–693] представляют собой значительный инструмент в конкретном и полном исследовании абстрактного строения теории полугрупп [103, 104, 204, 694]. Матричные пред-ставления [695–700] широко используются в изучении конечных полугрупп [701,702] и топологических полугрупп [703–707]. Обычно матричные полугруппы определяются над полем $\mathbb{K}$ [708–710]. Тем не менее, после обнаружения суперсимметрии физиками [34, 70] реалистичные объединенные теории частиц начали рассматриваться в суперпространстве (см., например, [711] и **Приложение Б**) — аналоге пространства, в котором все величины и функции определяются не над полем $\mathbb{K}$, но над грассман-банаховой супералгеброй над $\mathbb{K}$ [112, 174] (или их обобщениями [117,133]). Следовательно, представляется важным изучить различные представления полугрупп не матрицами, а суперматрицами [10].

В этом подразделе мы рассмотрим непрерывные суперматричные представления различных полугрупп связок, состоящих из идемпотентов [103,425,712]. Отметим, что исследование представлений полугрупп идемпотентов [704,713–715], с идемпотентно-генерированных полугрупп [716] и подмножеств идемпотентов [121, 717–720] и псевдоидемпотентов [721] в полугруппах, в особенности матричных полугрупп [722], является важным с абстрактно-алгебраической точки зрения. Идемпотенты также возникают и широко используются в приложениях случайных матричных полугрупп [699, 700, 723, 724].

Сначала рассмотрим возможные подполугруппы полугруппы редуцированных суперматриц (не множеств и не сэндвич, как в **Подразделе 4.1**). Множества нечетно-редуцированных матриц (см. **Определение 4.2**) образуют Г-полугруппы, которые определены в **Приложении В.6**. Рассмотрим сначала однопараметрические подполугруппы Г-полугрупп из (B.44)–(B.45).

**4.2.1.** Однопараметрические полугруппы редуцированных суперматриц. Наиболее элементарная однопараметрическая полугруппа суперматриц вида (B.44)–(B.45) предста-вляется антидиагональными нильпотентыми суперматрицами вида

$$\mathrm{Y}_{\alpha}(t) \stackrel{def}{=} \begin{pmatrix} 0 & \alpha t \\ \alpha & 0 \end{pmatrix}. \tag{4.39}$$

**Предложение 4.32.** *Суперматрицы* $\mathrm{Y}_{\alpha}(t)$ *наряду с нулевой суперматрицей*

$$\mathrm{Z} \stackrel{def}{=} \begin{pmatrix} 0 & 0 \\ 0 & 0 \end{pmatrix} \tag{4.40}$$

*образуют непрерывную полугруппу* $\mathbf{Z}_{\alpha} \stackrel{def}{=} \{\cup \mathrm{Y}_{\alpha}(t) \cup \mathrm{Z}; \cdot\}$ *с нулевым умножением*

$$\mathrm{Y}_{\alpha}(t) \cdot \mathrm{Y}_{\alpha}(u) = \mathrm{Z}. \tag{4.41}$$

*Доказательство.* Рассмотрим умножение двух элементов

$$Y_{\alpha}(t) \cdot Y_{\alpha}(u) = \begin{pmatrix} 0 & \alpha t \\ \alpha & 0 \end{pmatrix} \begin{pmatrix} 0 & \alpha u \\ \alpha & 0 \end{pmatrix} = \begin{pmatrix} \alpha^2 t & 0 \\ 0 & \alpha^2 u \end{pmatrix}.$$

Поскольку $\alpha$ — нильпотент второй степени $\alpha^2 = 0$, мы получаем необходимый результат — нулевое умножение (4.41). ∎

*Замечание* **4.33.** Это показывает, что здесь (как и во всех доказательствах ниже) нильпотентность играет решающую и обязательную роль, и, таким образом, эти построения возможны только для суперматриц и не имеют аналогов в обычном (несуперсимметричном) случае.

**Утверждение 4.34.** *Для любого фиксированного* $t = t_0 \in \Lambda^{1|0}$ *множество* $\{\mathrm{Y}_{\alpha}(t_0), \mathrm{Z}\}$ *представляет собой 0-минимальный идеал в полугруппе* $\mathbf{Z}_{\alpha}$.

Среди нетривиальных вариантов однопараметрических подполугруппы полугруппы $\mathbf{T}^{\Gamma}_{(L,R)}$ мы рассмотрим нечетно-редуцированные су-перматрицы следующего вида

$$\mathrm{P}_{\alpha}(t) \stackrel{def}{=} \begin{pmatrix} 0 & \alpha t \\ \alpha & 1 \end{pmatrix} \tag{4.42}$$

где $t \in \Lambda^{1|0}$ — четный параметр из $\Lambda$, который "нумерует" элементы $\mathrm{P}_{\alpha}(t)$, и $\alpha \in \Lambda^{0|1}$ представляет собой фиксированный нечетный элемент $\Lambda$, который "нумерует" множества $\bigcup\limits_{t} \mathrm{P}_{\alpha}(t)$.

*Замечание* **4.35.** Здесь мы исследуем однопараметрические подполугруппы полугруппы $\mathbf{T}^{\Gamma}_{(L,R)}$ как абстрактные полугруппы [102, 104], но не как полугруппы операторов [725, 726].

Сначала установим свойства умножения суперматриц $\mathrm{P}_{\alpha}(t)$. Из (4.42) видно, что

$$\begin{pmatrix} 0 & \alpha t \\ \alpha & 1 \end{pmatrix} \begin{pmatrix} 0 & \alpha u \\ \alpha & 1 \end{pmatrix} = \begin{pmatrix} \alpha^2 t & \alpha t \\ \alpha & 1 + \alpha^2 u \end{pmatrix} \stackrel{\alpha^2=0}{=} \begin{pmatrix} 0 & \alpha t \\ \alpha & 1 \end{pmatrix}, \tag{4.43}$$

и поэтому мы имеем

**Предложение 4.36.** *В случае* $\alpha^2 = 0$ *умножение суперматриц* $\mathrm{P}_{\alpha}(t)$ *имеет следующий вид*

$$\mathrm{P}_{\alpha}(t) \cdot \mathrm{P}_{\alpha}(u) = \mathrm{P}_{\alpha}(t). \tag{4.44}$$

**Следствие 4.37.** *Умножение* (4.44) *ассоциативно, поэтому множество суперматриц* $\mathrm{P}_{\alpha}(t)$ *представляет собой однопараметрическую полугруппу* $\mathbf{P}_{\alpha}$ *относительно умножения* $(\cdot)$.

**Следствие 4.38.** *Все суперматрицы* $\mathrm{P}_\alpha(t)$ *идемпотентны*

$$\begin{pmatrix} 0 & \alpha t \\ \alpha & 1 \end{pmatrix}^2 = \begin{pmatrix} \alpha^2 t & \alpha t \\ \alpha & 1+\alpha^2 t \end{pmatrix} \overset{\alpha^2=0}{=} \begin{pmatrix} 0 & \alpha t \\ \alpha & 1 \end{pmatrix}. \tag{4.45}$$

**Предложение 4.39.** *Если* $\mathrm{P}_\alpha(t) = \mathrm{P}_\alpha(u)$, *то*

$$t - u = \operatorname{Ann}\alpha. \tag{4.46}$$

*Доказательство.* Из определения (4.42) следует, что две суперматрицы $\mathrm{P}_\alpha(t)$ равны, если $\alpha t = \alpha u$, что дает искомое (4.46). ■

Аналогично мы можем ввести идемпотентные суперматрицы $\mathrm{Q}_\alpha(t)$ вида

$$\mathrm{Q}_\alpha(t) \overset{def}{=} \begin{pmatrix} 0 & \alpha \\ \alpha t & 1 \end{pmatrix}, \tag{4.47}$$

которые удовлетворяют

$$\begin{pmatrix} 0 & \alpha \\ \alpha t & 1 \end{pmatrix} \begin{pmatrix} 0 & \alpha \\ \alpha u & 1 \end{pmatrix} = \begin{pmatrix} 0 & \alpha \\ \alpha u & 1 \end{pmatrix} \tag{4.48}$$

или

$$\mathrm{Q}_\alpha(t) \cdot \mathrm{Q}_\alpha(u) = \mathrm{Q}_\alpha(u), \tag{4.49}$$

и поэтому суперматрицы $\mathrm{Q}_\alpha(t)$ также образуют полугруппу $\mathbf{Q}_\alpha$.

*Замечание* **4.40.** Полугруппы $\mathbf{P}_\alpha$ и $\mathbf{Q}_\alpha$ не содержат двусторонних нулей и единиц.

**Утверждение 4.41.** *Полугруппы* $\mathbf{P}_\alpha$ *и* $\mathbf{Q}_\alpha$ *— непрерывные объединения одноэлементных групп (соответствующие фиксированным* $t$*) с действиями* (4.44) *и* (4.49).

Соотношения (4.43)–(4.48) и

$$\begin{pmatrix} 0 & \alpha t \\ \alpha & 1 \end{pmatrix} \begin{pmatrix} 0 & \alpha \\ \alpha u & 1 \end{pmatrix} = \begin{pmatrix} 0 & \alpha t \\ \alpha u & 1 \end{pmatrix} \stackrel{def}{=} \mathrm{F}_{tu}, \qquad (4.50)$$

$$\begin{pmatrix} 0 & \alpha \\ \alpha u & 1 \end{pmatrix} \begin{pmatrix} 0 & \alpha t \\ \alpha & 1 \end{pmatrix} = \begin{pmatrix} 0 & \alpha \\ \alpha & 1 \end{pmatrix} \stackrel{def}{=} \mathrm{E} \qquad (4.51)$$

важны с абстрактной точки зрения и будут использоваться ниже.

*Замечание* **4.42.** В общем случае умножение суперматриц некоммутативно, необратимо, но ассоциативно, поэтому любые объекты, допускающие представление суперматрицами (с замкнутым умножением), автоматически будут полугруппами.

Так, непрерывные представления нулевых полугрупп, рассмотрены в **Приложении B.8**.

**4.2.2.** Скрученные прямоугольные связки. Теперь мы объединим полугруппы $\boldsymbol{P}_\alpha$ и $\boldsymbol{Q}_\alpha$ в некоторую нетривиальную полугруппу. Во-первых, мы рассмотрим объединенное множество элементов $\mathcal{P}_\alpha \cup \mathcal{Q}_\alpha$ и изучим их свойства умножения.

Используя (4.50) и (4.51), мы замечаем, что $\mathcal{P}_\alpha \cap \mathcal{Q}_\alpha = \boldsymbol{e}$, где $\varphi(\boldsymbol{e}) = \mathrm{E}$ из (4.51), и поэтому $\boldsymbol{e}\boldsymbol{\Delta}_\alpha \boldsymbol{p}_{t=1}$ и $\boldsymbol{e}\boldsymbol{\Delta}_\alpha \boldsymbol{q}_{t=1}$. Таким образом, мы вынуждены различать область $t = 1 + \operatorname{Ann}\alpha$ от других областей в суперпространстве параметра $t \in \Lambda^{1|0}$, и в дальнейшем для любых индексов в $\boldsymbol{p}_t$ и $\boldsymbol{q}_t$ мы подразумеваем $t \neq 1 + \operatorname{Ann}\alpha$.

**Утверждение 4.43.** *Элемент $\boldsymbol{e}$ представляет собой левый нуль и правую единицу для $\boldsymbol{p}_t$, и $\boldsymbol{e}$ представляет собой правый нуль и левую единицу для $\boldsymbol{q}_u$, т.е. $\boldsymbol{e} * \boldsymbol{p}_t = \boldsymbol{e}$, $\boldsymbol{p}_t * \boldsymbol{e} = \boldsymbol{p}_t$, и $\boldsymbol{q}_u * \boldsymbol{e} = \boldsymbol{e}$, $\boldsymbol{e} * \boldsymbol{q}_u = \boldsymbol{q}_u$.*

Используя (4.51), легко проверить, что $\boldsymbol{q}_u * \boldsymbol{p}_t = \boldsymbol{e}$, но обратное произведение требует рассмотрения дополнительных элементов, которые не содержатся в $\mathcal{P}_\alpha \cup \mathcal{Q}_\alpha$.

Из (4.50) мы получаем

$$\boldsymbol{r}_{tu} = \boldsymbol{p}_t * \boldsymbol{q}_u, \tag{4.52}$$

где $\varphi(\boldsymbol{r}_{tu}) = \mathrm{F}_{tu}$. Допустим, что $\mathcal{R}_\alpha \overset{def}{=} \bigcup\limits_{t,u \notin 1+\mathrm{Ann}\,\alpha} \boldsymbol{r}_{tu}$.

**Определение 4.44.** *Скрученная прямоугольная связка* $\boldsymbol{W}_\alpha$ *представляет собой объединение множеств идемпотентов* $\mathcal{P}_\alpha \cup \mathcal{Q}_\alpha \cup \mathcal{R}_\alpha$ *с* $*$*-произведением* (B.51)*, и следующей таблицей Кэли, представленной в Таблице* 4.1.

*Таблица* 4.1

*Таблица Кэли для непрерывной скрученной прямоугольной связки*

| $1 \setminus 2$ | $\boldsymbol{e}$ | $\boldsymbol{p}_t$ | $\boldsymbol{p}_u$ | $\boldsymbol{q}_t$ | $\boldsymbol{q}_u$ | $\boldsymbol{r}_{tu}$ | $\boldsymbol{r}_{ut}$ | $\boldsymbol{r}_{tw}$ | $\boldsymbol{r}_{vw}$ |
|---|---|---|---|---|---|---|---|---|---|
| $\boldsymbol{e}$ | $\boldsymbol{e}$ | $\boldsymbol{e}$ | $\boldsymbol{e}$ | $\boldsymbol{q}_t$ | $\boldsymbol{q}_u$ | $\boldsymbol{q}_u$ | $\boldsymbol{q}_t$ | $\boldsymbol{q}_w$ | $\boldsymbol{q}_w$ |
| $\boldsymbol{p}_t$ | $\boldsymbol{p}_t$ | $\boldsymbol{p}_t$ | $\boldsymbol{p}_t$ | $\boldsymbol{r}_{tt}$ | $\boldsymbol{r}_{tu}$ | $\boldsymbol{r}_{tu}$ | $\boldsymbol{r}_{tt}$ | $\boldsymbol{r}_{tw}$ | $\boldsymbol{r}_{tw}$ |
| $\boldsymbol{p}_u$ | $\boldsymbol{p}_u$ | $\boldsymbol{p}_u$ | $\boldsymbol{p}_u$ | $\boldsymbol{r}_{ut}$ | $\boldsymbol{r}_{uu}$ | $\boldsymbol{r}_{uu}$ | $\boldsymbol{r}_{ut}$ | $\boldsymbol{r}_{uw}$ | $\boldsymbol{r}_{uw}$ |
| $\boldsymbol{q}_t$ | $\boldsymbol{e}$ | $\boldsymbol{e}$ | $\boldsymbol{e}$ | $\boldsymbol{q}_t$ | $\boldsymbol{q}_u$ | $\boldsymbol{q}_u$ | $\boldsymbol{q}_t$ | $\boldsymbol{q}_w$ | $\boldsymbol{q}_w$ |
| $\boldsymbol{q}_u$ | $\boldsymbol{e}$ | $\boldsymbol{e}$ | $\boldsymbol{e}$ | $\boldsymbol{q}_t$ | $\boldsymbol{q}_u$ | $\boldsymbol{q}_u$ | $\boldsymbol{q}_t$ | $\boldsymbol{q}_w$ | $\boldsymbol{q}_w$ |
| $\boldsymbol{r}_{tu}$ | $\boldsymbol{p}_t$ | $\boldsymbol{p}_t$ | $\boldsymbol{p}_t$ | $\boldsymbol{r}_{tt}$ | $\boldsymbol{r}_{tu}$ | $\boldsymbol{r}_{tu}$ | $\boldsymbol{r}_{tt}$ | $\boldsymbol{r}_{tw}$ | $\boldsymbol{r}_{tw}$ |
| $\boldsymbol{r}_{ut}$ | $\boldsymbol{p}_u$ | $\boldsymbol{p}_u$ | $\boldsymbol{p}_u$ | $\boldsymbol{r}_{ut}$ | $\boldsymbol{r}_{uu}$ | $\boldsymbol{r}_{uu}$ | $\boldsymbol{r}_{ut}$ | $\boldsymbol{r}_{uw}$ | $\boldsymbol{r}_{uw}$ |
| $\boldsymbol{r}_{tw}$ | $\boldsymbol{p}_t$ | $\boldsymbol{p}_t$ | $\boldsymbol{p}_t$ | $\boldsymbol{r}_{tt}$ | $\boldsymbol{r}_{tu}$ | $\boldsymbol{r}_{tu}$ | $\boldsymbol{r}_{tt}$ | $\boldsymbol{r}_{tw}$ | $\boldsymbol{r}_{tw}$ |
| $\boldsymbol{r}_{vw}$ | $\boldsymbol{p}_v$ | $\boldsymbol{p}_v$ | $\boldsymbol{p}_v$ | $\boldsymbol{r}_{vt}$ | $\boldsymbol{r}_{vu}$ | $\boldsymbol{r}_{vu}$ | $\boldsymbol{r}_{vt}$ | $\boldsymbol{r}_{vw}$ | $\boldsymbol{r}_{vw}$ |

Из *Таблицы* 4.1 видно, что умножение в скрученной прямоугольной связке $\boldsymbol{W}_\alpha$ является ассоциативным *), как это и следовало ожидать.

---

*Примечание.* Для удобства мы показываем некоторые дополнительные соотношения.

Мы можем заметить из таблицы Кэли следующие непрерывные подполугруппы в скрученной прямоугольной связке:

- $\boldsymbol{e}$ – одноэлементная “почти тождественная” подполугруппа;
- $\tilde{\boldsymbol{P}}_\alpha = \left\{ \bigcup\limits_{t \neq 1 + \operatorname{Ann}\alpha} \boldsymbol{p}_t ; * \right\}$ – “приведенная” полугруппа левых нулей;
- $\boldsymbol{P}_\alpha = \left\{ \bigcup\limits_{t \neq 1 + \operatorname{Ann}\alpha} \boldsymbol{p}_t \cup \boldsymbol{e} ; * \right\}$ – полная полугруппа левых нулей;
- $\tilde{\boldsymbol{Q}}_\alpha = \left\{ \bigcup\limits_{t \neq 1 + \operatorname{Ann}\alpha} \boldsymbol{q}_t ; * \right\}$ – “приведенная” полугруппа правых нулей;
- $\boldsymbol{Q}_\alpha = \left\{ \bigcup\limits_{t \neq 1 + \operatorname{Ann}\alpha} \boldsymbol{q}_t \cup \boldsymbol{e} ; * \right\}$ – полная полугруппа правых нулей;
- $\tilde{\boldsymbol{F}}_\alpha^{(1|1)} = \left\{ \bigcup\limits_{t,u \neq 1 + \operatorname{Ann}\alpha} \boldsymbol{r}_{tu} ; * \right\}$ – “приведенная” прямоугольная связка;
- $\boldsymbol{F}_\alpha^{(1|1)} = \left\{ \bigcup\limits_{t,u \neq 1 + \operatorname{Ann}\alpha} \boldsymbol{r}_{tu} \cup \boldsymbol{e} ; * \right\}$ – полная прямоугольная связка;
- $\boldsymbol{V}_\alpha^L = \left\{ \bigcup\limits_{t,u \neq 1 + \operatorname{Ann}\alpha} \boldsymbol{r}_{tu} \cup \boldsymbol{p}_t ; * \right\}$ – “смешанная” левая прямоугольная связка;
- $\boldsymbol{V}_\alpha^R = \left\{ \bigcup\limits_{t,u \neq 1 + \operatorname{Ann}\alpha} \boldsymbol{r}_{tu} \cup \boldsymbol{q}_t ; * \right\}$ – “смешанная” правая прямоугольная связка.

Таким образом, мы получили непрерывное суперматричное представление для полугрупп левых и правых нулей и построили из них суперматричное представление прямоугольных связок. Хорошо известно, что любая прямоугольная связка изоморфна декартову произведению полугрупп левых и правых нулей [103, 433]. Здесь мы получили это в явном виде (см. (4.52)) и представили конкретную конструкцию (4.50). Кроме того, мы унифицировали все вышеупомянутые полугруппы в одном объекте, а именно в скрученной прямоугольной связке.

**4.2.3.** Представления прямоугольных связок. Умножение прямоугольных связок приводится в правом нижнем углу таблицы Кэли. Обычно [103,104] оно определяется одним соотношением

$$\boldsymbol{r}_{tu} * \boldsymbol{r}_{vw} = \boldsymbol{r}_{tw}. \tag{4.53}$$

В нашем случае индексы — это четные непрерывные грассмановы параметры из $\Lambda^{1|0}$. Что касается полугрупп нулей, это также приводит к некоторым особенностям в идеальном строении таких связок.

Другое отличие представляет собой отсутствие условия $u = v$ , что возникает в некотором приложениях из-за конечной природы индексов, рассматриваемых как некоторые величины, соответствующие строкам и столбцам в матрицах элементов (см. например, [727]). Поэтому, при поисках новых результатов в данном непрерывном суперсимметричном случае мы должны рассматривать и должны доказывать некоторые стандартные утверждения с самого начала.

Рассмотрим отношения Грина на $\boldsymbol{F}_{\alpha}^{(1|1)}$.

**Предложение 4.45.** *Любые два элемента в прямоугольной связке* $\boldsymbol{F}_{\alpha}^{(1|1)}$ *одновременно* $\mathscr{J}$ *- и* $\mathscr{D}$*-эквивалентны.*

*Доказательство.* Из (4.53) мы имеем

$$\boldsymbol{r}_{tu} * \boldsymbol{r}_{vw} * \boldsymbol{r}_{tu} = \boldsymbol{r}_{tw} * \boldsymbol{r}_{tu} = \boldsymbol{r}_{tu}, \tag{4.54}$$

$$\boldsymbol{r}_{vw} * \boldsymbol{r}_{tu} * \boldsymbol{r}_{vw} = \boldsymbol{r}_{vw} * \boldsymbol{r}_{tw} = \boldsymbol{r}_{vw} \tag{4.55}$$

для любого $t, u, v, w \in \Lambda^{1|0}$. Во-первых, мы обращаем внимание, что эти равенства совпадают с определением $\mathscr{J}$- классов [104], поэтому любые два элемента $\mathscr{J}$-эквивалентны, и таким образом $\mathscr{J}$ совпадает с универсальным отношением на $\mathscr{F}_{\alpha}^{(1|1)}$. Далее, используя (4.54), мы замечаем, что выполняются соотношения $\boldsymbol{r}_{tu} \mathscr{R} \boldsymbol{r}_{tu} * \boldsymbol{r}_{vw}$ и $\boldsymbol{r}_{tu} * \boldsymbol{r}_{vw} \mathscr{L} \boldsymbol{r}_{vw}$. Поскольку $\mathscr{D} = \mathscr{L} \circ \mathscr{R} = \mathscr{R} \circ \mathscr{L}$ (см., например, [103]), то $\boldsymbol{r}_{tu} \mathscr{D} \boldsymbol{r}_{vw}$. ■

**Утверждение 4.46.** *Каждый $\mathscr{R}$-класс $\mathsf{R}_{\boldsymbol{r}_{tu}}$ состоит из элементов $\boldsymbol{r}_{tu}$, которые $\boldsymbol{\Delta}_\alpha$-эквивалентны по первому индексу, т. е. $\boldsymbol{r}_{tu}\mathscr{R}\boldsymbol{r}_{vw} \Leftrightarrow t - v = \operatorname{Ann}\alpha$, и каждый $\mathscr{L}$-класс $\mathsf{L}_{\boldsymbol{r}_{tu}}$ состоит из элементов $\boldsymbol{r}_{tu}$, которые $\boldsymbol{\Delta}_\alpha$-эквивалентны по второму индексу, т. е. $\boldsymbol{r}_{tu}\mathscr{L}\boldsymbol{r}_{vw} \leftrightarrow u - w = \operatorname{Ann}\alpha$.*

*Доказательство.* Это следует из (4.54), явного разбиения прямоугольной связки (4.52) и **Теоремы B.36**. ■

Таким образом, пересечение $\mathscr{L}$- и $\mathscr{R}$-классов непусто. Для обыкновенных прямоугольных связок каждый $\mathscr{H}$- класс состоит из одного элемента [103, 104]. В нашем случае, однако, ситуация более сложная.

**Определение 4.47.** *Соотношение*

$$\boldsymbol{\Delta}_\alpha^{(1|1)} = \left\{ (\boldsymbol{r}_{tu}, \boldsymbol{r}_{vw}) \mid t - v = \operatorname{Ann}\alpha,\, u - w = \operatorname{Ann}\alpha,\, \boldsymbol{r}_{tu}, \boldsymbol{r}_{vw} \in \mathfrak{R}_\alpha \right\}. \tag{4.56}$$

*назовем двойным $\alpha$-отношением равенства .*

**Теорема 4.48.** *Каждый $\mathscr{H}$-класс в $\boldsymbol{F}_\alpha^{(1|1)}$ состоит из двойных $\boldsymbol{\Delta}_\alpha^{(1|1)}$-эквивалентных элементов, удовлетворяющих $\boldsymbol{r}_{tu}\boldsymbol{\Delta}_\alpha^{(2)}\boldsymbol{r}_{vw}$, и так $\mathscr{H} = \boldsymbol{\Delta}_\alpha^{(1|1)}$.*

*Доказательство.* Из (4.54) и **Определения 4.50** следует, что пересечение $\mathscr{L}$ - и $\mathscr{R}$-классов происходит, когда $\alpha t = \alpha v$ и $\alpha u = \alpha w$. Это дает $t = v + \operatorname{Ann}\alpha$, $u = w + \operatorname{Ann}\alpha$, что совмещается с двойным $\alpha$-отношением равенства (4.56). ■

Рассмотрим отображение $\psi : \boldsymbol{F}_\alpha^{(1|1)} \to \boldsymbol{F}_\alpha^{(1|1)}/\mathscr{R} \times \boldsymbol{F}_\alpha^{(1|1)}/\mathscr{L}$, которое отображает элемент $\boldsymbol{r}_{tu}$ в его $\mathscr{R}$ - и $\mathscr{L}$-классы

$$\psi(\boldsymbol{r}_{tu}) = \left\{ \mathsf{R}_{\boldsymbol{r}_{tu}}, \mathsf{L}_{\boldsymbol{r}_{tu}} \right\}. \tag{4.57}$$

В стандартном случае $\psi$ представляет собой биективное отображение

[103]. Теперь мы имеем

**Утверждение 4.49.** *Отображение $\psi$ представляет собой сюръекцию.*

*Доказательство.* Следует из **Теоремы B.36** и разложения (4.52). ■

Пусть декартово произведение $\boldsymbol{F}_{\alpha}^{(1|1)}/\mathscr{R} \times \boldsymbol{F}_{\alpha}^{(1|1)}/\mathscr{L}$ наделено $\diamond$-умножением прямоугольных связок его $\mathscr{R}$ - и $\mathscr{L}$-классов, аналогичных (4.53), т. е.

$$\left\{\mathsf{R}_{\boldsymbol{r}_{tu}}, \mathsf{L}_{\boldsymbol{r}_{tu}}\right\} \diamond \left\{\mathsf{R}_{\boldsymbol{r}_{vw}}, \mathsf{L}_{\boldsymbol{r}_{vw}}\right\} = \left\{\mathsf{R}_{\boldsymbol{r}_{tu}}, \mathsf{L}_{\boldsymbol{r}_{vw}}\right\}. \tag{4.58}$$

Для стандартных прямоугольных связок отображение $\psi$ является изоморфизмом [103]. В нашем случае мы имеем

**Теорема 4.50.** *Отображение $\psi$ — эпиморфизм.*

*Доказательство.* Во-первых, мы замечаем из (4.54), что

$$\mathsf{R}_{\boldsymbol{r}_{tu} * \boldsymbol{r}_{vw}} = \mathsf{R}_{\boldsymbol{r}_{tu}}, \tag{4.59}$$

$$\mathsf{L}_{\boldsymbol{r}_{tu} * \boldsymbol{r}_{vw}} = \mathsf{L}_{\boldsymbol{r}_{vw}}, \tag{4.60}$$

и таким образом, относительно $\diamond$-умножения (4.58) отображение $\psi$ представляет собой гомоморфизм, поскольку

$$\begin{aligned} \psi\left(\boldsymbol{r}_{tu} * \boldsymbol{r}_{vw}\right) &= \\ \left\{\mathsf{R}_{\boldsymbol{r}_{tu} * \boldsymbol{r}_{vw}}, \mathsf{L}_{\boldsymbol{r}_{tu} * \boldsymbol{r}_{vw}}\right\} &= \left\{\mathsf{R}_{\boldsymbol{r}_{tu}}, \mathsf{L}_{\boldsymbol{r}_{vw}}\right\} = \\ \left\{\mathsf{R}_{\boldsymbol{r}_{tu}}, \mathsf{L}_{\boldsymbol{r}_{tu}}\right\} \diamond \left\{\mathsf{R}_{\boldsymbol{r}_{vw}}, \mathsf{L}_{\boldsymbol{r}_{vw}}\right\} &= \\ &= \psi\left(\boldsymbol{r}_{tu}\right) * \psi\left(\boldsymbol{r}_{vw}\right). \end{aligned} \tag{4.61}$$

Далее, сюръективный гомоморфизм по определению является эпиморфизмом (см. [728]). ■

**4.2.4.** Непрерывные представления высших связок. Почти все полученные результаты могут быть обобщены для прямоугольных связок высшего порядка $(n|n)$, содержащие $2n$ непрерывных четных грассмановых параметров.

Соответствующая матричная конструкция имеет вид

$$\mathrm{F}_{t_1 t_2 \ldots t_n, u_1 u_2 \ldots u_n} \stackrel{def}{=} \begin{pmatrix} 0 & \alpha t_1 \; \alpha t_2 \; \ldots \; \alpha t_n \\ \alpha u_1 & \\ \alpha u_2 & \mathrm{I}\,(n \times n) \\ \vdots & \\ \alpha u_n & \end{pmatrix} \in \mathrm{RMat}_{\Lambda}^{odd}\,(1|n)\,, \tag{4.62}$$

где $t_1, t_2 \ldots t_n, u_1, u_2 \ldots u_n \in \Lambda^{1|0}$ четные параметры, $\alpha \in \Lambda^{1|0}$, $\mathrm{I}\,(n \times n)$ представляет собой единичную матрицу, и матричное умножение имеет вид

$$\mathrm{F}_{t_1 t_2 \ldots t_n, u_1 u_2 \ldots u_n} \mathrm{F}_{t'_1 t'_2 \ldots t'_n, u'_1 u'_2 \ldots u'_n} = \mathrm{F}_{t_1 t_2 \ldots t_n, u'_1 u'_2 \ldots u'_n}. \tag{4.63}$$

Таким образом, идемпотентные суперматрицы $\mathrm{F}_{t_1 t_2 \ldots t_n, u_1 u_2 \ldots u_n}$ образуют полугруппу $\boldsymbol{F}_{\alpha}^{(n|n)}$.

**Определение 4.51.** *Назовем* <u>*$(n|n)$-связкой высшего порядка*</u> *такую полугруппу* $\boldsymbol{F}_{\alpha}^{(n|n)} \ni \boldsymbol{f}_{t_1 t_2 \ldots t_n, u_1 u_2 \ldots u_n}$*, которая представляется суперматрицами* $\mathrm{F}_{t_1 t_2 \ldots t_n, u_1 u_2 \ldots u_n}$ *из* $\mathrm{RMat}_{\Lambda}^{odd}\,(1|n)$ *вида* (4.62).

Результаты, изложенные в **Приложении B.8**, с некоторыми незначительными отличиями справедливы также и для $\boldsymbol{F}_{\alpha}^{(n|n)}$.

**Определение 4.52.** *В* $\boldsymbol{F}_{\alpha}^{(n|n)}$ *соотношение*

$$\begin{gathered} \boldsymbol{\Delta}_{\alpha}^{(n|n)} \stackrel{def}{=} \left\{ \left( \boldsymbol{f}_{t_1 t_2 \ldots t_n, u_1 u_2 \ldots u_n}, \boldsymbol{f}_{t'_1 t'_2 \ldots t'_n, u'_1 u'_2 \ldots u'_n} \right) \mid t_k - t'_k = \mathrm{Ann}\,\alpha, \right. \\ \left. u_k - u'_k = \mathrm{Ann}\,\alpha,\; 1 \leq k \leq n,\; \boldsymbol{f}_{t_1 t_2 \ldots t_n, u_1 u_2 \ldots u_n}, \boldsymbol{f}_{t'_1 t'_2 \ldots t'_n, u'_1 u'_2 \ldots u'_n} \in \mathscr{F}_{\alpha}^{(2n)} \right\} \end{gathered} \tag{4.64}$$

*назовем $(n|n)$-ым $\alpha$-отношением равенства.*

*Замечание* **4.53.** Полугруппа $\boldsymbol{F}_{\alpha}^{(n|n)}$ эпиморфна полугруппе $\boldsymbol{F}_{\alpha}$, и два $\boldsymbol{\Delta}_{\alpha}^{(n|n)}$-эквивалентных элемента $\boldsymbol{F}_{\alpha}^{(n|n)}$ имеют тот же образ.

Рассмотрим идемпотентные суперматрицы из $\mathrm{RMat}_{\Lambda}^{odd}(k|m)$ вида

$$\mathrm{F}_{tu} \stackrel{def}{=} \begin{pmatrix} 0 & \alpha\mathrm{T} \\ \alpha\mathrm{U} & \mathrm{I} \end{pmatrix}, \tag{4.65}$$

где $\mathrm{T}\,(k \times m)$ и $\mathrm{U}\,(m \times k)$ представляют собой обыкновенные матрицы четных параметров связки, и $\mathrm{I}\,(m \times m)$ — единичная матрица. Данная связка содержит максимум $2km$ параметров из $\Lambda^{1|0}$. Умножение в этой связке есть

$$\begin{pmatrix} 0 & \alpha\mathrm{T} \\ \alpha\mathrm{U} & \mathrm{I} \end{pmatrix} \begin{pmatrix} 0 & \alpha\mathrm{T}' \\ \alpha\mathrm{U}' & \mathrm{I} \end{pmatrix} = \begin{pmatrix} 0 & \alpha\mathrm{T} \\ \alpha\mathrm{U}' & \mathrm{I} \end{pmatrix}, \tag{4.66}$$

что в блочном виде совпадает с умножением прямоугольной связки (4.53)

$$\mathrm{F}_{\mathrm{TU}}\mathrm{F}_{\mathrm{T}'\mathrm{U}'} = \mathrm{F}_{\mathrm{TU}'}. \tag{4.67}$$

**Теорема 4.54.** *Если $n = km$, то представления, заданные* (4.62) *и* (4.65)*, изоморфны.*

*Доказательство.* Поскольку в (4.63) и (4.67) не имеется перемножения между параметрами, то представления, заданные матрицами (4.62) и (4.65), отличаются перестановкой, если $n = km$. ■

**Следствие 4.55.** *Суперматрицы $\mathrm{F}_{t_1 t_2 \ldots t_n, u_1 u_2 \ldots u_n}$ из $\mathrm{RMat}_{\Lambda}^{odd}(1|n)$, имеющие вид* (4.62)*, исчерпывают все возможные непрерывные представления $(n|n)$-связок.*

*Замечание* **4.56.** Суперматрицы (4.62) представляют также $(k|m)$-связки, где $1 \leq k \leq n$, $1 \leq m \leq n$. В этом случае $t_{k+1} = 1 + \operatorname{Ann}\alpha, \ldots t_n = 1 + \operatorname{Ann}\alpha$, $u_{m+1} = 1 + \operatorname{Ann}\alpha, \ldots u_n = 1 + \operatorname{Ann}\alpha$. Таким образом, вышеупомянутый изоморфизм имеет место для различных связок, имеющих равное количество параметров. Поэтому мы будем рассматривать ниже в основном полные $(n|n)$-связки, подразумевая, что они содержат все частные и редуцированные случаи.

*Замечание* **4.57.** Для $k = 0$ и $m = 0$ они описывают $m$- правые полугруппы нулей $\boldsymbol{Q}_\alpha^{(m)}$ и $k$-левые полугруппы нулей $\boldsymbol{P}_\alpha^{(k)}$ соответственно, имеющие следующие законы умножения (ср. (B.51) и (B.54))

$$\begin{aligned} \boldsymbol{q}_{u_1 u_2 \ldots u_m} * \boldsymbol{q}_{u_1' u_2' \ldots u_m'} = \boldsymbol{q}_{u_1' u_2' \ldots u_m'}, \\ \boldsymbol{p}_{t_1 t_2 \ldots t_k} * \boldsymbol{p}_{t_1' t_2' \ldots t_k'} = \boldsymbol{p}_{t_1 t_2 \ldots t_k}. \end{aligned} \tag{4.68}$$

**Предложение 4.58.** *$m$-правые полугруппы нулей $\boldsymbol{Q}_\alpha^{(m)}$ и $k$-левые полугруппы нулей $\boldsymbol{P}_\alpha^{(k)}$ неприводимы в том смысле, что они не могут быть представлены в качестве прямого произведения “1-мерных” полугрупп правых нулей $\boldsymbol{Q}_\alpha$ и левых нулей $\boldsymbol{P}_\alpha$ соответственно.*

*Доказательство.* Следует непосредственно из сравнения структуры суперматриц (4.42), (4.47) и (4.62). ■

**Предложение 4.59.** *Для построения $(k|m)$-связки нельзя использовать “1-мерные” полугруппы правых нулей $\boldsymbol{Q}_\alpha$ и левых нулей $\boldsymbol{P}_\alpha$, потому, что они сводят его к обыкновенной “2-мерной” прямоугольной связке.*

*Доказательство.* В самом деле, пусть

$$\tilde{\boldsymbol{f}}_{t_1 t_2 \ldots t_k, u_1 u_2 \ldots u_m} = \boldsymbol{p}_{t_1} * \boldsymbol{p}_{t_2} \ldots * \boldsymbol{p}_{t_k} * \boldsymbol{q}_{u_1} * \boldsymbol{q}_{u_2} \ldots * \boldsymbol{q}_{u_m},$$

тогда, используя таблицу Кэли, имеем

$$\tilde{\boldsymbol{f}}_{t_1 t_2 \ldots t_k, u_1 u_2 \ldots u_m} = \boldsymbol{p}_{t_1} * \boldsymbol{q}_{u_m},$$

что тривиально совпадает с (4.52). Таким образом, любая комбинация элементов из "1-мерных" полугрупп правых и левых нулей не будет приводить к новым конструкциям, отличным от тех что перечислены в таблице Кэли. ■

Вместо этого мы имеем следующую декомпозицию $(k|m)$-связки в $k$-полугруппу левых нулей $\boldsymbol{P}_{\alpha}^{(k)}$ и $m$- полугруппу правых нулей $\boldsymbol{Q}_{\alpha}^{(m)}$

$$\boldsymbol{f}_{t_1 t_2 \ldots t_k, u'_1 u'_2 \ldots u'_m} = \boldsymbol{p}_{t_1 t_2 \ldots t_k} * \boldsymbol{q}_{u'_1 u'_2 \ldots u'_m}. \tag{4.69}$$

Несмотря на то, что эта формула аналогична (4.52), мы подчеркиваем, что увеличение числа суперпараметров не искусственный прием, а естественный путь к поиску новых построений, приводящих к обобщению отношений Грина и тонкого идеального строения $(n|n)$-связки, что не имеет аналогов в стандартном подходе [103, 104].

**4.2.5.** Тонкое идеальное строение высших связок. Рассмотрим отношения Грина для $(n|n)$-связки. Мы будем пытаться установить смысл свойств $\mathscr{R}, \mathscr{L}, \mathscr{D}, \mathscr{H}$-классов для суперматриц. Это позволит определить и изучить новые эквивалентности [10], обобщающие отношения Грина, равно как и прояснить предыдущие конструкции.

Мы строим искомое представление только для $(2|2)$-связок, имея ввиду то, что расширить все результаты на $(n|n)$-связки можно без труда простыми переобозначениями. Так, $\mathscr{R}$-эквивалентные элементы в этом частном случае рассмотрены в **Приложении B.9**. Продолжая его на общий случай $(n|n)$-связок, получаем общее

**Определение 4.60.** *$\mathscr{R}$-классы в $(n|n)$-связке состоят из элементов, имеющих все (!) $\alpha t_k$ фиксированными, где $1 \leq k \leq n$.*

Как дуальный аналог этого определения мы формулируем

**Определение 4.61.** *$\mathscr{L}$-классы в $(n|n)$-связке состоят из элементов, имеющих все (!) $\alpha u_k$ фиксированными, где $1 \leq k \leq n$.*

В такой картине очевидно, что объединение этих соотношений $\mathscr{D} = \mathscr{R} \vee \mathscr{L}$ покрывает все возможные элементы, и, следовательно, любые два элемента в $(n|n)$-связке $\mathscr{D}$-эквивалентны (см. предложение 4.45). Их пересечение $\mathscr{H} = \mathscr{R} \cap \mathscr{L}$ очевидно состоит из элементов со *всеми* (!) $\alpha t_k$ и $\alpha u_k$ фиксированными. Именно здесь — источник формулировки $(n|n)$-ых $\alpha$-отношений равенства (4.64).

**Предложение 4.62.** *В $(2|2)$-связке $\mathscr{J}$-отношение совпадает с универсальным отношением $\boldsymbol{\Delta}$.*

*Доказательство.* Умножая (B.58) на $\mathrm{F}_{t_1t_2,u_1u_2}$ справа и на $\mathrm{X}_{x_1x_2,y_1y_2}$ слева, мы получаем

$$\begin{aligned} \mathrm{F}_{t_1t_2,u_1u_2} \cdot \mathrm{X}_{x_1x_2,y_1y_2} \cdot \mathrm{F}_{t_1t_2,u_1u_2} = \mathrm{F}_{t_1t_2,u_1u_2}, \\ \mathrm{X}_{x_1x_2,y_1y_2} \cdot \mathrm{F}_{t_1t_2,u_1u_2} \cdot \mathrm{X}_{x_1x_2,y_1y_2} = \mathrm{X}_{x_1x_2,y_1y_2} \end{aligned} \tag{4.70}$$

для любых $t_1, t_2, u_1, u_2, x_1, x_2, y_1, y_2 \in \Lambda^{(1|0)}$, что совпадает с определением $\mathscr{J}$-отношения. Произвольность $\mathrm{F}_{t_1t_2,u_1u_2}$ и $\mathrm{X}_{x_1x_2,y_1y_2}$ доказывает утверждение. ■

Следовательно, мы имеем следующие абстрактные определения для $(2|2)$-связок

$$\boldsymbol{f}_{t_1t_2,u_1u_2} \mathscr{R} \boldsymbol{f}_{t'_1t'_2,u'_1u'_2} \iff \{\alpha t_1 = \alpha t'_1 \wedge \alpha t_2 = \alpha t'_2\}, \tag{4.71}$$

$$\boldsymbol{f}_{t_1t_2,u_1u_2} \mathscr{L} \boldsymbol{f}_{t'_1t'_2,u'_1u'_2} \iff \{\alpha u_1 = \alpha u'_1 \wedge \alpha u_2 = \alpha u'_2\}, \tag{4.72}$$

$$\boldsymbol{f}_{t_1t_2,u_1u_2}\mathscr{D}\boldsymbol{f}_{t'_1t'_2,u'_1u'_2} \iff \left\{ \begin{array}{l} (\alpha t_1 = \alpha t'_1 \wedge \alpha t_2 = \alpha t'_2) \vee \\ (\alpha u_1 = \alpha u'_1 \wedge \alpha u_2 = \alpha u'_2) \end{array} \right\}, \tag{4.73}$$

$$\boldsymbol{f}_{t_1t_2,u_1u_2}\mathscr{H}\boldsymbol{f}_{t'_1t'_2,u'_1u'_2} \iff \left\{ \begin{array}{l} (\alpha t_1 = \alpha t'_1 \wedge \alpha t_2 = \alpha t'_2) \wedge \\ (\alpha u_1 = \alpha u'_1 \wedge \alpha u_2 = \alpha u'_2) \end{array} \right\}. \tag{4.74}$$

Теперь мы разбиремся, что отсутствовало в стандартном подходе и введем обобщения отношений Грина.

Из (4.71) и (4.72) видно, что различные четыре возможности для удовлетворения равенств не исчерпываются ординарными отношениями $\mathscr{R}$ - и $\mathscr{L}$-эквивалентности. Ясно, почему мы писали выше восклицательные знаки: эти утверждения *будут исправляться*. Таким образом, мы вынуждены определить более общие отношения, "тонкие отношения эквивалентности". Они достаточны для описания всех возможных классов элементов в $(n|n)$-связках, пропущенных в стандартном подходе [104, 204].

Во-первых, мы определим их для использования в нашем частном случае.

**Определение 4.63.** *Тонкие $\mathscr{R}^{(k)}$- и $\mathscr{L}^{(k)}$-отношения в $(2|2)$-связке определяются следующим образом*

$$\boldsymbol{f}_{t_1t_2,u_1u_2}\mathscr{R}^{(1)}\boldsymbol{f}_{t'_1t'_2,u'_1u'_2} \iff \{\alpha t_1 = \alpha t'_1\}, \tag{4.75}$$

$$\boldsymbol{f}_{t_1t_2,u_1u_2}\mathscr{R}^{(2)}\boldsymbol{f}_{t'_1t'_2,u'_1u'_2} \iff \{\alpha t_2 = \alpha t'_2\}, \tag{4.76}$$

$$\boldsymbol{f}_{t_1t_2,u_1u_2}\mathscr{L}^{(1)}\boldsymbol{f}_{t'_1t'_2,u'_1u'_2} \iff \{\alpha u_1 = \alpha u'_1\}, \tag{4.77}$$

$$\boldsymbol{f}_{t_1t_2,u_1u_2}\mathscr{L}^{(2)}\boldsymbol{f}_{t'_1t'_2,u'_1u'_2} \iff \{\alpha u_2 = \alpha u'_2\}. \tag{4.78}$$

**Предложение 4.64.** *Тонкие $\mathscr{R}^{(k)}$- и $\mathscr{L}^{(k)}$-отношения являются отношениями эквивалентности.*

*Доказательство.* Следует из явного вида умножения (4.62) и (B.58)–

(B.59). ■

Поэтому они подразделяют связку $\boldsymbol{F}_{\alpha}^{(2|2)}$ на четыре тонких класса эквивалентности $\boldsymbol{F}_{\alpha}^{(2|2)}/\mathscr{R}^{(k)}$ и $\boldsymbol{F}_{\alpha}^{(2|2)}/\mathscr{L}^{(k)}$ следующим образом

$$\mathsf{R}_{\boldsymbol{f}}^{(1)} = \left\{ \boldsymbol{f}_{t_1 t_2, u_1 u_2} \in \boldsymbol{F}_{\alpha}^{(2|2)} \,|\, \alpha t_1 = const \right\}, \tag{4.79}$$

$$\mathsf{R}_{\boldsymbol{f}}^{(2)} = \left\{ \boldsymbol{f}_{t_1 t_2, u_1 u_2} \in \boldsymbol{F}_{\alpha}^{(2|2)} \,|\, \alpha t_2 = const \right\}, \tag{4.80}$$

$$\mathsf{L}_{\boldsymbol{f}}^{(1)} = \left\{ \boldsymbol{f}_{t_1 t_2, u_1 u_2} \in \boldsymbol{F}_{\alpha}^{(2|2)} \,|\, \alpha u_1 = const \right\}, \tag{4.81}$$

$$\mathsf{L}_{\boldsymbol{f}}^{(2)} = \left\{ \boldsymbol{f}_{t_1 t_2, u_1 u_2} \in \boldsymbol{F}_{\alpha}^{(2|2)} \,|\, \alpha u_2 = const \right\}. \tag{4.82}$$

Для прозрачности мы можем схематично представить

$$\begin{array}{cc} & \begin{array}{ccc} & \mathsf{R}_{\boldsymbol{f}}^{(1)} & \mathsf{R}_{\boldsymbol{f}}^{(2)} \\ & \updownarrow & \updownarrow \end{array} \\ \begin{array}{c} \\ \mathsf{L}_{\boldsymbol{f}}^{(1)} \leftrightarrow \\ \mathsf{L}_{\boldsymbol{f}}^{(2)} \leftrightarrow \end{array} & \begin{pmatrix} 0 & \alpha t_1 & \alpha t_2 \\ \alpha u_1 & 1 & 0 \\ \alpha u_2 & 0 & 1 \end{pmatrix}, \end{array} \tag{4.83}$$

где стрелки показывают, который элемент суперматрицы фиксируется согласно данному тонкому отношению эквивалентности.

Отсюда мы можем получать также и все известные отношения

$$\mathscr{R}^{(1)} \cap \mathscr{R}^{(2)} = \mathscr{R}, \tag{4.84}$$

$$\mathscr{L}^{(1)} \cap \mathscr{L}^{(2)} = \mathscr{L}, \tag{4.85}$$

$$\left(\mathscr{R}^{(1)} \cap \mathscr{R}^{(2)}\right) \cap \left(\mathscr{L}^{(1)} \cap \mathscr{L}^{(2)}\right) = \mathscr{H}, \tag{4.86}$$

$$\left(\mathscr{R}^{(1)} \cap \mathscr{R}^{(2)}\right) \vee \left(\mathscr{L}^{(1)} \cap \mathscr{L}^{(2)}\right) = \mathscr{D}. \tag{4.87}$$

Однако, кроме стандартных, имеется много других “смешанных” эквивалентностей [10], которые могут классифицироваться, используя

определения

$$\mathscr{H}^{(i|j)} = \mathscr{R}^{(i)} \cap \mathscr{L}^{(j)}, \tag{4.88}$$

$$\mathscr{D}^{(i|j)} = \mathscr{R}^{(i)} \vee \mathscr{L}^{(j)}, \tag{4.89}$$

$$\mathscr{H}^{(ij|k)} = \left(\mathscr{R}^{(i)} \cap \mathscr{R}^{(j)}\right) \cap \mathscr{L}^{(k)}, \tag{4.90}$$

$$\mathscr{H}^{(i|kl)} = \mathscr{R}^{(i)} \cap \left(\mathscr{L}^{(k)} \cap \mathscr{L}^{(l)}\right), \tag{4.91}$$

$$\mathscr{D}^{(ij|k)} = \left(\mathscr{R}^{(i)} \cap \mathscr{R}^{(j)}\right) \vee \mathscr{L}^{(k)}, \tag{4.92}$$

$$\mathscr{D}^{(i|kl)} = \mathscr{R}^{(i)} \vee \left(\mathscr{L}^{(k)} \cap \mathscr{L}^{(l)}\right). \tag{4.93}$$

Графическая интерпретация смешанных отношений эквивалентности дается диаграммой на *Рис.* 4.1.

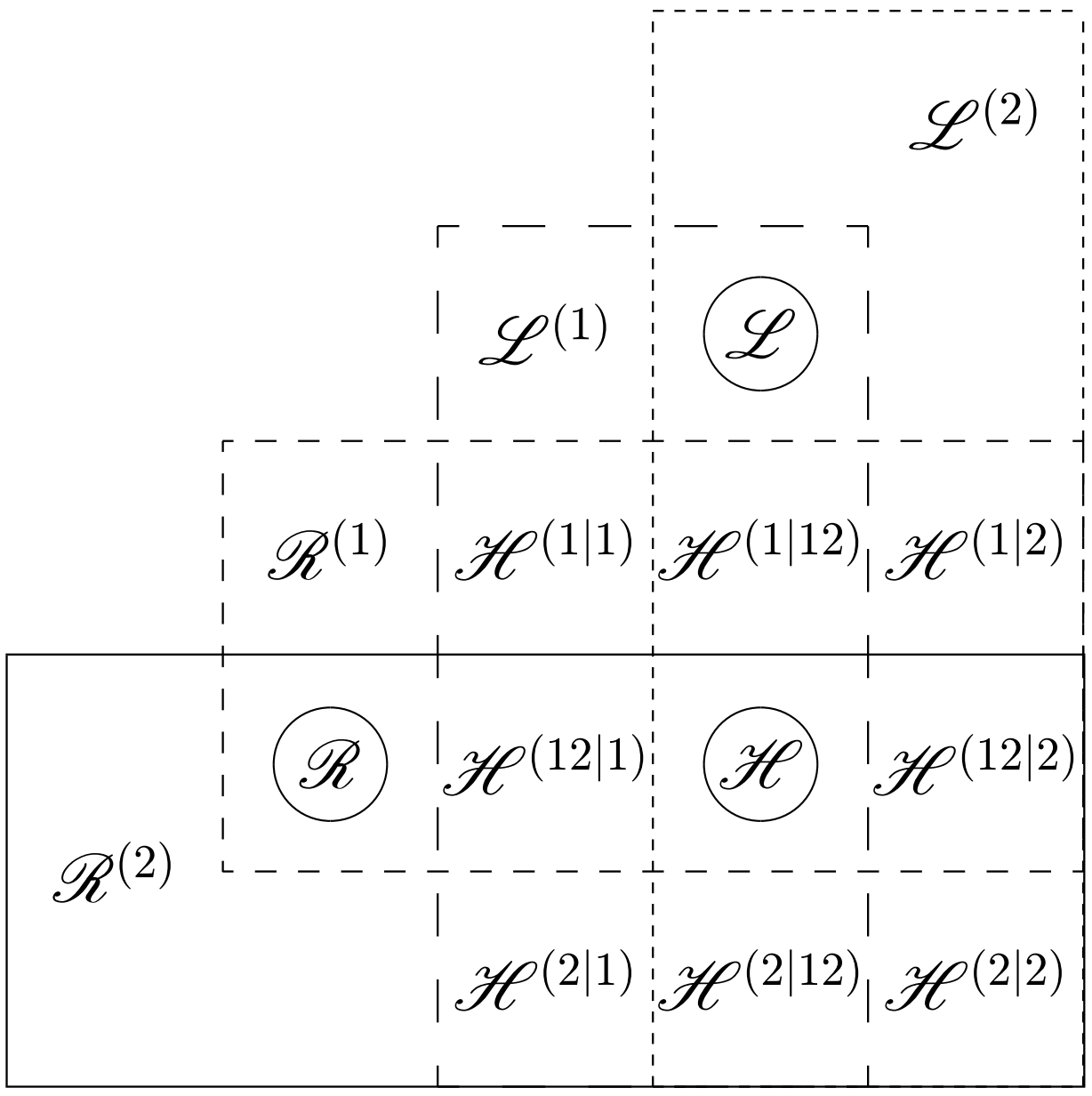

*Рис.* 4.1. *Тонкие отношения эквивалентности для (2|2)-связки (кружками отмечены стандартные отношения Грина)*

*Замечание* **4.65.** Стандартные $\mathscr{R}$ - и $\mathscr{L}$-отношения на *Рис.* 4.1 занимают 4 малых квадрата в длину, $\mathscr{H}^{(i|j)}$-отношения занимают 4 малых квадратов в квадрате, $\mathscr{H}^{(ij|k)}$ - и $\mathscr{H}^{(i|jk)}$-отношения занимают 2 малых квадрата, стандартное $\mathscr{H}$-отношение занимает 1 малый квадрат.

Мы замечаем, что смешанные отношения (4.88)–(4.93) в некотором смысле “шире”, чем стандартные (4.84)–(4.87). Поэтому, используя их, мы можем описать соответствующим образом все классы элементов из $(n|n)$-связки, включая те, что отсутствуют, если использовать только стандартные отношения Грина $^{*)}$.

Для каждого смешанного отношения мы можем определить соответствующий класс, используя очевидные определения. Тогда для каждого смешанного $\mathscr{D}$-класса мы можем построить смешанную eggbox диаграмму [104] тонких $\mathscr{R}, \mathscr{L}$-классов, которая будет такой размерности, сколько слагаемых имеет в своей правой части заданное смешанное отношение (4.89), (4.92) и (4.93). Например, eggbox диаграммы $\mathscr{D}^{(i|j)}$-классов двумерны, а диаграммы $\mathscr{D}^{(ij|k)}$ и $\mathscr{D}^{(i|jk)}$-классов должны быть трехмерны. В случае $(n|n)$-связки необходимо рассматривать все возможные $k$-размерные eggbox диаграммы, где $2 \leq k \leq n-1$.

Введенные тонкие отношения эквивалентности (4.75)–(4.78) допускают подполугрупповую интерпретацию.

**Лемма 4.66.** *Элементы из* $\boldsymbol{F}_{\alpha}^{(n|n)}$*, имеющие* $\alpha t_k = \beta_k$ *и* $\alpha u_k = \gamma_k$*, где* $\beta_k, \gamma_k \in \Lambda^{0|1}$ *фиксированы, и* $1 \leq k \leq m$*, образуют различные подполугруппы индекса* $m$*.*

*Доказательство.* Следует из явного вида матричного умножения суперматриц формы (4.62). ■

Рассмотрим различные подполугруппы индекса $(n-1)$ полугруппы $\boldsymbol{F}_{\alpha}^{(n|n)}$. Они состоят из элементов, имеющих все, кроме одного, $\alpha t_k$ и все, кроме одного, $\alpha u_k$ фиксированные.

---

*Примечание.* Для неотрицательных обычных матриц обобщенные (в ином смысле) отношения Грина были исследованы в [729].

Пусть такие элементы

$$\boldsymbol{U}_{\alpha}^{(k)} \stackrel{def}{=} \left\{ \boldsymbol{f}_{t_1 t_2 \ldots t_n, u_1 u_2 \ldots u_n} \in \boldsymbol{F}_{\alpha}^{(n|n)} \mid \bigwedge_{i \neq k} \alpha t_i = \beta_i \bigwedge_{i \neq k} \alpha u_i = \gamma_i \right\} \tag{4.94}$$

представляют собой подполугруппу индекса $(n-1)$, которая имеет только одну нефиксированную пару $\alpha t_k$, $\alpha u_k$. Стандартные отношения Грина [104] на подполугруппе $\boldsymbol{U}_{\alpha}^{(k)}$ следующие

$$\begin{aligned}
\boldsymbol{f}_{t_1 t_2 \ldots t_n, u_1 u_2 \ldots u_n} \mathscr{R}_{\boldsymbol{U}}^{(k)} \boldsymbol{f}_{t_1' t_2' \ldots t_n', u_1' u_2' \ldots u_n'} &\Leftrightarrow \left\{ \alpha t_k = \alpha t_k' \right\}, \qquad (4.95)\\
\boldsymbol{f}_{t_1 t_2 \ldots t_n, u_1 u_2 \ldots u_n} \mathscr{L}_{\boldsymbol{U}}^{(k)} \boldsymbol{f}_{t_1' t_2' \ldots t_n', u_1' u_2' \ldots u_n'} &\Leftrightarrow \left\{ \alpha u_k = \alpha u_k' \right\},\\
\boldsymbol{f}_{t_1 t_2 \ldots t_n, u_1 u_2 \ldots u_n} \mathscr{H}_{\boldsymbol{U}}^{(k)} \boldsymbol{f}_{t_1' t_2' \ldots t_n', u_1' u_2' \ldots u_n'} &\Leftrightarrow \left\{ \alpha t_k = \alpha t_k' \wedge \alpha u_k = \alpha u_k' \right\},\\
\boldsymbol{f}_{t_1 t_2 \ldots t_n, u_1 u_2 \ldots u_n} \mathscr{D}_{\boldsymbol{U}}^{(k)} \boldsymbol{f}_{t_1' t_2' \ldots t_n', u_1' u_2' \ldots u_n'} &\Leftrightarrow \left\{ \alpha t_k = \alpha t_k' \vee \alpha u_k = \alpha u_k' \right\},
\end{aligned}$$

где $\boldsymbol{f}_{t_1 t_2 \ldots t_n, u_1 u_2 \ldots u_n}, \boldsymbol{f}_{t_1' t_2' \ldots t_n', u_1' u_2' \ldots u_n'} \in \boldsymbol{U}_{\alpha}^{(k)} \subset \boldsymbol{F}_{\alpha}^{(n|n)}$.

**Теорема 4.67.** *Отношения Грина на $\boldsymbol{U}_{\alpha}^{(k)}$ представляют собой сужение соответствующих тонких отношений* (4.75)–(4.78) *на $\boldsymbol{F}_{\alpha}^{(n|n)}$ подполугруппу $\boldsymbol{U}_{\alpha}^{(k)}$*

$$\mathscr{R}_{\boldsymbol{U}}^{(k)} = \mathscr{R}^{(k)} \cap \left( \boldsymbol{U}_{\alpha}^{(k)} \times \boldsymbol{U}_{\alpha}^{(k)} \right), \tag{4.96}$$

$$\mathscr{L}_{\boldsymbol{U}}^{(k)} = \mathscr{L}^{(k)} \cap \left( \boldsymbol{U}_{\alpha}^{(k)} \times \boldsymbol{U}_{\alpha}^{(k)} \right), \tag{4.97}$$

$$\mathscr{H}_{\boldsymbol{U}}^{(k)} = \mathscr{H}^{(k|k)} \cap \left( \boldsymbol{U}_{\alpha}^{(k)} \times \boldsymbol{U}_{\alpha}^{(k)} \right), \tag{4.98}$$

$$\mathscr{D}_{\boldsymbol{U}}^{(k)} = \mathscr{D}^{(k|k)} \cap \left( \boldsymbol{U}_{\alpha}^{(k)} \times \boldsymbol{U}_{\alpha}^{(k)} \right). \tag{4.99}$$

*Доказательство.* Достаточно доказать утверждение для частного случая $\boldsymbol{F}_{\alpha}^{(2|2)}$ и $\boldsymbol{U}_{\alpha}^{(1)}$, а затем получить общее утверждение по индукции. Используя определение $\mathscr{R}$-класса в явном виде (B.58)–(B.59), мы заключаем, что условие $\alpha t_1 = \alpha t_1'$ общее для тонкого $\mathscr{R}^{(k)}$-класса и для подполугруппы $\mathscr{R}_{\boldsymbol{U}}^{(k)}$-классов. По аналогии можно доказать и остальные равенства. ■

*Замечание* **4.68.** Второе условие $\alpha t_2 = \alpha t_2'$ (что представляет собой вторую часть определения обыкновенного $\mathscr{R}$-отношения для $\boldsymbol{F}_{\alpha}^{(2|2)}$ (4.71)) выполняется также в $\boldsymbol{U}_{\alpha}^{(1)}$, но из-за собственного определения подполугруппы ($\alpha t_2 = \beta_2 = const$, $\alpha u_2 = \gamma_2 = const$), однако $\alpha t_2 = \alpha t_2'$ вообще не входят в тонкие отношения $\mathscr{R}^{(k)}$. Поэтому последнее представляет собой наиболее общее отношение среди рассматриваемых $\mathscr{R}$-отношений.

*Замечание* **4.69.** Можно рассматривать доказанную **Теорему 4.67** с точки зрения [238], где доказывались формулы, подобные (4.96)–(4.98), но с обычными отношениями Грина в правой части. Обращаясь к диаграмме на *Рис.* 4.1, мы делаем вывод, что наш результат содержит обычный случай [238] в качестве частного.

Кроме того, мы предполагаем, что **Теорема 4.67** имеет более глубокий смысл и дает другую общую трактовку тонким отношениям эквивалентности для абстрактных полугрупп.

**Предположение 4.70.** *Отношения Грина на подполугруппе* $\mathbf{U}$ *полугруппы* $\mathbf{S}$ *имеют как свой аналог продолженные образы в* $\mathbf{S}$, *а именно — тонкие отношения эквивалентности на* $\mathbf{S}$.

Мы доказали это утверждение для частного случая непрерывных представлений $(n|n)$-связок. Важно исследовать и другие алгебраические системы, где **Предположение 4.70** истинно.

## 4.3. Основные результаты и выводы

1. Построены новые суперматричные полугруппы и исследованы их идеальные свойства
2. Предложены нетривиальные редукции необратимых суперматриц, которые играют важную роль в приложениях.
3. Антитреугольные суперматрицы объединены с треугольными в различные сэндвич-полугруппы с необычными свойствами.
4. Получены новые типы нечетных супермодулей и антитранспонирования.
5. Найдены представления странной супералгебры Березина.
6. Введены нечетные аналоги собственных чисел, характеристических функций и сформулирована обобщенная теорема Гамильтона-Якоби.
7. Показано, что полугрупповые связки непрерывно представляются суперматричными полугруппами антитреугольного вида.
8. Построено непрерывное представление скрученной прямоугольной связки однопараметрическими суперматричными полугруппами, вычислены отношения Грина для различных подполугрупп.
9. Определен новый тип высших связок и для них введены обобщения отношений Грина — тонкие и смешанные отношения эквивалентности, которые трактуются как продолжения стандартных отношений Грина с подполугруппы на всю полугруппу.
10. Введены и изучены многомерные аналоги eggbox диаграмм для высших связок.

# <u>РАЗДЕЛ 5</u>

# ПЕРМАНЕНТЫ, SCF-МАТРИЦЫ И НЕОБРАТИМАЯ ГИПЕРБОЛИЧЕСКАЯ ГЕОМЕТРИЯ

В данном разделе исследуются необратимые свойства матриц, содержащих нильпотентные элементы и делители нуля, определенный тип которых возникает при анализе $N$-расширенных редуцированных преобразований. Показывается, что перманенты играют для них дуальную (по отношению к детерминантам) роль в большинстве принципиальных формул и утверждений (даже в нахождении обратной матрицы). Эти дуальные свойства изучаются в общем случае матриц содержащих нильпотентные элементы, что может быть применено во многих моделях элементарных частиц, использующих суперсимметрию в качестве основополагающего принципа.

Введенные матрицы используются для определения обратимых и необратимых дробно-линейных преобразований специального вида, для которых найден новый вид симметрии. Строится необратимая гиперболическая геометрия на четной части суперплоскости, в которой имеется два различно определенных инвариантных двойных отношения и два гиперболических расстояния, аналог производной Шварца и других классических формул.

## 5.1. Свойства scf-матриц и их перманентов

Свойства перманентов обычных матриц отличаются от свойств детерминантов (см. **Приложение Д**), что до сих пор существенно огра-ничивало их применение комбинаторными построениями и вероятностными задачами [626], а также теорией инвариантов [730] и перманентных идеалов [731]. Однако, если матрицы содержат нильпотентные элементы и делители нуля, то для некоторого типа матриц, возникающих при анализе $N$-расширенных суперконформных преобразований (см. **Раздел 3**), перманенты начинают играть дуальную (по отношению к детерминантам) роль [2]. Поэтому важно рассмотреть эти дуальные свойства в общем случае нильпотентных матриц, что может быть применено и в других моделях, использующих суперсимметрию в качестве основополагающего принципа.

**5.1.1.** $N = 2$ scf-матрицы. Рассмотрим сначала четные $2 \times 2$ матрицы с элементами из $\Lambda_0$, т. е. $\mathrm{A}=\begin{pmatrix} a & b \\ c & d \end{pmatrix} \in \mathrm{Mat}_{\Lambda_0}(2)$. Тогда из общей формулы (Д.3) следует, что

$$\mathrm{per}\,\mathrm{A} = ad + bc. \tag{5.1}$$

Если определить скалярное произведение стандартным образом

$$\mathrm{A} \times \mathrm{B} \stackrel{def}{=} \mathrm{tr}\,\mathrm{A}\mathrm{B}^T, \tag{5.2}$$

то для перманента суммы матриц получаем

$$\mathrm{per}\,(\mathrm{A} + \mathrm{B}) = \mathrm{per}\,\mathrm{A} + \mathrm{per}\,\mathrm{B} + \mathrm{A} \times \mathrm{B}^M, \tag{5.3}$$

где $\mathrm{B}^T$ — транспонированная матрица и $\mathrm{B}^M$ — матрица миноров.

Из (5.3) следуют важные частные случаи (см. (Д.6)), которые бу-дут использованы в дальнейших выкладках,

$$\operatorname{per}\,(\mathrm{A} - k\mathrm{I}) \;=\; k^2 - k \cdot \operatorname{tr}\mathrm{A} + \operatorname{per}\mathrm{A}, \tag{5.4}$$

$$\operatorname{per}\left(\mathrm{A} - \mathrm{A}^{MT}\right) \;=\; 2\operatorname{per}\mathrm{A} - \operatorname{tr}\mathrm{A}^2, \tag{5.5}$$

где $k \in \Lambda_0$ и I — единичная матрица.

Отсюда следует определение перманента $2 \times 2$ матрицы в терминах скалярного произведения

$$\operatorname{per}\mathrm{A} = \frac{1}{2}\operatorname{tr}\mathrm{A} \times \mathrm{A}^M \tag{5.6}$$

(ср. (Д.2)).

*Замечание* **5.1.** Если матрица A не содержит нильпотентных составляющих и положительна и $\operatorname{per}\mathrm{A} = 1$, то матрица $\mathrm{B} = \mathrm{A} - \mathrm{I}$ нильпотентна [732].

Введем в рассмотрение еще одну матричную функцию $\operatorname{scf}_{\pm}\mathrm{A}$, которая играет важную роль при рассмотрении свойств матриц, содержащих нильпотентные элементы, по формулам

$$\operatorname{scf}_{+}\mathrm{A} \overset{def}{=} ac, \;\; \operatorname{scf}_{-}\mathrm{A} \overset{def}{=} bd, \tag{5.7}$$

т. е. $\operatorname{scf}_{\pm}\mathrm{A}$ определяет степень ортогональности элементов первого и второго столбца матрицы A соответственно.

Необходимость введения функции $\operatorname{scf}_{\pm}\mathrm{A}$ видна из следующего ключевого соотношения

$$\mathrm{A}^{MT} \cdot \mathrm{A} = \begin{pmatrix} \operatorname{per}\mathrm{A} & 2\operatorname{scf}_{-}\mathrm{A} \\ 2\operatorname{scf}_{+}\mathrm{A} & \operatorname{per}\mathrm{A} \end{pmatrix}. \tag{5.8}$$

Сравним это соотношение с подобным для детерминанта

$$\mathrm{A}^{DT} \cdot \mathrm{A} = \begin{pmatrix} \det \mathrm{A} & 0 \\ 0 & \det \mathrm{A} \end{pmatrix} = \det \mathrm{A} \cdot \mathrm{I}, \tag{5.9}$$

где $\mathrm{A}^{D}$ — матрица алгебраических дополнений.

Тогда естественным является следующее

**Определение 5.2.** *$N = 2$ scf-матрица — это $2 \times 2$ четная матрица с элементами из $\Lambda_0$, у которой элементы столбцов ортогональны*

$$\mathrm{scf}_{\pm}\mathrm{A}_{\mathrm{scf}} = 0. \tag{5.10}$$

**Следствие 5.3.** *Для $N = 2$ scf-матриц выполняется соотношение, аналогичное* (5.9)

$$\mathrm{A}_{\mathrm{scf}}^{MT} \cdot \mathrm{A}_{\mathrm{scf}} = \begin{pmatrix} \operatorname{per} \mathrm{A}_{\mathrm{scf}} & 0 \\ 0 & \operatorname{per} \mathrm{A}_{\mathrm{scf}} \end{pmatrix} = \operatorname{per} \mathrm{A}_{\mathrm{scf}} \cdot \mathrm{I}, \tag{5.11}$$

*и, следовательно, имеет место дуальность*

$$\operatorname{per} \mathrm{A}_{\mathrm{scf}} \leftrightarrow \det \mathrm{A}_{\mathrm{scf}}, \quad \mathrm{A}_{\mathrm{scf}}^{M} \leftrightarrow \mathrm{A}_{\mathrm{scf}}^{D}. \tag{5.12}$$

Тогда понятно, что при $\epsilon\,[\operatorname{per} \mathrm{A}_{\mathrm{scf}}] \neq 0$ для scf-матриц можно ввести другое *дуальное* определение обратной матрицы, использующей не детерминант, а перманент [2].

**Определение 5.4.** *Для $N = 2$ scf-матрицы, удовлетворяющей условию $\epsilon\,[\operatorname{per} \mathrm{A}_{\mathrm{scf}}] \neq 0$, per-обратная матрица определяется следующей*

*формулой**)

$$\mathrm{A}_{\mathrm{scf}}^{-1per} \stackrel{def}{=} \frac{\mathrm{A}_{\mathrm{scf}}^{MT}}{\operatorname{per} \mathrm{A}_{\mathrm{scf}}}. \tag{5.13}$$

**Утверждение 5.5.** *Для* per*-обратной матрицы выполняется соотношение*

$$\mathrm{A}_{\mathrm{scf}}^{-1per} \cdot \mathrm{A}_{\mathrm{scf}} = \mathrm{I}. \tag{5.14}$$

*Доказательство.* Непосредственно получается из (5.11) при условии $\epsilon\,[\operatorname{per} \mathrm{A}_{\mathrm{scf}}] \neq 0$. ■

Отметим некоторые свойства $N = 2$ scf-матриц, следующие из их определения, которые, однако, не выполняются для обычных матриц. Например, для $n$-ой степени любой $N = 2$ scf-матрицы имеют место соотношения

$$\operatorname{tr} \mathrm{A}_{\mathrm{scf}}^{n} = a^n + d^n + [1 + (-1)^n]\,(bc)^{\frac{n}{2}}, \tag{5.15}$$

$$\begin{pmatrix} \operatorname{per} \\ \det \end{pmatrix}^{n} \mathrm{A}_{\mathrm{scf}} = \begin{pmatrix} \operatorname{per} \\ \det \end{pmatrix} \mathrm{A}_{\mathrm{scf}}^{n} = (ad)^n + (\pm 1)^n\,(bc)^n. \tag{5.16}$$

Отсюда, в частности, следуют связи между перманентом и детерминантом scf-матриц

$$\operatorname{per}^{2n} \mathrm{A}_{\mathrm{scf}} = \det^{2n} \mathrm{A}_{\mathrm{scf}}, \tag{5.17}$$

$$\operatorname{per} \mathrm{A}_{\mathrm{scf}}^{2n} = \det \mathrm{A}_{\mathrm{scf}}^{2n}. \tag{5.18}$$

**Утверждение 5.6.** *Если хотя бы один из элементов* scf*-матрицы на каждой из диагоналей нильпотентен и индекс нильпотентности равен 2 или элементы на каждой диагонали ортогональны, то произведение детерминанта на перманент равно нулю.*

---

*) *Примечание.* Ср. со стандартной формулой $A^{-1} = A^{DT} / \det A$ и (5.9).

*Доказательство.* Из определений детерминанта и перманента (5.1) получаем

$$\det \mathrm{A_{scf}} \cdot \mathrm{per}\, \mathrm{A_{scf}} = a^2 d^2 - b^2 c^2, \tag{5.19}$$

откуда и следует утверждение. ■

Кроме того, имеется нетривиальная связь между перманентом и следом scf-матрицы

$$\left(2\mathrm{per}\, \mathrm{A_{scf}} - \mathrm{tr}\, \mathrm{A^2_{scf}}\right)\left(2\mathrm{per}\, \mathrm{A_{scf}} + \mathrm{tr}\, \mathrm{A^2_{scf}} - \mathrm{tr}^{\,2} \mathrm{A_{scf}}\right) = 0, \tag{5.20}$$

где каждый из сомножителей отличен от нуля, а их ортогональность достигается за счет scf-условий (5.10).

По-видимому, одной из причин, почему перманенты не применялись широко в приложениях, как детерминанты, служит тот факт, что в общем случае перманент не мультипликативен, т. е. формула Бине-Коши $\det(\mathrm{AB}) = \det \mathrm{A} \cdot \det \mathrm{B}$ не выполняется*) без дополнительных условий для перманентов [626]. Замечательно, что именно уравнения (5.10) и являются требуемыми дополнительными условиями.

**Предложение 5.7.** (Формула Бине-Коши для перманентов) *Если* $\mathrm{A_{scf}}$ *и* $\mathrm{B_{scf}}$ *— любые* scf-*матрицы, то между их перманентами выполняется соотношения*

$$\mathrm{per}\, (\mathrm{A_{scf}} \cdot \mathrm{B_{scf}}) = \mathrm{per}\, \mathrm{A_{scf}} \cdot \mathrm{per}\, \mathrm{B_{scf}}. \tag{5.21}$$

*Доказательство.* Для $N = 2$ scf-матриц соотношение (5.21) следует из (5.1) непосредственным перемножением и затем применением scf-условий (5.10) . ■

Отметим также и другие важные формулы, справедливые для

---

*) *Примечание.* Также, как и инвариантность при линейных операциях над матрицами [626].

детерминантов и *только* для scf-матриц [2]

$$\mathrm{per}\,\left(\mathrm{A_{scf}}\cdot\mathrm{B_{scf}}\cdot\mathrm{A_{scf}^{-1}}\right) \;=\; \mathrm{per}\,\mathrm{B_{scf}}, \tag{5.22}$$
$$\mathrm{per}\,\mathrm{A_{scf}^{-1}} \;=\; \mathrm{per}^{\,-1}\mathrm{A_{scf}}, \tag{5.23}$$

где $\mathrm{A_{scf}^{-1}}$ — обратная матрица в обычном определении.

**5.1.2.** Ортогональные и scf-матрицы. Важным свойством scf-матриц является их связь с ортогональными матрицами при смене базиса [2], что использовалось нами при рассмотрении необратимых редуцированных $N=2$ и $N=4$ преобразований (см. **Раздел 3**).

Действительно, пусть

$$\mathrm{A_0}=\mathrm{U^{-1}}\cdot\mathrm{A}\cdot\mathrm{U},\quad \mathrm{B_0}=\mathrm{U^{-1}}\cdot\mathrm{B}\cdot\mathrm{U}, \tag{5.24}$$

где

$$\mathrm{U}=\frac{1}{\sqrt{2}}\begin{pmatrix}1 & i\\ 1 & -i\end{pmatrix} \tag{5.25}$$

— матрица перехода$^{*)}$ в комплексный базис, причем

$$\mathrm{U}^T\cdot\mathrm{U}=\begin{pmatrix}1 & 0\\ 0 & -1\end{pmatrix}=\sigma_3,\;\; \mathrm{U}\cdot\mathrm{U}^T=\begin{pmatrix}0 & 1\\ 1 & 0\end{pmatrix}=\sigma_1, \tag{5.26}$$

где $\sigma_i$ — матрицы Паули.

Тогда для произведения двух матриц в разных базисах можно получить

$$\mathrm{A}_0^T\cdot\mathrm{B_0}=\mathrm{U^{-1}}\cdot\mathrm{A}^{MT}\cdot\mathrm{B}\cdot\mathrm{U}. \tag{5.27}$$

Если выбрать $\mathrm{A_0}=\mathrm{B_0}$, то получим связь ортогональности в ко-

---

*Примечание.* С нулевым перманентом $\mathrm{per}\,\mathrm{U}=0$ и $\det\mathrm{U}=-i$.

ординатном базисе со свойствами scf-матриц в комплексном базисе [2]

$$\begin{aligned} \mathrm{A}_0^T \cdot \mathrm{A}_0 \;&=\; \mathrm{U}^{-1} \cdot \mathrm{A}^{MT} \cdot \mathrm{A} \cdot \mathrm{U} = \\ &\mathrm{per}\,\mathrm{A} \cdot \mathrm{I} + \mathrm{scf}_{+}\mathrm{A} \cdot \sigma^{+} + \mathrm{scf}_{-}\mathrm{A} \cdot \sigma^{-}, \end{aligned} \tag{5.28}$$

где I — единичная $2 \times 2$ матрица, $\sigma^{\pm} = \sigma_3 \pm i\sigma_1$ (см. (5.26)).

**Утверждение 5.8.** *В обратимом случае* $\epsilon\,[\mathrm{per}\,\mathrm{A}] \neq 0$ *нормированные на* $\sqrt{\mathrm{per}\,\mathrm{A}}$ scf-*матрицы подобны ортогональным матрицам.*

*Доказательство.* Используя scf-условия (5.10) $\mathrm{scf}_{\pm}\mathrm{A}_{\mathrm{scf}} = 0$, из (5.28) находим

$$\mathrm{A}_{0,\mathrm{scf}}^T \cdot \mathrm{A}_{0,\mathrm{scf}} = \mathrm{per}\,\mathrm{A}_{\mathrm{scf}} \cdot \mathrm{I}. \tag{5.29}$$

Обозначим $\mathrm{N}_{0,\mathrm{scf}} = \mathrm{A}_{0,\mathrm{scf}}/\sqrt{\mathrm{per}\,\mathrm{A}_{\mathrm{scf}}}$, тогда из (5.29) следует, что матрица $\mathrm{N}_{0,\mathrm{scf}}$ — ортогональная, т. е. $\mathrm{N}_{0,\mathrm{scf}}^T \cdot \mathrm{N}_{0,\mathrm{scf}} = \mathrm{I}$, следовательно $\mathrm{N}_{0,\mathrm{scf}} \in O_{\Lambda_0}(2)$. С другой стороны, пусть

$$\mathrm{N}_{\mathrm{scf}} = \frac{\mathrm{A}_{\mathrm{scf}}}{\sqrt{\mathrm{per}\,\mathrm{A}_{\mathrm{scf}}}}, \tag{5.30}$$

отсюда и из (5.24) получаем требуемую связь нормированных матриц в различных базисах $\mathrm{N}_{0,\mathrm{scf}} = \mathrm{U}^{-1} \cdot \mathrm{N}_{\mathrm{scf}} \cdot \mathrm{U}$. ■

**Следствие 5.9.** *Для нормированных* scf-*матриц ортогональность в одном базисе связана с* per-*обратимостью в другом*

$$\mathrm{N}_{0,\mathrm{scf}}^T \cdot \mathrm{N}_{0,\mathrm{scf}} = \mathrm{U}^{-1} \cdot \mathrm{N}_{\mathrm{scf}}^{-1per} \cdot \mathrm{N}_{\mathrm{scf}} \cdot \mathrm{U}, \tag{5.31}$$

*где* $\mathrm{N}_{\mathrm{scf}}^{-1per}$ *определено в* (5.13).

**5.1.3.** Обратимость и доопределенные scf-матрицы. Рассмотрим более подробно свойства обратимости scf-матриц.

**Утверждение 5.10.** *Для одной и той же матрицы$^{*)}$ числовые части детерминанта и перманента* (*отличных от нуля* $\operatorname{per} \mathrm{A} \neq 0$, $\det \mathrm{A} \neq 0$) *обращаются в нуль одновременно*

$$\epsilon\left[\operatorname{per} \mathrm{A}\right] = 0 \Leftrightarrow \epsilon\left[\det \mathrm{A}\right] = 0. \tag{5.32}$$

*Доказательство.* Следует из определений детерминанта и перманента (5.1) и разложения их в ряд по образующим $\Lambda_0$. ■

**Следствие 5.11.** *Для заданной* scf*-матрицы* $\mathrm{A}_{\mathrm{scf}}$ *при* $\operatorname{per} \mathrm{A}_{\mathrm{scf}} \neq 0$ *и* $\det \mathrm{A}_{\mathrm{scf}} \neq 0$ *обратная и per-обратная* (5.13) *матрицы определены или неопределены одновременно.*

Рассмотрим обратимый случай $\epsilon\left[\operatorname{per} \mathrm{A}_{\mathrm{scf}}\right] \neq 0$, $\epsilon\left[\det \mathrm{A}_{\mathrm{scf}}\right] \neq 0$, тогда единственным решением scf-условий (5.10) могут быть варианты, когда один из сомножителей обращается в нуль. Отсюда с очевидностью следует

**Утверждение 5.12.** *Обратимые* scf*-матрицы диагональны или антидиагональны.*

**Следствие 5.13.** *Для обратимых* scf*-матриц* per*-обратная матрица совпадает с обратной* $\mathrm{A}_{\mathrm{scf}}^{-1per} = \mathrm{A}_{\mathrm{scf}}^{-1}$.

В необратимом случае $\epsilon\left[\operatorname{per} \mathrm{A}_{\mathrm{scf}}\right] = 0$ нормировка, подобная (5.30), невозможна. Поэтому нужно непосредственно пользоваться scf-условиями (5.10) и ненормированными формулами (5.27)–(5.29). Тогда матрица $\mathrm{A}_{\mathrm{scf}}$ не обязательно будет диагональной или антидиагональной, как в **Утверждении 5.12**.

Для нахождения *доопределенной* per*-обратной* матрицы $\bar{\mathrm{A}}_{\mathrm{scf}}^{-1per}$ в

---

*Примечание.* Это утверждение справедливо для матриц любого порядка, состоящих из четных элементов.

этом случае необходимо избегать деления в (5.13) и решать уравнение

$$\bar{\mathrm{A}}_{\mathrm{scf}}^{-1per} \cdot \operatorname{per} \mathrm{A}_{\mathrm{scf}} = \mathrm{A}_{\mathrm{scf}}^{MT} \tag{5.33}$$

с нильпотентными обеими частями. Если аналогично ввести *доопределенную обратную* матрицу $\bar{\mathrm{A}}_{\mathrm{scf}}^{-1}$ по формуле

$$\bar{\mathrm{A}}_{\mathrm{scf}}^{-1} \cdot \det \mathrm{A}_{\mathrm{scf}} = \mathrm{A}_{\mathrm{scf}}^{DT}, \tag{5.34}$$

то в общем случае $\bar{\mathrm{A}}_{\mathrm{scf}}^{-1per} \neq \bar{\mathrm{A}}_{\mathrm{scf}}^{-1}$.

*Пример* **5.14.** Пусть

$$\mathrm{A}_{\mathrm{scf}} = \begin{pmatrix} \mu\nu & \alpha\beta \\ \mu\rho & \alpha\gamma \end{pmatrix} \tag{5.35}$$

— нильпотентная scf-матрица, для которой

$$\operatorname{per} \mathrm{A}_{\mathrm{scf}} = \mu\nu\alpha\gamma + \alpha\beta\mu\rho = \mu\alpha\left(\gamma\nu + \beta\rho\right), \tag{5.36}$$

$$\det \mathrm{A}_{\mathrm{scf}} = \mu\nu\alpha\gamma - \alpha\beta\mu\rho = \mu\alpha\left(\gamma\nu - \beta\rho\right). \tag{5.37}$$

Она необратима, поскольку $\epsilon\left[\operatorname{per} \mathrm{A}_{\mathrm{scf}}\right] = \epsilon\left[\det \mathrm{A}_{\mathrm{scf}}\right] = 0$. Пусть

$$\bar{\mathrm{A}}_{\mathrm{scf}}^{-1per} = \begin{pmatrix} x_1 & x_2 \\ x_3 & x_4 \end{pmatrix}, \quad \bar{\mathrm{A}}_{\mathrm{scf}}^{-1} = \begin{pmatrix} y_1 & y_2 \\ y_3 & y_4 \end{pmatrix}, \tag{5.38}$$

тогда из (5.33)–(5.34) и (5.35)–(5.37) имеем 2(!) *различные* системы урав-нений для определения элементов $x_i$ и $y_i$

$$\begin{cases} \mu\alpha\left(\gamma\nu+\beta\rho\right)x_1=\alpha\gamma, \\ \mu\alpha\left(\gamma\nu+\beta\rho\right)x_2=\alpha\beta, \\ \mu\alpha\left(\gamma\nu+\beta\rho\right)x_3=\mu\rho, \\ \mu\alpha\left(\gamma\nu+\beta\rho\right)x_3=\mu\nu, \end{cases} \begin{cases} \mu\alpha\left(\gamma\nu-\beta\rho\right)y_1=\alpha\gamma, \\ \mu\alpha\left(\gamma\nu-\beta\rho\right)y_2=-\alpha\beta, \\ \mu\alpha\left(\gamma\nu-\beta\rho\right)y_3=-\mu\rho, \\ \mu\alpha\left(\gamma\nu-\beta\rho\right)y_3=\mu\nu, \end{cases} \tag{5.39}$$

которые могут быть решены разложением по образующим $\Lambda$.

**5.1.4.** Полугруппа $N=2$ scf-матриц. Наряду с мультипликативностью перманента $N=2$ scf-матриц (5.21) важным также является поведение введенной матричной функции $\mathrm{scf}_{\pm}\mathrm{A}$ при умножении.

Рассмотрим функцию $\mathrm{scf}_{\pm}$ от произведения матриц A и $\mathrm{B}=\begin{pmatrix} p & q \\ r & s \end{pmatrix}$. Пользуясь определением (5.7), получаем

$$\mathrm{scf}_{+}(\mathrm{AB}) = p^2\cdot\mathrm{scf}_{+}\mathrm{A}+r^2\cdot\mathrm{scf}_{-}\mathrm{A}+2\mathrm{per}\,\mathrm{A}\cdot\mathrm{scf}_{+}\mathrm{B}, \tag{5.40}$$

$$\mathrm{scf}_{-}(\mathrm{AB}) = q^2\cdot\mathrm{scf}_{+}\mathrm{A}+s^2\cdot\mathrm{scf}_{-}\mathrm{A}+2\mathrm{per}\,\mathrm{A}\cdot\mathrm{scf}_{-}\mathrm{B}. \tag{5.41}$$

Обозначим множество $2\times 2$ четных матриц, удовлетворяющих условию (5.10), $\boldsymbol{\mathcal{A}}_{\mathrm{scf}}=\cup\,\mathrm{A}_{\mathrm{scf}}$. Тогда мы имеем

**Предложение 5.15.** *Множество $\boldsymbol{\mathcal{A}}_{\mathrm{scf}}$ образует подполугруппу полной линейной полугруппы $2\times 2$ четных матриц.*

*Доказательство.* Из (5.40)–(5.41) получаем отношения

$$\mathrm{scf}_{\pm}(\mathrm{AB})=0\Leftrightarrow\mathrm{scf}_{\pm}\mathrm{A}=0\wedge\mathrm{scf}_{\pm}\mathrm{B}=0, \tag{5.42}$$

что дает $\boldsymbol{\mathcal{A}}_{\mathrm{scf}}\star\boldsymbol{\mathcal{A}}_{\mathrm{scf}}\subseteq\boldsymbol{\mathcal{A}}_{\mathrm{scf}}$ и этим доказывает утверждение. ■

**Определение 5.16.** *Линейную полугруппу, изоморфную* $\{\boldsymbol{\mathcal{A}}_{\mathrm{scf}}|\cdot\}$, *где* $(\cdot)$ *— матричное умножение, назовем полугруппой* $N=2$ scf*-матриц* $SCF_{\Lambda_0}(2)$.

**Определение 5.17.** *Обратимые элементы из полугруппы* $SCF_{\Lambda_0}(2)$ *образуют группу* $N=2$ scf*-матриц* $GSCF_{\Lambda_0}(2)$.

**Определение 5.18.** *Небратимые элементы изполугруппы* $SCF_{\Lambda_0}(2)$ *образуют идеал* $ISCF_{\Lambda_0}(2)$.

Поскольку имеется соотношение подобия (5.24) и для scf-матриц выполняется ортогональность (5.29), в обратимом случае получаем

**Утверждение 5.19.** *Группа* $GSCF_{\Lambda_0}(2)$ *изоморфна ортогональной группе* $O_{\Lambda_0}(2)$.

Нетривиальным является необратимый случай $\epsilon\left[\mathrm{per}\,\mathrm{A}_{\mathrm{scf}}\right]=0$, когда scf-условия (5.7) выполняются не за счет зануления одного из сомножителей, а за счет ортогональности нильпотентных ненулевых сомножителей. Такие scf-матрицы принадлежат идеалу $ISCF_{\Lambda_0}(2)$ (см. *Пример* **5.14**).

**5.1.5.** $N=4$ scf-матрицы. Пусть

$$\mathrm{X}=\begin{pmatrix}\mathrm{A}_{11} & \mathrm{A}_{12}\\ \mathrm{A}_{21} & \mathrm{A}_{22}\end{pmatrix}\in\mathrm{Mat}_{\,\Lambda_0}(4) \tag{5.43}$$

— блочная $4\times4$ матрица, состоящая из четных элементов. Поставим ей в соответствие блочную $4\times4$ матрицу в координатном базисе

$$\mathrm{X}_0=\begin{pmatrix}\mathrm{U}^{-1}\mathrm{A}_{11}\mathrm{U} & \mathrm{U}^{-1}\mathrm{A}_{12}\mathrm{U}\\ \mathrm{U}^{-1}\mathrm{A}_{21}\mathrm{U} & \mathrm{U}^{-1}\mathrm{A}_{22}\mathrm{U}\end{pmatrix}, \tag{5.44}$$

где U — матрица подобия (5.25).

Найдем условия на матрицу X, при которых матрица $X_0$ будет ортогональной. Используя (5.27) и (5.28), получаем

$$X_0^T \cdot X_0 = \begin{pmatrix} P_1 & Q \\ Q & P_2 \end{pmatrix}, \tag{5.45}$$

$$\begin{aligned} P_i &= (\operatorname{per} A_{1i} + \operatorname{per} A_{2i}) \cdot I + \\ &\quad (\operatorname{scf}_+ A_{1i} + \operatorname{scf}_+ A_{2i})\, \sigma^+ + (\operatorname{scf}_- A_{1i} + \operatorname{scf}_- A_{2i})\, \sigma^-, \qquad (5.46) \\ Q &= U^{-1} \cdot \left( A_{11}^{MT} A_{12} + A_{21}^{MT} A_{22} \right) \cdot U, \qquad (5.47) \end{aligned}$$

где матричные функции $\operatorname{scf}_\pm A$ определены в (5.7).

Отсюда видно, что левая часть (5.45) будет пропорциональна единичной матрице

$$X_0^T \cdot X_0 = R \cdot I, \tag{5.48}$$

если занулить Q и слагаемые в (5.46), пропорциональные $\sigma$-матрицам, т. е. выбрать $Q = 0$ и $P_1 = P_2 = R \cdot I$. Таким образом, получаем

**Определение 5.20.** *Назовем* $N = 4$ scf-*матрицей такую матрицу* $X = X_{\text{scf}}^{N=4}$ (5.43), *блоки которой удовлетворяют уравнениям*

$$\begin{aligned} \operatorname{per} A_{11} + \operatorname{per} A_{21} &= \operatorname{per} A_{12} + \operatorname{per} A_{22} = R, \qquad (5.49) \\ \operatorname{scf}_\pm A_{11} + \operatorname{scf}_\pm A_{21i} &= \operatorname{scf}_\pm A_{12} + \operatorname{scf}_\pm A_{22} = 0, \qquad (5.50) \\ A_{11}^{MT} A_{12} + A_{21}^{MT} A_{22} &= 0. \qquad (5.51) \end{aligned}$$

Отметим, что для $N = 4$ scf-матриц $X_{\text{scf}}^{N=4}$ (и *только* для таких $4 \times 4$ матриц) также выполняется формула Бине-Коши (5.21) и сопутствующие тождества (5.22)–(5.23).

Чтобы определить детерминант $N=4$ scf-матрицы, используем следующее

**Утверждение 5.21.** *Для любых* $2\times 2$ *матриц* $\mathrm{E}_i$*, удовлетворяющих соотношению* $\mathrm{scf}_{\pm}\mathrm{E}_1+\mathrm{scf}_{\pm}\mathrm{E}_2=0$*, имеет место тождество*

$$\det\left(\mathrm{E}_1^{MT}\mathrm{E}_1+\mathrm{E}_2^{MT}\mathrm{E}_2\right)=\left(\mathrm{per}\,\mathrm{E}_1+\mathrm{per}\,\mathrm{E}_2\right)^2. \tag{5.52}$$

Кроме того, из (5.44) понятно, что $\det \mathrm{X}_{0,\mathrm{scf}}^{N=4}=\det^2\mathrm{U}^{-1}\cdot\det\mathrm{X}_{\mathrm{scf}}^{N=4}\cdot\det^2\mathrm{U}=\det\mathrm{X}_{\mathrm{scf}}^{N=4}$. Тогда из (5.48) следует, что $\det^2\mathrm{X}_{0,\mathrm{scf}}^{N=4}=R^4$, поэтому и $\det^2\mathrm{X}_{\mathrm{scf}}^{N=4}=R^4$, следовательно,

$$\det\mathrm{X}_{\mathrm{scf}}^{N=4}=kR^2, \tag{5.53}$$

где $k=\pm1$.

Отметим полезные соотношения между детерминантами блоков $N=4$ scf-матрицы

$$\det\mathrm{A}_{11}=k\det\mathrm{A}_{22},\quad \det\mathrm{A}_{21}=-k\det\mathrm{A}_{12}. \tag{5.54}$$

Перманент $N=4$ scf-матрицы также выражается через перманенты блоков по формуле

$$\mathrm{per}\,\mathrm{X}_{\mathrm{scf}}^{N=4}=\left(\mathrm{per}\,\mathrm{A}_{11}-\mathrm{per}\,\mathrm{A}_{21}\right)\left(\mathrm{per}\,\mathrm{A}_{22}-\mathrm{per}\,\mathrm{A}_{12}\right). \tag{5.55}$$

Примечательно, что при $k=+1$ формулы для детерминанта и перманента $N=4$ scf-матрицы можно объединить

$$\begin{pmatrix}\det\\ \mathrm{per}\end{pmatrix}\mathrm{X}_{\mathrm{scf}}^{N=4}=\left(\mathrm{per}\,\mathrm{A}_{11}\pm\mathrm{per}\,\mathrm{A}_{21}\right)\left(\mathrm{per}\,\mathrm{A}_{22}\pm\mathrm{per}\,\mathrm{A}_{12}\right), \tag{5.56}$$

что еще раз доказывает дуальность детерминанта и перманента при описании scf-матриц (см. (5.12)).

Из (5.53) следует, что классификация по необратимости $N = 4$ scf-матриц может быть проведена в терминах числовой части величины $R = \operatorname{per} \mathrm{A}_{11} + \operatorname{per} \mathrm{A}_{21} = \operatorname{per} \mathrm{A}_{22} + \operatorname{per} \mathrm{A}_{12}$. Обратимые $N = 4$ scf-матрицы имеют $\epsilon [R] \neq 0$, необратимые — $\epsilon [R] = 0$.

Обратимые $N = 4$ scf-матрицы можно отнормировать на $\sqrt{R}$ подобно (5.30), тогда соответствующая матрица $\mathrm{X}_{0,\mathrm{scf}}^{N=4}$ в координатном базисе будет $SO_{\Lambda_0}(4)$ матрицей при $k = +1$ и общей $O_{\Lambda_0}(4)$ матрицей при $k = -1$.

При умножении $N = 4$ scf-матриц соотношения (5.50) и (5.51) сохраняются, что можно показать, используя (5.40)–(5.41) и блочное умножение матриц вида (5.43). Поэтому, если обозначить множество $N = 4$ scf-матриц $\mathfrak{X}_{\mathrm{scf}}^{N=4} = \cup \mathrm{X}_{\mathrm{scf}}^{N=4}$, то $\mathfrak{X}_{\mathrm{scf}}^{N=4} \star \mathfrak{X}_{\mathrm{scf}}^{N=4} \subseteq \mathfrak{X}_{\mathrm{scf}}^{N=4}$, и получаем

**Предложение 5.22.** *Множество $N = 4$ scf-матриц $\mathfrak{X}_{\mathrm{scf}}^{N=4}$ образует подполугруппу полной линейной полугруппы $4 \times 4$ четных матриц.*

**Определение 5.23.** *Полугруппой $N = 4$ scf-матриц $SCF_{\Lambda_0}(4)$ назовем линейную полугруппу, изоморфную $\left\{\mathfrak{X}_{\mathrm{scf}}^{N=4} | \cdot \right\}$, где $(\cdot)$ — матричное умножение.*

**Определение 5.24.** *Обратимые элементы из полугруппы $SCF_{\Lambda_0}(4)$ образуют группу $N = 4$ scf-матриц $GSCF_{\Lambda_0}(4)$.*

**Определение 5.25.** *Небратимые элементы из полугруппы $SCF_{\Lambda_0}(4)$ образуют идеал $ISCF_{\Lambda_0}(4)$.*

Из-за соотношения (5.44) и ортогональности (5.48) в обратимом случае $\epsilon [R] \neq 0$ получаем

**Утверждение 5.26.** *Группа $GSCF_{\Lambda_0}(4)$ изоморфна ортогональной группе $O_{\Lambda_0}(4)$.*

Для $N = 4$ scf-матриц даже в обратимом случае scf-условия (5.7) могут выполняться и нетривиальным образом за счет ортогональности нильпотентных ненулевых сомножителей в $N = 4$ scf-условиях (5.50)–(5.51). Тогда среди блок-матриц $\mathrm{A}_{ij}$ могут появиться и не только диагональные и антидиагональные, как в стандартном варианте [563, 668], но и полные, состоящие из нильпотентных элементов (см., например, (Ж.83)–(Ж.84)).

Необратимые $N = 4$ scf-матрицы с $\epsilon[R] = 0$, принадлежащие идеалу $ISCF_{\Lambda_0}(4)$, имеют гораздо более богатую структуру и полугрупповые свойства относительно умножения [2].

## 5.2. Неэвклидова плоскость и scf-матрицы

Здесь мы рассмотрим некоторые необычные свойства дробно-линейных преобразований, к которым приводят $N = 2$ scf-матрицы.

Поставим в соответствие матрице $\mathrm{A} = \begin{pmatrix} a & b \\ c & d \end{pmatrix} \in \mathrm{Mat}_{\Lambda_0}(2)$ дробно-линейное преобразование $f : \mathbb{C}^{1|0} \to \mathbb{C}^{1|0}$ по формуле (см. например, [188, 733])

$$f_{\mathrm{A}}(z) = \frac{az+b}{cz+d}. \tag{5.57}$$

Доопределим $f_{\mathrm{A}}(z)$ на необратимый случай, когда $cz + d \neq 0$, но $\epsilon[cz + d] = 0$ по формуле

$$\bar{f}_{\mathrm{A}}(z)(cz+d) = az+b. \tag{5.58}$$

Будем обозначать равенства для доопределенных величин знаком $\stackrel{\circ}{=}$, а именно

$$\bar{f}_{\mathrm{A}}(z) \stackrel{\circ}{=} \frac{az+b}{cz+d}. \tag{5.59}$$

Пусть $\bar{\mathbf{F}}$ —полугруппа всех обратимых и необратимых доопре-деленных преобразований $\bar{f}_{\mathrm{A}}(z)$, а $L_{\Lambda_0}(2)$ — полугруппа матриц $\mathrm{A} \in \mathrm{Mat}_{\Lambda_0}(2)$. Поскольку для любых двух матриц A и B имеет место

$$\bar{f}_{\mathrm{A}} \circ \bar{f}_{\mathrm{B}} = \bar{f}_{\mathrm{AB}}, \tag{5.60}$$

то отображение полугрупп $\varphi : L_{\Lambda_0}(2) \to \bar{\mathbf{F}}$ есть гомоморфизм*) полугрупп.

**Предложение 5.27.** *Доопределенные дробно-линейные преобразования $\bar{f}_{\mathrm{A}}(z)$ имеют дополнительную неподвижную точку с нильпотентной координатой.*

*Доказательство.* Неподвижная точка $z_{\mathrm{fix}}$ отображения $\bar{f}_{\mathrm{A}}(z)$ определяется формулой $\bar{f}_{\mathrm{A}}(z_{\mathrm{fix}}) \mathrel{\mathring{=}} z_{\mathrm{fix}}$. Из (5.58) имеем $cz_{\mathrm{fix}}^2 + (d-a)\, z_{\mathrm{fix}} - b = 0$, откуда следует не одна, как в стандартном рассмотрении при $c \neq 0$ [734], а *две (!)* возможности:

1. $\epsilon[b] \neq 0$, $\epsilon[z_{\mathrm{fix}}] \neq 0$, тогда $z_{\mathrm{fix}}^{(\pm)} \mathrel{\mathring{=}} \dfrac{a - d \pm \sqrt{(a+d)^2 - 4\det \mathrm{A}}}{2c}$;
2. $\epsilon[b] = 0$, $\epsilon\left[z_{\mathrm{fix}}^{(0)}\right] = 0$, $\left(z_{\mathrm{fix}}^{(0)}\right)^2 = 0$, тогда $z_{\mathrm{fix}}^{(0)} \mathrel{\mathring{=}} \dfrac{b}{d-a}$.

■

Если выбрать в качестве матрицы A комплексную матрицу с единичным детерминантом, то $f_{\mathrm{A}}(z)$ — преобразование Мебиуса, играющее важную роль в теории струн и римановых поверхностей.

### 5.2.1. Определение и свойства per-отображений.

Выберем в качестве A введенные в **Подразделе 5.1** $N = 2$ scf-матрицы $\mathrm{A}_{\mathrm{scf}}$. Мы покажем, что наиболее ключевые соотношения будут иметь дуальные, где детерминант заменяется на перманент [2].

---

*Примечание.* Точнее — эпиморфизм с ненулевым ядром $a \cdot \mathrm{I}$, $a \in \Lambda_0$ (см. [733]).

Поскольку $N = 2$ scf-матрицы $\mathrm{A}_{\mathrm{scf}}$ образуют полугруппу $SCF_{\Lambda_0}(2)$ (см. **Предложение 5.15**), то соответствующие дробно-линейные преобразования $\bar{f}_{\mathrm{A}_{\mathrm{scf}}}(z)$ образуют полугруппу $\bar{\mathbf{F}}_{\mathrm{scf}} \subset \bar{\mathbf{F}}$ относительно композиции в силу (5.60), и отображение полугрупп $\varphi_{\mathrm{scf}} : SCF_{\Lambda_0}(2) \to \bar{\mathbf{F}}_{\mathrm{scf}}$ есть также гомоморфизм полугрупп.

**Определение 5.28.** *Назовем per-отображением дробно-линейное преобразование* (5.59) *с* $N = 2$ scf-*матрицей* $\mathrm{A} = \mathrm{A}_{\mathrm{scf}}$.

Основным для дальнейшего рассмотрения будет

**Утверждение 5.29.** Per-*отображения* (*и только они*) *удовлетворяют следующему тождеству*

$$n_1 \cdot \bar{f}_{\mathrm{A}_{\mathrm{scf}}}(z_1) \;+\; n_2 \cdot \bar{f}_{\mathrm{A}_{\mathrm{scf}}}(z_2) \,\mathring{=}$$

$$\frac{(n_1+n_2)(z_1+z_2)\cdot \operatorname{per}\mathrm{A}_{\mathrm{scf}}}{2(cz_1+d)(cz_2+d)} \;+\; \frac{(n_1-n_2)(z_1-z_2)\cdot \det\mathrm{A}_{\mathrm{scf}}}{2(cz_1+d)(cz_2+d)}. \quad (5.61)$$

*Доказательство.* Обозначим разность между левой и правой частями в (5.61) за $\Delta f(z_1,z_2)$. Для любой матрицы A непосредственно из (5.59) имеем

$$\Delta f(z_1,z_2) = (z_1 z_2 \cdot \mathrm{scf}_+\mathrm{A} + \mathrm{scf}_-\mathrm{A})(n_1+n_2), \quad (5.62)$$

тогда в силу того, что в нашем случае $\mathrm{A} = \mathrm{A}_{\mathrm{scf}}$ — $N = 2$ scf-матрица, из scf-условий (5.10) $\mathrm{scf}_\pm\mathrm{A}_{\mathrm{scf}} = 0$ получаем $\Delta f(z_1,z_2) = 0$. ■

Из тождества (5.61) явно прослеживается дуальная роль $\operatorname{per}\mathrm{A}_{\mathrm{scf}}$ и $\det\mathrm{A}_{\mathrm{scf}}$. Поэтому наряду с новыми формулами мы будем приводить и стандартные.

**Предложение 5.30.** *При* per-*отображениях разность координат преобразуется множителем, пропорциональным* $\det\mathrm{A}_{\mathrm{scf}}$, *а сумма координат преобразуется множителем, пропорциональным* $\operatorname{per}\mathrm{A}_{\mathrm{scf}}$.

*Доказательство.* Действительно, для суммы и разности (а не только для разности [733]) преобразованных координат из (5.61) получаем

$$\bar{f}_{\mathrm{A_{scf}}}(z_1) + \bar{f}_{\mathrm{A_{scf}}}(z_2) \doteqdot \frac{\operatorname{per} \mathrm{A_{scf}}}{(cz_1 + d)(cz_2 + d)}(z_1 + z_2), \tag{5.63}$$

$$\bar{f}_{\mathrm{A_{scf}}}(z_1) - \bar{f}_{\mathrm{A_{scf}}}(z_2) \doteqdot \frac{\det \mathrm{A_{scf}}}{(cz_1 + d)(cz_2 + d)}(z_1 - z_2). \tag{5.64}$$

■

*Замечание* **5.31.** Соотношение(5.64) выполняется не только для scf-матриц, но и для любых матриц A (см. например, [733]).

*Замечание* **5.32.** Соотношение (5.63) говорит о появлении на $\mathbb{C}^{1|0}$ новой симметрии, связанной с scf-матрицами и перманентами*).

Если элементы $\mathrm{A_{scf}}$ действительны, то из (5.63)–(5.64) находим дуальные формулы

$$\operatorname{Re} \bar{f}_{\mathrm{A_{scf}}}(z) \doteqdot \frac{\operatorname{per} \mathrm{A_{scf}}}{|cz + d|^2} \cdot \operatorname{Re} z, \tag{5.65}$$

$$\operatorname{Im} \bar{f}_{\mathrm{A_{scf}}}(z) \doteqdot \frac{\det \mathrm{A_{scf}}}{|cz + d|^2} \cdot \operatorname{Im} z, \tag{5.66}$$

откуда следует, что можно определить *два (!)* “единичных круга” на суперплоскости $\mathbb{C}^{1|0}$

$$\operatorname{Re} \bar{f}_{\mathrm{A_{scf}}}(z) = \operatorname{Re} z \Leftrightarrow |cz + d| = \sqrt{\operatorname{per} \mathrm{A_{scf}}}, \tag{5.67}$$

$$\operatorname{Im} \bar{f}_{\mathrm{A_{scf}}}(z) = \operatorname{Im} z \Leftrightarrow |cz + d| = \sqrt{\det \mathrm{A_{scf}}}. \tag{5.68}$$

---

*Примечание.* Что проясняет происхождение **Определения 5.28**.

Кроме того,

$$\frac{\left|\bar{f}_{\mathrm{A_{scf}}}\left(z_1\right)+\bar{f}_{\mathrm{A_{scf}}}\left(z_2\right)\right|^2}{\operatorname{Re}\bar{f}_{\mathrm{A_{scf}}}\left(z_1\right)\cdot\operatorname{Re}\bar{f}_{\mathrm{A_{scf}}}\left(z_2\right)} \mathrel{\mathring{=}} \frac{\left|z_1+z_2\right|^2}{\operatorname{Re}z_1\cdot\operatorname{Re}z_2}, \tag{5.69}$$

$$\frac{\left|\bar{f}_{\mathrm{A_{scf}}}\left(z_1\right)-\bar{f}_{\mathrm{A_{scf}}}\left(z_2\right)\right|^2}{\operatorname{Im}\bar{f}_{\mathrm{A_{scf}}}\left(z_1\right)\cdot\operatorname{Im}\bar{f}_{\mathrm{A_{scf}}}\left(z_2\right)} \mathrel{\mathring{=}} \frac{\left|z_1-z_2\right|^2}{\operatorname{Im}z_1\cdot\operatorname{Im}z_2}. \tag{5.70}$$

**5.2.2.** Правые и левые двойные отношения. Пусть $z_1, z_2, z_3, z_4 \in \mathbb{C}^{1|0}$ — четыре различные точки. Определим не одно (как в [733, 735]), а два двойных отношения следующим образом.

**Определение 5.33.** *Доопределенными правым и левым двойными отношениями назовем такие функции четырех точек*

$$\boldsymbol{D}^{\pm}\left(z_1, z_2, z_3, z_4\right) \mathrel{\mathring{=}} \frac{\left(z_1 \pm z_3\right)\left(z_2 \pm z_4\right)}{\left(z_1 \pm z_4\right)\left(z_2 \pm z_3\right)}. \tag{5.71}$$

*Замечание* **5.34.** В (5.71) прослеживается 2 отличия от стандартных определений [733]: **1)** наличие наряду с *левым* двойным отношением с разностями координат также и *правого* двойного отношения с их суммами; **2)** распространение определений (нового и известного) на нильпотентную область $\mathbb{C}^{1|0}$ с использованием доопределенного знака равенства $\mathring{=}$ (5.58)–(5.59).

Отметим, что, в частности,

$$\boldsymbol{D}^{\pm}\left(z, 1, 0, \infty\right) = z. \tag{5.72}$$

Относительно дробно-линейных преобразований общего вида (5.57) левое двойное отношение (5.71) инвариантно [733, 735] в силу (5.64) и *Замечания* **5.31**. Для per-отображений выполняются оба соотношения

(5.63) и (5.64), поэтому для них имеем

**Теорема 5.35.** *Как правое, так и левое двойные отношения* (5.71) *инвариантны относительно* per-*отображений*

$$\boldsymbol{D}^{\pm}\left(\bar{f}_{\mathrm{A_{scf}}}\left(z_1\right), \bar{f}_{\mathrm{A_{scf}}}\left(z_2\right), \bar{f}_{\mathrm{A_{scf}}}\left(z_3\right), \bar{f}_{\mathrm{A_{scf}}}\left(z_4\right)\right) \mathrel{\mathring{=}} \boldsymbol{D}^{\pm}\left(z_1, z_2, z_3, z_4\right) = \boldsymbol{r}^{\pm}. \tag{5.73}$$

*Доказательство.* Рассмотрим преобразованное правое двойное отношение с $\boldsymbol{D}^{+}\left(\bar{f}_{\mathrm{A_{scf}}}\left(z_1\right), \bar{f}_{\mathrm{A_{scf}}}\left(z_2\right), \bar{f}_{\mathrm{A_{scf}}}\left(z_3\right), \bar{f}_{\mathrm{A_{scf}}}\left(z_4\right)\right)$. Для различных сумм преобразованных координат воспользуемся (5.63)

$$\bar{f}_{\mathrm{A_{scf}}}\left(z_i\right) + \bar{f}_{\mathrm{A_{scf}}}\left(z_j\right) \mathrel{\mathring{=}} \frac{\operatorname{per} \mathrm{A_{scf}}}{\left(cz_i + d\right)\left(cz_j + d\right)}\left(z_i + z_j\right), \tag{5.74}$$

после чего в числителе и знаменателе (5.71) сократим на $\operatorname{per} \mathrm{A_{scf}}$ от каждой суммы и на общее выражение $\left(cz_1 + d\right)\left(cz_2 + d\right)\left(cz_3 + d\right)\left(cz_4 + d\right)$, тогда получим искомое непреобразованное правое двойное отношение $\boldsymbol{D}^{+}\left(z_1, z_2, z_3, z_4\right)$. Для левого двойного отношения доказательство проводится аналогично. ■

**Следствие 5.36.** *Если для двух четверок точек* $z_i$ *и* $w_i$ *левые* (*или правые*) *двойные отношения совпадают, то существует* per-*отображение, которое переводит одну четверку в другую* $w_i = \bar{f}_{\mathrm{A_{scf}}}\left(z_i\right)$.

Итак, мы доказали, что per-отображения имеют дополнительный инвариант $\boldsymbol{r}_{+}$ — правое двойное отношение $\boldsymbol{D}^{+}\left(z_1, z_2, z_3, z_4\right) = \boldsymbol{r}_{+}$, которое зависит не от конкретного значения координат $z_1, z_2, z_3, z_4$, а от их перестановок $z_{\sigma 1}, z_{\sigma 2}, z_{\sigma 3}, z_{\sigma 4}$, $\sigma \in S_4$, где $S_4$ — группа перестановок множества из 4 элементов. Введем правые и левые функции $p_{\sigma}^{\pm}\left(\boldsymbol{r}_{\pm}\right)$ на правых и левых двойных отношениях соответственно по формуле

$$\boldsymbol{D}^{\pm}\left(z_{\sigma 1}, z_{\sigma 2}, z_{\sigma 3}, z_{\sigma 4}\right) = p_{\sigma}^{\pm}\left(\boldsymbol{D}^{\pm}\left(z_1, z_2, z_3, z_4\right)\right). \tag{5.75}$$

**Утверждение 5.37.** *Отображения группы перестановок*

$$\omega^{\pm}:\sigma\rightarrow p_{\sigma}^{\pm} \tag{5.76}$$

*являются гомоморфизмами.*

*Доказательство.* Для двух последовательных перестановок из (5.75) имеем

$$\begin{aligned} p_{\pi}^{\pm}\left[p_{\sigma}^{\pm}\left(\boldsymbol{D}^{\pm}\left(z_1,z_2,z_3,z_4\right)\right)\right] &= p_{\pi}^{\pm}\left[\boldsymbol{D}^{\pm}\left(z_{\sigma 1},z_{\sigma 2},z_{\sigma 3},z_{\sigma 4}\right)\right] = \\ \boldsymbol{D}^{\pm}\left(z_{\pi\sigma 1},z_{\pi\sigma 2},z_{\pi\sigma 3},z_{\pi\sigma 4}\right) &= p_{\pi\sigma}^{\pm}\left(\boldsymbol{D}^{\pm}\left(z_1,z_2,z_3,z_4\right)\right), \end{aligned} \tag{5.77}$$

т. е. $p_{\pi}^{\pm}p_{\sigma}^{\pm}=p_{\pi\sigma}^{\pm}$, что и доказывает утверждение. ■

Найдем образ, например, транспозиции $\sigma=\sigma_{23}=(2,3)$ при гомоморфизме $\omega^{\pm}$ (5.76). Из (5.73) следует, что имеется *два (!)* per-отображения $h^{\pm}\left(z_1\right)$, переводящих точки $z_2,z_3,z_4$ в $1,0,\infty$ соответственно

$$\boldsymbol{D}^{\pm}\left(z_1,z_2,z_3,z_4\right)=\boldsymbol{D}^{\pm}\left(h^{\pm}\left(z_1\right),1,0,\infty\right). \tag{5.78}$$

С другой стороны, пользуясь (5.72), имеем $\boldsymbol{D}^{\pm}\left(h^{\pm}\left(z_1\right),1,0,\infty\right)=h^{\pm}\left(z_1\right)$. А из определения инвариантов (5.73) имеем

$$\boldsymbol{r}^{\pm}=\boldsymbol{D}^{\pm}\left(h^{\pm}\left(z_1\right),1,0,\infty\right)=h^{\pm}\left(z_1\right)=\boldsymbol{D}^{\pm}\left(\boldsymbol{r}^{\pm},1,0,\infty\right). \tag{5.79}$$

Применяем далее транспозицию $\sigma_{23}$ к (5.79) и получаем

$$p_{\sigma_{23}}^{\pm}\left(\boldsymbol{r}^{\pm}\right)=\boldsymbol{D}^{\pm}\left(\boldsymbol{r}^{\pm},0,1,\infty\right)=1\pm\boldsymbol{r}^{\pm}, \tag{5.80}$$

и, следовательно, $\sigma_{23}\overset{\omega^{\pm}}{\rightarrow}1\pm\boldsymbol{r}^{\pm}$. Проделывая подобные выкладки для остальных транспозиций, получаем следующее

**Утверждение 5.38.** *Образами группы перестановок $S_4$ при гомоморфизмах $\omega^+$ и $\omega^-$ (5.76) являются две (!) конечные группы, каждая из которых состоит из 6 элементов*

$$\omega^+(S_4) = \left\{\boldsymbol{r}^+, \frac{1}{\boldsymbol{r}^+}, 1+\boldsymbol{r}^+, \frac{1}{1+\boldsymbol{r}^+}, \frac{1+\boldsymbol{r}^+}{\boldsymbol{r}^+}, \frac{\boldsymbol{r}^+}{1+\boldsymbol{r}^+}\right\}, \tag{5.81}$$

$$\omega^-(S_4) = \left\{\boldsymbol{r}^-, \frac{1}{\boldsymbol{r}^-}, 1-\boldsymbol{r}^-, \frac{1}{1-\boldsymbol{r}^-}, \frac{1-\boldsymbol{r}^-}{\boldsymbol{r}^-}, \frac{\boldsymbol{r}^-}{1-\boldsymbol{r}^-}\right\} \tag{5.82}$$

*при* $\epsilon[\boldsymbol{r}^\pm] \neq 0$.

Если $\epsilon[\boldsymbol{r}^\pm] = 0$, то количество элементов в (5.81)–(5.82) уменьшается до 4

$$\omega^+(S_4) = \left\{\boldsymbol{r}^+, 1+\boldsymbol{r}^+, \frac{1}{1+\boldsymbol{r}^+}, \frac{\boldsymbol{r}^+}{1+\boldsymbol{r}^+}\right\}, \tag{5.83}$$

$$\omega^-(S_4) = \left\{\boldsymbol{r}^-, 1-\boldsymbol{r}^-, \frac{1}{1-\boldsymbol{r}^-}, \frac{\boldsymbol{r}^-}{1-\boldsymbol{r}^-}\right\}. \tag{5.84}$$

Однако другие критические значения инвариантов $\boldsymbol{r}^+$ и $\boldsymbol{r}^-$ не совпадают между собой. Из (5.81)–(5.82) следует, что образы отображений $\omega^\pm$ содержат по 3 элемента, если $1 \pm \boldsymbol{r}^\pm = 1/\boldsymbol{r}^\pm$, т. е. при различных значениях инвариантов

$$\boldsymbol{r}^+_{1,2} = \frac{-1 \pm \sqrt{5}}{2}, \tag{5.85}$$

$$\boldsymbol{r}^-_{1,2} = \frac{1 \pm i\sqrt{3}}{2} \tag{5.86}$$

соответственно. Если $\boldsymbol{r}^- = -1$, то говорят, что точки $z_1, z_2, z_3, z_4$ образуют гармоническую последовательность [733]. Соответствующее значение инварианта $\boldsymbol{r}^+$ равно $+1$, а такую последовательность точек можно назвать per-*гармонической.* При этом $\omega^+(S_4) = \omega^-(S_4) = \left\{\frac{1}{2}, 1, 2\right\}$.

**Утверждение 5.39.** *Для четырех точек $z_1, z_2, z_3, z_4$, лежащих на единичном круге, правое $\boldsymbol{D}^+(z_1, z_2, z_3, z_4)$ и левое $\boldsymbol{D}^-(z_1, z_2, z_3, z_4)$ двойные отношения действительны*).*

*Доказательство.* На единичном круге полагаем $z_i = e^{it_i}$, $t_i \in \mathbb{R}$, тогда из (5.46) получаем

$$\boldsymbol{D}^+\left(e^{it_1}, e^{it_2}, e^{it_3}, e^{it_4}\right) = \frac{\cos(t_1 - t_3)\cos(t_2 - t_4)}{\cos(t_1 - t_4)\cos(t_2 - t_3)}, \tag{5.87}$$

$$\boldsymbol{D}^-\left(e^{it_1}, e^{it_2}, e^{it_3}, e^{it_4}\right) = \frac{\sin(t_1 - t_3)\sin(t_2 - t_4)}{\sin(t_1 - t_4)\sin(t_2 - t_3)}. \tag{5.88}$$

Следовательно $\boldsymbol{D}^\pm\left(e^{it_1}, e^{it_2}, e^{it_3}, e^{it_4}\right) \in \mathbb{R}$. ■

Имеется также per-аналог формулы Лаггера [736], позволяющий выразить правое двойное отношение через “угол” $\vartheta$ между “прямыми”. Действительно, пусть $\mathrm{tg}\vartheta = \frac{m_1 - m_2}{1 + m_1 m_2}$, тогда из (5.71) можно получить

$$\boldsymbol{D}^\pm(m_1, m_2, +i, -i) = e^{\pm i\vartheta}, \tag{5.89}$$

где выражение с нижним знаком представляет собой классическую формулу Лаггера [736].

Если A — матрица, соответствующая дробно-линейному преобразованию $f_{\mathrm{A}}(z)$ (см. (5.57)), то для левого двойного отношения можно вывести формулу

$$\boldsymbol{D}^-\left(z, f_{\mathrm{A}}^{\circ 3}(z), f_{\mathrm{A}}^{\circ 2}(z), f_{\mathrm{A}}(z)\right) = \frac{\mathrm{tr}^2\mathrm{A}}{\det \mathrm{A}}, \tag{5.90}$$

---

*Примечание.* Для левого двойного отношения см., например, [736].

где $f_{\mathrm{A}}^{\circ n}(z)$ — композиция из $n$ преобразований.

Подобная формула для правого двойного отношения (если $\mathrm{A} = \mathrm{A}_{\mathrm{scf}}$) имеет вид

$$\boldsymbol{D}^{+}\left(z, f_{\mathrm{A}_{\mathrm{scf}}}^{\circ 3}(z), f_{\mathrm{A}_{\mathrm{scf}}}^{\circ 2}(z), f_{\mathrm{A}_{\mathrm{scf}}}(z)\right) = z\left(ad^3 + 2b^2c^2 + a^3d\right) =$$
$$z\left[\operatorname{per}\mathrm{A}_{\mathrm{scf}}\left(\operatorname{tr}\mathrm{A}_{\mathrm{scf}}^2 + \frac{1}{2}\operatorname{tr}^2\mathrm{A}_{\mathrm{scf}}\right) + \frac{1}{4}\operatorname{tr}^2\mathrm{A}_{\mathrm{scf}}\left(\operatorname{tr}^2\mathrm{A}_{\mathrm{scf}} - \operatorname{tr}\mathrm{A}_{\mathrm{scf}}^2\right)\right], \quad (5.91)$$

где $f_{\mathrm{A}_{\mathrm{scf}}}^{\circ n}(z)$ — композиция $n$ per-отображений.

Отметим, что имеется тесная связь между левым двойным отношением и производной Шварца [737]. Действительно, для любой функции $f(z)$ из (5.71) получаем

$$\boldsymbol{D}^{-}\left(f(z+ta), f(z+tb), f(z+tc), f(z+td)\right) =$$
$$\boldsymbol{D}^{-}(a,b,c,d)\left[1 + \frac{t^2}{6}(a-b)(c-d)\,\boldsymbol{S}_f^{-}(z)\right] + O\left(t^3\right), \quad (5.92)$$

где $a, b, c, d, t \in \Lambda_0$, и производная Шварца определяется формулой

$$\boldsymbol{S}_f^{-}(z) = \frac{f'''(z)}{f'(z)} - \frac{3}{2}\left(\frac{f''(z)}{f'(z)}\right)^2. \quad (5.93)$$

Аналогичная формула для правого двойного отношения имеет следующий вид

$$\boldsymbol{D}^{+}\left(f(z+ta), f(z+tb), f(z+tc), f(z+td)\right) =$$
$$1 + \frac{t^2}{6}(a-b)(c-d)\,\boldsymbol{S}_f^{+}(z) +$$
$$\frac{t^3}{8}(a-b)(c-d)(a+b+c+d)\,\boldsymbol{S}_f^{(3)+}(z) + O\left(t^4\right), \quad (5.94)$$

где функции $\boldsymbol{S}_f^+(z)$ и $\boldsymbol{S}_f^{(3)+}(z)$ равны

$$\boldsymbol{S}_f^+(z) = -\frac{3}{2}\left(\frac{f'(z)}{f(z)}\right)^2, \tag{5.95}$$

$$\boldsymbol{S}_f^{(3)+}(z) = -\frac{f'(z)}{f(z)}\left(\frac{f'(z)}{f(z)}\right)'. \tag{5.96}$$

Из сравнения выражения в квадратных скобках (5.92) и второй строки в (5.94) следует, что функцию $\boldsymbol{S}_f^+(z)$ (5.95) можно трактовать как per-аналог производной Шварца [188, 738].

**5.2.3.** Per-аналог гиперболического расстояния на суперплоскости. Пусть точки $z_1, z_2, z_3, z_4$ лежат на одной и той же "геодезической", определяемой лишь точками $z_1, z_2$, в то время, как точки $z_3, z_4$ лежат на "единичном круге" (5.67)–(5.68). Тогда можно определить вместо одного [734, 737] два гиперболических расстояния.

**Определение 5.40.** *Правое и левое гиперболические расстояния в "единичном круге" определяются формулами* [2]

$$\boldsymbol{d}^{\pm}(z_1, z_2) \stackrel{def}{=} \ln \boldsymbol{D}^{\pm}(z_1, z_2, z_3, z_4). \tag{5.97}$$

Из **Утверждения 5.39** следует, что, если точки $z_1, z_2$ лежат на "единичном круге" (5.67)–(5.68), то расстояния $\boldsymbol{d}^{\pm}(z_1, z_2)$ действительны (см. [736, 739]). Аддитивность расстояния $\boldsymbol{d}^{\pm}(z_1, z_2)$ (5.97) обеспечивается мультипликативностью правого и левого двойных отношений

$$\boldsymbol{D}^{\pm}(z_1, z_2, z, z')\,\boldsymbol{D}^{\pm}(z_2, z_3, z, z') = \boldsymbol{D}^{\pm}(z_1, z_3, z, z'). \tag{5.98}$$

Имеются и другие формулы для расстояния*) [733, 737], которые учитывают явно условие $\operatorname{Im} z \geq 0$, определяющее верхнюю полуплоскость $\mathbb{H}^2_{im}$ [734, 740]. Например, из (5.70) следует, что выражение [733]

$$\boldsymbol{d}^-_{im}(z_1, z_2) = \operatorname{Arch}\left(1 + \frac{|z_1 - z_2|^2}{\operatorname{Im} z_1 \operatorname{Im} z_2}\right) \tag{5.99}$$

инвариантно относительно дробно-линейных преобразований. Однако в случае per-отображений (см. **Определение 5.28**) мы имеем другую инвариантность (5.69), что приводит к необходимости рассмотрения “правой полуплоскости” $\mathbb{H}^2_{re}$, определяемой условием $\operatorname{Re} z \geq 0$. Тогда по аналогии с (5.99) можно определить правое расстояние [2]

$$\boldsymbol{d}^+_{re}(z_1, z_2) = \operatorname{Arch}\left(1 + \frac{|z_1 + z_2|^2}{\operatorname{Re} z_1 \operatorname{Re} z_2}\right), \tag{5.100}$$

инвариантное относительно per-отображений вследствие (5.69).

В терминах правого двойного отношения $\boldsymbol{D}^+(z_1, z_2, z_3, z_4)$ и правого расстояния $\boldsymbol{d}^+(z_1, z_2)$ (5.97) (или $\boldsymbol{d}^+_{re}(z_1, z_2)$) можно последовательно построить per-аналог гиперболической геометрии [733, 741], тригонометрии [737, 739] на комплексной суперплоскости или в многомерных комплексных суперпространствах [742].

---

*) *Примечание.* Левого в нашем определении.

## 5.3. Основные результаты и выводы

**1.** Определены и подробно изучены специальные необратимые матрицы с нильпотентными элементами — scf-матрицы, возникающие в $N$-расширенной суперконформной геометрии.

**2.** Обнаружена дуальность между перманентом и детерминантом и между минорами и алгебраическими дополнениями.

**3.** Предложена новая формула для обратной матрицы через перманент и миноры.

**4.** Получена новая формула Бине-Коши для перманентов.

**5.** Построены полугруппы $N = 2$ и $N = 4$ scf-матриц.

**6.** Приведены дробно-линейные преобразования специального вида, для которых найден новый вид симметрии.

**7.** Сформулирована необратимая гиперболическая геометрия на суперплоскости, в которой имеется два различных инвариантных двойных отношения и, соответственно, два расстояния.

**8.** Получены аналоги производной Шварца и различных классических формул на суперплоскости.

# ЗАКЛЮЧЕНИЕ

**1.** Построена теория необратимых супермногообразий — полусупермногообразий, необратимых расслоений и гомотопий, что является основой математического аппарата суперсимметричных моделей элементарных частиц.

**а)** Получена формулировка полусупермногообразий в терминах функций перехода, найдены обобщенные условия коцикла, новый тип ориентируемости.

**б)** Предложен общий принцип полукоммутативности для необратимых морфизмов.

**в)** Сформулированы необратимые аналоги расслоений — полурасслоения — в терминах уравнений на функции перехода, изучены морфизмы полурасслоений.

**г)** Введены полугомотопии с необратимым четным или нечетным суперпараметром.

**2.** Построена и исследована в терминах теории полугрупп необратимая суперконформная геометрия на суперплоскости, необходимая для формулировки суперструнных теорий элементарных частиц.

**а)** Построена супераналитическая полугруппа и дано определение супераналитических полусупермногообразий.

**б)** Рассмотрены дополнительные редукции касательного суперпространства с учетом необратимости. Они приводят к обобщению понятия комплексной структуры на необратимый случай.

**в)** Найдены новые необратимые преобразования — сплетающие четность преобразования, которые дуальны суперконформным в смысле полученной формулы сложения березинианов и являются необратимым супераналогом антиголоморфных преобразований. Они вращают четность в касательном суперпространстве и приводят к появлению нового типа коциклов — сплетенных коциклов. Единым образом описаны оба типа редуцированных преобразований с помощью альтернативной параметризации, в которой переключение между ними производится введенным спином редукции, равным половина и $N/2$ для расширенных $N$-преобразований.

**г)** Построена $N = 1$ суперконформная полугруппа, принадлежащая к новому абстрактному типу полугрупп, которые имеют необычную идеальную структуру. Определены обобщенные векторные и тензорные отношения Грина.

**д)** Исследованы дробно-линейные необратимые редуцированные преобразования в терминах полуминоров и полуматриц, для которых определены функции полуперманента и полудетерминанта (дуального корню из обычного детерминанта). Найдена четно-нечетная симметрия дробно-линейных $N = 1$ суперконформных преобразований, которая состоит в симметрии относительно одновременной замены детерминанта на полудетерминант и четных координат на нечетные.

**е)** Найдены необратимые супераналоги расстояния в $(1|1)$-мерном суперпространстве. Введен необратимый аналог метрики и показана ее полуинвариантность.

**ж)** Изучены нелинейные реализации $N = 1$ редуцированных преобразований и найден новый тип голдстино как решение, соответствующее сплетающим четность преобразованиям.

**з)** Исследована необратимая геометрия на $N = 2$ и $N = 4$ расширенной суперплоскости. Построены $N = 2$ и $N = 4$ суперконформные полугруппы в альтернативной параметризации. Обобщается на произвольное $N$ понятие комплексной структуры на суперплоскости.

**3.** Построены суперматричные полугруппы и исследованы их идеальные свойства и нетривиальные редукции, применяемые в феноменологии суперсимметричных моделей элементарных частиц.

**а)** Найдено несколько возможностей объединить антитреугольные суперматрицы с треугольными в сэндвич-полугруппы с необычными свойствами.

**б)** Получены новые типы нечетных супермодулей и антитранспонирования, представления странной супералгебры Березина.

**в)** Введены нечетные аналоги собственных чисел, характеристических функций и сформулирована обобщенная теорема Гамильтона-Якоби.

**4.** Обнаружено, что полугрупповые связки непрерывно представляются суперматричными полугруппами антитреугольного вида.

**а)** Получено объединение однопараметрических полугрупп в некоторую нетривиальную полугруппу — скрученную прямоугольную связку.

**б)** Определены высшие связки и для них введены обобщения отношений Грина — тонкие и смешанные отношения эквивалентности. Для них построены многомерные диаграммы.

**5.** Исследованы необратимые свойства матриц, содержащих нильпотентные элементы и делители нуля и возникающих в $N$-расширенной суперконформной геометрии.

а) Найдена дуальность между перманентом и детерминантом и между минорами и алгебраическими дополнениями, предложена новая формула для обратной матрицы через перманент и миноры.

б) Изучены обратимые и необратимые дробно-линейные преобразования специального вида, для которых найден новый вид симметрии.

в) Построена необратимая гиперболическая геометрия на суперплоскости, в которой имеется два различных инвариантных двойных отношения и, соответственно, два расстояния.

Таким образом, проведенные исследования геометрических и симметрийных аспектов суперсимметричных и суперструнных моделей элементарных частиц, полученные конкретные аналитические и общенаучные результаты можно квалифицировать как новое научное направление, состоящее в построении новой модели элементарных частиц на основе более абстрактных категорных понятий и базовых внутренних структур.

К перспективам дальнейшего развития этого направления можно отнести поиск новых проявлений необратимых и полугрупповых свойств в современной теории суперструн и супермембран, продвижение в сторону конкретных расчетов возможных дополнительных вкладов в фермионные амплитуды и наблюдаемые, а также разработка общих принципов построения суперсимметричных моделей элементарных частиц на основе соответствующих теорий полугрупп.

Различные вопросы, связанные с материалом диссертации, интенсивно и многократно обсуждались на конференциях и семинарах с такими учеными, как Ader J.-P., Aspinwall P., Cariñena J. F., Ćirić M., Comtét A., de Wit B., Deligne P., Duff M. J., Gates J., Grisaru M. T., Howie J. M., Kelarev A., Kupsch J., Lawson M. V., Lukierski J., Nieuwenhuizen P., O'Raifeartaigh L., Preston G. B., Rühl W., Rabin J. M., Schein B. M., Sezgin E., Stasheff J., Tucker R. W., Umble R., Wess J., Wightman A. S., Акулов В. П., Алексеевский Д. В., Аринкин Д., Бережной Ю. А., Бессмертный М. Ф., Ваксман Л. Л., Воронов А. А., Громов Н. А. Демичев А. П., Дринфельд В. Г., Зима В. Г., Капустников А. А., Куренной Г. Ч., Лейтес Д. А., Манин Ю. И., Меренков Н. П., Натанзон С. М., Новиков Б. В., Пашнев А. И., Синельщиков С. С., Смилга А. В., Степановский Ю. П., Фаддеев Л. Д., Фомин П. И., Хренников А. Ю., которым автор выражает искреннюю и глубокую благодарность за плодотворные дискуссии.

# СПИСОК ИСПОЛЬЗОВАННЫХ ИСТОЧНИКОВ

# СПИСОК РИСУНКОВ

# СПИСОК ТАБЛИЦ

# <u>Приложение A</u>

# Теория абстрактных полугрупп

Введем понятия алгебраической теории полугрупп [102–104], необходимые для понимания основного текста.

## A.1. Группоиды, полугруппы и идеалы

*Бинарной операцией* на множестве $\mathcal{S}$ называется отображение $\mathcal{S} \times \mathcal{S}$ в $\mathcal{S}$, где $\mathcal{S} \times \mathcal{S}$ есть множество всех упорядоченных пар элементов из $\mathcal{S}$. Если это отображение обозначается $(*)$, то образ в $\mathcal{S}$ элемента $(\mathbf{a}, \mathbf{b}) \in \mathcal{S} \times \mathcal{S}$ будет обозначаться через $\mathbf{a} * \mathbf{b}$. *Частичной бинарной операцией* на множестве $\mathcal{S}$ называется отображение непустого подмножества множества $\mathcal{S} \times \mathcal{S}$ в $\mathcal{S}$. Под *частичным группоидом* мы будем понимать систему $\{\mathcal{S}; *\}$, состоящую из непустого множества $\mathcal{S}$ и частичной бинарной операции $(*)$ на нем. Бинарная операция $(*)$ на множестве $\mathcal{S}$ называется *ассоциативной*, если $\mathbf{a} * (\mathbf{b} * \mathbf{c}) = (\mathbf{a} * \mathbf{b}) * \mathbf{c}$ для всех $\mathbf{a}, \mathbf{b}, \mathbf{c}$, из $\mathcal{S}$. *Полугруппа* $\mathbf{S}$ - это такой группоид $\{\mathcal{S}; *\}$, в котором операция $(*)$ ассоциативна.

Отображение $\alpha$ множества $\mathcal{X}$ в множество $\mathcal{Y}$ есть отображение *на*, если каждый элемент из $\mathcal{Y}$ является образом по крайней мере одного элемента из $\mathcal{X}$. Отображение $\alpha$ множества $\mathcal{X}$ в $\mathcal{Y}$ *взаимно однозначно*, если различные элементы из $\mathcal{Y}$ отображаются посредством $\alpha$ в различные элементы из $\mathcal{Y}$. Взаимно однозначное отображение множества $\mathcal{X}$ на себя будет называться *подстановкой* множества $\mathcal{X}$, даже если $\mathcal{X}$ конечно. Множество $\mathbf{T}_{\mathcal{X}}$ всех подстановок множества $\mathcal{X}$ с операцией суперпозиции называется *симметрической группой на* $\mathcal{X}$.

Для любого положительного целого числа $n$ назовем *$n$-й степенью*

$\mathbf{a}^{*n}$ элемента $\mathbf{a}$ полугруппы $\mathbf{S}$ элемент $\mathbf{a}_1 * \mathbf{a}_2 * ...\mathbf{a}_n$ при $\mathbf{a}_1 = \mathbf{a}_2 = ... = \mathbf{a}_n = \mathbf{a}$. Следующие два "закона показателей" $\mathbf{a}^{*m+n} = \mathbf{a}^{*m}\mathbf{a}^{*n}$, $(\mathbf{a}^{*m})^{*n} = \mathbf{a}^{*mn}$ очевидно, выполняются для любого $\mathbf{a} \in \mathbf{S}$ и для любых положительных чисел $m$ и $n$.

Непустое подмножество $\mathcal{T}$ группоида $\mathcal{S}$ называется его *подгруппоидом (подполугруппой*, если $(*)$ ассоциативно), если из включений $\mathbf{a} \in \mathcal{T}$ и $\mathbf{b} \in \mathcal{T}$ следует, что $\mathbf{a} * \mathbf{b} \in \mathcal{T}$. Пересечение любого семейства подгруппоидов, очевидно, либо пусто, либо является подгруппоидом. Если $\mathcal{A}$ - непустое подмножество группоида $\mathcal{S}$, то пересечение всех группоидов из $\mathcal{S}$, содержащих $\mathcal{A}$ ($\mathcal{S}$ само является одним из таких подгруппоидов), есть подгруппоид $< \mathcal{A} >$ группоида $\mathcal{S}$, содержащий $\mathcal{A}$ и содержащийся в каждом подгруппоиде из $\mathcal{S}$, содержащем $\mathcal{A}$. Если $\mathbf{S}$ - полугруппа, то любой подгруппоид из $\mathbf{S}$ является подполугруппой.

Если $\mathcal{S}$ - группоид, то мощность $|\mathcal{S}|$ множества $\mathcal{S}$ называется *порядком* $\mathcal{S}$. Если этот порядок конечен, то мы можем задать бинарную операцию в $\mathcal{S}$ посредством ее *таблицы умножения(таблицы Кэли)* так же, как и для конечных групп; часто такой наглядный способ задания полезен даже для бесконечного $\mathcal{S}$. Таблица Кэли это квадратная матрица, состоящая из элементов полугруппы $\mathcal{S}$, строки и столбцы которой занумерованы элементами из $\mathcal{S}$ таким образом, что элемент, находящийся в $\mathbf{a}$- строке и $\mathbf{b}$-столбце $(\mathbf{a}, \mathbf{b} \in \mathcal{S})$, равен произведению $\mathbf{a} * \mathbf{b}$.

Элемент $\mathbf{a}$ группоида $\mathcal{S}$ *сократим слева (справа)*, если для любых $\mathbf{x}, \mathbf{y} \in \mathcal{S}$ из соотношения $\mathbf{a} * \mathbf{x} = \mathbf{a} * \mathbf{y}$ $(\mathbf{x} * \mathbf{a} = \mathbf{y} * \mathbf{a})$ следует равенство $\mathbf{x} = \mathbf{y}$. Группоид $\mathcal{S}$ называется *группоидом с левым (правым) сокращением,* если каждый элемент из $\mathcal{S}$ сократим слева (справа). Таким образом, $\mathcal{S}$ - *группоид с сокращениями*, если $\mathcal{S}$ есть группоид и с левым, и с правым сокращением.

Два элемента $\mathbf{a}$ и $\mathbf{b}$ группоида $\mathcal{S}$ *коммутируют*, если $\mathbf{a} * \mathbf{b} = \mathbf{b} * \mathbf{a}$. В этом случае выполняется еще один "закон показателей": $(\mathbf{a} * \mathbf{b})^{*n} = \mathbf{a}^{*n}\mathbf{b}^{*n}$. Группоид $\mathcal{S}$ называется *коммутативным*, если любые два его элемента коммутируют. Элемент группоида $\mathcal{S}$, коммутирующий с каждым элементом из $\mathcal{S}$, называется *центральным элементом.* Для произвольного подмножества $\mathcal{X}$ группоида $\mathcal{S}$ множество

$$\mathrm{Cent}\,(\mathcal{X}) = \{\mathbf{a} \in \mathcal{S} \mid \mathbf{a} * \mathbf{x} = \mathbf{x} * \mathbf{a},\ \ \forall \mathbf{x} \in \mathcal{X}\} \tag{A.1}$$

называется *централизатором* подмножества $\mathcal{X}$.

Если $\mathbf{S}$ — полугруппа, множество всех центральных элементов $\mathbf{S}$, либо пусто, либо является подполугруппой. В последнем случае $\mathrm{Cent}\,(\mathcal{X})$ называется *центром* полугруппы $\mathbf{S}$.

Элемент $\mathbf{e}$ полугруппы $\mathbf{S}$ называется *левой (правой) единицей*, если $\mathbf{e} * \mathbf{a} = \mathbf{a}$ ($\mathbf{a} * \mathbf{e} = \mathbf{a}$) для всех $\mathbf{a} \in \mathbf{S}$. Элемент $\mathbf{e}$ полугруппы $\mathbf{S}$ называется *двусторонней единицей* (или просто *единицей*), если $\mathbf{e}$ — и левая, и правая единица. Заметим, что если $\mathbf{S}$ содержит левую единицу $\mathbf{e}$ и правую единицу $\mathbf{f}$, то $\mathbf{e} = \mathbf{f}$; действительно, $\mathbf{e} * \mathbf{f} = \mathbf{f}$, так как $\mathbf{e}$ — левая единица, и $\mathbf{e} * \mathbf{f} = \mathbf{e}$, так как $\mathbf{f}$ - правая единица.

Как следствие этого факта получаем, что для полугруппы $\mathbf{S}$ выполняется в точности одно из следующих утверждений:

**Утверждение A.1.** 1. $\mathbf{S}$ *не имеет ни левых, ни правых единиц;*

2. $\mathbf{S}$ *обладает по крайней мере одной левой единицей, но не имеет правых единиц;*

3. $\mathbf{S}$ *обладает по крайней мере одной правой единице, но не имеет левых единиц;*

4. $\mathbf{S}$ *обладает единственной двусторонней единицей и не имеет других левых или правых единиц.*

Элемент $\mathbf{z}$ полугруппы $\mathbf{S}$ называется *левым (правым) нулем,* если $\mathbf{z} * \mathbf{a} = \mathbf{z}$ ($\mathbf{a} * \mathbf{z} = \mathbf{z}$) для любого $\mathbf{a} \in \mathbf{S}$. Элемент $\mathbf{z}$ полугруппы $\mathbf{S}$ называется *нулем*, если $\mathbf{z}$ — и левый, и правый нуль. Если полугруппа $\mathbf{S}$ обладает левым нулем $\mathbf{z}_1$ и правым нулем $\mathbf{z}_2$, то $\mathbf{z}_1 = \mathbf{z}_2$. Следовательно, для любой полугруппы $\mathbf{S}$ выполняется в точности одно из предыдущих четырех утверждений с заменой в них слова *"единица"* на слово *"нуль"*.

Пусть $\mathcal{X}$ - произвольное множество. Определим бинарную операцию $(\circledast_R)$ в $\mathcal{X}$, полагая $\mathbf{x} \circledast_R \mathbf{y} = \mathbf{y}$ для всех $\mathbf{x}, \mathbf{y} \in \mathcal{X}$. Ассоциативность легко проверяется. Назовем $\mathbf{X}_R = \{\mathcal{X}; \circledast_R\}$ *полугруппой правых нулей.* Каждый элемент из $\mathbf{X}_R$ является правым нулем и левый единицей одновременно. *Полугруппа левых нулей* $\mathbf{X}_L = \{\mathcal{X}; \circledast_L\}$ определяется двойственным образом ($\mathbf{x} \circledast_L \mathbf{y} = \mathbf{x}$ для всех $\mathbf{x}, \mathbf{y} \in \mathcal{X}$). Несмотря на кажущуюся их тривиальность, эти полугруппы естественным образом появляются в ряде исследований.

Полугруппу $\mathbf{S}$ с нулем $\mathbf{z}$ будем называть *полугруппой с нулевым умножением,* если $\mathbf{a} * \mathbf{b} = \mathbf{z}$ для всех $\mathbf{a}, \mathbf{b}, \in \mathbf{S}$. Пусть $\mathbf{s}$ — произвольная полугруппа, и пусть $\mathbf{1} \notin \mathbf{S}$ — символ, не являющийся элементом из $\mathbf{S}$. Распространим бинарную операцию, заданную в $\mathbf{S}$, на множество $\mathbf{S}^1 = \mathbf{S} \bigcup \mathbf{1}$, полагая, $\mathbf{1} * \mathbf{1} = \mathbf{1}$ и $\mathbf{1} * \mathbf{a} = \mathbf{a} * \mathbf{1} = \mathbf{a}$ для любого $\mathbf{a} \in \mathbf{S}$. Легко проверить, что $\mathbf{S}^1$ есть полугруппа с единицей $\mathbf{1}$. Аналогичным образом можно присоединить нуль $\mathbf{0}$ к $\mathbf{S}$, а именно $\mathbf{S}^0 = \mathbf{S} \bigcup \mathbf{0}$, $\mathbf{0} * \mathbf{0} = \mathbf{0} * \mathbf{a} = \mathbf{a} * \mathbf{0} = \mathbf{0}$ для всех $\mathbf{a} \in \mathbf{S}$.

Элемент $\mathbf{e}$ полугруппы $\mathbf{S}$ называется *идемпотентом*, если $\mathbf{e} * \mathbf{e} = \mathbf{e}$. Односторонние единицы и нули суть идемпотентны. Если каждый элемент полугруппы $\mathbf{S}$ есть идемпотент, то будем говорить, что $\mathbf{S}$ является *полугруппой идемпотентов* или *связкой.*

Умножение множеств определяется формулой

$$\mathbf{A} \star \mathbf{B} \stackrel{def}{=} \bigcup \{\mathbf{a} * \mathbf{b} \mid \mathbf{a} \in \mathbf{A}, \quad \mathbf{b} \in \mathbf{B}\}\,. \tag{A.2}$$

Подмножество $\mathbf{L}$ полугруппы $\mathbf{S}$ называют *левым идеалом*, если $\mathbf{S} \star \mathbf{L} \subseteq \mathbf{L}$. Двойственно определяется правый идеал; так что $\mathbf{R}$ — *правый идеал* полугруппы $\mathbf{S}$. если $\mathbf{R} \star \mathbf{S} \subseteq \mathbf{R}$. Левые и правые идеалы вме-сте обычно называются *односторонними.* Подмножество полугруппы, являющееся как левым, так и правым идеалом, называется *двусторонним идеалом* или просто *идеалом.* Если $\mathbf{I}$ есть левый (правый, двусторонний) идеал полугруппы $\mathbf{S}$, то пишут $\mathbf{I} \trianglelefteq_l \mathbf{S}$ ($\mathbf{I} \trianglelefteq_r \mathbf{S}, \mathbf{I} \trianglelefteq \mathbf{S}$), опуская черточку внизу, если идеал собственный. Всякий односторонний идеал является подполугруппой. Для любого подмножества $\mathbf{A}$ полугруппы $\mathbf{S}$ множество $\mathbf{S} \star \mathbf{A}$ ($\mathbf{A} \star \mathbf{S}$, $\mathbf{S}{\star}\mathbf{A}{\star}\mathbf{S}$) будет левым (правым, двусторонним) идеалом; в частности, таковым будет множество $\mathbf{S} \star \mathbf{a}$ ($\mathbf{a} \star \mathbf{S}$, $\mathbf{S} \star \mathbf{a} \star \mathbf{S}$) для любого элемента $\mathbf{a} \in \mathbf{S}$. Для любого $n$ полугруппа $\mathbf{S}^{\star n}$ есть идеал в $\mathbf{S}$. Если $\mathbf{S}^{\star n} = \mathbf{S}^{\star n+k}$ для некоторого $k$, то $\mathbf{S}^{\star n} = \mathbf{S}^{\star m}$ для любого $m \geq n$; если $\mathbf{S}^{\star 2} = \mathbf{S}$, то полугруппа $\mathbf{S}$ называется *глобально идемпотентной.* Для любого $\mathbf{a} \in \mathbf{S}$ множество $\mathbf{L}(\mathbf{a}) = \{\mathbf{a}\} \cup \mathbf{S} \star \mathbf{a}$ ($\mathbf{R}(\mathbf{a}) = \{\mathbf{a}\} \cup \mathbf{a} \star \mathbf{S}$, $\mathbf{J}(\mathbf{a}) = \{\mathbf{a}\} \cup \mathbf{S} \star \mathbf{a} \cup \mathbf{a} \star \mathbf{S} \cup \mathbf{a} \star \mathbf{S} \star \mathbf{a}$) будет левым (правым, двусторонним) идеалом, содержащим $\mathbf{a}$ и содержащимся в любом левом (правом, двустороннем) идеале $\mathbf{I}$ таком, что $\mathbf{a} \in \mathbf{I}$, идеал $\mathbf{L}(\mathbf{a})$ ($\mathbf{R}(\mathbf{a}), \mathbf{J}(\mathbf{a})$) называют *главным левым* (*правым, двусторонним*) *идеалом*, порожденным элементом $\mathbf{a}$. Подполугруппу $\mathbf{T}$ полугруппы $\mathbf{S}$ называют *изолированной* (*вполне изолированной*), если для любого $\mathbf{a} \in \mathbf{S}$ и любого натурального $n$ (любых $\mathbf{a}, \mathbf{b} \in \mathbf{S}$) из того, что $\mathbf{a}^{*n} \in \mathbf{T}$ ($\mathbf{a}, \mathbf{b} \in \mathbf{T}$), следует, что $\mathbf{a} \in \mathbf{T}$ (хотя бы один из элементов $\mathbf{a}, \mathbf{b}$ принадлежит $\mathbf{T}$); если это условие выполняется тогда и только тогда, когда $\mathbf{S} \setminus \mathbf{T}$ есть объединение подполугрупп (подполугруппа) или $\mathbf{T} = \mathbf{S}$. Вполне изолированный идеал называется также *вполне первичным или простым.*

Подполугруппа $\mathbf{T}$ полугруппы $\mathbf{S}$ называется *выпуклой* (или *фильтром*, если для любых $\mathbf{a}, \mathbf{b} \in \mathbf{S}$ из того, что $\mathbf{a} * \mathbf{b} \in \mathbf{T}$, следует $\mathbf{a} \in \mathbf{T}$ и $\mathbf{b} \in \mathbf{T}$; это условие выполняется, очевидно, тогда и только тогда, когда $\mathbf{T} = \mathbf{S} \setminus \mathbf{I}$ для некоторого (необходимо вполне изолированного) идеала $\mathbf{I}$ или $\mathbf{T} = \mathbf{I}$. Всякое множество попарно не пересекающихся подполугрупп $\mathbf{T}_i$ полугруппы будем называть *россыпью.* Типичный пример — россыпь максимальных подгрупп. Если $\{\mathbf{T}_i\}_{i \in I}$ — россыпь полугруппы $\mathbf{S}$ такая, что $\bigcup\limits_{i\in I}\mathbf{T}_i = \mathbf{S}$ (т. е. компоненты россыпи образуют разбиение $\mathbf{S}$), то будем говорить, что данная россыпь *покрывает* $\mathbf{S}$. Если $\bigcup\limits_{i\in I}\mathbf{T}_i$ является порождающим множеством полугруппы $\mathbf{S}$, то будем говорить, что россыпь $\{\mathbf{T}_i\}_{i\in I}$ *порождает* $\mathbf{S}$.

## A.2. Полугруппы и преобразования

Один из важнейших примеров полугрупп доставляет множество $\mathcal{T}(\mathfrak{X})$ всех преобразований (отображений в себя) произвольного множества $\mathfrak{X}$. Образ элемента $x \in \mathfrak{X}$ при преобразовании $\alpha \in \mathcal{T}$ будем обозначать через $x\alpha$. *Произведение (суперпозиция, композиция)* $\alpha \circ \beta$ преобразований $\alpha$ и $\beta$ задается тогда формулой $x(\alpha\beta) = (x\alpha)\beta$. Введенная операция ассоциативна, так что $\mathcal{T}(\mathfrak{X})$ превращается в полугруппу, которая называется *симметрической полугруппой* или *полной полугруппой преобразований* на множестве $\mathfrak{X}$.

Принципиальная важность симметрических полугрупп состоит в том, что справедлив следующий аналог известной теоремы Кэли для групп: любая полугруппа вложима в подходящую симметрическую полугруппу; или, другими словами, любая полугруппа изоморфна некоторой полугруппе преобразований. Говорят также, что любая полугруппа изоморфно *представима* преобразованиями. Обсуждаемое сейчас утверждение может быть уточнено: полугруппа $\mathbf{S}$ вложима в $\mathcal{T}(\mathfrak{X})$, где множество $\mathfrak{X}$ либо совпадает с $\mathbf{S}$, либо получается из $\mathbf{S}$ с добавлением одного элемента. Умножение в множестве $\mathcal{T}(\mathfrak{X})$ можно определить и "справа налево" (записывая символы отображений слева от соответствующих элементов $\mathfrak{X}$); положим для любого $x \in \mathfrak{X}$

$$\alpha \circ \beta(x) \stackrel{def}{=} \alpha(\beta(x)) \tag{A.3}$$

Полученная таким образом полугруппа (ее также называют симметри-ческой) двойственна введеной выше полугруппе $\mathcal{T}(X)$. Мы будем пользоваться в тексте определением (A.3).

Многие изучаемые полугруппы преобразований оказываются подполугруппами каких-либо из перечисленных выше полугрупп. Наиболее типична ситуация, когда множество $\mathfrak{X}$ наделено той или иной математической структурой, и рассматриваются ее эндоморфизмы, т. е. преобразования, согласованные с этой структурой — сохраняющие соответствующие отношения и (или) операции, заданные на $\mathfrak{X}$. Совокупность $\mathrm{End}\,\mathfrak{X}$ всех эндоморфизмов данной структуры является подполугруппой в $\mathcal{T}(\mathfrak{X})$ — это *полугруппа эндоморфизмов.* Классический пример такой ситуации — полугруппа $\mathrm{End}_{\mathbb{F}}V$ линейных операторов векторного пространства $V$ над телом $\mathbb{F}$.

## A.3. Обратимость, нильпотентность и регулярность

*Моноидом* называется полугруппа с единицей $\mathbf{1}$. Элемент $\mathbf{a}$ моноида $\mathbf{S}$ называется *обратимым справа* (*слева*), если существует такой элемент $\mathbf{b} \in \mathbf{S}$, что $\mathbf{a} * \mathbf{b} = \mathbf{1}$ ($\mathbf{b} * \mathbf{a} = \mathbf{1}$). Элемент, обратимый слева и справа, называется двусторонне обратимым или просто *обратимым.* Множество $\mathbf{G}_r(\mathbf{S})$ (множество $\mathbf{G}_l(\mathbf{S})$) всех обратимых справа (слева) элементов моноида $\mathbf{S}$ является подмоноидом с правым (левым) сокращением; множество $\mathbf{G}(\mathbf{S}) = \mathbf{G}_r(\mathbf{S}) \cap \mathbf{G}_l(\mathbf{S})$ всех обратимых элементов является (максимальной) подгруппой в $\mathbf{S}$, называется *группой обратимых элементов*моноида $\mathbf{S}$. Группа $\mathbf{G}(\mathbf{S})$ тогда и только тогда включает в себя все односторонние обратимые элементы (т. е. верно равенство $\mathbf{G}_r(\mathbf{S}) = \mathbf{G}_l(\mathbf{S})$), когда $\mathbf{G}(\mathbf{S})$ выпукла в $\mathbf{S}$; при этом множество $\mathbf{S} \setminus \mathbf{G}(\mathbf{S})$, если оно не пусто, является наибольшим отличным от $\mathbf{S}$ идеалом в $\mathbf{S}$. Полугруппа $\mathbf{S}$ с таким свойством называется полугруп-пой с *отделяющейся групповой частью.* Полугруппами с отделяющейся групповой частью будут всякий конечный и всякий комутативный моноид, всякий моноид с сокращением, всякий моноид матриц над полем, а также полугруппы,рассматриваемы в основном тексте.

Элемент $\mathbf{a}$ полугруппы $\mathbf{S} = \mathbf{S}^0$ с нулем $\mathbf{0}$ называют *левым* (*правым*) *делителем нуля,* если $\mathbf{a} \neq \mathbf{0}$ и в $\mathbf{S}$ существует такой элемент $\mathbf{b} \neq \mathbf{0}$,что $\mathbf{a} * \mathbf{b} = \mathbf{0}$ ( $\mathbf{b} * \mathbf{a} = \mathbf{0}$). Элемент $\mathbf{a}$ из $\mathbf{S} = \mathbf{S}^0$называется *нильэлементом* (или *нильпотентным* элементом), если $\mathbf{a}^{*n} = \mathbf{0}$ для некоторого натурального $n$ ; наименьшее $n$ с таким свойством называется *индексом* элемента $\mathbf{a}$. Нильэлемент индекса $> 1$ является, очевидно, делителем нуля (левым и правым) . Множество нильэлеменов полугруппы $\mathbf{S} = \mathbf{S}^0$ обозначается $\mathrm{Nil}\,\mathbf{S}$. Элемент $\mathbf{a}$ *аннулирует слева* (*справа*) подмножество $\mathcal{X} \subseteq \mathbf{S}$, если $\mathbf{a} * \mathcal{X} = \mathbf{0}$ ($\mathcal{X} * \mathbf{a} = \mathbf{0}$). Множество $\mathrm{Ann}_L\mathcal{X} = \{\mathbf{a} \mid \mathbf{a} * \mathcal{X} = \mathbf{0}\}$ называется *левым аннулятором* множества $\mathcal{X}$; двойственно определяется *правый аннулятор* $\mathrm{Ann}_R\mathcal{X}$. Множество $\mathrm{Ann}\,\mathcal{X} = \mathrm{Ann}_L\mathcal{X} \cap \mathrm{Ann}_R\mathcal{X}$ называется (двусторонним) *аннулятором* множества $\mathcal{X}$. Свойства аннуляторов в полугруппах с нулем параллельны свойствам аннуляторов в кольцах; в частности, если $\mathcal{X}$ есть левый (правый) идеал, то $\mathrm{Ann}_L\mathcal{X}$ ($\mathrm{Ann}_R\mathcal{X}$) является двусторонним идеалом.

Если аннулятор содержит ненулевые элементы, то его называют *нетривиальным,* в противном случае — *тривильаным.*

Для полугруппы $\mathbf{S}$ через $\mathcal{E}(\mathbf{S})$ обозначают множество всех ее идемпотентов, определяемых $\mathbf{e}^{*2}= \mathbf{e}$. Во многих рассмотрениях полезную роль играет отношение *естественного частичного порядка* на $\mathcal{E}(\mathbf{S})$ заданное условием:

$$\mathbf{e} \leq \mathbf{f} \Leftrightarrow \mathbf{e} * \mathbf{f} = \mathbf{f} * \mathbf{e} = \mathbf{e}. \tag{A.4}$$

В этом смысле можно, например, говорить о цепях и антицепях в $\mathcal{E}(\mathbf{S})$. Очевидно, что единица (нуль) полугруппы $\mathbf{S}$ будет наибольшим (наименьшим) элементом в $\mathcal{E}(\mathbf{S})$. Идемпотент $\mathbf{e} \neq \mathbf{0}$ называется *примитивным,* если $\mathbf{e}$ является минимальным элементом в множестве ненулевых идемпотентов из $\mathcal{E}(\mathbf{S})$. В частности, всякий односторонний нуль полугруппы, не являющейся двусторонним нулем, будет примитивным. В полугруппе с левым (правым) сокращением всякий идемпотент является левой (правой) единицей. Следовательно, в полугруппе с сокращением может быть не более одного идемпотента, и если таковой есть, то это единица. Идемпотент $\mathbf{e}$ полугруппы $\mathbf{S}$ называется *центральным,* если $\mathbf{e} \in \mathrm{Cent}\,(\mathbf{S})$, т. е. $\mathbf{e} * \mathbf{x} = \mathbf{x} * \mathbf{e}$ для любого $\mathbf{e} \in \mathbf{S}$. Полугруппу, содержащую единственный идемпотент, называют *унипотентной.* Полугруппу, каждый элемент которой является идемпотентом, называют *полугруппой идемпотентов (или идемпотентной полугруппой),* а также *связкой.* Коммутативная связка называется *полурешеткой.* Последний термин оправдан, если рассмотреть на полурешетке $\mathbf{S}$ отношение естественного частичного порядка, заданное формулой (A.4), то для любых $\mathbf{a}, \mathbf{b} \in \mathbf{S}$ произведение $\mathbf{a} * \mathbf{b}$ будет равно $\inf(\mathbf{a}, \mathbf{b})$; и обратно, если $\mathcal{P}$— частично упорядоченное множество, в котором любые два элемента имеют точную нижнюю грань, то операция $(\circledcirc)$, заданная условием $a \circledcirc b = \inf(a, b)$, превращает $\mathcal{P}$ в коммутативную связку.

Простейшие примеры некоммутативных связок представляют *полугруппы левых (правых) нулей,* удовлетворяющие, по определению, тождеству $\mathbf{x} * \mathbf{y} = \mathbf{x}$ $(\mathbf{x} * \mathbf{y} = \mathbf{y})$. Полугруппу левых (правых) нулей называют также *левосингулярной (правосингулярной)*; полугруппа, являющаяся левосингулярной или правосингулярной, называется *сингулярной.* Сингулярная полугруппа не только некоммутативна, она обладает следующим свойством "антикоммутативности": $\mathbf{a} * \mathbf{b} \neq \mathbf{b} * \mathbf{a}$ для любых различных элементов $\mathbf{a}$ и $\mathbf{b}$. Произвольная полугруппа с указанным свойством, очевидно, является связкой и удовлетворяет тождеству $\mathbf{x} * \mathbf{y} * \mathbf{x} = \mathbf{x}$; такие полугруппы называются *прямоугольными* (или *прямоугольными связками*).

Элемент $\mathbf{a}$ полугруппы $\mathbf{S}$ называется регулярным, если имеет место включение $\mathbf{a} \in \mathbf{a} \star \mathbf{S} \star \mathbf{a}$, т. е., если в $\mathbf{S}$ существует такой элемент $\mathbf{x}$, что $\mathbf{a} = \mathbf{a} * \mathbf{x} * \mathbf{a}$. Из последнего равенства вытекает, что элементы $\mathbf{e} = \mathbf{a} * \mathbf{x}$ и $\mathbf{f} = \mathbf{x} * \mathbf{a}$ — идемпотенты, причем элемент $\mathbf{e}$ (элемент $\mathbf{f}$) служит для $\mathbf{a}$ левой (правой) единицей; если при этом $\mathbf{e} = \mathbf{f}$, то $\mathbf{a}$ будет групповым элементом. Обратно, если элемент $\mathbf{a} \in \mathbf{S}$ обладает левой (правой) единицей, принадлежащей множеству $\mathbf{a} \star \mathbf{S}$ (множеству $\mathbf{S} \star \mathbf{a}$) то $\mathbf{a}$, очевидно, регулярен. Элемент $\mathbf{a}$ регулярен тогда и только тогда, когда главный левый идеал $\mathbf{L}(\mathbf{a})$ (главный правый идеал $\mathbf{R}(\mathbf{a})$) порождается некоторым идемпотентом. Элементы $\mathbf{a}$ и $\mathbf{b}$ называются *инверсными* друг к другу (*обобщеннообратными*, *регулярносопряженными*), если $\mathbf{a} * \mathbf{b} * \mathbf{a} = \mathbf{a}$ и $\mathbf{b} * \mathbf{a} * \mathbf{b} = \mathbf{b}$. Всякий регулярный элемент обладает хотя бы одним инверсным к нему элементом. Всякий группой элемент $\mathbf{g}$ будет регулярным, обратный к нему в соответствующей максимальной подгруппе $\mathbf{G}$ элемент $\mathbf{g}^{-1}$будет инверсным к $\mathbf{g}$ (подчеркнем, что вне $\mathbf{G}$ могут существовать и другие инверсные к $\mathbf{g}$ элементы), и кроме того, $\mathbf{g}$ и $\mathbf{g}^{-1}$ перестановочны. Обратно, два перестановочных инверсных друг к другу элемента будут групповыми и взаимно обратными в соответствующей подгруппе $\mathbf{G}_e$. Групповые элементы называют также *вполне регулярными.*

Для элемента $\mathbf{a}$ произвольной полугруппы среди степеней $\mathbf{a}$, $\mathbf{a}^{*2} \ldots$ будет лишь конечное число различных тогда и только тогда, когда некоторая степень $\mathbf{a}$ равна идемпотенту; элемент $\mathbf{a}$ с таким свойством называется элементом *конечного порядка,* в противном случае $\mathbf{a}$ называется элементом *бесконечного порядка.* Полугруппа, все элементы которой иметют конечный порядок, называется *периодической.* Периодическая полугруппа с законом сокращения будет группой. Полярный к группам класс унипотентных периодических полугрупп составляют *нильполу-* *группы* — полугруппы с нулем $\mathbf{0}$, все элементы которых суть нильэлементы. Полугруппа $\mathbf{S} = \mathbf{S}^0$ называется *нильпотентной*, если $\mathbf{S}^{\star n} = \mathbf{0}$ для некоторого $n$; при желании указать $n$ говорят о *$n$-ступенно-нильпотентной* (или *$n$-нильпотентной*) полугруппе, наименьшее $n$ с таким свойством называют *ступенью нильпотентности.* Всякая нильпотентная полугруппа будет, очевидно, нильполугруппой с *нулевым умножением.* Полугруппу называют *левой (правой) нильполугруппой*, если некоторая степень каждого ее элемента есть левый (правый) нуль. Полугруппу $\mathbf{S}$ называют *нильпотентной слева(справа)*, если для некоторого $n$ множество $\mathbf{S}^{\star n}$ состоит из левых (правых) нулей.

## A.4. Отношения и гомоморфизмы

Бинарное отношение $\rho$ на полугруппе $\mathbf{S}$ назывется *стабильным* (или *устойчивым*) *слева,* если для любых $\mathbf{a}, \mathbf{b}, \mathbf{c} \in \mathbf{S}$ из $\mathbf{a}\rho\mathbf{b}$ следует $(\mathbf{c} * \mathbf{a})\,\rho\mathbf{b}$. Двойственно определяется *стабильность справа.* Отношение, стабильное слева и справа, называется (двусторонне) *стабильным.* Стабильная эквивалентность на полугруппе называется *конгруэнцией.*

Если $\rho$ — конгруэнция на полугруппе $\mathbf{S}$, то факторомножество $\mathbf{S}/\rho$ превращается в полугруппу заданием на нем операции $(\bullet)$, определяемой формулой $\rho\,(\mathbf{x}) \bullet \rho\,(\mathbf{y}) = \rho\,(\mathbf{x} * \mathbf{y})$. Эта полугруппа называется *факторполугруппой* полугруппы $\mathbf{S}$ по конгуэнции $\rho$.

Отображение $\rho^{\#} : \mathbf{S} \to \mathbf{S}/\rho$, ставящее в соответствие каждому элементу содержащий его $\rho$-класс

$$\rho\,(\mathbf{x}) \overset{def}{=} \{\mathbf{y} \in \mathbf{S} \mid \mathbf{x}\rho\mathbf{y}\}\,, \tag{A.5}$$

является сюръективным гомоморфизмом, он называется *естественным* (или *каноническим*) *гомоморфизмом* $\mathbf{S}$ на $\mathbf{S}/\rho$. Для произвольного гомоморфизма $\varphi : \mathbf{S} \to \mathbf{T}$ отношение $\ker\varphi = \{(\mathbf{a}, \mathbf{b}) \in \mathbf{S} \times \mathbf{S} \mid \mathbf{a}\varphi = \mathbf{b}\varphi\}$, называемое *ядром гомоморфизма*$\varphi$, есть конгруэнция на $\mathbf{S}$, причем факторполугруппа $\mathbf{S}/\ker\varphi$ изоморфна $\mathbf{T}$; более точно, существует изоморфизм $\psi$ полугруппы $\mathbf{S}/\ker\varphi$ на $\mathbf{T}$ такой, что $\psi = (\ker\varphi)^{\#}\,\psi$. Приведенные утверждения представляют собой конкретную версию *теоремы о гомоморфизмах*, верной для любых универсальных алгебр. Если $\rho, \tau$ — конгруэнции на полугруппе S, причем $\rho \subseteq \tau$, то существует (единственный) сюръективный гомоморфизм $\chi : \mathbf{S}/\rho \to \mathbf{S}/\tau$ такой, что $\tau^{\#} = \rho^{\#} \circ \chi$.

**Утверждение А.2.** *Следующие условия для непустого подмножества* $\mathfrak{N}$ *полугруппы* $\mathbf{S}$ *эквивалентны:*

1. $N$ *является классом некоторой конгруэнции на* $S$.

2. *Для любых* $\mathbf{a}$, $\mathbf{b} \in \mathfrak{N}$ *и любых* $\mathbf{x}, \mathbf{y}, \in \mathbf{S}$ *из* $\mathbf{x} * \mathbf{a} * \mathbf{y} \in \mathfrak{N}$ *следует* $\mathbf{x} * \mathbf{b} * \mathbf{y} \in \mathfrak{N}$.

Подмножество $\mathfrak{N}$ удовлетворяющее этим условиям, называется *нормальным комплексом.* Нормальный комплекс $\mathfrak{N}$, содержащий подполугруппу, будет подполугруппой (конкретная версия общеалгебраического факта). В частности, $\mathfrak{N}$ будет подполугруппой, если $\mathfrak{N}$ содержит идемпотент. Для регулярных полугрупп и эпигрупп справедливо обратное утверждение: всякий нормальный комплекс, являющийся подполугруппой, содержит идемпотент.

Специальный случай нормального комплекса $\mathfrak{N}$ представляет собой *нормальная подполугруппа* $\mathbf{N}$ — так называют полный прообраз единицы при некотором гомоморфизме данной полугруппы на моноид. Подполугруппа $\mathbf{N}$ полугруппы $\mathbf{S}$ будет нормальной тогда и только тогда, когда для любого $\mathbf{a} \in N$ и любых $\mathbf{x}, \mathbf{y} \in \mathbf{S}$ таких, что $\mathbf{x} * \mathbf{y} \in \mathbf{S}$, каждое из включений $\mathbf{x} * \mathbf{y} \in \mathbf{N}$ и $\mathbf{x} * \mathbf{a} * \mathbf{y} \in \mathbf{N}$ влечет за собой другое. Нормальные подполугруппы группы — это в точности ее нормальные подгруппы. В отличие от групп и колец, произвольная конгруэнция на полугруппе не определяется, вообще говоря, каким-либо одним из своих классов; это обусловливает специфику и сложность изучения конгруэн-ций на полугруппах.

Важный пример — рисовские конгруэции на произвольной полугруппе. Пусть $\mathbf{I}$ — идеал полугруппы $\mathbf{S}$. Определим отношение $\rho_{\mathbf{I}}$ на $\mathbf{S}$, полагая

$$\rho_{\mathbf{I}} = \{(\mathbf{a}, \mathbf{b}) \in \mathbf{S} \times \mathbf{S} \,|\, \mathbf{a}, \mathbf{b} \in \mathbf{I} \text{или } \mathbf{a} = \mathbf{b}\}\,. \tag{A.6}$$

Легко видеть,что $\rho_{\mathbf{I}}$— конгруэция; ее называют *идеальной* или *рисовской конгруэнцией* (или конгруэнцией Риса), соответствующей идеалу $\mathbf{I}$. Классы конгруэции $\rho_{\mathbf{I}}$ — это идеал $\mathbf{I}$ и (если $\mathbf{I} \neq \mathbf{S}$ ) одноэлементные подмножества $\{\mathbf{a}\}$, где $\mathbf{a} \in \mathbf{S} \setminus \mathbf{I}$. Фактор полугруппу $\mathbf{S}/\rho_{\mathbf{I}}$, как правило обозначают $\mathbf{S}/\mathbf{I}$ и называют *факторполугруппой Риса* полугруппы $\mathbf{S}$ по идеалу $I$. Факторполугруппа Риса всегда есть полугруппа с нулем. Образно говоря, $\mathbf{S}/\mathbf{I}$ получается из $\mathbf{S}$ "склеиванием" всех элементов идеала $\mathbf{I}$ и превращением их в нуль. Таким образом, идеалы представляют собой полярный по отношению к нормальным подполугруппам тип нормальных комплексов: они (и, очевидно, только они) являются полными прообразами нуля при гомоморфизмах данной полугруппы на полугруппу с нулем. Если в определении рисовской конгруэнции идеал $\mathbf{I}$ заменить произвольным левым(правым) идеалом, то введенное отношение $\rho_{\mathbf{I}}$ будет левой (правой) конгруэнцией.

Всякий гомоморфный образ $\mathbf{A}$ произвольной подполугруппы $\mathbf{T}$ полугруппы $\mathbf{S}$ называется *фактором* или *делителем* полугруппы $\mathbf{S}$; говорят, также, что $\mathbf{A}$ *делит* $\mathbf{S}$, и пишут $\mathbf{A}|\mathbf{S}$. Если $\mathbf{A} \simeq \mathbf{T}/\mathbf{I}$ ,где $\mathbf{I} \trianglelefteq \mathbf{T}$, то $\mathbf{A}$ называют *рисовским фактором.* Если $\mathbf{T}$, есть эпигруппа, то такой (рисовский) фактор будем называть (рисовским) эпифактором. Для любых подполугруппы $\mathbf{T}$ и идеала $\mathbf{J}$ из $\mathbf{S}$ множество $\mathbf{T} \cup \mathbf{J}$ будет подполугруппой; если при этом $\mathbf{T} \cap \mathbf{J} \neq \varnothing$, то $\mathbf{T} \cap \mathbf{J} \trianglelefteq \mathbf{T}$ и $\mathbf{T} \cup \mathbf{J}/\mathbf{J} \simeq \mathbf{T}/\mathbf{T} \cap \mathbf{J}$. Если $\mathbf{J}$ и $\mathbf{K}$ — идеалы из $\mathbf{S}$, причем $\mathbf{J} \subseteq \mathbf{K}$, то $\mathbf{K}/\mathbf{J} \trianglelefteq \mathbf{S}/\mathbf{J}$ и $(\mathbf{S}/\mathbf{J})\,/\,(\mathbf{K}/\mathbf{J}) \simeq \mathbf{S}/\mathbf{K}$. Если $\mathbf{J} \trianglelefteq \mathbf{S}$, то полугруппа $\mathbf{S}$ назы-вается *идеальным расширением полугруппы* $\mathbf{J}$ при помощи полугруппы $\mathbf{S}/\mathbf{J}$. Класс $\mathcal{K}$ называется *замкнутым относительно идеальных расширений* (*идеалов, факторполугрупп Риса*), если идеальное расширение $\mathcal{K}$-полугруппы при помощи $\mathcal{K}$-полугруппы (идеал $\mathcal{K}$-полугруппы, факторполугруппа Риса $\mathcal{K}$-полугруппы) будет $\mathcal{K}$-полугруппой. Идеальное расширение данной полугруппы при помощи нильпотентной полугруппы(нильполугруппы) называют ее *нильпотентным расширением (нильрасширением).*

Подмножество $\mathfrak{T}$ из $\mathbf{S}$, содержащее в точности один элемент из каждого $\rho$-класса, называется *трансверсалом.* Таким образом, подмножество $\mathfrak{T}$ и отношение $\rho$ *трансверсальны* (и каждое из них трансверсально другому).

Любое отображение $\theta : \mathbf{S} \to \mathbf{S}'$ определяет на $\mathbf{S}$ эквивалентность

$$\ker\theta = \{(\mathbf{x},\mathbf{y}) \in \mathbf{S}\times\mathbf{S} : \theta(\mathbf{x}) = \theta(\mathbf{y})\}\,. \tag{A.7}$$

Если $\theta$ сюръективно, то отображение $\hat{\theta} : \mathbf{S}/\ker\theta \to \mathbf{S}'$, определяемое равенством $\hat{\theta}(\rho(\mathbf{x})) = \theta(x)$ для любого $\mathbf{x} \in \mathbf{S}$, является биекцией (теорема об изоморфизме для множеств). Существует также биекция произвольного трансверсала множества $S$ по модулю $\ker\theta$ на $\mathbf{S}'$.

Частичное упорядочение бинарных отношений индуцирует структуру решетки на множестве всех эквивалентностей на $\mathbf{S}$. В самом деле, для произвольных эквивалентностей $\rho_1$ и $\rho_2$на $\mathbf{S}$ мы имеем

$$\inf(\rho_1,\rho_2) = \rho_1 \cap \rho_2, \quad \sup(\rho_1,\rho_2) = \bigcup_{n\in N} (\rho_1 \cup \rho_2)^n\,. \tag{A.8}$$

Если дан сюръективный гомоморфизм $\theta : \mathbf{S} \to \mathbf{S}'$, определим на $\mathbf{S}$

конгруэнцию $\mathrm{Ker}\,\theta$ (ядерную конгруэнцию отображения $\theta$) :

$$\mathrm{Ker}\,\theta = \{(\mathbf{x},\mathbf{y}) \in \mathbf{S}\times\mathbf{S} : \theta(\mathbf{x}) = \theta(\mathbf{y})\}\,.$$

Единственная возможность удовлетворяющей равенству $\hat{\theta}\circ\chi_{\mathrm{Ker}\,\theta} = \theta$, это положить $\hat{\theta}(\tilde{\mathbf{x}}) = \theta(\mathbf{x})$ для любого $x \in S$,где $\tilde{\mathbf{x}} = \chi_{\mathrm{Ker}\,\theta}(\mathbf{x})$ . Этот результат является частным случаем следующей теоремы.

**Теорема А.3.** *Пусть $\theta : \mathbf{S} \to \mathbf{S}'$ — гомоморфизм полугруппы $S$ на полугруппу $\mathbf{S}'$, $\rho$ — конгруэнция на $S$, для которой $\mathrm{Ker}\,\theta \subseteq \rho$. Определим бинарное отношение $\rho'$ на $S'$, положив*

$$\rho' = \{(\mathbf{x}',\mathbf{y}') \in \mathbf{S}'\times\mathbf{S}' \mid \exists \mathbf{x},\mathbf{y} \in \mathbf{S},\, \mathbf{x}\rho\mathbf{y},\, \theta(\mathbf{x}) = \mathbf{x}', \theta(\mathbf{y}) = \mathbf{y}'\}\,.$$

*Тогда:*

1. *Отношение $\rho'$ есть конгруэнция на $\mathbf{S}'$.*

2. *Существует единственное отображение $\hat{\theta}$ из $\mathbf{S}/\rho$ в $\mathbf{S}'/\rho'$, такое, что $\hat{\theta}\circ\chi_\rho = \chi_{\rho'}\circ\theta$, причем $\hat{\theta}$ — изоморфизм.*

3. *Отображение $\rho' \longmapsto \rho$ определяет изоморфизм решетки всех конгруэнций на $\mathbf{S}$, содержащих $\mathrm{Ker}\,\theta$,на решетку всех конгруэнций на $\mathbf{S}'$.*

## А.5. Теория идеалов

Двусторонний идеал $\mathbf{I}$ полугруппы $\mathbf{S}$ называется *минимальным идеалом,* если для любого идела $\mathbf{J} \subseteq \mathbf{S}$ из $\mathbf{J} \subseteq I$ следует $\mathbf{J} = \mathbf{I}$. Если $\mathbf{I}$ — минимальный идеал и $\mathbf{J}$ —любой другой идеал, то пересечение $\mathbf{I}\cap\mathbf{J}$ непусто, поскольку $\mathbf{I}\star\mathbf{J} \subseteq \mathbf{I}\cap\mathbf{J}$; кроме того, включение $\mathbf{I}\cap\mathbf{J} \subseteq \mathbf{I}$ влечет за собой равенство $\mathbf{I}\cap\mathbf{J} = \mathbf{I}$ и поэтому $\mathbf{J} \subseteq \mathbf{I}$. Таким образом, минимальный идеал является универсально минимальным (т. е. наименьшим) и, следовательно, *единственным.* По этой причине минимальный идеал полугруппы (если он существует) называют ее ядром Сушкевича. Любая конечная полугруппа обладает минимальным идеалом.

Полугруппа называется *простой*, если она не содержит идеалов, отличных от самой себя. Если полугруппа $\mathbf{S}$ проста, то для любого $\mathbf{a} \in S$ выполняется равенство $\mathbf{S} \star \mathbf{a} \star \mathbf{S} = \mathbf{S}$. Обратно, если $\mathbf{S} \star \mathbf{a} \star \mathbf{S} = \mathbf{S}$ для любого $\mathbf{a} \in \mathbf{S}$, то, взяв идеал $\mathbf{I}$ в $\mathbf{S}$ и элемент $\mathbf{a} \in \mathbf{I}$, получим $\mathbf{S} = \mathbf{S} \star \mathbf{a} \star \mathbf{S} \subseteq \mathbf{I}$, т. е. $\mathbf{S} = \mathbf{I}$, и потому $\mathbf{S}$ проста. Следовательно, чтобы доказать простоту полугруппы $\mathbf{S}$, достаточно предъявить по крайней мере одну пару $\mathbf{x}, \mathbf{y}$ — решение уравнения $\mathbf{x} * \mathbf{a} * \mathbf{y} = \mathbf{b}$ — для любых $\mathbf{a}, \mathbf{b} \in \mathbf{S}$.

**Определение A.4.** *Главным идеальным рядом полугруппы* $\mathbf{S}$ *называется конечная цепь* $\mathbf{I}_1 \subset \mathbf{I}_2 \subset \ldots \subset \mathbf{I}_n = \mathbf{S}$ *идеалов из* $\mathbf{S}$*, где* $\mathbf{I}_1$ *— мимальныйй идеал, а* $\mathbf{I}_k$ *является максимальным среди идеалов из* $\mathbf{S}$*, содержащихся в* $\mathbf{I}_{k+1}$*,* $k = 1, 2, \ldots, n-1$*. Факторполугруппы Риса* $\mathbf{I}_{k+1}/\mathbf{I}_k$ *и идеал* $\mathbf{J}_1$ *называются факторами этого ряда.*

*Пример* **A.5.** Полугруппами с главными идеальными рядами являются, например, конечные полугруппы и полугруппа $\mathrm{End}_{\mathbb{K}} V$ всех линейных преобразований конечномероного векторного пространства $V$ над полем $\mathbb{K}$.

**Лемма A.6.** *Пусть* $\mathbf{I}$ *— идеал полугруппы* $\mathbf{S}$ *и* $\mathbf{J}$ *— максимальнный идеал из* $\mathbf{S}$*, строго содержащийся в* $\mathbf{I}$*. Для произвольного* $\mathbf{a} \in \mathbf{I} \setminus \mathbf{J}$ *обозначим через* $\mathbf{I}(\mathbf{a})$ *множество всех таких* $\mathbf{x} \in \mathbf{S}^1 \star \mathbf{a} \star \mathbf{S}^1$*, что* $\mathbf{S}^1 \star \mathbf{x} \star \mathbf{S}^1 \subset \mathbf{S}^1 \star \mathbf{a} \star \mathbf{S}^1$*. Тогда* $\mathbf{I}(\mathbf{a})$ *является идеалом в* $S$ *и фактор полугруппы Риса* $\mathbf{S}^1 \star \mathbf{a} \star \mathbf{S}^1 / \mathbf{I}(\mathbf{a})$ *и* $\mathbf{I}/\mathbf{J}$ *изоморфны.*

Множество $\mathbf{S}^1 \star \mathbf{a} \star \mathbf{S}^1 \setminus \mathbf{I}(\mathbf{a})$ есть $\mathscr{J}$-класс элемента $\mathbf{a}$. Заметим, что множество $\mathbf{I}(\mathbf{a})$, если оно непусто, является максимальным идеалом из $\mathbf{S}$, содержащимся в $\mathbf{S}^1 \star \mathbf{a} \star \mathbf{S}^1$. В самом деле, если $\mathbf{I}$ — такой идеал из $\mathbf{S}$, что $\mathbf{I}(\mathbf{a}) \subset \mathbf{I} \subseteq \mathbf{S}^1 \star \mathbf{a} \star \mathbf{S}^1$, то, взяв $\mathbf{x} \in \mathbf{I}$, $\mathbf{x} \notin \mathbf{I}(\mathbf{a})$, получим $\mathbf{S}^1 \star \mathbf{a} \star \mathbf{S}^1 = \mathbf{S}^1 \star \mathbf{x} \star \mathbf{S}^1 \subseteq \mathbf{I}$ и следовательно, $\mathbf{I} = \mathbf{S}^1 \star \mathbf{a} \star \mathbf{S}^1$. Из этого замечания вытекает, что полугруппа $\mathbf{S}^1 \star \mathbf{a} \star \mathbf{S}^1/\mathbf{I}(\mathbf{a})$ является 0-минимальным идеалом полугруппы $\mathbf{S}/\mathbf{I}(\mathbf{a})$.

## A.6. Свойства отношений Грина

Отношения Грина на полугруппе $\mathbf{S}$ определяются формулами

$$\mathbf{a}\mathscr{R}\mathbf{b} \iff \mathbf{a} \star \mathbf{S}^1 = \mathbf{b} \star \mathbf{S}^1, \tag{A.9}$$

$$\mathbf{a}\mathscr{L}\mathbf{b} \iff \mathbf{S}^1 \star \mathbf{a} = \mathbf{S}^1 \star \mathbf{b}, \tag{A.10}$$

$$\mathbf{a}\mathscr{J}\mathbf{b} \iff \mathbf{S}^1 \star \mathbf{a} \star \mathbf{S}^1, \tag{A.11}$$

$$\mathscr{D} = \mathcal{R} \vee \mathscr{L}, \tag{A.12}$$

$$\mathscr{H} = \mathscr{R} \cap \mathscr{L}. \tag{A.13}$$

Из определения видно, что $\mathscr{R}$ (соответственно $\mathscr{L}$) есть левая (соответственно правая) конгруэнция. Остальные отношения являются вообще говоря, просто эквивалентностями. Класс элемента $\mathbf{a} \in \mathbf{S}$ обозначается латинской буквой, соответствующей валентности, с индексом $\mathbf{a}$ : $\mathsf{R}_\mathbf{a}$ обозначает $\mathscr{R}$-класс элемента $\mathbf{a}$, $\mathsf{L}_\mathbf{a}$ — соответственно $\mathscr{L}$-класс и т.д. Отметим, что $\mathsf{J}_\mathbf{a} = \mathbf{S}^1 \star \mathbf{a} \star \mathbf{S}^1 \setminus \mathbf{I}(\mathbf{a})$, где идеал $\mathbf{I}(\mathbf{a})$ определен в Лемме A.6.

**Предложение A.7.** *Любая правая конгруэнция, содержащаяся в $\mathscr{L}$, коммутирует с любой левой конгруэнцией, содержащейся в $\mathscr{R}$.*

**Следствие A.8.** $\mathscr{D} = \mathscr{R} \vee \mathscr{L} = \mathscr{R} \circ \mathscr{L} = \mathscr{L} \circ \mathscr{R}$

*Доказательство.* Так как $(\mathscr{R} \circ \mathscr{L}) \circ (\mathscr{R} \circ \mathscr{L}) = \mathscr{R} \circ \mathscr{R} \circ \mathscr{L} \circ \mathscr{L} \subseteq \mathscr{R} \circ \mathscr{L}$ , отношение $\mathscr{R} \circ \mathscr{L}$ — эквивалентность на $\mathbf{S}$. Учитывая включение $\mathscr{R} \circ \mathscr{L} \supseteq \mathscr{R} \cup \mathscr{L}$ и определение отношения $\mathscr{R} \vee \mathscr{L}$, получаем $\mathscr{R} \vee \mathscr{L} =$

$\mathscr{R} \circ \mathscr{L}$. ■

**Предложение A.9.** *Если полугруппа* $\mathbf{S}$ *конечна, то* $\mathscr{D} = \mathscr{J}$.

Существует естественный частичный порядок на множестве классов каждого из отношений $\mathscr{H}, \mathscr{R}, \mathscr{L}, \mathscr{J}$. Напрмер, частичный порядок на множестве $\mathscr{R}$-классов определяется условием: $\mathsf{R}_{\mathbf{a}} \trianglelefteq \mathsf{R}_{\mathbf{b}}$, если и только если $\mathbf{a} \star \mathbf{S}^1 \subseteq \mathbf{b} \star \mathbf{S}^1$. Для глобального описания полугруппы $\mathbf{S}$ наиболее важен частичный порядок на множестве $\mathscr{J}$-классов, определяемый условием: $\mathsf{J}_{\mathbf{a}} \trianglelefteq \mathsf{J}_{\mathbf{b}}$, если и только если $\mathbf{S}^1 \star \mathbf{a} \star \mathbf{S}^1 \subseteq \mathbf{S}^1 \star \mathbf{b} \star \mathbf{S}^1$. Частично упорядоченное множество $\mathscr{J}$-классов мы назовем *остовом* полугруппы $\mathbf{S}$. Полугруппы, в которых $\mathscr{D} = \mathscr{J}$, могут быть описаны в терминах их остова и локального строения различных $\mathscr{D}$-классов.

# Приложение Б

# Суперпространства, супермногообразия и их типы

Существуют две основные математические концепции супермногообразия. Первая, разработанная Березиным [22,27,30,47,68,107], Лейтесом [29, 106, 186, 275, 743, 744] и Костантом [185], называемая алгебраической, состоит в расширении пучка вещественных функций на действительном многообразии до пучка $\mathbb{Z}_2$-градуированных коммутативных алгебр [108,109,113,679,745–754]. Второй подход, функциональный, развитый в работах Роджерс [94,112,127,755–757], ДеВитта [174] и Владимирова-Воловича [117, 758] (см. обзор в [91]), сводится к модификации определения самого многообразия [116,128,175,187,226,256, 389,759]. В работе [760] делалась попытка объединить эти два подхода и рассмотреть бесконечномерные супермногообразия алгебраического подхода с мультиградуировкой. Сравнительный анализ различных подходов к определению супермногообразий проводился в [91,115,268,761–765].

## Б.1. Алгебраический подход к супермногообразиям

Супермногообразие *алгебраического подхода* — это пара $(\mathfrak{X}, \mathcal{O}_{\mathfrak{X}})$, где $\mathfrak{X}$ — $C^\infty$-многообразие и $\mathcal{O}_{\mathfrak{X}}$-пучок $\mathbb{Z}_2$-градуированных коммутативных алгебр, удовлетворяющих следующим условиям: 1) существует сюръективное отображение пучков $\sigma : \mathcal{O}_{\mathfrak{X}} \to C^\infty$, где $C^\infty$ — пучок гладких действительных функций на $\mathfrak{X}$; 2) существует открытое покрытие

$\{\mathscr{U}_i\}$ на $\mathfrak{X}$ и изоморфизмы $\mathbb{Z}_2$-градуированных коммутативных алгебр:

$$\varphi_i : \mathcal{O}_{\mathfrak{X}}|_{\mathscr{U}_i} \to \Lambda \otimes C^\infty|_{\mathscr{U}_i}, \tag{Б.1}$$

где $\Lambda$ — алгебра Грассмана, имеющая конечное число $K$ канонических антикоммутирующих образующих $\{\xi_1, \dots \xi_K\}$.

Такое определение супермногообразий обобщает алгебраическое определение действительного гладкого многообразия [766, 767].

Изоморфизмы $\varphi_i$ означают, что $\mathcal{O}_{\mathfrak{X}}$ локально можно рассматривать как пучок ростков функций на $\mathfrak{X}$ со значением в алгебре Грассмана $\Lambda$. Любая такая функция $f$ полностью определяется семейством $2^K$ действительных функций $f_{i_1 \dots i_r}$, входящих в разложение (В.1) при $n = K$. Если рассматривать $f$ как суперполе, ему будет соответствовать супермультиплет, все элементы которого — функции $\mathfrak{X} \to \mathbb{R}$, и, следовательно, эти функции могут рассматриваться как классические поля.

Система координат на области $\mathscr{U}_i$ тривиализации супермногообразия $(\mathfrak{X}, \mathcal{O}_{\mathfrak{X}})$ состоит из координат $\{x_i\}$ на многообразии $\mathfrak{X}$ и образующих $\mathbb{Z}_2$-градуированной алгебры $\mathcal{O}_{\mathfrak{X}}|_{\mathscr{U}_i}$. В качестве таких образующих можно выбрать координатные функции $[x_i] : \mathscr{U}_i \to \mathbb{R}$ и локально постоянные на $\mathscr{U}_i$ функции $[\xi_A]$ со значением в образующих алгебрах Грассмана. Таким образом нечетные координаты появляются как нечетные образующие $\mathbb{Z}_2$-градуированной алгебры функций.

В качестве суперкоординатных преобразований выступают автоморфизмы пучка $(\mathfrak{X}, \mathcal{O}_{\mathfrak{X}})$. Существует однозначное соответствие между такими автоморфизмами и семейством $\{y_i([x_i], [\xi_A]), \eta_A([x_i], [\xi_A])\}$ локальных сечений пучка $\mathcal{O}_{\mathfrak{X}}$. По заданному сечению автоморфизм определяется как: $x_i \to y_i(x_i, 0)$, $f([y_i], [\eta_A]) \to f([x_i], [\xi_A])$. Таким образом, четные образующие пучка $[x_i]$ отождествляются с координатами в обычном смысле. Однако такое отождествление нарушается при суперкоординатном преобразовании, поскольку четные образующие $[y_i]$ пучка в отличие от новых координат содержат нильпотентную часть.

Нечетные образующие пучка $[\xi_A]$ не являются координатами в обычном смысле. Например, в алгебраическом подходе не существует глобальных трансляций $[\xi_A] \to [\xi_A + \alpha]$,где $\alpha$ — нечетный элемент $\Lambda$,не зависящий от $\xi_A$. Следовательно, координаты $[x_i], [\xi_A]$ супермногообразия и суперполя $f([x_i], [\xi_A])$ допускают представления супералгебры Пуанкаре, но не супергруппы Пуанкаре.

Необходимо отметить, что элементы супергрупп Ли параметризуются определенным набором четных и нечетных элементов $\Lambda$ [27, 30, 186]. И, таким образом, нечетные координаты группового пространства отличаются от функциональных нечетных координат $[\xi_A]$. Поэтому групповое пространство супергрупп Ли не является супермногообразием в рамках алгебраического подхода.

## Б.2. Функциональный подход

На алгебре Грассмана может быть задана структура банаховой алгебры. Это можно сделать, например, с помощью нормы вида [112]

$$\|\xi\| = \sum \left|a^{A_1 \ldots A_J}\right|, \; \xi = \sum_{J=0}^{K} a^{A_1 \ldots A_J} \xi_{A_1} \ldots \xi_{A_J}. \tag{Б.2}$$

Суперпространство *функционального подхода* [91] размерности $(n|m)$ определяется как прямое произведение $n$ экземпляров четной части $\Lambda$ и $m$ экземпляров нечетной части $\Lambda$: $B^{n|m} = \Lambda_0^n \times \Lambda_1^m$. С одной стороны, такое суперпространство может рассматриваться как $\Lambda$-оболочка $\mathbb{Z}_2$-градуированного векторного пространства $L^{n|m} = L_0 \oplus L_1 = \mathbb{R}^n \oplus \mathbb{R}^m$, которая получается умножением четных (нечетных) эле-ментов $L$ на четные (нечетные) элементы $\Lambda$. При таком подходе в качестве базиса $B^{n|m}$ выступают $(n+m)$ базисных векторов пространства $L:\{l_i, i=1,\ldots,n; l_j, j=1,\ldots,m\}$, а в качестве координат — элементы $\{x_i,\theta_j\}$ из $B^{n|m}$. С другой стороны, $B^{n|m}$ является $R^{2^{L-1}(n+m)}$— мерным действительным векторным пространством [190, 768].

На суперпространстве $B^{n|m}$ рассматриваются $\Lambda$-значные функции $f\left(x_i,\theta_j\right)$. Их дифференцирование по грассмановым координатам определяется аналогично обычному дифференцированию на банаховых пространствах с учетом специфики, связанной с антикоммутированием нечетных координат [769, 770].

Для задания как четной, так и нечетной координаты в функциональном подходе необходимо и достаточно задать $2^{K-1}$ действительных коэффициентов разложения ее по базису $\Lambda$. Существует аналогия с комплексным анализом, где переменная $z=x+iy$ содержит две действительные переменные $x$ и $y$. Эта аналогия может быть расширена. Например, условие супердифференцируемости ведет к уравнениям для производных по действительным координатам, аналогичным условиям Коши-Римана [117]. На суперпространстве $\mathscr{M}^{n|m}$ может быть построена теория контурного интегрирования [771–775], в том числе в нечетном секторе [776, 777] и в некомпактном случае [778].

Супермногообразием $\mathscr{M}^{n|m}$ размерности $(n|m)$ называется банахово многообразие, допускающее атлас $\left\{\mathscr{U}_i,\psi_i:\mathscr{U}_i\to B^{n|m}\right\}$, функции перехода которого — супергладкие. Можно построить касательное суперрасслоение $T\mathscr{M}^{n|m}$ над многообразием $\mathscr{M}^{n|m}$. Типичным слоем его будет суперпространство $B^{n|m}$, и структурной группой будет супергруппа Ли $L\left(n|m\right)$ автоморфизмов $B^{n|m}$.

Понятия суперпространства, супермногообразия, суперрасслоения в функциональном подходе являются непосредственными градуированными обобщениями понятий обычной дифференциальной геометрии.

## Б.3. Различия между алгебраическим и функциональным подходами

Нечетные координаты $[\xi_A]$ в алгебраическом подходе являются образующими алгебры Грассмана, в то время как координаты $\theta_j$ принимают значения во всей нечетной части $\Lambda$. Индексы $i$ — это индексы $\mathbb{Z}_2$-градуированного векторного пространства $L$, а не индексы образующих $\Lambda$. Поскольку при строгом рассмотрении нечетные координаты $[\xi_A]$ алгебраического подхода выступают как образующие алгебры функций со значениями в $\Lambda$, то в функциональном подходе не существует понятия нечетных переменных, а запись вида $f(\xi)$ лишь уточняет по какой системе образующих производится разложение. Напротив, в подходе функциональном нечетные величины $\theta_j$ могут рассматриваться как нечетные переменные, каждая из которых содержит $2^{K-1}$ "скрытых" индексов $\Lambda$. Четные, координаты супермногообразия функционального подхода $x_i$ не являются вещественными, а принимают значения в четной части $\Lambda$. Следовательно,в подходе функциональном необходима особая процедура для придания физического смысла мультиплетам, отвечающим суперполям.

Функции в алгебраическом и функциональном подходах, принимая значения в $\Lambda$, определены на совершенно разных множествах ( $\mathbb{R}$ и $B^{n|m}$ соответственно). В алгебраическом подходе разложение функции по нечетным образующим $[\xi_A]$ есть ее разложение по базису области значений, в то время как в функциональном подходе разложение функции по нечетным переменным $\theta_j$ аналогично разложению в ряд Тейлора. Кроме того, при нечетном числе генераторов в $\Lambda$ существует неоднозначнсть в определении производных $\partial/\partial\theta$, которая отсутствует при алгебраическом определении $\partial/\partial\xi_A$ в алгебраическом подходе [779].

Для придания физического смысла суперполевым моделям необходимо с каждым супермногообразием $\mathscr{M}^{n|m}$ связать действительное многообразие и установить соответствие между функциями на них. При рассмотрении тривиального супермногообразия — области в $B^{n|m}$ — трудностей не возникает, так как $\mathbb{R}^n \subset B^{n|m}$ и существует взаимнооднозначное соответствие между супергладкими функциями на $B^{n|0}$ (т.е. элементами супермультиплета $f_{i_1 \ldots i_r}(x_i)$ для суперполя $f(x_i, \theta_j)$) и гладкими функциями на $\mathbb{R}^n$ (физическими полями $f_{i_1 \ldots i_r}(x_0)$). Хотя переход от $f_{i_1 \ldots i_r}(x_i)$ к $f_{i_1 \ldots i_r}(x_0)$ не оговаривается в физических работах, он необходим, так как физические поля — функции вещественных переменных.

Сложнее дело обстоит в общем случае. Нетривиальная склейка областей $B^{n|m}$ затрудняет выделение $\mathbb{R}^n$. Для этого обычно используется отображение $\epsilon$, ставящее в соответствие элементу $\Lambda$ его действительную часть. На супермногообразии определяется отношение эквивалентности: $x \sim y \left(x, y \in \mathscr{M}^{n|m}\right)$, если существует карта $(\mathscr{U}, \psi) : x \in \mathscr{U}, y \in \mathscr{U}$ и $\epsilon\psi(x) = \epsilon\psi(y)$. Однако это отношение не выявляется в общем случае независимым от выбора карты (если карты имеют несвязное пересечение). На супермногообразии можно задать такой атлас, что $\epsilon$-эквивалентность будет определена глобальнона многообразии [181, 225, 276]. Однако фактор супермногообразия по такому отношению эквивалентности не всегда будет даже топологическим многообразием.

Эта проблема является общей для всех моделей, использующих формализм суперполей, и остается принципиальной для физической интерпретации супергеометрического формализма [374, 780, 781].

## Б.4. Суперконформные многообразия

Одним из важейших классов супермногообразий являются суперконформные многообразия [341–343], введенные для вычисления суперструнных амплитуд [327, 330, 331, 344].

При рассмотрении струны в формализме Полякова [333] анализ взаимодействия струн сводится к изучению топологически нетривиальных мировых поверхностей струны. Для бозонной струны в силу конформной инвариантности вычисление струнных амплитуд в критической размерности ($d = 26$) сводится к интегрированию по так называемому конформному пространству модулей — множеству конформных классов двумерных римановых многообразий [349, 782, 783]. Этот факт был впервые установлен в работе Полякова [333] и подробно проанализирован в работах Книжника [280, 326, 328, 336, 784–786]. Аналогичные утверждения имеет место и для фермионной струны [277, 334, 335]. При этом конформное пространство модулей следует заменить на суперконформное пространство модулей — классов суперконформной эквивалентности $(1|1)$-мерных комплексных многообразий с суперконформной структурой [353, 356, 362].

Понятие суперриманова многообразия, появлялось ранее в работах, в которых исследовалалсь связь двумерной супергравитации и суперструны [320, 323]. В этих работах рассматривались $(2|2)$-мерные римановы супермногообразия с определенными ограничениями на кривизну и кручение. Можно проверить, что суперримановы многообразия в определенном выше смысле находятся во взаимно однозначном соотвестсвии с многообразиями, исследовавшимися в этих работах (это вытекает из доказанной в них возможности выбора супертетрады в конформно-плоском виде).

*Суперконформным многообразием* будем называть многообразие, склеенное из $(1|1)$-мерных суперобластей с помощью суперконформных преобразований. Примером суперконфомного многообразия является суперпроективное пространство $\mathbb{C}P^{1,1}$, получающееся из комплексного суперпространства $\mathbb{C}^{2|1}$ с координатами $(w_1, w_2, \eta)$ при отождествлении $(w_1, w_2, \eta) \sim (\lambda w_1, \lambda w_2, \lambda\eta)$ (здесь $w_1, w_2, \lambda$ — четные элементы, $\eta$ — нечетный). Вводя координаты $(z, \theta) = \left(w_1, w_2^{-1}, \eta w_2^{-1}\right)$ и $(z', \theta') = \left(w_1, w_2^{-1}, \eta w_2^{-1}\right)$, мы можем представлять себе $\mathbb{C}P^{1,1}$ как суперпространство, склеенное из двух суперобластей с помощью суперконформного преобразования $z' = z^{-1}$, $\theta' = \theta z^{-1}$.

Более общий пример суперконформного многообразия можно построить, исходя из произвольного комплексного многообразия размерности 1. Если исходное многообразие склеено из областей $\mathscr{U}_i$, локальные координаты которых связаны аналитическими преобразованиями $z_i = f_{ij}(z_j)$, то соответствующее суперконформное многообразие склено из суперобластей с помощью суперконформных преобразований $z_i = f_{ij}(z_j)$, $\theta_i = \sqrt{f'_{ij}(z_j)}\theta_j$. Отметим, что описанная процедура не вполне однозначна из-за свободы в выборе ветви квадратного корня.

Можно дать инвариантное определение суперконформного многообразия. Для этого заметим, что в каждой точке суперконформного-многообразия выделено $(0,1)$-мерное комплексное подпространство касательного пространства, порожденное в локальных координатх $Z = (z, \theta)$ вектором $D$ (как обычно дифференциальный оператор первого порядка отождествляется в векторным полем). В силу того, что $\tilde{D} = F^{-1}D$, где $F$ — конформный фактор, это подпространство не зависит от выбора локальной системы координат. Мы можем определить суперконформное многообразие как $(1|1)$-мерное комплексное многообразие, в каждой точке которого выделено $(0|1)$-мерное подпространство касательного пространства, аналитически зависящее от точки. При этом должно быть выполнено следующее условие невырожденности: если $\hat{e}$ — четное векторное поле, касающееся в каждой точке выделенного $(0|1)$-мерного подпространства, то векторные поля $\hat{e}$ и $\hat{E} = \frac{1}{2}\{\hat{e}, \hat{e}\}$ должны составлять базис комплексного касательного пространства к рассматриваемому многообразию. Можно доказать, что при этом условии в окрестности каждой точки найдется такая система координат $(z, \theta)$, что $\hat{e} = \Phi\left(z, \theta, \tilde{z}, \tilde{\theta}\right) D$ . Суперконформное преобразование $Z \to \tilde{Z}$ характеризуется как аналитическое преобразование, переводящее выделенное $(0|1)$-мерное подпространство в точке $Z$ в аналогичное подпростран-ство в точке $\tilde{Z}$.

Два суперриманова многообразия суперконформно-эквиалентны, если эквивалентны соответствующие суперконформные структуры. Другими словами, если на супермногообразии заданы суперримановы структуры с помощью векторных полей $\hat{e}$ и $\hat{e}'$, то эти структуры эквивалентны в случае, когда существует преобразование многообразия $\mathscr{M}$ (суперконформное пространство модулей) может быть получено с помощью факторизации пространства всех суперримановых структур на $\mathscr{M}$. Мы воспользуемся этим замечанием, чтобы описать касательное пространство $T$ к суперпространству модулей T.

Всякое комплексное многообразие может быть получено из односвязного комплексного многообразия с помощью факторизации по дискретной подгруппе его группы автоморфизмов (аналитических преобразований). Этот факт следует из замечания, что всякое многообразие получается с помощью факторизации своего универсального накрывающего. Поскольку односвязные одномерные комплексные многообразия и их группы автоморфизмов хорошо известны (см., например, [188, 733, 734]), этот факт позволяет дать некоторое описание всех одномерных компактных комплексных многообразий с точностью до аналитической эквивалентности (или, что то же самое, описание всех конформных структур на ориентируемом двумерном компактном многообразии). Задача сводится к описанию всех таких подгрупп группы автоморфизмов, при факторизации по которым снова получается многообразие. Для того, чтобы подгруппа обладала этим свойством, нужно, чтобы она действовала свободно, т. е. чтобы отличные от тождественного преобразования не имели неподвижных точек. Подгруппа, по которой происходит факторизация, изоморфна фундаментальной группе профакторизованного многообразия. Фундаментальные группы компактных двумерных поверхностей хорошо известны. Для поверхности рода 0 (сферы) фундаменталная группа равна нулю. Это означает, что все компакт-ные комплексные многообразия рода 0 совпадают со сферой Римана. Если компактное комплексное многообразие имеет род 1 (топологически эквивалентно тору), то его фундаментальная группа яаляется абелевой группой с двумя образующими. Отсюда следует, что оно может быть получено из $\mathbb{C}^1$ факторизацией по решетке (т. е. с помощью отождествления $z \sim z + m_1e_1 + m_2e_2$, где $m_1, m_2$ — произвольные целые числа, $e_1, e_2$ — фиксированные комплексные числа $e_1 \neq \lambda e_2, \lambda = \tilde{\lambda}$ . Две разные решетки приводят к эквивалентным многообразиям, если одну можно перевести в другую с помощью автоморфизма.

Во всяком односвязном одномерном комплексном многообразии можно ввести метрику, имеющую постоянную кривизну. Автоморфизмы многообразия реализуются как движения (преобразования, сохраняющие метрику). Это позволяет ввести метрику постоянной кривизны на любом компактном комлексном многообразии; метрику можно нормировать так, чтобы площадь была единичной. Тем самым в каждом классе конформно-эквивалентных метрик выбирается единственный представитель. Классификация суперконформных многообразий может быть проведена с помощью аналогичных рассуждений [341, 342].

# Приложение B

# Суперматрицы и необратимость

Здесь мы вначале изложим необходимые сведения из линейной супералгебры и теории суперматриц [30, 106, 186], а затем рассмотрим некоторые новые свойства и явления, связанные с необратимостью суперматриц [8, 10].

## B.1. Линейная супералгебра

*Линейным суперпространством* называется $\mathbb{Z}_2$-градуированное линейное пространство $\mathbf{\Lambda}$, разложенное в прямую сумму $\mathbf{\Lambda} = \Lambda_0 \oplus \Lambda_1$. Элементы из $\Lambda_0$ и $\Lambda_1$ называются *однородными* (*четными* и *нечетными* соответственно) элементами. Если $a \in \Lambda_i$, где $i \in \mathbb{Z}_2$, то будем писать $\mathrm{p}(a) = \mathrm{i}$ и называть $\mathrm{p}(a)$ четностью элемента $a$. Любой элемент (за исключением нуля) может быть единственным образом представлен в виде $a = a_0 + a_1$, где $a_i \in \Lambda_i$. *Линейное подсуперпространство* — это такое $\mathbb{Z}_2$-градуированное подпространство $\mathrm{L} \subset \mathbf{\Lambda}$, что $\mathrm{L}_i = \mathrm{L} \cap \Lambda_i$. *Размерностью* $\mathbb{Z}_2$-градуированного линейного пространства называется пара $(p|q)$, где $p$ — размерность четного и $q$ — размерность нечетного подпространств. Будем обозначать $\mathbb{Z}_2$-градуированное линейное пространство с фиксированной четность как $\mathbf{\Lambda}^{p|q}$. Тогда четные и нечетные подсуперпространства будут обозначаться $\mathbf{\Lambda}^{p|0}$ и $\mathbf{\Lambda}^{0|q}$ соответственно. Отметим, что размерность $(p|q)$ не связана с числом образующих $\Lambda$.

Пусть $\mathbf{\Lambda}^{p|q}$ и $\mathbf{\Lambda}^{m|n}$ — линейные суперпространства. На $\mathbf{\Lambda}^{p|q} \oplus \mathbf{\Lambda}^{m|n}$, $\mathbf{\Lambda}^{p|q} \otimes \mathbf{\Lambda}^{m|n}$ и $\mathrm{Hom}\left(\mathbf{\Lambda}^{p|q}, \mathbf{\Lambda}^{m|n}\right)$ структура суперпространства вводится естественным образом. Элементы суперпространства $\mathrm{Hom}\left(\mathbf{\Lambda}^{p|q}, \mathbf{\Lambda}^{m|n}\right)$ называются *гомоморфизмами* из $\mathbf{\Lambda}^{p|q}$ в $\mathbf{\Lambda}^{m|n}$. Четные гомоморфизмы,

т. е. элементы из $\mathrm{Hom}_0\left(\mathbf{\Lambda}^{p|q}, \mathbf{\Lambda}^{m|n}\right)$, называются *морфизмами суперпространств.* Обозначим через $\mathbf{\Pi}(\mathbf{\Lambda})$ суперпространство, определенное формулами $\mathbf{\Pi}(\Lambda_0) = \Lambda_1$, $\mathbf{\Pi}(\Lambda_1) = \Lambda_0$, т. е. $\mathbf{\Pi}$ — оператор смены четности, а гомоморфизм $\mathbf{\Pi} : \mathbf{\Lambda} \to \Pi(\mathbf{\Lambda})$ по назовем *каноническим нечетным гомоморфизмом суперпространства* $\mathbf{\Lambda}$ в $\mathbf{\Pi}(\mathbf{\Lambda})$.

*Супералгеброй* называется суперпространство $\mathrm{A}$ вместе с морфизмом суперпространств $\mathrm{A} \oplus \mathrm{A} \to \mathrm{A}$. Отметим, что каждая супералгебра является алгеброй. *Идеал* в супералгебре $\mathrm{A}$ — идеал алгебры $\mathrm{A}$, являющийся одновременно подсуперпространством. *Подсупералгеброй* в $A$, являющаяся подсуперпространством.

Пусть $\mathrm{A}$ и $\mathrm{B}$ — супералгебры. Гомоморфизм алгебр $\varphi : \mathrm{A} \to \mathrm{B}$ называется *морфизмом супералгебр*, если $\mathrm{p}(\varphi) = 0$. Для любой супералгебры $\mathrm{A}$ определим *коммутирование* (или *скобку*) $[,] : \mathrm{A} \oplus \mathrm{A} \to \mathrm{A}$ по правилу о знаках, положив $[\mathrm{a}, \mathrm{b}] = \mathrm{ab} - (-1)^{\mathrm{p(a)p(b)}}\,\mathrm{ba}$. Элементы $\mathrm{a}, \mathrm{b} \in \mathrm{A}$ называются *коммутирующими*, если $[\mathrm{a}, \mathrm{b}] = 0$. Супералгебра называется *коммутативной*, если любые два ее элемента коммутируют.

*Централизатором* множества $\mathrm{S}$ однородных элементов из $\mathrm{A}$ называется множество $C(\mathrm{S}) = \{\mathrm{a} \in \mathrm{A} \mid [\mathrm{a}, \mathrm{s}] = 0, \mathrm{s} \in \mathrm{S}\}$. *Нормализатором* такого множества $\mathrm{S}$ называется $N(\mathrm{S}) = \{\mathrm{a} \in \mathrm{A} \mid \mathrm{aS} = \mathrm{Sa}\}$. *Центром* супералгебры $\mathrm{A}$ называется множество $Z(\mathrm{A}) = \{\mathrm{a} \in \mathrm{A} \mid [\mathrm{a}, \mathrm{A}] = 0\}$. Множества $C(\mathrm{S})$ и $Z(\mathrm{A})$ являются коммутативными супералгебрами, а $N(\mathrm{S})$ — супералгеброй.

Обозначим через $\Lambda(n)$ внешнюю (грассманову) алгебру от $n$ переменных $\xi_1, \ldots, \xi_n$ — образующих, которые удовлетворяют соотношениям $\xi_1\xi_j + \xi_j\xi_i = 0$, $1 \le i,\, j \le n$. В частности $\xi_i^2 = 0$. Произвольный элемент $f \in \Lambda(n)$ можно единственным образом представить в виде

$$f = f_0 + \sum_{1 \le r \le n} \sum_{1 < i_1 < \ldots < i_r \le n} f_{i_1 \ldots i_r} \xi_{i_1} \ldots \xi_{i_r}. \tag{B.1}$$

Определим на $\Lambda(n)$ структуру супералгебры, полагая $\mathrm{p}(\xi_i) = 1$. Очевидно, что супералгебра $\Lambda(n)$ коммутативна. В дальнейшем $\Lambda(n)$ называется *супералгеброй Грассмана.*

*Тензорным произведением* супералгебр A и B называется суперпространство $\mathrm{A} \otimes \mathrm{B}$, на котором задана структура супералгебры по формуле

$$(\mathrm{a} \otimes \mathrm{b})(\mathrm{a}_1 \otimes \mathrm{b}_1) = (-1)^{\mathrm{p}(a_1)\mathrm{p}(b)}\, \mathrm{aa}_1 \otimes \mathrm{bb}_1, \tag{B.2}$$

где $\mathrm{a}, \mathrm{a}_1 \in \mathrm{A}$, $\mathrm{b}, \mathrm{b}_1 \in \mathrm{B}$.

Тензорное произведение коммутативных супералгебр является коммутативной супералгеброй. В частности, $\Lambda(n) \otimes \Lambda(m) \cong \Lambda(n+m)$.

Каждой коммутативной супералгебре $\mathrm{C} = \mathrm{C}_0 \oplus \mathrm{C}_1$ соответствует *каноническая проекция* $\epsilon : \mathrm{C} \to \mathrm{C}/\mathrm{id}\,\mathrm{C}_1 = \mathrm{C}_0/(\mathrm{id}\,\mathrm{C}_1)^2$, где $\mathrm{id}\,\mathrm{X}$ обзначает идеал, порожденный множеством X.

**Лемма B.1.** *Пусть* C — *коммутативная супералгебра. Тогда элемент* $\mathrm{c} \in \mathrm{C}$ *обратим в том случае, когда обратим* $\epsilon[\mathrm{c}]$.

Пусть A — супералгебра с единицей, $M$ — некоторое суперпространство. *Левым действием* супералгебры A на $M$,или *левым* A-*действием*, называется морфизм суперпространств $\mathrm{A} \otimes M \to M$, удовлетворяющий условиям: $\mathrm{a}(\mathrm{b}m) = (\mathrm{ab})m$, $\mathrm{a}, \mathrm{b}, 1 \in \mathrm{A}$, $m \in M, 1m = m$.

*Левым модулем* над A, или левым A-модулем, называется суперпространство $M$, на котором задано левое A-действие. Понятие *правого* A-модуля вводится аналогично.

Пусть C-коммутативноая супералгебра. Тогда каждый левый C-модуль можно превратить в правый C-модуль (и наооборот) p:

$$m\mathrm{c} = \begin{cases} (-1)^{\mathrm{p}(m)p(\mathrm{c})}\, \mathrm{c}m \\ (-1)^{(\mathrm{p}(m)+1)p(\mathrm{c})}\, \mathrm{c}m \end{cases}, \tag{B.3}$$

где $\mathrm{c} \in \mathrm{C}, m \in M$.

Структуры левого и правого модуля на $M$ согласованы в следующем смысле:

$$(\mathrm{a}m)\,\mathrm{b} = \mathrm{a}\,(m\mathrm{b})\,, \quad \mathrm{a}, \mathrm{b} \in \mathrm{C},\ m \in M. \tag{B.4}$$

Множество C-гомоморфизмов из $M$ в $N$ является подсуперпространством в $\mathrm{Hom}\,(M, N)$ ,которое обозначается через $\mathrm{Hom}_{\,\mathrm{C}}\,(M, N)$. Когда $M = N$, суперпространство $\mathrm{Hom}_{\,\mathrm{C}}\,(M, N)$ обозначается через $\mathrm{End}_{\,\mathrm{C}}\,(M)$ называются *автоморфизмами* $M$, и они образуют группу $GL_{\mathrm{C}}\,(M)$.

Определим на суперпространстве $\mathrm{Hom}_{\,\mathrm{C}}\,(M, \mathrm{C})$ структуру C-модуля, полагая

$$(\mathrm{cF})\,(m) = c\,(\mathrm{F}\,(m))\,;\ (\mathrm{F}c)\,(m) = \mathrm{F}\,(cm)\,, \tag{B.5}$$

где $\mathrm{F} \in \mathrm{Hom}_{\,\mathrm{C}}\,(M, \mathrm{C})$.

Из формул (B.3) и (B.5) немедленно следует, что

$$c\mathrm{F} = (-1)^{\mathrm{p(c)p(F)}}\,\mathrm{Fc} \tag{B.6}$$

*Тензорным произведением* C-модулей $M$ и $N$ называется тензорное произведение суперпространств $M \otimes N$, профакторизированное по соотношениям $m\mathrm{c} \otimes n = m \otimes \mathrm{c}n$, где $m \in M$, $n \in N$, $\mathrm{c} \in \mathrm{C}$. Обозначим факторпространство через $M \otimes_{\mathrm{C}} N$. На суперпространстве $M \otimes_C N$ структрура C-модуля вводится по формулам $\mathrm{c}\,(m \otimes n) = \mathrm{c}m \otimes n$, $(m \otimes n)\,\mathrm{c} = m \otimes n\mathrm{c}$. Легко проверяется, что C-модули $(L \otimes_{\mathrm{C}} M) \otimes_{\mathrm{C}} N$ и $L \otimes_{\mathrm{C}} (M \otimes_{\mathrm{C}} N)$ естественно изоморфны. Следовательно, такой C-модуль можно обозначить через $L \otimes_{\mathrm{C}} M \otimes_{\mathrm{C}} N$.

Если A и B суть C-алгебры, то на C-модуле $\mathrm{A} \otimes_{\mathrm{C}} \mathrm{B}$ можно ввести структуру C-алгебры, полагая

$$c\,(a \otimes b) = ca \otimes b,\ c \in C,\ a \in A,\ b \in B. \tag{B.7}$$

Легко проверить, что изоморфизм C-модулей $T : A \otimes_C B \to B \otimes_C A$ согласован с умножением и является поэтому изоморфизмом C-алгебр.

Пусть $I$-множество, представленное в виде объединения непересекающихся подмножеств $I_0$ и $I_1$.

*Базисом* C-модуля $M$ называется набор однородных элементов $m_i \in M$, где $i \in I$, такой, что $p\,(m_i) = 0$ при $i \in I_0$ и $p\,(m_i) = 1$ при $i \in I_1$, причем каждый элемент $m$ однозначно записывается в виде суммы $\sum\limits_i c_i m_i$, где все $c_i \in C$, кроме конечного числа, равны нулю. C-модуль называется *свободным*, если в нем можно выбрать базис, соответствующий некоторому набору индексов.

## B.2. Суперматричная алгебра

*Суперматричной структурой* назовем матричную структуру с приписанной каждой строке и каждому столбцу четностью. Четность $i$-й строки обозначим $p_{row}\,(i)$, четность $j$-столбца — $p_{col}\,(j)$. Обычно суперматричная структура будет выбираться так, чтобы все четные строки и столбцы шли сначала, а нечетные — потом. Такая суперматричная структура будет называться *стандартной**). Стандартную суперматричную структуру можно записывать в блочном $2 \times 2$ виде:

$$M = \begin{pmatrix} R & S \\ T & U \end{pmatrix}, \tag{B.8}$$

---

*Примечание.* Нестандартные суперматричные структуры (когда нечетные элементы располагаются не блоками, а по диагоналям) рассматривались в [787].

где $\mathrm{R}, \mathrm{S}, \mathrm{T}, \mathrm{U}$ — матричные структуры, согласованные с делением строк и столбцов на четные и нечетные. В случае обобщенной $\mathbb{Z}_3$ суперсимметрии [788] суперматричная структура описывается блочной $3 \times 3$ матрицей [789].

Если суперматричная структура содержит $p$ четных и $q$ нечетных строк и $m$ четных и $n$ нечетных столбцов, то размер этой структуры равен $(p|q) \times (m|n)$. *Порядком* суперматричной структуры размера $(p|q) \times (p|q)$ называется пара натуральных чисел $(p|q)$. Суперматричные структуры порядка $(p|q)$ соответствуют элементам $\mathrm{Hom}\left(\boldsymbol{\Lambda}^{p|q}, \boldsymbol{\Lambda}^{p|q}\right)$.

Пусть задана суперматричная структура $\mathrm{M}$ и некоторое суперпространство $\Lambda$. *Матрицей* с элементами из $\boldsymbol{\Lambda}$ называется множество $\{\mathrm{X}_{ij} \mid \mathrm{X}_{ij} \in \boldsymbol{\Lambda}\}$, соответствующее клеткам суперматричной структуры $\mathrm{M}$. Определим на линейном пространстве матриц с элементами из $\boldsymbol{\Lambda}$ четность следующим образом: $\mathrm{p}\left(\mathrm{M}\right) = 0$, если $\mathrm{p}\left(\mathrm{X}_{ij}\right) + \mathrm{p}_{row}\left(i\right) + \mathrm{p}_{col}\left(j\right) = 0$, и $p\left(\mathrm{X}\right) = 1$, если $\mathrm{p}\left(\mathrm{X}_{ij}\right) + \mathrm{p}_{row}\left(i\right) + \mathrm{p}_{col}\left(j\right) = 1$, для всех $i, j$.

Легко проверить, что относительно таким образом введенной четности линейное пространство матриц превращается в суперпространство. Если суперматричная структура стандартна, то определение четности матриц (B.8) можно переписать в виде $\mathrm{p}\left(\mathrm{M}\right) = 0$, если $\mathrm{p}\left(\mathrm{R}_{ij}\right) = \mathrm{p}\left(\mathrm{U}_{kl}\right) = 0$, $\mathrm{p}\left(\mathrm{S}_{il}\right) = \mathrm{p}\left(\mathrm{T}_{kj}\right) = 1$, и $\mathrm{p}\left(\mathrm{X}\right) = 1$, если $\mathrm{p}\left(\mathrm{R}_{ij}\right) = \mathrm{p}\left(\mathrm{U}_{kl}\right) = 1$, $\mathrm{p}\left(\mathrm{S}_{il}\right) = \mathrm{p}\left(\mathrm{T}_{kj}\right) = 0$.

Введем на суперпространстве матриц размера $(p|q) \times (m|n)$ с элементами из коммутативной супералгебры $\mathrm{C}$ структуру $\mathrm{C}$-модуля, полагая

$$(\mathrm{Mc})_{ij} = (-1)^{\mathrm{p(c)p}_{col}(j)}\, \mathrm{X}_{ij}\mathrm{c}, \quad (\mathrm{cX})_{ij} = (-1)^{\mathrm{p(c)p}_{row}(i)}\, \mathrm{cX}_{ij}. \tag{B.9}$$

Эту структуру можно задать и другим, эквивалентным образом, а именно, определив для каждой пары целых чисел $(p|q)$ гомоморфизм супералгебр $\mathrm{C} \to \mathrm{Mat}_{\mathrm{C}}(p|q)$, который каждому элементу $\mathrm{c} \in \mathrm{C}$ ставит в соответствие диагональную матрицу

$$\mathrm{scalar}_{p|q}(\mathrm{C}) = \mathrm{diag}\left(\mathrm{c}, \ldots, \mathrm{c}, (-1)^{\mathrm{p(c)}}\mathrm{c}, \ldots, (-1)^{\mathrm{p(c)}}\mathrm{c}\right) \tag{B.10}$$

со стандартной суперматричной структурой. Теперь структуру C-модуля на суперпространстве матриц размера $(p,q) \times (m,n)$ можно ввести по формуле

$$\mathrm{cM} = \mathrm{scalar}_{p|q}(\mathrm{C}) \cdot \mathrm{M} = \mathrm{M} \cdot \mathrm{scalar}_{m|n}(\mathrm{C}). \tag{B.11}$$

Из ассоциативности матричного умножения следует, что

$$(\mathrm{cX})\,\mathrm{Y} = \mathrm{c}\,(\mathrm{XY}), \;\; (\mathrm{Xc})\,\mathrm{Y} = \mathrm{X}\,(\mathrm{cY}), \;\; \mathrm{X}\,(\mathrm{Yc}) = (\mathrm{XY})\,\mathrm{c} \tag{B.12}$$

при $\mathrm{X}, \mathrm{Y} \in \mathrm{Mat}\,(p|q;\mathrm{C})$. Следовательно, супералгебра $\mathrm{Mat}_{\mathrm{C}}(p|q)$ является C-алгеброй.

Пусть $\mathrm{M}{=}(\mathrm{X}_{ij})$ — матрица размера $(p,q) \times (m,n)$ с элементами из суперпространства $\mathbf{\Lambda}$. *Супертранспонированной* к ней назовем матрицу размера $(m,n) \times (p,q)$ ,элементы которой имеют вид:

$$\left(\mathrm{M}^{\mathrm{st}}\right)_{ij} = (-1)^{(\mathrm{p}_{row}(i)+\mathrm{p}_{col}(j))(\mathrm{p}(X)+\mathrm{p}_{row}(i))}\,\mathrm{X}_{ji} =$$

$$(-1)^{(\mathrm{p}_{row}(i)+\mathrm{p}_{col}(j))(\mathrm{p}(X)+\mathrm{p}_{col}(j))}\,\mathrm{X}_{ji}, \tag{B.13}$$

а суперматричная структура определяется естественным образом. В формуле (B.13) четности $\mathrm{p}_{row}(i)$, $\mathrm{p}_{col}(j)$ берутся согласовано с суперматричной структурой матрицы M. Как увидим, при таком выборе знаков переход от матрицы C-линейного оператора осуществляется с помощью супертранспонирования. Если суперматричная структура имеет стан-дарный вид, то формула (B.13) дает

$$M^{\mathrm{st}} = \begin{pmatrix} R & S \\ T & U \end{pmatrix}^{\mathrm{st}} = \begin{pmatrix} R^T & T^T \\ -S^T & U^T \end{pmatrix}, \tag{B.14}$$

если $p(M) = 0$, и

$$M^{st} = \begin{pmatrix} R & S \\ T & U \end{pmatrix}^{st} = \begin{pmatrix} R^T & -T^T \\ S^T & U^T \end{pmatrix}, \tag{B.15}$$

если $p(M) = 1$.

Дважды транспонированная матрица имеет вид:

$$\left(\begin{pmatrix} R & S \\ T & U \end{pmatrix}^{\mathrm{st}}\right)^{\mathrm{st}} = \begin{pmatrix} R & -S \\ -T & U \end{pmatrix}. \tag{B.16}$$

Следовательно, порядок супертранспонирования равен 4.

## B.3. Суперслед и супердетерминант

Пусть C — коммутативная супералгебра. Аналогом обычной матричной алгебры является супералгебра $\mathrm{Mat}_C(p|q)$. *Суперследом* называется гомоморфизм супералгебр $\mathrm{str} : \mathrm{Mat}_C(p|q) \to C$, определяемый по формуле

$$\mathrm{str}\, M = \sum_{i,j} (-1)^{(p(X)+1)p_{row}(i)} X_{ij} = \sum_{i,j} (-1)^{(p(X_{ij})+1)p_{row}(i)} X_{ij}. \tag{B.17}$$

Иначе говоря, если $\mathrm{M}=\begin{pmatrix} \mathrm{R} & \mathrm{S} \\ \mathrm{T} & \mathrm{U} \end{pmatrix}$, то

$$\operatorname{str}\mathrm{M} = \operatorname{tr}\mathrm{R} - \operatorname{tr}\mathrm{U} \tag{B.18}$$

для четных матриц $\mathrm{p}(\mathrm{M}) = 0$ и

$$\operatorname{str}\mathrm{M} = \operatorname{tr}\mathrm{R} + \operatorname{tr}\mathrm{U} \tag{B.19}$$

для нечетных матриц $\mathrm{p}(\mathrm{M}) = 1$.

**Определение B.2.** *Отображение* $\operatorname{str} : \mathrm{Mat}_{\mathrm{C}}(p|q) \to \mathrm{C}$ *является* C-*линейным гомоморфизмом.*

$$\operatorname{str}\mathrm{M}^{\mathrm{st}} = \operatorname{str}\mathrm{M}, \ \ \operatorname{str}\mathbf{\Pi}(\mathrm{M}) = (-1)^{\mathrm{p}(X)+1}\operatorname{str}\mathrm{M}. \tag{B.20}$$

Пусть X-матрица размера $(p|q)\times(m|n)$, Y-матрица размера $(m|n)\times(p|q)$. Тогда

$$\operatorname{str}\mathrm{XY} = (-1)^{\mathrm{p(X)p(Y)}}\operatorname{str}\mathrm{YX}. \tag{B.21}$$

В частности, если $\mathrm{X}, \mathrm{D} \in \mathrm{Mat}_{\mathrm{C}}(p|q)$, где D есть четная обратимая матрица,то

$$\operatorname{str}\mathrm{DXD}^{-1} = \operatorname{str}\mathrm{X}. \tag{B.22}$$

Будем обозначать через $GL(p|q; \mathrm{C})$ мультипликативную группу четных обратимых элементов из $\mathrm{Mat}_{\mathrm{C}}(p|q)$. Эта группа является аналогом обычной общей линейной группы. Определим гомоморфизм группы $GL(p|q; \mathrm{C})$ в группу $\mathrm{C}_0^*$ обратимых элементов из $\mathrm{C}_0$ — аналог обычного определителя, положив

$$\operatorname{Ber}\mathrm{M} = \det\left(\mathrm{R} - \mathrm{SU}^{-1}\mathrm{T}\right)\det\mathrm{U}^{-1}. \tag{B.23}$$

**Лемма B.3.** *Пусть* C*-некоторая коммутативная супералгебра, отображение* $\epsilon : \mathrm{C} \to \mathrm{C}/\mathrm{idC}_1$ *— каноническая проекция и* $\epsilon : \mathrm{Mat}_{\mathrm{C}}(p|q) \to \mathrm{Mat}_{\epsilon[\mathrm{C}]}(p|q)$ *— соответствующий гомоморфизм матричных алгебр. Матрица* $\mathrm{M} \in \mathrm{Mat}_{\mathrm{C}}(p|q)$ *обратима тогда и только тогда, когда обратим элемент* $\epsilon[\mathrm{M}] \in \mathrm{Mat}_{\epsilon[\mathrm{C}]}(p|q)$.

В силу **Леммы B.3** матрица U обратима, и элементы матриц $\mathrm{R} - \mathrm{SU}^{-1}\mathrm{T}$ и U лежат в коммутативной алгебре $\mathrm{C}_0$. Поэтому все определители имеют смысл и $\mathrm{Ber}\,\mathrm{M} \in \mathrm{C}_0^*$ . Функция Ber называется *березианином.* Суперслед и березиниан связаны точно так же, и обычный след и детерминант. А именно:

$$\mathrm{Ber}\,\mathrm{M} = \exp \mathrm{str}\, \ln \mathrm{M}, \tag{B.24}$$

если правая и левая части (B.24) определены.

В отличие от детерминанта березиниан определен не на всем множестве $\mathrm{Mat}_{\mathrm{C}}(p|q)$ (см. **Приложение B.5**).

**Теорема B.4.** *Пусть* $\mathrm{XY} \in GL_{\mathrm{C}}(p|q)$, *тогда*

$$\mathrm{Ber}\,\mathrm{XY} = \mathrm{Ber}\,\mathrm{X} \cdot \mathrm{Ber}\,\mathrm{Y}. \tag{B.25}$$

**Определение B.5.** *Суперматричная функция* str *является гомоморфизмом* C*-модулей, а* Ber *— гомоморфизмом групп.*

Пусть $M, N$-свободные C-модули, $\mathrm{X} \in \mathrm{End}_{\mathrm{C}}(M)$ и $\mathrm{Y} \in \mathrm{End}_{\mathrm{C}}(N)$, а $\mathrm{V} \in GL_{\mathrm{C}}(M), \mathrm{W} \in GL_{\mathrm{C}}(N)$. Тогда

$$\mathrm{str}\,(\mathrm{X} \oplus \mathrm{Y}) = \mathrm{str}\,\mathrm{X} + \mathrm{str}\,\mathrm{Y}, \tag{B.26}$$

$$\mathrm{Ber}\,\mathrm{V} \oplus \mathrm{W} = \mathrm{Ber}\,\mathrm{V} \cdot \mathrm{Ber}\,\mathrm{W}. \tag{B.27}$$

Если $\mathrm{X} \in \mathrm{Hom}_{\mathrm{C}}(M, N)$ и $\mathrm{Y} \in \mathrm{Hom}_{\mathrm{C}}(M, N)$, то

$$\mathrm{str}\, XY = (-1)^{\mathrm{p}(X)\mathrm{p}(Y)}\, \mathrm{str}\, YX. \tag{B.28}$$

## B.4. Странные супералгебра, след и детерминант

Определим супералгебру $Q(n)$, которая является еще одним аналогом матричной алгебры. Интерес к ней в последнее время возобновился в связи с новыми свойствами инвариантных функций на супералгебрах Ли [790–796] и их квантованых версиях [797, 798].

Пусть A — произвольная супералгебра. Алгебра $\mathrm{Q}(\mathrm{A})$ состоит из выражений вида $\mathrm{a} + \varepsilon\mathrm{b}$, где $\mathrm{a}, \mathrm{b} \in \mathrm{A}$. Сложение выражений $\mathrm{a} + \varepsilon\mathrm{b}$ определяется естественным образом, а структуры суперпространства и супералгебры вводятся по формулам

$$(\mathrm{Q}(\mathrm{A}))_i = \mathrm{A}_i + \varepsilon \mathrm{A}_{i-1}, \tag{B.29}$$

$$(\mathrm{a} + \varepsilon\mathrm{b})(\mathrm{c} + \varepsilon\mathrm{d}) = \left(\mathrm{ac} - (-1)^{\mathrm{p(b)}}\,\mathrm{bd}\right) + \varepsilon\left(\mathrm{bc} + (-1)^{\mathrm{p(a)}}\,\mathrm{ad}\right). \tag{B.30}$$

Пусть C-коммутативная супералгебра. Тогда $\mathrm{Q}(\mathrm{C})$ можно определить как расширение супералгебры C с помощью одного элемента $\varepsilon$ такого, что $\mathrm{p}(\varepsilon) = 1, \varepsilon^2 = -1$ и $[\varepsilon, \mathrm{c}] = 0$ для любого $\mathrm{c} \in \mathrm{C}$.

Если супералгебра A является C-алгеброй, то $\mathrm{Q}(\mathrm{A}) \cong \mathrm{Q}(\mathrm{C}) \otimes_{\mathrm{C}} \mathrm{A}$. Тем самым $\mathrm{Q}(\mathrm{A})$ тоже является C-алгеброй. Из определения супералгебры $\mathrm{Q}(\mathrm{A})$ следует, что $(\mathrm{Q}(\mathrm{A}))_0$ является алгеброй. Имеется изоморфизм алгебр (но не супералгебр!) $(\mathrm{Q}(\mathrm{A}))_0 \to \mathrm{A}$, задаваемый формулой $\mathrm{a} + \varepsilon\mathrm{b} \longmapsto \mathrm{a} + \mathrm{b}$, где $\mathrm{a} \in \mathrm{A}_0$, $\mathrm{b} \in \mathrm{A}_1$.

Если супералгебра A ассоциативна, то ассоциативна и супералгебра $\mathrm{Q}(\mathrm{A})$. Если A — алгебра с единицей, то $\mathrm{Q}(\mathrm{A})$ — некоммутативна.

Положим $Q_{\mathrm{C}}(n) = \mathrm{Q}(\mathrm{Mat}_{\mathrm{C}}(n|0))$, где C — коммутативная суперапгебра. Мы будем часто рассматривать $Q_{\mathrm{C}}(n)$ как подалгебру в $\mathrm{Mat}_{\mathrm{C}}(n|n)$, причем вложение $Q_{\mathrm{C}}(n) \to \mathrm{Mat}_{\mathrm{C}}(n|n)$ проводится по формуле

$$\mathrm{A} + \varepsilon\mathrm{B} \longmapsto \begin{pmatrix} \mathrm{A}\,(-1)^{\mathrm{p(B)}+1}\,\mathrm{B} \\ \mathrm{B}\,(-1)^{\mathrm{p(A)}}\,\mathrm{A} \end{pmatrix}, \tag{B.31}$$

где $\mathrm{A}, \mathrm{B} \in \mathrm{Mat}_{\mathrm{C}}(n|0)$.

На подсуперапгебре $Q_{\mathrm{C}}(n) \subset \mathrm{Mat}_{\mathrm{C}}(n|n)$ суперслед тождественно равен нулю. Мы определим на $Q_{\mathrm{C}}(n)$ otr $: Q_{\mathrm{C}}(n) \to \mathrm{C}$ (otr — нечетный, "странный"), полагая

$$\mathrm{otr}\,(A + \varepsilon B) = \mathrm{str}\, B. \tag{B.32}$$

**Определение B.6.** *Отображение* otr $: Q_{\mathrm{C}}(n) \to \mathrm{C}$ *есть нечетный гомоморфизм* C*-модулей.*

$$\mathrm{otr}\,\mathrm{XY} = (-1)^{\mathrm{p(X)p(Y)}}\,\mathrm{otr}\,\mathrm{YX},$$

*если* $\mathrm{X}, \mathrm{Y} \in Q_{\mathrm{C}}(n)$.

Обозначим через $GQ_{\mathrm{C}}(n)$ группу четных обратимых элементов из $Q_{\mathrm{C}}(n)$. Она является нечетным ("странным") аналогом полной линейной группы. "Странные" аналоги для других групп получены в [794].

Группу $GQ_{\mathrm{C}}(n)$ можно рассматривать как подгруппу в $GL_{\mathrm{C}}(n|n)$, соотоящую из матриц M вида

$$\mathrm{M} = \begin{pmatrix} \mathrm{A} & -\mathrm{B} \\ \mathrm{B} & \mathrm{A} \end{pmatrix}. \tag{B.33}$$

Другая реализация: группа $GQ_{\mathrm{C}}(n)$ изоморфна группе всех обра-тимых матриц из $\mathrm{Mat}_{\mathrm{C}}(n|0)$, причем изоморфизм задается формулой $\mathrm{A}+\varepsilon\mathrm{B}\to\mathrm{A}+\mathrm{B}$.

**Лемма B.7.** *На всех матрицах подгруппы $GQ_{\mathrm{C}}(n)\subset GL_{\mathrm{C}}(n|n)$ березиниан тождественно равен* 1.

Определим теперь *нечетный детерминант* $\mathrm{qet}: GQ_{\mathrm{C}}(n)\to\mathrm{C}_1$. Для каждой нечетной матрицы $\mathrm{M}\in\mathrm{Mat}_{\mathrm{C}}(n|0)$ положим

$$\mathrm{F}(\mathrm{M})=\sum_{i\geq 0}\frac{1}{2i+1}\mathrm{str}\,\mathrm{M}^{2i+1}. \tag{B.34}$$

На самом деле эта сумма конечна, так как $X^k=0$ при $k>n^2$. Отметим также, что $F_{2k}=0$.

Определим отображение $\mathrm{qet}: GQ_{\mathrm{C}}(n)\to\mathrm{C}_1$ ( qet — queer, "странный" детерминант) формулой

$$\mathrm{qet}\,(\mathrm{A}+\varepsilon\mathrm{B})=F\left(\mathrm{A}^{-1}\mathrm{B}\right)=\sum_{i\geq 0}\frac{1}{2i+1}\mathrm{str}\,\left(\mathrm{A}^{-1}\mathrm{B}\right)^{2i+1}. \tag{B.35}$$

Происхождение этой формулы таково. Мы хотим определить гомоморфизм $\mathrm{qet}: GQ_{\mathrm{C}}(n)\to\mathrm{C}_1$, соответствующий нечетному следу $\mathrm{otr}$. Кроме того, qet должен равняться нулю на элементах из $GL_{\mathrm{C}}(n|0)$. Если $\mathrm{M}=\varepsilon\mathrm{B}$, то естественно положить

$$\mathrm{qet}\,(1+\mathrm{M})=\mathrm{otr}\,\ln(1+\mathrm{M})=\sum_{i\geq 0}\frac{1}{2i+1}\mathrm{str}\,\mathrm{B}^{2i+1}, \tag{B.36}$$

что и приводит к данному определению [186, 743, 799].

**Определение B.8.** *Пусть* $\mathrm{I}\subset\mathrm{C}$ — *идеал и* $\mathrm{M}\in\mathrm{Q}_{\mathrm{I}}(n)$. *Тогда*

$$\mathrm{qet}\,(1+\mathrm{M})=\mathrm{otr}\,\mathrm{M}\mathrm{mod}\,\mathrm{I}^2. \tag{B.37}$$

**Теорема B.9.** *Если* $\mathrm{X}, \mathrm{Y} \in GQ_{\mathrm{C}}(n)$, *то*

$$\operatorname{qet} \mathrm{XY} = \operatorname{qet} \mathrm{X} + \operatorname{qet} \mathrm{Y}. \tag{B.38}$$

Иначе говоря, qet есть гомоморфизм групп.

## B.5. Идеалы $(1|1) \times (1|1)$ суперматриц

Рассмотрим подробнее полугрупповую структуру суперматриц из $\mathrm{Mat}_{\Lambda}(1|1)$. Обозначим

$$\mathfrak{M}' = \{\mathrm{M} \in \mathfrak{M} | \, \epsilon(a) \neq 0\}, \tag{B.39}$$

$$\mathfrak{M}'' = \{\mathrm{M} \in \mathfrak{M} | \, \epsilon(b) \neq 0\}, \tag{B.40}$$

$$\mathfrak{J}' = \{\mathrm{M} \in \mathfrak{M} | \, \epsilon(a) = 0\}, \tag{B.41}$$

$$\mathfrak{J}'' = \{\mathrm{M} \in \mathfrak{M} | \, \epsilon(b) = 0\}. \tag{B.42}$$

Тогда $\mathfrak{M} = \mathfrak{M}' \cup \mathfrak{J}' = \mathfrak{M}'' \cup \mathfrak{J}''$ и $\mathfrak{M}' \cap \mathfrak{J}' = \varnothing$, $\mathfrak{M}'' \cap \mathfrak{J}'' = \varnothing$, поэтому $\mathfrak{M}^{inv} = \mathfrak{M}' \cap \mathfrak{M}''$ и $\mathfrak{T} \subset \mathfrak{M}''$.

Березиниан $\operatorname{Ber} \mathrm{M}$ хорошо определен только для суперматриц из $\mathfrak{M}''$ и обратим, когда $\mathrm{M} \in \mathfrak{M}^{inv}$. Но для суперматриц из $\mathfrak{M}'$ обратный элемент $(\operatorname{Ber} \mathfrak{M})^{-1}$ хорошо определен и обратим, если $\mathrm{M} \in \mathfrak{M}^{inv}$ [30]. Относительно умножения суперматриц $(\cdot)$ множество $\mathfrak{M}$ представляет собой полугруппу $\mathbf{M} \stackrel{def}{=} \{\mathfrak{M}; \cdot\}$ всех $(1|1)$-мерных суперматриц, и множество $\mathfrak{M}^{inv}$ представляет подгруппу $\mathbf{G} \stackrel{def}{=} \left\{\mathfrak{M}^{inv}; \cdot\right\} \subseteq \mathbf{M}$. В стандартном базисе $\mathfrak{M}^{inv}$ представляет полную линейную группу $GL_{\Lambda}(1|1)$ [30]. Подмножество $\mathfrak{J} \subset \mathfrak{M}$ представляет идеал полугруппы $\mathbf{M}$ [104].

**Предложение B.10.** 1) *Множества* $\mathfrak{J}$, $\mathfrak{J}'$ *и* $\mathfrak{J}''$ *представляют изолированные идеалы полугруппы* $\mathbf{M}$.

2) *Множества* $\mathfrak{M}^{inv}$, $\mathfrak{M}'$ *и* $\mathfrak{M}''$ *— фильтры полугруппы* $\mathbf{M}$.

3) *Множества* $\mathfrak{M}'$ *и* $\mathfrak{M}''$ *представляют подполугруппы полугруппы* $\mathbf{M}$, *при этом* $\mathfrak{M}' = \mathfrak{M}^{inv} \cup \mathfrak{J}'$ *и* $\mathfrak{M}'' = \mathfrak{M}^{inv} \cup \mathfrak{J}''$, *где соответствующие изолированные идеалы* $\mathfrak{K}' = \mathfrak{M}' \setminus \mathfrak{M}^{inv} = \mathfrak{M}' \cap \mathfrak{J}''$ *и* $\mathfrak{K}'' = \mathfrak{M}'' \setminus \mathfrak{M}^{inv} = \mathfrak{M}'' \cap \mathfrak{J}'$.

*4) Идеал* $\mathbf{I}$ *полугруппы* $\mathbf{M}$ *представлен множеством* $\mathfrak{J} = \mathfrak{J}' \cup \mathfrak{K}' = \mathfrak{J}'' \cup \mathfrak{K}''$.

*Доказательство.* Допустим $M_3 = M_1 M_2$, тогда $a_3 = a_1 a_2 + \alpha_1 \beta_2$ и $b_3 = b_1 b_2 + \beta_1 \alpha_2$. Взяв числовую часть, мы выводим $\epsilon(a_3) = \epsilon(a_1)\,\epsilon(a_2)$, и $\epsilon(b_3) = \epsilon(b_1)\,\epsilon(b_2)$. Далее используем определения подполугрупп и идеалов из **Приложения А**. ■

## B.6. Правые и левые Γ-матрицы

В общем случае, нечетно-редуцированные матрицы $T \in \mathfrak{T}$ (см. (4.4) и **Подраздел 4.1**) не образуют полугруппу, поскольку их умножение не замкнуто (4.9). Однако, некоторое подмножество в $\mathfrak{T}$ может образовать полугруппу $\mathbf{T}^{SG}$, именно то, в котором $(1|1)$-элемент в результирующей суперматрице (4.9) обращается в нуль (см. **Предложение 4.8**).

Чтобы определить класс полугрупп такого типа, мы рассмотрим некоторые обобщения. Пусть $\alpha, \beta \in \Gamma$, где $\Gamma \subset \Lambda_1$ — нечетная подсуперобласть. Мы обозначим

$$\operatorname{Ann}\alpha \overset{def}{=} \left\{\Gamma \in \Lambda_1 \,|\, \Gamma \cdot \alpha = 0\right\}, \ \operatorname{Ann}\Gamma = \bigcap_{\alpha \in \Gamma} \operatorname{Ann}\alpha, \qquad (B.43)$$

В последнем определении пересечение множеств является решающим.

*Замечание* **B.11.** Нильпотентность*) $\alpha$ приводит к $\alpha \in \operatorname{Ann}\alpha$ и как следствие $\Gamma \cdot \operatorname{Ann}\Gamma = 0$.

**Определение B.12.** *Определим левые и правые Г-матрицы следующим образом*

$$\mathrm{T}^{\Gamma}_{(L)} \stackrel{def}{=} \begin{pmatrix} 0 & \Gamma \\ \operatorname{Ann}\Gamma & b \end{pmatrix}, \tag{B.44}$$

$$\mathrm{T}^{\Gamma}_{(R)} \stackrel{def}{=} \begin{pmatrix} 0 & \operatorname{Ann}\Gamma \\ \Gamma & b \end{pmatrix}. \tag{B.45}$$

**Предложение B.13.** *Г-матрицы* $\mathrm{T}^{\Gamma}_{(L,R)} \subset \mathfrak{T}$ *образуют подполугруппы* $\mathbf{T}^{\Gamma}_{(L,R)}$ *относительно умножения суперматриц.*

*Доказательство.* Рассмотрим аналог умножения (4.9) для множеств в случае левых Г-матриц $\mathrm{T}^{\Gamma}_{(L)}$ следующим образом

$$\begin{pmatrix} 0 & \Gamma \\ \operatorname{Ann}\Gamma & b_1 \end{pmatrix} \begin{pmatrix} 0 & \Gamma \\ \operatorname{Ann}\Gamma & b_2 \end{pmatrix} = \begin{pmatrix} \Gamma \cdot \operatorname{Ann}\Gamma & \Gamma \cdot b_2 \\ b_1 \cdot \operatorname{Ann}\Gamma & b_1 \cdot b_2 + \operatorname{Ann}\Gamma \cdot \Gamma \end{pmatrix}.$$

Таким образом, условие $\Gamma \cdot \operatorname{Ann}\Gamma = 0$ и доказывает утверждение. ■

*Замечание* **B.14.** В полугруппах $\mathbf{T}^{\Gamma}_{(L,R)}$ подмножество матриц с $\beta = 0$ представляет собой левый идеал, и с $\alpha = 0$ представляет собой правый идеал, матрицы с $b = 0$ образуют двусторонний идеал.

**Определение B.15.** *Назовем Г-полугруппами введенные в* B.13 *полугруппы* $\mathbf{T}^{\Gamma}_{(L,R)}$.

*Замечание* **B.16.** Г-полугруппы $\mathbf{T}^{\Gamma}_{(L,R)}$ не содержат единицу.

Сущность Г-полугрупп $\mathbf{T}^{\Gamma}_{(L,R)}$ может быть выяснена из следую-

---

*Примечание.* Здесь мы рассматриваем только тот случай, когда индекс нильпотентности 2 и $\alpha^2 = 0$.

щей аналогии с биидеалами [451, 452, 800]. Напомним, что биидеал в полугруппе $\mathbf{M}$ может быть введен как множество $\mathbf{B}$ суперматриц, удовлетворяющих $\mathbf{B} * \mathbf{M} * \mathbf{B} \subseteq \mathbf{B}$ [451]. Для $\Gamma$-полугрупп $\mathbf{T}^{\Gamma}_{(L,R)}$ это соотношение слишком сильное и может не выполняться. Тем не менее, некоторый более общий аналог его может быть найден.

**Предложение B.17.** *Для любого заданного* $\Gamma \subset \Lambda_1$ *полугруппы* $\mathbf{T}^{\Gamma}_{(L,R)}$ *являются одновременно слабыми биидеалами*), которые удовлетворяют соотношениям*

$$\mathbf{T}^{\Gamma}_{(L,R)} * \mathbf{M} * \mathbf{T}^{\Gamma}_{(L,R)} \subseteq \mathbf{T}^{\Gamma_1}_{(L,R)}, \tag{B.46}$$

*где* $\Gamma_1(\Gamma) \subset \Lambda_1$ *- суперобласть в нечетном секторе* $\Lambda$.

*Доказательство.* Рассматрим аналог (B.46) для множеств в виде

$$\begin{pmatrix} 0 & \Gamma \\ \operatorname{Ann}\Gamma & b_1 \end{pmatrix} \begin{pmatrix} a & \alpha \\ \beta & b \end{pmatrix} \begin{pmatrix} 0 & \Gamma \\ \operatorname{Ann}\Gamma & b_2 \end{pmatrix} =$$

$$\begin{pmatrix} \Gamma \cdot \operatorname{Ann}\Gamma \cdot b & \Gamma \cdot bd - \Gamma^2 \cdot \beta \\ \operatorname{Ann}\Gamma \cdot cb - (\operatorname{Ann}\Gamma)^2 \cdot \alpha & \Gamma \cdot \operatorname{Ann}\Gamma \cdot a + c\beta \cdot \Gamma + \operatorname{Ann}\Gamma \cdot \alpha \cdot d + cbd \end{pmatrix}. \tag{B.47}$$

Мы видим, что условие $\Gamma \cdot \operatorname{Ann}\Gamma = 0$ снова дает нечетно-редуцированную суперматрицу в правой части, за счет исчезновения $(1|1)$-слагаемого. Тогда произведение $(2|1)$ и $(1|2)$-элементов равно нулю по той же причине, и мы имеем $\Gamma$-матрицу, однако, определенной над иной суперобластью $\Gamma_1(\Gamma) \subset \Lambda_1$. ■

---

*) *Примечание.* Слово "обобщенный биидеал" резервировано для другой конструкции в [451].

## B.7. Полугруппа множеств редуцированных матриц

Чтобы объединить введенные множества суперматриц (4.12), мы рассмотрим тройные произведения

$$\begin{aligned} \mathcal{S} \star \mathcal{A} \star \mathcal{T} &= \mathcal{S}, \\ \mathcal{T} \star \mathcal{A} \star \mathcal{T} &= \mathcal{T}, \\ \mathcal{S} \star \mathcal{D} \star \mathcal{S} &= \mathcal{S}, \\ \mathcal{T} \star \mathcal{D} \star \mathcal{S} &= \mathcal{T}. \end{aligned} \tag{B.48}$$

Здесь мы замечаем, что множества суперматриц $\mathcal{A}$ и $\mathcal{D}$ играем роль "сэндвич" элементов в особом $\mathcal{S}$ и $\mathcal{T}$ умножении. Более того, сэндвич элементы находятся во взаимооднозначном соответствии с правыми множествами, на которых они действуют, и таким образом они "чувствительны справа". Следовательно, вполне естественно ввести следующее

**Определение B.18.** *Сэндвич-произведение множеств редуцированных суперматриц* $\mathcal{R} = \mathcal{S}, \mathcal{T}$

$$\mathcal{R}_1 \odot \mathcal{R}_2 \stackrel{def}{=} \begin{cases} \mathcal{R}_1 \star \mathcal{D} \star \mathcal{R}_2, & \mathcal{R}_2 = \mathcal{S}, \\ \mathcal{R}_1 \star \mathcal{A} \star \mathcal{R}_2, & \mathcal{R}_2 = \mathcal{T}. \end{cases} \tag{B.49}$$

В терминах сэндвич-произведения из (B.48) мы получаем

$$\begin{aligned} \mathcal{S} \odot \mathcal{T} &= \mathcal{S}, \\ \mathcal{T} \odot \mathcal{T} &= \mathcal{T}, \\ \mathcal{S} \odot \mathcal{S} &= \mathcal{S}, \\ \mathcal{T} \odot \mathcal{S} &= \mathcal{T}. \end{aligned} \tag{B.50}$$

**Предложение B.19.** *$\odot$-умножение ассоциативно.*

*Доказательство.* Перебор всех тройных произведений с различной расстановкой скобок и использование таблицы умножения (B.50). ■

**Определение B.20.** *Элементы $\mathcal{S}$ и $\mathcal{T}$ образовывают полугруппу множеств относительно $\odot$- умножения* (B.49)*, которую мы будем называть полугруппой множеств редуцированных матриц и обозначим* $\mathbf{RMS}_{set}$ (**R***educed super***M***atrix* **S***emigroup of sets*)*.*

Из (B.50) видно, что $\mathbf{RMS}_{set}$ есть полугруппа идемпотентов, причем каждый элемент является правым нулем, поэтому мы можем сформулировать следующую теорему.

**Теорема B.21.** *Полугруппа* $\mathbf{RMS}_{set}$ *изоморфна особой полугруппе правых нулей, т. е.* $\mathbf{RMS}_{set} \cong \mathbf{Z}_R = \{\mathcal{R} = \mathcal{S} \cup \mathcal{T}; \odot\}$.

## B.8. Непрерывное суперматричное представление нулевых полугрупп

Рассматрим абстрактное множество $\mathcal{P}_\alpha$ (которое "нумеруется" нечетным параметром $\alpha \in \Lambda^{0|1}$), состоящее из элементов $\boldsymbol{p}_t \in \mathcal{P}_\alpha$ ($t \in \Lambda^{1|0}$ представляет собой непрерывный четный суперпараметр), которые подчиняются закону умножения

$$\boldsymbol{p}_t * \boldsymbol{p}_u = \boldsymbol{p}_t. \qquad \text{(B.51)}$$

**Утверждение B.22.** *Умножение* (B.51) *ассоциативно и следовательно множество $\mathcal{P}$ является полугруппой* $\boldsymbol{P}_\alpha \stackrel{def}{=} \{\mathcal{P}_\alpha; *\}$.

**Утверждение B.23.** *Полугруппа $\boldsymbol{P}_\alpha$ представляет собой непрерывное однопараметрическое представление полугруппы левых нулей* [104], *в которой каждый элемент одновременно — и левый нуль и правая единица.*

**Предложение B.24.** *Полугруппа $\boldsymbol{P}_\alpha$ эпиморфна (не изоморфна!) полугруппе $\mathbf{P}_\alpha$.*

*Доказательство.* Сравнивая (4.43) и (B.51), мы замечаем, что отображение $\varphi : \boldsymbol{P}_\alpha \to \mathbf{P}$ представляет собой гомоморфизм. Видно, что два элемента $\boldsymbol{p}_t$ и $\boldsymbol{p}_u$, удовлетворяющих (4.46), имеют один и то же образ

$$\varphi(\boldsymbol{p}_t) = \varphi(\boldsymbol{p}_u) \leftrightarrow t - u = \operatorname{Ann}\alpha, \ \boldsymbol{p}_t, \boldsymbol{p}_u \in \mathcal{P}_\alpha. \tag{B.52}$$

■

**Определение B.25.** *Соотношение*

$$\boldsymbol{\Delta}_\alpha = \{(\boldsymbol{p}_t, \boldsymbol{p}_u) \mid t - u = \operatorname{Ann}\alpha, \boldsymbol{p}_t, \boldsymbol{p}_u \in \mathcal{P}_\alpha\}. \tag{B.53}$$

*назовем $\alpha$-отношением равенства.*

*Замечание* **B.26.** Если суперпараметры $t$ и $\alpha$ принимают значения в различных алгебрах Грассмана, которые не содержат взаимно уничтожающихся элементов кроме нуля, тогда $\operatorname{Ann}\alpha = 0$ и $\boldsymbol{\Delta}_\alpha = \boldsymbol{\Delta}$, где $\boldsymbol{\Delta}$ — стандартное отношение равенства [104].

Теперь мы можем сформулировать более общее высказывание.

**Утверждение B.27.** *В суперсимметричном случае аналог стандартного отношения равенства $\boldsymbol{\Delta}$ представляет собой $\alpha$-отношение равенства $\boldsymbol{\Delta}_\alpha$ (B.53).*

Тот факт, что $\boldsymbol{\Delta} \neq \boldsymbol{\Delta}_\alpha$ приводит к некоторые новым абстрактным алгебраическим структурам в суперматричной теории и нетривиальным результатам. Среди последних имеется следующий

**Следствие B.28.** *Ядро гомоморфизма* $\varphi$ *определяется следующей формулой* $\ker\varphi = \bigcup_{t\in\operatorname{Ann}\alpha} \boldsymbol{p}_t$.

Напомним, что в несуперсимметричном случае $\ker\varphi = \boldsymbol{p}_{t=0}$.

*Замечание* **B.29.** Вне $\ker\varphi$ полугруппа $\mathbf{P}_\alpha$ непрерывна и супергладка, что может быть показано посредством стандартных методов суперанализа [30, 174].

**Утверждение B.30.** *Полугруппа* $\mathbf{P}_\alpha$ *нередуктивна и несократима, поскольку* $\boldsymbol{p} * \boldsymbol{p}_t = \boldsymbol{p} * \boldsymbol{p}_u \rightarrow \boldsymbol{p}_t \boldsymbol{\Delta}_\alpha \boldsymbol{p}_u$, *но не* $\boldsymbol{p}_t = \boldsymbol{p}_u$ (*или* $\boldsymbol{p}_t \boldsymbol{\Delta} \boldsymbol{p}_u$) *для всех* $\boldsymbol{p}\in\mathcal{P}_\alpha$. *Следовательно, суперматричное представление, заданное* $\varphi$, *не является точным.*

**Следствие B.31.** *Если* $t + \operatorname{Ann}\alpha \cap u + \operatorname{Ann}\alpha \neq \varnothing$, *тогда* $\boldsymbol{p}_t \boldsymbol{\Delta}_\alpha \boldsymbol{p}_u$ (*а не* $\boldsymbol{p}_t \Delta \boldsymbol{p}_u$ *как в обычном случае).*

Аналогично, полугруппа $\boldsymbol{Q}_\alpha$ с умножением

$$\boldsymbol{q}_t * \boldsymbol{q}_u = \boldsymbol{q}_u \tag{B.54}$$

изоморфна полугруппе правых нулей, в которой каждый элемент является одновременно и правым нулем, и левой единицей, и, кроме того, полугруппа $\boldsymbol{Q}_\alpha$ эпиморфна полугруппе $\mathbf{Q}_\alpha$.

**Определение B.32.** *Полугруппы левых и правых нулей* $\boldsymbol{P}_\alpha$ *и* $\boldsymbol{Q}_\alpha$ *могут быть названы* *почти антикоммутативными**), *поскольку для них* $\boldsymbol{p}_t * \boldsymbol{p}_u = \boldsymbol{p}_u * \boldsymbol{p}_t$ *или* $\boldsymbol{q}_t * \boldsymbol{q}_u = \boldsymbol{q}_u * \boldsymbol{q}_t$ *дает* $\alpha t = \alpha u$ *и* $t = u + \operatorname{Ann}\alpha$.

Нетривиальность данного определения и его отличие от случая абстрактных полугрупп левых и правых нулей основана на том факте, что суперматричное представление, заданное $\varphi$, не является точным

---

*Примечание.* По аналогии с антикоммутативными прямоугольными связками [104].

согласно **Утверждению B.30**.

**Предложение B.33.** *Полугруппы $\boldsymbol{P}_\alpha$ и $\boldsymbol{Q}_\alpha$ регулярны, но не инверсны.*

*Доказательство.* Для любых двух элементов $\boldsymbol{p}_t$ и $\boldsymbol{p}_u$, используя (B.51), мы имеем $\boldsymbol{p}_t * \boldsymbol{p}_u * \boldsymbol{p}_t = (\boldsymbol{p}_t * \boldsymbol{p}_u) * \boldsymbol{p}_t = \boldsymbol{p}_t * \boldsymbol{p}_t = \boldsymbol{p}_t$.

Аналогично, и для $\boldsymbol{q}_t$ и $\boldsymbol{q}_u$. Тогда $\boldsymbol{p}_t$ имеет хотя бы обратный элемент $\boldsymbol{p}_u * \boldsymbol{p}_t * \boldsymbol{p}_u = \boldsymbol{p}_u$. Но $\boldsymbol{p}_u$ произвольно выбран, поэтому в полугруппах $\boldsymbol{P}_\alpha$ и $\boldsymbol{Q}_\alpha$ любые два элемента взаимноинверсны. Однако, $\boldsymbol{P}_\alpha$ и $\boldsymbol{Q}_\alpha$ не инверсные полугруппы, в которых каждый элемент имеет единственный инверсный [104]. ■

Важно подчеркнуть, что идеальное строение $\boldsymbol{P}_\alpha$ и $\boldsymbol{Q}_\alpha$ не полностью совпадают (хотя имеет много общего) с полугруппами левых и правых нулей в следующем смысле.

**Предложение B.34.** *Каждый элемент из $\boldsymbol{P}_\alpha$ образовывает изолированный главный правый идеал, каждый элемент из $\boldsymbol{Q}_\alpha$ образовывает главный левый идеал, и поэтому каждый главный правый и левый идеал в $\boldsymbol{P}_\alpha$ и $\boldsymbol{Q}_\alpha$ соответственно имеют идемпотентный генератор.*

*Доказательство.* Из (B.51) и (B.54) следует, что $\boldsymbol{p}_t = \boldsymbol{p}_t * \boldsymbol{\mathcal{P}}_\alpha$ и $\boldsymbol{q}_u = \boldsymbol{\mathcal{Q}}_\alpha * \boldsymbol{q}_u$. ■

**Предложение B.35.** *Полугруппы $\boldsymbol{P}_\alpha$ и $\boldsymbol{Q}_\alpha$ просты слева и справа соответственно.*

*Доказательство.* Из (B.51) и (B.54) видно, что $\boldsymbol{\mathcal{P}}_\alpha = \boldsymbol{\mathcal{P}}_\alpha * \boldsymbol{p}_t$ и $\boldsymbol{\mathcal{Q}}_\alpha = \boldsymbol{q}_u * \boldsymbol{\mathcal{Q}}_\alpha$. ■

Отношения Грина в стандартных полугруппах левых нулей следующие: $\mathscr{L}$-эквивалентность совпадает с универсальным отношением, и $\mathscr{R}$-эквивалентность совпадает с отношением равенства [104]. В нашем случае первое утверждение то же самое, но вместо последнего мы имеем

**Теорема В.36.** *В* $\boldsymbol{P}_\alpha$ *и* $\boldsymbol{Q}_\alpha$ *соответственно* $\mathscr{R}$*-эквивалентность и* $\mathscr{L}$*-эквивалентность совпадает с* $\alpha$*-отношением равенства* (В.53).

*Доказательство.* Рассмотрим $\mathscr{R}$-эквивалентность в $\boldsymbol{P}_\alpha$. Два элемента $\boldsymbol{p}_t, \boldsymbol{p}_u \in \boldsymbol{P}_\alpha$ $\mathscr{R}$-эквивалентны тогда и только тогда, если $\boldsymbol{p}_t * \boldsymbol{\mathcal{P}}_\alpha = \boldsymbol{p}_u * \boldsymbol{\mathcal{P}}_\alpha$. В терминах элементов матриц это выглядит, как $\alpha t = \alpha u$, что дает $t - u = \operatorname{Ann}\alpha$. По определению (В.53) это приводит к $\boldsymbol{p}_t \boldsymbol{\Delta}_\alpha \boldsymbol{p}_u$, и мы получаем $\mathscr{R} = \boldsymbol{\Delta}_\alpha$, и аналогично для $\mathscr{L}$-эквивалентности. ■

## В.9. Отношение $\mathscr{R}$-эквивалентности для прямоугольной $(2|2)$-связки

Явный вид $(2|2)$-связки $\boldsymbol{F}_\alpha^{(2|2)} \ni \boldsymbol{f}_{t_1 t_2, u_1 u_2}$ в суперматричном представлении есть

$$\mathrm{F}_{t_1 t_2, u_1 u_2} = \begin{pmatrix} 0 & \alpha t_1 & \alpha t_2 \\ \alpha u_1 & 1 & 0 \\ \alpha u_2 & 0 & 1 \end{pmatrix}. \tag{В.55}$$

Согласно определению $\mathscr{R}$-классов [104], два элемента $\mathrm{F}_{t_1 t_2, u_1 u_2}$ и $\mathrm{F}_{t'_1 t'_2, u'_1 u'_2}$ в связке $\mathscr{R}$-эквивалентны тогда и только тогда, если существует два других элемента $\mathrm{X}_{x_1 x_2, y_1 y_2}$, $\mathrm{W}_{v_1 v_2, w_1 w_2}$ таких, что

$$\mathrm{F}_{t_1 t_2, u_1 u_2} \cdot \mathrm{X}_{x_1 x_2, y_1 y_2} = \mathrm{F}_{t'_1 t'_2, u'_1 u'_2}, \tag{В.56}$$

$$\mathrm{F}_{t'_1 t'_2, u'_1 u'_2} \cdot \mathrm{W}_{v_1 v_2, w_1 w_2} = \mathrm{F}_{t_1 t_2, u_1 u_2} \tag{В.57}$$

одновременно. Или в явном виде

$$\begin{pmatrix} 0 & \alpha t_1 & \alpha t_2 \\ \alpha y_1 & 1 & 0 \\ \alpha y_2 & 0 & 1 \end{pmatrix} = \begin{pmatrix} 0 & \alpha t'_1 & \alpha t'_2 \\ \alpha u'_1 & 1 & 0 \\ \alpha u'_2 & 0 & 1 \end{pmatrix}, \tag{В.58}$$

$$\begin{pmatrix} 0 & \alpha t'_1 & \alpha t'_2 \\ \alpha w_1 & 1 & 0 \\ \alpha w_2 & 0 & 1 \end{pmatrix} = \begin{pmatrix} 0 & \alpha t_1 & \alpha t_2 \\ \alpha u_1 & 1 & 0 \\ \alpha u_2 & 0 & 1 \end{pmatrix}. \tag{B.59}$$

Чтобы удовлетворить последнему равенству в (B.58) и (B.59) мы должны выбрать

$$\alpha y_1 = \alpha u'_1, \ \alpha y_1 = \alpha u'_1, \tag{B.60}$$

$$\alpha w_1 = \alpha u_1, \ \alpha w_2 = \alpha u_2, \tag{B.61}$$

$$\alpha t_1 = \alpha t'_1, \ \alpha t_2 = \alpha t'_2. \tag{B.62}$$

Из-за произвольности $\mathrm{X}_{x_1 x_2, y_1 y_2}$ и $\mathrm{W}_{v_1 v_2, w_1 w_2}$ первые уравнения в (B.60)–(B.61) всегда могут быть решены возможностью выбора параметра. Вторые уравнения в (B.62) представляют собой определение $\mathscr{R}$-класса в $(2|2)$-связке в суперматричной интерпретации [10].

# Приложение Д

# Перманенты и их обобщения для матриц с нильпотентными элементами

Перманенты представляют собой объект математического исследования, в настоящее время весьма распространенный, прежде всего, в комбинаторике и линейной алгебре [626]. Теория перманентов дважды стохастических матриц и $(0,1)$-матриц стала сейчас существенной и неотъемлемой частью комбинаторной математики, а именно того ее раздела, где рассматриваются матричные комбинаторные задачи. Интересные сами по себе проблемы, связанные с перманентами, приобрели актуальность также в связи с многообразными их приложениями — как математическими (например, в алгебре и теории вероятностей), так и в других отраслях знания (в квантовой теории поля, физической химии, статистической физике).

В своем знаменитом мемуаре 1812 г. Коши развивал теорию детерминантов как специального вида знакопеременных симметрических функций, которые он отличал от обычных симметрических функций, называя последние "перманентными симметрическими функциями". Он ввел также некоторый подкласс симметрических функций, которые были позднее Мюиром названы *перманентами*.

Интересно, что еще в 1872 г. рассматривались соотношения между перманентами и детерминантами матриц, элементами которых являлись суть “альтернирующие” (alternate) числа, т. е. антикоммутирующие (!) [801].

С появлением суперматематики роль перманентов принципиально меняется, поскольку элементами матриц могут быть нильпотентные и антикоммутирующие числа и функции, и поэтому многие классические теоремы становятся неприменимыми или модифицируются (см. **Разделы 3** и **5**, а также [2, 3, 13]).

## Д.1. Перманенты и детерминанты

Пусть $\boldsymbol{V}$ есть $n-$мерное пространство со скалярным произведением [626]. Тогда $\mathbb{Z}$-градуированное контравариантное тензорное пространство над $\boldsymbol{V}$, т. е. пространство $\boldsymbol{T}_0(\boldsymbol{V}) = C\dot{+}\boldsymbol{V}\dot{+}\boldsymbol{V}\otimes\boldsymbol{V}\dot{+}\boldsymbol{V}\otimes\boldsymbol{V}\otimes\boldsymbol{V}\dot{+}\ldots$ наследует от $\boldsymbol{V}$ скалярное произведение, определяемое формулой

$$(\boldsymbol{x}_1\otimes\ldots\otimes\boldsymbol{x}_p,\,\boldsymbol{y}_1\otimes\ldots\boldsymbol{y}_p) = \prod_{t=1}^{p}(\boldsymbol{x}_t,\boldsymbol{y}_t) \tag{Д.1}$$

для однородных степени $p$ разложимых элементов. Симметрическое пространство $\boldsymbol{V}$ есть область значений определенного на $\boldsymbol{T}_0(\boldsymbol{V})$ оператора симметрии $\sum\limits_{p=0} S_p$, где $S_p = \dfrac{1}{p!}\sum\sigma$, и суммирование производится по элементам симметрической группы степени $p$ (действие перестановки $\sigma$ на разложимом тензоре определяется как $\sigma(\boldsymbol{x}_1\otimes\boldsymbol{x}_2\ldots\otimes\boldsymbol{x}_p) = \boldsymbol{x}_{\sigma(1)}\otimes\boldsymbol{x}_{\sigma(2)}\ldots\otimes\boldsymbol{x}_{\sigma(p)}$). Каждый $S_p$ эрмитово идемпотентен, так что, если $\boldsymbol{x}_1\ldots\boldsymbol{x}_p = S_p\boldsymbol{x}_1\otimes\boldsymbol{x}_2\ldots\otimes\boldsymbol{x}_p$, то

$$\begin{aligned}(\boldsymbol{x}_1\ldots\boldsymbol{x}_p,\,\boldsymbol{y}_1\ldots\boldsymbol{y}_p) &= (\boldsymbol{x}_1\otimes\ldots\otimes\boldsymbol{x}_p, S_p\boldsymbol{y}_1\otimes\ldots\otimes\boldsymbol{y}_p) = \\ \frac{1}{p!}\sum_{\sigma}\prod_{t=1}^{p}\left(\boldsymbol{x}_t,\boldsymbol{y}_{\sigma(t)}\right) &= \frac{1}{p!}\mathrm{per}\left((\boldsymbol{x}_i,\boldsymbol{y}_j)\right).\end{aligned} \tag{Д.2}$$

Таким образом, функция перманента естественно возникает как аналитическое выражение для скалярного произведения в $\boldsymbol{V}^{(p)} = \mathrm{im}\,S_p$ точно таким же образом, как детерминант в $p$-м внешнем произведении $\wedge^p\boldsymbol{V}$. Это означает, что унитарную геометрию $\boldsymbol{V}^{(p)}$ можно применить для исследования $\mathrm{per}\,A$, и это наблюдение привело к значительному прогрессу в обращении с этой функцией.

Пусть $\mathrm{A} = (a_{ij})$ — матрица размера $m \times n$ над коммутативным кольцом, $m \trianglelefteq n$ . Перманент матрицы $\mathrm{A}$, обозначаемый $\mathrm{Per}\,\mathrm{A}$, определяется как

$$\mathrm{Per}\,\mathrm{A} = \sum_{\sigma} a_{1\sigma(1)} a_{2\sigma(2)} \dots a_{m\sigma(m)}, \tag{Д.3}$$

где суммирование распространяется на все взаимно однозначные отображения из $\{1, 2, \dots, m\}$ в $\{1, 2, \dots n\}$.

Последовательность $\left(a_{1\sigma(1)}, \dots, a_{m\sigma(m)}\right)$ называется *диагональю*, а произведение $a_{1\sigma(1)}, \dots, a_{m\sigma(m)}$ — *диагональным произведением матрицы* $\mathrm{A}$. Таким образом, $\mathrm{Per}\,\mathrm{A}$ есть сумма диагональных произведений матрицы. Другими словами, $\mathrm{Per}\,\mathrm{A}$ есть сумма всех произведений $m$ таких элементов $\mathrm{A}$,что никакие два из них не находятся в одной строке или одном столбце. Отсюда следует, что все члены $\mathrm{Per}\,\mathrm{A}$ наряду с другими содержатся в множестве членов, получающихся при перемножении сумм по строке матрицы $\mathrm{A}$.

Особенно важен случай $m = n$. Перманент квадратной матрицы $\mathrm{A}$ обозначается через $\mathrm{per}\,\mathrm{A}$ вместо $\mathrm{Per}\,\mathrm{A}$. В большинстве случаев употребление термина "перманент" фактически ограничивает случаем квадратных матриц.

Пусть $\mathrm{A} = (a_{ij})$ — матрица порядка $n$. Тогда

$$\mathrm{per}\,\mathrm{A} \det \mathrm{A} = \left(\sum_{\sigma \in E} \prod_{i=1}^{n} a_{i\sigma(i)}\right)^2 - \left(\sum_{\sigma \in F} \prod_{i=1}^{n} a_{i\sigma(i)}\right)^2, \tag{Д.4}$$

где $E$ и $F$ — множества всех четных и нечетных перестановок соответственно.

**Теорема Д.1.** *Пусть* $\mathrm{A} = (a_{ij})$ *и* $\mathrm{X} = (x_{ij})$ *— квадратные матрицы порядка* $n$. *Тогда*

$$\mathrm{per}\,\mathrm{A} \det \mathrm{X} = \sum_{\sigma \in S_n} \varepsilon(\sigma) \det(\mathrm{A} * \mathrm{X}_\sigma), \tag{Д.5}$$

*где* $X_\sigma$ *— матрица,* $i$*-строка которой есть* $\sigma(i)$*-я строка матрицы* $X$*,* $A \bigcirc X_\sigma$ *— произведение Адамара, а* $\varepsilon(\sigma)$ *обозначает знак подстановки* $\sigma$ *из симметрической группы* $S_n$*.*

**Определение Д.2.** *Произведение Адамара двух матриц* $P = (p_{ij})$ *и* $Q = (q_{ij})$ *порядка* $n$ *есть* $P \bigcirc Q = R$*, где матрица* $R = (p_{ij}q_{ij})$*.*

Для матрицы A порядка $n$ имеем

$$\mathrm{per}\,(A - \lambda I_n) = \lambda^n + \sum_{k=1}^{n} c_k \lambda^{n-k}, \tag{Д.6}$$

где $c_k = (-1)^k \sum_{\omega \in Q_{k,n}} \mathrm{per}\,A[\omega]$. При этом $\mathrm{per}\,(A - \lambda I_n)$ называется *перманентным характеристическим многочленом* A (см. [626]). Если A — квадратная матрица, то

$$|\mathrm{per}\,A|^2 \le \mathrm{per}\,(AA^*). \tag{Д.7}$$

Равенство получается в том и только в том случае, когда A имеет нулевую строку или A есть обобщенная матрица перестановки.

Если U — унитарная матрица, то

$$|\mathrm{per}\,U| \ge \det U \tag{Д.8}$$

с равенством в том и только в том случае, когда A диагональна или имеет нулевую строку.

**Теорема Д.3.** (Теорема Шура) *Если* A *— положительно полуопределенная эрмитова матрица, то*

$$\mathrm{per}\,A \ge \det A \tag{Д.9}$$

*с равенством в том и только в том случае, когда* A *диагональна или имеет нулевую строку.*

Пусть A есть матрица размера $m \times n$, а D и G — диагональные матрицы порядков $m$ и $n$ соответственно. Тогда

$$\mathrm{Per}\ (\mathrm{DAG}) \neq \mathrm{Per}\,\mathrm{D} \cdot \mathrm{Per}\,\mathrm{A} \cdot \mathrm{Per}\,\mathrm{G}. \qquad (Д.10)$$

Пусть $\mathrm{A} = (a_{ij}) \in M_n$ есть $(0,1)$-матрица, т. е. матрица, составленная из $0$ и $1$. Пусть $\mathrm{B} = (b_{ij})$ — матрица "перманентных дополнений" для $\mathrm{A}$, т. е. $b_{ij} = \mathrm{per}\,\mathrm{A}\,(j|i)$. Отсюда можно вывести что

$$(\mathrm{per}\,\mathrm{A})^2 \leq k\mathrm{tr}\ (\mathrm{BB}^*)\,, \qquad (Д.11)$$

где $k = \sum_{i,j} a_{ij}/n^2$.

## Д.2. Полуминоры и полуматрицы

Введем в рассмотрение супераналоги миноров в матрице M — “полуминоры”

$$\begin{aligned} &\mathrm{M}_a = \begin{pmatrix} d & \beta \\ \delta & e \end{pmatrix}, \quad \mathrm{M}_b = \begin{pmatrix} c & \beta \\ \gamma & e \end{pmatrix}, \quad \mathrm{M}_c = \begin{pmatrix} b & \alpha \\ \delta & e \end{pmatrix}, \\ &\mathrm{M}_d = \begin{pmatrix} a & \alpha \\ \gamma & e \end{pmatrix}, \quad \mathrm{M}_e = \begin{pmatrix} a & b \\ c & d \end{pmatrix}, \quad \mathcal{M}_\alpha = \begin{pmatrix} c & d \\ \gamma & \delta \end{pmatrix}, \\ &\mathcal{M}_\beta = \begin{pmatrix} a & b \\ \gamma & \delta \end{pmatrix}, \quad \mathcal{M}_\gamma = \begin{pmatrix} b & \alpha \\ d & \beta \end{pmatrix}, \quad \mathcal{M}_\delta = \begin{pmatrix} a & \alpha \\ c & \beta \end{pmatrix}. \end{aligned} \qquad (Д.12)$$

Не все полуминоры (Д.12) являются суперматрицами в обычном смысле [30], а лишь $\mathrm{M}_a, \mathrm{M}_b, \mathrm{M}_c, \mathrm{M}_d, \mathrm{M}_e$, т. е. полуминоры четных эле-ментов, причем $\mathrm{M}_e$ - обычная (не супер) матрица.

**Определение Д.4.** *Назовем полуминоры $\mathcal{M}_\alpha, \mathcal{M}_\beta, \mathcal{M}_\gamma, \mathcal{M}_\delta$ нечетных элементов полуматрицами.*

По аналогии с суперматрицами (см. [30] и **Подраздел 4.1**) обозначим множество $2 \times 2$ полуматриц $\mathcal{M}\mathrm{at}\,(1|1)$.

Тогда можно сформулировать общее утверждение.

**Предположение Д.5.** *В $(p+q) \times (p+q)$-суперматрице общего положения $\mathrm{M} \in \mathrm{Mat}\ (p|q)$ полуминоры четных элементов $a_i$ являются суперматрицами $\mathrm{M}_{a_i} \in \mathrm{Mat}\ (p-1|q-1)$, а полуминоры нечетных элементов $\alpha_i$ являются полуматрицами $\mathcal{M}_{\alpha_i} \in \mathcal{M}\mathrm{at}\,(p-1|q-1)$.*

**Определение Д.6.** *Назовем горизонтальными полуматрицы $\mathcal{M}_\alpha$, $\mathcal{M}_\beta$, а полуматрицы $\mathcal{M}_\gamma, \mathcal{M}_\delta$ - вертикальными (в зависимости от расположения нечетных элементов).*

Обозначим $\mathcal{M}_\alpha, \mathcal{M}_\beta \in \mathcal{M}\mathrm{at}^H\,(1|1)$ и $\mathcal{M}_\gamma, \mathcal{M}_\delta \in \mathcal{M}\mathrm{at}^V\,(1|1)$. Тогда легко получить следующее

**Утверждение Д.7.** *Произведение горизонтальной и вертикальной полуматриц дает суперматрицу общего положения, а произведение вертикальной и горизонтальной полурматриц дает обычную (не супер) матрицу.*

В общем случае полуматрицы не образуют полугруппу относительно обычного умножения матриц. Они отличаются от суперматриц перестановкой элементов **только в одном** столбце или строке.

*Замечание* **Д.8.** Полуматрицы следует отличать от нестандартных (точнее, диагональных [787]) форматов суперматриц, применяемых в $N=2$ суперконформной теории поля [485, 802] и бесконечномерных суперпредставлениях [803].

По аналогии с супертранспонированием [30] и **Π**-транспонированием [186, 679] (см. также **Пункт 4.1.3**) введем

**Определение Д.9.** *Определим вертикальное* $\mathbf{\Theta}_V$ *и горизонтальное* $\mathbf{\Theta}_H$ *полутранспонирования как перестановку элементов второго столбца или строки соответственно*

$$\begin{pmatrix} a_1 & a_2 \\ a_3 & a_4 \end{pmatrix}^{\mathbf{\Theta}_V} = \begin{pmatrix} a_1 & a_4 \\ a_3 & a_2 \end{pmatrix}, \tag{Д.13}$$

$$\begin{pmatrix} a_1 & a_2 \\ a_3 & a_4 \end{pmatrix}^{\mathbf{\Theta}_H} = \begin{pmatrix} a_1 & a_2 \\ a_4 & a_3 \end{pmatrix} \tag{Д.14}$$

*независимо от четности элементов.*

**Утверждение Д.10.** *Полутранспонирования являются идемпотентами, поскольку* $\mathbf{\Theta}_V^2 = \mathbf{\Theta}_V$ *и* $\mathbf{\Theta}_H^2 = \mathbf{\Theta}_H$.

Кроме того, они превращают полуматрицы в суперматрицы и наоборот по формулам

$$\begin{aligned} \mathcal{M}\mathrm{at}^H(1|1) &\overset{\mathbf{\Theta}_V}{\leftrightarrow} \mathrm{Mat}\,(1|1)\,, \\ \mathcal{M}\mathrm{at}^V(1|1) &\overset{\mathbf{\Theta}_H}{\leftrightarrow} \mathrm{Mat}\,(1|1)\,, \end{aligned} \tag{Д.15}$$

а их произведение переводит горизонтальные полуматрицы в вертикальные и наоборот

$$\mathcal{M}\mathrm{at}^H(1|1) \overset{\mathbf{\Theta}_V\mathbf{\Theta}_H}{\longleftrightarrow} \mathcal{M}\mathrm{at}^V(1|1)\,. \tag{Д.16}$$

Однако, $\mathbf{\Theta}_V$ для вертикальных полуматриц и $\mathbf{\Theta}_H$ для горизон-тальных полуматриц являются автоморфизмами

$$\begin{aligned} \mathcal{M}\mathrm{at}^{H}(1|1) &\overset{\mathbf{\Theta}_H}{\leftrightarrow} \mathcal{M}\mathrm{at}^{H}(1|1)\,, \\ \mathcal{M}\mathrm{at}^{V}(1|1) &\overset{\mathbf{\Theta}_V}{\leftrightarrow} \mathcal{M}\mathrm{at}^{V}(1|1)\,. \end{aligned} \tag{Д.17}$$

**Утверждение Д.11.** *Произведение полутранспонирований дает* П-*транспонирование* (*из* [186, 679])

$$\mathbf{\Theta}_V \mathbf{\Theta}_H = \mathbf{\Pi}. \tag{Д.18}$$

Поэтому полутранспонирование можно трактовать как извлечение квадратного корня из **П**-транспонирования (см. также (4.20)).

Горизонтальные и вертикальные полуматрицы описывают вращающие четность отображения линейных двумерных суперпространств

$$\begin{aligned} \mathcal{M}_\alpha, \mathcal{M}_\beta &: \mathbf{\Lambda}^{2|0} \to \mathbf{\Lambda}^{1|1}, \\ \mathcal{M}_\gamma, \mathcal{M}_\delta &: \mathbf{\Lambda}^{1|1} \to \mathbf{\Lambda}^{2|0} \end{aligned} \tag{Д.19}$$

соответственно. Тогда, как суперматрицы действуют в суперпространстве $\mathbf{\Lambda}^{1|1}$

$$\mathrm{M}_a, \mathrm{M}_b, \mathrm{M}_c, \mathrm{M}_d : \mathbf{\Lambda}^{1|1} \to \mathbf{\Lambda}^{1|1}, \tag{Д.20}$$

а обычная матрица $\mathrm{M}_e$ действует в четном пространстве $\mathbf{\Lambda}^{2|0}$

$$\mathrm{M}_e : \mathbf{\Lambda}^{2|0} \to \mathbf{\Lambda}^{2|0}. \tag{Д.21}$$

**Следствие Д.12.** *Полуматрицы, в отличие от обычных матриц и суперматриц, меняют тип пространства, в котором они действуют и вращают четность одной из координат.*

Это легко видеть из следующей диаграммы

$$\begin{array}{ccc} \mathbf{\Lambda}^{1|1} & \xrightarrow[\text{susy}]{\text{M}} & \mathbf{\Lambda}^{1|1} \\ \mathcal{M}\uparrow & & \uparrow\mathcal{M} \\ \mathbf{\Lambda}^{2|0} & \xrightarrow[\text{nonsusy}]{\text{M}} & \mathbf{\Lambda}^{2|0} \end{array} \tag{Д.22}$$

где полуматрицы действуют по вертикальным стрелкам, изменяя четно-нечетную сигнатуру пространства, в то время, как (супер)матрицы действуют по горизонтальным, оставляя четно-нечетную сигнатуру неизменной.

*Замечание* **Д.13.** Интересно сравнить и проследить аналогии рассматриваемого изменения четно-нечетной сигнатуры суперпространства с возможными эффектами изменения пространственно-временной сигнатуры обычного пространства [197, 804–806].

Для полуматриц из (Д.12) можно ввести нечетные аналоги обычного (не супер) детерминанта и перманента. Различные свойства перманентов [626] и матриц, содержащих нильпотентные элементы, приведены в **Разделе 5**.

**Определение Д.14.** *Полудетерминант горизонтальной полуматрицы $\mathcal{M}_\alpha$ определяется формулой*

$$\delta\mathrm{et}\mathcal{M}_\alpha = \delta\mathrm{et}\begin{pmatrix} c & d \\ \gamma & \delta \end{pmatrix} \overset{def}{=} c\delta - d\gamma. \tag{Д.23}$$

**Определение Д.15.** *Полуперманент горизонтальной полуматрицы $\mathcal{M}_\alpha$ определяется формулой*

$$\pi\mathrm{er}\mathcal{M}_\alpha = \pi\mathrm{er}\begin{pmatrix} c & d \\ \gamma & \delta \end{pmatrix} \overset{def}{=} c\delta + d\gamma. \tag{Д.24}$$

Аналогичные определения справедливы и для вертикальных полу-матриц. Кроме того, в случае суперматриц кроме березиниана мы будем пользоваться и обычными детерминантом и перманентом, например,

$$\det \mathrm{M}_a = \det \begin{pmatrix} d & \beta \\ \delta & e \end{pmatrix} = ed - \delta\beta. \tag{Д.25}$$

$$\operatorname{per} \mathrm{M}_a = \operatorname{per} \begin{pmatrix} d & \beta \\ \delta & e \end{pmatrix} = ed + \delta\beta. \tag{Д.26}$$

Такую же формулу будем применять и для матриц со всеми нечетными элементами

$$\det \begin{pmatrix} \alpha & \beta \\ \gamma & \delta \end{pmatrix} = \alpha\delta - \gamma\beta. \tag{Д.27}$$

$$\operatorname{per} \begin{pmatrix} \alpha & \beta \\ \gamma & \delta \end{pmatrix} = \alpha\delta + \gamma\beta. \tag{Д.28}$$

*Замечание* **Д.16.** Полудетерминанты и полуперманенты не связаны с квазидетерминантами [807–810], которые применяются для матриц с некоммутирующими элементами и решают некоторые проблемы с обратимостью при изучении систем линейных уравнений над грассмановой алгеброй (см. [811] и приложения в [269]).

Приведем некоторые свойства полудетерминантов и полуперманентов.

Очевидно, что они нильпотентны, т. е. для любой полуматрицы $\mathcal{M}$ имеем

$$(\delta\mathrm{et}\mathcal{M})^2 = (\pi\mathrm{er}\mathcal{M})^2 = 0.$$

Кроме того,

$$\delta\text{et}\begin{pmatrix} \det \mathrm{M}_e & \operatorname{per} \mathrm{M}_e \\ \delta\text{et}\mathcal{M}_\alpha & \pi\text{er}\mathcal{M}_\alpha \end{pmatrix} = 2cd \cdot \delta\text{et}\mathcal{M}_\beta, \tag{Д.29}$$

$$\delta\text{et}\mathcal{M}_\alpha \cdot \pi\text{er}\mathcal{M}_\alpha = 2cd\delta\gamma. \tag{Д.30}$$

Последнее соотношение интересно сравнить с аналогичным соотношением для обычных (и супер) матриц

$$\det \mathrm{M}_e \cdot \operatorname{per} \mathrm{M}_e = a^2 b^2 \tag{Д.31}$$

(см. [626] и **Раздел 5**).

Приведем также некоторые полезные и используемые в дальнейшем соотношения между полудетерминантами полуминорами

$$\begin{aligned} b \cdot \delta\text{et}\mathcal{M}_\delta \pm a \cdot \delta\text{et}\mathcal{M}_\gamma &= \alpha \cdot \binom{\text{per}}{\det}\mathrm{M}_e, \\ d \cdot \delta\text{et}\mathcal{M}_\delta \pm c \cdot \delta\text{et}\mathcal{M}_\gamma &= \beta \cdot \binom{\text{per}}{\det}\mathrm{M}_e, \\ c \cdot \delta\text{et}\mathcal{M}_\beta \pm a \cdot \delta\text{et}\mathcal{M}_\alpha &= \gamma \cdot \binom{\text{per}}{\det}\mathrm{M}_e, \\ d \cdot \delta\text{et}\mathcal{M}_\beta \pm b \cdot \delta\text{et}\mathcal{M}_\alpha &= \delta \cdot \binom{\text{per}}{\det}\mathrm{M}_e. \end{aligned} \tag{Д.32}$$

Другие свойства матриц с нильпотентными элементами можно найти в **Разделе 5**.

# Приложение E

# $N$-расширенные суперпространства и необратимые якобианы

Здесь мы рассмотрим обобщенные супераналитические преобразования в $N = 2$ и $N = 4$ суперпространствах и их необратимые якобианы.

## E.1. $N = 1$ суперякобиан

Здесь мы вводим аналог березиниана для необратимых преобразований. Запишем супераналитическое преобразование (2.1) в виде композиции

$$1)\begin{cases} \tilde{z} &= F\left(z,\tilde{\theta}\right), \\ \tilde{\theta} &= \tilde{\theta}, \end{cases} \qquad 2)\begin{cases} z &= z, \\ \tilde{\theta} &= \tilde{\theta}\left(z,\theta\right), \end{cases} \tag{E.1}$$

где $F\left(z,\tilde{\theta}\right) = \tilde{z}\left(z,\theta\right)$.

Суперякобиан первого преобразования есть просто $J_1 = \partial F/\partial z$. Если

$$\epsilon\left[\frac{\partial\tilde{\theta}}{\partial\theta}\right] \neq 0, \tag{E.2}$$

тогда, учитывая, что $\theta$ – нечетное, мы находим $J_2 = \left(\partial\tilde{\theta}/\partial\theta\right)^{-1}$ [30].

Таким образом полный суперякобиан есть

$$\boldsymbol{J}_{SA} = J_1 J_2 = \frac{\partial F}{\partial z}\cdot\left(\frac{\partial\tilde{\theta}}{\partial\theta}\right)^{-1}. \tag{E.3}$$

Чтобы получить $J_1$, мы запишем $J\left(z,\tilde{\theta}\right) = \tilde{z}\left(z,\theta\left(z,\tilde{\theta}\right)\right)$, тогда мы дифференцируем $\tilde{z}\left(z, \theta\left(z, \tilde{\theta}\right)\right)$ как сложную функцию

$$\frac{\partial F}{\partial z} = \frac{\partial \tilde{z}}{\partial z} + \frac{\partial \tilde{z}}{\partial \theta} \cdot \frac{\partial \theta}{\partial \tilde{\theta}} \cdot \frac{\partial \tilde{\theta}}{\partial z}. \tag{E.4}$$

Таким образом, мы получаем полный супер Якобиан

$$\boldsymbol{J}_{SA} = \frac{\dfrac{\partial \tilde{z}}{\partial z} - \dfrac{\partial \tilde{\theta}}{\partial z} \cdot \dfrac{\partial \theta}{\partial \tilde{\theta}} \cdot \dfrac{\partial \tilde{z}}{\partial \theta}}{\dfrac{\partial \tilde{\theta}}{\partial \theta}} \tag{E.5}$$

без условия обратимости всего преобразования, т.е. без стандартного требования $\epsilon\left[\partial \tilde{z}/\partial z\right] \neq 0$ [106].

Тем не менее, в [30] было показано, что выражение вида (E.5) (в алгебре матриц) может расширяться в случае $\epsilon\left[\partial \tilde{z}/\partial z\right] = 0$ (полунеобратимый случай (2.4) в нашей классификации).

**Предложение E.1.** *Формула* (E.5) *дает суперякобиан для обратимого и полунеобратимого суpeраналитических преобразований.*

*Доказательство.* Из (2.2) мы получаем

$$\frac{\partial \tilde{z}}{\partial z} = f'(z) + \theta \cdot \chi'(z), \tag{E.6}$$

$$\frac{\partial \tilde{\theta}}{\partial \theta} = g(z), \tag{E.7}$$

поэтому

$$\epsilon\left[\frac{\partial \tilde{z}}{\partial z}\right] = \epsilon\left[f'(z)\right] = \epsilon\left[f(z)\right],$$

$$\epsilon\left[\frac{\partial \tilde{\theta}}{\partial \theta}\right] = \epsilon\left[g(z)\right],$$

и, таким образом, согласно определениям (2.3) и (2.4), условие (E.2)

охватывает обратимые и полунеобратимые преобразования. ■

**Следствие E.2.** *Для обратимых и полунеобратимых супераналитических преобразований мы имеем*

$$\boldsymbol{J}_{SA}^{inv,halfinv} = \mathrm{Ber}\left(\tilde{Z}/Z\right) \tag{E.8}$$

*с*

$$\mathrm{Ber}\left(\tilde{Z}/Z\right) = \mathrm{Ber}\,\mathrm{P}_{SA}^{0}, \tag{E.9}$$

*где*

$$\mathrm{P}_{SA}^{0} = \begin{pmatrix} \dfrac{\partial \tilde{z}}{\partial z} & \dfrac{\partial \tilde{\theta}}{\partial z} \\ \dfrac{\partial \tilde{\theta}}{\partial z} & \dfrac{\partial \tilde{\theta}}{\partial \theta} \end{pmatrix}. \tag{E.10}$$

В необратимом случае, когда (E.2) не удовлетворяется, мы не можем использовать (E.3) и (E.4), и соотношение (E.8) более не применимо. Так, что мы вынуждены расширять определения. Якобиан $J_1$ должен вычисляться из

$$J_1^{noninv} \cdot \frac{\partial \tilde{\theta}}{\partial \theta} = \frac{\partial \tilde{z}}{\partial z} \cdot \frac{\partial \tilde{\theta}}{\partial \theta} + \frac{\partial \tilde{z}}{\partial \theta} \cdot \frac{\partial \tilde{\theta}}{\partial z}, \tag{E.11}$$

и поэтому вместо (E.5) и (E.8) мы имеем

**Определение E.3.** *Суперякобиан необратимого супераналитического преобразования $\mathcal{T}_{SA}$ определяется формулой*

$$\boldsymbol{J}_{SA}^{noninv} \cdot \left(\frac{\partial \tilde{\theta}}{\partial \theta}\right)^2 = \frac{\partial \tilde{z}}{\partial z} \cdot \frac{\partial \tilde{\theta}}{\partial \theta} + \frac{\partial \tilde{z}}{\partial \theta} \cdot \frac{\partial \tilde{\theta}}{\partial z}. \tag{E.12}$$

Здесь условие (E.2) больше не является необходимым. Чтобы вычислять $J_1^{noninv}$ и $\boldsymbol{J}_{SA}^{noninv}$, нужно решить уравнения (E.11) и (E.12) (т.е. раскладывая обе части в ряд по генераторам алгебры Грассмана).

В зависимости от компонентных функций суперякобиан полунеобратимого супераналитического преобразования (т.е. при $\epsilon\left[g\left(z\right)\right]\neq 0$) имеет вид

$$\boldsymbol{J}_{SA}=\frac{f'\left(z\right)}{g\left(z\right)}+\frac{\chi\left(z\right)\cdot\psi'\left(z\right)}{g^{2}\left(z\right)}+\theta\left(\frac{\chi\left(z\right)}{g\left(z\right)}\right)', \tag{E.13}$$

который совпадает с березинианом для обратимого и полунеобратимого преобразования.

В случае необратимого преобразования мы должны использовать следующее уравнение

$$\begin{aligned}\boldsymbol{J}_{SA}^{noninv}\cdot g^{2}\left(z\right) &= f'\left(z\right)\cdot g\left(z\right)+\chi\left(z\right)\cdot\psi'\left(z\right)\\ &+\theta\left(\chi'\left(z\right)\cdot g\left(z\right)-\chi\left(z\right)\cdot g'\left(z\right)\right)\end{aligned} \tag{E.14}$$

которое можно решить специальными методами вычислений с нильпотентами [120, 812].

**Следствие E.4.** *Для обратимых супераналитических преобразований березиниан существует и обратим* $\left(\epsilon\left[f\left(z\right)\right]\neq 0,\ \epsilon\left[g\left(z\right)\right]\neq 0\right)$, *для полунеобратимых преобразований березиниан существует и необратим, в то время, как для необратимых супераналитических преобразований* $\left(\epsilon\left[f\left(z\right)\right]=0\right)$ *мы можем использовать только суперякобиан* $\boldsymbol{J}_{SA}^{noninv}$ (E.14).

Чтобы классифицировать все супераналитические преобразования, мы должны ввести некоторую числовую характеристику необратимости.

**Определение E.5.** *Индекс необратимости супераналитического преобразования определяется формулой*

$$\operatorname{ind}\boldsymbol{J}_{SA}\stackrel{def}{=}\left\{n\in\mathbb{N}\,|\,\boldsymbol{J}_{SA}^{n}=0,\ \boldsymbol{J}_{SA}^{n-1}\neq 0\right\}. \tag{E.15}$$

*Замечание* **E.6.** Мы исключаем из рассмотрения тривиальный случай нулевого суперякобиана $\boldsymbol{J}_{SA}=0$.

Очевидно, что числовая мера необратимости на самом деле задается обратной величиной.

**Определение E.7.** *Степень необратимости супераналитического преобразования есть*

$$\boldsymbol{m}_{SA}\stackrel{def}{=}\frac{1}{\operatorname{ind}\boldsymbol{J}_{SA}}. \tag{E.16}$$

**Следствие E.8.** *Обратимые супераналитические преобразования обладают бесконечным индексом* $\operatorname{ind}\boldsymbol{J}_{SA}=\infty$ *и нулевой степенью необратимости* $\boldsymbol{m}_{SA}=0$.

**Следствие E.9.** *“Наиболее необратимые”* (*кроме тривиальных с нулевым якобианом* $\boldsymbol{J}_{SA}=0$) *супераналитические преобразования имеют* $\operatorname{ind}\boldsymbol{J}_{SA}=2$ *и* $\boldsymbol{m}_{SA}=1/2$.

## E.2. $(1|N)$-мерное суперпространство

Рассмотрим $(1|N)$-мерное суперпространство $\mathbb{C}^{1|N}$ с комплексными четной $z\in\mathbb{C}^{1|0}$ и нечетными $\theta^i\in\mathbb{C}^{0|1}$ координтами (обозначим $Z=\left(z,\theta^1,\theta^2,\ldots,\theta^N\right)$), где $\left\{\theta^i,\theta^j\right\}=0$.

Произвольная голоморфная суперфункция от $Z$ раскладывается в ряд

$$F\left(z,\theta^1,\theta^2,\ldots,\theta^N\right)=F_0\left(z\right)+\sum_i\theta^iF_i\left(z\right)+\sum_{i<j}\theta^i\theta^jF_{ij}\left(z\right)+\ldots, \tag{E.17}$$

который конечен вследствие нильпотентности $\theta^i$, причем последнее слагаемое пропорционально произведению всех нечетных координат, т. е. $\theta^1\theta^2\ldots\theta^N$.

В общем случае суперпроизводные определяются формулами [403]

$$D_i = \partial_i + u_{ij}\theta^j\partial, \tag{E.18}$$

где $\partial_i = \partial/\partial\theta^i$ и по повторяющимся индексам подразумевается суммирование.

Если в (E.18) $u_{ij} = \delta_{ij}$, то это означает $O(N)$ симметрию в нечетном секторе [563,565]. Другие обратимые варианты обсуждались в [403].

Таким образом, касательное суперпространство в $\mathbb{C}^{1|N}$ определяется вектором $(\partial, D_1, \ldots, D_N)^T$, где

$$\{D_i, D_j\} = 2\delta_{ij}\partial. \tag{E.19}$$

*Замечание* **E.10.** При $N = 1$ , когда $D_1^2 = \partial$, единственный нечетный дифференциальный оператор $D_1$ рассматривался как “квадратный корень” из $\partial$, что приводило в суперструнных приложениях к обыкновенным дифференциальным уравнениям. Тогда, как в случае $N > 1$ необходимо рассматриваить дифференциальные уравнения в частных производных [485].

При супераналитических преобразованиях $\mathcal{T}_{SA} : \mathbb{C}^{1|N} \to \mathbb{C}^{1|N}$ и $Z \to \tilde{Z}$ имеем закон преобразования

$$\begin{pmatrix} \partial \\ D_1 \\ \vdots \\ D_N \end{pmatrix} = \mathrm{P}_{SA}^{(N)} \cdot \begin{pmatrix} \tilde{\partial} \\ \tilde{D}_1 \\ \vdots \\ \tilde{D}_N \end{pmatrix}, \tag{E.20}$$

где суперматрица касательного пространства имеет вид

$$\mathrm{P}_{SA}^{(N)} = \begin{pmatrix} \partial\tilde{z} - \partial\tilde{\theta}^i \cdot \tilde{\theta}_i & \partial\tilde{\theta}^1 & \cdots & \partial\tilde{\theta}^N \\ D_1\tilde{z} - D_1\tilde{\theta}^j \cdot \tilde{\theta}_j & D_1\tilde{\theta}^1 & \cdots & D_1\tilde{\theta}^N \\ \vdots & \vdots & \ddots & \vdots \\ D_N\tilde{z} - D_N\tilde{\theta}^j \cdot \tilde{\theta}_j & D_N\tilde{\theta}^1 & \cdots & D_N\tilde{\theta}^N \end{pmatrix}. \tag{E.21}$$

Тогда предполагается выполнение $N$ суперконформных условий [563, 565, 624, 665]

$$D_i\tilde{z} - D_i\tilde{\theta}^j \cdot \tilde{\theta}_j = 0 \tag{E.22}$$

(ср. (2.37)) как требование однородности преобразования суперпроизводных

$$D_i = D_i\tilde{\theta}^j \cdot \tilde{D}_j. \tag{E.23}$$

(ср. (2.43)). Отсюда делается вывод, что композиция суперконформных преобразований снова дает суперконформное преобразование [563]. При этом стандартным образом редуцированная к суперконформному виду суперматрица $\mathrm{P}_{SCf}^{(N)}$ имеет блочно-треугольную форму, аналогичную (2.51)

$$\mathrm{P}_{SCf}^{(N)} = \begin{pmatrix} \partial\tilde{z} - \partial\tilde{\theta}^k \cdot \tilde{\theta}_k & \partial\tilde{\theta}^1 & \cdots & \partial\tilde{\theta}^N \\ 0 & D_1\tilde{\theta}^1 & \cdots & D_1\tilde{\theta}^N \\ \vdots & \vdots & \ddots & \vdots \\ 0 & D_N\tilde{\theta}^1 & \cdots & D_N\tilde{\theta}^N \end{pmatrix}. \tag{E.24}$$

Определяются также $N$-обобщения дифференциалов $d\theta^i$ и

$$dZ = dz + \theta_k \cdot d\theta^k. \tag{E.25}$$

При супераналитических преобразованиях

$$\begin{pmatrix} d\tilde{Z} \\ d\tilde{\theta}^1 \\ \vdots \\ d\tilde{\theta}^N \end{pmatrix} = \begin{pmatrix} dZ \\ d\theta^1 \\ \vdots \\ d\theta^N \end{pmatrix} \cdot \mathrm{P}_{SA}^{(N)}. \tag{E.26}$$

В обратимом суперконформном случае $dZ$ преобразуются по формулам

$$d\tilde{Z} = dZ \left( \partial\tilde{z} - \partial\tilde{\theta}^k \cdot \tilde{\theta}_k \right). \tag{E.27}$$

При четном $N$ можно применить дополнительное дифференцирование к (E.23) и симметризовать, тогда получим

$$\delta_{ij} \left( \partial\tilde{z} - \partial\tilde{\theta}^k \cdot \tilde{\theta}_k \right) = D_i\tilde{\theta}_k \cdot D_j\tilde{\theta}^k, \tag{E.28}$$

что можно сравнить с (2.58).

Подставляя (E.28) в (E.27), получаем

$$d\tilde{Z} = dZ \cdot D_i\tilde{\theta}_k \cdot D_j\tilde{\theta}^k, \tag{E.29}$$

что в стандартном случае [563, 565] трактуется как $N$-обобщение соотношения $d\tilde{z} = (\partial\tilde{z}/\partial z)\, dz$.

Соотношение (E.28) в обратимом случае после нормировки на множитель в левой части приводит к обычной $O(N)$ матрице, составленной из $D_i\tilde{\theta}_k$ (правый нижний угол в (E.24)). Детерминант этой матрицы, равный по модулю единице, различает между собой два топологически отделимых случая $SO(N)$ преобразований с тривиальным расслоением и общих $O(N)$ преобразований с твистом [563, 565].

Приведенные рассуждения, однако, справедливы лишь в случае инфинитезимальных и обратимых преобразований, а также при стандартной суперконформной редукции суперматрицы $\mathrm{P}_{SA}^{(N)} \to \mathrm{P}_{SCf}^{(N)}$ (E.24). С учетом возможной необратимости преобразований, нильпотентной левой части в (E.28) и наличия нильпотентных компонентных функций в $Z \to \tilde{Z}$ стандартные методы можно существенно видоизменить и расширить число различных типов преобразований [12, 17].

## E.3. $N = 2$ березиниан

Рассмотрим общие $N = 2$ супераналитические преобразования $Z\left(z, \theta^{+}, \theta^{-}\right) \to \tilde{Z}\left(\tilde{z}, \tilde{\theta}^{+}, \tilde{\theta}^{-}\right)$. Их действие в касательном $(1|2)$ суперпространстве имеет следующий вид

$$\begin{pmatrix} \partial \\ D^{-} \\ D^{+} \end{pmatrix} = \mathrm{P}_{SA}^{(N=2)} \cdot \begin{pmatrix} \tilde{\partial} \\ \tilde{D}^{-} \\ \tilde{D}^{+} \end{pmatrix}, \tag{E.30}$$

$$\begin{pmatrix} d\tilde{Z} & d\tilde{\theta}^{+} & d\tilde{\theta}^{-} \end{pmatrix} = \begin{pmatrix} dZ & d\theta^{+} & d\theta^{-} \end{pmatrix} \cdot \mathrm{P}_{SA}^{(N=2)}, \tag{E.31}$$

где

$$\mathrm{P}_{SA}^{(N=2)} = \begin{pmatrix} \partial\tilde{z} - \partial\tilde{\theta}^{+} \cdot \theta^{-} - \partial\tilde{\theta}^{-} \cdot \theta^{+} & \partial\tilde{\theta}^{+} & \partial\tilde{\theta}^{-} \\ D^{-}\tilde{z} - D^{-}\tilde{\theta}^{-} \cdot \tilde{\theta}^{+} - D^{-}\tilde{\theta}^{+} \cdot \tilde{\theta}^{-} & D^{-}\tilde{\theta}^{+} & D^{-}\tilde{\theta}^{-} \\ D^{+}\tilde{z} - D^{+}\tilde{\theta}^{-} \cdot \tilde{\theta}^{+} - D^{+}\tilde{\theta}^{+} \cdot \tilde{\theta}^{-} & D^{+}\tilde{\theta}^{+} & D^{+}\tilde{\theta}^{-} \end{pmatrix}. \tag{E.32}$$

**Предложение E.11.** *Внешний $N = 2$ дифференциальный оператор де Рама* [211]

$$\mathrm{d}^{(N=2)} = dz\partial + d\theta^{+}\partial_{-} + d\theta^{-}\partial_{+} \tag{E.33}$$

*инвариантен относительно общих $N = 2$ супераналитических преобразований $Z\left(z, \theta^{+}, \theta^{-}\right) \to \tilde{Z}\left(\tilde{z}, \tilde{\theta}^{+}, \tilde{\theta}^{-}\right)$.*

*Доказательство.* Пользуясь определениями, запишем (E.33) в виде

$$\begin{aligned} \mathrm{d}^{(N=2)} &= \left(dz - d\theta^+\theta^- - d\theta^-\theta^+\right)\partial + d\theta^+\left(\partial_- + \theta^-\partial\right) + \\ d\theta^-\left(\partial_+ + \theta^+\partial\right) &= dZ\partial + d\theta^+D^- + d\theta^-D^+. \end{aligned} \tag{E.34}$$

Тогда из (E.30) и (E.31) следует

$$\mathrm{d}^{(N=2)} = \begin{pmatrix} dZ & d\theta^+ & d\theta^- \end{pmatrix} \begin{pmatrix} \partial \\ D^- \\ D^+ \end{pmatrix} =$$

$$\begin{pmatrix} dZ & d\theta^+ & d\theta^- \end{pmatrix} \cdot \mathrm{P}_{SA}^{(N=2)} \cdot \begin{pmatrix} \tilde{\partial} \\ \tilde{D}^- \\ \tilde{D}^+ \end{pmatrix} =$$

$$\begin{pmatrix} d\tilde{Z} & d\tilde{\theta}^+ & d\tilde{\theta}^- \end{pmatrix} \begin{pmatrix} \tilde{\partial} \\ \tilde{D}^- \\ \tilde{D}^+ \end{pmatrix} = \tilde{\mathrm{d}}^{(N=2)}.$$

■

Найдем связь между березинианом и суперматрицей $\mathrm{P}_{SA}^{(N=2)}$ (E.32). Березиниан $N = 2$ супераналитических преобразований $Z\left(z, \theta^+, \theta^-\right) \to \tilde{Z}\left(\tilde{z}, \tilde{\theta}^+, \tilde{\theta}^-\right)$ определяется формулой [30]

$$\mathrm{Ber}^{\,N=2}\left(\tilde{Z}/Z\right) = \mathrm{Ber}\,\mathrm{P}_0^{(N=2)}, \tag{E.35}$$

где

$$\mathrm{P}_0^{(N=2)} = \begin{pmatrix} \dfrac{\partial \tilde{z}}{\partial z} & \dfrac{\partial \tilde{\theta}^+}{\partial z} & \dfrac{\partial \tilde{\theta}^-}{\partial z} \\ \dfrac{\partial \tilde{z}}{\partial \theta^+} & \dfrac{\partial \tilde{\theta}^+}{\partial \theta^+} & \dfrac{\partial \tilde{\theta}^-}{\partial \theta^+} \\ \dfrac{\partial \tilde{z}}{\partial \theta^-} & \dfrac{\partial \tilde{\theta}^+}{\partial \theta^-} & \dfrac{\partial \tilde{\theta}^-}{\partial \theta^-} \end{pmatrix}. \tag{E.36}$$

**Предложение E.12.** *Березиниан общих $N = 2$ супераналитических преобразований равен березиниану суперматрицы* $\mathrm{P}_{SA}^{(N=2)}$ (E.32)

$$\mathrm{Ber}\,\left(\tilde{Z}/Z\right) = \mathrm{Ber}\,\mathrm{P}_{SA}^{(N=2)}. \tag{E.37}$$

*Доказательство.* Разложим суперматрицу $P_0^{(N=2)}$ на произведение

$$\mathrm{P}_0^{(N=2)} = \mathrm{K} \cdot \mathrm{P}_{SA}^{(N=2)} \cdot \tilde{\mathrm{K}}, \tag{E.38}$$

где

$$\mathrm{K} = \begin{pmatrix} 1 & 0 & 0 \\ -\theta^- & 1 & 0 \\ -\theta^+ & 0 & 1 \end{pmatrix}, \quad \tilde{\mathrm{K}} = \begin{pmatrix} 1 & 0 & 0 \\ -\tilde{\theta}^- & 1 & 0 \\ -\theta^+ & 0 & 1 \end{pmatrix}.$$

Легко заметить, что $\mathrm{Ber}\,\mathrm{K} = \mathrm{Ber}\,\tilde{\mathrm{K}} = 1$. Пользуясь мультипликативностью березиниана [30], имеем

$$\begin{aligned} \mathrm{Ber}\,\mathrm{P}_0^{(N=2)} &= \mathrm{Ber}\,\mathrm{K} \cdot \mathrm{P}_{SA}^{(N=2)} \cdot \tilde{\mathrm{K}} = \mathrm{Ber}\,\mathrm{K} \cdot \mathrm{Ber}\,\mathrm{P}_{SA}^{(N=2)} \cdot \mathrm{Ber}\,\tilde{\mathrm{K}} = \\ 1 \cdot \mathrm{Ber}\,\mathrm{P}_{SA}^{(N=2)} \cdot 1 &= \mathrm{Ber}\,\mathrm{P}_{SA}^{(N=2)}. \end{aligned}$$

Тогда из (E.35) получаем

$$\mathrm{Ber}\,\left(\tilde{Z}/Z\right) = \mathrm{Ber}\,\mathrm{P}_0^{(N=2)} = \mathrm{Ber}\,\mathrm{P}_{SA}^{(N=2)}.$$

■

## E.4. Березинианы $N = 4$ редуцированных преобразований

Здесь мы найдем полный березиниан в каждом случае из $N = 4$ редукций (3.184)–(3.197).

Для этого, по аналогии с $N = 2$ **Утверждением 3.18**, докажем

**Утверждение E.13.** *Условия редукции* (3.184)–(3.197)*, примененные в обратном порядке, дают вырожденную суперматрицу* $\mathrm{P}^{(N=4)}$*, имеющую нулевой березиниан.*

*Доказательство.* Применяя $\mathsf{TPt}_i^{\pm}$ условия к суперматрице $\mathrm{P}_S^{(N=4)}$ (3.173), получаем

$$\mathrm{P}_{D_i^{\pm}}^{(N=4)} = \mathrm{P}_S^{(N=4)}|_{TPt_i^{\pm}} =$$

$$\begin{pmatrix} 0 & \partial\tilde{\theta}^{+}_{1(TPt_i^{\pm})} & \partial\tilde{\theta}^{-}_{1(TPt_i^{\pm})} & \partial\tilde{\theta}^{+}_{2(TPt_i^{\pm})} & \partial\tilde{\theta}^{-}_{2(TPt_i^{\pm})} \\ 0 & & & & \\ 0 & & & & \\ 0 & & \mathrm{H}_{TPt_i^{\pm}} & & \\ 0 & & & & \end{pmatrix}. \tag{E.39}$$

И аналогично для $\mathsf{TPt}_2^{\pm}$. С другой стороны, применяя $\mathsf{SCf}$ условие к $\mathrm{P}_{TPt_i^{\pm}}^{(N=4)}$ (3.174)–(3.177), имеем

$$\mathrm{P}_D^{(N=4)} = \mathrm{P}_S^{(N=4)}|_{\Delta_i^{\pm}\left(z,\theta_i^{+},\theta_i^{-}\right)=0} =$$

$$\begin{pmatrix} 0 & \partial\tilde{\theta}^{+}_{1(SCf)} & \partial\tilde{\theta}^{-}_{1(SCf)} & \partial\tilde{\theta}^{+}_{2(SCf)} & \partial\tilde{\theta}^{-}_{2(SCf)} \\ 0 & & & & \\ 0 & & & & \\ 0 & & \mathrm{H}_{SCf} & & \\ 0 & & & & \end{pmatrix}. \tag{E.40}$$

Очевидно, что $\operatorname{Ber}\mathrm{P}_D^{(N=4)} = \operatorname{Ber}\mathrm{P}_{D_i^{\pm}}^{(N=4)} = 0.$ ■

*Замечание* **E.14.** Все 5 вырожденных суперматриц (E.39)–(E.40), несмотря на подобный внешний вид, не совпадают между собой

$$\mathrm{P}_{D_1^+}^{(N=4)} \neq \mathrm{P}_{D_1^-}^{(N=4)} \neq \mathrm{P}_{D_2^+}^{(N=4)} \neq \mathrm{P}_{D_2^-}^{(N=4)} \neq \mathrm{P}_{D}^{(N=4)}, \tag{E.41}$$

поскольку на их оставшиеся ненулевые элементы $\partial\tilde{\theta}_i^{\pm}$ и H наложены различные условия.

Чтобы найти березиниан редуцированных преобразований для каждого типа редукций, спроектируем формулу сложения $N = 4$ березинианов (3.172) и воспользуемся **Предложением 3.38** и **Утверждением E.13**, а затем формулами (3.178)–(3.179) для березинианов введенных матриц $\mathrm{P}_S^{(N=4)}$ и $\mathrm{P}_{T_i^{\pm}}^{(N=4)}$. Тогда для $\operatorname{Ber}^{N=4}\left(\tilde{Z}/Z\right)$ получаем (ср. $N = 1$ (2.48) и $N = 2$ (3.63))

$$\operatorname{Ber}^{N=4}\left(\tilde{Z}/Z\right) = \operatorname{Ber}\mathrm{P}_{SA}^{(N=4)} =$$

$$\begin{cases} \left(\operatorname{Ber}\mathrm{P}_S^{(N=4)} + \operatorname{Ber}\mathrm{P}_{T_1^+}^{(N=4)} + \operatorname{Ber}\mathrm{P}_{T_1^-}^{(N=4)} + \operatorname{Ber}\mathrm{P}_{T_2^+}^{(N=4)} + \operatorname{Ber}\mathrm{P}_{T_2^-}^{(N=4)}\right)|_{SCf} \\ \left(\operatorname{Ber}\mathrm{P}_S^{(N=4)} + \operatorname{Ber}\mathrm{P}_{T_1^+}^{(N=4)} + \operatorname{Ber}\mathrm{P}_{T_1^-}^{(N=4)} + \operatorname{Ber}\mathrm{P}_{T_2^+}^{(N=4)} + \operatorname{Ber}\mathrm{P}_{T_2^-}^{(N=4)}\right)|_{TPt_1^+} \\ \left(\operatorname{Ber}\mathrm{P}_S^{(N=4)} + \operatorname{Ber}\mathrm{P}_{T_1^+}^{(N=4)} + \operatorname{Ber}\mathrm{P}_{T_1^-}^{(N=4)} + \operatorname{Ber}\mathrm{P}_{T_2^+}^{(N=4)} + \operatorname{Ber}\mathrm{P}_{T_2^-}^{(N=4)}\right)|_{TPt_1^-} \\ \left(\operatorname{Ber}\mathrm{P}_S^{(N=4)} + \operatorname{Ber}\mathrm{P}_{T_1^+}^{(N=4)} + \operatorname{Ber}\mathrm{P}_{T_1^-}^{(N=4)} + \operatorname{Ber}\mathrm{P}_{T_2^+}^{(N=4)} + \operatorname{Ber}\mathrm{P}_{T_2^-}^{(N=4)}\right)|_{TPt_2^+} \\ \left(\operatorname{Ber}\mathrm{P}_S^{(N=4)} + \operatorname{Ber}\mathrm{P}_{T_1^+}^{(N=4)} + \operatorname{Ber}\mathrm{P}_{T_1^-}^{(N=4)} + \operatorname{Ber}\mathrm{P}_{T_2^+}^{(N=4)} + \operatorname{Ber}\mathrm{P}_{T_2^-}^{(N=4)}\right)|_{TPt_2^-} \end{cases} =$$

$$\begin{cases} \operatorname{Ber} \mathrm{P}_{SCf}^{(N=4)} + 0 + 0 + 0 + 0 \\ 0 + \operatorname{Ber} \mathrm{P}_{TPt_1^+}^{(N=4)} + 0 + 0 + 0 \\ 0 + 0 + \operatorname{Ber} \mathrm{P}_{TPt_1^-}^{(N=4)} + 0 + 0 \\ 0 + 0 + 0 + \operatorname{Ber} \mathrm{P}_{TPt_2^+}^{(N=4)} + 0 \\ 0 + 0 + 0 + 0 + \operatorname{Ber} \mathrm{P}_{TPt_2^-}^{(N=4)} \end{cases} = \begin{cases} \operatorname{Ber} \mathrm{P}_{SCf}^{(N=4)} & (\mathsf{SCf}) \\ \operatorname{Ber} \mathrm{P}_{TPt_1^+}^{(N=4)} & (\mathsf{TPt}\,{}_1^+) \\ \operatorname{Ber} \mathrm{P}_{TPt_1^-}^{(N=4)} & (\mathsf{TPt}\,{}_1^-) \\ \operatorname{Ber} \mathrm{P}_{TPt_2^+}^{(N=4)} & (\mathsf{TPt}\,{}_2^+) \\ \operatorname{Ber} \mathrm{P}_{TPt_2^-}^{(N=4)} & (\mathsf{TPt}\,{}_2^-) \end{cases} =$$

$$\begin{cases} \dfrac{Q_{SCf}\left(z, \theta_i^+, \theta_i^-\right)}{\det \mathrm{H}_{SCf}} & (\mathsf{SCf}) \\ \dfrac{\varDelta^-_{1\left(TPt_1^+\right)}\left(z, \theta_i^+, \theta_i^-\right)}{\det^2 \mathrm{H}_{TPt_1^+}} \cdot \boldsymbol{K}_{TPt_1^+} & (\mathsf{TPt}\,{}_1^+) \\ \dfrac{\varDelta^+_{1\left(TPt_1^-\right)}\left(z, \theta_i^+, \theta_i^-\right)}{\det^2 \mathrm{H}_{TPt_1^-}} \cdot \boldsymbol{K}_{TPt_1^-} & (\mathsf{TPt}\,{}_1^-) \\ \dfrac{\varDelta^-_{2\left(TPt_2^+\right)}\left(z, \theta_i^+, \theta_i^-\right)}{\det^2 \mathrm{H}_{TPt_2^+}} \cdot \boldsymbol{K}_{TPt_2^+} & (\mathsf{TPt}\,{}_2^+) \\ \dfrac{\varDelta^+_{2\left(TPt_2^-\right)}\left(z, \theta_i^+, \theta_i^-\right)}{\det^2 \mathrm{H}_{TPt_2^-}} \cdot \boldsymbol{K}_{TPt_2^-} & (\mathsf{TPt}\,{}_2^-) \end{cases}, \qquad \text{(E.42)}$$

где $\boldsymbol{K}_{TPt_i^\pm}$ определено в (3.180).

# Приложение Ж

# Частные случаи редуцированных преобразований

Здесь рассматриваются частные случаи редуцированных $N = 1$, $N = 2$ и $N = 4$ преобразований, расщепленные и дробно-линейные преобразования.

## Ж.1. $\rho$-суперконформные преобразования и нильпотентные суперполя

Существует несколько различных определений суперконформных преобразований [343, 354, 491]. В одном из них [341, 342] утверждается, что $N = 1$ преобразование $Z \to \tilde{Z}$ является суперконформным, если множитель, с которым преобразуются производные равен березиниану $\mathrm{Ber}\, \mathrm{P}_{SCf} = D\tilde{\theta}_{SCf}$ (см. (2.52)). Здесь мы рассмотрим в общем случае преобразования, для которых выполняется соотношение

$$\mathrm{Ber}\left(\tilde{Z}/Z\right) = D\tilde{\theta}. \tag{Ж.1}$$

В компонентах это уравнение (при $\epsilon\left[g\left(z\right)\right] \neq 0$) приводит к системе (см. (2.2))

$$f'\left(z\right) g\left(z\right) + \chi\left(z\right) \psi'\left(z\right) = g^2\left(z\right), \tag{Ж.2}$$

$$\left(\frac{\chi\left(z\right)}{g\left(z\right)}\right)' = \psi'\left(z\right). \tag{Ж.3}$$

Отсюда получаем общий вид пребразований в стандартной параметри-

зации [111,176,405]

$$\tilde{z} = f(z) + \theta \cdot (\psi(z) + \rho)\sqrt{f'(z) + \psi(z)\psi'(z)}, \tag{Ж.4}$$

$$\tilde{\theta} = \psi(z) + \theta \cdot \sqrt{f'(z) + \psi(z)\psi'(z) + \rho\psi(z)}. \tag{Ж.5}$$

По сравнению со стандартными суперконформными преобразованиями [343,354,491] новыми в (Ж.4)–(Ж.5) являются слагаемые с нечетным параметром $\rho$, который появляется из-за наличия производных*) в обеих частях уравнения (Ж.3).

**Определение Ж.1.** *$\rho$-суперконформными преобразованиями назовем преобразования* (Ж.4)–(Ж.5)*, а супермногообразия, склееенные с помощью таких преобразований — $\rho$-суперримановыми поверхностями.*

Суперматрица касательного расслоения $\mathrm{P} = \mathrm{P}_\rho$ из (2.22) имеет вид

$$\mathrm{P}_\rho = \mathrm{P}_{SCf} + \rho \cdot \begin{pmatrix} \partial\tilde{\theta} & 0 \\ D\tilde{\theta} & 0 \end{pmatrix}, \tag{Ж.6}$$

где $\mathrm{P}_{SCf}$ определяется в (2.39).

Из (Ж.6) видно, что в общем случае суперпроизводная $D$ преобразуется неоднородно при $\rho$-суперконформных преобразованиях, а именно,

$$D = D\tilde{\theta} \cdot \tilde{D} + \rho \cdot D\tilde{\theta} \cdot \tilde{\partial} \tag{Ж.7}$$

или

$$D = \mathrm{Ber}\left(\tilde{Z}/Z\right) \cdot \tilde{D} + \rho \cdot \frac{\partial\tilde{\theta}}{\partial\theta} \cdot \tilde{\partial} \tag{Ж.8}$$

Кроме того, для $\rho$-аналога SCf супердифференциала $d\tau = dZD + d\theta$

---

*Примечание.* В системе уравнений, следующей из $D\tilde{z} - D\tilde{\theta} \cdot \tilde{\theta} = 0$ (2.29), равны сами выражения под знаком производных в (Ж.3).

выполняется

$$d\tilde{\tau} = d\tau \cdot D\tilde{\theta} \cdot (1 + \rho D)\,. \qquad (Ж.9)$$

Из (Ж.7)–(Ж.9) следует

**Определение Ж.2.** *Суперполя, позволяющие выделение нечетного множителя $\rho$, назовем $\rho$-суперполями.*

**Предложение Ж.3.** *При $\rho$-суперконформных преобразованиях $\rho$-суперполя преобразуются ковариантно.*

*Доказательство.* Следует из нильпотентности $\rho$. ■

**Следствие Ж.4.** *$\rho$-суперполя на $\rho$-суперримановых поверхностях обладают всеми "хорошими" свойствами обычных суперполей на суперримановых поверхностях* [359–361].

## Ж.2. Полугруппа расщепленных $N = 2$ SCf преобразований

Важным частным случаем $N = 2$ суперконформных преобразований являются *расщепленные (split)* $N = 2$ преобразования $\mathcal{T}_{SCf}^{(N=2)}$ [563], которые не содержат нечетных функций в разложении (3.14)

$$\begin{cases} \tilde{z} = f(z)\,, \\ \tilde{\theta}^{\pm} = \theta^{\pm} g_{\pm\mp}(z) + \theta^{\mp} g_{\pm\pm}(z)\,. \end{cases} \qquad (Ж.10)$$

Такие преобразования могут быть функциями перехода на обычных римановых поверхностях со спиновой структурой [563].

Применение $N = 2$ SCf условий (3.39)–(3.40) дает систему уравне-

ний

$$\operatorname{per} \mathrm{G}_{split} = f'(z), \qquad (Ж.11)$$

$$\operatorname{scf}_{\pm} \mathrm{G}_{split} = 0, \qquad (Ж.12)$$

или в явном виде

$$g_{+-}(z)\, g_{-+}(z) + g_{++}(z)\, g_{--}(z) = f'(z), \qquad (Ж.13)$$

$$g_{+-}(z)\, g_{--}(z) = 0, \qquad (Ж.14)$$

$$g_{-+}(z)\, g_{++}(z) = 0. \qquad (Ж.15)$$

Из (Ж.12) следует, что матрица $\mathrm{G}_{split}$ (3.17) является scf-матрицей (см. **Подраздел 5.1**), параметризованной двумя четными функциями из $\mathrm{G}_{split}$, а уравнение (Ж.12) получается из (3.80) занулением нечетных функций. Это означает, что при $\epsilon\left[\operatorname{per} \mathrm{G}_{split}\right] \neq 0$ матрица $\mathrm{G}_0$ в координатном базисе, соответствующая $\mathrm{G}_{split}$ (связанная соотношением, подобным (3.79)), после перенормировки на $\sqrt{\operatorname{per} \mathrm{G}_{split}}$ будет $O_{\Lambda_0}(2)$ матрицей, причем условие $\epsilon\left[\operatorname{per} \mathrm{G}_{split}\right] \neq 0$ оставляет лишь две возможности (подобно (3.77)–(3.78)): $\mathrm{G}_{split}$ — диагональная и антидиагональная матрица

$$\mathrm{G}_{U(1)} = \begin{pmatrix} g_{+-}(z) & 0 \\ 0 & g_{-+}(z) \end{pmatrix} - U_{\Lambda_0}(1), \qquad (Ж.16)$$

$$\mathrm{G}_{O(2)} = \begin{pmatrix} 0 & g_{++}(z) \\ g_{--}(z) & 0 \end{pmatrix} - O_{\Lambda_0}(2). \qquad (Ж.17)$$

**Утверждение Ж.5.** *“Таблица умножения” матриц* G

$$\mathrm{G}_{U(1)}\mathrm{G}_{U(1)} = \mathrm{G}_{U(1)}, \qquad (Ж.18)$$

$$\mathrm{G}_{U(1)}\mathrm{G}_{O(2)} = \mathrm{G}_{O(2)}, \qquad (Ж.19)$$

$$G_{O(2)}G_{O(2)} = G_{U(1)}. \qquad (Ж.20)$$

*совпадает с таблицей умножения типов $N = 2$ расщепленных преобразований.*

Соответствующие (Ж.16)–(Ж.17) преобразования имеют вид

$$\begin{cases} \tilde{z} = f(z), \quad f'(z) = g_{+-}(z)\, g_{-+}(z), \\ \tilde{\theta}^{\pm} = \theta^{\pm} g_{\pm\mp}(z), \end{cases} \quad - U_{\Lambda_0}(1) \qquad (Ж.21)$$

$$\begin{cases} \tilde{z} = f(z), \quad f'(z) = g_{++}(z)\, g_{--}(z), \\ \tilde{\theta}^{\pm} = \theta^{\mp} g_{\pm\pm}(z), \end{cases} \quad - O_{\Lambda_0}(2) \qquad (Ж.22)$$

откуда следует, что только $U_{\Lambda_0}(1)$ преобразования образуют подгруппу (или подполугруппу в необратимом случае), поскольку отсутствует переворот киральности в $\theta$ секторе. Именно такие функции перехода (но в другой параметризации) описывают произвольное линейное расслоение над обычными римановыми поверхностями [563].

В необратимом случае $\epsilon\,[\operatorname{per} G_{split}] = 0$ ситуация не столь прозрачна, поскольку SCf условия (Ж.14)–(Ж.15) могут быть выполнены не только занулением сомножителей, но и за счет возможных делителей нуля в функциях $g_{ab}(z)$ $(a, b = \pm)$. Это может случиться, например, когда $g_{ab}(z)$ являются произведениями нечетных функций, и тогда для параметризации необратимого преобразования необходимо выбрать не четные, а нечетные функции.

*Пример* **Ж.6.** Действительно, пусть

$$G_{split} = \begin{pmatrix} \mu_{+}(z)\, \nu_{-}(z) & \mu_{+}(z)\, \nu_{+}(z) \\ \mu_{-}(z)\, \nu_{-}(z) & -\mu_{-}(z)\, \nu_{+}(z) \end{pmatrix} \qquad (Ж.23)$$

где $\mu_a, \nu_b : \mathrm{C}^{1|0} \to \mathrm{C}^{0|1}$ и $\mu_a^2(z) = \nu_b^2(z) = 0$, тогда

$$\begin{cases} \tilde{z} = f(z), \quad f'(z) = \mu_+(z)\,\nu_+(z)\,\mu_-(z)\,\nu_-(z) \\ \tilde{\theta}^{\pm} = \pm\theta^{\pm}\mu_{\pm}(z)\,\nu_{\mp}(z) + \theta^{\mp}\mu_{\pm}(z)\,\nu_{\pm}(z), \end{cases} \tag{Ж.24}$$

причем SCf условия (Ж.14)–(Ж.15) выполняются вследствие нильпотентности нечетных функций $\mu_{\pm}(z)$ и $\nu_{\pm}(z)$, а матрица $\mathrm{G}_{split}$ не (анти) диагонализуется (как в (Ж.16)–(Ж.17), а представляет собой scf-матрицу с нильпотентными элементами (см. **Подраздел 5.1**).

Из сравнения (3.14) и (Ж.10) следует, что расщепленные $N=2$ преобразования образуют подполугруппу общей полугруппы $N=2$ супераналитических преобразований, которая характеризуется только лишь элементами матрицы $\mathrm{G}_{split}$ (3.17). Поэтому представление расщепленной $N=2$ SCf полугруппы функциональными матрицами (см. **Определение 3.9**) будет сужением представления (3.18) на элементы матрицы $\mathrm{G}_{split}$, т.е.

$$\left\{\begin{matrix} f & h & \chi_- & \chi_+ \\ \psi_+ & \lambda_+ & g_{+-} & g_{++} \\ \psi_- & \lambda_- & g_{--} & g_{-+} \end{matrix}\right\}\Big|_{split} \longrightarrow \left\{\begin{matrix} g_{+-} & g_{++} \\ g_{--} & g_{-+} \end{matrix}\right\}. \tag{Ж.25}$$

Отсюда следует

**Определение Ж.7.** *Элемент* $\mathbf{s}$ *расщепленной* $N=2$ *суперконформной полугруппы* $\mathbf{S}_{SCf(split)}^{(N=2)}$ *параметризуется функциональной матрицей*

$$\left\{\begin{matrix} g_{+-} & g_{++} \\ g_{--} & g_{-+} \end{matrix}\right\}\Big|_{g_{\mp\pm}(z)g_{\pm\pm}(z)=0} \stackrel{def}{=} \mathbf{s}_{split} \in \mathbf{S}_{SCf(split)}^{(N=2)}, \tag{Ж.26}$$

*а действие*

$$\mathbf{s}^{(1)}_{split} *_s \mathbf{s}^{(2)}_{split} = \mathbf{s}^{(3)}_{split} \qquad (Ж.27)$$

*определяется композицией расщепленных преобразований* $Z \to \tilde{Z} \to \tilde{\tilde{Z}}$ *и имееет следующий вид*

$$\left\{ \begin{array}{cc} g^{(3)}_{+-} & g^{(3)}_{++} \\ g^{(3)}_{--} & g^{(3)}_{-+} \end{array} \right\} = \left\{ \begin{array}{cc} g^{(1)}_{+-} & g^{(1)}_{++} \\ g^{(1)}_{--} & g^{(1)}_{-+} \end{array} \right\} *_s \left\{ \begin{array}{cc} g^{(2)}_{+-} & g^{(2)}_{++} \\ g^{(2)}_{--} & g^{(2)}_{-+} \end{array} \right\} = \qquad (Ж.28)$$

$$\left\{ \begin{array}{cc} g^{(1)}_{+-} \circ f^{(2)} \cdot g^{(2)}_{+-} + g^{(1)}_{++} \circ f^{(2)} \cdot g^{(2)}_{--} & g^{(1)}_{+-} \circ f^{(2)} \cdot g^{(2)}_{++} + g^{(1)}_{++} \circ f^{(2)} \cdot g^{(2)}_{-+} \\ g^{(1)}_{--} \circ f^{(2)} \cdot g^{(2)}_{+-} + g^{(1)}_{-+} \circ f^{(2)} \cdot g^{(2)}_{--} & g^{(1)}_{--} \circ f^{(2)} \cdot g^{(2)}_{++} + g^{(1)}_{-+} \circ f^{(2)} \cdot g^{(2)}_{-+} \end{array} \right\},$$

*где*

$$f^{(2)\prime}(z) = \operatorname{per} \mathrm{G}^{(2)}_{split} = g^{(2)}_{+-}(z)\, g^{(2)}_{-+}(z) + g^{(2)}_{++}(z)\, g^{(2)}_{--}(z)\,,$$

$$g^{(1)}_{\mp\pm}(z)\, g^{(1)}_{\pm\pm}(z) = 0, \quad g^{(2)}_{\mp\pm}(z)\, g^{(2)}_{\pm\pm}(z) = 0.$$

Ассоциативность действия $*_s$ (Ж.27) следует из ассоциативности композиции расщепленных преобразований.

**Утверждение Ж.8.** *Ортогональность элементов столбца* (Ж.14)–(Ж.15) *или* scf-*свойство матрицы* G (Ж.12) *при действии* $*_s$ *сохраняется, т.е.* $g^{(3)}_{\mp\pm}(z)\, g^{(3)}_{\pm\pm}(z) = 0$ *в* (Ж.28).

Очевидно, что необратимые преобразования соответствуют идеалу $\mathbf{I}^{(N=2)}_{SCf(split)}$ полугруппы $\mathbf{S}^{(N=2)}_{SCf(split)}$, а обратимые преобразования — ее подгруппе $\mathbf{G}^{(N=2)}_{SCf(split)}$.

Двусторонняя единица в полугруппе $\mathbf{S}^{(N=2)}_{SCf(split)}$ определяется как

$$\mathbf{e}_{split} = \left\{ \begin{array}{cc} 1 & 0 \\ 0 & 1 \end{array} \right\}, \qquad (Ж.29)$$

а двусторонний нуль представляется нулевой матрицей в (Ж.26).

## Ж.3. Вложение $N=1 \hookrightarrow N=2$

Ранее частый случай вложения $N=1 \hookrightarrow N=2$ использовалось в [637, 813] при вычислении суперструнных амплитуд методом функционального интегрирования [814]. Мы рассмотрим общий случай суперконформного $N=1 \hookrightarrow N=2$ вложения [3] с учетом необратимых преобразований.

Опишем погружение $N=1$ мирового листа $W=(w,\eta)$ в $N=2$ суперплоскость $Z=(z,\theta^+,\theta^-)$ тремя четными и тремя нечетными функциями и следующим преобразованием общего вида

$$z = f(w)+\eta\cdot\chi(w), \quad (Ж.30)$$

$$\theta^{\pm} = \psi_{\pm}(w)+\eta\cdot g_{\pm}(w), \quad (Ж.31)$$

где $f, g_{\pm}: \mathbb{C}^{1,0}\to\mathbb{C}^{1,0}$, $\chi,\psi_{\pm}:\mathbb{C}^{1,0}\to\mathbb{C}^{0,1}$.

При супераналитических $N=1 \hookrightarrow N=2$ преобразованиях (Ж.30)–(Ж.31) $N=1$ суперпроизводная $D=\partial_\eta+\eta\cdot\partial_w$ $\left(D^2=\partial_w\right)$ переходит в

$$D = D\theta^+\cdot D^- + D\theta^-\cdot D^+ + \left(Dz-\theta^+\cdot D\theta^- - \theta^-\cdot D\theta^+\right)\cdot\partial_w, \quad (Ж.32)$$

где $D^{\pm}$ определены в (3.1), поэтому суперконформные условия в данном случае имеют вид

$$Dz = \theta^+\cdot D\theta^- + \theta^-\cdot D\theta^+. \quad (Ж.33)$$

Применяя к (Ж.33) оператор $D$, получаем

$$\partial_w z + \theta^+\cdot\partial_w\theta^- + \theta^-\cdot\partial_w\theta^+ = 2\cdot D\theta^+\cdot D\theta^-. \quad (Ж.34)$$

Условие того, что два погружения $(z,\theta^+,\theta^-)$ и $\left(\tilde{z},\tilde{\theta}^+,\tilde{\theta}^-\right)$ параме-тризуют один и тот же мировой лист, приводят к соотношениям

$$D^+\tilde{\theta}^- \cdot D^-\tilde{\theta}^+ + D^+\tilde{\theta}^+ \cdot D^-\tilde{\theta}^- = \frac{D\tilde{\theta}^+ \cdot D\tilde{\theta}^-}{D\theta^+ \cdot D\theta^-}. \qquad (\text{Ж}.35)$$

По аналогии с (3.9) введем в рассмотрение матрицу

$$\mathrm{H}_w = \begin{pmatrix} D\theta^+ & D\theta^- \\ D\theta^+ & D\theta^- \end{pmatrix}, \qquad (\text{Ж}.36)$$

тогда условие (Ж.35) запишется в виде [3]

$$\operatorname{per}\tilde{\mathrm{H}}_w = \operatorname{per}\mathrm{H} \cdot \operatorname{per}\mathrm{H}_w, \qquad (\text{Ж}.37)$$

где H определена в (3.9).

Классификацию вложений $N = 1 \hookrightarrow N = 2$ по необратимости можно провести в полной аналогии с классификацией $N = 2$ редуцированных преобразований (см. **Пункт 2.1.3**).

Приведем пример вложения $N = 1 \hookrightarrow N = 2$ при $\epsilon\left[\operatorname{per}\mathrm{H}_w\right] \neq 0$ [3]

$$\begin{aligned} z &= f(w) + \frac{\eta}{\sqrt{2}} e^{q(w)} \psi_-(w) \sqrt{f'(w) + \psi_+(w)\,\psi_-'(w)} + \\ &\quad \frac{\eta}{\sqrt{2}} e^{-q(w)} \psi_+(w) \sqrt{f'(w) + \psi_-(w)\,\psi_+'(w)}, \qquad (\text{Ж}.38) \\ \theta^\pm &= \psi_\pm(w) + \qquad (\text{Ж}.39) \\ &\quad \frac{\eta}{\sqrt{2}} e^{\pm q(w)} \sqrt{f'(w) + \psi_-(w)\,\psi_+'(w) + \psi_+(w)\,\psi_-'(w)}. \end{aligned}$$

Среди необратимых преобразований с $\epsilon\left[\operatorname{per}\mathrm{H}_w\right] = 0$ приведем следующее [3]

$$z = f_N(w) + \eta\left[\psi_+(w)\,\rho_-(w) + \psi_-(w)\,\rho_+(w)\right]\sigma(w), \qquad (\text{Ж}.40)$$

$$\theta^{\pm} \;=\; \psi_{\pm}(w) + \eta\rho_{\pm}(w)\,\sigma(w)\,, \tag{Ж.41}$$

где $f'_N(w) = \psi'_+(w)\,\psi_-(w) + \psi'_-(w)\,\psi_+(w)$.

Отметим, что многие формулы и соответствующие выводы можно перенести с $N=2$ преобразований на вложения $N=1 \hookrightarrow N=2$, если в первых положить $\theta^{\pm} = \eta/\sqrt{2}$ (см. **Пункт 2.1.3**).

## Ж.4. Расщепленные $N=4$ SCf преобразования

Среди $N = 4$ преобразований, удовлетворяющих SCf условиям (3.184)–(3.185), наиболее простыми оказываются так называемые *расщепленные (split)* $N = 4$ SCf преобразования (см. $N = 2$ в **Приложении Ж.2**). Они не содержат нечетных компонентных функций в разложении (3.235) и (обратимые) могут служить функциями перехода на специальных (несупер) римановых поверхностях со спиновой структурой [563].

Общий вид расщепленных $N = 4$ SCf преобразований (ср. (Ж.10))

$$\begin{cases} \tilde{z} = f(z)\,, \\ \tilde{\theta}_i^{\pm} = \theta_j^{\pm} g_{ij}^{\pm\mp}(z) + \theta_j^{\mp} g_{ij}^{\pm\pm}(z) \end{cases} \tag{Ж.42}$$

задается 17 четными функциями $f$, $g_{ij}^{ab} : \mathbb{C}^{1|0} \to \mathbb{C}^{1|0}$, на которые налагаются SCf условия [2], следующие из (3.213)–(3.215),

$$\mathrm{per}\,\mathrm{G}_{11} + \mathrm{per}\,\mathrm{G}_{21} \;=\; \mathrm{per}\,\mathrm{G}_{12} + \mathrm{per}\,\mathrm{G}_{22} = f'(z) \tag{Ж.43}$$

$$\mathrm{scf}_{\pm}\mathrm{G}_{11} + \mathrm{scf}_{\pm}\mathrm{G}_{21} \;=\; \mathrm{scf}_{\pm}\mathrm{G}_{12} + \mathrm{scf}_{\pm}\mathrm{G}_{22} = 0, \tag{Ж.44}$$

$$\mathrm{G}_{12}^{T}\cdot\mathrm{G}_{11}^{M} + \mathrm{G}_{22}^{T}\cdot\mathrm{G}_{21}^{M} \;=\; 0, \tag{Ж.45}$$

$$\begin{aligned} \mathrm{G}_{11}^{T\prime}\cdot\mathrm{G}_{11}^{M} + \mathrm{G}_{21}^{T\prime}\cdot\mathrm{G}_{21}^{M} \;&=\; \mathrm{G}_{11}^{T}\cdot\mathrm{G}_{11}^{M\prime} + \mathrm{G}_{21}^{T}\cdot\mathrm{G}_{21}^{M\prime} = \\ \mathrm{G}_{12}^{T\prime}\cdot\mathrm{G}_{12}^{M} + \mathrm{G}_{22}^{T\prime}\cdot\mathrm{G}_{22}^{M} \;&=\; \mathrm{G}_{12}^{T}\cdot\mathrm{G}_{12}^{M\prime} + \mathrm{G}_{22}^{T}\cdot\mathrm{G}_{22}^{M\prime}, \end{aligned} \tag{Ж.46}$$

где 4 матрицы $\mathrm{G}_{ij}$ определяются*) аналогично (3.17)

$$\mathrm{G}_{ij} = \begin{pmatrix} g_{ij}^{+-}(z) & g_{ij}^{++}(z) \\ g_{ij}^{--}(z) & g_{ij}^{-+}(z) \end{pmatrix}, \qquad (Ж.47)$$

$$\mathrm{scf}_{\pm}\mathrm{G}_{ij} = g_{ij}^{+\mp}(z)\, g_{ij}^{-\mp}(z). \qquad (Ж.48)$$

Условия (Ж.43)–(Ж.46) свидетельствуют о том, что $4\times4$ матрица

$$\mathrm{G} = \begin{pmatrix} \mathrm{G}_{11} & \mathrm{G}_{12} \\ \mathrm{G}_{21} & \mathrm{G}_{22} \end{pmatrix} \qquad (Ж.49)$$

является $N = 4$ scf-матрицей, т. е. $\mathrm{G} \in SCF_{\Lambda_0}(4)$ (см. **Пункт 5.1**), удовлетворяющей дополнительному дифференциальному ограничению (Ж.46) (ср. [563]).

*Замечание* **Ж.9.** Из дифференциального условия (Ж.46) после интегрирования следует линейное

$$\mathrm{G}_{11}^{T}\cdot\mathrm{G}_{11}^{M} + \mathrm{G}_{21}^{T}\cdot\mathrm{G}_{21}^{M} = \mathrm{G}_{12}^{T}\cdot\mathrm{G}_{12}^{M} + \mathrm{G}_{22}^{T}\cdot\mathrm{G}_{22}^{M} + c, \qquad (Ж.50)$$

где $c$ — четная константа.

Отметим также соотношение между G и $\mathrm{H}_{SCf}$ матрицами

$$\mathrm{G} = \mathrm{H}_{SCf}^{T}\big|_{\theta_i^{\pm}=0} \qquad (Ж.51)$$

(см. (3.165)).

Из уравнения (Ж.43) видно, что именно матрица G определяет тип и обратимость $N = 4$ расщепленных преобразований. Так, расщеп-

---

*Примечание.* Штрих означает матрицу, каждый элемент которой продифференцирован по $z$.

ленные $N = 4$ преобразования образуют подполугруппу общей полугруппы $N = 4$ супераналитических преобразований, которая характеризуется только элементами матрицы G (Ж.49).

Представление расщепленной $N = 4$ SCf полугруппы функциональными матрицами (см. для $N = 2$ **Определение 3.9**) будет сужением представления $N = 4$ супераналитической полугруппы функциональными матрицами, содержащими в качестве элементов все 80 функций (аналогичного (3.18)) на элементы только матрицы G, состоящие из 16 элементов (как в (Ж.25)).

**Определение Ж.10.** *Элемент* $\mathbf{s}$ *расщепленной* $N = 4$ *суперконформной полугруппы* $\mathbf{S}^{(N=4)}_{SCf(split)}$ *параметризуется функциональной* $4 \times 4$ *матрицей*

$$\left\{ \begin{array}{cc} \mathrm{G}_{11} & \mathrm{G}_{12} \\ \mathrm{G}_{21} & \mathrm{G}_{22} \end{array} \right\} |_{SCf} \stackrel{def}{=} \mathbf{s}_{split} \in \mathbf{S}^{(N=4)}_{SCf(split)}, \qquad (Ж.52)$$

*а действие*

$$\mathbf{s}^{(1)}_{split} *_s \mathbf{s}^{(2)}_{split} = \mathbf{s}^{(3)}_{split} \qquad (Ж.53)$$

*определяется композицией расщепленных преобразований* $Z \to \tilde{Z} \to \tilde{\tilde{Z}}$ *(аналогично, как в* $N = 2$ *(Ж.28)).*

*Замечание* **Ж.11.** Ассоциативность действия $*_s$ (Ж.53) следует из ассоциативности композиции $N = 4$ расщепленных преобразований.

Очевидно, что необратимые преобразования соответствуют идеалу полугруппы $\mathbf{I}^{(N=4)}_{SCf(split)} \trianglelefteq \mathbf{S}^{(N=4)}_{SCf(split)}$, а обратимые преобразования — ее подгруппе $\mathbf{G}^{(N=4)}_{SCf(split)} \subset \mathbf{S}^{(N=4)}_{SCf(split)}$.

Двусторонняя единица в полугруппе $\mathbf{S}_{SCf(split)}^{(N=4)}$ определяется как

$$\mathbf{e}_{split}=\begin{Bmatrix}1&0&0&0\\0&1&0&0\\0&0&1&0\\0&0&0&1\end{Bmatrix}, \tag{Ж.54}$$

а двусторонний нуль представляется нулевой матрицей в (Ж.52).

*Замечание* **Ж.12.** Функциональная матрица (Ж.52) совпадает с G только по внешнему виду*), поскольку умножение в расщепленной $N=4$ суперконформной полугруппе $\mathbf{S}_{SCf(split)}^{(N=4)}$ не связано с обычным умножением матриц G, а определяется композицией $N=4$ SCf преобразований (см. *Замечание* **3.10** и (Ж.28)).

В обратимом случае при $\epsilon\left[\det \mathrm{G}\right]\neq 0$ березиниан расщепленных $N=4$ SCf преобразований определяется формулой

$$\operatorname{Ber}_{split}^{N=4}\left(\tilde{Z}/Z\right)=\frac{f'\left(z\right)}{\det \mathrm{G}}, \tag{Ж.55}$$

которая следует непосредственно из (Ж.42).

Применение формул (3.220) и (Ж.43) дает

$$\begin{aligned}\det \mathrm{G}=k_{split}\left(\operatorname{per}\mathrm{G}_{11}+\operatorname{per}\mathrm{G}_{12}\right)^2 &=\\ k_{split}\left(\operatorname{per}\mathrm{G}_{21}+\operatorname{per}\mathrm{G}_{22}\right)^2 &= k_{split}\left[f'\left(z\right)\right]^2,\end{aligned} \tag{Ж.56}$$

откуда

$$\operatorname{Ber}_{split}^{N=4}\left(\tilde{Z}/Z\right)=\frac{k_{split}}{f'\left(z\right)}, \tag{Ж.57}$$

где $k_{split}=+1$ для $SO_{\Lambda_0}\left(4\right)$ преобразований и $k_{split}=-1$ для общих

---

*Примечание.* Фигурные скобки призваны их отличить.

$O_{\Lambda_0}(4)$ преобразований$^{*)}$.

Рассмотрим более подробно различные типы расщепленных обратимых преобразований. Для этого отметим, что SCf условия (Ж.44)–(Ж.46) налагают жесткие ограничения на вид обратимых матриц $\mathrm{G}_{ij}$: они могут быть либо диагональными ($k_{ij} = +1$, $U(1)$-матрица), либо антидиагональными ($k_{ij} = -1$, $O(2)$-матрица), т. е. $\mathrm{G}_{ij} = \mathrm{G}_{ij}^{k_{ij}} = \mathrm{G}_{ij}^{\pm} \in GSCF_{\Lambda_0}(2)$ (см. **Пункт 5.1**).

Сначала рассмотрим случай, матрица G (Ж.47) является или блочно диагональной $\mathrm{G}_{12} = \mathrm{G}_{21} = 0$ ($k_G = +1$), или блочно антидиагональной $\mathrm{G}_{11} = \mathrm{G}_{22} = 0$ ($k_G = -1$). Тогда условия (Ж.44)–(Ж.46) выполняются тождественно, и расщепленные $N = 4$ SCf преобразования $\mathcal{T}_{SCf}^{N=4}$ характеризуются 3 индексами, т. е. $\mathcal{T}_{SCf}^{N=4} = \mathcal{T}_{k_1,k_2}^{k_G} = \mathcal{T}_{\pm\pm}^{\pm}$, где $k_1, k_2 = k_{11}, k_{22}$ при $k_G = +1$ и $k_1, k_2 = k_{12}, k_{21}$ при $k_G = -1$. Отсюда получаем общую формулу для типа преобразования через типы составляющих (всего 8 вариантов)

$$k_{split} = k_G k_1 k_2. \tag{Ж.58}$$

В том случае, когда все 4 матрицы $\mathrm{G}_{ij}$ отличны от нуля, по аналогии с (Ж.58) имеем $\mathcal{T}_{SCf}^{N=4} = \mathcal{T}_{k_{21}k_{22}}^{k_{11}k_{12}}$ (16 вариантов) и

$$k_{split} = k_{11} k_{12} k_{21} k_{22}. \tag{Ж.59}$$

*Замечание* **Ж.13.** Поскольку для $U(1)$ матриц $\mathrm{G}_{ij}$ $k_{ij} = +1$, то

$$k_{split} = (-1)^{n_{O(2)}}, \tag{Ж.60}$$

---

*Примечание.* Другими словами, нормированная матрица G является $SO_{\Lambda_0}(4)$ (при $k_{split} = +1$) или $O_{\Lambda_0}(4)$ (при $k_{split} = -1$) матрицей.

где $n_{O(2)}$ — число $O(2)$ матриц в G.

При композиции расщепленных преобразований (Ж.53) $k_{split}^{(3)} = k_{split}^{(1)} k_{split}^{(2)}$. Отсюда следует очевидное

*Замечание* **Ж.14.** Подгруппу в $O(4)$ составляют только те преобразования, для которых $k_{split} = +1$.

Покажем, каким образом в нашем формализме возникает подгруппа глобальных вращений $SU_{global}(2) \cong SO_{global}(4)$ [668]. Для этого рассмотрим, например, преобразование $\mathcal{T}_{++}^{+}$

$$\tilde{z} = f(z), \quad \tilde{\theta}_1^{\pm} = \theta_1^{\pm} g_{11}^{\pm\mp}(z), \quad \tilde{\theta}_2^{\pm} = \theta_2^{\pm} g_{22}^{\pm\mp}(z), \tag{Ж.61}$$

где $f'(z) = g_{11}^{+-}(z)\, g_{11}^{-+}(z)$ (см. (Ж.42)).

Перепараметризуем функции $g_{ij}^{\pm\mp}(z)$ по формулам*)

$$g_{11}^{\pm\mp}(z) = u_1(z)\, e^{\pm q_1(z)}, \tag{Ж.62}$$

$$g_{22}^{\pm\mp}(z) = u_2(z)\, e^{\pm q_2(z)}. \tag{Ж.63}$$

Тогда из уравнений (Ж.46) следует соотношения

$$g_{11}^{+-}(z)\, g_{11}^{-+\prime}(z) = g_{11}^{+-\prime}(z)\, g_{11}^{-+}(z) \Rightarrow u_1^2(z)\, q_1'(z) = 0, \tag{Ж.64}$$

$$g_{22}^{+-}(z)\, g_{22}^{-+\prime}(z) = g_{22}^{+-\prime}(z)\, g_{22}^{-+}(z) \Rightarrow u_2^2(z)\, q_2'(z) = 0. \tag{Ж.65}$$

В силу обратимости $\epsilon\left[g_{ij}^{\pm\mp}(z)\right] \neq 0$, и, следовательно $\epsilon\left[u_i(z)\right] \neq 0$, поэтому $u_i^2(z) \neq 0$, значит $q_i'(z) = 0$ и $q_i(z) = q_i = const$, и получаем

$$\tilde{z} = f(z), \quad \tilde{\theta}_1^{\pm} = \hat{\theta}_1^{\pm} u_1(z), \quad \tilde{\theta}_2^{\pm} = \hat{\theta}_2^{\pm} u_2(z), \tag{Ж.66}$$

---

*Примечание.* В силу того, что в обратимом случае $\epsilon\left[g_{ij}^{\pm\mp}(z)\right] \neq 0$, это возможно.

где $f'(z) = u_1^2(z) = u_2^2(z)$ и, следовательно, в нечетном секторе имеются искомые глобальные вращения

$$\hat{\theta}_1^{\pm} = \theta_1^{\pm} e^{\pm q_1}, \quad \hat{\theta}_2^{\pm} = \theta_2^{\pm} e^{\pm q_2}. \tag{Ж.67}$$

В случае преобразований $\mathcal{T}_{++}^{+}$ имеем дальнейшие упрощения. Из (Ж.46) и (Ж.50) можно получить также, что $u_1^2(z) = u_2^2(z) = f'(z)$, т. е. $u_1(z) = ku_2(z) = u(z)$, где $k = \pm 1$. Окончательно

$$\tilde{z} = f(z), \quad \tilde{\theta}_1^{\pm} = \hat{\theta}_1^{\pm} u(z), \quad \tilde{\theta}_2^{\pm} = k\hat{\theta}_2^{\pm} u(z), \tag{Ж.68}$$

где $f'(z) = u^2(z)$.

Более нетривиальными являются преобразования $\mathcal{T}_{k_{21}k_{22}}^{k_{11}k_{12}}$ (Ж.59), в которых все 4 матрицы $\mathrm{G}_{ij}$ отличны от нуля. Рассмотрим подробнее преобразования $\mathcal{T}_{++}^{++}$, для которых все $\mathrm{G}_{ij}$ диагональны и все $k_{ij} = +1$. Снова параметризуем ненулевые $g_{ij}^{\pm\mp}(z)$, как в (Ж.62)–(Ж.63), тогда SCf условия (Ж.44)–(Ж.46) приводят к следующим $N = 4$ расщепленным преобразованиям [2]

$$\tilde{z} = f(z), \tag{Ж.69}$$

$$\tilde{\theta}_1^{\pm} = \hat{\theta}_1^{\pm} u_1^{\pm}(z) + \hat{\theta}_2^{\pm} u_2^{\pm}(z), \tag{Ж.70}$$

$$\tilde{\theta}_2^{\pm} = -\hat{\theta}_1^{\pm} u_2^{\mp}(z) + \hat{\theta}_2^{\pm} u_1^{\mp}(z), \tag{Ж.71}$$

где $\hat{\theta}_i^{\pm}$ определены в (Ж.67), $f'(z) = u_1^{+}(z)\, u_1^{-}(z) + u_2^{+}(z)\, u_2^{-}(z)$ и

$$u_1^{+}(z)\, u_1^{-\prime}(z) + u_2^{+}(z)\, u_2^{-\prime}(z) = u_1^{+\prime}(z)\, u_1^{-}(z) + u_2^{+\prime}(z)\, u_2^{-}(z). \tag{Ж.72}$$

*Замечание* **Ж.15.** Очевидно, что преобразования с $u_2^{\pm}(z) = 0$ образуют подгруппу расщепленных преобразований (Ж.69)–(Ж.71).

Локальные $SU_{\Lambda_0}(2)$ вращения возникают, если перепараметризовать $u_i^{\pm}(z)$ в виде

$$u_1^{\pm}(z) = v_1(z)\, e^{\pm p_1(z)}, \quad (Ж.73)$$

$$u_2^{\pm}(z) = v_2(z)\, e^{\pm p_2(z)}, \quad (Ж.74)$$

где функции $v_i(z)$ и $p_i(z)$ удовлетворяют соотношениям

$$v_1^2(z) + v_2^2(z) = f'(z), \quad (Ж.75)$$

$$p_1'(z)\, v_1^2(z) = p_2'(z)\, v_2^2(z). \quad (Ж.76)$$

Наиболее симметричным*) решением является

$$v_1(z) = v_2(z) = \sqrt{\frac{f'(z)}{2}}, \quad (Ж.77)$$

$$p_1(z) = p_2(z). \quad (Ж.78)$$

Отметим, что среди обратимых расщепленных преобразований, удовлетворяющих (Ж.44)–(Ж.46), имеется дополнительный промежуточный тип, который описывается двумя обратимыми матрицами $\mathrm{G}_{ij}$ и двумя необратимыми [2]. Действительно, пусть scf-матрицы $\mathrm{G}_{11}, \mathrm{G}_{22} \in GSCF_{\Lambda_0}(2)$ и $\mathrm{G}_{12}, \mathrm{G}_{21} \in ISCF_{\Lambda_0}(2)$ (см. **Пункт 5.1**), т. е. $\mathrm{per}\, \mathrm{G}_{ij} \neq 0$, $\epsilon[\mathrm{per}\, \mathrm{G}_{11}] \neq 0$, $\epsilon[\mathrm{per}\, \mathrm{G}_{22}] \neq 0$, $\epsilon[\mathrm{per}\, \mathrm{G}_{12}] = 0$, $\epsilon[\mathrm{per}\, \mathrm{G}_{21}] = 0$. Матрицы $\mathrm{G}_{11}$ и $\mathrm{G}_{22}$ возьмем диагональными $k_{11} = k_{22} = +1$ (см. (Ж.59)). Параметризуем scf-матрицы $\mathrm{G}_{12}$ и $\mathrm{G}_{21}$ нечетными функциями $\mu_i^{\pm}(z)$ следу-

---

*Примечание.* Хотя и не единственным даже из обратимых.

ющим образом (см. *Пример* **Ж.6**)

$$\mathrm{G}_{ij} = \begin{pmatrix} \mu_i^+(z)\,\mu_j^-(z) & \mu_i^+(z)\,\mu_j^+(z) \\ \mu_i^-(z)\,\mu_j^-(z) & -\mu_i^-(z)\,\mu_j^+(z) \end{pmatrix}, \qquad (\text{Ж}.79)$$

где

$$\mathrm{per}\,\mathrm{G}_{ij} = 2\rho(z) \neq 0 \quad (i \neq j), \qquad (\text{Ж}.80)$$

$$\rho(z) = \mu_1^+(z)\,\mu_1^+(z)\,\mu_2^-(z)\,\mu_2^-(z). \qquad (\text{Ж}.81)$$

Тогда SCf условия (Ж.44) выполняются за счет нильпотентности нечетных функций $\mu_i^{\pm}(z)$, условие (Ж.45) приводит к $g_{11}^{\pm\mp}(z) = g_{22}^{\mp\pm}(z)$, а остальные дают систему уравнений для функций $\mu_i^{\pm}(z)$, которая может быть решена многими способами (также из-за нильпотентности $\mu_i^{\pm}(z)$). Приведем один пример из этой серии решений при $\mu_1^{\pm}(z) = \mu_2^{\pm}(z)$

$$\tilde{z} = f(z), \qquad (\text{Ж}.82)$$

$$\tilde{\theta}_1^{\pm} = \hat{\theta}_1^{\pm} u(z) \pm \hat{\theta}_2^{\pm} \mu_1^{\pm}(z)\,\mu_1^{\mp}(z), \qquad (\text{Ж}.83)$$

$$\tilde{\theta}_2^{\pm} = \hat{\theta}_2^{\pm} u(z) \mp \hat{\theta}_1^{\pm} \mu_1^{\mp}(z)\,\mu_1^{\pm}(z), \qquad (\text{Ж}.84)$$

где $f'(z) = u(z)$.

В необратимом случае $\epsilon\,[\mathrm{per}\,\mathrm{G}] = 0$ матрицы $\mathrm{G}_{ij} \in ISCF_{\Lambda_0}(2)$ не обязательно должны быть диагональными. Если параметризовать их 4 нечетными функциями $\mu_i^{\pm}(z)$ и выбрать $\mathrm{G}_{12}$ и $\mathrm{G}_{21}$ подобно (Ж.79), а для остальных матриц выбрать $\mathrm{G}_{11} = \mathrm{G}_{12}^{TM}$, $\mathrm{G}_{22} = \mathrm{G}_{21}^{TM}$, то получаем необратимые расщепленные преобразования, принадлежащие иде-алу $N=4$ SCf полугруппы (см. **Определение 3.58**)

$$\begin{aligned}\tilde{z} &= f(z), && (Ж.85)\\ \tilde{\theta}_1^{\pm} &= \mp\hat{\theta}_1^{\pm}\mu_1^{\pm}(z)\,\mu_2^{\mp}(z) \pm \hat{\theta}_2^{\pm}\mu_1^{\mp}(z)\,\mu_2^{\pm}(z) + \\ &\quad \hat{\theta}_1^{\mp}\mu_1^{\pm}(z)\,\mu_2^{\pm}(z) + \hat{\theta}_2^{\mp}\mu_1^{\pm}(z)\,\mu_2^{\pm}(z), && (Ж.86)\\ \tilde{\theta}_2^{\pm} &= \mp\hat{\theta}_2^{\pm}\mu_2^{\pm}(z)\,\mu_1^{\mp}(z) \pm \hat{\theta}_1^{\pm}\mu_2^{\mp}(z)\,\mu_1^{\pm}(z) + \\ &\quad \hat{\theta}_2^{\mp}\mu_2^{\pm}(z)\,\mu_1^{\pm}(z) + \hat{\theta}_1^{\mp}\mu_2^{\pm}(z)\,\mu_1^{\pm}(z), && (Ж.87)\end{aligned}$$

где $f'(z)=4\rho(z)$ (см. (Ж.81)).

## Ж.5. Дробно-линейные $N=4$ преобразований и полуматрицы

В обратимом случае дробно-линейные $N=4$ преобразования можно получить непосредственно из (3.250)–(3.252), если выбрать в качестве компонентных функций дробно-линейные

$$f(z)=\frac{az+b}{cz+d},\quad \psi_i^{\pm}=\frac{\gamma_i^{\pm}z+\delta_i^{\pm}}{cz+d}, \qquad (Ж.88)$$

где $a,b,c,d\in\Lambda_0$ — четные константы, $\delta_i^{\pm}\in\Lambda_1$ — нечетные.

Тогда получаем

$$\begin{aligned}\tilde{z} &= \frac{az+b}{cz+d}+\frac{Y\hat{\theta}_1^{+}}{(cZ^{+}+d)^2}\left[Z^{+}\cdot\delta\mathrm{et}\mathcal{Y}^{--}+\delta\mathrm{et}\mathcal{X}^{--}\right]+\\ &\quad \frac{Y\hat{\theta}_1^{-}}{(cZ^{-}+d)^2}\left[Z^{-}\cdot\delta\mathrm{et}\mathcal{Y}^{++}+\delta\mathrm{et}\mathcal{X}^{++}\right]+\\ &\quad \frac{Y\hat{\theta}_2^{+}}{(cZ^{+}+d)^2}\left[Z^{+}\cdot\pi\mathrm{er}\mathcal{Y}^{-+}+\pi\mathrm{er}\mathcal{X}^{-+}\right]+\end{aligned}$$

$$\frac{Y\hat{\theta}_2^-}{(cZ^- + d)^2}\left[Z^- \cdot \pi\mathrm{er}\mathcal{Y}^{+-} + \pi\mathrm{er}\mathcal{X}^{+-}\right] +$$
$$\left(\hat{\theta}_1^+\hat{\theta}_1^- + \hat{\theta}_2^+\hat{\theta}_2^-\right) \cdot \frac{(d - cz)\left(\det\Gamma_{11}^{+-} + \det\Gamma_{22}^{+-}\right) + 2d\mathrm{per}\,\Omega - 2c\mathrm{per}\,\Upsilon}{(cz + d)^3} +$$
$$2\hat{\theta}_1^+\hat{\theta}_2^+ \cdot \frac{(d - cz)\det\Gamma_{21}^{--} - 2c\delta_2^-\delta_1^-}{(cz + d)^3} +$$
$$2\hat{\theta}_1^-\hat{\theta}_2^- \cdot \frac{(d - cz)\det\Gamma_{21}^{++} - 2c\delta_2^+\delta_1^+}{(cz + d)^3} - \hat{\theta}_1^+\hat{\theta}_1^-\hat{\theta}_2^+\hat{\theta}_2^-\frac{2c}{(cz + d)^3}, \qquad (Ж.89)$$
$$\tilde{\theta}_1^\pm = \frac{\gamma_1^\pm z + \delta_1^\pm}{cz + d} + \hat{\theta}_1^\pm\frac{Ye^{\pm p_2}}{cZ^\pm + d} + \hat{\theta}_2^\pm\frac{Ye^{\pm p_1}}{cZ^\pm + d} + 2\hat{\theta}_1^\pm\hat{\theta}_2^\pm\frac{\delta\mathrm{et}\mathcal{V}_2^\mp}{(cz + d)^2}, \qquad (Ж.90)$$
$$\tilde{\theta}_2^\pm = \frac{\gamma_2^\pm z + \delta_2^\pm}{cz + d} - \hat{\theta}_1^\pm\frac{Ye^{\mp p_1}}{cZ^\pm + d} + \hat{\theta}_2^\pm\frac{Ye^{\mp p_2}}{cZ^\pm + d} + 2\hat{\theta}_2^\pm\hat{\theta}_1^\pm\frac{\delta\mathrm{et}\mathcal{V}_1^\mp}{(cz + d)^2}, \qquad (Ж.91)$$

где $Y = \det\begin{pmatrix} a & b \\ c & d \end{pmatrix}$, $\Gamma_{ij}^{ab}, \Upsilon, \Omega$ — матрицы со всеми нечетными элементами

$$\Gamma_{ij}^{ab} = \begin{pmatrix} \delta_i^a & \delta_j^b \\ \gamma_i^a & \gamma_j^b \end{pmatrix}, \ \Upsilon = \begin{pmatrix} \delta_1^+ & \delta_2^+ \\ \delta_2^- & \delta_1^- \end{pmatrix}, \ \Omega = \begin{pmatrix} \gamma_1^+ & \gamma_2^+ \\ \gamma_2^- & \gamma_1^- \end{pmatrix}, \qquad (Ж.92)$$

и горизонтальные полуматрицы (см. **Пункт Д.2**) $\mathcal{Y}^{ab}, \mathcal{X}^{ab}$ и $\mathcal{V}_i^a$ определяются

$$\mathcal{Y}^{ab} = \begin{pmatrix} \gamma_1^a & \gamma_2^a \\ e^{bp_1} & e^{-bp_2} \end{pmatrix}, \ \mathcal{X}^{ab} = \begin{pmatrix} \delta_1^a & \delta_2^a \\ e^{-bp_1} & e^{bp_2} \end{pmatrix}, \ \mathcal{V}_i^a = \begin{pmatrix} \gamma_i^a & \delta_i^a \\ c & d \end{pmatrix} \qquad (Ж.93)$$

где $\hat{\theta}_i^\pm$ определены в (Ж.67), $a, b = \pm$ и $p_1, p_2$ отвечают различным $SU(2, \Lambda_0)$ вращениям.

*Замечание* **Ж.16.** Матрицы $Y, \Gamma_{ij}^{ab}, \Upsilon, \Omega$ и полуматрицы $\mathcal{V}_i^a$ представляют собой миноры и полуминоры (см. **Пункт Д.2**) соответствующих матриц и полуматриц в суперматрице дробно-линейных $N = 4$ SCf пре-образований в однородных координатах, аналогично $N = 1$ в (2.205).

Отметим, что в необратимом случае решение уравнения (3.258) для дробно-линейных функций $\psi_i^{\pm}$ (Ж.88) имеет вид

$$f(z) = \frac{\pi \mathrm{er} \Gamma_{11}^{+-} + \pi \mathrm{er} \Gamma_{22}^{+-}}{c(cz+d)}. \tag{Ж.94}$$

# Приложение З

# Сплетающие четность преобразования, нечетные коциклы и деформации

Теория деформаций супермногообразий [746, 753, 815] с одной стороны представляет собой необходимую составляющую анализа суперструн и суперконформных теорий поля в терминах суперримановых поверхностей [111, 816], а с другой стороны интересна и с математической точки зрения [346, 357, 624, 817] как суперобобщение соответствующей теории для обычных комплексных многообразий [818–821].

Здесь мы рассмотрим некоторые особенности координатного описания и деформаций полусупермногообразий (см. **Раздел 1**), возникающие при учете сплетающих четность преобразований (см. **Подраздел 2.3**). Проследим подробно, каким образом возникают новые типы условий согласованности и коциклов.

## З.1. Смешанные условия согласованности и нечетные аналоги коциклов

Пусть имеется $(1|0)$-мерное комплексное полусупермногообразие $\mathscr{M}$ (в смысле **Определения 1.3**), представленное в виде полуатласа $\mathscr{M} = \bigcup\limits_{\alpha} \{\mathscr{U}_\alpha\}$ с локальными координатами $z_\alpha$. Тогда основные формулы и теоремы будут повторять соответствующие формулы несуперсимметричного случая [819]. Единственное добавление состоит в учете наряду с обратимыми полунеобратимых преобразований (2.4) в качестве функций перехода $z_\alpha = f_{\alpha\beta}(z_\beta)$ с ненулевым, но необратимым нильпотентным якобианом $J_{\alpha\beta} = \partial z_\alpha / \partial z_\beta$, т. е. $J_{\alpha\beta} \neq 0$, но $\epsilon[J_{\alpha\beta}] = 0$. Этот случай является промежуточным между стандартным обратимым, когда $J_{\alpha\beta} \neq 0$, и предельным необратимым, когда $J_{\alpha\beta} = 0$.

*Замечание* **З.1.** Преобразования с нулевым якобианом рассматривались в [812] для комплексных афинных пространств, а также для векторных пространств [382, 383] при исследовании контракций различных алгебраических структур [381, 822, 823].

На пересечении трех суперобластей $\mathscr{U}_\alpha \cap \mathscr{U}_\beta \cap \mathscr{U}_\gamma$ для поледовательных переходов $z_\gamma \to z_\beta \to z_\alpha$ имеем условие согласования

$$f_{\alpha\gamma} = f_{\alpha\beta} \circ f_{\beta\gamma} \tag{З.1}$$

или в локальных координатах $f_{\alpha\gamma}(z_\gamma) = f_{\alpha\beta}(f_{\beta\gamma}(z_\gamma))$.

При этом соответствующие якобианы преобразуются мультипликативно (с поточечным умножением)

$$J_{\alpha\gamma} = J_{\alpha\beta} \cdot J_{\beta\gamma}, \tag{З.2}$$

что отвечает касательному расслоению на $\mathscr{M}$ [260, 819, 824].

В случае $(1|1)$-мерного полусупермногообразия с локальными координатами $Z_\alpha = (z_\alpha, \theta_\alpha)$ роль якобиана в обратимом суперконформном случае играет березиниан перехода $Z_\beta \to Z_\alpha$ (см. **Приложение Е.1**). Однако для выполнения условия коцикличности, аналогичного (З.2), необходимо рассматривать редуцированные преобразования (см. **Пункт 2.1.3**). Здесь мы покажем, что при ослаблении обратимости возникает не один вариант суперобобщения условия коцикличности (З.2) [405,491], а два [9] в соответствие с двумя типами редуцированных преобразований [7, 13].

Для этого запишем общее преобразование $(1|1)$-мерного касатель-ного вектора $T\mathscr{M}|_\beta \to T\mathscr{M}|_\alpha$ в матричном виде (см. 2.24)

$$\begin{pmatrix} \partial_\beta \\ D_\beta \end{pmatrix} = \mathrm{P}^{SA}_{\alpha\beta} \cdot \begin{pmatrix} \partial_\alpha \\ D_\alpha \end{pmatrix}, \tag{3.3}$$

$$\mathrm{P}^{SA}_{\alpha\beta} = \begin{pmatrix} Q_{\alpha\beta} & \partial_\beta\theta_\alpha \\ \varDelta_{\alpha\beta} & D_\beta\theta_\alpha \end{pmatrix}, \tag{3.4}$$

$$Q_{\alpha\beta} = \partial_\beta z_\alpha - \partial_\beta\theta_\alpha \cdot \theta_\alpha, \tag{3.5}$$

$$\varDelta_{\alpha\beta} = D_\beta z_\alpha - D_\beta\theta_\alpha \cdot \theta_\alpha, \tag{3.6}$$

где $D_\alpha = \partial/\partial\theta_\alpha + \theta_\alpha\partial_\alpha, \partial_\alpha = \partial/\partial z_\alpha$ (нет суммирования).

При двух последовательных преобразованиях $Z_\gamma \to Z_\beta \to Z_\alpha$ на $\mathscr{U}_\alpha \cap \mathscr{U}_\beta \cap \mathscr{U}_\gamma$ для суперматриц $\mathrm{P}^{SA}_{\alpha\beta}$ из (3.3) имеем условие коцикличности, аналогичное (3.2)

$$\mathrm{P}^{SA}_{\alpha\gamma} = \mathrm{P}^{SA}_{\beta\gamma} \cdot \mathrm{P}^{SA}_{\alpha\beta}. \tag{3.7}$$

Отсюда следуют выражения для нечетной и четной производных конечной нечетной координаты

$$D_\gamma\theta_\alpha = D_\gamma\theta_\beta \cdot D_\beta\theta_\alpha + \varDelta_{\beta\gamma} \cdot \partial_\beta\theta_\alpha, \tag{3.8}$$

$$\partial_\gamma\theta_\alpha = \partial_\gamma\theta_\beta \cdot D_\beta\theta_\alpha + Q_{\beta\gamma} \cdot \partial_\beta\theta_\alpha. \tag{3.9}$$

Легко видеть, что зануление вторых слагаемых в (3.9)–(3.9)

$$\varDelta_{\beta\gamma} = 0, \quad \text{(SCf)} \tag{3.10}$$

$$Q_{\beta\gamma} = 0, \quad \text{(TPt)} \tag{3.11}$$

приводит к двум (!), а не к одному, как в стандартном случае [491], условиям коцикла [7,9] и соответствующим двум редукциям суперматрицы

$\mathrm{P}^{SA}_{\alpha\beta}$ (см. **Пункт 2.1.3**).

Уравнения (3.10)–(3.11) снова, как и (2.37)–(2.38), определяют суперконформные (SCf) и сплетающие четность (TPt) преобразования соответственно (см. **Пункт 2.1.3** и **Подраздел 2.3**). Тогда вместо одного условия коцикличности для суперматриц $\mathrm{P}^{SA}_{\alpha\beta}$ (3.7) имеем два условия

$$\mathrm{P}^{SCf}_{\alpha\gamma} = \mathrm{P}^{SCf}_{\beta\gamma} \cdot \mathrm{P}^{SCf}_{\alpha\beta}, \tag{3.12}$$

$$\mathrm{P}^{TPt}_{\alpha\gamma} = \mathrm{P}^{TPt}_{\beta\gamma} \cdot \mathrm{P}^{SCf}_{\alpha\beta}. \tag{3.13}$$

для редуцированных различным образом суперматриц

$$\mathrm{P}^{SCf}_{\alpha\beta} = \begin{pmatrix} Q^{SCf}_{\alpha\beta} & \partial_\beta \theta^{SCf}_\alpha \\ 0 & D_\beta \theta^{SCf}_\alpha \end{pmatrix}, \tag{3.14}$$

$$\mathrm{P}^{TPt}_{\alpha\beta} = \begin{pmatrix} 0 & \partial_\beta \theta^{TPt}_\alpha \\ \mathit{\Delta}^{TPt}_{\alpha\beta} & D_\beta \theta^{TPt}_\alpha \end{pmatrix}, \tag{3.15}$$

где $Q^{SCf}_{\alpha\beta} \stackrel{def}{=} Q_{\alpha\beta}|_{\mathit{\Delta}_{\alpha\beta}=0}$, $\mathit{\Delta}^{TPt}_{\alpha\beta} \stackrel{def}{=} \mathit{\Delta}_{\alpha\beta}|_{Q_{\alpha\beta}=0}$.

Таким образом, из (3.8)–(3.13) следует

**Утверждение 3.2.** *При ослаблении обратимости для условия коцикличности* (3.2) *имеется два возможных суперобобщения — четное и нечетное*

$$\boldsymbol{J}^{SCf}_{\alpha\gamma} = \boldsymbol{J}^{SCf}_{\beta\gamma} \cdot \boldsymbol{J}^{SCf}_{\alpha\beta} \tag{3.16}$$

$$\mathcal{J}^{TPt}_{\alpha\gamma} = \mathcal{J}^{TPt}_{\beta\gamma} \cdot \boldsymbol{J}^{SCf}_{\alpha\beta} \tag{3.17}$$

*где*

$$\boldsymbol{J}^{SCf}_{\alpha\beta} \stackrel{def}{=} D_\beta \theta^{SCf}_\alpha, \tag{3.18}$$

$$\mathcal{J}_{\alpha\beta}^{TPt} \stackrel{def}{=} \partial_\beta \theta_\alpha^{TPt}. \tag{3.19}$$

*Замечание* **3.3.** Из (3.19) следует, что $\mathcal{J}_{\alpha\beta}^{TPt}$ является нечетным и, следовательно, нильпотентным.

Отсюда естественно вытекает

**Определение 3.4.** *Назовем* $\boldsymbol{J}_{\alpha\beta}^{SCf}$ *и* $\mathcal{J}_{\alpha\beta}^{TPt}$ *четным и нечетным коциклом соответственно, а условие* (3.17) — *смешанным условием согласованности (условием коцикла).*

Все рассмотренные условия согласованности можно представить также в более наглядном виде, отражающем нетривиальную четно-нечетную симметрию между коциклами,

$$\partial_\gamma z_\alpha = \partial_\gamma z_\beta \cdot \partial_\beta z_\alpha \stackrel{SUSY}{\Longrightarrow} \begin{cases} D_\gamma \theta_\alpha^{SCf} = D_\gamma \theta_\beta^{SCf} \cdot D_\beta \theta_\alpha^{SCf}, & (\mathsf{SCf}) \\ \partial_\gamma \theta_\alpha^{TPt} = \partial_\gamma \theta_\beta^{TPt} \cdot D_\beta \theta_\alpha^{SCf}, & (\mathsf{TPt}) \end{cases} \tag{3.20}$$

где индексы $SCf$ и $TPt$ отвечают типу редуцированного преобразования $\mathcal{T}_{\alpha\beta}$ между соответствующими суперобластями $\mathcal{U}_\alpha$ и $\mathcal{U}_\beta$ (см. также *Замечание* **2.29**). Следовательно, в рамках категории редуцированных преобразований ( а не суперконформных) мы имеем две коммутативные диаграммы

$$\begin{array}{ccc} \mathcal{U}_\gamma & \xrightarrow{\mathcal{T}_{\beta\gamma}^{SCf}} & \mathcal{U}_\beta \\ & {\scriptstyle \mathcal{T}_{\alpha\gamma}^{SCf}} \searrow & \downarrow {\scriptstyle \mathcal{T}_{\alpha\beta}^{SCf}} \\ & & \mathcal{U}_\alpha \end{array} \qquad \begin{array}{ccc} \mathcal{U}_\gamma & \xrightarrow{\mathcal{T}_{\beta\gamma}^{TPt}} & \mathcal{U}_\beta \\ & {\scriptstyle \mathcal{T}_{\alpha\gamma}^{TPt}} \searrow & \downarrow {\scriptstyle \mathcal{T}_{\alpha\beta}^{SCf}} \\ & & \mathcal{U}_\alpha \end{array} \tag{3.21}$$

соответствующие условиям согласованности (3.16) и (3.17) (ср. (4.13)).

*Замечание* **3.5.** По терминологии [825] коциклы, удовлетворяющие соотношениям типа (3.2) и (3.16)–(3.17) называются склеивающими коциклами соответствующего расслоения (в данном случае касательного).

Существенным для суперструнных приложений (см., например, [327, 343, 362, 826]) фактом является

**Предложение 3.6.** *Четный коцикл* $\boldsymbol{J}_{\alpha\beta}^{SCf}$ (3.18) *совпадает с березинианом — четным супераналогом якобиана — суперконформного* (SCf) *преобразования* $Z_\beta \to Z_\alpha$

$$\boldsymbol{J}_{\alpha\beta}^{SCf} = \mathrm{Ber}\,\mathrm{P}_{\alpha\beta}^{SCf}. \tag{3.22}$$

*Доказательство.* Следует непосредственно из (3.14) и (2.52). ■

Это позволяет построить каноническое расслоение с функциями перехода (3.22), а также соответствующее линейное расслоение [183, 343,365,405]. Сопоставляя (3.1)–(3.2) и **Предложение 3.6**, можно придать похожий смысл также и нечетному коциклу*) (3.19).

**Предположение 3.7.** *Нечетный коцикл* $\mathcal{J}_{\alpha\beta}^{TPt}$ *можно трактовать как нечетный супераналог якобиана для сплетающих четность* (TPt) *преобразований* $Z_\beta \to Z_\alpha$ *(см.* **Определение 2.55***).*

*Замечание* **3.8.** Формула (3.17) может рассматриваться не только как условие коцикличности, но и как закон умножения четного и нечетного супераналогов якобинана.

Тогда соответствующие аналоги канонического и линейного расслоений будут обладать необычными свойствами, например, кручение четности и нильпотентность коциклов (см. подробнее **Пункт 2.3.2**).

---

*) *Примечание.* Введенные нечетные коциклы не связаны с $\mathbb{Z}_2$-градуированными коциклами, возникающими при суперсимметризации швингеровского слагаемого для нейтральной частицы [827, 828].

## 3.2. Деформации и TPt преобразования

Возникновение дополнительного условия согласования (3.13) и нечетного условия коцикличности (3.17) приводит к соответствующей модификации стандартных условий деформации в локальном подходе [357, 816, 817, 829]. Это, в свою очередь, играет важную роль в суперструнных вычислениях [324,348,362] для определения свойств пространства супермодулей [347, 353, 356, 404, 566] и формулировки суперобобщения фундаментальной теоремы Римана-Роха [342, 343, 346, 352, 830].

Здесь мы переформулируем стандартный подход, используя альтернативной параметризацию (см. **Пункт 2.1.5**), что позволит учесть также и нечетные условия коцикличности (3.13) и (3.17).

В несуперсимметричном случае [818–820] деформация условия согласованности (3.1)

$$z_\alpha = f_{\alpha\beta}(z_\beta) + t b_{\alpha\beta}(z_\beta) \tag{3.23}$$

приводит к тому же условию (3.1) для недеформинрованных функций $f_{\alpha\beta}(z_\beta)$ и к уравнению для деформаций $b_{\alpha\beta}(z_\beta)$

$$b_{\alpha\gamma}(z_\gamma) = b_{\alpha\beta}(f_{\beta\gamma}(z_\gamma)) + f'_{\alpha\beta}(f_{\beta\gamma}(z_\gamma)) \cdot b_{\beta\gamma}(z_\gamma). \tag{3.24}$$

Умножим это соотношение тензорно на $\partial/\partial z_\alpha$ и воспользуемся $f'_{\alpha\beta} = \partial z_\alpha/\partial z_\beta$, тогда получаем условие согласованности в виде

$$b_{\alpha\beta}\frac{\partial}{\partial z_\alpha} + b_{\beta\gamma}\frac{\partial}{\partial z_\beta} - b_{\alpha\gamma}\frac{\partial}{\partial z_\alpha} = 0, \tag{3.25}$$

которое показывает, что $\left\{b_{\alpha\beta}\dfrac{\partial}{\partial z_\alpha}\right\}$ действительно является коциклом. При инфинитезимальных преобразованиях $z_\alpha \longmapsto z_\alpha + t s_\alpha(z_\alpha)$ коцикл

(З.25) изменяется на кограницу

$$\left\{ b_{\alpha\beta}\frac{\partial}{\partial z_\alpha}\right\} \longmapsto \left\{ b_{\alpha\beta}\frac{\partial}{\partial z_\alpha} + s_\alpha \frac{\partial}{\partial z_\alpha} - s_\beta \frac{\partial}{\partial z_\beta}\right\}, \qquad (З.26)$$

что определяет когомологический класс (Кодайры-Спенсера) деформаций первого порядка [819].

В суперконформном случае [357,816] рассматриваются недеформированные расщепленные преобразования, имеющие в стандартной параметризации [341] вид

$$\mathsf{SCf}_{split}\colon \begin{cases} z_\alpha = f_{\alpha\beta}\left(z_\beta\right), \\ \theta_\alpha = \theta_\beta \cdot \sqrt{f'_{\alpha\beta}\left(z_\beta\right)}, \end{cases} \qquad (З.27)$$

которые не содержат никаких нечетных параметров, кроме $\theta_\alpha$. Поэтому расщепленные суперримановы поверхности, имеющие преобразования (З.27) в качестве функций склейки содержат ту же информацию, что и обычные римановы поверхности, наделенные спиновой структурой, которая определяется знаком квадратного корня [327, 330, 826, 831].

Теперь суперконформные деформации определяются двумя параметрами*), четным $t$ и нечетным $\tau$ [357,816] и двумя четными функциями $b_{\alpha\beta}\left(z_\beta\right)$ и $c_{\alpha\beta}\left(z_\beta\right)$ следующим образом

$$z_\alpha^{SCf}(t,\tau) = f_{\alpha\beta}\left(z_\beta\right) + tb_{\alpha\beta}\left(z_\beta\right) + \theta_\beta \cdot \tau c_{\alpha\beta}\left(z_\beta\right) \cdot F_{\alpha\beta}\left(z_\beta, t\right), \qquad (З.28)$$

$$\theta_\alpha^{SCf}(t,\tau) = \tau c_{\alpha\beta}\left(z_\beta\right) + \theta_\beta \cdot F_{\alpha\beta}\left(z_\beta, t\right), \qquad (З.29)$$

где $F_{\alpha\beta}\left(z_\beta, t\right) = \sqrt{f'_{\alpha\beta}\left(z_\beta\right) + tb'_{\alpha\beta}\left(z_\beta\right)}$.

Четное условие согласованности (см. первую диаграмму в (З.21))

---

*Примечание.* Точнее, (1|1)-суперпространством параметров $\mathbb{P}^{(1|1)}$.

на тройных пересечениях $\mathscr{U}_\alpha \cap \mathscr{U}_\beta \cap \mathscr{U}_\gamma$ записывается в виде

$$z_\alpha^{SCf}(z_\gamma,\theta_\gamma) = z_\alpha^{SCf}\left(z_\beta^{SCf}(z_\gamma,\theta_\gamma),\theta_\beta^{SCf}(z_\gamma,\theta_\gamma)\right), \quad (3.30)$$

$$\theta_\alpha^{SCf}(z_\gamma,\theta_\gamma) = \theta_\alpha^{SCf}\left(z_\beta^{SCf}(z_\gamma,\theta_\gamma),\theta_\beta^{SCf}(z_\gamma,\theta_\gamma)\right), \quad (3.31)$$

что в первом порядке по*) $t,\tau$ приводит к уравнениям (3.1) и (3.24) плюс дополнительное уравнение на функцию $c_{\alpha\beta}(z_\beta)$

$$c_{\alpha\gamma}(z_\gamma) = c_{\alpha\beta}(f_{\beta\gamma}(z_\gamma)) + c_{\beta\gamma}(z_\gamma)\cdot\sqrt{f'_{\alpha\beta}(f_{\beta\gamma}(z_\gamma))}. \quad (3.32)$$

Тензорное умножение на $\dfrac{\partial}{\partial z_\alpha}$ в дополнение к (3.26) и использование (3.27) дает

$$c_{\alpha\beta}\theta_\alpha\frac{\partial}{\partial z_\alpha} + c_{\beta\gamma}\theta_\beta\frac{\partial}{\partial z_\beta} - c_{\alpha\gamma}\theta_\alpha\frac{\partial}{\partial z_\alpha} = 0. \quad (3.33)$$

Уравнения (3.25) и (3.33) свидетельствуют о том, что чисто суперконформные деформации описываются двумя коциклами $\left\{b_{\alpha\beta}\dfrac{\partial}{\partial z_\alpha}\right\}$ и $\left\{c_{\alpha\beta}\theta_\alpha\dfrac{\partial}{\partial z_\alpha}\right\}$, которые при суперконформных репараметризациях

$$z_\alpha \overset{SCf}{\longmapsto} z_\alpha + ts_\alpha(z_\alpha) + \theta_\alpha\cdot\tau r_\alpha(z_\alpha)\cdot\sqrt{1+ts'_\alpha(z_\alpha)}, \quad (3.34)$$

$$\theta_\alpha \overset{SCf}{\longmapsto} \tau r_\alpha(z_\alpha) + \theta_\alpha\cdot\sqrt{1+ts'_\alpha(z_\alpha)}, \quad (3.35)$$

изменяются на кограницы (3.26) и

$$\left\{c_{\alpha\beta}\theta_\alpha\frac{\partial}{\partial z_\alpha}\right\} \longmapsto \left\{c_{\alpha\beta}\theta_\alpha\frac{\partial}{\partial z_\alpha} + r_\alpha\theta_\alpha\frac{\partial}{\partial z_\alpha} - r_\beta\theta_\beta\frac{\partial}{\partial z_\beta}\right\}, \quad (3.36)$$

---

*Примечание.* В (3.30)–(3.31) эти дополнительные аргументы опущены, но подразумеваются.

что определяет соответствующие когомологические классы [357, 511] и пространство супермодулей [566, 829].

Переформулируем теперь супердеформации таким образом, чтобы можно было учесть также и нечетные условия согласованности (3.17). Для этого воспользуемся альтернативной параметризацией (см. **Пункт 2.1.5**) и запишем редуцированные SCf и TPt преобразования на $\mathcal{U}_\alpha \cap \mathcal{U}_\beta$ в едином виде (см. (2.81))

$$z_\alpha = f_{\alpha\beta}(z_\beta) + \theta_\alpha \cdot \chi_{\alpha\beta}(z_\beta), \tag{3.37}$$

$$\theta_\alpha = \psi_{\alpha\beta}(z_\beta) + \theta_\alpha \cdot g_{\alpha\beta}(z_\beta), \tag{3.38}$$

где независимыми являются функции $g_{\alpha\beta}(z_\beta), \psi_{\alpha\beta}(z_\beta)$ (в отличие от стандартной параметризации функциями $f_{\alpha\beta}(z_\beta), \psi_{\alpha\beta}(z_\beta)$ [111, 566]), через которые выражаются отстальные по формулам

$$\mathsf{SCf} \;:\; \begin{cases} f_{\alpha\beta}^{SCf\,\prime}(z_\beta) = g_{\alpha\beta}^2(z_\beta) + \psi_{\alpha\beta}'(z_\beta) \cdot \psi_{\alpha\beta}(z_\beta), \\ \chi_{\alpha\beta}^{SCf}(z_\beta) = g_{\alpha\beta}(z_\beta) \cdot \psi_{\alpha\beta}(z_\beta), \end{cases} \tag{3.39}$$

$$\mathsf{TPt} \;:\; \begin{cases} f_{\alpha\beta}^{TPt\,\prime}(z_\beta) = \psi_{\alpha\beta}'(z_\beta) \cdot \psi_{\alpha\beta}(z_\beta), \\ \chi_{\alpha\beta}^{TPt\,\prime}(z_\beta) = g_{\alpha\beta}'(z_\beta) \cdot \psi_{\alpha\beta}(z_\beta) - g_{\alpha\beta}(z_\beta) \cdot \psi_{\alpha\beta}'(z_\beta). \end{cases} \tag{3.40}$$

Отсюда следует, расщепленное SCf преобразование в альтернативной параметризации (3.27) имеет вид

$$\mathsf{SCf}_{split} : \begin{cases} z_\alpha = \int g_{\alpha\beta}^2(z_\beta)\, dz_\beta, \\ \theta_\alpha = \theta_\beta \cdot g_{\alpha\beta}(z_\beta), \end{cases} \tag{3.41}$$

в то время как TPt аналогом (3.41) является вложение $2 \hookrightarrow 1$ [423], т. е.

$$\mathsf{TPt}_{split}\colon \begin{cases} z_\alpha = 0, \\ \theta_\alpha = \theta_\beta \cdot g_{\alpha\beta}(z_\beta)\,. \end{cases} \tag{3.42}$$

Теперь смешанные (в смысле **Определения 3.4**) как SCf, так и TPt деформации будут определяться теми же параметрами $t, \tau$, но уже парой четных функций $p_{\alpha\beta}, c_{\alpha\beta}$ (вместо $b_{\alpha\beta}, c_{\alpha\beta}$ в (3.28)–(3.29)) следующим образом

$$z_\alpha(t,\tau) = f_{\alpha\beta}(z_\beta, t, \tau) + \theta_\beta \cdot \chi_{\alpha\beta}(z_\beta, t, \tau)\,, \tag{3.43}$$

$$\theta_\alpha(t,\tau) = \tau c_{\alpha\beta}(z_\beta) + \theta_\beta \cdot (g_{\alpha\beta}(z_\beta) + t p_{\alpha\beta}(z_\beta))\,, \tag{3.44}$$

т. е. вместо $f_{\alpha\beta}(z_\beta)$ изначально деформируется $g_{\alpha\beta}(z_\beta)$, а остальные функции $f_{\alpha\beta}(z_\beta, t, \tau), \chi_{\alpha\beta}(z_\beta, t, \tau)$ находятся из соответствующих уравнений (2.81).

Теперь, с учетом (3.16)–(3.17) и диаграммы (3.21), наряду с четными (3.30)–(3.31) получаем нечетные условия согласованности для деформированных функций (дополнительные аргументы $t, \tau$ снова опущены)

$$z_\alpha^{TPt}(z_\gamma, \theta_\gamma) = z_\alpha^{SCf}\left(z_\beta^{TPt}(z_\gamma, \theta_\gamma), \theta_\beta^{TPt}(z_\gamma, \theta_\gamma)\right), \tag{3.45}$$

$$\theta_\alpha^{TPt}(z_\gamma, \theta_\gamma) = \theta_\alpha^{SCf}\left(z_\beta^{TPt}(z_\gamma, \theta_\gamma), \theta_\beta^{TPt}(z_\gamma, \theta_\gamma)\right). \tag{3.46}$$

Разложение этих уравнений по $t, \tau$, аналогичное четному случаю (3.30)–(3.31), дает

$$g_{\alpha\gamma}^{TPt}(z_\gamma) = g_{\alpha\beta}^{SCf}(z_\beta) \cdot g_{\beta\gamma}^{TPt}(z_\gamma)\,, \tag{3.47}$$

$$c_{\alpha\gamma}^{TPt}(z_\gamma) = c_{\alpha\beta}^{SCf}\left(f_{\beta\gamma}^{TPt}(z_\gamma)\right) + g_{\alpha\beta}^{SCf}(z_\beta)\cdot c_{\beta\gamma}^{TPt}(z_\gamma), \qquad (3.48)$$

$$p_{\alpha\gamma}^{TPt}(z_\gamma) = p_{\alpha\beta}^{SCf}\left(f_{\beta\gamma}^{TPt}(z_\gamma)\right)\cdot g_{\beta\gamma}^{TPt}(z_\gamma) + g_{\alpha\beta}^{SCf}(z_\beta)\cdot p_{\beta\gamma}^{TPt}(z_\gamma). \qquad (3.49)$$

Первое уравнение (3.47) является условием коцикличности для функций $g_{\alpha\beta}(z_\beta)$ и говорит о том, что эти функции реализуют соответствующий смешанный (несимметричный) аналог линейного расслоения над суперримановыми поверхностями [361, 365, 405]. Уравнение (3.48) аналогично уравнению (3.32), если учесть, что преобразование $z_\beta \to z_\alpha$ как для четного условия согласованности, так и для нечетного (3.45)–(3.46) является SCf преобразованием, в котором выполняется соотношение

$$g_{\alpha\beta}^{SCf\,2}(z_\beta) = f_{\alpha\beta}^{SCf\,\prime}(z_\beta) \qquad (3.50)$$

(см. также (3.27) и (2.81)).

В четном случае (когда все три преобразования $z_\gamma \to z_\beta \to z_\alpha$ являются SCf преобразованиями) из уравнения (3.49) при $\epsilon\left[g_{\alpha\beta}^{SCf}(z_\beta)\right] \neq 0$, если для всех трех переходов воспользоваться подстановкой

$$p_{\alpha\beta}^{SCf}(z_\beta) = \frac{b_{\alpha\beta}^{SCf\,\prime}(z_\beta)}{2g_{\alpha\beta}^{SCf}(z_\beta)}, \qquad (3.51)$$

после интегрирования можно получить

$$b_{\alpha\gamma}^{SCf}(z_\gamma) = b_{\alpha\beta}^{SCf}\left(f_{\beta\gamma}^{SCf}(z_\gamma)\right) + g_{\alpha\beta}^{SCf}\left(f_{\beta\gamma}^{SCf}(z_\gamma)\right)\cdot b_{\beta\gamma}^{SCf}(z_\gamma), \qquad (3.52)$$

что совпадает с (3.24) при учете (3.50).

Применяя полученные соотношения можно построить TPt-аналоги сперктральных последовательностей и соответствующих комплексов со сплетением четности по аналогии со стандартными SCf [269, 332, 511] (см. однако *Замечание* **2.62**).

## 3.3. Нечетные аналоги препятствий и смешанные $\theta$-коциклы

Препятствия [214–218, 220] играют важную роль в пониманиии внутренней структуры супермногообразий [30, 221] и суперконформных многообразий [108, 832].

Стандартное препятствие [214, 220] можно вычислить как отклонение левой части соответствующей формулы согласованности (например, (3.25), (3.33)) от нуля [824]. Для функций $b_{\alpha\beta}(z_\alpha)$ (3.25) и $c_{\alpha\beta}(z_\alpha)$ (3.33) имеем

$$\hat{\mathsf{D}}_{\alpha\beta\gamma}(b) = b_{\alpha\beta}\frac{\partial}{\partial z_\alpha} + b_{\beta\gamma}\frac{\partial}{\partial z_\beta} - b_{\alpha\gamma}\frac{\partial}{\partial z_\alpha}, \tag{3.53}$$

$$\hat{\mathsf{D}}_{\alpha\beta\gamma}(c) = c_{\alpha\beta}\theta_\alpha\frac{\partial}{\partial z_\alpha} + c_{\beta\gamma}\theta_\beta\frac{\partial}{\partial z_\beta} - c_{\alpha\gamma}\theta_\alpha\frac{\partial}{\partial z_\alpha}. \tag{3.54}$$

Например, в суперконформном случае для $b_{\alpha\beta}^{SCf}(z_\beta)$ (3.52) тогда получаем

$$\hat{\mathsf{D}}_{\alpha\beta\gamma}^{SCf}(b) = \left(g_{\alpha\beta}^{SCf\,2}(z_\alpha)\frac{\partial z_\beta}{\partial z_\alpha} - 1\right)\cdot b_{\beta\gamma}^{SCf}(z_\alpha)\frac{\partial}{\partial z_\beta}. \tag{3.55}$$

Отсюда следует

**Утверждение 3.9.** *Если преобразование $z_\beta \to z_\alpha$ является обратимым* SCf *преобразованием, то препятствие $\hat{\mathsf{D}}_{\alpha\beta\gamma}^{SCf}(b)$ равно нулю.*

*Доказательство.* Используем (3.50), тогда для выражения в скобках (3.55) имеем $g_{\alpha\beta}^{SCf\,2}(z_\alpha)\dfrac{\partial z_\beta}{\partial z_\alpha} = f_{\alpha\beta}^{SCf\,\prime}(z_\beta)\dfrac{\partial z_\beta}{\partial z_\alpha} = \dfrac{\partial z_\alpha}{\partial z_\beta}\dfrac{\partial z_\beta}{\partial z_\alpha} = 1$. ■

Рассмотрение редуцированных преобразований (SCf и TPt единым образом) в альтернативной параметризации приводит к возможности определения наряду с коциклами по четной переменной $z$ (например, (3.25) и (3.33)) также коциклов по нечетной переменной $\theta$.

**Определение 3.10.** *Назовем $\theta$-коциклом конструкцию, аналогичную четному коциклу, в которой тензорное умножение производится на нечетное векторное поле $\partial/\partial\theta_\alpha$ вместо $\partial/\partial z_\alpha$.*

Рассмотрим условия согласованности, связанные с деформациями $c_{\alpha\beta}(z_\alpha)$ и $p_{\alpha\beta}(z_\alpha)$ (3.48)–(3.49) в альтернативной параметризации, не конкретизируя вид редуцированного преобразования. Умножим тензорно уравнение (3.48) на $\partial/\partial\theta_\alpha$ и воспользуемся соотношением

$$\frac{\partial}{\partial\theta_\beta} = g_{\alpha\beta}(z_\beta)\frac{\partial}{\partial\theta_\alpha}, \tag{3.56}$$

которое следует из вторых уравнений в (3.41)–(3.42), тогда получим

$$c_{\alpha\gamma}\frac{\partial}{\partial\theta_\alpha} = c_{\alpha\beta}\frac{\partial}{\partial\theta_\alpha} + c_{\beta\gamma}\frac{\partial}{\partial\theta_\beta}. \tag{3.57}$$

**Утверждение 3.11.** $\left\{c_{\alpha\beta}\dfrac{\partial}{\partial\theta_\alpha}\right\}$ *является $\theta$-коциклом.*

*Доказательство.* Следует непосредственно из (3.57). ∎

Аналогично, умножив (3.49) на $\theta_\alpha\partial/\partial\theta_\alpha$, получаем

$$p_{\alpha\gamma}\theta_\alpha\frac{\partial}{\partial\theta_\alpha} = g_{\beta\gamma}\cdot p_{\alpha\beta}\theta_\alpha\frac{\partial}{\partial\theta_\alpha} + g_{\alpha\beta}\cdot p_{\beta\gamma}\theta_\beta\frac{\partial}{\partial\theta_\beta}. \tag{3.58}$$

*Замечание* **3.12.** $\left\{p_{\alpha\beta}\theta_\alpha\dfrac{\partial}{\partial\theta_\alpha}\right\}$ не является $\theta$-коциклом из-за подкручивающих множителей $g_{\beta\gamma}$ и $g_{\alpha\beta}$ в (3.58).

Для характеризации отличия набора функций на пересечениях $\mathscr{U}_\alpha\cap\mathscr{U}_\beta\cap\mathscr{U}_\gamma$ от $\theta$-коцикла, введем $\theta$-аналог препятствий (3.53)–(3.54).

**Определение 3.13.** *Назовем $\theta$-препятствием степень незамкнуто-*

*сти набора соответствующих функций (с нечетным векторным полем $\partial/\partial\theta_\alpha$) на пересечениях $\mathscr{U}_\alpha \cap \mathscr{U}_\beta \cap \mathscr{U}_\gamma$.*

Тогда для $\left\{c_{\alpha\beta}\dfrac{\partial}{\partial\theta_\alpha}\right\}$ и $\left\{p_{\alpha\beta}\theta_\alpha\dfrac{\partial}{\partial\theta_\alpha}\right\}$ имеем $\theta$-препятствия

$$\hat{\mathbf{\Delta}}_{\alpha\beta\gamma}(c) = c_{\alpha\beta}\frac{\partial}{\partial\theta_\alpha} + c_{\beta\gamma}\frac{\partial}{\partial\theta_\beta} - c_{\alpha\gamma}\frac{\partial}{\partial\theta_\alpha}, \tag{3.59}$$

$$\hat{\mathbf{\Delta}}_{\alpha\beta\gamma}(p) = p_{\alpha\beta}\theta_\alpha\frac{\partial}{\partial\theta_\alpha} + p_{\beta\gamma}\theta_\beta\frac{\partial}{\partial\theta_\beta} - p_{\alpha\gamma}\theta_\alpha\frac{\partial}{\partial\theta_\alpha}. \tag{3.60}$$

**Утверждение 3.14.** *$\theta$-препятствие $\hat{\mathbf{\Delta}}_{\alpha\beta\gamma}(c)$ равно нулю.*

*Доказательство.* Следует из **Утверждения 3.11** и (3.57). ■

Вычислим $\theta$-препятствие $\hat{\mathbf{\Delta}}_{\alpha\beta\gamma}(p)$. Для этого воспользуемся (3.56) и получим

$$\hat{\mathbf{\Delta}}_{\alpha\beta\gamma}(p) = \left[p_{\alpha\beta}(z_\beta)\cdot(g_{\beta\gamma}(z_\gamma)-1) + p_{\beta\gamma}(z_\gamma)\cdot(g_{\alpha\beta}(z_\beta)-1)\right]\cdot\theta_\beta\frac{\partial}{\partial\theta_\beta}. \tag{3.61}$$

Тогда в силу произвольности $p_{\alpha\beta}(z_\beta)$ справедливо

**Утверждение 3.15.** *$\theta$-препятствие $\hat{\mathbf{\Delta}}_{\alpha\beta\gamma}(p)$ обращается в нуль для преобразований, не меняющих нечетную координату, т. е. для которых выполняется $g_{\alpha\beta}(z_\beta) = 1$.*

Таким образом, введенные $\theta$-препятствия и $\theta$-коциклы являются дополнительными характеристиками полусупермногообразий, для которых функциями склейки служат редуцированные преобразования.